%% file: borowka_thesis.tex
\documentclass[twoside,openright,english,a4paper,fleqn,numbers=noenddot]{scrreprt}
\usepackage[latin1]{inputenc}
\usepackage[T1]{fontenc}
\usepackage{lmodern}
\usepackage{titlesec, blindtext,color}
\usepackage{babel}
\usepackage{feynmp}
\usepackage{rotating}
\usepackage{slashed}
\usepackage{amsmath}
\usepackage{amssymb}
\usepackage{graphicx, subfigure}
\usepackage{ulem}
\usepackage[amssymb]{SIunits}
\usepackage{fancyhdr}
\usepackage{booktabs} 
\usepackage{array} 
\usepackage{paralist} 
\usepackage{verbatim} 
\usepackage{feynmp}
\usepackage{rotating}
\usepackage{slashed}
\usepackage{colortbl}
\usepackage{cite}
\usepackage{abstract}
\input{definitions.tex}
\definecolor{gray75}{gray}{0.75}
\newcommand{\hsp}{\hspace{20pt}}
\titleformat{\chapter}[hang]{\Huge\bfseries}{\thechapter\hsp\textcolor{gray75}{|}\hsp}{0pt}{\Huge\bfseries}
\begin{document}
\DeclareGraphicsRule{*}{mps}{*}{} 
\setlength{\unitlength}{1mm} 
%

\addtolength{\oddsidemargin}{0.75cm}
\thispagestyle{empty}
\begin{center}

\Large{
Technische Universit\"at M\"unchen\\
\vspace*{\stretch{3}}
Max-Planck-Institut f\"ur Physik\\
(Werner-Heisenberg-Institut)\\
}

\vspace*{\stretch{5}}
\noindent
  \vspace*{\stretch{1}}
  {\bfseries\Huge \\Evaluation of multi-loop multi-scale integrals and phenomenological two-loop applications\\}

  \vspace*{\stretch{5}}

  {\bfseries\LARGE Sophia Carola Borowka\\}
  \vspace*{\stretch{1}}

\vspace*{\stretch{5}}

\large{
 Vollst\"andiger Abdruck der von der Fakult\"at f\"ur Physik\\
 der Technischen Universit\"at M\"unchen zur Erlangung des akademischen Grades eines\\
 \begin{bfseries} Doktors der Naturwissenschaften (Dr. rer. nat.) \end{bfseries}\\
 genehmigten Dissertation.

\vspace*{\stretch{5}}

\begin{tabular}{llll}
Vorsitzender:               &     & Univ.-Prof.~Dr.~L.~Oberauer \\
Pr\"ufer der Dissertation:  &  1. & Hon.-Prof.~Dr.~W.~F.~L.~Hollik\\
                           &  2. & Univ.-Prof.~Dr.~A.~Ibarra
\end{tabular}

\vspace*{\stretch{5}}

\makeatletter
 \def\vhrulefill#1{\leavevmode\leaders\hrule\@height#1\hfill \kern\z@}
\makeatother

Die Dissertation wurde am 04.06.2014 \\
bei der Technischen Universit\"at M\"unchen eingereicht und\\
durch die Fakult\"at f\"ur Physik am 03.07.2014 angenommen. 

}

\end{center}
\pagebreak
\addtolength{\oddsidemargin}{-0.75cm}
 \addtolength{\oddsidemargin}{0.6cm} 
 \addtolength{\evensidemargin}{-0.6cm} 

\mbox{}
\thispagestyle{empty}
\newpage

This thesis is based on the author's work conducted at the Max Planck Institute 
for Physics (Werner-Heisenberg-Institut) in Munich. 
Parts of this work have already been published in Refs.~\cite{Borowka:2012yc,Borowka:2013cma,
Borowka:2014wla,Borowka:2012ii,Borowka:2012rt,Borowka:2013lda,
Borowka:2013uea}.
~\\
~\\
\textbf{Articles}
\begin{itemize}
\item[\cite{Borowka:2012yc}]
Borowka, Sophia and Carter, Jonathon and Heinrich, Gudrun \\
``Numerical Evaluation of Multi-Loop Integrals for Arbitrary Kinematics \\with SecDec 2.0'', 
{\it Comput.Phys.Commun.} \textbf{184} (2013) 396-408, \\{\tt arXiv:1204.4152 [hep-ph]}
\item[\cite{Borowka:2013cma}] 
Borowka, Sophia and Heinrich, Gudrun \\
``Massive non-planar two-loop four-point integrals with SecDec 2.1'', \\
{\it Comput.Phys.Commun.} \textbf{184} (2013) 2552-2561, {\tt arXiv:1303.1157 [hep-ph]}
\item[\cite{Borowka:2014wla}] 
Borowka, S. and Hahn, T. and Heinemeyer, S. and Heinrich, G. and Hollik, W. \\
``Momentum-dependent two-loop QCD corrections to the neutral Higgs-boson \\masses in the MSSM'', {\it Eur.Phys.J} \textbf{C74} (2014) 2994, %
{\tt arXiv:1404.7074 [hep-ph]}
\end{itemize}
~\\
\textbf{Proceedings}
\begin{itemize}
\item[\cite{Borowka:2012ii}]
Borowka, Sophia and Carter, Jonathon and Heinrich, Gudrun \\
``SecDec: A tool for numerical multi-loop calculations'',\\
{\it J.Phys.Conf.Ser.} \textbf{368} (2012), {\tt arXiv:1206.4908 [hep-ph]}
\item[\cite{Borowka:2012rt}]
Borowka, Sophia and Heinrich, Gudrun \\
``Numerical evaluation of massive multi-loop integrals with SecDec'',\\
{\it PoS} \textbf{LL2012} (2012) 038, {\tt arXiv:1209.6345 [hep-ph]}
\item[\cite{Borowka:2013lda}]
Borowka, Sophia and Heinrich, Gudrun \\
``Numerical multi-loop calculations with SecDec'',\\
{\it C13-05-16}, {\tt arXiv:1309.3492 [hep-ph]}
\item[\cite{Borowka:2013uea}] 
Borowka, Sophia and Heinrich, Gudrun \\
``Two-loop applications of the program SecDec'',\\
{\it PoS} \textbf{RADCOR2013} (2014) 009, {\tt arXiv:1311.6476 [hep-ph]}
\end{itemize}

\newpage
\begin{abstract}
In this thesis, major new developments in the program \secdec{} are presented. 
SecDec is a publicly available program for the numerical evaluation of 
multi-loop multi-scale integrals and in this thesis it has been extended from 
Euclidean to physical kinematics.

\medskip 

The program \secdec{} is based on sector decomposition to 
extract dimensionally regulated singularities. To deal with integrable 
singularities due to mass thresholds, the integrand is analytically continued 
to the complex plane. 
Further improvements are shown, proving invaluable in the two applications 
within this thesis. 

In the first application, numerical results for several massive two-loop four-point functions 
are presented. In particular, results for two of the most complicated massive non-planar 
two-loop box integrals entering the heavy-quark pair production at next-to-next-to leading 
order in QCD are shown. A mixed analytical and numerical approach proves beneficial 
in the evaluation of the most complicated diagram. 
It is shown that the program can deal not only with scalar integrals, 
but also with tensor integrals of in principle arbitrary rank. 

In its second application within this thesis, the neutral MSSM Higgs-boson spectrum 
is discussed. In particular, the calculation of the leading momentum-dependent 
order ${\cal O} (\alpha_s \alpha_t)$ corrections using a mixed on-shell/$\overline{DR}$ 
renormalization scheme is presented. 
Integrals which are available in analytic form have been implemented in a way allowing 
for a stable numerical evaluation.
Analytically inaccessible integrals 
are evaluated numerically using the program \secdec{}.
The combination of the new momentum-dependent two-loop contribution with
the existing one- and two-loop corrections in the
on-shell/\DRbar\ scheme leads to an improved prediction of the light
MSSM Higgs-boson mass and a correspondingly reduced theoretical
uncertainty. 
The effect of the newly included momentum-dependent terms on the 
neutral $\cal{ CP}$-even Higgs-boson masses is discussed. 
The corresponding shifts in the lightest Higgs-boson mass $\Mh$
are below~$1 \gev$ in all scenarios considered, but can extend up
to the level of the current experimental accuracy.
The results are included in the code \fh, a publicly available program 
to calculate parameters related to the Higgs-boson sector in the 
framework of the MSSM. 

\end{abstract}

\tableofcontents

\cleardoublepage

\input{introduction/introduction.tex}
\input{higgsintro/higgsintro.tex}
\input{twoloopfrontier/twoloopfrontier.tex}
\input{sectordecompo/sectordecompo.tex}
\input{contourdef/contourdef.tex}
\input{theprogram/theprogram.tex}
\input{application1/application1.tex}
\input{application2/application2.tex}
\input{conclusion/conclusion.tex}

\appendix
\input{appendix/appendix.tex}
\bibliography{borowka_thesis.bib}
\listoffigures
\chapter*{Acknowledgement} 

First and foremost, I would like to express my deep gratitude to Dr. Gudrun Heinrich 
and Prof. Dr. Wolfgang Hollik, my research supervisors, for their patient guidance and 
enlightening discussions. 
In particular, I am grateful to Gudrun Heinrich for the continuous support, openness to 
discussions and detailed explanations, and Wolfgang Hollik for the 
opportunity to join the institute and for giving me the freedom and support 
to pursue the proposed project in my own way. 
%

\medskip

I gratefully acknowledge collaboration with Jonathon Carter, Thomas Hahn and Sven Heinemeyer. 

\medskip 

I would also like to extend my thanks to Thomas Hahn and Peter Breitenlohner for the computer 
administration and their constructive comments and advice on good programming practices.

\medskip

Furthermore, I wish to thank all current and former PhD students of the theory group, especially Andreas Deser, 
Stephan Steinfurt, Hansj{\"o}rg Zeller, Migael Strydom, Sebastian Halter, Federico Bonetti, 
Sebastian Passehr and Jonas Lindert 
for interesting discussions. I want to thank my office mates Thi Nhung Dao, Davide Pagani, Tiziano Peraro and 
Johannes Schlenk for interesting discussions and a productive environment. 

\medskip 

I am grateful to Scott Mandry and Raphaela Borowka for proofreading (parts of) this thesis.  

\medskip

Importantly, the easy access to literature, knowledgeable surrounding, the up-to-date equipment, 
very well working IT-support, travel opportunities and good overall management of 
the institute must be acknowledged, as it made working at the institute 
particularly fun. 
%
Moreover, I would like to thank all people working at the institute for contributing to such a 
friendly atmosphere, especially the secretaries of the theory department Rosita Jurgeleit and 
Monika Goldammer, Silke Zollinger from the PR department and Karim Kannouf at the reception desk. 

\medskip

I am furthermore very grateful to my whole family and friends for their enduring moral support. 
Particular thanks go to my mother, father and sister for their support, 
and to my grandfather for his invaluable advice.
%
%
%
%
%
%
%
%
\end{document}

%% file: definitions.tex
\newcommand{\bq}{\begin{align}}
\newcommand{\eq}{\end{align}}

\newcommand{\secdec}{{\textsc{SecDec}}}
\newcommand{\idel}{i\,\delta}
\newcommand{\eps}{\epsilon}
\newcommand{\rd}{{\mathrm{d}}}

\def\refeq#1{\mbox{Eq.~(\ref{#1})}}

\def\reffi#1{\mbox{Fig.~\ref{#1}}}

\def\citere#1{\mbox{Ref.~\cite{#1}}}

\newcommand{\mste}{m_{\tilde{t}_1}}
\newcommand{\mstz}{m_{\tilde{t}_2}}

\newcommand{\MstL}{M_{\tilde{t}_L}}
\newcommand{\MstR}{M_{\tilde{t}_R}}

\newcommand{\Xt}{X_t}

\newcommand{\msusy}{M_{\rm SUSY}}

\newcommand{\cHe}{\cH_1}
\newcommand{\cHz}{\cH_2}

\newcommand{\Pe}{\phi_1^0}
\newcommand{\Pz}{\phi_2^0}

\newcommand{\PePz}{\phi_1^0\phi_2^0}

\newcommand{\drbar}{$\overline{\rm{DR}}$}

\newcommand{\DRbar}{\ensuremath{\overline{\mathrm{DR}}}}

\def\order#1{\ensuremath{{\cal O}(#1)}}
\newcommand{\cp}{{\cal CP}}

\newcommand{\cH}{{\cal H}}

\newcommand{\cL}{{\cal L}}

\newcommand{\fh}{{\sc FeynHiggs}}

\newcommand{\MA}{m_{A^0}}
\newcommand{\mh}{m_h}

\newcommand{\Mh}{M_h}
\newcommand{\MH}{M_H}

\newcommand{\mhmax}{$\mh^{\rm max}$}

\newcommand{\mt}{m_t}

\newcommand{\mgl}{m_{\tilde{g}}}

\newcommand{\Stop}{\tilde{t}}

\newcommand{\tsf}{\theta\kern-.20em_{\tilde{f}}}
\newcommand{\tsfp}{\theta\kern-.20em_{\tilde{f}\prime}}
\newcommand{\tsq}{\theta\kern-.15em_{\tilde{q}}}

\newcommand{\KL}{\left(}
\newcommand{\KR}{\right)}

\newcommand{\VL}{\left( \begin{array}{c}}
\newcommand{\VR}{\end{array} \right)}
\newcommand{\ML}{\left( \begin{array}{cc}}
\newcommand{\MLd}{\left( \begin{array}{ccc}}
\newcommand{\MLv}{\left( \begin{array}{cccc}}
\newcommand{\MR}{\end{array} \right)}
\newcommand{\re}{\mathop{\mathrm{Re}}}

\newcommand{\tb}{\tan \beta}

\newcommand{\CTb}{\cot \beta}

\newcommand{\Sbe}{\sin \beta}

\newcommand{\SQb}{\sin^2\!\beta}

\newcommand{\Cb}{\cos \beta}

\newcommand{\CQb}{\cos^2\!\beta}

\newcommand{\CZb}{\cos 2\beta}

\newcommand{\tev}{\,\, \mathrm{TeV}}
\newcommand{\gev}{\,\, \mathrm{GeV}}
\newcommand{\mev}{\,\, \mathrm{MeV}}

\newcommand{\BC}{\begin{center}}
\newcommand{\EC}{\end{center}}
\newcommand{\BE}{\begin{equation}}
\newcommand{\BEA}{\begin{eqnarray}}
\newcommand{\EEA}{\end{eqnarray}}
\newcommand{\non}{\nonumber}
\newcommand{\id}{{\rm 1\kern-.12em
\rule{0.3pt}{1.5ex}\raisebox{0.0ex}{\rule{0.1em}{0.3pt}}}}
\newcommand{\lsim}
{\;\raisebox{-.3em}{$\stackrel{\displaystyle <}{\sim}$}\;}
\newcommand{\gsim}
{\;\raisebox{-.3em}{$\stackrel{\displaystyle >}{\sim}$}\;}

\newcommand{\tanb}{\tan \beta\,}

\newcommand{\dMAsqt}{\delta \MA^{2(2)}}

\newcommand{\dZ}[1]{\delta Z_{#1}}

\def\als{\alpha_s}
\def\alt{\alpha_t}

\def\de{\delta}

\def\eps{\varepsilon}

\def\De{\Delta}

\def\Si{\Sigma}

\def\hSi{\hat{\Sigma}}
\newcommand{\se}[1]{\Sigma_{#1}}

\newcommand{\ser}[1]{\hat{\Sigma}_{#1}}

 \newcommand{\deVez}{\de V_{\Pe\Pe}^{(2)}}
 \newcommand{\deVzz}{\de V_{\Pz\Pz}^{(2)}}
 \newcommand{\deVezz}{\de V_{\PePz}^{(2)}}

\definecolor{Lightblue}{cmyk}{0.9,0.1,0.1,0.3}
\definecolor{dgelborange}{cmyk}{0.,0.3,0.5, 0.}
\definecolor{Lila}{rgb}{0.5,0.,1}

%% file: introduction/introduction.tex
\chapter{Introduction}
\label{sec:introduction}%
%
%
%
Hadron colliders have set the stage to a whole new era of discovery. 
With the increasing wealth of high energy collision data, physics up to 
the TeV scale is being explored. 
Within the theoretical framework of the Standard Model of particle 
physics (SM)~\cite{Glashow:1961tr,Weinberg:1967tq,Salam:1968rm,
Fritzsch:1973pi,Gross:1973ju,Gross:1973id,Politzer:1973fx} most of the observations 
made by past and present collider experiments 
can successfully be described. 
Its predictive power has lead to the discovery of almost all of its constituents. 
These are three families of quarks and leptons, four 
gauge bosons mediating the electroweak and strong interaction, and
the simplest manifestation of the 
Brout-Englert-Higgs 
mechanism~\cite{Guralnik:1964eu,Englert:1964et,
Higgs:1964ia,Higgs:1966ev} - the Higgs-boson. 
Although the discovery of the latter is still not 
fully confirmed, 
a particle behaving like the Standard Model Higgs-boson has recently 
been observed~\cite{Aad:2012tfa,Chatrchyan:2012ufa} 
in the ATLAS and CMS experiments
at the Large Hadron Collider (LHC). 
The characteristics of this new 
boson with a mass around $125$ GeV
have been determined already rather 
accurately~\cite{Aad:2013xqa,
ATLAS-CONF-2014-009,Chatrchyan:2013mxa,
Chatrchyan:2013iaa}. 
If deviations with respect to the SM characteristics 
are found with 
the collation of more data, 
this particle must be interpreted 
within a different model. 
%
There are already several other reasons to 
search for an embedding of the Standard Model 
as an effective theory into a more general 
theoretical framework. 
Apart from the fact that gravity is not incorporated, 
the indirect observation of dark matter~\cite{Begeman:1991iy,
Clowe:2006eq,Ade:2013zuv} does not find a description in 
the Standard Model either. 
Furthermore, the predicted violation of the ${\cal CP}$ 
symmetry is not large enough as to explain the observed 
excess of matter over antimatter in the universe. 
More peculiarities are related to the newly found boson. 
If it indeed is the SM Higgs-boson, it is discussed~\cite{Cabibbo:1979ay,
Hung:1979dn,Lindner:1985uk,Degrassi:2012ry} that the 
electroweak vacuum of the Standard Model may not be 
absolutely stable and its low 
mass can only be accommodated for by assuming an 
unnatural amount of fine-tuning~\cite{Weinberg:1979bn}. 
Ideas for models beyond the Standard Model in 
which the newly found boson is realized range from interpreting 
it as a dilaton~\cite{Gildener:1976ih,Cao:2013cfa} or in the 
framework of a composite Higgs model~\cite{Kaplan:1983fs,Giudice:2007fh}. 
A different proposal for a new framework is 
formulated as a supersymmetric extension to the Standard 
Model, in particular the Minimal Supersymmetric Standard 
Model (MSSM)~\cite{Nilles:1983ge,Haber:1984rc,Barbieri:1987xf}. 
It has been broadly discussed over the last 
few decades. 

The motivation for supersymmetry (SUSY)~\cite{Ramond:1971gb,
Neveu:1971rx,Gervais:1971ji,Golfand:1971iw,
Volkov:1973ix,Wess:1973kz,Wess:1974tw} is twofold. On the one 
hand, %
%
it can provide for a solution to the fine-tuning and the 
hierarchy problem, achieve a gauge coupling unification 
and moreover accommodate for a dark matter candidate. 
On the other hand, it allows for the embedding of 
present observations into a more generalized 
mathematical framework.
Supersymmetry 
arises as the only possible extension to the Poincar\'{e} 
algebra~\cite{Haag:1974qh}, evading the 
no-go-theorem found by Coleman and 
Mandula~\cite{Coleman:1967ad}.
In supersymmetric theories, all known fermionic particles of 
the Standard Model are assigned a scalar superpartner and 
all bosonic SM particles a fermionic one. 
The Standard Model contains one scalar doublet. 
In renormalizable supersymmetric models, the necessity 
for an even number 
of scalar doublets arises. 
At least two Higgs
doublets are required, to give mass to respectively both, up-type 
and down-type particles and scalar particles 
(sparticles)~\cite{Fayet:1974pd,Witten:1981nf,
Dimopoulos:1981zb,Sakai:1981gr,Inoue:1982ej,
Inoue:1982pi,Inoue:1983pp}. 
The MSSM 
contains two scalar doublets which 
conserve hypercharge gauge invariance.
Due to this invariance, all 
up-type particles and sparticles
couple exclusively to one 
scalar doublet, while all down-type
(s)particles couple to the other doublet. 
This evades constraints from 
flavor-changing neutral currents (FCNCs), as was 
pointed out by Glashow and 
Weinberg~\cite{Glashow:1976nt}.
The two scalar doublets of the MSSM give rise to 
five physical Higgs-bosons. In
lowest order, these are the light and heavy $\cp$-even, the $\cp$-odd, 
and the charged Higgs-bosons.
While the mass of the 
Higgs-boson remains a free input parameter in the Standard Model, it is 
predicted within the MSSM. 
%
%
%
%
%
Associating the newly observed boson with the 
lightest $\cp$-even Higgs-boson $h^0$, 
the upper bound on its predicted mass $m_{h^0}$ at leading order (LO) 
is given by the $Z$ gauge boson mass. This would already have lead 
to the exclusion of the MSSM at past collider experiments. 
%
%
Yet, higher-order quantum corrections to the MSSM Higgs-boson masses lead to 
a shift in the upper limit towards 
$m_{h^0} \lesssim 135$ GeV.

Higher-order corrections are not only decisive in the precise prediction 
of physics beyond the Standard Model, but are of proven importance 
in the understanding of SM processes at colliders. 
The state of the art of higher-order corrections to Standard Model 
processes and a future wish list is summarized in the 
proceedings of the 2013 Les Houches workshop~\cite{Butterworth:2014efa}. 
%
The more accurate predictions are desired, the more involved the 
calculations become. 
%
%
%
%
%
%
%
%
Leading order theoretical predictions can most commonly not 
meet the current experimental precision. 
The calculation of perturbative corrections at next-to-leading order (NLO) 
in the strong or electroweak coupling constant has 
reached an impressive level of
automation meanwhile. 
Corrections beyond NLO accuracy still require 
quite some effort, both on the 
conceptual and on the technical side before they can be performed in 
a largely automated way.
There are a few processes measured at the LHC where 
the need for next-to-next-to leading order (NNLO) QCD predictions 
arises. 
One of them is top-quark pair production. 
Top-quark pair production is vital for the precise measurements of 
the top-quark properties but also enters into other measurements, e.g., 
of parton distributions. 
At the LHC top quarks are produced so numerously that 
they also constitute a significant background to new physics signals. 
It is therefore crucial to understand this background properly to be 
able to discriminate the signal. 
A full NNLO prediction for the total cross section of top-quark pair production 
is known in a semi-numerical form~\cite{Czakon:2013goa} along with 
many partial results in semi-numerical and analytical form~\cite{Korner:2008bn,Bonciani:2008az,
Bonciani:2009nb,Bonciani:2010mn,Aliev:2010zk,Abelof:2011ap,Czakon:2011ve,
Czakon:2012pz,Czakon:2012zr,Baernreuther:2012ws,vonManteuffel:2012je,
Abelof:2012he,Abelof:2012rv}. 
Soft gluon and Coulomb effects also have been taken into account 
beyond the next-to-leading logarithmic accuracy
and have been combined with fixed order results to come up 
with predictions as precise as possible~\cite{Czakon:2013vfa,Moch:2012mk,
Cacciari:2011hy,Czakon:2011xx,Beneke:2011mq,Beneke:2012wb,
Ahrens:2011px,Kidonakis:2010dk}. 
Among the key ingredients of the full NNLO calculation are 
complicated two-loop integrals entering the virtual corrections. 
Analytical expressions for these 
are known for diagrams dependent on relatively 
few mass scales~\cite{Bonciani:2010mn,
Bonciani:2009nb,Bonciani:2008az,Bonciani:2013ywa,
vonManteuffel:2012je,vonManteuffel:2013uoa}. 
As soon as several mass scales are involved, numerical 
methods to calculate multi-loop integrals 
become increasingly important. 

\medskip

The brief outline of this thesis is as follows: 
In Chapters~\ref{chap:higgsintro}-\ref{chap:contourdef}, 
the basic concepts of the author's work presented in this thesis are established. 
In Chapter~\ref{chap:theprogram}, the developed version 2 of the program \secdec{} 
is described, laying the foundation for 
two applications presented in Chapters~\ref{chap:application1} 
and~\ref{chap:application2}. 

More comprehensively, Chapter~\ref{chap:higgsintro} covers an introduction 
of the tree-level Higgs-boson sector of the MSSM. The scalar quark (squark) 
sector is discussed as well, focussing on strong and Yukawa-type 
interactions. Afterwards, a motivation for higher-order corrections 
to the Higgs-boson masses is discussed, along with an introduction to their 
computation. 
The latter involves the evaluation of two-loop integrals with 
multiple scales, leading to mass thresholds. 
Different methods to approach multi-loop multi-scale 
integrals are reviewed in Chapter~\ref{chap:twoloopfrontier}, before 
motivating the pursuit of a universal numerical approach using 
Feynman parameterization. 
The method of sector decomposition is used for the 
disentanglement of overlapping ultraviolet (UV), collinear 
and infrared (IR) singularities, as discussed in 
Chapter~\ref{chap:sectordecompo}.
Various algorithms performing differently with respect to this task 
are also reviewed. 

In Chapter~\ref{chap:contourdef}, the appearance of thresholds 
is discussed. To compute integrable thresholds, the integrand 
needs to be analytically continued to the complex plane. 
Towards this aim, a deformation of the integration contour, 
applicable in numerical calculations, is explained.  
Finally, studies by the author 
are presented which tune the analytical continuation 
further towards a stable evaluation of integrals containing 
thresholds. 
%
%
%

In Chapter~\ref{chap:theprogram}, 
the features incorporated in an upgrade 
of the open-source program \secdec{} are presented. 
Based on the concepts introduced in 
Chapters~\ref{chap:twoloopfrontier}-\ref{chap:contourdef}, 
\secdec{} allows the automated numerical 
computation of multi-loop multi-scale integrals, in addition to 
an evaluation of more general parametric integrals. 
Restricted to non-physical kinematics in version 1, the extension 
to physical kinematics including thresholds is 
achieved in version 2 of the program. 
The upgraded features are presented along with 
diverse other improvements. 

In Chapter~\ref{chap:application1}, the full power 
of the program \secdec{} is shown in an application 
to massive 
non-planar two-loop four-point functions, among them various 
ones where analytical results are unknown. 
Several of the topologies shown are 
computed in a fully automated way. 
For one topology which is of particular 
interest in the top-quark pair production at 
NNLO,
analytical transformations beforehand are shown, 
improving on the numerical stability. 
In particular, the integration of one Feynman parameter 
of a sub-loop is found to be beneficial.
Furthermore, a transformation first introduced by 
the author and collaborators, proves to 
allow for a simplification of the singularity 
structure, leading to a reduction in the number of 
sub-functions to be integrated. The transformation is 
presented in detail. 
%
%
%

In Chapter~\ref{chap:application2}, the calculation of the 
dominant neutral ${\cal CP}$-even MSSM Higgs-boson mass 
corrections at the two-loop order including momentum dependence 
is presented. 
This requires the calculation of two-loop self-energies with 
a proper renormalization at the two-loop level, 
using an overall mixed on-shell and $\overline{DR}$ scheme for 
the renormalization. 
%
The program \secdec{} is used in the evaluation of analytically 
unaccessible integrals.  
The mass shifts resulting from the additional momentum-dependent 
contributions are presented. 

The conclusions are given in Chapter~\ref{chap:conclusion}.

%% file: higgsintro/higgsintro.tex
\chapter{Higgs-bosons in the MSSM}
\label{chap:higgsintro}%
In the following, the Higgs-boson, quark and 
scalar quark (squark) sector of the MSSM are introduced. 
The tree-level mass matrices are derived from the MSSM Lagrangian.  
All interactions appearing in the 
two-loop corrections to the MSSM Higgs-boson masses discussed 
in Chap.~\ref{chap:application2} are shown as well. 
This includes supersymmetric QCD (SCQD) interactions. 
The following two sections, Sec.~\ref{sec:higgstreelevel} and~\ref{sec:scalarquarktreelevel}, 
are based on Refs.~\cite{Gunion:1984yn,
Gunion:1986nh,Fayet:1976cr,Gunion:1989we,Haber:1984rc,Rzehak:2005zz}. 
Afterwards, the current status of higher-order 
Higgs-boson mass corrections in the MSSM is 
reviewed in Sec.~\ref{sec:highorderreview}.
%
%
\section{The Higgs-boson sector of the MSSM at tree-level}
\label{sec:higgstreelevel}
%
The Higgs-boson sector of the MSSM with real parameters (rMSSM) is part of the full 
MSSM Lagrangian and consists of the following four components
\begin{align}
 \mathcal{L}_{\rm{MSSM}} \supset \mathcal{L}_{\rm{H_{free}}} + 
 \mathcal{L}_{\rm{H_V}} + \mathcal{L}_{\rm{H_{int}}} + 
 \mathcal{L}_{\rm{H_{fix}}} +  \mathcal{L}_{\rm{H_{ghost}}} \text{ ,}
 \label{eq:mssmlag}
 \end{align}
where 
$\mathcal{L}_{\rm{H_{free}}}$ contains the free-field kinetic terms, 
$\mathcal{L}_{\rm{H_V}}$ is derived from the Higgs-boson potential, and 
$ \mathcal{L}_{\rm{H_{fix}}} $ is the gauge-fixing term. 
With the introduction of a gauge-fixing term, unphysical 
degrees of freedom arise which are compensated by 
Faddeev-Popov ghost terms~\cite{Faddeev:1967fc} in 
$ \mathcal{L}_{\rm{H_{ghost}}} $. 
The interaction part of the Higgs-boson sector Lagrangian
can be summarized as 
\begin{align}
\non \mathcal{L}_{\rm{H_{int}}} =&\hspace{4pt} \mathcal{L}_{\rm{H}\rm{H}\rm{H}} + \mathcal{L}_{\rm{H}\rm{H}\rm{H}\rm{H}}\\
\non &+ \mathcal{L}_{\rm{H}\rm{H} \rm{V}} + \mathcal{L}_{\rm{HVV}} + \mathcal{L}_{\rm{HHVV}} \\
\non &+ \mathcal{L}_{\rm{H}\psi\bar{\psi}} + \mathcal{L}_{\rm{H}\tilde{s}\tilde{s}} + \mathcal{L}_{\rm{H}\rm{H}\tilde{s}\tilde{s}}\\
&+ \mathcal{L}_{\rm{H}\tilde{\chi}\tilde{\chi}}\text{ .} 
 \label{eq:MSSMHint}
\end{align}
All physical neutral and charged Higgs-boson 
fields are referred to with the index H, the index 
V is short for vector boson fields, $\psi$ and $\bar{\psi}$ 
denote the Standard Model
quarks and leptons, $\tilde{s}$ denotes 
all squarks and scalar leptons (sleptons) and $\tilde{\chi}$
symbolizes the neutralinos and charginos. 
Adopting the Feynman-'t Hooft gauge, all 
ghost contributions vanish. 

\medskip 

The MSSM requires two doublets  $\mathcal{H}_1$ and 
$\mathcal{H}_2$ of complex scalar fields, 
which are conventionally written in terms of 
their components as follows, 
\begin{align}
\mathcal{H}_1 &= \left( \begin{matrix} \mathcal{H}_1^0 \\ \mathcal{H}_1^- \end{matrix} \right) = 
\left( \begin{matrix} v_1 + \frac{1}{\sqrt{2}}(\phi_1^0 - \textrm{i} \chi_1^0) \\ -\phi_1^- \end{matrix} \right) \text{, } \\
\mathcal{H}_2 &= \left( \begin{matrix} \mathcal{H}_2^+ \\ \mathcal{H}_2^0 \end{matrix} \right) = 
\left( \begin{matrix}  \phi_2^+ \\ v_2 + \frac{1}{\sqrt{2}}(\phi_2^0 + \textrm{i} \chi_2^0) \end{matrix} \right) \text{, }
\end{align}
with an associated hypercharge $Y_1=-1$ 
and $Y_2=+1$, respectively. Their vacuum expectation 
values are given by $v_1$ and $v_2$, respectively. The fields 
$\phi_i$ and $\chi_i$ are still unphysical, but are
brought into the physical basis
\begin{align}
\hspace{-20pt}\VL H^0 \\[0.5ex] h^0 \VR = A(\alpha) \VL \phi_1^0 \\[0.5ex] \phi_2^0 \VR \!\text{ ,\;\;} 
 \VL G^0 \\[0.5ex] A^0 \VR = A(\beta) \VL \chi_1^0 \\[0.5ex] \chi_2^0 \VR \!\text{ ,\;\;} 
 \VL\! G^{\pm} \!\!\\[0.5ex]\! H^{\pm} \!\!\VR  = A(\beta) \VL \phi_1^{\pm} \\[0.5ex] \phi_2^{\pm} \VR 
 \label{eq:physbasis} 
\end{align}
via orthogonal transformations of the type 
\begin{align}
A(x) = \ML \cos(x) & \sin(x) \\[0.5ex] -\sin(x) & \cos(x) \MR \text{, }
\label{eq:orthotrafo}
\end{align}
giving rise to the particle spectrum of physical Higgs- and unphysical 
Goldstone-bosons, compare Tab.~\ref{tab:higgspartspectr}.
\begin{table}[ht]
\begin{center}
    \begin{tabular}{ !{\color{gray75}\vrule} l !{\color{gray75}\vrule} l !{\color{gray75}\vrule} }
    \arrayrulecolor{gray75} \hline
2 neutral ${\cal CP}$-even Higgs-bosons & $h^0$, $H^0$ \\ 
1 neutral ${\cal CP}$-odd Higgs-boson & $A^0$ \\ 
1 neutral ${\cal CP}$-odd Goldstone boson & $G^0$ \\ 
2 charged Higgs-bosons & $H^+$, $H^-$ \\ 
2 charged Goldstone bosons & $G^+$, $G^-$ \\ \arrayrulecolor{gray75} \hline 
\end{tabular}
\end{center}
\label{tab:higgspartspectr}
\caption{Higgs- and Goldstone boson particle spectrum.} 
\end{table}
\\The kinetic part of the MSSM Higgs-boson sector 
Lagrangian reads 
\begin{align}
  \mathcal{L}_{\phi_{\rm{free}}} =& \sum_{a=1}^2 \partial_\mu \mathcal{H}_a^\dagger \partial^\mu \mathcal{H}_a \text{ .} \label{eq:MSSMHfree}
\end{align}
Note the index $\phi$ instead of $H$ in Eq.~(\ref{eq:MSSMHfree}). It is introduced to distinguish 
between the $\phi_1^0$-$\phi_2^0$ and the $h^0$-$H^0$ basis. 

\medskip

The potential part of the rMSSM 
Higgs-boson sector Lagrangian $\mathcal{V}_{\phi}$ 
can be written in terms of the supersymmetric
F- and D-term contributions 
\begin{align}
 \mathcal{V}_{\phi_{\rm{F}}} = |\mu|^2 (\mathcal{H}_1^{i\dagger}\mathcal{H}_1^{i}+\mathcal{H}_2^{i\dagger}\mathcal{H}_2^{i}) \text{, }
 \label{eq:VHF}
\end{align}
and 
\begin{align}
 \mathcal{V}_{\phi_{\rm{D}}} = \frac18 (g^2 + g'^2) (\mathcal{H}_1^{i\dagger}\mathcal{H}_1^{i}-\mathcal{H}_2^{i\dagger}\mathcal{H}_2^{i})^2 + \frac12 g^2 |H_1^{i\dagger}\mathcal{H}_2^{i}|^2 \text{ ,}
  \label{eq:VHD}
\end{align}
where $ i=1,2$ from now on. In contrast to the Standard Model, the Higgs-boson self-couplings in the 
MSSM, resulting from Eq.~(\ref{eq:VHD}), are determined through the gauge coupling constants.  
The dagger in Eqs.~(\ref{eq:VHF}) and~(\ref{eq:VHD}) indicates Hermitian adjoints, 
$\mu$ is the higgsino (fermionic superpartner of the Higgs-boson) mass parameter, $g$ 
is the SU(2)$_L$ and $g'$ the U(1)$_Y$ coupling 
constant. 
The coupling constants $g$ and $g'$ are related
to the electric charge $e$ and the electro-weak mixing
angle $\theta_W$ of the Standard Model by
\begin{align}
\hspace{60pt}g = \frac{e}{\sin\theta_W} \text{ ,} \hspace{60pt} 
g' = \frac{e}{\cos\theta_W} 
\label{eq:coupldefs}\text{ .}
\end{align}
Due to the non-observation of supersymmetric partners to 
the Standard Model particles, supersymmetry must 
be broken. 
Various supersymmetry breaking mechanisms can be 
considered~\cite{Fayet:1974pd,Witten:1981nf,Dimopoulos:1981zb,
Girardello:1981wz,Sakai:1981gr,Inoue:1982ej,Inoue:1982pi,
Inoue:1983pp}. In the MSSM, explicit breaking 
terms~\cite{Dimopoulos:1981zb,Girardello:1981wz} parameterize 
the effect of SUSY breaking. In order to accommodate 
for a solution to the hierarchy problem, these terms may not 
introduce additional quadratic divergences. They must 
have mass dimension less than four. 
These so called soft supersymmetry breaking terms 
are added to the MSSM Higgs-boson potential 
\begin{align}
\mathcal{V}_{\phi_{\rm{soft}}} = m_1^2 (\mathcal{H}_1^{i\dagger}\mathcal{H}_1^{i}) + m_2^2 (\mathcal{H}_2^{i\dagger}\mathcal{H}_2^{i}) - 
m_{12}^2 (\eps_{ij} \mathcal{H}_1^{i}\mathcal{H}_2^{j} + h.c.)\text{, }
 \label{eq:VHsoft}
\end{align}
where $ j=1,2$ from now on, and with $\eps_{ij}$\footnote{Note the convention: 
$\eps_{ij}=-\eps^{ij}$, with $\eps_{12}=-1$ and $\eps_{21}=1$.} being totally antisymmetric,   
resulting in an overall MSSM Higgs-boson potential of
\begin{align}
 \mathcal{V}_{\phi} &= \mathcal{V}_{\phi_{\rm{F}}} + \mathcal{V}_{\phi_{\rm{D}}}+ \mathcal{V}_{\phi_{\rm{soft}}}\label{eq:fullHiggspotential} \text{ .}
\end{align}
Relating the MSSM Higgs-boson potential to the MSSM Lagrangian by
\begin{align}
 \mathcal{L}_{\phi_{\rm{V}}} &= - \left( \int \text{d}^2 \theta\, \mathcal{V}_{\phi_{\rm{F}}} + 
 \iint \text{d}^2 \theta \text{d}^2 \bar{\theta}\, \mathcal{V}_{\phi_{\rm{D}}} + 
 \mathcal{V}_{\phi_{\rm{soft}}} \right) \label{eq:PotMSSMtoLag} \text{,}
\end{align}
where the F-term part of the potential is integrated over the auxiliary superspace 
component $\theta$, while the D-term part of the potential is integrated
over both superspace components $\theta$ and $\bar{\theta}$. In the
following, the notation
\begin{align}
 \int \text{d}^2 \theta \; \mathcal{V} = \left[ \mathcal{V} \right]_F \hspace{20pt}\text{ and }\hspace{20pt}
\iint \text{d}^2 \theta \text{d}^2 \bar{\theta} \;\mathcal{V} = \left[ \mathcal{V} \right]_D \text{ }
\end{align}
is adopted. 

\medskip

Until now, the MSSM Higgs-boson potential, Eq.~(\ref{eq:fullHiggspotential}), contains 
four free parameters: $m_1$, $m_2$, $m_{12}$ and $\mu$.
Exploiting the fact that the two vacuum expectation values $v_1$ and $v_2$
need to minimize the potential and be nonzero at the same time, 
the following necessary minimization conditions
\begin{align}
\frac{\partial}{\partial \mathcal{H}_a^i} \,  \mathcal{V}_{\phi} \Big|^{\text{lin}}= 0\text{ , } \hspace{20pt} a=1,2
\label{eq:minlinpartcond}
\end{align}
are required to hold. The 
minimization conditions originate 
from the equations of motion, 
compare Eq.~(\ref{eq:eoms}).
In Eq.~(\ref{eq:minlinpartcond}), the linear part
of the rMSSM potential reads
\begin{align}
\non \mathcal{V}_{\phi} \Big|^{\text{lin}} =& \sqrt{2} \tilde{m}_1^2 v_1 \phi_1^0 + \sqrt{2} \tilde{m}_2^2 v_2  \phi_2^0 + \sqrt{2} m_{12}^2 (v_2 \phi_1^0+ v_1\phi_2^0 )\\
&+\frac{1}{2 \sqrt{2}} (g^2+g'^2)(v_1^2-v_2^2)(v_1 \phi_1^0 -v_2 \phi_2^0 )\text{ ,}
\label{eq:linVpart}
\end{align}
where $\tilde{m}_i^2\equiv m_i + |\mu|^2 $. It should be 
noted that there are no contributions from the fields $\chi_a^0$
and $\phi_a^{\pm}$ to the linear part of the potential. This
is due to the fact that the MSSM potential is 
$\mathcal{CP}$-conserving, meaning that it is invariant under 
the consecutive application of a charge conjugation ${\cal C}$ 
and a parity transformation ${\cal P}$. 
Without imposing the 
minimization condition just yet, and
writing the coefficients to the fields 
$\phi_1^0, \phi_2^0$ in Eq.~(\ref{eq:linVpart})
as tadpole parameters $T_a$ instead, the parameters 
$\tilde{m}_1$ and $\tilde{m}_2$ can 
be expressed in terms of experimentally accessible 
quantities ($g$, $g'$, $v_1$, $v_2$) 
and the parameter $m_{12}$
\begin{align}
 \tilde{m}_1^2 &= \frac{1}{\sqrt{2} v_1} T_1 - \frac{v_2}{v_1} m_{12}^2 - \frac{1}{4} (g^2+g'^2)(v_1^2-v_2^2) \text{ ,} \label{eq:mtilde1}\\
 \tilde{m}_2^2 &= \frac{1}{\sqrt{2} v_2} T_2 - \frac{v_1}{v_2} m_{12}^2 + \frac{1}{4} (g^2+g'^2)(v_1^2-v_2^2)  \text{ .} \label{eq:mtilde2}
\end{align}
Consequently, after making use of Eq.~(\ref{eq:minlinpartcond}),  
the tadpole parameters $T_a$ vanish
\begin{align}
T_a=0 \text{ .}
\label{eq:tadszero}
\end{align}
Turning to the part of the MSSM Higgs-boson potential which is 
bilinear in the fields, the mass matrices $M^2_{\phi^0}$, 
$M^2_{\chi^0}$ and $M^2_{\phi^\pm}$ of the scalar
fields of the neutral $\mathcal{CP}$-even, the
neutral $\mathcal{CP}$-odd and, respectively, the charged
Higgs- and Goldstone-bosons can be identified 
\begin{align}
\hspace{-20pt} \mathcal{V}_{\phi} \supset \frac12 \left( \begin{matrix} \phi_1^0 & \phi_2^0 \end{matrix} \right) M^2_{\phi^0}
\left( \begin{matrix} \phi_1^0 \\ \phi_2^0 \end{matrix} \right) +
 \frac12 \left( \begin{matrix} \chi_1^0 & \chi_2^0 \end{matrix} \right) M^2_{\chi^0}
 \left( \begin{matrix} \chi_1^0 \\ \chi_2^0 \end{matrix} \right) +
 \left( \begin{matrix} \phi_1^{+} & \phi_2^{+} \end{matrix} \right) M^2_{\phi^\pm}
  \left( \begin{matrix} \phi_1^{-} \\ \phi_2^{-} \end{matrix} \right) \text{,}
 \label{eq:MSSMmassmatrices}
\end{align}
where $\phi_2^{-}=(\phi_2^{+})^*$ and $\phi_1^{+}=(\phi_1^{-})^*$. 
The tree-level mass matrix of 
the neutral $\mathcal{CP}$-odd bosons reads 
\begin{subequations}
\begin{align}
M^2_{\chi^0} &= 
\left( 
\begin{matrix} 
\tilde{m}_1^2 + \frac14 (g^2 + g'^2)(v_1^2-v_2^2) & m_{12}^2  \\ 
 m_{12}^2 & \tilde{m}_2^2 + \frac14 (g^2 + g'^2)( v_2^2-v_1^2)  
\end{matrix} 
\right) \\
 &= m_{12}^2
 \left( 
\begin{matrix} 
-\frac{v2}{v1} &  1  \\ 
 1 & - \frac{v1}{v2}  
\end{matrix} 
\right) \text{,}
\end{align}
\end{subequations}%
where in the last step, the relations of Eqs.~(\ref{eq:mtilde1}-\ref{eq:tadszero}) were used.
Afterwards, the mass matrix can be made diagonal 
\begin{subequations}
\begin{align}
\text{diag}(m_{G^0}^2,m_{A^0}^2) &= A(\beta)^T M^2_{\chi^0} A(\beta) \\
 &= \left( 
\begin{matrix} 
0 & 0  \\ 
0 & -m_{12}^2 (\tan \beta + \cot \beta)  
\end{matrix} 
\right) \text{,}
\end{align}
\end{subequations}%
resulting with the tree-level relation for 
the physical neutral $\mathcal{CP}$-odd Higgs-boson
mass $m_{A^0}^2$. Note that the
tree-level $A^0$-boson mass $m_{A^0}^2$ does not attain any
dependence on the Standard Model vector-boson masses $m_W$ or $m_Z$.
Defining the vacuum expectation values $v_1$ and $v_2$ as
\begin{align}
v_1 \equiv \frac{\sqrt{2} m_W \cos\beta}{g} \text{ ,} \hspace{20pt} v_2 \equiv \frac{\sqrt{2} m_W \sin\beta}{g} \text{ ,}
\label{eq:vevsdef}
\end{align}
the following relation results 
\begin{align}
\textrm{tan}\beta = \frac{v_2}{v_1} \text{ ,} \hspace{20pt} 0 \leq \beta \leq \frac{\pi}{2} \text{ .}
\label{eq:tanb}
\end{align}
The lower and upper bound on the angle $\beta$ result from the 
assumption that $v_1$ and $v_2$ are real and 
positive, compare Ref.~\cite{Gunion:1984yn}.\\
Hereby, all free mass parameters are 
expressed in terms of physical observables and the
MSSM Higgs-boson potential is fixed by the parameters 
$v_1$, $v_2$, $m_A$, $g'$ and $g$.
With this knowledge in 
mind, the masses of the neutral $\mathcal{CP}$-even
Higgs-bosons can be derived from the mass matrix
\begin{subequations}%
\begin{align}
M^2_{\phi^0} &= \ML M^2_{\phi_1^0\phi_1^0} & M^2_{\phi_1^0\phi_2^0} \\ M^2_{\phi_2^0\phi_1^0} & M^2_{\phi_2^0\phi_2^0} \MR \\
&= \left( 
\begin{matrix} 
m_{A^0}^2 \textrm{sin}^2 \beta + m_Z^2 \textrm{cos}^2 \beta  & -(m_{A^0}^2 + m_Z^2) \textrm{sin} \beta  \textrm{cos} \beta \\ 
-(m_{A^0}^2 + m_Z^2) \textrm{sin} \beta  \textrm{cos} \beta & m_{A^0}^2 \textrm{cos}^2 \beta + m_Z^2 \textrm{sin}^2 \beta  
\end{matrix} 
\right) \text {,}
\label{eq:mphi0tree}
\end{align}
\end{subequations}%
being written in terms of the three parameters $m_{A^0}$, $m_Z$ and the angle $\beta$.
After bringing the mass matrix into diagonal form, the physical
$\mathcal{CP}$-even Higgs-boson masses read
\begin{align}
 m_{H^0,h^0}^2=\frac12 \left( m_{A^0}^2 + m_Z^2 \pm \sqrt{(m_{A^0}^2 + m_Z^2)^2-4 m_{A^0}^2 m_Z^2\cos^2(2\beta)} \right) \text{.}
 \label{eq:mHmhtree}
\end{align}
The mass of the light CP-even Higgs-boson 
is therefore bound from above through 
the relation $m_{h^0} \leq \text{min}(m_Z,m_{A^0}) \,|\text{cos}(2 \beta)|$.\\
For completeness, the masses of the 
charged Higgs-bosons can be derived 
via the mass matrix of the 
charged-boson components 
\begin{align}
\hspace{-20pt} M^2_{\phi^\pm} &= 
\left( 
\begin{matrix} 
\tilde{m}_1^2 + \frac14 (g^2 + g'^2) (v_1^2-v_2^2) + \frac12 g^2 v_2^2 &   m_{12}^2 - \frac12 g^2 v_1 v_2  \\ 
 m_{12}^2 - \frac12 g^2 v_1 v_2 & \tilde{m}_2^2 + \frac14 (g^2 + g'^2) (v_2^2-v_1^2) + \frac12 g^2 v_1^2 
\end{matrix} 
\right) 
\end{align}
as
\begin{subequations}%
\begin{align}
\text{diag}(m_{G^\pm}^2,m_{H^\pm}^2) &= A(\beta)^T M^2_{\phi^\pm} A(\beta) \\
 &= \left( 
\begin{matrix} 
0 & 0  \\ 
0 & -m_{12}^2 (\tan \beta + \cot \beta) + m_W^2  
\end{matrix} 
\right) \text{ ,}
\end{align}
\end{subequations}%
where relations Eqs.~(\ref{eq:mtilde1}-\ref{eq:mtilde2}) are again useful.
Similar to the SM relations, the gauge-boson masses are 
given by 
\begin{align}
 m_W^2 = \frac12 g^2 (v_1^2+v_2^2) \text{ ,} \hspace{10pt} m_Z^2 = \frac12 (g^2 + g'^2) (v_1^2+v_2^2) 
\text{ ,} \hspace{10pt} m_\gamma^2 = 0 \text{ .}
\end{align}
The rich phenomenological 
implications of the real MSSM can be explored 
further, when studying the dependence on the 
angles $\alpha$ and $\beta$. The angle $\beta$ is linked
to the vacuum expectation values through Eq.~(\ref{eq:tanb}).
In turn, the angle $\alpha$ can be determined from 
the rotation of Eq.~(\ref{eq:mphi0tree}) into the physical basis. 
The following basic relation among the two angles holds~\cite{Gunion:1986nh}
\begin{align}
 \textrm{tan} (2 \alpha) =  \textrm{tan} (2 \beta) \frac{m_{A^0}^2 + m_Z^2}{m_{A^0}^2-m_Z^2} \text{ ,} 
 \hspace{20pt} -\frac{\pi}{2} < \alpha < 0 \text{ .}
 \label{eq:tan2alphatan2beta}
\end{align}
Many more relations among the angles can be found, compare 
Ref.~\cite{Gunion:1984yn,Gunion:1986nh}. 
When expressing the couplings in terms of these, 
they can be formulated as angle 
suppression factors with respect to 
Standard Model Higgs-boson couplings to, e.g., 
vector bosons
\begin{align}
 \frac{g_{h^0 V V}}{g_{h V V}} \propto \sin(\beta-\alpha) \text{ ,}\\
  \frac{g_{H^0 V V}}{g_{h V V}} \propto \cos(\beta-\alpha) \text{ ,}
\end{align}
where $h$ denotes the Standard Model 
Higgs-boson and $h^0, H^0$ the MSSM Higgs-bosons. 

\medskip

Besides, in order not to be a toy model, the features of 
the Standard Model must be reproduced in the 
MSSM, at least in certain parametric limits. 
This is fulfilled in the decoupling limit, 
taking the limit $m_{A^0} \rightarrow \infty$. 
Then, the physical Higgs-bosons $A^0$, $H^0$ and $H^\pm$
decouple from the theory and the Standard Model Higgs-boson sector
consisting of a single physical 
$\mathcal{CP}$-even scalar $h^0$ results. 
Additionally, a SUSY mass scale much larger than the 
electro-weak scale can be assumed. 
Then, $h^0$ becomes 
indistinguishable from the Higgs-boson $h$ of the Standard 
Model, since all Standard Model tree-level 
and loop-induced couplings to 
Standard Model gauge 
bosons and fermions are reproduced. 
A decoupling may also occur independent of the 
$A^0$ boson mass in other regions of the 
MSSM parameter space. For a discussion, see 
Refs.~\cite{Gunion:1989we,Carena:2001bg}. 
%
%
\section{The scalar quark sector and SQCD at tree-level}
\label{sec:scalarquarktreelevel}
In light of the calculation to be discussed in
detail in Chap.~\ref{chap:application2}, 
an introduction of the tree-level quark and
squark interactions is necessary, 
including those of supersymmetric 
Quantum Chromodynamics (SQCD). 
Interactions with the Higgs-boson sector are 
also discussed. 

\medskip

The relevant parts of the
MSSM Lagrangian regarding SQCD, the fermion $\psi$ 
and the scalar fermion (sfermion) $\tilde{s}$ sector, reads
\begin{align}
 \mathcal{L}_{\rm{MSSM}} \supset  \mathcal{L}_{\tilde{s}_{\rm{free}}} +
 \mathcal{L}_{\psi_{\rm{free}}} + 
 \mathcal{L}_{\tilde{s}_{\rm{int}}} +
  \mathcal{L}_{\psi_{\rm{int}}} +
 \mathcal{L}_{G} + \mathcal{L}_{\tilde{g}}\text{ ,}
 \label{eq:squarkLag}
 \end{align}
where the first two terms on the righthand side of 
Eq.~(\ref{eq:squarkLag}) are the free field equations
for the squarks and sleptons and the quarks and 
leptons, respectively. The last two terms are
the QCD and SQCD gauge field contributions. They 
can be combined as follows
\begin{align}
\mathcal{L}_{G} + \mathcal{L}_{\tilde{g}} = 
\left[ \frac{1}{16 g_s^2} W_s^{a \alpha} W_{s  \alpha}^{a} + h.c.\right]_F 
+ \mathcal{L}_{\tilde{g}_{\rm{soft}}}\text{ ,}
 \label{eq:QCDandSQCD}
\end{align}
where $W_s^a$ in the first term on the righthand side is the supersymmetric
SU(3) Yang-Mills field-strength tensor defined as
\begin{align}
 W_{s \alpha}^a= - \frac14 \bar{D}_\alpha \bar{D}^\alpha (e^{-2 g_s t_s^a G^a} D_\alpha e^{2 g_s t_s^a G^a }) \text{ ,}
\end{align}
compare Ref.~\cite{Ferrara:1974pu}. 
Here, the $t_s^a$ denote the generators of SU(3)$_c$, $g_s$ is the strong 
coupling constant and the $G^a$ represent the 
gluon and gluino fields, compare Tab.~\ref{tab:superfieldcontent}. 
\begin{table}
\begin{center}
    \begin{tabular}{ 
    !{\color{gray75}\vrule}!{\color{gray75}\vrule} c 
    !{\color{gray75}\vrule}!{\color{gray75}\vrule} c 
    !{\color{gray75}\vrule} c
    !{\color{gray75}\vrule}!{\color{gray75}\vrule} c 
    !{\color{gray75}\vrule} c
    !{\color{gray75}\vrule}!{\color{gray75}\vrule} c 
    !{\color{gray75}\vrule}!{\color{gray75}\vrule}
    }
    \arrayrulecolor{gray75} \hline
   superfield & particle content & spin & sparticle content & spin & $Y$\\
    \arrayrulecolor{gray75} \hline
$\hat{Q}$ & $Q=\begin{pmatrix} {u_{L}}\\ {d_L} \end{pmatrix}$ & $\frac12$ &
$\tilde{Q}=\begin{pmatrix} {\tilde{u}_{L}}\\ {\tilde{d}_L} \end{pmatrix}$ & $0$ & $1/3$\\ 
$\hat{U}$ & $U=u_R^\dagger$ & $\frac12$ & $\tilde{U}=\tilde{u}_R^*$ & $0$& $-4/3$\;\;\;\\ 
$\hat{D}$ & $D=d_R^\dagger$ & $\frac12$ & $\tilde{D}=\tilde{d}_R^*$ & $0$& $2/3$\\ 
$\hat{L}$ & $\begin{pmatrix} {\nu_{L}}\\ {e_L} \end{pmatrix}$ & $\frac12$ &
$\begin{pmatrix} {\tilde{\nu}_{L}}\\ {\tilde{e}_L} \end{pmatrix}$ & $0$& \hspace{-12pt} $-1$\\ 
$\hat{E}$ & $e_R^\dagger$ & $\frac12$ & $\tilde{e}_R^*$ & $0$ & $2$\\ 
$G^a$   & $G_\mu^a$ & $1$ & $\tilde{g}^a$ & $\frac12$ & $0$\\ 
$W$ & $W_\mu^i$ & $1$ & $\tilde{W}^i$ & $\frac12$ & $2$\\ 
$B$ & $B_\mu$ & $1$ & $\tilde{B}$ & $\frac12$ & $2$\\ 
$H_{1}$ & $H_{1}$ & $0$ & $\tilde{H}_{1}$ & $\frac12$ & $-1$\\ 
$H_{2}$ & $H_{2}$ & $0$ & $\tilde{H}_{2}$ & $\frac12$ & $+1$\\ 
\arrayrulecolor{gray75} \hline 
\end{tabular}
\end{center}
\label{tab:superfieldcontent}
\caption{The superfield content of the MSSM, the respective spin of the particles and sparticles 
	and the corresponding weak hypercharge $Y$. The index $a$ labels the different components. 
	Neutralinos and charginos are formed from linear combinations of the 
	gauginos, $\tilde{B}$ and $\tilde{W}^i$, and the higgsinos $\tilde{H}_{1,2}$. }
%
\end{table}
The 
$D_\alpha$ and $\bar{D}_\alpha$ are the covariant derivatives with
respect to the superspace coordinates and are defined as 
\begin{align}
 D_\alpha=\partial_\alpha - i \sigma^\mu_{\alpha\beta} \bar{\theta}^\beta \partial_\mu \text{,} 
 \hspace{20pt} \bar{D}^\alpha=\bar{\partial}^\alpha - i \bar{\sigma}^{\mu \alpha\beta} \theta_\beta \partial_\mu \text{ .}
\end{align}
While the gluon is massless, the gluino acquires
a mass term from explicit but soft supersymmetry
breaking
\begin{align}
\mathcal{L}_{\tilde{g}_{\rm{soft}}}  = - ( \frac12 m_{\tilde{g}} \tilde{g}^\dagger\tilde{g} + h.c.) \text{ .}
\end{align}
This type of mass term can rather generically be introduced for 
all fermions of a supergauge multiplet, compare 
Refs.~\cite{Dimopoulos:1981zb,Girardello:1981wz}. \\
Following Ref.~\cite{Gunion:1984yn}, the third and fourth 
term in Eq.~(\ref{eq:squarkLag}) 
can be split 
into F-terms, D-terms and soft breaking terms. 
Additionally, interactions between the 
gluino $\tilde{g}$, a quark and a squark must
be taken into account. 
Hence, the full potential reads
\begin{align}
 \mathcal{L}_{\tilde{s}_{\rm{int}}} + 
 \mathcal{L}_{\psi_{\rm{int}}} &= 
 -\left[\mathcal{V}_{\tilde{s}_{\rm{F}}}\right]_F - \left[\mathcal{V}_{\tilde{s}_{\rm{D}}}\right]_D - 
 \mathcal{V}_{\tilde{s}_{\rm{soft}}} + 
 \mathcal{L}_{\psi} +
  \mathcal{L}_{qqG} + 
 \mathcal{L}_{\tilde{q}\tilde{q}\tilde{q}\tilde{q}}^s +
  \mathcal{L}_{\tilde{q}\tilde{q}G} +
 \mathcal{L}_{q\tilde{q}\tilde{g}} 
 \label{eq:fullQuarkSquarkpotential} \text{ .}
\end{align}
The F-term contributions are
\begin{align}
\nonumber\mathcal{V}_{\tilde{s}_{\rm{F}}} =&\,
|-\mu^\dagger \mathcal{H}_1^{i\dagger} + y_u \tilde{Q}^{i\dagger}\tilde{U}|^2 
 + |\mu^\dagger \mathcal{H}_2^{i\dagger} + y_d \tilde{Q}^{i\dagger}\tilde{D}|^2 \\
\nonumber &+\, y_u^2 \,|\eps_{ij} \mathcal{H}_2^i \tilde{Q}^{j}|^2 + y_d^2\, |\eps_{ij} \mathcal{H}_1^i \tilde{Q}^{j}|^2 \\
 &+\,(y_u \mathcal{H}_2^{i\dagger}\tilde{U}^{\dagger}-y_d \mathcal{H}_1^{i\dagger}\tilde{D}^{\dagger})(y_u \mathcal{H}_2^{i} \tilde{U} - y_d \mathcal{H}_1^i \tilde{D})
\end{align}
where Yukawa interactions of the type $H\tilde{s}\tilde{s}$, 
$HH\tilde{s}\tilde{s}$ and $\tilde{s}\tilde{s}\tilde{s}\tilde{s}$ 
can be read off. While the $\tilde{Q},\,\tilde{U}$ and $\tilde{D}$  
denote the scalar superfields, see Tab.~\ref{tab:superfieldcontent}, 
the Yukawa couplings read
\begin{align}
 y_u = \frac{g \,m_u}{\sqrt{2} m_W \sin\beta}\text{,} \hspace{20pt} 
 y_d = \frac{g\, m_d}{\sqrt{2} m_W \cos\beta} 
 \label{eq:yukdefs}\text{ .}
\end{align}
The D-terms contributing to the scalar potential Eq.~(\ref{eq:fullQuarkSquarkpotential}) read
\begin{align}
\nonumber \mathcal{V}_{\tilde{s}_{\rm{D}}} =& \frac18 g^2 \left( 
4 |\mathcal{H}_1^{i\dagger} \tilde{Q}^{i} |^2 + 
4 |\mathcal{H}_2^{i\dagger} \tilde{Q}^{i} |^2 - 
2 (\tilde{Q}^{i\dagger}\tilde{Q}^{i}) (\mathcal{H}_1^{i\dagger} \mathcal{H}_1^{i} + \mathcal{H}_2^{i\dagger} \mathcal{H}_2^{i}) +
(\tilde{Q}^{i\dagger}\tilde{Q}^{i})^2\right) \\
\nonumber +& \frac14 g'^2 (\mathcal{H}_2^{i\dagger} \mathcal{H}_2^{i} - \mathcal{H}_1^{i\dagger} \mathcal{H}_1^{i}) 
(Y_q \tilde{Q}^{i\dagger}\tilde{Q}^{i} + Y_u \tilde{U}^{\dagger}\tilde{U} + Y_d \tilde{D}^{\dagger}\tilde{D})\\
 +& \frac18 g'^2 \left(
Y_q \tilde{Q}^{i\dagger}\tilde{Q}^{i} + Y_u \tilde{U}^{\dagger}\tilde{U} + Y_d \tilde{D}^{\dagger}\tilde{D} \right)^2
\label{eq:vdscalar} \text{ ,}
\end{align}
where $Y_q=\frac13$, $Y_u=-\frac{4}{3}$ and $Y_d=\frac{2}{3}$ 
are the hypercharges of the respective superfield.
The squarks couple not only weakly to each other, but also strongly via
\begin{align}
\mathcal{L}_{\tilde{q}\tilde{q}\tilde{q}\tilde{q}}^s = - \frac12 g_s^2 \sum_{a} 
(\tilde{q}_L^\dagger t_s^a \tilde{q}_L -  \tilde{q}_R^\dagger t_s^{*a} \tilde{q}_R)^2 \text{ ,}
\end{align}
compare e.g. Ref.~\cite{Drees:2004jm}.
The soft-breaking part of the potential reads
\begin{align}
\nonumber \mathcal{V}_{\tilde{s}_{\rm{soft}}} =& 
M_{\tilde{Q}}^2 \,\tilde{Q}^{i\dagger}\tilde{Q}^{i} + M_{\tilde{u}}^2 \,\tilde{U}^{\dagger}\tilde{U}  + 
M_{\tilde{d}}^2 \,\tilde{D}^{\dagger}\tilde{D}\\
+& m_6 (-\eps^{ij} y_u A_u H_2^i \tilde{Q}^{j} \tilde{U}^{\dagger} + \eps^{ij} y_d A_d H_1^i \tilde{Q}^{j} \tilde{D}^{\dagger} + h.c.) 
\label{eq:vsoftscalar}\text{ ,}
\end{align}
where the products $m_6\, A_{u,d}$ denote the trilinear couplings 
of the Higgs-bosons to the squarks. In the following, $m_6=1$ is 
chosen by convention. $M_{\tilde{Q}}$ is the 
mass parameter of the left-hand sparticles, 
$M_{\tilde{u}}$ and $M_{\tilde{d}}$ are the
mass parameters of the righthand up-type and 
down-type sparticles, respectively. The soft breaking
terms lead at most to a logarithmically divergent 
behavior and gauge invariance is ensured, 
see Refs.~\cite{Dimopoulos:1981zb,Girardello:1981wz}. 
The slepton fields $\tilde{L},\,\tilde{E}$ are omitted here 
but can be included as well with an appropriate choice of hypercharges.
The Yukawa interaction of the quarks and leptons reads
\begin{align}
\mathcal{L}_{\psi} = - \eps_{ij} (y_d \mathcal{H}_1^i Q^j D - y_u \mathcal{H}_2^i Q^j U + y_e \mathcal{H}_1^i L^j E) + h.c. \text{ ,}
\end{align}
where $y_e$ is the Yukawa coupling to the leptons. The
quark and lepton masses purely arise from the Yukawa 
interactions, therefore the quark and lepton mass matrices
can be directly deducted from the Yukawa terms. 
The interaction of two quarks with a gluon reads
\begin{align}
\mathcal{L}_{qqG} = 
- g_s G_\mu^a \sum_{i=u,d} \bar{q}_{i}^j \gamma^\mu (t_s^a)_{jk} q_{i}^k \text{ ,}
\end{align}
where $j$ and $k$ are color indices.
Now, the squark-squark gluon interaction is
\begin{align}
 \mathcal{L}_{\tilde{q}\tilde{q}G} = -i g_s G_\mu^a \sum_{i=u,d} \tilde{q}_{i}^{j\dagger} 
 (t_s^a)_{jk} \partial^\mu \tilde{q}_{i}^{k} \text{ ,}
\end{align}
where the sums run 
over both left- and righthanded components, compare Ref.~\cite{Haber:1984rc}.
Finally, the squark-quark-gluino interaction reads
\begin{align}
\hspace{-20pt} \mathcal{L}_{q\tilde{q}\tilde{g}} = -\sqrt{2} g_s (t_s^a)_{jk} \sum_{i=u,d}(
 \tilde{g}_a^\dagger P_L q_i^k \tilde{q}_{iL}^{j\dagger} + 
 q_i^{j\dagger} P_R \tilde{g}_a \tilde{q}_{iL}^{k} - 
 \tilde{g}_a^\dagger P_R q_i^k \tilde{q}_{iR}^{j\dagger} -
 q_i^{j\dagger} P_L \tilde{g}_a \tilde{q}_{iR}^{k} ) \text{ ,}
\end{align}
where the relative minus sign comes in with the negative sign of
the color generator $t_s^{a*}$ of the color antitriplets.
Likewise, there are electroweak quark-squark-gaugino 
interactions. They are not listed here because they are 
not needed in the calculation of Chap.~\ref{chap:application2}. 

\medskip

The squark masses are composed 
of the 
soft breaking terms, but also 
the F- and D-terms of the squark potential, 
when the Higgs-bosons acquire vacuum expectation 
values.
Altogether, the massive part of the squark sector 
in the MSSM reads
\begin{align}
\cL_{\tilde{q}, \text{mass}} &= - 
\begin{pmatrix}
{{\tilde{u}}_{L}}^{\dagger} & {{\tilde{u}}_{R}}^{\dagger} 
\end{pmatrix}
M^2_{\tilde{u}_{LR}}
\begin{pmatrix}
{\tilde{u}}_{L}\\{\tilde{u}}_{R}
\end{pmatrix} -
\begin{pmatrix}
{{\tilde{d}}_{L}}^{\dagger} & {{\tilde{d}}_{R}}^{\dagger} 
\end{pmatrix}
M^2_{\tilde{d}_{LR}}
\begin{pmatrix}
{\tilde{d}}_{L}\\{\tilde{d}}_{R}
\end{pmatrix} 
\text{,}
\end{align}
where the up-type squark mass matrix is
given by 
\begin{align}
\hspace{-20pt} M^2_{\tilde{u}_{LR}} =
\begin{pmatrix}  
 M_{\tilde{Q}}^2 + y_u^2 v_2^2 + \frac14 (g^2-Y_q g'^2) (v_1^2-v_2^2) & 
 y_u v_2 ( A_u^\dagger - \mu \cot \beta)   \\[.2em]
 y_u v_2 ( A_u - \mu^\dagger \cot \beta) &
 M_{\tilde{u}}^2 + y_u^2 v_2^2  + \frac14 g'^2 Y_u (v_2^2-v_1^2)
\end{pmatrix}  \text{,}
\label{eq:upmassmatrix}
\end{align}
with $y_u v_2=m_u$ from previous definitions in 
Eq.~(\ref{eq:vevsdef}) and Eq.~(\ref{eq:yukdefs}). 
The parameter $\mu$ is taken to be real in the rMSSM. 
A similar mass matrix can be set up for the down-type squarks
from the previously described parts of the Lagrangian. Using 
further the definitions of Eq.~(\ref{eq:coupldefs})
and of the hypercharges below Eq.~(\ref{eq:vdscalar}), the
up-type squark mass matrix can be written as
\begin{align}
\hspace{-25pt}M^2_{\tilde{u}_{LR}} =
\begin{pmatrix} 
 M_{\tilde{Q}}^2 + m_u^2 + m_Z^2 \CZb \,(I_u^3 - Q_u \sin^2\theta_W) & 
 m_u X_{\tilde{u}} \\[.6em]
 m_u X_{\tilde{u}} &
 \hspace{-15pt} M_{\tilde{u}}^2 + m_u^2 + m_Z^2 \CZb \, Q_u  \sin^2\theta_W
 \end{pmatrix} \text{,}
 \label{eq:upmassmatrixnicer}
 \end{align}
with $X_{\tilde{u}} = \,A_u - \mu\,\CTb$ and where $Q_u$ denotes their
charge and $I_u^3$ the third component of the isospin 
of the up-type squark, respectively.
For completeness, the down-type squark mass matrix is also given, 
\begin{align}
\hspace{-25pt}M^2_{\tilde{d}_{LR}} =
\begin{pmatrix} 
 M_{\tilde{Q}}^2 + m_d^2 + m_Z^2 \CZb \,(I_d^3 - Q_d \sin^2\theta_W) & 
 m_d X_{\tilde{d}} \\[.6em]
 m_d X_{\tilde{d}} &
 \hspace{-15pt} M_{\tilde{d}}^2 + m_d^2 + m_Z^2 \CZb \, Q_d  \sin^2\theta_W
 \end{pmatrix} \text{,}
 \label{eq:downmassmatrixnicer}
\end{align}
with $X_{\tilde{d}} = \,A_d - \mu \tanb$. 
The squark mass matrices can be rotated into
the physical basis
\begin{align}
\label{eq:orthosquarkmass}
\cL_{\tilde{q}, \text{mass}} &= - 
\begin{pmatrix}
{{\tilde{q}}_{1}}^{\dagger} & {{\tilde{q}}_{2}}^{\dagger} 
\end{pmatrix}
M^2_{\tilde{q}_{12}}
\begin{pmatrix}
{\tilde{q}}_{1}\\[.2em]{\tilde{q}}_{2} 
\end{pmatrix} \text{,}
\end{align}
with the physical squark mass
eigenstates $m_{\tilde{q}_1}^2$ and 
$m_{\tilde{q}_2}^2$. The new
mass eigenstates are related to the unphysical masses via 
an orthogonal transformation
\begin{subequations}
\begin{align}
 M^2_{\tilde{q}_{LR}} &= U^T_{\tilde{q}} \,M^2_{\tilde{q}_{12}} \,U_{\tilde{q}} \\
  &=
  \begin{pmatrix} 
  \cos^2\!\theta_{\tilde{q}} \,m_{\tilde{q}_1}^2 + \sin^2\!\theta_{\tilde{q}} \,m_{\tilde{q}_2}^2 & 
  \sin\theta_{\tilde{q}} \cos\theta_{\tilde{q}} (m_{\tilde{q}_1}^2-m_{\tilde{q}_2}^2) \\ 
  \sin\theta_{\tilde{q}} \cos\theta_{\tilde{q}} (m_{\tilde{q}_1}^2-m_{\tilde{q}_2}^2) & 
  \sin^2\!\theta_{\tilde{q}} \,m_{\tilde{q}_1}^2 + \cos^2\!\theta_{\tilde{q}} \,m_{\tilde{q}_2}^2
  \end{pmatrix} \text{,}
  \label{eq:msquark12}
\end{align}%
\label{alleqs:msquark12}%
\end{subequations}%
where the unitarity matrix
\begin{align}
 U_{\tilde{q}}= 
 \begin{pmatrix} 
  \cos {\theta}_{\tilde{q}} & \sin {\theta}_{\tilde{q}} \\ 
  -\sin {\theta}_{\tilde{q}} & \cos {\theta}_{\tilde{q}} \end{pmatrix} \text{ }
  \label{eq:utransformtheta}
\end{align}
is parametrized by the mixing angle $ {\theta}_{\tilde{q}}$. \\
Matching the two mass matrices in Eq.~(\ref{eq:upmassmatrixnicer}) 
and Eq.~(\ref{eq:msquark12}), an expression for the parameter $X_{\tilde{q}}$
can be formulated as follows
\begin{align}
 X_{\tilde{q}} = \frac{\sin\theta_{\tilde{q}} \cos\theta_{\tilde{q}} 
 (m_{\tilde{q}_1}^2-m_{\tilde{q}_2}^2)}{m_q} \text{ .}
 \label{eq:definitionXt}
\end{align}

%
%
\section{Higher-order Higgs-boson mass corrections within the real MSSM}
\label{sec:highorderreview}
With the light neutral $\mathcal{CP}$-even Higgs-boson tree-level 
mass $m_{h^0}$ being
limited to $m_Z$ at most, compare Eq.~(\ref{eq:mHmhtree}), the 
MSSM could already have been excluded at LEP due to the 
lack of its observation. 
Yet, higher-order self-energy corrections shifted the upper bound on the 
light Higgs-boson mass considerably. 
There are mainly three different methods 
to approach higher-order mass corrections. They 
can be combined as well. 
An exact calculation, invariant under different gauge-fixing 
terms, 
is achieved using the 
Feynman-diagrammatic  (FD) approach, 
where the self-energy diagrams are
evaluated explicitly.
The second method uses an
effective potential approximation
further developed
for higher loop calculations, compare
Ref.~\cite{Jackiw:1974cv}. In this approach, 
the scalar Higgs-boson fields are expanded
around their vacuum expectation values. 
This allows for the computation of 
the higher-loop effective 
potential, involving vacuum diagrams 
of the given loop order. The results
are compact, but also 
of limited accuracy. As this 
approach expands around a constant
value of the fields, the
momentum dependence of the two-loop self-energies 
cannot be taken into account.
A third approach uses effective
Lagrangians capturing the
dynamics and symmetries of 
a system in generic terms,
while the phenomenology is
contained in their coefficients.
Effective Lagrangians are often
used in model-independent analyses.

\medskip

In the calculation 
presented in Chap.~\ref{chap:application2} of this thesis, 
the Feynman-diagrammatic
approach is adopted. In this approach, higher order mass 
corrections are computed by 
allowing for perturbations to the 
propagators of the fields $\boldsymbol{\phi}_a^0,\;\boldsymbol{\chi}_a^0,\;\boldsymbol{\phi}_a^{\pm}$
which result from the solutions to the equations of motion.
For completeness 
\begin{align}
\partial_\mu \left( \frac{\partial \mathcal{L}}{\partial(\partial_\mu x_a)}\right) - 
\frac{\partial \mathcal{L}}{\partial x_a} = 0 \text{ ,}
\label{eq:eoms}
\end{align}
where $x=\boldsymbol{\phi}^0,\;\boldsymbol{\chi}^0,\;\boldsymbol{\phi}^{\pm}$ and where $a=1,2$. 
Firstly, the equations of motion require that the 
terms of the MSSM potential part of the Lagrangian
linear in the 
$\mathcal{CP}$-even Higgs-boson fields 
$\boldsymbol{\phi}_a^0$, see Eq.~(\ref{eq:linVpart}), 
must vanish. For this condition to be met, 
all higher-order corrections 
up to $l^{\text{th}}$ order need to be 
seen as canceling. At tree level, the 
terms linear in the fields are tadpole 
coefficients. Higher orders include
additional propagators to tadpole lines
in terms of loops, where the coefficients
are termed $T_a^{(l)}$.
In conclusion, the statement reads 
\begin{align}
\sum_{a=1}^2 \,(T_a + T_a^{(1)}  + T_a^{(2)} +\dots ) = 0 \text{ .}
\label{eq:linLaghigherorders}
\end{align}
Secondly, the bilinear free field and potential parts of the Lagrangian, 
Eq.~(\ref{eq:MSSMHfree}) and Eqs.~(\ref{eq:VHF})-(\ref{eq:VHsoft}),
lead to the following contributions to 
the equations of motion
\begin{align}
(\partial_\mu \partial^\mu + M_{x_ax_a}^2) \; x_a 
+ \eps_{ab} M_{x_ax_b}^2 \;x_b  = 0  
\hspace{20pt}\text{ with }\hspace{5pt} a,b=1,2 \text{; } \hspace{10pt} a \neq b \text{ .}
\label{eq:eomshigherorders}
\end{align}
Hence, those terms bilinear in the same field 
get solutions to the equations of motion in 
terms of causal Green functions including a 
momentum and a massive part, while the solutions
to terms bilinear in two different fields
contain only a massive part. 
Computing higher orders in perturbation theory 
corresponds to adding one-particle irreducible 
terms $\Sigma(p^2)$ to the propagators~\cite{Peskin:1995ev}
\begin{align}
& \frac{i}{\boldsymbol{p^2-M_{x_ax_a}^2}} \\
\nonumber &= \frac{i}{p^2-M_{x_ax_a}^2} + \frac{i}{p^2-M_{x_ax_a}^2}(-i \Sigma_{x_ax_a}(p^2))\frac{i}{p^2-M_{x_ax_a}^2}\\
  &+ \frac{i}{p^2-M_{x_ax_a}^2}(-i \Sigma_{x_ax_a}(p^2))\frac{i}{p^2-M_{x_ax_a}^2}
  (-i \Sigma_{x_ax_a}(p^2))\frac{i}{p^2-M_{x_ax_a}^2} + \dots \label{eq:allpartirredcomp} \\
  &=\frac{i}{p^2-M_{x_ax_a}^2- \Sigma_{x_ax_a}(p^2)} \text{ ,} \label{eq:usegeometricseries}
\end{align}
where the bold-faced propagator is the one corrected 
to all orders in perturbation theory and where each 
one-particle irreducible term can be split into 
its different orders, 
\begin{align}
\Sigma_{x_{a}x_b} &=\Sigma_{x_{a}x_b}^{(1)} + \Sigma_{x_ax_b}^{(2)} + ... \text{ .} \label{eq:sigmaxixj}
\end{align} 
A geometric series relation is used
in the last step from Eq.~(\ref{eq:allpartirredcomp}) to 
Eq.~(\ref{eq:usegeometricseries}). The cases where different fields 
enter the time ordered two-point correlation functions 
can be treated similarly.
Assuming that the self-energy corrections in 
Eq.~(\ref{eq:sigmaxixj}) can be renormalized, 
the renormalized self-energies 
$\hat{\Sigma}_{x_{a}x_b}(p^2)$ enter as
corrections to the inverse propagator matrix of the field $x$, 
\begin{align}
(\Delta_{\text{x}})^{-1} = -\text{i}
\left( \begin{matrix} 
p^2 - m_{x_1}^2 + \hat{\Sigma}_{x_1x_1}(p^2) & -m_{x_1x_2}^2 + \hat{\Sigma}_{x_1x_2}(p^2)\\ 
-m_{x_1x_2}^2 + \hat{\Sigma}_{x_1x_2}(p^2) & p^2 - m_{x_2}^2 + \hat{\Sigma}_{x_2x_2}(p^2) 
\end{matrix} \right) \text{ .}
\label{eq:invpropmatrixx}
\end{align}
The loop-corrected masses $M_{x_1}$ and $M_{x_2}$ are determined by the
real parts of the propagator matrix of the field $x$. 
This is equivalent to solving the equation
\begin{equation}
\left[p^2 - m_{x_1}^2 + \hat{\Sigma}_{x_1x_1}(p^2)\right]
\left[p^2 - m_{x_2}^2 + \hat{\Sigma}_{x_2x_2}(p^2)\right] -
\left[-m_{x_1x_2}^2 + \hat{\Sigma}_{x_1x_2}(p^2)\right]^2 = 0\,.
\label{eq:proppolex}
\end{equation}

\medskip

The status of the available self-energy corrections 
to the real MSSM
can be summarized as follows. 

\medskip

At the one-loop level, the full corrections 
to the MSSM Higgs-boson masses are
known, including gauge bosons contributions and 
momentum dependence, see Refs.~\cite{Ellis:1990nz,Haber:1990aw,
Okada:1990vk,Brignole:1992uf,Chankowski:1992ej,
Chankowski:1992er,Dabelstein:1994hb,Dabelstein:1995js}.

\medskip

At two loops, a full result using an effective potential
approach is known~\cite{Martin:2002iu,Martin:2002wn}. 
Preceding works used a two-loop renormalization 
group equation (RGE) improved one-loop effective potential
approach~\cite{Kodaira:1993yt,
Gladyshev:1994iw,Carena:1995bx,
Casas:1994us,Carena:1995wu}, or a two-loop 
effective potential approach~\cite{Hempfling:1993qq,Zhang:1998bm,
Espinosa:1999zm,Espinosa:2000df,
Degrassi:2001yf,Brignole:2001jy,
Brignole:2002bz,Dedes:2003km}.
Furthermore, explicit computations have been
performed in the Feynman-dia-grammatic
approach, neglecting gauge contributions and 
assuming vanishing external momentum \cite{Heinemeyer:1998jw,
Heinemeyer:1998kz,Heinemeyer:1998np,
Heinemeyer:1999be}.
The latter cover the dominant 
corrections of the order $\mathcal{O}(\alpha_s \alpha_t)$ 
and $\mathcal{O}(\alpha_t^2)$ and the sub-dominant 
two-loop contributions of the order
$\mathcal{O}(\alpha_s \alpha_b)$, 
$\mathcal{O}(\alpha_t \alpha_b)$ and 
$\mathcal{O}(\alpha_b^2)$. The orders are given in
terms of the coupling factors entering the vertices of 
the loop diagrams. These are the strong 
coupling constant $\alpha_s= \frac{g_s^2}{4 \pi}$
and the Yukawa couplings $\alpha_{t}= \frac{y^2_{t}}{4 \pi}$ and 
$\alpha_{b}= \frac{y^2_{b}}{4 \pi}$ 
of the top and the bottom quark, respectively. 
The relative size of a correction can be estimated a priori by 
assessing the relative size of its couplings. 
Due to supersymmetry, the Yukawa couplings for the 
quarks and squarks are equivalent 
$\alpha_{\tilde{t},\tilde{b}}=\alpha_{t,b}$. 
The soft supersymmetry \textbf{breaking} terms 
contributing in the coupling of the Higgs-bosons to 
the squarks are proportional to the Standard Model 
Yukawa couplings as well, compare 
Eq.~(\ref{eq:vsoftscalar}). 
Therefore, no distinction between $\alpha_{\tilde{q}}$ and
$\alpha_{q}$ is needed. 

Complementary to the Feynman-diagrammatic approach, 
higher-order corrections to the Higgs-boson masses 
have been found using the effective Lagrangian 
approach~\cite{Espinosa:1991fc,Hempfling:1993kv,
Hall:1993gn,Carena:1994bv,Carena:1999py,Espinosa:2001mm}.
Other studies aim towards a combination of the 
existing two-loop results obtained in different approaches, 
see Refs.~\cite{Zhang:1998bm,
Espinosa:1999zm,Espinosa:2000df,
Carena:2000dp,Degrassi:2002fi}.

\medskip

Regarding the third order, the dominant corrections
of the order $\mathcal{O}(\alpha_s^2 \alpha_t)$ 
are available~\cite{Martin:2007pg,Harlander:2008ju,Kant:2010tf}, 
where gauge contributions and a non-vanishing external momentum 
still need to be incorporated in future calculations. 
Also, third and higher order resummation effects 
have recently been taken into account~\cite{Hahn:2013ria,Draper:2013oza}.

\medskip

Remarkably, all these
higher-order corrections lead to an approximate 
upper bound of $m_{h^0} \lesssim 135\,$GeV, where the 
maximal value for the light Higgs-boson mass depends 
sensitively on the precise value of the top-quark mass, 
compare Refs.~\cite{Heinemeyer:1998np,Degrassi:2002fi}.\\
While the one-loop corrections relocate 
the upper bound towards higher masses, the two-loop corrections 
enter with competing signs and the 
three-loop corrections further stabilize the mass, entering with
both signs. The overall dominant corrections
come from those self-energies involving the top quark and 
the scalar top (stop) quarks. 

The remaining contributions of higher orders can be 
estimated based on the already available results. To reach 
a theoretical precision matching or even surpassing the 
experimental one, corrections beyond the above mentioned ones 
must be taken into account. 
At two loops, the remaining uncertainties originate from 
neglecting gauge contributions and momentum 
dependence~\cite{Heinemeyer:2004ms}. 
Therefore, the 
calculation of the momentum dependent two-loop 
corrections to the MSSM Higgs-boson self-energies 
is of interest. 
The momentum dependence at the 
one-loop level is known to generally amount to
less than 2 GeV, compare Ref.~\cite{Frank:2006yh}. 
%
The dominant corrections at the two-loop level have 
been calculated adopting 
a full $\overline{DR}$ renormalization 
scheme, see Refs.~\cite{Martin:2003it,Martin:2004kr,Martin:2005qm}. 
Higher-order corrections 
to the tree-level MSSM Higgs-boson masses can only
be applied consistently if they are computed 
within the same renormalization scheme.
It is therefore interesting to analyze the momentum-dependent 
contribution again, but
using an on-shell renormalization for all masses entering 
the calculation. Although the calculation
becomes more involved with this renormalization scheme choice, 
the benefit of being able to incorporate those
corrections into the public program 
\textsc{FeynHiggs}~\cite{Heinemeyer:1998yj,Frank:2002qf,
Hahn:2009zz,Hahn:2010te} and 
thereby making the corrections readily 
available, is outweighing.

%
%

%% file: twoloopfrontier/twoloopfrontier.tex
%
\chapter{Multi-scale integrals beyond one loop}
\label{chap:twoloopfrontier}
This chapter explores diverse techniques regarding the computation
of multi-scale integrals beyond one loop. In particular, a focus is laid on 
two-loop calculations, where fully differential phenomenological predictions 
start to emerge for a variety of processes. 
The chapter culminates in a motivation for the usage of Feynman 
parameterization in a tool to compute multi-loop multi-scale integrals 
in an automated way.
%
%
\section{The two loop frontier}
\label{sec:twoloopfrontier}
%
%
Due to the high energies at present hadron colliders, 
processes at very small distances can be resolved. At the Tevatron and the LHC, 
these dominantly involve quarks and gluons (partons). 
While the Tevatron is a proton-antiproton collider, the dominant production channel 
has $q\bar{q}$ pairs in the initial state, 
at the LHC the $gg$ channel is the dominant initial state. 
Compared to $e^+e^-$ annihilation processes at past colliders this 
is one additional external leg when 
higher orders are computed. This greatly increases the 
complexity of such a computation. 

\medskip

At next-to-leading order %
\footnote{It should be noted 
that the leading order usually encompasses tree-level diagrams, but there 
are LO processes (e.g. Higgs production in gluon fusion) which start with 
loop diagrams.} in perturbative QCD, 
the frontier is looking at final states with many 
particles and matching them with a parton shower. 
The motivation to go to NNLO accuracy is at least twofold. 
First, the dependence on the renormalization scale is expected to be reduced. 
Second, a process has more partons in the final state. This 
initiates the reconstruction of the parton shower, thus approaching 
an experimental jet reconstruction. 

\medskip

The forefront at NNLO is the computation of four-point processes, 
where various complications can arise, one being a complicated 
singularity structure of individual diagrams, and the other the 
involvement of internal and external masses leading to 
multi-scale problems. 
This can render the evaluation of sometimes just single diagrams 
a highly non-trivial task. 
While multiple legs and scales can already be very complicated 
at NLO, at NNLO completely new challenges arise in addition 
when generalizing the techniques employed at NLO to NNLO 
and diverse conceptual differences must be acknowledged. 

\medskip

Beyond NLO, the renormalization procedure 
for the cancellation of ultraviolet (UV) singularities involves not 
only counter-terms of the loop order to be considered but 
also counter-terms of lower loop order with insertions. 
Finding the full but minimal set of diagrams is non trivial 
but was solved and further developed by  
Bogoliubov, Parasiuk, Hepp and Zimmermann, summarized 
in the (BPHZ) theorem, see 
Refs.~\cite{Bogoliubov:1957gp,Hepp:1966eg,Zimmermann:1968mu}. 

\medskip

\begin{figure}[htb!]
\begin{center}
\subfigure[]{
\includegraphics[width=0.31\textwidth]{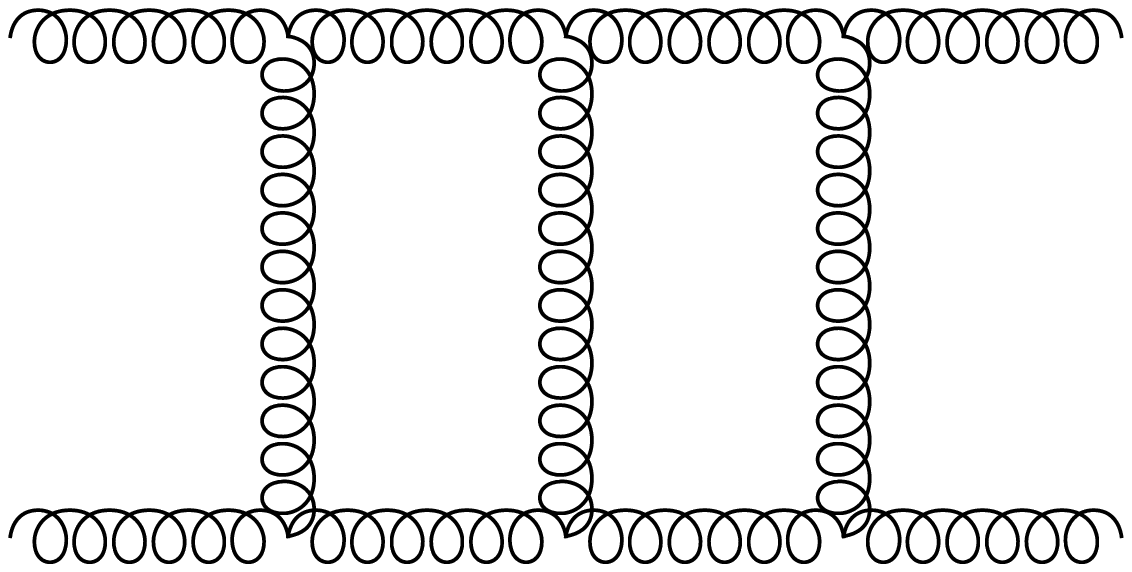}
                \label{subfig:virtualvirtual}}%
\subfigure[]{
\includegraphics[width=0.31\textwidth]{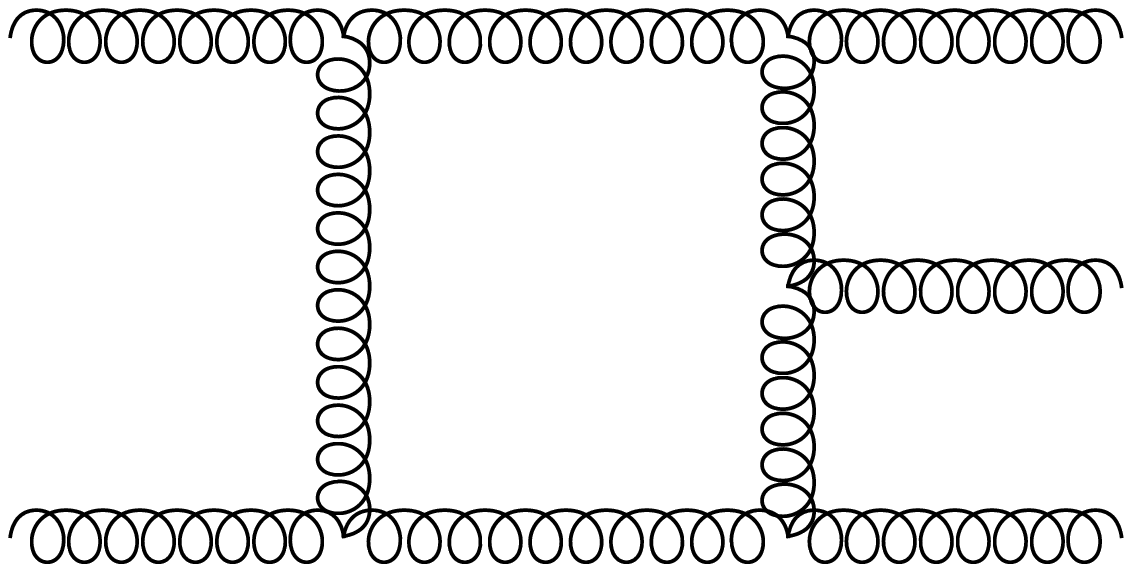}
                \label{subfig:realvirtual}}%
\subfigure[]{
\includegraphics[width=0.31\textwidth]{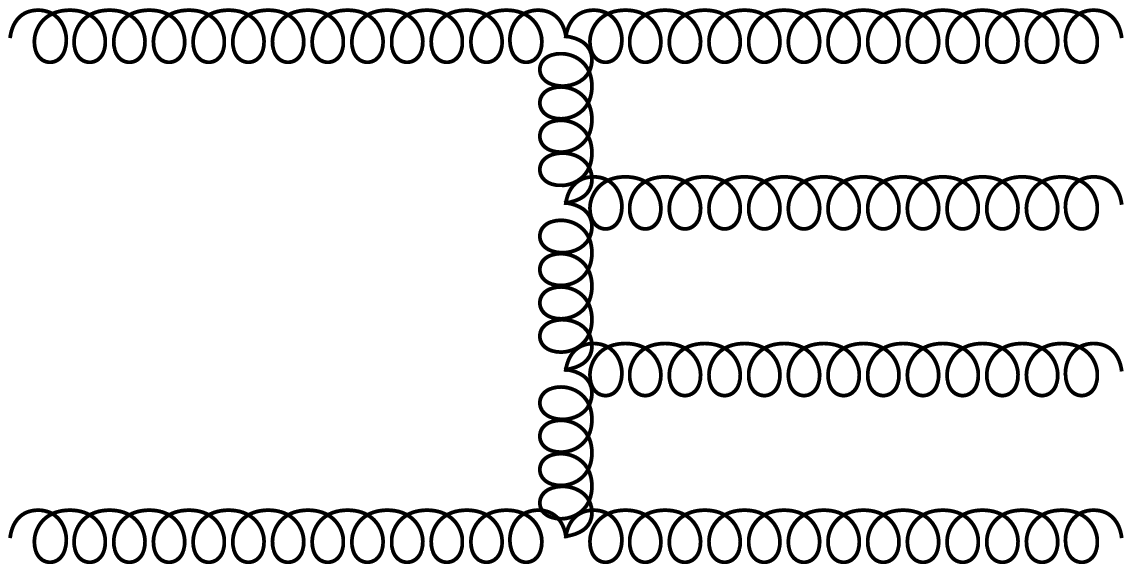}
                \label{subfig:realreal}}%
\end{center}
\caption{Example diagrams for double virtual \subref{subfig:virtualvirtual}, real 
virtual \subref{subfig:realvirtual} and double real radiation \subref{subfig:realreal} 
contributions entering an NNLO calculation are shown.}
\label{fig:twolooprealvirtual}
\end{figure} 
Furthermore, a mixture of real radiation and virtual 
corrections enter a cross section at NNLO, in addition
to double real and double virtual contributions. 
%
%
%
The typical ingredients of an NNLO calculation are depicted in 
Fig.~\ref{fig:twolooprealvirtual}. 
Given the fact that now three pieces enter the calculation which are 
all part of different phase space dimensions, the need for more 
discriminating and refined subtraction schemes emerges. \\ 
At NLO, subtraction schemes are already well established, 
see Refs.~\cite{Frixione:1995ms,Frixione:1997np,Catani:1996vz}. 
At NNLO there are various subtraction 
schemes available, 
all with different aims and capabilities. After the 
introduction of $q_T$ subtraction by Catani and 
Grazzini~\cite{Catani:1998nv,Catani:1999ss}, 
the idea of using the sector decomposition algorithm 
\cite{Binoth:2000ps,Heinrich:2008si} for a complete NNLO 
calculation was originally proposed by 
Heinrich~\cite{Heinrich:2002rc}. It was taken up and further 
developed into a full subtraction scheme in Refs.~\cite{GehrmannDeRidder:2003bm,
Anastasiou:2003gr,Binoth:2004jv}, and first applied to a full 
process in Ref.~\cite{Anastasiou:2004qd}. 
The idea proposed by Czakon to combine sector decomposition 
applied to real emission integrals with phase-space partitioning 
from FKS subtractions, lead to the sector improved residue 
subtraction~\cite{Czakon:2010td}, successfully applied, e.g., in 
Refs.~\cite{Czakon:2011ve,Boughezal:2011jf}. 
Further, the antenna factorization introduced in Refs.~\cite{Kosower:1997zr,
Campbell:1998nn,Kosower:2003bh} 
was explicitly worked out and applied to full NNLO 
processes~\cite{GehrmannDeRidder:2004tv,
GehrmannDeRidder:2005cm,GehrmannDeRidder:2007bj,Currie:2013vh}. 
Lastly, a direct generalization of dipole subtraction to NNLO 
processes was presented by 
Somogyi, Trocsanyi et al.~\cite{Somogyi:2006cz,Somogyi:2006da}.

\medskip

While for the real radiation the computation of more and more legs is 
of interest, for the virtual contributions higher-order loop integrals 
need to be solved. 
In light of the fact that the number of diagrams contributing to 
higher-order processes increases tremendously from one 
order to the next, and the diagrams themselves become more and 
more complicated, it is desirable to find highly automatable 
procedures to tackle these. At NLO, tools towards this 
aim are already highly developed and sophisticated. 
The procedure of generating the real radiation and loop amplitudes 
contributing to a full process is automated to a large extent by programs 
like \textsc{aMC@NLO}~\cite{Alwall:2014hca}, 
\textsc{BlackHat}~\cite{Berger:2008sj}, 
\textsc{FeynArts, FormCalc \& LoopTools}~\cite{Hahn:2010zi}, 
\textsc{GoSam}~\cite{Cullen:2014yla}, 
\textsc{HELAC-NLO}~\cite{Bevilacqua:2011xh}, 
\textsc{HERWIG++}~\cite{Bahr:2008pv}, 
\textsc{Matchbox}~\cite{Platzer:2011bc}, 
\textsc{MCFM}~\cite{Campbell:2011bn}, 
\textsc{NJet}~\cite{Badger:2012pg}
\textsc{Open Loops}~\cite{Cascioli:2011va}, 
\textsc{POWHEG}~\cite{Alioli:2010xd}, 
\textsc{RECOLA}~\cite{Actis:2012qn}, 
\textsc{Sherpa}~\cite{Gleisberg:2008ta}, 
\textsc{VBFNLO}~\cite{Arnold:2011wj}. 
There are diverse tools allowing for an automated generation of the pure 
Feynman amplitudes. The programs with loop capabilities are 
{\sc FeynArts}~\cite{Kublbeck:1990xc,Hahn:2000kx} or {\sc QGRAF}~\cite{Nogueira:1991ex}. 

\medskip

The full basis of irreducible master integrals is known at one-loop order. 
It comprises scalar pentagon, box, triangle, 
bubble and tadpole diagrams which are known analytically or can be 
computed numerically~\cite{Hahn:1998yk,Ellis:2007qk,Binoth:2008uq,
vanHameren:2010cp}. 
Beyond one loop, the set of irreducible integrals, so called master integrals, 
is not known which makes the decomposition more difficult. 
Master integrals beyond one loop can have irreducible numerators 
which need to be evaluated in addition. In the reduction of multi-loop 
amplitudes to 
a set of resulting master integrals, integration by parts relations 
\cite{'tHooft:1972fi,Tkachov:1981wb,Chetyrkin:1981qh} 
and identities resulting from Lorentz invariance \cite{Gehrmann:1999as} 
are indispensable. 
The former are based on the fact that the integral over the total 
derivative with respect to any loop momentum $k_l$ vanishes in 
dimensional regularization
\begin{align}
0 = \int \text{d}^D k_l \frac{\partial}{\partial k_l^\mu} f(k_l,\dots) \text{ ,}
\end{align}
where the integrand $f$ may contain any combination of propagators, 
scalar products and loop momentum vectors. 
Together with the exploitation of Lorentz invariance by
\begin{align}
(p_1^\nu \frac{\partial}{\partial p_{1\mu}} -  p_1^\mu \frac{\partial}{\partial p_{1\nu}} 
+\dots + p_n^\nu \frac{\partial}{\partial p_{1\mu}} -  p_n^\mu \frac{\partial}{\partial p_{1\nu}} ) 
\, G(\{p\},\{m\}) = 0 \text{ ,}
\end{align}
where $n$ is the number of external momenta $p_i$
and $m$ the internal masses, the reduction to master 
integrals can be achieved. The full reduction into 
less complicated integrals can be done if the number of constraints 
matches or exceeds the number of unknown integrals. 

\medskip 

In the following, the general structure of the resulting master 
integrals is shown and different methods to solve them 
are described. 
\section{Two and more loop integrals with multiple scales}
\label{sec:generalintegraldefinition}
%
%
The difficulty of such multi-loop integrals has led to their extensive study 
and the development of various specialized integration techniques. 
In the following, a general multi-loop integral is introduced before 
presenting a variety of techniques to tackle these. 

\medskip

A general Feynman loop integral $G$ at $L$ loops with $N$ propagators, where 
the propagators $P_j$ can have in principle arbitrary powers $\nu_j$ and mass $m_j$,  
has the following representation in momentum space
\begin{align}
G_{l_1 \dots l_R}^{\mu_1 \dots \mu_R} (\{p\},\{m\}) =
& \prod\limits_{l=1}^{L} \int \rd^D\kappa_l\; 
\frac{k_{l_1}^{\mu_1} \cdots k_{l_R}^{\mu_R}} {\prod\limits_{j=1}^{N} P_{j}^{\nu_j}(\{k\},\{p\},m_j^2)} \label{eq:genfeynintegral} \\
\rd^D\kappa_l=&\frac{\mu_r^{4-D}}{i\pi^{\frac{D}{2}}}\,\rd^D k_l\;,\; P_j(\{k\},\{p\},m_j^2)=q_j^2-m_j^2+i\delta\;,
\label{eq:propagatordefinition}
\end{align}
where the $q_j$ are linear combinations of external momenta $p_i$ 
and loop momenta $k_l$. 
While the rank $R$ of the integral is indicated by the number 
of loop momenta appearing in the 
numerator, the 
indices $l_i$ denote which of the $L$ loop momenta belongs to 
which Lorentz index $\mu_i$. 
The factor of  $i\pi^{\frac{D}{2}}$ in $\kappa$ is chosen by convention to remove any 
dependence on $\pi$ after the integration over the loop momenta within Feynman 
parameterization. The renormalization scale is denoted by $\mu_r$. 
The $+i\delta$ in Eq.~(\ref{eq:propagatordefinition}) results from the solutions 
of the field equations in terms of causal Green functions.

\medskip

The integral is regulated dimensionally meaning that the integer dimension 
number $d$ is shifted by an 
infinitesimal quantity $\eps$ to the new dimension variable  $D= d - 2 \,\eps$. Infrared or 
ultraviolet poles then appear as poles in the regulator $\eps$. Careful 
distinction has to be made between ultraviolet ($\eps > 0$) and infrared 
regulators ($\eps < 0$). 
Formally, the theory is first renormalized dimensionally by going a little 
below the integer number $d$ of space-time dimensions and afterwards analytically 
continued to a little value above $d$ to regulate the mass singularities, see 
Ref.~\cite{Bardin:1999ak} for a comprehensive discussion. 
A modified version of the dimensional regularization is the dimensional 
reduction (DR). It is of particular 
interest in SUSY calculations, since 
it preserves global gauge invariance and supersymmetry, also at 
the two loop level~\cite{Hollik:2005nn}. Adopting 
this regularization scheme first introduced and applied 
in Refs.~\cite{Siegel:1979wq,Capper:1979ns}, 
all external particles and all gamma matrices $\{\gamma^\mu,\gamma^5\}$ 
appearing in the couplings are treated as quasi four-dimensional, 
while the loop integrals are computed
in $D=4-2\,\eps$ dimensions. 
For comparison, the 
introduction of a cutoff $\Lambda$ to the propagators to regulate UV 
divergences corresponds to the 
Pauli-Villars regularization. 

\medskip

The complexity of the evaluation of loop integrals generally increases with the 
number of loops and the number of legs. Massive internal lines 
further increase the level of 
complexity by raising the number of involved scales. 
All masses and invariants formed from external momenta are summarized in 
the term ``kinematic invariants''. The sum of independent kinematic variables 
corresponds to the number of scales involved in a diagram. 
Already at one-loop many-scale integrals are hard to compute. 
Multi-loop integrals involving multiple scales are particularly demanding. 
Further complexity arises with the non-planarity of graphs. Such 
diagrams only appear beyond one loop. 
These additional complications make the evaluation of multi-loop 
integrals an extremely non-trivial task and shrewd and refined 
techniques need to be employed to tackle these. Analytical techniques 
are very advanced, but when it comes to automation they very often 
still reach their limit. 
Numerical methods are in general easier to automate but issues here 
are the speed and the accuracy. 

\medskip 

In the following, the main technical approaches towards 
the evaluation of multi-loop and multi-scale integrals are 
reviewed. 
The main recent developments are sketched before 
concluding with a motivation for the method chosen to 
be investigated in this thesis. 
\section{Introduction of Feynman parametrization}
\label{sec:feynparametrization}
%
%
A first method to deal with Feynman loop integrals is to 
introduce Feynman parameters to every propagator. 
To prove that this technique indeed works, it is descriptive 
to have a look at a two-propagator example \cite{Peskin:1995ev}
\begin{align}
 \frac{1}{AB} = \int_0^\infty \text{d} x_1 \text{d} x_2 \;\delta(1-x_1-x_2)\frac{1}{(x_1 A + x_2 B)^2}  \text{ ,}
\label{eq:feynmantrick}
\end{align}
with the propagators $A$ and $B$ and the Feynman parameters $x$ and $y$.
Generalizing this idea to multiple propagators is straightforward  \cite{Heinrich:2008si}
\begin{align}
\frac{1}{\prod\limits_{j=1}^{N} P_j^{\nu_j}} = 
\frac{\Gamma(N_\nu) }{\prod\limits_{j=1}^{N} \Gamma(\nu_j)} 
 \,\prod\limits_{j=1}^{N}\,\int_0^\infty \rd x_j\,\,x_j^{\nu_j-1} \delta\big(1-\sum_{i=1}^N x_i\big) \left[ \sum_{j=1}^N x_j P_j \right]^{-N_\nu} \text{ ,}
\end{align}
where $N_\nu=\sum_{j=1}^N \nu_j$.
An equivalent representation was derived by Schwinger.  
Parameters are usually referred to as Schwinger variables when 
they are assumed to have values between zero and infinity, while 
values between zero and one are associated with Feynman 
parameterization. 
The general form of a multi-loop integral reads
\begin{align}
\non G_{l_1 \dots l_R}^{\mu_1 \dots \mu_R}=
& \frac{\Gamma(N_\nu)}{\prod_{j=1}^{N}\Gamma(\nu_j)}
\int_0^\infty \,\prod\limits_{j=1}^{N}\rd x_j\,\,x_j^{\nu_j-1}\, 
\delta\big(1-\sum_{i=1}^N x_i\big)\int \rd^D\kappa_1\ldots\rd^D\kappa_L\\
&\times k_{l_1}^{\mu_1} \cdots k_{l_R}^{\mu_R} \left[ 
       \sum\limits_{i,j=1}^{L} k_i^{\rm{T}}\, M_{ij}\, k_j  - 
       2\sum\limits_{j=1}^{L} k_j^{\rm{T}}\cdot Q_j +J +\idel
                             \right]^{-N_\nu} \text{ ,}
\end{align}
where the propagators are written in terms  
bilinear in the loop momenta with coefficients contained 
in the matrix $M$, terms linear in the loop 
momenta with coefficients $Q_j$ and remaining terms 
included in $J$ depending on the masses, external 
momenta and the Feynman parameters only. 
To be able to integrate out the loop momenta, the terms 
bilinear and linear in the loop momenta need to be 
brought into a quadratic form by a shift 
\begin{align}
k_l = \tilde{k}_l + v_l \text{ ,} \quad v_l = \sum_{i=1}^L M_{li}^{-1} Q_i \text{ .}
\end{align}
The integration of the quadratic form is then straightforward. 
After also integrating out the radial coordinates the general 
loop integral reads
\begin{align}                           
\non G_{l_1 \dots l_R}^{\mu_1 \dots \mu_R} =&
\frac{(-1)^{N_{\nu}}}{\prod_{j=1}^{N}\Gamma(\nu_j)}
\,\prod\limits_{j=1}^{N}\,  \int \limits_{0}^{\infty}  
dx_j\,\,x_j^{\nu_j-1}\,\delta(1-\sum_{l=1}^N x_l) 
\\ \non & \times
\sum_{m=0}^{\lfloor R/2\rfloor} \left( -\frac{1}{2} \right)^m 
\Gamma(N_{\nu}-m - LD/2) \left[ (\tilde{M}^{-1} \otimes g)^{(m)} \tilde{l}^{(R-2 m)} \right]^{\Gamma_1,\dots,\Gamma_R}
\\ & \times
\frac{{\cal U}^{N_{\nu}-(L+1) D/2 - R}}
{{\cal F}^{N_\nu-L D/2 - m}} \text{ ,}
\label{eq:thefeynmanloopintegral}
\end{align}
where 
\begin{align}
{\cal F}(\vec x) =& 
\det (M) \left[ \sum\limits_{j,l=1}^{L} Q_j \, M^{-1}_{jl}\, Q_l
-J -\idel\right] \label{eq:defF}\\
{\cal U}(\vec x) =& \det (M), \quad \tilde{M}^{-1}= {\cal U} M^{-1} \text{ ,} \quad \tilde{l}=  {\cal U} v\text{ .}
\label{eq:defU}
\end{align}
Note the sign of the infinitesimal imaginary part in Eq.~(\ref{eq:defF}). It 
results from factoring an overall minus sign into the prefactor during Wick 
rotation. 
Each loop momentum $k_{l_i}$ in the numerator is associated with a fixed 
Lorentz index $\mu_i$. The $\Gamma_i$ refer to the combination of both 
indices $ \Gamma_i = (l_i, \mu_i)$. For rank $R=0$, the second line of 
Eq.~(\ref{eq:thefeynmanloopintegral}) reduces to the factor $\Gamma(N_{\nu} - LD/2)$, 
containing overall UV divergences if present. In the case of $R=1$, 
the product $(\tilde{M}^{-1} \otimes g)^{(0)}$ does not contribute and 
$\tilde{l}^{(1)}= \tilde{l}_{l_1}^{\mu_1} $. For loop integrals of higher 
rank $R>1$, products of the matrix element $\tilde{M}^{-1}_{ij} $ with the metric tensor $g^{\mu\nu}$ 
contribute as a sum of all possible combinations with the vectors $\tilde{l}$ in 
the double indices $\Gamma_i$. As an example of how to read the notation involving the 
metric tensors, for rank $R=2$ and correspondingly $m=1$ it is 
\begin{align}
(\tilde{M}^{-1} \otimes g)^{(1)} =  (\tilde{M}^{-1} \otimes g)^{\mu_1\mu_2}_{l_1 l_2}   = 
\tilde{M}^{-1}_{l_1 l_2} g^{\mu_1\mu_2} \text{ ,}  \quad l_1,l_2 \in \{1,2\} \text{ .}
\end{align}
The functions ${\cal U}$ and ${\cal F}$ in Eq.~(\ref{eq:thefeynmanloopintegral}) are the 
first and second Symanzik polynomial, respectively, 
and are homogeneous (in the Feynman parameters). 
${\cal U}$ is positive semi-definite and 
${\cal F}$ is negative semi-definite when all propagators 
are massless. 
\begin{figure}[ht!]
	\begin{center}
\includegraphics[width=0.4\textwidth]{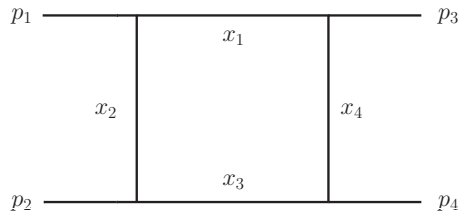} 	      
	\end{center}
\caption{The one-loop box diagram with massless propagators.}
\label{fig:1loopallmasslessbox} 
\end{figure}
The two functions can also be obtained using a 
graph-theoretical method, where the polynomials 
are constructed from topological cuts of the corresponding 
Feynman graph. For the construction of the function ${\cal U}$, 
$L$ lines of the graph are cut, whereas $L+1$ lines are cut to 
arrive at the function $\mathcal{F}$, see 
Refs.~\cite{Tarasov:1996br,Smirnov:2006ry,Heinrich:2008si}. 
\pagebreak
To illustrate this for a simple example, assume a massless one-loop box with all 
external legs being light-like, compare Fig.~\ref{fig:1loopallmasslessbox}. 
Then, the first Symanzik polynomial ${\cal U}$ is constructed from adding up all 
possible $L$ line cuts of propagators which lead to a tree-level diagram. 
Each cut propagator contributes with its 
Feynman parameter, if more than one propagator needs to 
be cut (which happens for $L >1$), 
products of Feynman parameters, all of the same degree, enter the function ${\cal U}$.
In the case of the one-loop box with massless propagators, this reads\\
\begin{figure}[ht!]
\hspace{20pt} 
\raisebox{-19pt} {%
\includegraphics[width=0.2\textwidth]{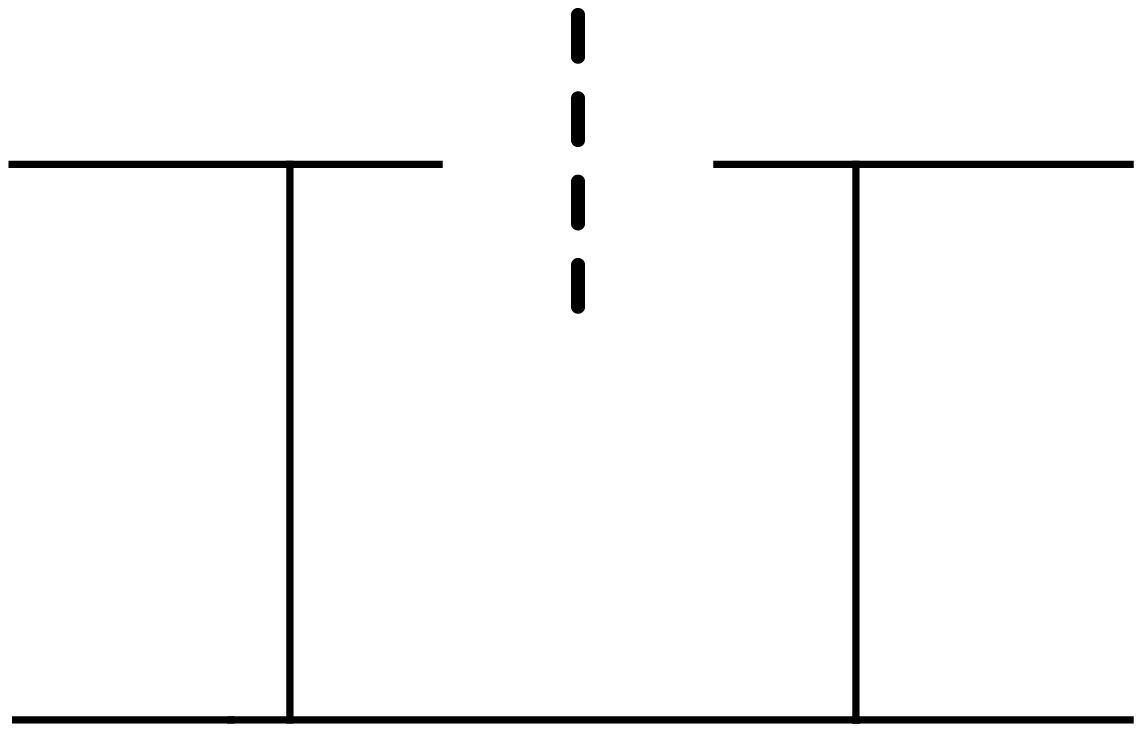} 
} $+$     
\raisebox{-19pt} {%
\includegraphics[width=0.2\textwidth]{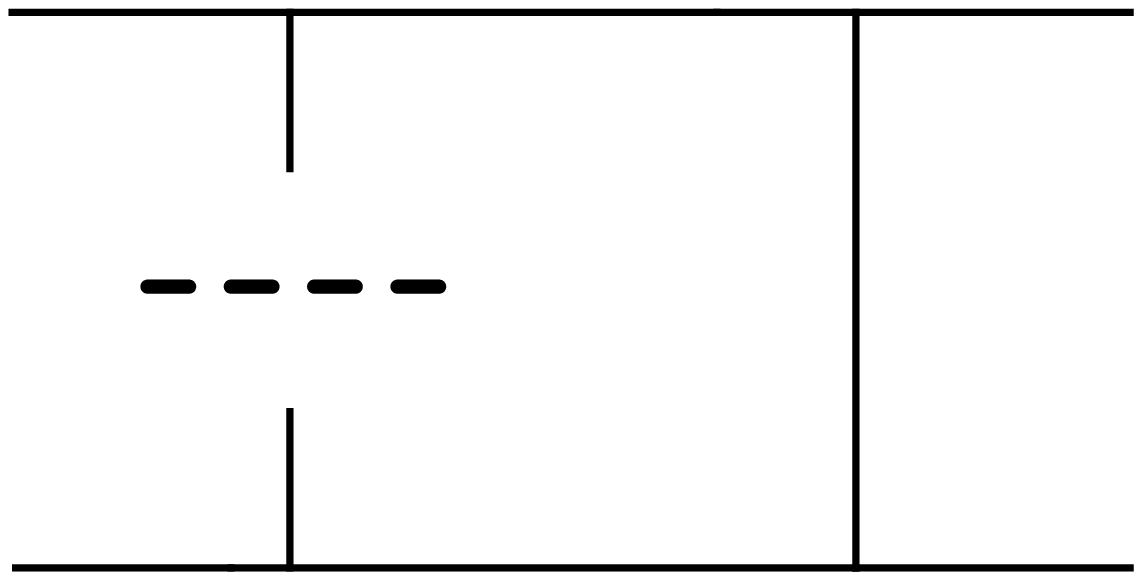} 	
}  $+$	      
\raisebox{-30pt} {%
\includegraphics[width=0.2\textwidth, angle=180,origin=c]{twoloopfrontier/twoloopfrontierpics/1l_boxmasslesscut1}
} $+$     
\raisebox{-19pt} {%
\includegraphics[width=0.2\textwidth,angle=180,origin=c]{twoloopfrontier/twoloopfrontierpics/1l_boxmasslesscut2}
} \vspace{8pt}\\
${\cal U}= \hspace{39pt} x_1  \hspace{39pt} + \hspace{39pt}  x_2  \hspace{39pt} + 
 \hspace{39pt} x_3  \hspace{39pt} + \hspace{39pt}  x_4$
\label{fig:constructU} 
\end{figure}
\\The second Symanzik polynomial ${\cal F}$ is constructed from adding up all 
possible $L+1$ line cuts of propagators. All $L+1$ cut propagators contribute with their 
Feynman parameter and the squared momentum flowing into the resulting tree. 
In the case of the one-loop box with massless propagators, the 
function ${\cal F}$ then reads \\
\begin{figure}[ht!]
\raisebox{-19pt} {%
\includegraphics[width=0.13\textwidth]{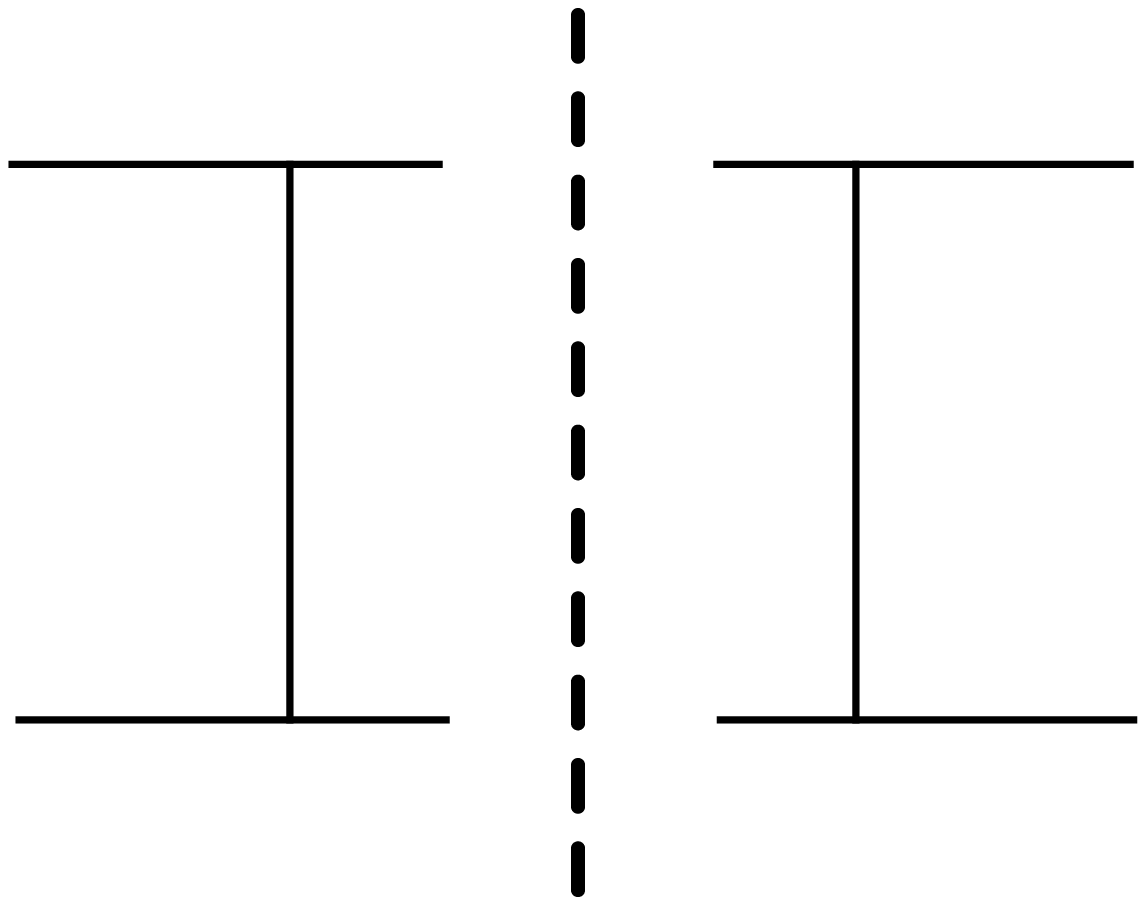} 
} $+$     
\raisebox{-11pt} {%
\includegraphics[width=0.13\textwidth]{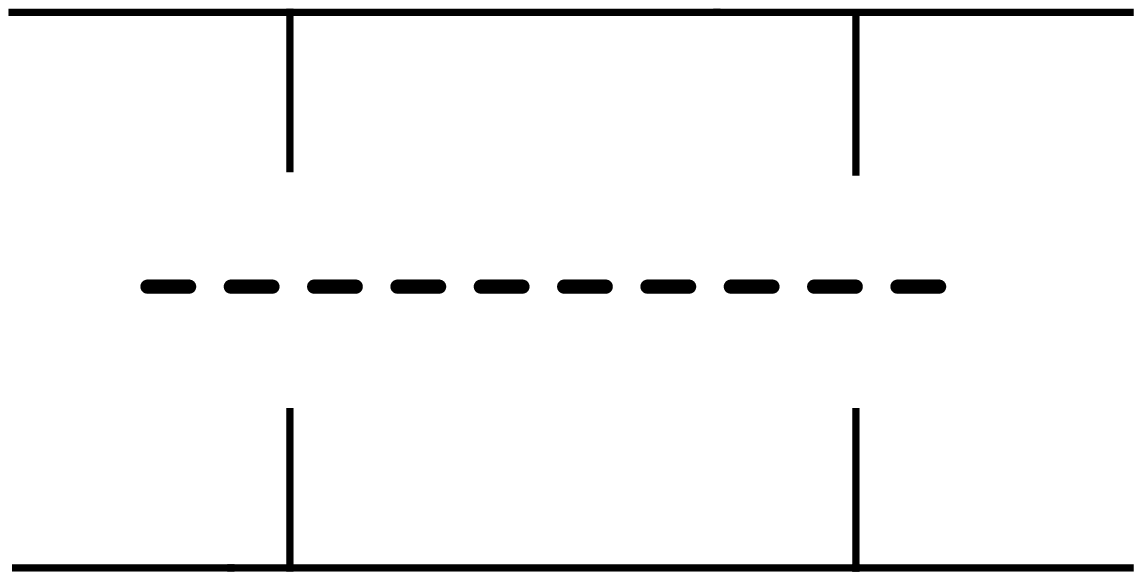} 	
}  $+$	      
\raisebox{-11pt} {%
\includegraphics[width=0.13\textwidth]{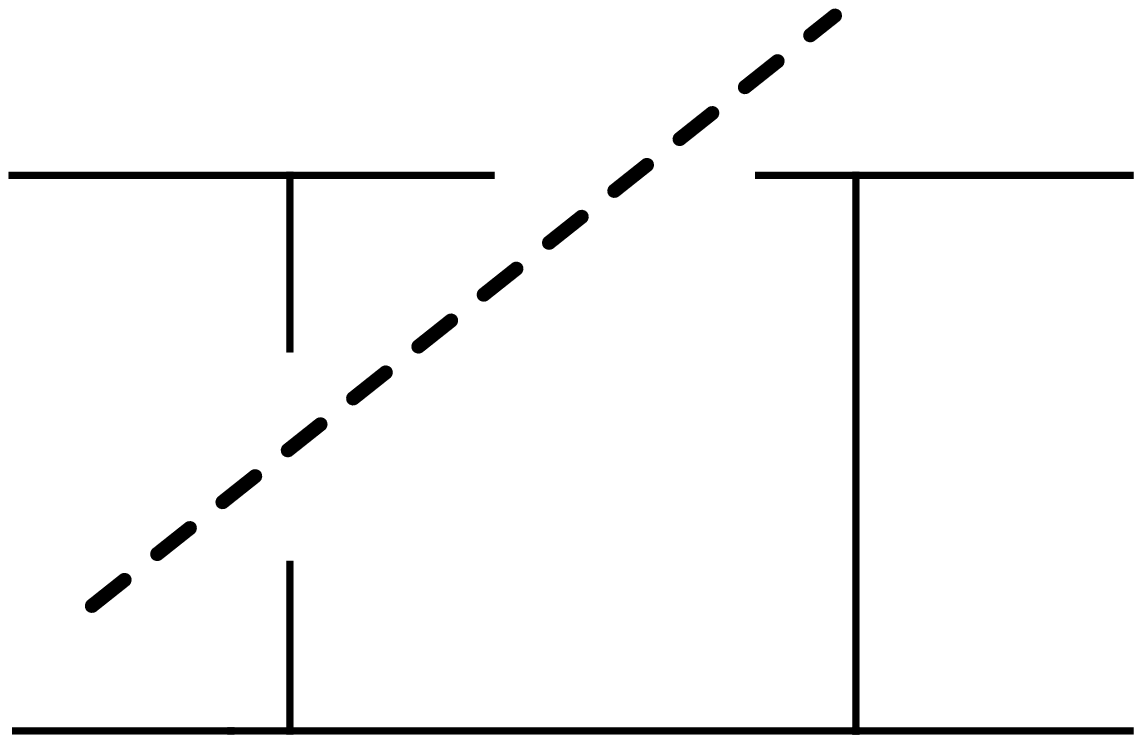}
} $+$ 
\raisebox{-19pt} {%
\reflectbox{%
\includegraphics[width=0.13\textwidth,angle=180,origin=c]{twoloopfrontier/twoloopfrontierpics/1l_boxmasslesscut5}
}}  $+$ 
\raisebox{-19pt} {%
\includegraphics[width=0.13\textwidth,angle=180,origin=c]{twoloopfrontier/twoloopfrontierpics/1l_boxmasslesscut5}
} $+$ 
\raisebox{-11pt} {%
\reflectbox{%
\includegraphics[width=0.13\textwidth]{twoloopfrontier/twoloopfrontierpics/1l_boxmasslesscut5}
}} \vspace{8pt}\\
${\cal F}= 
\hspace{0pt} - s_{12} \; x_1x_3  \hspace{0pt} 
- \hspace{0pt} s_{23}\; x_2x_4   \hspace{5pt} 
- \hspace{10pt} p_1^2\; x_1x_2   \hspace{15pt} 
- \hspace{13pt}  p_2^2\; x_2x_3  \hspace{10pt} 
- \hspace{10pt} p_3^2\; x_3x_4   \hspace{15pt} 
- \hspace{15pt} p_4^2\; x_4x_1  $ \vspace{8pt} .
\label{fig:constructF} 
\end{figure}

\medskip

While the prefactor $\Gamma(N_{\nu}-m-LD/2)$ in 
Eq.~(\ref{eq:thefeynmanloopintegral}) contains  
overall ultraviolet divergences if present, the vanishing of the function 
${\cal U}$ is related to ultraviolet sub-divergences of the graph. 
The second Symanzik polynomial contains the 
occurring infrared singularities. 
The occurrence of these depends not only 
on the topology as in the UV case, but 
on the kinematics as well. If some of the kinematic 
invariants are zero, e.g. when some external momenta are 
light-like, the vanishing of $\mathcal{F}$
may induce an infrared (IR) divergence. Therefore
general theorems about the IR singularity structure 
of multi-loop integrals are sparse. 
For practical purposes sector decomposition can provide information about the 
singularity structure and numerical results, because it offers 
a constructive algorithm to extract the poles in $1/\eps$.
When generalizing the kinematic invariants to physical space, the
second Symanzik polynomial can also vanish when linear combinations 
of Feynman parameters and kinematic invariants vanish. A clever 
deformation of the integration 
contour to the complex plane helps dealing with these physical 
poles and the integration over thresholds.
~\\
For a diagram with only massless propagators, the 
function ${\cal F}$ does not contain any 
squares in the Feynman parameters. 
If massive internal lines are present, 
terms of the type 
\begin{align}
{\cal F}(\vec x) \propto {\cal U}(\vec x) \sum\limits_{j=1}^{N} x_j m_j^2
\label{eq:squaresinF}
\end{align}
appear. These are the source of 
complexity when it comes to the 
analytical evaluation of multi-scale integrals, as many 
methods used for the simplifcation of an integrand 
can no longer be applied. \\
This opens the field for a numerical treatment 
of multi-scale integrals, where squared 
Feynman parameters are not a bottleneck to 
the calculation. Additionally, the introduction 
of Feynman parameters is highly automatable 
paving the way towards a very general solution to 
a large class of multi-scale integrals with arbitrary 
kinematics. 
~\\
The problems occurring with this method and their 
solution will be explained in the next two chapters. 
Beforehand, several other approaches towards 
solving multi-loop integrals will be reviewed. 
%
%
\section{The virtues of a Mellin-Barnes representation}
In analogy to Eq.~(\ref{eq:feynmantrick}), the main transformation 
to arrive at a Mellin-Barnes representation can be summarized 
in one line
\begin{align}
 \frac{1}{(A+B)^\lambda} = \frac{1}{\Gamma(\lambda)} \frac{1}{2 \pi i}
\int_{-i\infty}^{+i\infty} \text{d} z\, \Gamma(\lambda + z)\, \Gamma(- z) 
\frac{B^z}{A^{\lambda + z}} \text{ ,}
\label{eq:mellinbarnestrick}
\end{align}
with the difference, that now a sum in the denominator is 
transformed into a product. 
The sum on the left-hand side can either be a massive 
propagator or a sum of two propagators after Feynman 
parametrization. 
In the first case, massive propagators are 
expressible as a continuous 
superposition of massless propagators. Considering that the 
massive propagators 
introduce squares in the Feynman parameters, see 
Sec.~\ref{sec:feynparametrization}, this transformation 
can be very beneficial. 

In general, a factorization of the type Eq.~(\ref{eq:mellinbarnestrick}) 
can be used to achieve a representation 
of loop integrals in terms of gamma functions, which are in 
general easier to evaluate. This benefit comes at the 
cost of extra Mellin integrations. Within their integration 
domain, poles in the variable $z$ can occur. 
%
%
Taking these into account, the integration contour 
must always be chosen such that 
the poles with a $\Gamma(\dots +z)$ dependence 
are placed left of the contour and the poles with a 
$\Gamma(\dots - z)$ dependence are situated on the 
right-hand side with respect to the contour. 
Closing the contour to the right and taking a series 
of residues, the integral can be evaluated. Yet, finding 
the appropriate contour is non-trivial. 

With the computation of the planar~\cite{Smirnov:1999gc} and 
non-planar~\cite{Tausk:1999vh} massless two-loop 
four-point diagram, Smirnov and Tausk pioneered the 
utilization of a Mellin-Barnes representation finding an 
appropriate choice of contours for physical kinematics 
including thresholds. 
Several software packages became available automating the 
analytical procedure where possible, see Refs.~\cite{Czakon:2005rk,
Smirnov:2009up,Gluza:2010rn}. The more scales 
are involved, the less easy it is to arrive at a fully analytical result. 
Numerical approaches have also been considered~\cite{Anastasiou:2005cb,
Freitas:2010nx,Smirnov:2013eza}, putting much effort in the  
automation of a proper analytical continuation of the integrand.  
Very recently an idea by Pilipp~\cite{Pilipp:2008ef}
was implemented with in combination with Feynman parametrization 
to treat such contours in an automated 
way, see Ref.~\cite{Smirnov:2013eza}. 
It works as follows: The second Symanzik polynomial ${\cal F}$ 
is decomposed as
\begin{align}
 \mathcal{F}(x_1,\dots ,x_n)= \rho\; F_1(x_i) + F_2(x_i) \text{ ,}
\end{align}
where a small coefficient to terms in $\cal{F}$ is extracted into the parameter $\rho$ 
and where all terms contained in $F_1$ and $F_2$ are sufficiently large 
as not to contribute to a singularity.
The Mellin-Barnes representation is then 
introduced for the product
\begin{align}
\frac{\Gamma(N_{\nu}-LD/2)}{ (\rho F_1 + F_2)^{N_{\nu}-LD/2} } =
 \frac{1}{2 \pi i} &\int_{-i \infty}^{i \infty} \text{d}z \;\rho^z 
 \frac{\Gamma(N_{\nu}-LD/2 + z) \Gamma(-z)}{F_1^{-z} F_2^{N_{\nu}-LD/2+z}} \text{ ,}
\end{align}
After the application of sector decomposition which will be described 
in detail in Chap.~\ref{chap:sectordecompo}, the functions $F_i$ and $\cal{U}$ are
constant in the Feynman parameters $x_i$, so that the singularity 
structure is revealed in the exponents of the factorized $x_i$ 
\begin{align}
x_i^{n_i -1 + b_i \eps + c_i z} \text{ ,}
\end{align}
where the $-1$ enters with the Feynman parametrization, and where
$n_i$, $b_i$ and $c_i$ are some coefficients resulting from the 
division into sectors. Allowing only those integrals where $c_i<0$, 
the integration contour of the integral over the variable $z$ 
can be closed to the right, allowing 
for the application of Cauchy's integral theorem. The residues 
arising from terms of the type $\Gamma(n_i + b_i \eps + c_i z)$ 
after $x_i$ integration need to be taken into account.
Afterwards, an expansion in the parameter $\rho$ can be performed. 

The usage of a Mellin-Barnes representation can 
be very beneficial in diverse contexts and can even be applied in 
an automated way to the physical region with the computation of 
asymptotic expansions in the Feynman integrals. Yet, a fully 
automated approach in all regions of phase space is difficult.  
%
%
%
\section{The method of differential equations}
As it turns out, a representation for the master integrals resulting from the reduction can 
also be found by setting up differential equations in all kinematic 
invariants and solving them with the appropriate boundary conditions. 
The method was first introduced by Kotikov who related massive loop 
integrals to massless ones, see Ref.~\cite{Kotikov:1990kg} and then generalized to 
differential equations in all kinematic invariants, see Refs.~\cite{Remiddi:1997ny,Caffo:1998yd}. \\
Taking the derivative of an integral with respect to one of its invariants yields 
linear combinations of integrals with at most one additional propagator in 
the denominator and 
one additional scalar product in the numerator. 
The derivatives of the invariants $s_{ij}=(p_i+p_j)^2$ 
can be expressed in terms of derivatives in the external momenta, e.g. for 
box diagrams 
\begin{align}
s_{ij}\frac{\partial}{\partial s_{ij}} = 
\frac12 \left(    
p_i^\mu \frac{\partial}{\partial p_i^\mu} + 
p_j^\mu \frac{\partial}{\partial p_j^\mu} -
p_k^\mu \frac{\partial}{\partial p_k^\mu} 
\right) \text{ , } i \ne j \ne k  \text{ , } i,j,k=1,2,3 \text{ .}
\end{align}
This generates similar expressions as those resulting from IBP relations mentioned 
in Sec. ~\ref{sec:twoloopfrontier} and Lorenz invariance identities (LI) introduced 
by Gehrmann and Remiddi, see Ref.~\cite{Gehrmann:1999as}. 
Using the IBP and LI relations, 
the integrals which received an additional propagator or scalar product can 
be reduced again to such an extent as to result in a system of 
differential equations. In the example of box diagrams, one of the equations reads 
\begin{align}
\non s_{ij} \frac{\partial}{\partial s_{ij}} I_{t,t,0}(s_{ij},s_{jk},s_{ki},D) =& 
A(s_{ij},s_{jk},s_{ki},D)\; I_{t,t,0}(s_{ij},s_{jk},s_{ki},D) \\
&+ F(s_{ij},s_{jk},s_{ki},D, I_{t-1,N_\nu,R}(s_{ij},s_{jk},s_{ki},D))
\end{align}
where $ I_{t,N_\nu,R}$ is an integral of a diagram with four external legs of $t$ 
different propagators and $R$ scalar products and where the function $A$ is 
rational. The function $F$ plays the role of an inhomogeneous term and 
the integral contained in it,  $I_{t-1,N_\nu,R}$, is one differing propagator 
short compared to the $I_{t,N_\nu,R}$ integral.
The boundary conditions can be derived from kinematical limits, e.g. a 
vanishing invariant $s_{ij}$, 
\begin{align}
\non I_{t,t,0}(0,s_{jk},s_{ki},D) =& - A(0,s_{jk},s_{ki},D)^{-1} \\
&\times F(0,s_{jk},s_{ki},D, I_{t-1,N_\nu,R}(0,s_{jk},s_{ki},D)) \text{ ,}
\end{align}
where $A(0,s_{jk},s_{ki},D) \ne 0$. 
Afterwards, the differential equations can be solved by 
introducing an integrating factor $M$ of the type
\begin{align}
M(s_{ij}) = e^{\int  \text{d} s_{ij} \;A(s_{ij},s_{jk},s_{ki},D)} \text{ ,}
\end{align}
yielding solutions to the inhomogeneous equation 
\begin{align}
\non I_{t,t,0}(s_{ij},&s_{jk},s_{ki},D) = \frac{1}{M(s_{ij}) } \\
&\times \left( \int  \text{d} s_{ij}
F(s_{ij},s_{jk},s_{ki},D, I_{t-1,N_\nu,R}(s_{ij},s_{jk},s_{ki},D)) \;M(s_{ij}) + C
\right) \text{ ,}
\end{align}
where the integral over the function $F$ and $M$ is either 
known or relatively easy to integrate and where the constant $C$ is 
chosen such that it matches the boundary conditions. \\
The nice feature of this technique is that it can be applied to arbitrary 
multi-loop integrals with arbitrary scales. However, the current bottleneck is 
related to the appearance of elliptic integrals. 
\begin{figure}[ht]
	\begin{center}
\includegraphics[width=0.45\textwidth]{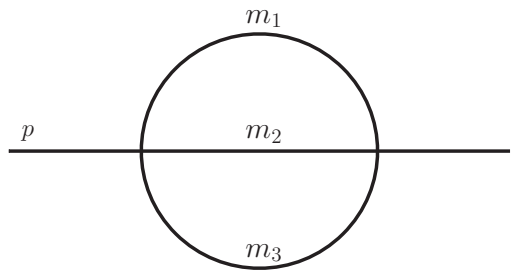} 
	\end{center}
\caption{Two-loop two-point massive bubble diagram, also termed the 
	"sunrise" topology.}
\label{fig:sunrisediag}
\end{figure}
These already appear in the rather simple but all-massive two-loop 
bubble with different masses, see Fig.~\ref{fig:sunrisediag}. After the developments 
summarized in this thesis, such an integral is easily treated numerically for in principle 
arbitrary kinematics and in a fully automated way.
%
%
\section{Further analytic developments}
%
%
%
Apart from the multi-purpose techniques 
already mentioned in the previous sections, there are many more 
specialized tricks to attack a special class of multi-loop integrals 
on the one hand, and other ideas based on long-known mathematical 
concepts to simplify the result on the other hand. 
Furthermore, intensive exploration of diverse mathematical concepts uncovered 
new criteria and underlying structures to easier access scattering 
amplitudes. 

\medskip

A presentation of analytic results in terms of generalized 
hypergeometric functions has been found to yield very compact results. 
The evaluation of special cases of hypergeometric functions, namely 
generalized Lauricella functions 
involving elliptic integrals are not accessible by present analytical 
techniques. In contrast, for all results expressible 
in terms of generalized harmonic polylogarithms (GHPLs) 
a fast, accurate and stable numerical evaluation of the analytical 
expressions can be found. 
GHPLs are generalizations of harmonic polylogarithms~\cite{Remiddi:1999ew}, 
introduced in Refs.~\cite{Goncharov:1998kja,Bonciani:2003cj} and 
applied in innumerable phenomenological applications. 
They are not all independent and relations among them can 
become very complicated. A systematic approach to govern the 
complexity of such relations is therefore highly desirable in the study 
of multi-loop integrals. One such approach is the formulation 
of results in terms of symbols. The concept, introduced by Zagier and 
Gonachrov in Refs.~\cite{0728.11062,0863.19004,2009arXiv0908.2238G}, 
allows for particularly simple and elegant expressions. 
After the symbol calculus was applied in the context of ${\cal N} = 4$ 
supersymmetric Yang-Mills (SYM) theory, see 
Refs.~\cite{Goncharov:2010jf,Alday:2010jz,Gaiotto:2011dt,DelDuca:2011ne,
Dixon:2011ng,DelDuca:2011jm,DelDuca:2011wh,CaronHuot:2011ky,
Dixon:2011pw,Heslop:2011hv,Brandhuber:2012vm}, 
it was found to be also applicable to diverse phenomenological problems, see 
Refs.~\cite{Buehler:2011ev,Duhr:2011zq,Duhr:2012fh,Gehrmann:2013vga,
Anastasiou:2013srw,vonManteuffel:2013uoa,Gehrmann:2013cxs,Bonciani:2013ywa,
Abreu:2014cla,Gehrmann:2014bfa}. 
The coproduct, as a generalization of the symbol, allowed for the 
conservation of information on constants with an associated weight, 
as was pointed out in Refs.~\cite{Goncharov:2002math,Brown:1102.1310B,
Duhr:2012fh}. 

As already mentioned in Sec.~\ref{sec:twoloopfrontier}, beyond one-loop 
the basis of master integrals is not fixed. Finding criteria for an optimal basis 
was therefore a major breakthrough in the computation of multi-loop 
amplitudes. These were introduced by Henn, see Ref.~\cite{Henn:2013pwa}, 
and further explored and applied in Refs.~\cite{Henn:2013tua,Henn:2013woa,
Henn:2013nsa,Argeri:2014qva,Henn:2014lfa,Caron-Huot:2014lda}.
They lead to a straightforward iterative solution of the differential equations in 
the dimensional regulator $\eps$. 

\medskip 

While the formulation of results in terms of much simpler representations 
is of vital importance in pushing the frontier towards the computation of 
higher loop integrals, the introduction of completely different approaches 
to the computation of the master integrals forms the second pillar in 
multi-loop computations. 
Conceptually, every mathematical object constituting a loop integral can 
be reformulated using a different approach which is more suitable in a 
specific calculation. 
Master integrals entering at higher order in perturbation theory can be 
approached from a graph theoretical point of view, 
an algebraic or a geometric point of view, to just name a few of the many 
areas of interplay between physics and mathematics that lead to 
attractive solutions to yet unsolved integral representations. 
The liberation from calculations in strictly 4 dimensions, for example, lead to plural 
ingenious approaches. The old concept of an infinitesimal $\eps$ shift used 
in dimensional regularization is predominantly used in the 
computation of loop integrals. It is also an old concept to shift the dimension 
by positive or negative integer numbers, see Refs.~\cite{Bern:1993kr,
Tarasov:1996br,Tarasov:1996bz,Campbell:1996zw,
Tarasov:1998nx,Fleischer:1999hq}, but one may 
also benefit from an integral representation adopting a negative dimension. 
The latter was developed in 
Refs.~\cite{Halliday:1987an,Dunne:1987am,Dunne:1987qb,Ricotta:1989ia} 
and successfully applied in the computation of a 
massless two-loop five-propagator diagram, where the expression of 
a sub-loop is derived using negative dimensions, see 
Ref.~\cite{Broadhurst:1987tv}. An abstraction to scalar one-loop 
vertex functions including internal masses, off-shell legs and 
arbitrary propagator powers was achieved for general dimensions, 
see Ref.~\cite{Anastasiou:1999ui}. \\
Another concept centers around the analysis of discontinuities across 
the branch cuts of Feynman integrals. 
In the traditional approach, the integral might be reconstructed 
directly from one of its discontinuities using a dispersion relation, see 
Refs.~\cite{Landau:1959fi,Cutkosky:1960sp,'tHooft:1973pz,Remiddi:1981hn}. 
This technique can be generalized to the application of sequential 
unitarity cuts in different channels, reconstructing one- and multi-loop 
integrals, see Ref.~\cite{Abreu:2014cla}. \\
Furthermore, the criterion of linear reducibility of a 
graph has been studied over the past few years and 
recently used in the computation of diverse examples, compare 
Refs.~\cite{Brown:2004svm,Brown:2009ta,
Brown:2009qja,Brown:2008um,Bogner:2013tia,
Panzer:2013cha,Panzer:2014gra,Panzer:2014caa}. 
The examples, all being linearly reducible in the Feynman 
parameters, can be integrated sequentially and analytical 
results can be given in terms of multiple polylogarithms. 
%
%
\section{Motivation for adopting a numerical approach}
%
%
One may realize that the quality of an employed technique to tackle 
multi-leg, -loop and -scale integrals on the one hand lies 
in its applicability to very generic cases of loop integrals, and on 
the other hand in the achieved accuracy within a given time span 
in addition to control over the parametric 
dependences. 
A fully automated elegant analytical approach to compute all 
possibly existing loop integrals would therefore be the perfect 
solution. Yet, analytical methods are still struggling with the 
appearance of elliptic integrals, entering already in rather simple 
two-loop diagrams, while numerical approaches need a better 
ratio of speed to accuracy to compete with the elegance of analytical 
results.  
The two main pillars therefore mutually enrich each other and 
methods including analytical and numerical approaches 
push the boundaries of what is computable with present 
techniques.

\medskip

While the achievements using analytical methods were analyzed in 
the previous sections, there are diverse groups who contributed highly 
non-trivial results 
to significant phenomenological applications 
taking up a numerical approach, compare e.g. Refs.~\cite{Passarino:2001wv,
Passarino:2001jd,Passarino:2006gv,Passarino:2007fp,Actis:2008ts,
Anastasiou:2007qb,Beerli:2008zz,Anastasiou:2008rm,Freitas:2012iu,
Czakon:2013goa,Boughezal:2011jf,Boughezal:2013uia}. 

\medskip

%
In the work summarized in this thesis, a highly 
automated numerical approach is adopted, filling the gap of the 
missing automated evaluation of multi-loop multi-scale integrals including 
thresholds. 
To this end, a representation of the integrals of interest in terms of plain 
Feynman parameters is used.
Due to its generality, the Feynman parametrization can serve 
as the most universal approach to a numerical treatment 
of integrals with arbitrary kinematics. 
Divergences are regulated dimensionally and are factorized 
using the method of sector decomposition. 
The program \secdec{} version 1 already implemented the 
automated formulation of integrals in terms of Feynman 
parameterization, integration of loop momenta and a series 
expansion in the dimensional regulator $\eps$, where the 
coefficients to each order in $\eps$ are integrated numerically. 
The upgrade of this program to be able 
to deal with mass thresholds within the integration region 
is one of the main achievements of the work presented in this 
thesis. The advancement is accomplished by an automated 
analytical continuation of the Feynman parametrized integrand, 
building on work presented in Refs.~\cite{Soper:1999xk,
Binoth:2005ff,Nagy:2006xy,Anastasiou:2007qb,Beerli:2008zz}. 
With the resulting version 2 of \secdec, valuable predictions and 
checks can be done, regardless of the number of scales 
involved. Contrary to analytical methods, it can even be 
beneficial to include more 
scales. While purely finite integrals with multiple scales are hard 
to access with analytical methods, it is comparatively easy 
using the numerical approach. 
Finally, it turns out that not only is the developed tool useful for checks 
against analytical results, but it has also proven powerful in 
computing analytically unaccessible integrals for phenomenological 
applications. 
%
%

%% file: sectordecompo/sectordecompo.tex
\chapter[The method of sector decomposition]{The method of \\sector decomposition}
\label{chap:sectordecompo}%
%
%
As described in the introduction, a theory can be ultra-violet and infrared divergent,  
The idea of renormalization is to subtract the divergent parts and thereby 
make the theory finite. Finding the right subtraction terms for the UV divergent parts was a 
long standing problem whose solution resulted in the BPHZ theorem. To this end 
Hepp used a decomposition of higher order loop diagrams into sectors to 
disentangle overlapping UV divergences~\cite{Hepp:1966eg}. 
Thirty years later, the idea was taken up by Denner and Roth for a 
disentanglement of UV divergences~\cite{Roth:1996pd}. 

\smallskip

The application to infrared singularities and a systematic 
treatment of these to arbitrary loop order using 
sector decomposition was pointed out by Binoth and 
Heinrich~\cite{Binoth:2000ps}. 
%
It serves as a local subtraction procedure to separate infrared 
divergences from individual diagrams. 
%
\section{Conceptual idea}
\label{sec:secdecconcept}
%
%
%
The algorithm to find the subtraction terms of individual graphs works in three main 
steps. In the first step, 
the singular components of an integral are disentangled by iteratively 
decomposing the integral into sectors. In a second step, the pole 
coefficients to each order in the poles of the dimensional regulator $\eps$ 
are extracted. 
In the last step, the coefficients containing kinematic invariants and 
Feynman parameters 
are integrated analytically or numerically if an analytical treatment is 
not accessible. 
\begin{figure}[htb]
	\begin{center}
	      \includegraphics[width=0.7\textwidth]{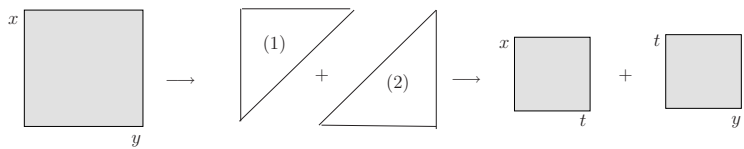}
	\end{center}
	\caption{The basic idea of sector decomposition.}
	\label{fig:basicsecdec} 
\end{figure}
\\The idea of sector decomposition is essential for 
the first step of the algorithm. 
It is based on splitting the integration region to achieve 
a disentanglement of the singularities. 
As a simple example, consider the following integral
\begin{subequations}
\begin{align}
\hspace{-20pt}\int_0^1 \!\!\!\!\text{d}x_1  \!\! &\int_0^1 \!\!\!\! \text{d}x_2 \,\, \frac{1}{(x_1+x_2)^{2+\eps}} \\
 &= \int_0^1 \!\!\!\!\text{d}x_1  \!\! \int_0^1 \!\!\!\! \text{d}x_2 \,\, \frac{1}{(x_1+x_2)^{2+\eps}} ( \Theta(x_1-x_2) + \Theta(x_2-x_1) ) \label{eq:showsecdecline2}\\ 
 &=\int_0^1 \!\!\!\! \text{d}x_1 \!\! \int_0^{x_1} \!\!\!\!\!\! \text{d}x_2 \,\,
 \frac{1}{(x_1+x_2)^{2+\eps}} \,\,+\int_0^1 \!\!\!\! \text{d}x_2 \!\! \int_0^{x_2} \!\!\!\! \text{d}x_1 \,\, \frac{1}
{(x_1+x_2)^{2+\eps}} \label{eq:showsecdecline3}\\
 &=\int_0^1 \!\!\!\! \text{d}x_1 \!\! \int_0^1 \!\!\!\! \text{d}\tilde{x}_2 \,\, \frac{x_1}{(x_1+ x_1 \tilde{x}_2)^{2+\eps}} 
 + \int_0^1 \!\!\!\! \text{d}x_2 \!\! \int_0^1 \!\!\!\! \text{d} \tilde{x}_1 \,\, \frac{x_2}{(x_2 \tilde{x}_1+x_2)^{2+\eps}} 
 \label{eq:showsecdecline4} \text{ ,} \\
  &=\int_0^1 \!\!\!\! \text{d}x_1 \!\! \int_0^1 \!\!\!\! \text{d}\tilde{x}_2 \,\, \frac{1}{x_	1^{1+\eps}(1+  \tilde{x}_2)^{2+\eps}} 
  + \int_0^1 \!\!\!\! \text{d}x_2 \!\! \int_0^1 \!\!\!\! \text{d} \tilde{x}_1 \,\, \frac{1}{x_2^{1+\eps}( \tilde{x}_1+1)^{2+\eps}} 
 \label{eq:showsecdecline5} \text{ ,}
\end{align}%
\label{eq:secdecexample}%
\end{subequations}%
where overlapping divergences appear in the limit of a vanishing of 
both Feynman parameters $x_1,x_2 \rightarrow 0$ in the first line. 
In the second line, see Eq.~(\ref{eq:showsecdecline2}), the integration 
region is split into one part where $x_1$ is always bigger than $x_2$ 
and a second part where the hierarchy is reversed. The splitting 
can be translated into a change of the integration boundaries, compare 
Eq.~(\ref{eq:showsecdecline3}). Using the transformation 
\begin{subequations}
\begin{align}
x_1 &\rightarrow x_1 \\
x_2 &\rightarrow x_1\tilde{x}_2
\end{align}%
\label{eq:blowuptrafo1}%
\end{subequations}%
in the first integral on the righthand side of Eq.~(\ref{eq:showsecdecline3}) and 
the transformation 
\begin{subequations}
\begin{align}
x_1 &\rightarrow x_2 \tilde{x}_1 \\
x_2 &\rightarrow x_2 
\end{align}%
\label{eq:blowuptrafo2}%
\end{subequations}%
in the second integral on the righthand side of Eq.~(\ref{eq:showsecdecline3}), 
both integrals are remapped onto 
the unit hypercube, compare Eq.~(\ref{eq:showsecdecline4}). 
The transformations in Eqs.~(\ref{eq:blowuptrafo1}) and (\ref{eq:blowuptrafo2}) 
are known in the mathematical literature as blowing-up 
an affine $N$-dimensional space, 
compare e.g. Ref.~\cite{Hartshorne:1977}. The number of variables 
participating in this blowing-up is two, therefore $N=2$ in this example. 
The blowing-up leads to two integrals with disentangled (non-overlapping) 
singularities, compare Eq.~(\ref{eq:showsecdecline5}). 

\medskip

In the following, all three steps of the algorithm are discussed in more detail.  
When treating Feynman loop integrals, a decomposition into primary 
sectors is beneficial and performed before the iterated decomposition into 
sub-sectors. 
The description of the algorithm is restricted to scalar multi-loop integrals 
for better readability, but 
the extension to multi-loop tensor integrals is straightforward. 
The discussion of the algorithm is based on Refs.~\cite{Binoth:2000ps,Heinrich:2008si}. 
%
%
\subsection{Generation of primary sectors}
\label{subsec:primarysecdec}
A general loop integral in Feynman parametrization, compare 
Eq.~(\ref{sec:feynparametrization}), contains a $\delta$-distribution, 
which can be formulated in various ways. To arrive at the 
simplest representation for a subsequent iterated sector 
decomposition, the representation of the $\delta$-distribution 
is chosen such that a definite hierarchy is introduced after 
integration, where one Feynman parameter $x_l$ out of $N$ 
is always larger than the rest. To this end, the 
$N$ dimensional unit 
hypercube is split into $N$ sectors. 
In each of these so called primary sectors, one Feynman 
parameter $x_l$ is chosen to be 
larger than all others
\begin{align}
\prod\limits_{j=1}^{N}\,\int_0^\infty  \rd x_j = 
\prod\limits_{j=1}^{N}\,\int_0^\infty  \rd x_j \sum_{l=1}^N \Theta(x_l \geq x_j \geq 0) \text{ ,}
\end{align}
where $\Theta$ is the Heaviside step function 
with values 
\begin{align}
\Theta(x  - y) = 
\begin{cases}
  1,  & x>y\text{ ,}\\
  0, & x \leq y\text{ .}
\end{cases}
\end{align}
In the distribution sense, it is a generalized function defined as 
\begin{align}
\Theta(x-y) = \int_y^\infty \rd x\; \varphi(x) \text{ ,}
\end{align}
where the derivative of $\varphi(x)$ with respect to $x$ gives the Dirac $\delta$-distribution.
After the decomposition into primary sectors, the 
integral $G$ is split into $N$ integrals $G_l$ with $x_l$ the upper 
integration boundary of all integrals over $x_i\, (\forall i \neq l)$, compare 
Eq.~(\ref{eq:showsecdecline3}). 
In the next step, all integrals are remapped to the unit hypercube 
by using a blowing up transformation on the Feynman parameters 
\begin{align}
x_j = 
\begin{cases}
  x_l t_j,  & j \neq l \text{ ,}\\
  x_l ,  & j=l \text{ .}
\end{cases}
\end{align}
The homogeneity of the functions $\cal{U}$ and $\cal{F}$ 
lead to a scaling behavior of ${\cal{U}} \propto x_l^{L}$ and 
${\cal{F}} \propto x_l^{L+1}$, where $L$ is the number of loops 
of the diagram, see Sec.~\ref{sec:feynparametrization}. 
Taking into consideration all powers in $x_l$ appearing in 
one sector 
\begin{align}
G_l &\propto \int_0^\infty \prod\limits_{j=1 \atop j \neq l}^{N} 
(x_l^{N-1} \rd t_j)\, \rd x_l\,  (x_l t_j)^{\nu_j-1} x_l^{\nu_l-1} 
x_l^{L(N_{\nu}-(L+1) D/2)-(L+1)(N_\nu-L D/2)} \\
&\propto \int_0^\infty \prod\limits_{j=1 \atop j \neq l}^{N} 
(t_j^{\nu-1} \rd t_j) \, \rd x_l\, x_l^{\left( \sum_{k=1}^N (\nu_k -1) +N-1 \right)} x_l^{-N_\nu} \\
&\propto \int_0^\infty \prod\limits_{j=1 \atop j \neq l}^{N} 
(t_j^{\nu-1} \rd t_j)\, \rd x_l\, x_l^{-1} \text{ ,}
\end{align}
an overall factor of $x_l^{-1}$ remains. 
The integration of the $\delta$-distribution 
\begin{align}
\int_0^\infty  \frac{\rd x_l}{x_l} \delta(1- x_l \,(1 + \sum_{k=1}^{N-1} t_k )) = 1
\end{align}
then yields primary sectors of the type
\begin{align}
G_{l} = \prod\limits_{j=1 \atop j \neq l}^{N}\,  \int \limits_{0}^{1}  
\rd t_j\,\,t_j^{\nu_j-1} 
\frac{{\cal U}_l^{N_{\nu}-(L+1) D/2}(\vec{t})}
{{\cal F}_l^{N_\nu-L D/2}(\vec{t})} \text{ ,}
\end{align}
where the $G_l$ are connected with the full integral $G$ by 
\begin{align}
G = \frac{(-1)^{N_{\nu}}}{\prod_{j=1}^{N}\Gamma(\nu_j)} \Gamma(N_{\nu} - LD/2) \sum_{l=1}^{N} G_l \text{ .}
\end{align}
\subsection{Iterated sector decomposition}
In the one-loop case, the first 
Symanzik sub-sector polynomials ${\cal U}_l$ are already brought into the form 
$\mathcal{U}_l = 1 + \sum\limits_{j=1,\, j \neq l}^{N} t_j$ after the decomposition 
into primary sectors. 
This is different at higher loop order $L>1$ and also for the second 
Symanzik sub-sector polynomials ${\cal F}_l$. 
An iterative procedure allows for the successive disentanglement of 
all singularities. It follows three steps which are 
performed until completion. 

\bigskip

At first, a minimal set of parameters ${\cal S} = \{ t_{\alpha_1}, \dots , t_{\alpha_r}\}$ 
is assigned which leads to a vanishing of the primary sector functions 
${\cal U}_l$ and ${\cal F}_l$ in the limit of vanishing elements of ${\cal S}$. 
The success of the 
decomposition is dependent on the choice of ${\cal S}$ which is by no 
means unique. 

\medskip

Then, the defined $r$-dimensional cube is split into sub-sectors 
\begin{align}
\prod\limits_{j=1}^{r} \Theta(1 \geq t_{\alpha_j} \geq 0) = 
\sum_{k=1}^r \prod\limits_{j=1 \atop j \neq k}^{r}  \Theta(t_{\alpha_k} \geq t_{\alpha_j} \geq 0) \text{ .}
\end{align}

Next, the integration boundaries are transformed back to the 
unit cube by applying a blowing up once more, leading to the following 
transformation rules for the Feynman parameters 
\begin{align}
t_{\alpha_j} = 
\begin{cases}
  t_{\alpha_k} t_{\alpha_j},  & j \neq k \text{ ,}\\
  t_{\alpha_k} ,  & j=k \text{ .}
\end{cases}
\end{align}
At least one of the functions ${\cal U}_l$ and ${\cal F}_l$ factorize 
in the parameter $ t_{\alpha_k} $ with the exponent of ${\cal U}_l$ 
or ${\cal F}_l$, respectively. Taking the additional Jacobian factor 
of $t_{\alpha_k}^{r-1}$ into account, exponents of the type $A_k - B_k\, \eps$ 
result for each integration parameter $t_{\alpha_k}$. $A_k$ and $B_k$ are 
numbers independent of the regulator $\eps$. 
The resulting sub-sector integrals are of the form 
\begin{align}
G_{lk} = \prod\limits_{j=1  \atop j \neq k}^{N}\,  \int \limits_{0}^{1}  
\rd t_{\alpha_j}\,\,\left( t_{\alpha_k}^{A_k-B_k \eps}  \right) 
\frac{{\cal U}_{lk}^{N_{\nu}-(L+1) D/2}(\vec{t}_{\alpha_j})}
{{\cal F}_{lk}^{N_\nu-L D/2}(\vec{t}_{\alpha_j}) } \text{ ,} \quad k=1, \dots, r \text{ .}
\end{align}
The three steps are repeated creating further sub-sectors 
${\cal U}_{lk_1 k_2 \dots k_c}$ and ${\cal F}_{lk_1 k_2 \dots k_c}$, 
until no further set ${\cal S}$ can be found 
after $c$ iterations which leads to a vanishing of 
the sub-sector functions. This is the case when they contain 
a constant term in form of a $1$ in the case of the 
first, and in form of a kinematic invariant in the 
case of the second Symanzik polynomial
\begin{align}
U_{lk_1k_2\dots} =& 1 + u(\vec{t}_{\alpha_j}) \text{ ,}\\
F_{lk_1k_2\dots} =& s_1 + \sum_{\beta} (s_\beta) f_{\beta}(\vec{t}_{\alpha_j}) \text{ ,}
\end{align}
where $u(\vec{t}_{\alpha_j})$ and $f_{\beta}(\vec{t}_{\alpha_j})$ are polynomials 
in the Feynman parameters and where kinematic invariants including masses 
are termed $s_1$ and $s_\beta$. \\
The singular behavior is now contained in the exponent $A_k$ and 
all overlapping divergences are disentangled. 
\subsection{Extraction of the poles}
\label{subsec:extractpoles}
It is now possible to find subtraction terms to extract poles in a Laurent series in 
the regulator $\eps$. 
Each obtained sub-sector integrand and all variables $t_{\alpha_j}$ with exponents 
$A_j - B_j \,\eps$ can be written in the general form
\begin{align}
I_j =\int_0^1 \rd t_{\alpha_j} t_{\alpha_j}^{A_j - B_j \eps} 
\mathcal{I}(t_{\alpha_j},\{ t_{\alpha_i \neq \alpha_j} \},\eps) \text{ ,}
\label{eq:subsectorforpoleextraction}
\end{align}
where $\mathcal{I}$ is a function of the decomposed sub-sector functions 
${\cal U}_{lk_1 k_2 \dots k_c}$ and ${\cal F}_{lk_1 k_2 \dots k_c}$. If the 
Feynman parameter is of positive or vanishing exponent, $A_j \geq 0$, 
the integration is finite in the regulator $\eps$ and no subtraction is needed. 
In all other cases, the integration will lead to a logarithmic pole for $A_j = -1$ or 
a higher pole if $A_j <-1$ and in the limit of a vanishing Feynman 
parameter $t_{\alpha_j}$. To expand around the pole, an expansion into a 
Taylor series around $t_{\alpha_j}=0$ can be performed 
\begin{align}
\mathcal{I}(t_{\alpha_j},\{ t_{\alpha_i \neq \alpha_j} \},\eps)=
\sum_{p=0}^{|A_j|-1}
\mathcal{I}_j^{(p)}(0, \{ t_{\alpha_i \neq \alpha_j} \},\eps) 
\frac{t_{\alpha_j}^p}{p!}+ R(\vec{t},\eps) \text{ ,}
\label{eq:extractpoletaylorseries}
\end{align}
where $R(\vec{t},\eps)$ denotes the remainder term which does not 
contain any poles in the parameter $t_{\alpha_j}$ by construction and where 
\begin{align}
\mathcal{I}^{(p)}(0,\{ t_{i \neq j} \},\eps)=\frac{\partial^p}{\partial t_j^p} 
\mathcal{I}_j(t_j, \{ t_{i \neq j} \},\eps) \Big|_{t_j=0} \text{ .}
 \end{align}
Reinserting Eq.~(\ref{eq:extractpoletaylorseries}) into 
Eq.~(\ref{eq:subsectorforpoleextraction}) the only terms 
depending on the variable $t_{\alpha_j}$ are powers of it and 
the remainder polynomial $R$. Expanding the whole 
integrand into plus-distributions using the identity 
\begin{align}
x^{-1 + \kappa \, \eps} = \frac{1}{\kappa \, \eps} \delta(x)
+ \sum_{n=0}^{\infty} \frac{(\kappa \, \eps)^n}{n!} \Big [ \frac{\text{ln}^n(x)}{x} \Big]_+ \text{ ,}
\end{align}
where 
\begin{align}
\int_0^1 \rd x\, f(x)  \Big [ \frac{g(x)}{x} \Big]_+ = 
 \int_0^1 \rd x\, g(x) \left[ \frac{f(x)-f(0)}{x} \right] \text{ ,}
\label{eq:subtractiongeneral}
\end{align}
the integration over $t_{\alpha_j}$ can be performed 
straightforwardly for the first term on the righthand side of 
Eq.~(\ref{eq:extractpoletaylorseries}), resulting with only the 
integration left in the finite remainder term
\begin{align}
I_j =\sum_{p=0}^{|A_j|-1} \frac{1}{A_j - B_j \, \eps + p + 1 } 
\frac{\mathcal{I}_j^{(p)}(0, \{ t_{i \neq j} \},\eps)}{p!}
+ \int_0^1 \text{d}t_j t_j^{A_j-B_j \eps} R(\vec{t},\eps) \text{ .}
\label{eq:poleextraction}
\end{align}
In the case of a logarithmic divergence, the sub-sector integrand with poles 
subtracted in the variable $t_{\alpha_j}$ would read 
\begin{align}
\non I_j =& \int_0^1 \rd t_{\alpha_j} t_{\alpha_j}^{-1-B_j \eps}
\mathcal{I}(t_{\alpha_j},\{ t_{\alpha_i \neq \alpha_j} \},\eps)  \\
\non=& -\frac{\mathcal{I}(0,\{  t_{\alpha_i \neq \alpha_j} \},\eps)}{B_j \eps} \\
& + \int_0^1 \rd t_{\alpha_j} t_{\alpha_j}^{-1 - B_j \eps} 
 (\mathcal{I}(t_{\alpha_j},\{  t_{\alpha_i \neq \alpha_j} \},\eps) -
 \mathcal{I}(0,\{ t_{\alpha_i \neq \alpha_j} \},\eps) ) \text{ .}
 \label{eq:subtractionexample}
\end{align}
\subsection{Calculation of the pole coefficients}
After repetition of these subtraction steps for all variables $t_{\alpha_j} \forall \, j$  
and all obtained sub-sectors, nested sums result where each summand can 
be dependent on the regulator $\eps$. The whole 
expression can be expanded in $\eps$ yielding a Laurent series with 
coefficients $C_{lk_1 k_2 \dots k_c, m}$ for each of the $c(l)$ sub-sector integrals 
of the $l$-th primary sector  
\begin{align}
G_{lk_1 k_2 \dots k_c} = \sum_{m=-2L}^{n} C_{lk_1 k_2 \dots k_c, m} \eps^m + \mathcal{O}(\eps^{n+1})
\end{align}
which again enter the full result for a (scalar) loop diagram as 
\begin{align}
G = (-1)^N \Gamma(N_\nu - LD/2) \sum_{l=1}^{N} \sum_{k=1}^{c(l)} G_{lk_1 k_2 \dots k_c} \text{ .}
\end{align}
\section{The choice of algorithm}
\subsection{Goals}
\label{subsec:decompogoals}
The best suited sector decomposition algorithm 
may differ in view of the two aspects, applicability 
and simplicity of the result.  
One algorithm may be applicable to every multi-loop 
diagram, but result in very complicated expressions. 
Another one may lead to relatively simple expressions, 
but is not guaranteed to stop.  

\medskip

A sector decomposition algorithm does not stop, as soon 
as it runs into an infinite recursion. 
The appearance of such can be exemplified assuming the 
following function 
\begin{align}
f ( x_1 , x_2 , x_3 ) = x^2_1 + x^2_2 x_3 \text{ .}
\end{align}
When decomposing it first in the variables $x_1$ and 
$x_3$, two sub-sectors with opposite hierarchy 
\footnote{For an equal sign in Eqs.~(\ref{eq:f1insec42})-(\ref{eq:f12insec42}), the 
Jacobian determinants need to be considered as well.}
\begin{align}
f_1 ( t_1 , x_2 , x_3 ) \propto & \, x_3\, (x_3 t_1^2 + x^2_2)\label{eq:f1insec42}\\
f_2 ( x_1 , x_2 , t_3 ) \propto & \, x_1\, (x_1 + x_2^2 t^2_3)
\end{align}
are created by rescaling the Feynman parameter 
$x_1 = x_3 t_1$ in the first sub-sector of function $f$, and 
$x_3 = x_1 t_3$ in the second sub-sector. 
Choosing the sub-sector $f_1$ and the set $S=\{ 2,3 \}$ of Feynman 
parameters, the initial functional results 
\begin{align}
f_{11} ( t_1 , t_2 , x_3 ) \propto & \, x_3^2 \, (t_1^2 + x_3 t^2_2)\\
f_{12} ( t_1 , x_2 , t_3 ) \propto & \, x_2^2\, t_3 \, (t_3 t_1^2 + x_2) \label{eq:f12insec42} \text{ ,}
\end{align}
augmented by an additional factor of Feynman parameters. 
The set $S=\{ 1,2 \}$ would instead lead to a termination 
of the algorithm. 
If chosen in an inconvenient way, decomposition sequences 
can complicate integrand functions 
or even lead to an infinite recursion. 
The occurrence of the latter limits the applicability. 
Manifestations of the former have a direct influence 
on the numerical convergence. 

\medskip

The fewer decomposition steps are needed in an iterative 
algorithm, the fewer sub-sectors are produced and the smaller 
the powers of factorized integration parameters are.  
%
%
%
%
%
%
%
\subsection{A heuristic algorithm for slim results}
\label{subsec:heuristicstrategies}
A first algorithm, which is also the one employed in the 
program \secdec, is completely heuristic. 
First, the primary sector decomposition is performed 
as described in 
Sec.~\ref{subsec:primarysecdec}. 
Then, each individual primary sector is iteratively 
decomposed into sub-sectors until both Symanzik 
polynomials are finite for vanishing Feynman 
parameters. The procedure works as follows:
\begin{itemize}
\item[1)] Determine which of the two polynomials ${\cal U}_l$ and 
${\cal F}_l$ of the primary sector $l$ 
turns zero in the limit of 
vanishing Feynman parameters. Find the best 
decomposition set ${\cal S}$ for this function. 
If both polynomials nullify, 
find the best decomposition set for ${\cal U}_l$.
\item[2)] Compute all possible subsets of Feynman 
parameters contained in one primary sector. 
\item[3)] Find the smallest set ${\cal S}_{\text{min}}$ that nullifies the function 
detected in 1). If two or more such sets have equal but 
minimal length, proceed with step 4), otherwise continue 
with step 5). The smallest set must contain more than one Feynman 
parameter, otherwise it would factorize from the polynomials 
${\cal U}_l$ and/or ${\cal F}_l$.
\item[4)] Discover which of the minimal sets maximizes the number of 
vanishing sub-sector polynomials. If there are several minimal sets 
leading to the same maximal number of vanishing polynomials, 
analyze the powers in the Feynman parameters 
of each nullifying set for the function detected in 1). Choose 
the set with lowest powers in the vanishing 
Feynman parameters.  
%
\item[5)] Divide both, ${\cal U}_l$ and ${\cal F}_l$ into sub-sectors 
using the encountered best minimal set. 
\end{itemize}
The procedure is iterated and more sub-sectors are 
produced as 
long as there is a set that nullifies ${\cal U}_{l k_1 \dots k_c}$ or/and 
${\cal F}_{l k_1 \dots k_c}$. \\
This heuristic strategy is found to produce 
the least sectors compared to strategies described separately 
by Bogner, Weinzierl and Smirnov, Tentyukov, compare Ref.~\cite{Bogner:2007cr} 
and Ref.~\cite{Smirnov:2008py}, respectively. 
Yet, the algorithm is not guaranteed to stop. The 
probability for running into an infinite recursion as 
described in the previous section 
can be reduced by introducing an additional heuristic 
strategy to the algorithm. 
In the latter, a pre-decomposition is carried out for all 
those Feynman parameters appearing quadratically or in 
higher powers in the primary sectors. No selection of 
a subset of Feynman parameters is performed. 
This treatment has proven to be very beneficial in many 
cases, especially in the computation of two-loop integrals 
with massive internal lines, but can 
lead to an increase in the number of 
produced sectors if it is always carried out. Applied to 
inconveniently chosen sectors of complicated integrals, this 
additional strategy can even introduce higher spurious 
negative powers in the factorized Feynman parameters. 
\subsection{Algorithms guaranteed to stop}
\label{subsec:stoppingalgorithm}
There are examples of three-loop diagrams which cannot be treated 
with the heuristic algorithms described in the previous chapter 
due to the occurrence of infinite recursion. 
To this end, it is interesting to find algorithms which are 
guaranteed to stop, regardless of the numbers of sectors 
produced. 

The possibly simplest algorithm to that matter is the one 
introduced by Hepp \cite{Hepp:1966eg}, 
where $n!$ sectors are produced due to the fact that each sector 
is split in all Feynman variables $x_n$, thereby always choosing the 
maximal decomposition set. Although this strategy will eventually 
terminate, the amount of sectors produced is by far too large. 
The problem can be solved differently by formulating it in terms of the 
polyhedra game introduced by Hironaka where the player A is supposed 
to win over player B after a finite number of moves and 
independent of the reaction of player B, see Ref.~\cite{Hironaka1964}. 
The relation to sector decomposition was found by Bogner and 
Weinzierl who also analyzed three strategies leading to the 
termination of the sector decomposition algorithm, see 
Ref.~\cite{Bogner:2007cr}. 
The first strategy analyzed there is based on 
work by Zeilinger~\cite{1245.14052}, the second on a strategy by 
the mathematician Spivakovsky~\cite{0531.14009}, and the third strategy is 
inspired by a proof of Encinas and Hauser, see Ref.~\cite{1059.14022}.
The strategies are all based on enforcing a sequence of 
decreasing decomposition sets of Feynman parameters 
used for each step 
in the iterated decomposition into sub-sectors. It was found, that the 
heuristic strategy always wins over the terminating algorithms in terms 
of the numbers of sectors produced, see Ref.~\cite{Bogner:2007cr,Bogner:2008ry}. \\
This situation does not change with the introduction of another 
strategy $S$ found by Smirnov and Tentyukov, although it results with less 
sectors than the previously mentioned terminating strategies, see 
Ref.~\cite{Smirnov:2008py}. The strategy involves the computation 
of normal vectors to facets of the convex hull of all weights, where the 
weights are found by the exponents in the Feynman parameters of each 
monomial in the sub-sector polynomial. If there is no facet which would 
lead to a, with respect to the lexicographical ordering, smaller set of Feynman 
parameters to be decomposed in the next step, the decomposition is 
finished using strategy which is guaranteed to stop, e.g. the one based on 
work by Zeilinger. 
It was also found that strategy $S$ produces the same number of sectors 
as the strategy based on Speer sectors \cite{Speer:1975dc}, which can process 
more information about the graph to be computed. The introduction of 
Speer sectors leads to a higher efficiency in the sector decomposition, 
regarding the speed and the memory intensity, see 
Refs.~\cite{Smirnov:2008aw,Smirnov:2009pb}. 

\medskip

Having introduced all these strategies, it would be nice to have a 
strategy which produces comparatively few sectors with regard to the 
heuristic strategy and 
is guaranteed to terminate in a finite number of steps. 
Such a strategy was introduced by Kaneko and 
Ueda~\cite{Kaneko:2009qx,Kaneko:2010kj}. They take a deterministic 
approach and reformulate the primary sectors using 
convex and combinatorial geometry. By the construction of 
intersections of dual cones to convex polyhedral cones a unique 
decomposition of the integration region can be found for each 
polynomial. Some of the cones may still be too complicated for integration, 
therefore they can be cut into simplices using 
triangulation. The total number of 
sectors produced depends on the triangulation 
algorithm. For the latter, there are many implementations 
available in the literature, 
see e.g. Refs.~\cite{Ueda:2009xx,Barber96thequickhull}. Using the first 
of the two, the resulting number of sectors is found to be even smaller 
compared to the heuristic strategy and the algorithm is, by construction, 
always guaranteed to stop, see Ref.~\cite{Kaneko:2010kj}. 
The drawback is that the resulting functions are more 
complicated compared to the integrands resulting from the 
heuristic strategy. This is due to 
the fact that it is not an iterative algorithm.  
While currently 
only the heuristic strategy, augmented by the option of 
applying a pre-decomposition, is 
implemented in the program \secdec, the algorithm of Kaneko and 
Ueda will be included in the next improved version of the program, 
see Sec.~\ref{sec:secdecfuture}. 
%
%

%% file: contourdef/contourdef.tex
\chapter[Singularity structure of Feynman integrals]{Singularity structure of \\Feynman integrals}
\label{chap:contourdef}
\section{Euclidean vs. physical kinematics}
\label{sec:euclidvsphyskinem}
In Sec.~\ref{sec:feynparametrization} it was already pointed out 
that both Symanzik polynomials are of definite sign when 
computing integrals in the Euclidean region. This implies that 
the energy component of the external momenta lies on the 
imaginary axis, leading to 
negative values in the kinematic invariants $s_{ij} < 0$ and $p_i^2<0$ 
and an overall positive contribution in the second 
Symanzik polynomial $\cal{F}$. To verify this, compare e.g. with the 
one-loop box example in Sec.~\ref{sec:feynparametrization}. 
Then, together with masses entering with a positive sign, 
$\cal{F}$ is positive semi-definite. 
After the application of sector decomposition, all possible singularities 
appearing in $\cal{F}$ are factorized leading to only positive definite 
integrands. 

\medskip

Switching to physical kinematics, the invariants formed from external 
momenta can be real and four-momentum conservation 
\begin{align}
\left( \sum_{i}^{n} p_{i} \right)^2 = 0
\end{align}
must hold, where $n$ is the number of external legs 
of which $n-1$ are linearly independent. 
Due to this fact, $\cal{F}$ is no longer definite and further 
singularities, though integrable, can occur within the integration 
region.
\begin{figure}[htb!]
\begin{center}
\includegraphics[width=0.4\textwidth]{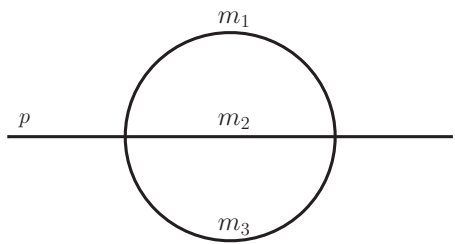}
\caption{Two-loop two-point "sunrise" graph with three internal masses.}
\label{fig:sunriseallgemein}
\end{center}
\end{figure}

An intuitive example are production thresholds 
which appear as internal particles go on-shell. This means that 
the overall incoming external momentum reaches 
any sum of masses of internal propagators potentially leading to 
physical final states.  
A simple example to demonstrate this 
is a two-loop two-point function with three internal masses, see 
Fig.~\ref{fig:sunriseallgemein}. The three-particle-cut discontinuity 
occurs for 
\begin{align}
p^2 = (m_1+m_2+m_3)^2 \text{ .}
\end{align}
The analytical determination of the threshold locations gets more 
and more complicated, the more external legs and propagators 
are involved. 
Writing an integral in Feynman parametrization, 
thresholds may be parametrized at the integrand 
level, as a combination 
of kinematic invariants and Feynman parameters. 
The full set of thresholds can be determined solving the 
Landau equations of an integrand. 
%
%
%
\section{Landau equations}
\label{sec:landauequations}
The Landau equations 
\begin{subequations}
\begin{align}
 x_j\,(q_j^2(\{k \},\{p\})-m_j^2) =& 0 \quad \forall \,j \in \{1,\dots,N\} \label{eq:landau1}\\
\frac{\partial}{\partial k_i^\mu}\sum_{j=1}^N x_j\,\left(q_j^2(\{k \},\{p\})-m_j^2\right) =& 0 
 \quad \forall \,i \in \{1,\dots,L\} \text{ ,} \label{eq:landau2}
\end{align}
\end{subequations}
give the necessary (but not sufficient) conditions for a divergence, 
see Refs.~\cite{Landau:1959fi,ELOP,Nakanishi}. 
In accordance with the notation of Sec.~\ref{sec:generalintegraldefinition}, 
the $q_j$ are linear combinations of 
external momenta $p_i$ and loop momenta $k_i$, $N$ is the number 
of propagators and $L$ the number of loops. 
Paraphrasing Eqs.~(\ref{eq:landau1}), either the propagator 
$q_j^2-m_j^2$ or their respective Feynman parameter 
$x_j$ must vanish to potentially contribute to a singularity. 
Only if Eqs.~(\ref{eq:landau2}), involving a 
derivative by the loop momenta, vanish 
simultaneously, the conditions for a 
Landau singularity are fulfilled. 

\medskip

A solution to the system with $x_j \neq 0 \;\forall \,j$ gives the 
leading Landau singularity, which is not
integrable for $N>2$ when $D=4-2\eps$, and real values of 
masses and momenta. 
Those singularities 
where the vanishing of one Feynman parameter leads to a 
singular behavior are termed sub-leading Landau singularities. 
These correspond to the thresholds of a subgraph as a vanishing 
Feynman parameter can be associated with the removal of one 
propagator and the junction of two vertices. These singularities 
are integrable and of logarithmic or square-root type. 

\medskip

The Landau equations can be solved by contracting the momenta 
of Eq.~(\ref{eq:landau2}) with those loop and external momenta the 
equation depends on, to 
get a system of equations which can be solved by using the 
constraints arising from Eq.~(\ref{eq:landau1}).
A nice example analysis can be found in Ref.~\cite{Passarino:2006gv}. 
Another example is shown in Sec.~\ref{sec:expectedthresholds}. \\
The Landau equations can also be formulated as 
\begin{subequations}
\begin{align}
{\cal F}(\vec{x},\{p, m^2\}) &= 0 \text{ ,} \label{eq:landaualtern1}\\
\frac{\partial}{\partial x_j}{\cal F}(\vec{x},\{p, m^2\}) &= 0  \quad \forall \,j \in \{1,\dots,N\} 
 \text{ ,} \label{eq:landaualtern2}
\end{align}%
\label{eq:landaualtern}%
\end{subequations}%
after having integrated out the loop momenta, see 
Ref.~\cite{Bjorken:1965zz}.
The leading Landau singularity is again given by the solution to 
the system of equations assuming an empty set of vanishing Feynman 
parameters. \\
How we deal with these singularities will be described in the following section.
\section{Deformation of the integration contour}
\subsection{Cauchy theorem}
\label{subsec:cauchy}
\begin{figure}[ht]
	\begin{center}
	      \includegraphics[width=0.5\textwidth]{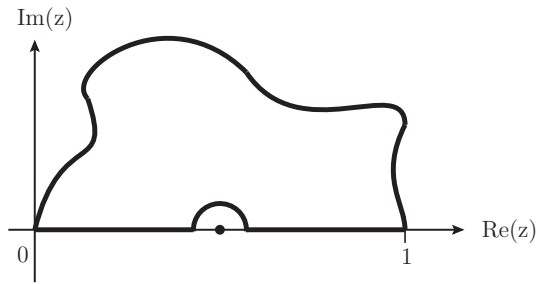}
	\end{center}
	\caption{Schematic picture of the closed contour avoiding poles on the real axis.}
	\label{fig:cauchypicture} 
\end{figure}
Unless the function ${\cal F}$ is of definite sign for
all possible values of invariants and Feynman parameters, 
the  denominator of a multi-loop integral will vanish within the integration 
region on a hypersurface given by the solutions of the Landau equations. 
To avoid the non-physical poles on the real axis, the Cauchy theorem  
\begin{align}
\oint_c  \prod\limits_{j=1}^{N}\rd z_j {\cal I}(\vec{z}) =
\int_0^1 \prod\limits_{j=1}^{N}\rd x_j {\cal I}(\vec{x}) + 
\int_1^0 \prod\limits_{j=1}^{N}\rd z_j {\cal I}(\vec{z})  
= 0
\label{eq:cauchysentence}
\end{align}
can be exploited, where $\text{Re}(\vec{z})=\vec{x}$. 
To be able to use the theorem, the original integrand, depending only on 
the real coordinates $x_j$, is analytically continued to 
the complex plane. The coordinate transformation reads
\begin{align}
\int_0^1 \prod\limits_{j=1}^{N}\rd x_j {\cal I}(\vec{x}) = 
\int_0^1 \prod\limits_{j=1}^{N}\rd x_j 
\left\vert \left(\frac{\partial z_k(\vec{x})}{\partial x_l}\right)\right\vert 
{\cal I}(\vec{z}(\vec{x})) \text{ ,}
\label{eq:cauchyformula}
\end{align}
where the new complex coordinates $\vec{z}$ describe a path parametrized by 
the variables $\vec{x}$. 
With a given description of the coordinates $\vec{z}$, the Cauchy theorem in Eq.~(\ref{eq:cauchysentence}) can be formulated.
It is valid in this form as long as the deformation is in 
accordance with the causal $i\delta$ prescription 
of the Feynman propagators, as the region enclosed by the integration 
contour then does not contain any singular points, compare Fig.~\ref{fig:cauchypicture}.
It is important to keep in mind, that no poles should be 
crossed while changing the integration
path, otherwise Eq.~(\ref{eq:cauchysentence}) is no longer valid. \\
Finding the right analytical continuation to the coordinates $\vec{z}$ is 
equivalent to finding the proper deformation to the integration contour. 
It is the crucial step for the success of this method and will be 
treated in the following.
\subsection{Deformation}
\label{subsec:deformation}
The aim is to find a clever deformation which is well suited 
for an automated application in numerical calculations. 
For its realization, a good parametrization of the complex variables $z_i$ in Eq.~(\ref{eq:cauchyformula}) 
must be found which on the one hand preserves the causal $i\delta$ 
prescription, and on the 
other ensures all physical thresholds to appear as such in the result. 
As the latter are contained in the Landau equations, an inclusion of these 
in the deformation is desirable. 
It is therefore required that all Landau equations, 
Eqs.~(\ref{eq:landaualtern}), are realized when the deformed function 
${\cal F}(\vec{z}(\vec{x}))$ vanishes. 
Furthermore, the $i\delta$ prescription for the Feynman propagators 
requires that the contour deformation to the complex plane is 
chosen such that the infinitesimal imaginary part is conserved. 
The negative sign of the imaginary part of the second Symanzik 
polynomial ${\cal F}$ was discussed in 
Sec.~\ref{sec:feynparametrization}. 
For real masses and Mandelstam invariants $s_{ij}$, the following Ansatz~\cite{Soper:1999xk,Nagy:2006xy,Binoth:2005ff}
is therefore convenient
\begin{align}
\vec{z}( \vec x) &= \vec{x} - i\;  \vec{\tau}(\vec{x})\nonumber\\
\tau_k ( \vec x) &= \lambda\, x_k (1-x_k)
 \, \frac{\partial {\cal F}(\vec{x})}{\partial x_k}  \text{ ,}
 \label{eq:condef}
\end{align}
where $\lambda$ is an arbitrary real and positive parameter. 
A closed integration contour is guaranteed by the factors $x_k$ and $(1-x_k)$, 
keeping the endpoints fixed. 
From Eq.~(\ref{eq:condef}), the negative sign of the imaginary part is only 
guaranteed if the derivative by ${\cal F}(\vec{x})$ is not negative. 
Assuming the overall deformation to be small, the analytic continuation of the 
integrand can be expanded into a series
\begin{equation}
{\cal F}(\vec{x}) \rightarrow {\cal F}(\vec{z}(\vec{x}))={\cal F}(\vec{x})
-i \,\sum\limits_{k} \, \lambda\, x_k (1-x_k) \left(\frac{\partial {\cal F}}{\partial x_k}  \right)^2 + {\cal O}(\tau_k ( \vec x)^2)\;,
\label{eq:newF}
\end{equation}
where the expansion is done individually in each component $k$. 
The physically motivated requirement that all Landau equations, 
Eqs.~(\ref{eq:landaualtern}), be fulfilled is met in Eq.~(\ref{eq:newF}), 
when the deformed integrand ${\cal F}(\vec{z}(\vec{x}))$ vanishes. 
Furthermore, the imaginary part of ${\cal F}(\vec{z}(\vec{x}))$ 
is always negative 
due to an ever positive $(\frac{\partial {\cal F}}{\partial x_k} )^2$ term. While the 
absolute size of the derivative parts are determined by the 
diagrams to be computed, 
$\lambda$ is chosen to be a free parameter determining the 
scale of the deformation. 
Following the analysis of Ref.~\cite{Soper:1999xk}, the Ansatz for the analytical 
continuation must guarantee a full cancellation of singularities in 
subtraction terms present in the remainder term of Eq.~(\ref{eq:poleextraction}), see 
also Eq.~(\ref{eq:subtractionexample}). If the analytic continuation is done only after 
computing the subtraction terms, one Feynman parameter is deformed, while 
the one of the subtraction term is not. Assume the deformation of a function $I$
depending on one Feynman parameter
\begin{align}
I = \int_0^1 \rd t_{\alpha}\, t_{\alpha}^{-1 + \eps} 
 (\mathcal{I}(t_{\alpha},\eps) - \mathcal{I}(0,\eps) ) \text{ .}
\end{align}
Analytic continuation of the parameter 
$t_\alpha \to z_\alpha=t_\alpha + i \tau (t_\alpha)$ yields  
\begin{align}
I = \int_0^1 \rd t_{\alpha}\, 
\left( 1+ i \frac{\partial \tau(t_\alpha)}{\partial t_\alpha} \right) 
\frac{ \mathcal{I}(t_{\alpha}+ i \tau (t_\alpha),\eps) - \mathcal{I}(0,\eps) } 
{ (t_{\alpha}+ i \tau (t_\alpha))^{1 - \eps} } \text{ .}
\end{align}
The subtraction term $ \mathcal{I}(0,\eps)$ was set up to cancel 
the soft singularity in the real part. This parameterization 
can introduce spurious poles in the imaginary part, which are 
not taken care of in the limit of $t_{\alpha} \to 0$, unless $\tau (t_\alpha)$ 
vanishes faster than linear in the Feynman parameter $t_{\alpha}$. If 
the deformation vanishes faster than linear in the Feynman parameter 
$t_{\alpha}$, the imaginary part vanishes faster than the real part, resulting 
in the original subtraction term.  
This condition is no longer necessary, when the analytic continuation is 
done prior to the construction of the subtraction terms. It is due to this 
analysis that the analytic continuation of each Feynman parameter is 
done right after the iterated sector decomposition procedure. 

\medskip

In summary, unless a kinematic point fulfills all Landau equations, where 
both ${\cal F}$ and its derivatives with respect to $x_i$ vanish, 
the deformation of the integration contour leads to a well behaved 
integral at the points where only the function ${\cal F}$ vanishes. 

\medskip

An implementation and further analysis of this deformation for numerical 
calculations has already been worked out in 
Refs.~\cite{Anastasiou:2007qb,Beerli:2008zz}. 
To assure a high numerical stability of the evaluation of multi-loop integrals, 
necessary to make the implementation publicly available, 
supplementary studies of the deformation are necessary which are 
presented in the following. 
\subsubsection{Deformation studies}
\label{subsubsec:deformationstudies}
The aim of these deformation studies is to find an optimal procedure for 
an optimal choice for the 
parameter $\lambda$ which guarantees a good behavior of the integrand. 
To this end, the terms of order ${\cal O}(\tau_k ( \vec x))$, ${\cal O}(\tau_k ( \vec x))^2$ 
and ${\cal O}(\tau_k ( \vec x))^3$ are analyzed, assuming a decreasing effect in 
higher orders, as is expected from a convergent Taylor series expansion. 

\medskip

The analytic continuation of ${\cal F}(\vec{z}(\vec{x}))$ to the third power in the deformation reads
\begin{align}
\non {\cal F}(\vec{z}(\vec{x})) =& 
\mathcal{F}(x) - 
i \,\lambda \,\sum\limits_{j} \, x_j (1-x_j) \left(\frac{\partial {\cal F}}{\partial x_j}  \right)^2 \\
 &- \frac12 \lambda^2 \sum\limits_{j} \sum\limits_{k}  x_j x_k (1-x_j) (1-x_k) 
\frac{\partial {\cal F}}{\partial x_j} \frac{\partial {\cal F}}{\partial x_k}  
\left( \frac{\partial^{2} {\cal F}}{\partial x_j \partial x_k} \right) \\
& + \frac i6 \lambda^3 \sum\limits_{j} \sum\limits_{k}\sum\limits_{l} \frac{ \tau_j\, \tau_k\, \tau_l }{\lambda^3} 
\left( \frac{\partial^{3} {\cal F}}{\partial x_j \partial x_k\partial x_l} \right) \label{eq:deformationplusterm}
\end{align}
which uncovers two non-trivial aspects of the deformation. One leads to the 
fact that the term proportional to $\lambda^2$ contributes to the \textbf{real} 
part of ${\cal F}(\vec{z}(\vec{x}))$ and the other to an ambiguity in the 
sign of the imaginary part. 

\pagebreak
\begin{figure}[htb!]
	\begin{center}
	      \includegraphics[width=0.6\textwidth]{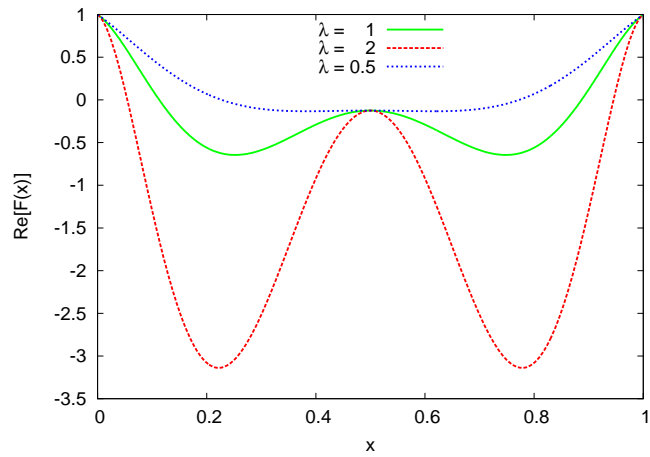}
	\end{center}
	\caption{Influence of the deformation on the real part for the one-loop bubble 
	and $m=1$, $s=4.5$.}
	\label{fig:realbubblepartlambda} 
\end{figure}
\begin{figure}[htb!]
	\begin{center}
	      \includegraphics[width=0.6\textwidth]{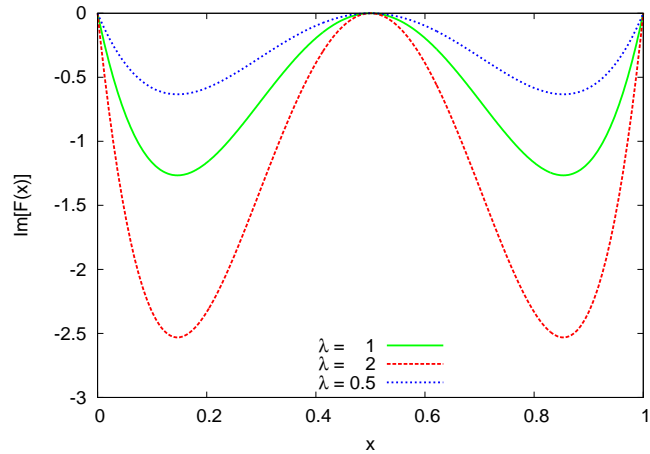}
	\end{center}
	\caption{Influence of the deformation on the imaginary part for the one-loop bubble 
	and $m=1$, $s=4.5$.}
	\label{fig:imagbubblepartlambda} 
\end{figure}
To show the effect on the real part, it is descriptive to look at the 
specific but simple example of the massive one-loop bubble, where 
the leading Landau singularity is well known to be situated at 
$s=4 \,m^2$ when $x=\frac12$. 
The function ${\cal F}(x)$ of the one-loop bubble reads
\begin{align}
\mathcal{F}_{\text{1L-bubble}}(x)= - s \,x\, (1-x)  + m^2- i \delta  \text{ .}
\label{eq:Foneloopbubble}
\end{align}
The real part of $\mathcal{F}$ after the analytical continuation is shown in 
Fig.~\ref{fig:realbubblepartlambda}, where a point above threshold was 
chosen, with a mass $m=1$ and $s=4.5$ and assuming arbitrary units. 
From its basic geometric properties, it is known that the derivative of ${\cal F}$ in Eq.~(\ref{eq:condef}) is smallest in the extrema and largest where 
the slope is maximal. 
Around $x=0.5$, the function $\mathcal{F}$ is almost, but not exactly, vanishing. 
The size of the deformation coming from the derivative of $\mathcal{F}$ is 
shown for $\lambda=1$. One can notice that choosing a rather small $\lambda=0.5$ 
the function $\mathcal{F}$ never vanishes except at the endpoints of the integration 
region $x=0,1$, while for the cases of $\lambda=1,2$ the 
function additionally vanishes in four points. 
In principle, this should not be a problem, 
as the imaginary part is not vanishing in any point beyond the end-points, see 
Fig.~\ref{fig:imagbubblepartlambda}. 
But the larger the value for $\lambda$ is chosen, the closer the points where the 
real part is zero, get to the endpoints, where also the imaginary part is small. 
This can easily lead to numerical instabilities, so the parameter $\lambda$ should not 
be chosen too large. 
\begin{figure}[htb!]
	\begin{center}
	      \includegraphics[width=0.6\textwidth]{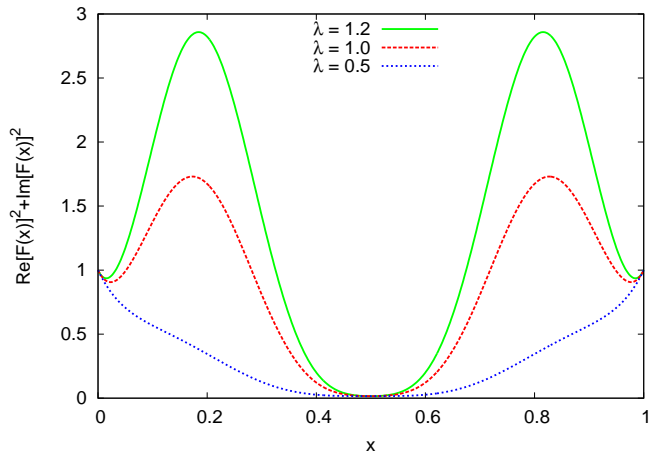}
	\end{center}
	\caption{Implications for the best choice of $\lambda$ from the modulus of 
	${\cal F}(z(x))$ for the one-loop bubble and $m=1$, $s=4.5$.}
	\label{fig:minmod} 
\end{figure}

\medskip

It should be noted that a value for $\lambda > 1$ is still viable, as long as the overall 
deformation is small. Otherwise, the series in the deformation Eq.~(\ref{eq:newF}) is 
no longer converging, leading to a wrong sign of the imaginary part. 

\medskip

After having settled that the deformation parameter should not be chosen too large, 
it must be found that it should neither be chosen to small, see Fig.~\ref{fig:minmod}. 
Though never striking zero, the modulus is extremely small for $\lambda=0.5$, in particular 
in the vicinity of $x=0.5$, which is bad for the numerical convergence. A maximization 
of the modulus of the function ${\cal F}(\vec{z}(\vec{x}))$ close to the critical points where it 
becomes minimal stabilizes the numerical evaluation.

\medskip

In the case of the one-loop bubble, the term of order $\lambda^3$ in 
Eq.~(\ref{eq:deformationplusterm}) is zero. This is not the case for more complicated 
integrals. In order for this term not to grow dominant and by that spoil the overall minus 
sign of the imaginary part, either lambda must be chosen below one or the terms 
proportional to the derivative must be $ |\frac{\partial {\cal F}(\vec{x})}{\partial x_k} | < 1$. 
This task can be accomplished by performing a small sampling of the derivative 
terms for each Feynman parameter in various values and divide the parameter $\lambda$ 
by the maximally achieved value for the derivative
\begin{align}
\lambda \rightarrow \tilde{\lambda} = \frac{\lambda}{\text{max}(|\partial {\cal F}(\vec{x})/{\partial x_1}|,\dots,|\partial {\cal F}(\vec{x})/{\partial x_N}|)} \text{ .}
\end{align}
Then, the derivative parts are   
roughly normalized to one and the scale of the deformation is again dominated by 
the value for the parameter $\lambda$. 
\begin{figure}[htb!]
	\begin{center}
	      \includegraphics[width=0.6\textwidth]{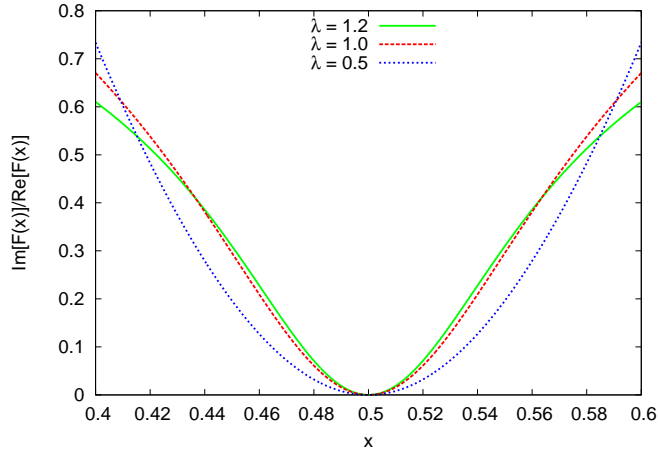}
	\end{center}
	\caption{Implications for the best choice of $\lambda$ from the minimization of 
	the complex argument of ${\cal F}(z(x))$ for the one-loop bubble and $m=1$, $s=4.5$.}
	\label{fig:minarg} 
\end{figure}

\medskip

In order to further prevent the deformation to become too large, valuable information on 
the right choice of $\lambda$ can be extracted from the analysis of the 
complex argument 
\begin{align}
\varphi_{{\cal F}(\vec{z}(\vec{x}))}=\left| \frac{- \,\sum\limits_{k} \, \tau_k ( \vec x) \left(\frac{\partial {\cal F}}{\partial x_k}  \right)^2}{{\cal F}(\vec{x})} \right|
 \text{ ,}
\end{align} 
where the numerator contains the coefficient of the imaginary part of order 
$\mathcal{O}(\lambda)$, see Fig.~\ref{fig:minarg}.
Minimizing the complex argument $\varphi_{{\cal F}(\vec{z}(\vec{x}))}$ improves the 
numerical convergence in the whole integration region when kinematically far from 
a critical point. When the imaginary part is relatively small 
compared to the real part, the terms of order $\lambda^2$ contributing 
to the real part cannot become too large and those terms going with $\lambda^3$ 
cannot spoil the overall sign of the imaginary part. 
This can be advantageous for speeding up a calculation and is especially 
interesting in the case of highly fluctuating integrands. Close to a threshold, this 
additional check is however not advisable because it clashes with the maximization 
of the modulus of ${\cal F}(\vec{z}(\vec{x}))$.

\medskip

For more technical details about the implementation of the deformation, see 
Sec.~ \ref{subsec:program:contourdef}.
\subsubsection{Pinch singularities}
%
%
%
If a singularity falls together with an endpoint of the integration 
region, it is trapped and no proper deformation of the 
integration contour is possible. 
The same applies to the case 
where two singularities fall together, where one singular point 
could only be bypassed deforming the contour into the 
direction of the other singularity. The result is a pinch in 
the integration contour. Both, pinch singularities and 
singularities at the endpoint of the integration region, 
are described by the Landau equations, compare 
Eqs.~(\ref{eq:landaualtern}). 

\medskip

With the introduction of the analytical continuation to 
the complex plane and the deformation of the integration contour, 
integrable singularities do not appear as divergences in the 
coefficients of the Laurent series in $\eps$. 
Nevertheless, the method leads to numerical instabilities 
in the vicinity of either a pinch singularity or a 
singularity at the endpoint of the integration region. 
This is due to the fact that the deformation 
of the integration contour becomes negligible. 
Returning to the 
one-loop bubble of Eq.~(\ref{eq:Foneloopbubble}), this 
behavior can be observed in Fig.~\ref{fig:minmod}, where the modulus of the 
function ${\cal F}$ can become very small. The 
evaluation time and accuracy of an integral close to such a 
singular point heavily depends on the chosen value for $\lambda$ 
and the numerical integrator. 

%% file: theprogram/theprogram.tex
\chapter[Extension of the program {\sc SecDec} to physical kinematics]{Extension of the program \\{\sc SecDec} to physical \\kinematics}
\label{chap:theprogram}%
In the following, the public program \secdec{} version 2~\cite{Borowka:2012yc,
Borowka:2012ii,Borowka:2012rt,Borowka:2013cma,Borowka:2013lda,
Borowka:2013uea} is presented. 
\secdec{} is a program for the numerical evaluation of 
multi-scale multi-loop and 
multi-dimensional polynomial parameter integrals. 
It is based on the sector decomposition algorithm described 
in Chap.~\ref{chap:sectordecompo}, where dimensionally 
regulated singularities are extracted. 
Even though their coefficients are available in algebraic form,             
they are usually too complicated to be integrated analytically.                                                                     
Therefore the final computation of the coefficients to each order 
in the regulator $\eps$ is done by Monte Carlo integration. 
To deal with 
integrable singularities due to mass thresholds, the integration 
contour is deformed to the complex plane. 
Before the extension to arbitrary kinematics was achieved with \secdec{} 
version 2, a first version of the program was publicly available~\cite{Carter:2010hi}. 
Other public implementations of the sector decomposition algorithm 
working in the Euclidean space are introduced in 
Refs.~\cite{Bogner:2007cr,Smirnov:2008py,
Smirnov:2009pb,Gluza:2010rn,Ueda:2009xx}. 
Recently, a new version of the program \textsc{Fiesta} has become 
available~\cite{Smirnov:2013eza}, 
including interesting and valuable new features. 
The extension to arbitrary kinematics was also achieved there, though their 
approach is heavily based on the one employed in \secdec, as 
mentioned in their publication. 

\bigskip 

The structure of this chapter is as follows: 
in Sec.~\ref{sec:programfunc}, the functionality of the program \secdec{} 
version 2 is reviewed. Its characteristic features are 
explained in Sec.~\ref{sec:charsecdecfeatures}, further 
capabilities are elaborated in Sec.~\ref{sec:additionalsecdeccaps}. 
The operational sequence of the program is shown in 
Sec.~\ref{sec:op_sequence}, before studying two examples and 
discussing the computation times in Sec.~\ref{sec:checkssecdec}. 
Prospective future developments are discussed in Sec.~\ref{sec:secdecfuture}.
\section{Functionality}
\label{sec:programfunc}
In this section, the functionality of the program \secdec{} is described. 
The program has two main branches, one where the computation of 
any loop integral or integral with a similar structure is possible. The 
user can start from a diagram knowing the 
propagators involved, or can even feed some of their own functions into 
the program. All other steps including the output of the final result are 
performed in a fully automated way. 
In the other branch, more 
general parametric functions can be computed, including the special 
feature that additional finite functions can be left symbolic until shortly before 
numerical integration. 

\medskip

Up to the step of the final integration, the coefficients are computed 
in a fully analytical way, where all kinematic invariants are left symbolic by 
default. This feature allows for a fast evaluation of multiple kinematic 
points, as only the integration step is left to be done if any of the 
kinematic invariants change. 
If a user is interested in just the result for one particular diagram for 
one set of kinematics, it is also possible to insert kinematic values 
in the beginning. 

\medskip 
%
%
%
%

Version 2 of  \secdec{} contains the following new features, 
which will be described in detail in the next sections. 
\subsubsection{Loop integrals and integrals of similar structure}
\begin{itemize}
\item Multi-scale loop integrals can be evaluated without restricting the kinematics 
to the Euclidean region. This has been achieved by performing a (numerical) 
contour integration in the complex plane. The program automatically tries to 
find an optimal deformation of the integration path. 
In addition, a kinematic threshold can be defined symbolically. Above this 
threshold, a complex contribution is expected and the deformation of the 
integration contour is automatically switched on. 
\item For scalar multi-loop integrals, the integrand can be constructed 
from the topological cuts of the diagram. The user only has to provide 
the vertices and the propagator masses, but does not have to provide 
the momentum flow. \footnotemark\footnotetext{This new feature has been implemented 
in collaboration with J. Carter.}
\item Tensor integrals can be evaluated with (in principle) no limitation 
on the rank. 
This means that a numerical approach in certain cases can help to 
alleviate or even avoid the procedure of amplitude reduction 
to master integrals. \footnotemark[1]
\item Another new feature is the option to apply the sector decomposition 
algorithm and subsequent contour deformation on user-defined functions 
which do not necessarily have the form of standard loop integrals, but have 
a simliar structure. 
\item The files for the numerical integration of functions amenable to contour 
deformation (multi-scale multi-loop integrals, user-defined functions) 
are written in C$^{++}$ rather than Fortran. 
For integrations in Euclidean space, the user can choose between using 
Fortran or C$^{++}$. 
\item A parallelization of the algebraic part for Mathematica versions 7 
and higher is possible if multiple cores are available. This is of special 
interest when computing very complicated multi-scale multi-loop integrals, 
see e.g. Chap.~\ref{chap:application1}.
\item A rescaling of all kinematic invariants by the absolute value of the largest 
invariant can be chosen to achieve a faster convergence of the numerical result. 
\item Looping over ranges of parameter values is automated, allowing 
scans over different kinematic configurations within one topology. 
\footnotemark[1]
\item A stable and recent version of the 
\textsc{Cuba library}~\cite{Hahn:2004fe,Agrawal:2011tm}, \textsc{Cuba}-3.2, allowing 
for a parallelized numerical integration is added to the program and used by default. 
\end{itemize}
\subsubsection{General parametric integrals}
\begin{itemize}
\item The user can define additional (finite) functions at a symbolic level. These can be 
specified later, after the integrand has been transformed 
into a set of finite parameter integrals, for each order in 
$\eps$. \footnotemark[1]
\item Looping over ranges of parameter values is automated, allowing 
scans over parameter sets for more general polynomial functions. \footnotemark[1]
\end{itemize}
%
%

Below, these new features are described in more detail, but also see 
Appendix~\ref{sec:appendix:usermanual} for a user manual. 
Comprehensive documentation can be found with the code itself, 
available at {\tt http://secdec.hepforge.org}.
\section{Characteristic features}
\label{sec:charsecdecfeatures}
\subsection{Loop integrals}
The program is capable of integrating general loop and 
multi-loop diagrams, including kinematic thresholds, using Feynman parameters. 
In accordance with Sec.~\ref{sec:feynparametrization}, such 
a loop integral is composed of five parts, the two Symanzik polynomials 
${\cal U}$ and  ${\cal F}$, the numerator, the $\delta$-distribution 
and the powers of factorizing Feynman parameters. 
While the numerator in a scalar integral is equal to unity, it 
can contain contractions of loop momenta with each other 
or with external momenta when the integral is of higher rank. 
While any kinematic invariant or scalar factor is treated as a
constant in the numerator, loop 
momenta appearing as contractions in the numerator influence 
the singularity structure of the integrand, compare 
Eq.~(\ref{eq:thefeynmanloopintegral}). 

Tensor integrals up to in principle arbitrary rank can be 
computed with \secdec{} by evaluating 
the coefficient functions of the Lorentz decomposed tensors. 
Take for example, the Lorentz decomposition of a 
one-loop two-point (bubble) integral $B_\mu$ of rank $R=1$ with 
one external momentum $p$
\begin{align}
B_\mu = p_\mu B_1 \text{ ,}
\end{align}
where the coefficient function $B_1$ reads
\begin{align}
p^2 B_1= P  \int \text{d}^D q 
\frac{q_\mu p^\mu}{(q^2-m_1^2)((q+p_1)^2-m_2^2)} \text{ }
\label{eq:b1mu}
\end{align}
with $P=\left( \frac{(2 \pi \mu_r)^{(4-D)}}{i \pi^2} \right)$. In the 
case of $B_\mu$, the coefficient function $B_1$ can be 
computed with \secdec, thereby delivering a result for the whole tensor integral. 

\medskip

The algorithms in \secdec{} are not restricted by any 
loop order, tensor rank or the number of scales. 
Provided with the information on the diagram to be computed 
\secdec{} calculates the Laurent series up to the desired order 
in the regulator $\eps$ in a fully automated way. 
For the iterated sector decomposition two heuristic strategies, 
described in Sec.~\ref{subsec:heuristicstrategies}, are available.
A diagram is 
specified by its propagators, loop momenta and irreducible 
numerators, and by the 
number of external legs and their on-shell conditions. The 
on-shell conditions become of special importance when 
external legs are light-like. 
This is the minimal information needed. Yet, \secdec{} has several  
options allowing for a more efficient evaluation tailored to 
specific integrals and/or a customization of the output of the results. 
One of the features of \secdec{} is that all kinematic 
invariants are left symbolic up to the numerical integration. 
This allows for a fast evaluation of integrals of the same topology 
for different sets of kinematic values. If only one 
kinematic point is of interest, it can be beneficial to 
set the values for the invariants already in the beginning to 
allow for an additional simplification of the integrands prior to 
numerical integration. This feature is included in \secdec{} as 
well, by abuse of the on-shell conditions, see 
App.~\ref{subsec:graphm}. 
Furthermore, \secdec{} allows for the choice of the 
desired prefactor and the maximal order in the 
regulator $\eps$ to be computed. 
Among further options regarding the numerical integration, see 
Sec.~\ref{subsec:program:cubaparameters}, \secdec{} 
arranges for a user-adjustment of the contour-deformation 
parameters for calculations in the physical region, see 
Sec.~\ref{subsec:program:contourdef}. A removal of spurious 
divergences can be achieved using integration by 
parts, see Sec.~\ref{subsec:ibprelations}. 
%
%
%
\subsection{Parametric integrals}
The program \secdec{} can also factorise singularities from parameter integrals which are  more 
general than those in multi-loop integrals.
The only restrictions are firstly that the integration domain 
should be a unit hypercube, 
and secondly singularities should reside only at the upper and/or lower integration 
boundary, i.e. at zero or one. Contour deformation is not available for more 
general parametric functions, because it requires 
the sign of the imaginary part to be known a priori in order not to give a 
wrong result. 
Currently the singularities are assumed to be regulated 
by non-integer powers of the integration parameters, e.g. the 
$\eps$ of dimensional regularization, or some other regulator.
The general form of such integrals is 
\begin{align}
I=\int_0^1 dx_1 \ldots \int_0^1 dx_N \prod_{i=1}^m P_i(\vec{x},\{\alpha\})^{\nu_i} \text{ ,}
\label{eq:general}
\end{align}
where $P_i(\vec{x},\{\alpha\})$ are polynomial functions of the parameters $x_j$,
which can also contain a set of symbolic constants $\{\alpha\}$. 
The user can leave the parameters $\{\alpha\}$ symbolic during the decomposition,
assigning values only for the numerical integration step.
This way the decomposition and subtraction steps do not have to be redone if 
the values for the constants are changed.
In Eq.~(\ref{eq:general}), the indices $\nu_i$ are of the 
form $\nu_i=a_i+b_i \, \eps$, with $a_i$ such 
that the integral is convergent.
Note that half integer powers are also possible. 
\subsection{Implementation of Contour deformation}
\label{subsec:program:contourdef}
As explained in Sec.~\ref{subsec:cauchy}, singularities on the real axis can be avoided 
by a deformation of the integration contour to the complex plane.
The scale of the deformation is controlled by the parameter
 $\lambda$ defined in Eq.~(\ref{eq:condef}).
The convergence of the numerical integration can be improved significantly 
by choosing an ``optimal'' value for $\lambda$. 
As analyzed in Sec.~\ref{subsubsec:deformationstudies}, values of 
$\lambda$ which are too small lead to contours which are too close 
to the poles on the real axis and therefore lead to bad convergence.
Values of $\lambda$ which are too large can modify the real part of the function 
to an unacceptable extent, and could even change the sign of the
imaginary part if the terms of order $\lambda^3$ become larger than those 
terms linear in $\lambda$. This would lead to a wrong result.
Therefore a four-step procedure is implemented in \secdec{} to 
optimize the value of $\lambda$. These are
\begin{itemize}
\item Ratio check: 
To make sure that terms of order $\lambda^3$ in
Eq.~(\ref{eq:newF}) do not spoil the sign of the imaginary part, the 
ratio of the terms linear and cubic in $\lambda$ are evaluated 
for a quasi-randomly chosen set of samples to 
determine $\lambda_{\text{max}}$. 
The size of the set can be chosen by the user. 
\item Modulus check: The imaginary part is vital at the points where
  the real part of  ${\cal F}$ is vanishing. In these regions,
  the deformation should be large enough to avoid large numerical
  fluctuations due to a highly peaked integrand. Therefore the 
  modulus of each sub-sector function ${\cal F}_i$ is checked at a number 
  of sample points. At the points where the modulus is close to vanishing, 
  the fraction of the value $\lambda_{\text{max}}$ is picked which 
  maximizes the modulus of ${\cal F}_i$. Hereby, the value of $\lambda$ 
  which keeps ${\mathcal F}_i$ furthest from zero is chosen. 
\item Individual $\lambda(i,j)$ adjustments: If the discrepancy of the values of
  $\frac{\partial {\cal F}_i}{\partial x_j}$ for different $x_j$ is very large among 
  one sub-sector $i$, it can be convenient to have an individual 
  parameter $\lambda(i,j)$ for each sub-sector function ${\cal F}_i$ and 
  each Feynman parameter $x_j$. As was shown in 
  Sec.~\ref{subsubsec:deformationstudies}, it is beneficial to have small 
  overall deformations of the integration contour. 
  Therefore each individual parameter $\lambda(i,j)$ is divided by the 
  largest value of $\frac{\partial {\cal F}_i}{\partial x_j}$ for all $x_j$ in 
  one sub-sector $i$, decreasing the overall size of the deformation. 
  If the largest deformations is smaller or equal to one, the $\lambda(i,j)$ are 
  left unchanged.
\item Further optional $\lambda(i,j)$ adjustments: 
\begin{itemize}
\item[1)] If the integrand is expected to be oscillatory and hence 
sensitive to small changes in the deformation parameter $\lambda$, 
\secdec{} can minimize the argument 
of each sub-sector function ${\cal F}_i$ by varying $\lambda(i,j)$. 
The effect is shown in Sec.~\ref{subsubsec:deformationstudies}. 
\item[2)] If the integrand is expected to have (integrable) 
singularities close to the endpoints of the integration ($x_j=0,1$), the 
deformation should be as large as possible in order 
to move the contour away 
from the problematic region. To this end, each individual parameter 
$\lambda(i,j)$ is multiplied by the largest value of 
$\frac{\partial {\cal F}_i}{\partial x_j}$ for all $x_j$ in one sub-sector $i$. 
\end{itemize}
\item Sign check: After the above adjustments to $\lambda$ have been made, 
the sign of Im$(\mathcal F)$ is again checked for a number of sample
points. If the sign is ever positive, this value of $\lambda$ is disallowed.
\end{itemize}
The contour deformation can be switched on or off, see 
App.~\ref{subsec:programinputparameters}. 
The calculation takes longer if a deformation 
of the integration contour is performed, so if the integrand is known to be 
positive definite, the contour deformation option should be switched off. 
It must also be emphasized that for integrands with a complicated singularity 
structure, the success of the numerical integration can critically depend 
on the parameters which tune the deformation 
and on the settings for the Monte Carlo integration. 
\section{Additional capabilities}
\label{sec:additionalsecdeccaps}
\subsection{Evaluation of user-defined functions with arbitrary kinematics}
\label{subsec:userdefinedfuncs}
To calculate a ``standard'' loop integral, it is sufficient 
to specify the numerator, the loop momenta and the propagators. 
The program will construct the integrand in terms of 
Feynman parameters automatically. 
It can also be desirable to take a mixed approach of computing an integral 
numerically after having manipulated it analytically. An example approach 
is given in Chap.~\ref{chap:application1}, where the numerical efficiency 
is shown to be improved by a clever analytical preparation of the 
integrand for the subsequent Monte Carlo integration. In this example, the 
preparation includes the analytical integration of one sub-loop. 
This implies that the constraint $\delta(1-\sum_i x_i)$ has been already used 
to achieve a convenient parametrization, and therefore no primary sector 
decomposition is needed anymore to eliminate the $\delta$-constraint.
In such a case, the user can skip this step in \secdec{} and 
insert the functions to be factorized directly into the Mathematica 
template file, using the favored parametrization.
More generally, this option offers more 
flexibility regarding the functions to be integrated, such as expressions 
for loop integrals which are not in the ``standard form''. For example, 
analytic manipulations which have already been performed 
on the integral can be dealt with as well.
This includes the possibility to perform a deformation of the integration 
contour to the complex plane. \\
To better understand the types of function a user could insert, the reader is 
invited to look back to Eq.~(\ref{eq:thefeynmanloopintegral}). 
A general loop integral in Feynman parametrization thus contains  
a numerator function ${\cal N}$ non-divergent by construction, 
two Symanzik polynomials ${\cal U}$ and ${\cal F}$, whose exponent can 
have either sign and therefore singular points. Furthermore, it contains 
a fully analytical, but arbitrary prefactor ${\cal P}$ allowed to contain singularities 
in the regulator $\eps$, factorizing powers of Feynman parameters, and 
a $\delta$-distribution constraint. 
A user-defined function, may 
contain any of the previously mentioned components or none, with 
the only exception that the $\delta$-constraint needs to either not 
exist or have been integrated out already. 
In a more general form, such a user-defined integral $G_{\text{user}}$ may have 
any of the following components
\begin{align}
G_{\text{user}} = {\cal P} (\eps) 
\prod_{j=1}^{N} \{ \int_0^1 \rd x_j \;x_j^{a_j(\eps)}  \} \;
\mathcal{N}(\eps) \; 
\mathcal{U}^{\text{ex}{\cal U}}(\vec{x},\{ m\},\{ p\}) \; 
\mathcal{F}^{\text{ex}{\cal F}} (\vec{x},\{ m\},\{ p\}) \text{ .}
\label{eq:userdefinedint}
\end{align}
The functions ${\cal N}$, ${\cal U}$ and ${\cal F}$ can be polynomials 
or products of polynomials with the only condition that they 
share a common exponent, $\text{ex}{\cal U}$ 
and $\text{ex}{\cal F}$. The 
factorizing Feynman parameters can appear with exponents $a_j$ dependent 
or independent of $\eps$. 
Any of the exponents may in principle also contain fractional numbers. 
Details about the usage of this option are given in the user manual, see 
App.~\ref{subsec:graphuserdef}.
\subsection{Topology-based construction of the integrand}
\label{subsec:cutconstruct}
As already mentioned in Sec.~\ref{sec:feynparametrization}, the functions ${\cal U}$ and 
${\cal F}$ can be constructed from the topology of the corresponding 
Feynman graph, without the need to assign the momenta for each propagator explicitly. 
The implementation in \secdec{} is such that the 
user only has to label the external momenta, the vertices and the masses 
of a graph. 
An example is given in Sec.~\ref{sec:checkssecdec} and more examples 
can be found in \secdec. 
This feature of constructing the graph topologically is only implemented 
for scalar integrals so far. The syntax is explained in App.~\ref{subsec:graphtopo}.
\subsection{Looping over ranges of parameters}
\label{subsec:multinumerics}
In phenomenological applications usually not just one kinematical 
point is of interest, but looping over ranges of parameters becomes 
necessary. To this end, it is beneficial to decrease the computation 
time where possible.  
The algebraic part of \secdec{} can deal with symbolic expressions for 
the kinematic invariants, or other parameters contained in the integrand. 
Consequently, the decomposition and subtraction need only be done 
once, producing functions which contain general kinematics. 
The generality of these functions allows for the computation of 
many sets of different values for the invariants. 
\secdec{} allows for an automated calculation
of many numerical points, minimizing the effort for the user. 
Scripts are provided for ``standard'' loop, user-defined and 
more general parametric integrals, see App.~\ref{subsec:appendixmultinumerics}.
%
%
\subsection{Integration-by-parts relations}
\label{subsec:ibprelations}
After the iterated sector decomposition has been performed, 
poles of the type 
\begin{align}
G_l(A_j,B_j,\vec{t},\eps) \propto  I (A_j,B_j,\vec{t},\eps) =
 \int_0^1 \rd t_j \; t_j^{A_j+ B_j \eps} R(\vec{t},\eps)
\end{align}
can arise in an integral $G_l$ of the sub-sector $l$, where $I$ is 
a sub-sector integrand. An exponent 
$A_j =-2$ is associated with a spurious linear pole, powers $A_j <-2$ 
correspond to spurious poles of higher order. These terms must 
be artificial because a renormalizable gauge theory must be integrable. 
The function $R(\vec{t},\eps)$ 
denotes the residue integrand after subtraction, compare 
Eq.~(\ref{eq:extractpoletaylorseries}). Choosing $A_j =-2$, it 
would read
\begin{align}
R(\vec{t},\eps) = 
\mathcal{I}(t_{\alpha_j},\{ t_{\alpha_i \neq \alpha_j} \},\eps) - 
\mathcal{I} (0,\{ t_{i \neq j} \},\eps) - 
t_j \Big [ \mathcal{I}^{(1)}(t_j,\{ t_{i \neq j} \},\eps) \Big ]_{t_j=0} \text{ .}
\end{align}
Even in the limit of a vanishing $t_j$, the integrand will remain integrable 
and finite assuming the decomposition into plus-distributions has 
already been performed. While approaching $t_j=0$, both the numerator 
and the denominator will become very small. This can introduce numerical 
instabilities resulting, e.g., from rounding errors. To mitigate this, the 
sub-sector integrand $I$ can be integrated by parts
\begin{align}
\non I (A_j,B_j,\vec{t},\eps) =& \;
\Big [ \frac{t_j^{1+ A_j+ B_j \eps}}{1+ A_j+ B_j \eps} R(\vec{t},\eps) \Big ]^1_0  - \\
&\quad \; \frac{1}{1+ A_j+ B_j \eps} \int_0^1 \rd t_j \; t_j^{1+ A_j+ B_j \eps} 
\frac{\partial}{\partial t_j}R(\vec{t},\eps) \\
\non =&\; \frac{1}{1+ A_j+ B_j \eps} \\
&\quad \times \Big [ R(1,\{ t_{i \neq j} \},\eps)  - 
\int_0^1 \rd t_j \; t_j^{1+ A_j+ B_j \eps} 
\frac{\partial}{\partial t_j}R(\vec{t},\eps) \Big ] \text{ ,}
\end{align}
thereby reducing the negative power in the Feynman parameter $t_j$ 
by one and enhancing numerical stability, see Ref.~\cite{Carter:2011uza} 
for a more detailed description of the implementation. This procedure 
is automated for arbitrary pole order.  
\subsection{Leaving functions implicit during the algebraic part}
\label{subsec:dummy}
When evaluating general parametric functions, the user 
may wish to introduce a ``dummy'' function depending 
on (some of) the integration parameters, specifying 
the actual form of the function later at the 
numerical integration stage.
There are a number of reasons why one might want to leave
functions implicit during the algebraic stage: 
for example, squared matrix elements typically contain
large but finite functions of the phase space variables in the numerator, so
the algebraic part of the calculation will be
quicker and produce much smaller intermediate files if these functions are left
implicit. Also, one might like to use a number of measurement
functions and be able to specify or change them without 
having to perform the decomposition more than once. 
Note that one may use more than one implicit function at a time, 
and that these functions can have any number of arguments. 
The syntax and usage are described in App.~\ref{subsec:appendixdummy}.
\subsection{Assessing the reliability of the numerical result} 
\label{subsec:reliabilityofnumericalresult}
When dealing with numerical techniques, the knowledge of 
the reliability of the result is of major importance. 
Although the integrands of all sectors are stored analytically, 
it may, especially when dealing with complicated integrals, be 
time consuming to analyze these, 
either due to the abundance of Feynman parameters appearing 
in one function, or simply due to a large number of functions. 
And even then, the numerical integrator may still 
appear to be a black box. 
It is therefore appealing to get an estimate for the 
correctness of the stated uncertainty. 
The numerical integrators contained in the {\sc Cuba 
library}~\cite{Hahn:2004fe,Agrawal:2011tm} 
return a probability for an estimated numerical 
uncertainty to be erroneous. A maximal probability of $1$ 
therefore means that the stated uncertainty of a result cannot be trusted. 
The program collects the maximal probability for each computed 
order in the dimensional regulator $\eps$. 
The probability can be reduced by 
increasing the number of sampling points used in an evaluation. 
More options tuning the numerical integration parameters are 
given in the user manual, see 
App.~\ref{sec:appendix:usermanual}. 
To prevent a suffering from underestimation of the true 
error given by the numerical integrators, it is beneficial 
to check a result with different integrators when dealing 
with complicated integrands. 
\subsection{Automated remapping to one endpoint} 
\label{subsec:endpointremapping}
The program is capable of remapping singularities in 
a Feynman parameter $x_j$  
appearing at the endpoint $1$ of the integration region to zero.
A remapping of the singularity to zero becomes necessary, 
if one of the sub-sectors of ${\cal U}$ or ${\cal F}$ in the case of 
loop integrals, or one of the $P_i$ of Eq.~(\ref{eq:general}) in the 
case of more general parametric functions, 
diverge in the limit of one or more $t_j \to1$. 
It works as follows. 
The integration region is split into two parts at the point $a$, where $a$ 
is chosen rather arbitrarily 
\begin{align}
\int_0^1 f(x) \text{d}x =&  \int_0^a f(x) \text{d}x + \int_a^1 f(x) \text{d}x \label{eq:remap1}\\
=& \int_0^1\frac1a f(\frac{\tilde{x}}{a}) \text{d}\tilde{x} + 
 \int_0^1\frac1a f(1-\frac{x'}{a}) \text{d}x' \text{ .}\label{eq:remap2}
\end{align}
From Eq.~(\ref{eq:remap1}) to Eq.~(\ref{eq:remap2}) the substitutions $x \to \frac{\tilde{x}}{a}$
and $x \to (1-\frac{\tilde{x}}{a})$ are applied to the first and 
the second integral of the right-hand side, respectively. Hereby, 
a remapping of the integration boundaries to the unit 
hypercube is achieved. 
The resulting functions $f$ either vanish for $\lim_{\tilde{x}\to 0}$ or 
do not vanish at all. 

\medskip

%
%
%
%
%
%
In \secdec, the integration region of those 
integrals over Feynman 
parameters $t_{j_i}$ leading to a divergence at the 
upper integration boundary, is split at $1/2$ and the resulting 
two integrals are remapped to the unit interval.  \\
This splitting of the integration region is performed before the 
iterated sector decomposition. 
It increases the number of primary 
sectors by $2 \, n$, where $n$ is 
the number of split integration variables, 
in favor of improved numerical convergence. 
After all singularities at the endpoint are remapped to the 
lower integration boundary, an iterated sector 
decomposition can be applied. 
The occurrence of singularities at both endpoints is typically 
encountered in massless diagrams. In 
Sec.~\ref{subsec:subloopintegration}, the integration 
of one sub-loop prior to the treatment of the full integral serves as 
an example. 
%
%
\subsection{A word on the numerical integration}
\label{subsec:program:cubaparameters}
The numerical integration forms a crucial part in the calculation of any 
type of integrand function resulting from a Laurent series expansion in $\eps$. 
\secdec{} contains interfaces to six different numerical integrators, 
{\sc Bases}~\cite{Kawabata:1995th}, 
Vegas, Suave, Divonne and Cuhre contained in 
the {\sc Cuba library}~\cite{Hahn:2004fe,Agrawal:2011tm}, and {\sc NIntegrate} 
contained in Mathematica~\cite{Wolfram}. The user 
is offered to choose one of these in the input files, see 
App.~\ref{subsec:programinputparameters}. 
It is crucial to have the parameters for the numerical 
integrator under control. 
\secdec{} incorporates several options, allowing for a good 
adjustment of these parameters. 
Two of them are the desired relative  
and absolute accuracy, where the desired absolute 
accuracy is necessary for integrals close to zero. 
If the real or imaginary part of the integral 
tends to zero, the relative accuracy can never be reached. 
The numerical integrators therefore attempt to find an estimate $\hat{I}$ 
of the integral $I$ that fulfills 
\begin{align}
\hat{I}-I \leq \text{max}(\eps_{\text{abs}},\eps_{\text{rel}} I) \text{ ,}
\end{align} 
see e.g. Ref.~\cite{Hahn:2004fe}. 
For vanishing values of the integral, the integration time 
then heavily depends on the value chosen for the 
desired absolute accuracy. 
When looping over ranges of kinematic invariants, 
points below threshold have a zero imaginary part. 
This can artificially increase the computation times, 
if the absolute accuracy goal was set to a reasonable, 
but small value. This artifact can be circumvented by 
setting the lowest threshold symbolically when 
specifying the integrand. 
This new feature is incorporated in \secdec{}, 
switching off the contour deformation below the 
user-defined threshold.
Hence, unnecessarily long calculations are 
avoided in kinematic regions where 
the imaginary part is known to be zero. 

\medskip

The other selectable parameters 
are described in the manuals of the different numerical integrators, 
{\sc Bases}~\cite{Kawabata:1995th} and the 
{\sc Cuba library}~\cite{Hahn:2004fe,Agrawal:2011tm}, 
and in the user manual in App.~\ref{sec:appendix:usermanual}. 
Regarding the advantages of these integrators, 
{\sc Bases} is a Fortran compatible Monte Carlo integrator that 
allows for a sequential evaluation only. In the sequential mode  
it is fast, producing reliable results. 
The four integrators included in the  {\sc Cuba library} can run in parallel mode 
and are usable with Fortran and C/C$^{++}$. While Vegas gives 
very stable results and tends to overestimate the numerical integration 
uncertainty, Divonne is extremely adaptive but occasionally underestimates 
integration uncertainties. The latter is very useful for very complicated integrands 
and is especially good in regions close to a threshold. 
Suave is heuristically found to converge slowly, but gives very stable 
results. While Vegas, Suave and Divonne are mainly Monte Carlo integrators 
which can sample pseudo and quasi-random numbers, Cuhre is a fully 
deterministic integrator, able to reach high accuracy if the integrand 
is comparatively simple. 

\medskip 

In the integration phase, \secdec{} allows for different choices regarding 
the number of integrands to be summed before integration. 
Setting {\it togetherflag=0} and {\it grouping=0} at the same time 
leads to the separate integration of each sector in all different pole 
structures $\text{P}i\text{l}j\text{h}k$ and orders $ord$ in the regulator $\eps^{ord}$.  
Here $i$ denotes the number of logarithmic poles, $j$ the number 
of linear poles and $k$ the number of higher poles in $\eps$. 
The summation of some integrand files before 
integration is enabled by entering the allowed summed size 
of the grouped files in bytes, e.g. {\it grouping=2000000} corresponds 
to a grouping of integrands with a maximal total file size of around 2 MB. 
Switching on {\it togetherflag=1}, all integrands leading to 
the full coefficient of a certain order in the regulator 
are first summed up and then integrated numerically. \\
The uncertainty resulting from the Monte Carlo 
integration is expected to be bigger when each sector 
is integrated individually before summing up all sectors. Yet, the 
difference turns out not to be large, as the numerical integrator 
can tackle single functions much better and yield more accurate 
results. 
Grouping files before integration can also lead to a 
faster convergence if the integrand contains highly oscillating terms 
which cancel each other out, but will in general  
slow down the numerical calculation if the polynomial structure is 
rather smooth. 
\section{Operational sequence}
\label{sec:op_sequence}
%
%
\begin{figure}[htb!]
	\begin{center}
	      \includegraphics[width=1.0\textwidth]{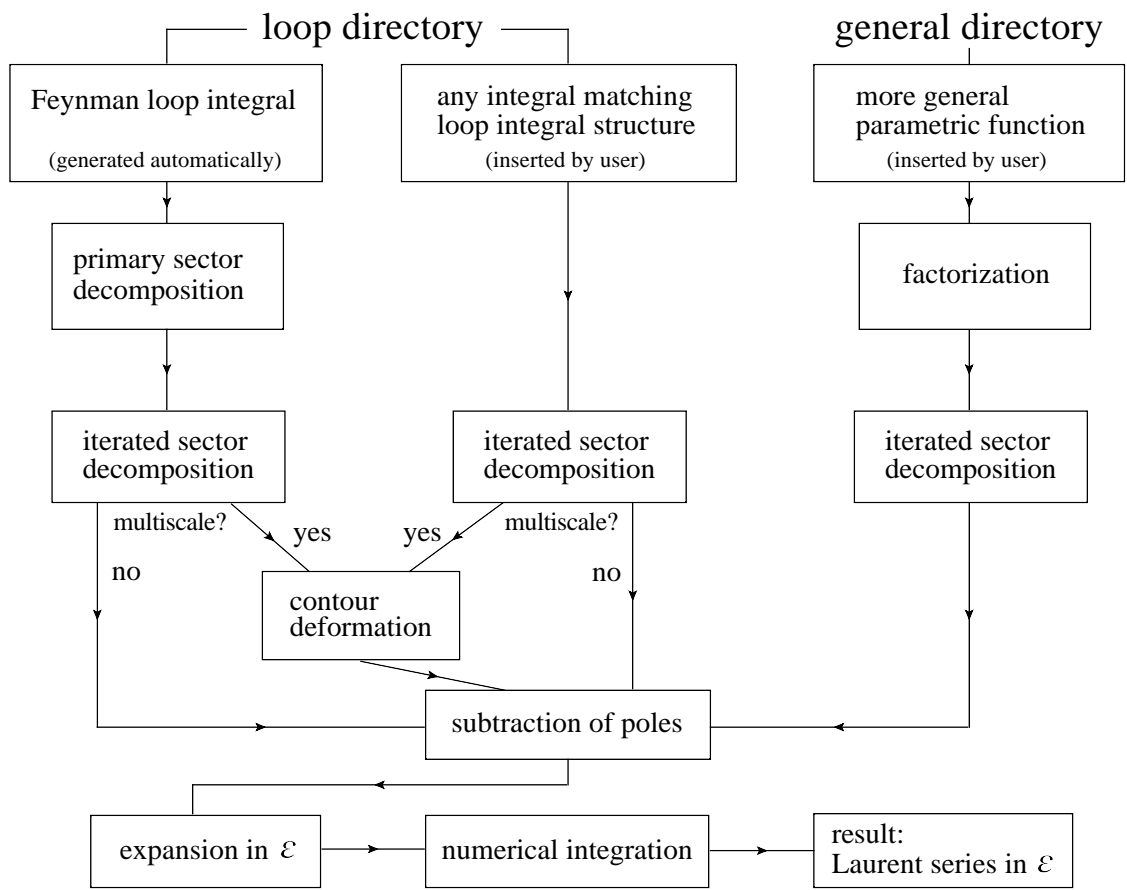}
	\end{center}
	\caption{The operational sequence of \secdec. This flowchart shows the main 
	steps the program performs to produce the result as a Laurent series in the regulator 
	$\eps$.}
	\label{fig:structure2} 
\end{figure}
The program is divided into two main directories, {\tt loop} and {\tt general}, 
respectively. 
They have the same global structure, only individual files 
are specific to either loop integrals and integrals of loop structure or to 
more general parametric functions. 
The directories contain a number of Perl scripts steering 
Mathematica~\cite{Wolfram} modules (located in the subdirectory  {\tt src}), 
writing the files necessary for numerical integration and executing them. 
The scripts use further Perl modules contained in the 
subdirectory {\tt perlsrc}. 

\medskip

To calculate a loop integral within the standard or user-defined setup, 
the user needs to enter the {\tt loop} directory in \secdec, while for the 
computation of parameter integrals the user is referred to the 
{\tt general} directory. 
When operating \secdec{} in either of the two directories, 
the program works through algebraic and numerical parts.
The algebraic part uses Mathematica code and performs the 
sector decomposition, the subtraction of the 
singularities, the expansion in $\eps$ and the generation of the 
files necessary for the numerical integration.
In the numerical part,  Fortran or C$^{++}$ functions forming the coefficient 
of each Laurent series term in $\eps$ are integrated using the 
Monte Carlo integrators contained in the
\textsc{Cuba library}~\cite{Hahn:2004fe,Agrawal:2011tm},  
\textsc{Bases}~\cite{Kawabata:1995th} or {\sc NIntegrate}~\cite{Wolfram}. 

\medskip

The flowchart of \secdec{} shows the basic 
building blocks to calculate multi-loop or more general parametric 
integrals, see Fig.~\ref{fig:structure2}. 
The Mathematica source files 
are located in the subdirectories {\tt src/deco} (files used for the decomposition), 
{\tt src/subexp} (files used for the pole subtraction and expansion in $\eps$) and 
{\tt src/util} (miscellaneous useful functions).  
The Robodoc~\cite{Robodoc} documentation 
is located in the subdirectory {\tt doc}. It contains an index to look 
up a documentation of the source code in html format by loading 
{\tt masterindex.html} into a browser.
%
The program comes with example 
input and template files in the subdirectories {\tt loop/demos} 
and {\tt general/demos}, respectively. 
Further details on the installation and operation 
are found in App.~\ref{subsec:install} and App.~\ref{subsec:operation}. 
\section{Selection of checks and examples}
\label{sec:checkssecdec}
In the following, two loop topologies are looked 
at in more detail. The first example is a two-loop three-point function 
which serves as the default graph when running \secdec{} without 
an alteration of the {\tt myparamfile.input, mytemplatefile.m} in the 
{\tt loop} directory. The second example is the two-point two-loop 
function with five massive propagators. Different cases of 
numerators and numbers of mass scales are discussed. 
Further highly non-trivial two-loop examples are discussed in 
Chap.~\ref{chap:application1}.
For a detailed description of examples in 
the {\tt general} directory, see the directory {\tt SecDec/general/demos} 
and Refs.~\cite{Carter:2010hi,Carter:2011uza,Borowka:2012yc}. 
\subsection{A two-loop three-point function}
\label{subsec:p126}
\begin{figure}[htb!]
\begin{center}
\includegraphics[width=5.cm]{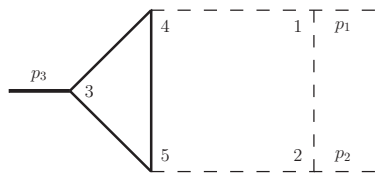}
\end{center}
\caption{Two-loop vertex graph $P_{126}$, containing a massive triangle loop.  
Solid lines are massive, dashed lines are massless. The vertices are labeled to match the 
construction of the integrand from the topology as explained in the text.}
\label{fig:P126}
\end{figure}
In this example, three of the new features of \secdec{} version 2 
are demonstrated: the construction of the two Symanzik polynomials 
$\mathcal U$ and $\mathcal F$ from the
topology of the graph and the evaluation of the graph in
the physical region. Finally, it is shown how results for a whole set of 
different numerical values for the kinematic invariants can be produced 
and plotted in an automated way. 
The diagram under discussion is a two-loop three-point function, 
see Fig.~\ref{fig:P126}. It has been studied extensively adopting either 
an analytical~\cite{Davydychev:2003mv}, or a 
numerical~\cite{Ferroglia:2003yj,Bonciani:2003hc} approach. 
The name for the diagram, $P_{126}$, was introduced in Ref.~\cite{Davydychev:2003mv}. 

\medskip

\begin{figure}[htb!]
\begin{center}
\includegraphics[width=0.7\textwidth]{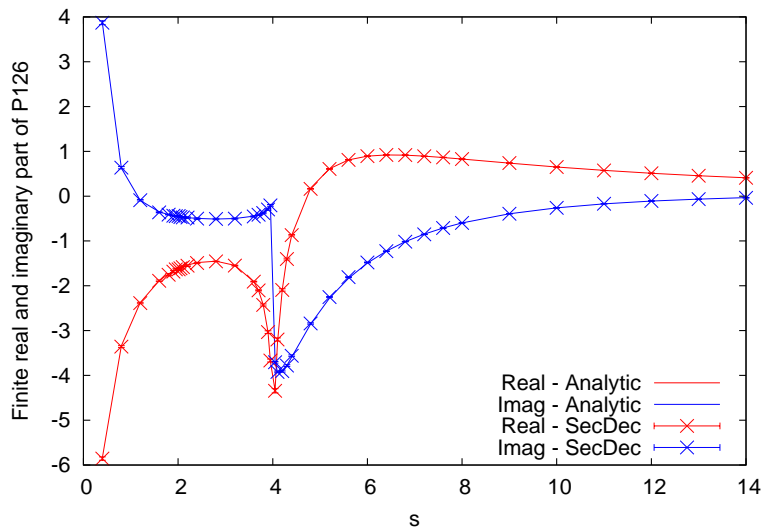}
\caption{\label{fig:P126result}
Comparison of analytic and numerical results for the diagram $P_{126}$ using $m^2=1$ 
and varying $s$. Due to the high accuracy of the numerical integration, error 
bars are barely seen. }
\end{center}
\end{figure}

The template file {\tt templateP126.m} in the {\tt loop/demos} subdirectory 
contains the following lines \\
{\it proplist} =
{\small\{\{ms[1],\{3,4\}\},\{ms[1],\{4,5\}\},\{ms[1],\{5,3\}\},\{0,\{1,2\}\},\{0,\{1,4\}\},\{0,\{2,5\}\}\}};\\
{\it onshell} = \{ssp[1] $\to$ 0,\,ssp[2] $\to$ 0,\,ssp[3] $\to$ sp[1,2]\};\\
where each entry in {\tt proplist} corresponds to a propagator of 
the diagram; the first entry is the mass of the
propagator and the second entry contains the labels of the 
two vertices which the propagator connects. 
The labels for the vertices are as shown in Fig.~\ref{fig:P126}.
For vertices containing only internal propagators the labeling is arbitrary.
The {\textit onshell} conditions in the above example state that
$p_1^2=p_2^2=0,\,p_3^2=s_{12}=s$.
Results for the finite $\eps^0$ part of graph $P_{126}$ agree very 
accurately with the analytic prediction, compare 
Fig.~\ref{fig:P126result}. 
The calculation time for the numerical integration of the finite part 
is around 5 secs, for a relative accuracy of 
about 1\% and an absolute 
accuracy of $10^{-5}$ using the integrator Divonne of the \textsc{Cuba library}.  
This is for a typical point far from the 
$s=4\,m^2$ threshold on an 
Intel{\small \textregistered}\,Core\texttrademark\,i7-2600 CPU at 3.40GHz using two cores. 
For a point close to threshold ($s/m^2=3.9$), the timings are similar.
To run this example, from the {\tt loop} directory, issue the command
`{\it ./launch -d demos -p paramP126.input -t templateP126.m}'. 
The diagram $P_{126}$ can also be computed using the 
user-defined setup, see App.~\ref{subsec:graphuserdef} for a detailed 
explanation. 
\subsubsection{Producing data files for sets of numerical values}
To loop over a set of numerical values for the invariants  $s$ and $m^2$
once the C++ files are created, issue the command \\
`{\it perl multinumericsloop.pl -d demos -p multiparamP126.input}'. \\
This will run the numerical integrations for the values of $s$ and $m^2$ specified
in the file {\tt demos/multiparamP126.input}. 
The files containing the results will be found in {\tt demos/2loop/P126}, and the
files {\tt p-2.gpdat, p-1.gpdat} and {\tt p0.gpdat} will contain the  
data files for each point, corresponding to the coefficients of 
$\eps^{-2},\eps^{-1}$ and $\eps^0$, respectively. 
These files can be used to plot the numerical results using gnuplot, see 
Fig.~\ref{fig:P126result} for an exemplary result. \\
The same steps can be performed using the user-defined option, where the command 
reads `{\it perl multinumericsuserdefined.pl -p multiparamuserdefined.input}' 
and where the values for $s$ and $m^2$ are given in the 
{\tt multiparamuserdefined.input} file. 
\subsection{Massive tensor two-loop two-point functions}
\label{subsec:bubblerankintegrals}
In this subsection the option to evaluate integrals with a non-trivial numerator 
is demonstrated by calculating two-loop two-point functions involving different 
mass scales. This implies that a reduction to only scalar (master) integrals is 
not necessarily needed.
\begin{figure}[htb]
\begin{center}
\includegraphics[width=5.2cm]{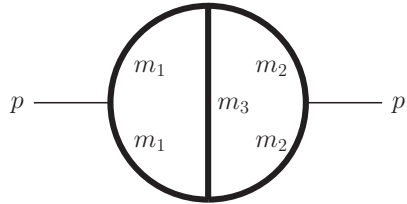} 
\caption{Two-loop bubble diagram with  different masses}
\label{fig:bubble2m}
\end{center}
\end{figure}
The exemplary diagram considered here contains up to four different scales, 
see Fig.~\ref{fig:bubble2m}. 
As an example, a non-trivial numerator a coefficient function 
resulting from a rank three integral Lorentz decomposition 
is chosen
\begin{align}
&\mathcal{G}_{B} = \left( \frac{1}{\mathrm{i} \pi^{\frac{\mathrm{D}}{2}}}\right)^2 
 \int \mathrm{d^D}k_1\, \mathrm{d^D}k_2 \frac{ (k_1\cdot k_2)\,(k_1\cdot p_1)}{D_1\ldots D_5} \text{ ,}
\label{eq:bubbleint}\\
& D_1=k_1^2-m_1^2, \quad D_2=(k_1+p_1)^2-m_1^2, \quad D_3=(k_1-k_2)^2-m_3^2, \\
& D_4=(k_2+p_1)^2-m_2^2, \quad D_5=k_2^2-m_2^2 \;, 
\end{align}
where the causal $i \delta$ prescription is omitted. 
The fact that this tensor integral is reducible does not play a role 
here, as the purpose is to demonstrate that a reduction may become 
obsolete, when considering the short integration times for the tensors. \\
\begin{figure}[htb!]
\begin{center}
\subfigure[scalar integral] {
\includegraphics[width=0.7\textwidth]{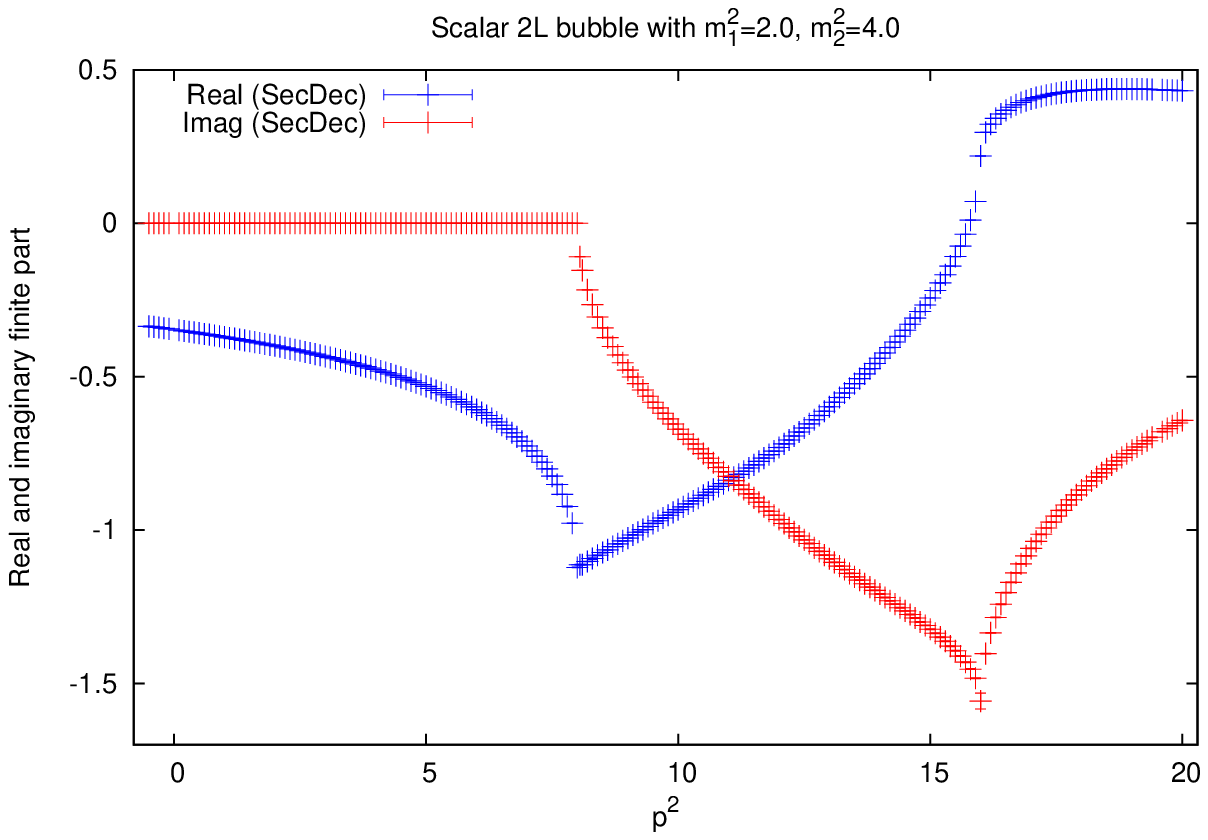} 
\label{fig:bubblescalar2m} }\\
\subfigure[rank 3 tensor integral] {
\includegraphics[width=0.7\textwidth]{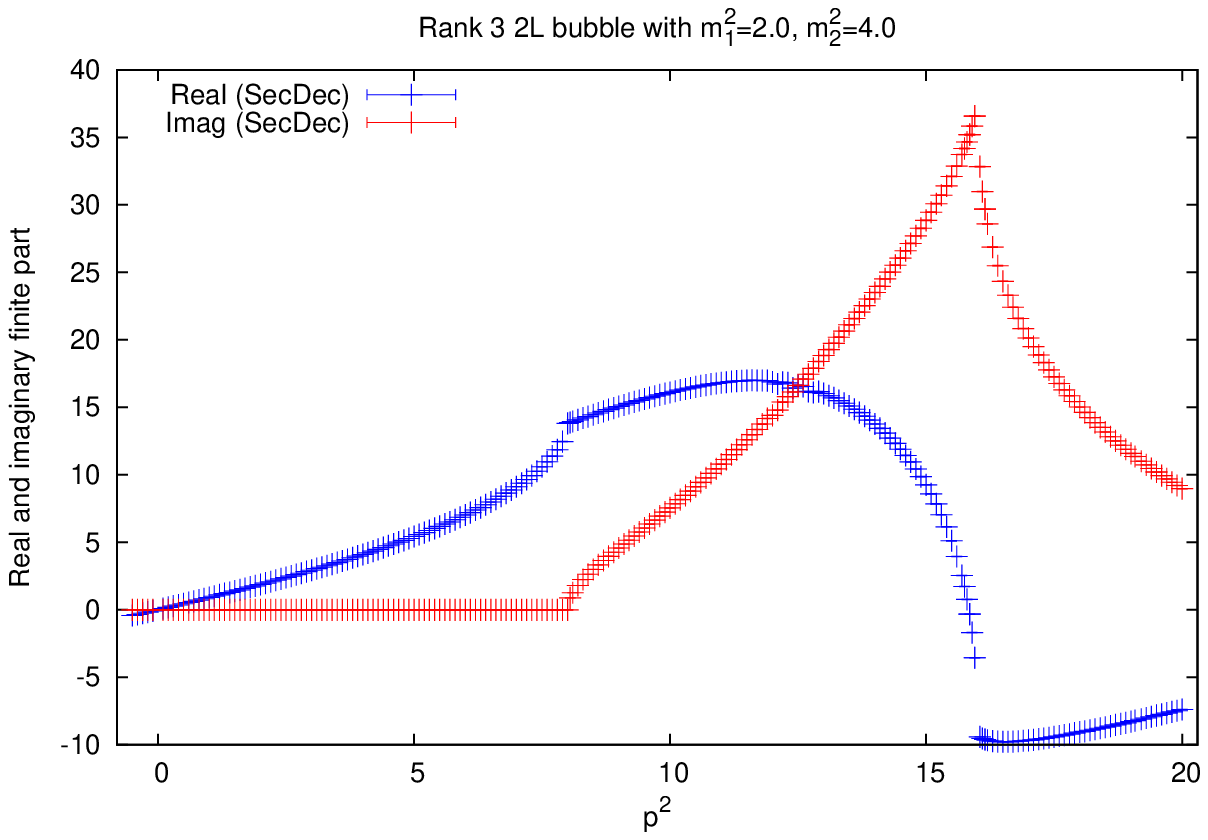} 
\label{fig:bubblerank32m} }
\end{center}
\caption{Results for the real (blue) and imaginary (red) parts of 
the two-loop two-point topology of Fig.~\ref{fig:bubble2m} are shown 
in the scalar case~\subref{fig:bubblescalar2m} and for the 
the rank three coefficient function $\mathcal{G}_{B}$~\subref{fig:bubblerank32m} 
with numerator $(k_1\cdot k_2)\,(k_1\cdot p_1)$. 
The masses are $m_1^2=2, m_2^2=4, m_3=0$.}
\label{fig:bub_m3zero}
\end{figure}
\\%
\begin{figure}[ht!]
\begin{center}
\subfigure[scalar integral] {
\includegraphics[width=0.7\textwidth]{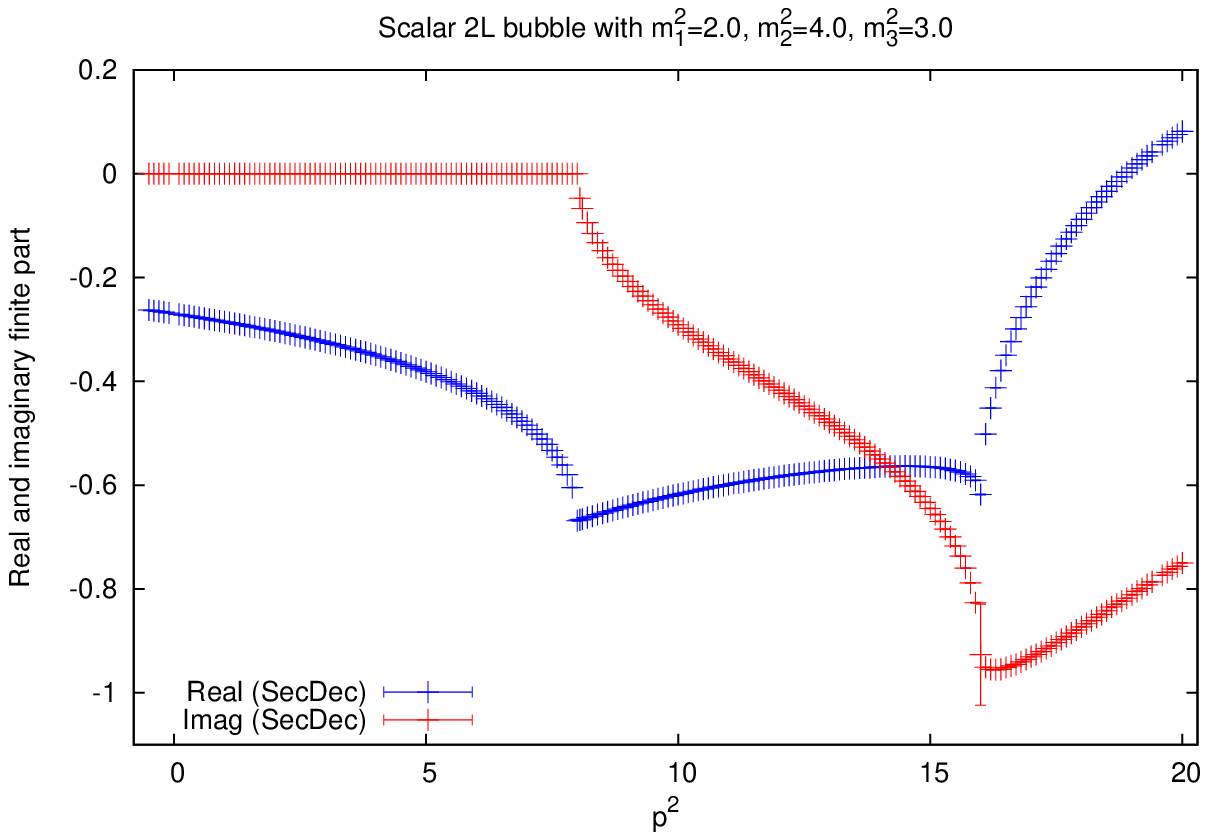}
\label{fig:bubblescalar3m} } \\
\subfigure[rank 3 tensor integral] {
\includegraphics[width=0.7\textwidth]{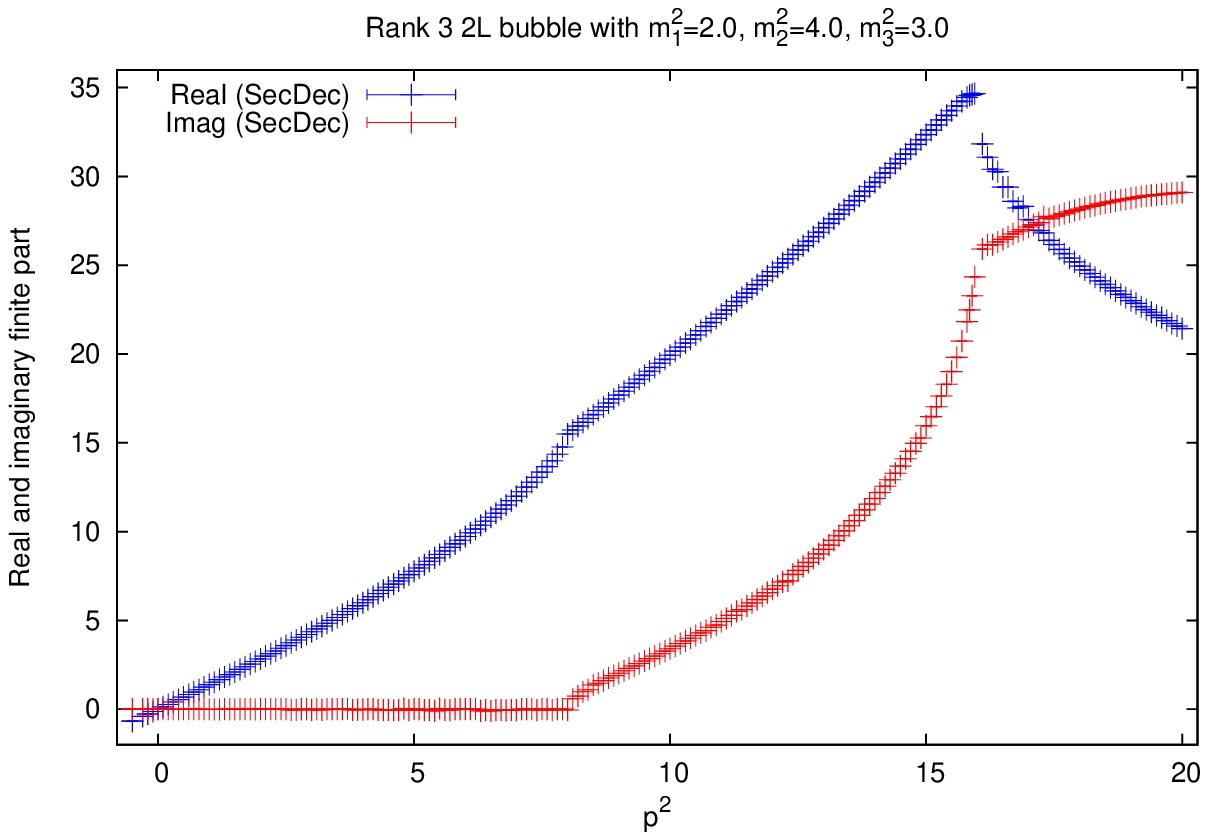} 
\label{fig:bubblerank33m} }
\end{center}
\caption{Results for the real (blue) and imaginary (red) parts of 
the two-loop two-point topology of Fig.~\ref{fig:bubble2m} are shown 
for three non-vanishing masses 
in the scalar case~\subref{fig:bubblescalar3m} and for the 
the rank three coefficient function $\mathcal{G}_{B}$~\subref{fig:bubblerank33m} 
with numerator $(k_1\cdot k_2)\,(k_1\cdot p_1)$.
The masses are $m_1^2=2, m_2^2=4, m_3^2=3$.} 
\label{fig:bubble3m} 
\end{figure}
\\Results for the scalar and tensor integrals with $m_3=0$ are shown in 
Fig.~\ref{fig:bub_m3zero}, while results for $m_3^2=3$ are shown in 
Fig.~\ref{fig:bubble3m}. 
\begin{figure}[htb!]
\begin{center}
\subfigure[] {
\includegraphics[width=0.7\textwidth]{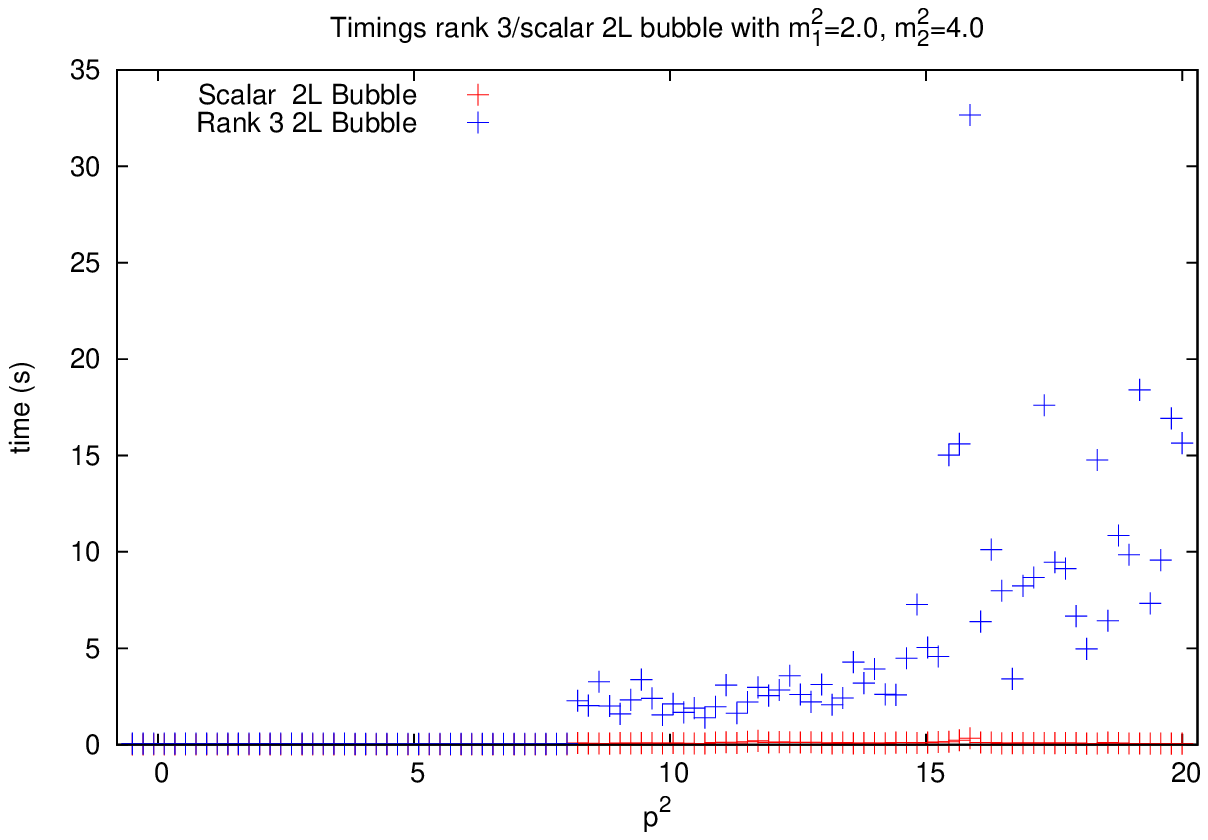} 
\label{fig:timingsbubbles2m}} \\
\subfigure[] {
\includegraphics[width=0.7\textwidth]{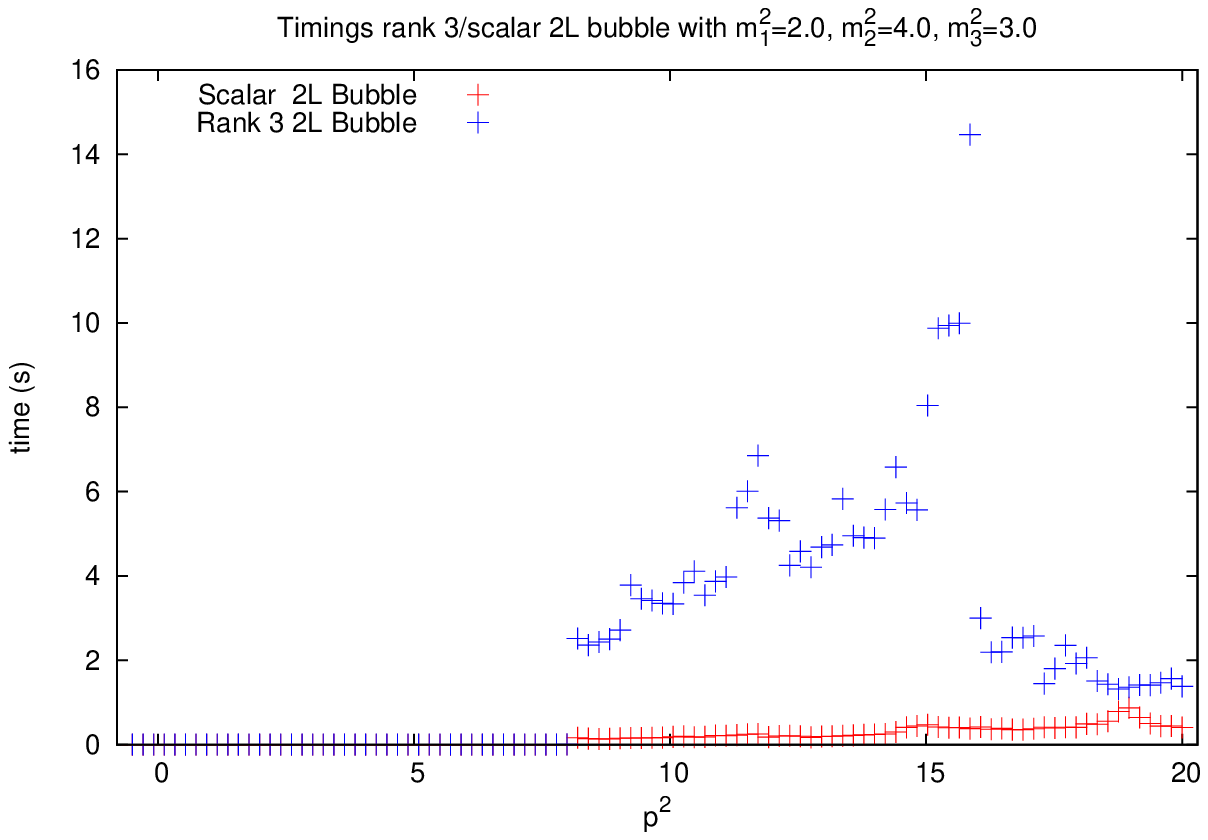} 
\label{fig:timingsbubbles3m}}
\end{center}
\caption{Comparison of evaluation times between  scalar and rank three tensor integrals
corresponding to a two-loop two-point function with \subref{fig:timingsbubbles2m} two, 
and \subref{fig:timingsbubbles3m} three masses.
The red points are the evaluation times in seconds for the scalar integral at a given 
kinematic point, the timings for the rank three tensor integral are marked in blue. 
The timings were obtained on computers with Intel i7 processors and 4 cores.}
\label{fig:bubble3m_timings}
\end{figure}

\medskip

The timings are expected to be higher for the rank three coefficient 
function because its leading pole is of order ${\cal O}(\eps^{-2})$, while 
the scalar integral is finite. In the case of finite integrals, less functions  
need to be integrated, leading to a faster result. 
A comparison of the timings of the scalar massive two-point integrals  
with the rank three two-loop two-point coefficient function shows, 
that for values of $p^2$ above the mass threshold at $p^2=4m_2^2$, 
the timings for the rank three integral coefficient function do not differ 
much from the ones for the scalar integrals, 
compare Fig.~\ref{fig:timingsbubbles2m} for the $m_3^2=0$ case 
and Fig.~\ref{fig:timingsbubbles3m} for $m_3^2=3$. 
A relative accuracy of 0.1\% and an absolute accuracy of $10^{-5}$ 
was required for the Monte Carlo integration in scalar and rank 3 
integrals alike. 
With the new feature 
of symbolically defining a threshold below which the 
contour deformation is switched off, 
unnecessarily long calculations are avoided in kinematic regions where 
the imaginary part is known to be zero. 
In the case of the rank three two-loop bubble with two masses, 
no threshold appears in the ${\cal O}(\eps^{-2})$ coefficient, 
a threshold at $p^2=4 m_2^2$ appears in the sub-leading pole 
and the lowest threshold of the finite part is located at 
$p^2=4 m_1^2$. As only the lowest threshold is incorporated 
in the user-definition, the computing times of this integral are 
largest, where the imaginary part in the 
pole coefficients is zero, compare Fig.~\ref{fig:timingsbubbles2m} and 
Sec.~\ref{subsec:program:cubaparameters}.
%
%
%
The minimal numerical integration time for a kinematic point 
below threshold is $0.6$ms for the scalar two-loop bubble with $m_3^2=0$. 
Above threshold, the scalar two-loop bubble integrals minimally take 0.1\,secs 
for $m_3^2 \neq 0$ and 0.03\,secs for $m_3^2=0$. 
In Fig.~\ref{fig:timingsbubbles3m}, a small peak in the timings can be 
observed for $p^2=12$. When solving the Landau equations for the 
two-loop two-point function with two masses, a sub-threshold can be 
detected at exactly $p^2=12$. Given this observation, it may sometimes 
be of interest to have a closer look at the timings and possibly learn 
something about the singularity structure of the integrand. 
%
%
\section{Future developments}
\label{sec:secdecfuture}
%
%
The upgraded version 2 of the program \secdec{} was 
presented. The main new feature, an implementation of an 
automated deformation of the integration contour to analytically 
continue the integrand, was described along with other new 
capabilities. It allows for the computation of multi-loop integrals 
with in principle no limitation on the number of scales involved. 
Examples to show its full power are discussed in 
Chap.~\ref{chap:application1}, 
where numerical results for diverse two-loop four-point topologies 
are shown.

Although comparatively fast, 
the numerical evaluation of multi-loop diagrams can, in general, still not 
compete with the evaluation time of analytical results. 
Including more analytical calculations would therefore be
very beneficial. 
A future version of the program 
could therefore integrate functions analytically where 
possible. Especially for the pole parts, the integration 
of several Feynman parameters prior to numerical 
integration is feasible. This can lead to a significant 
decrease in the numerical integration times, as became 
apparent in the discussion of the timings in 
Sec.~\ref{subsec:bubblerankintegrals}.

Furthermore, a decrease in the numerical integration times 
can be achieved using the 
fastest Monte Carlo integrator available for one-dimensional 
parameter representations, \textsc{Quadpack} which 
is included in the GNU scientific library. Currently, 
the integrators Vegas or Suave are used in such cases, 
although they only rise to their full power when multiple 
integration parameters are involved.

Another missing piece, as mentioned in Sec.~\ref{subsec:stoppingalgorithm}, 
is the implementation of an algorithm which is guaranteed to 
stop. It should set in automatically (if not already chosen in the 
beginning) when the heuristic algorithm runs into an 
infinite recursion. 

\medskip

Although some of the mentioned new features are already 
implemented in a private version of the code, 
they will be made publicly available as a whole in version 3 of \secdec. 

\medskip

Besides new features accelerating \secdec, further technological 
developments are needed, e.g., for kinematic configurations 
very close to pinch singularities. An extensive treatment of 
line singularities within the integration region or singularities 
appearing close to the endpoints 
of the integration region is still to be found.

Beyond improvements on the computation of the multi-loop 
integrals, interfaces 
to other programs are highly desirable. 
For example, the construction of interfaces to existing 
amplitude reduction programs would mean 
another cornerstone towards the long-term goal of 
assembling a highly automated program for 
the computation of processes beyond one loop. 
%
%
%
%
%
%
%
%

%% file: application1/application1.tex
\chapter[Non-planar two-loop four-point integrals with external or internal masses]{Non-planar two-loop \\four-point integrals with \\external or internal masses}
\label{chap:application1}%
\begin{figure}[htb!]
\begin{center}
\subfigure[$ggtt_2$]{\raisebox{-2pt}{\includegraphics[width=0.4\textwidth]{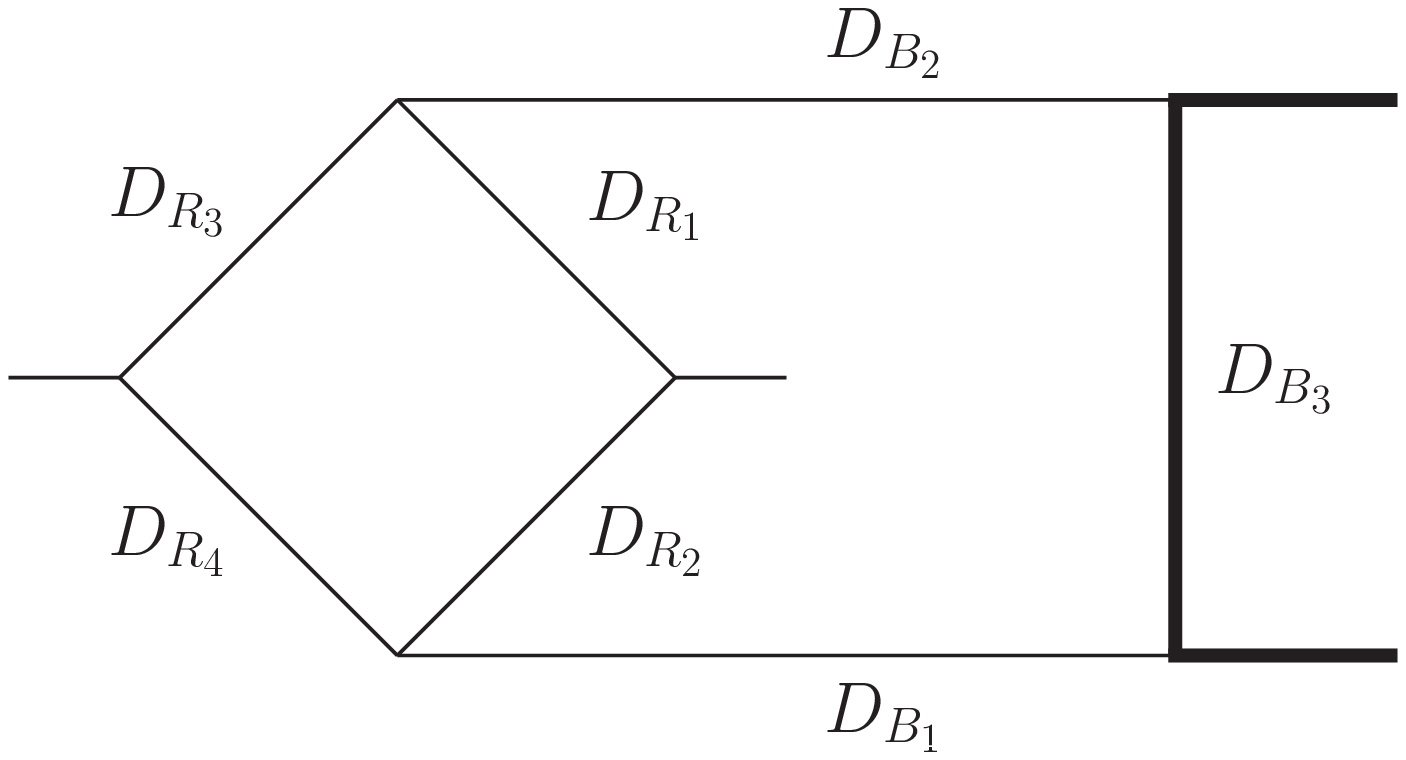}}
\label{fig:ggtt2diagram}} \hspace{13pt}
\subfigure[$ggtt_1$]{\raisebox{8pt}{\includegraphics[width=0.45\textwidth]{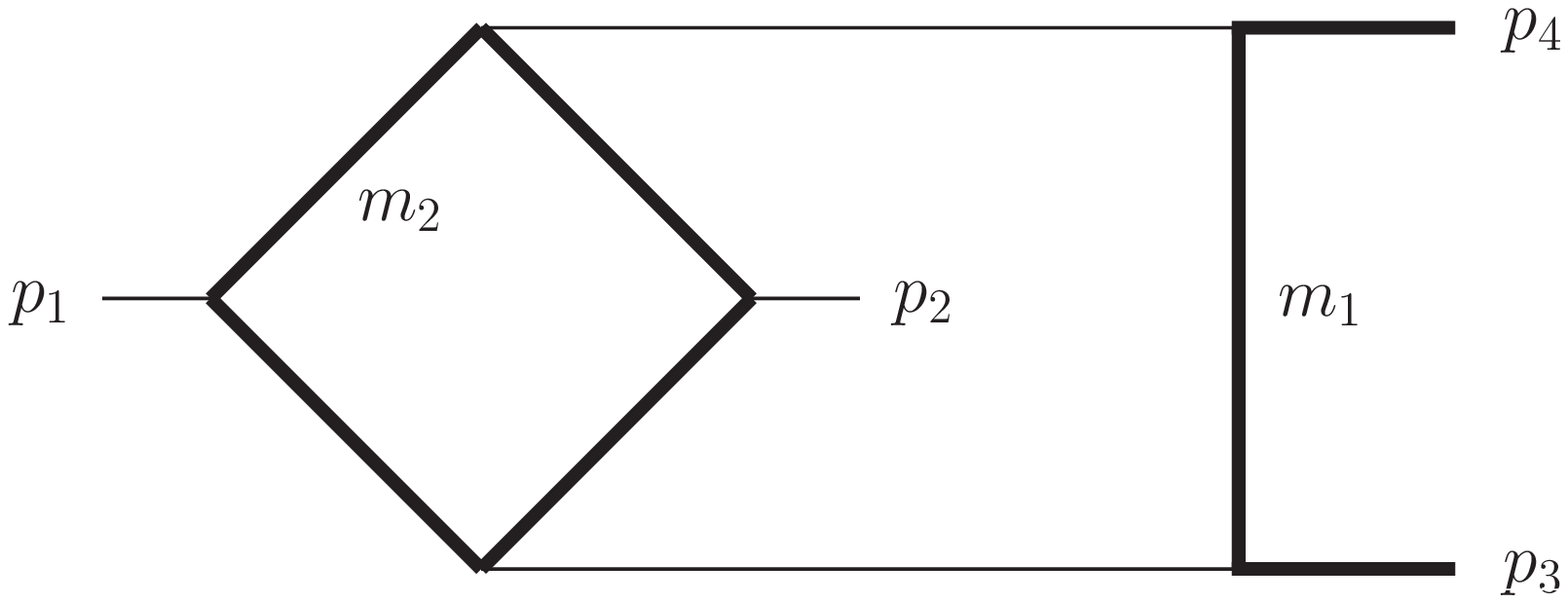}} \label{fig:ggtt1diagram}} \\
\subfigure[$M_7^{\text{P}}$]{\includegraphics[width=0.45\textwidth]{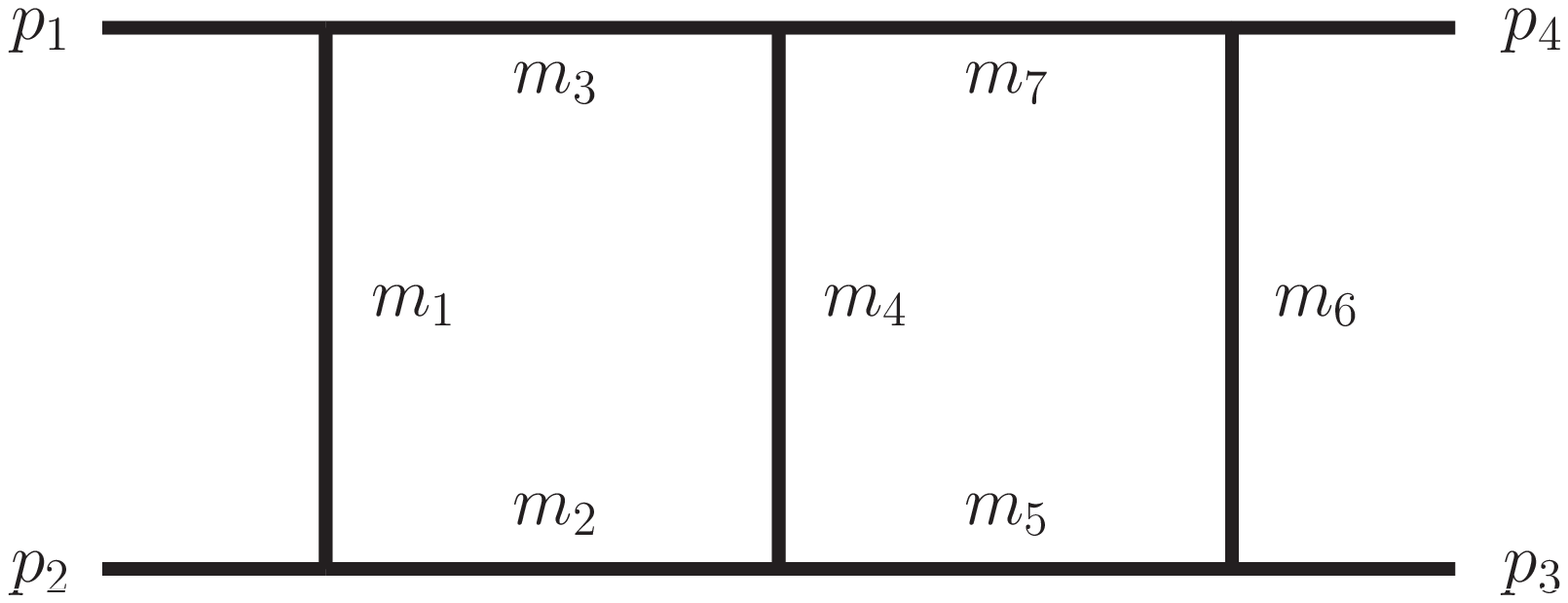}
\label{fig:JapPdiagram}}\hspace{13pt}
\subfigure[$M_7^{\text{NP}}$]{\includegraphics[width=0.45\textwidth]{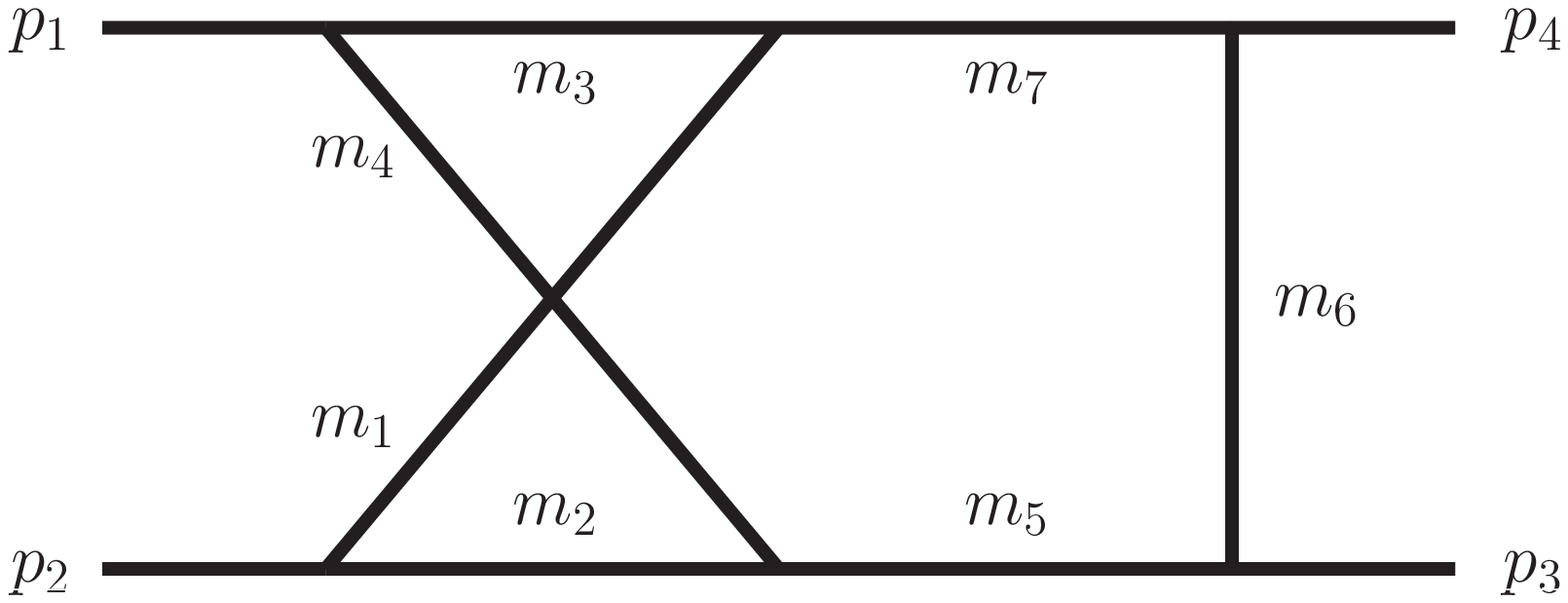}
\label{fig:JapNPdiagram}}\\
\subfigure[$B_6^{NP}$]{\includegraphics[width=5.5cm]{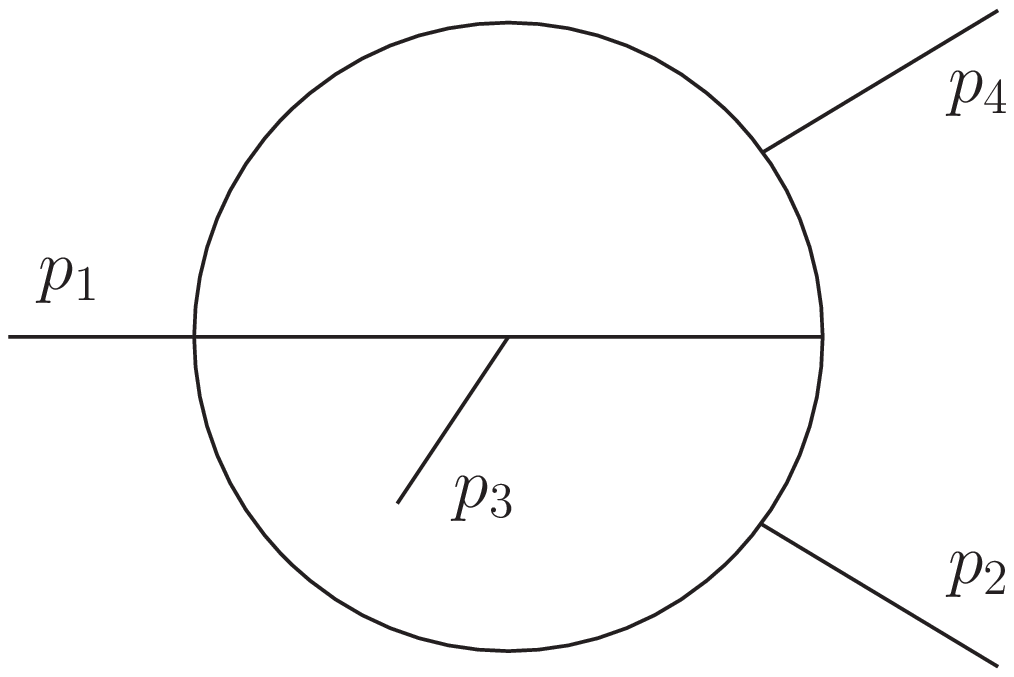}
\label{fig:Bnp6diagram}}
\end{center}
\caption{All diagrams considered in this chapter: the 
	massive non-planar two-loop box diagrams entering the light \subref{fig:ggtt2diagram} and 
	heavy \subref{fig:ggtt1diagram} fermionic correction to the $gg$ channel, 
	the all-massive planar $M_7^{\text{P}}$ \subref{fig:JapPdiagram} and 
	non-planar $M_7^{\text{NP}}$ \subref{fig:JapNPdiagram} seven-propagator diagram and 
	the non-planar six-propagator diagram $B_6^{\text{NP}}$
	 \subref{fig:Bnp6diagram}. The thick lines denote massive particles.} 
\label{fig:alldiagschapter6} 
\end{figure}

In this chapter, the program \secdec{} is used to make predictions for 
two-loop four-point integrals, with massive external legs and/or massive 
internal propagators, see Fig.~\ref{fig:alldiagschapter6}. 
The diagrams considered are non-planar with a single planar case. 
The focus lies on the non-planar diagrams as they are usually more complicated. In this way 
the power of the program \secdec{} can be explored and its limits extended. 
Contrary to analytical calculations, where several mass scales lead to unaccessible elliptic 
integrals, the limits for numerical integration are rather reached when 
spurious higher than logarithmic divergences occur. Such spurious divergences lead to 
more complicated subtraction terms which in turn 
may lead to numerical instabilities if the divergence canceling terms
are not fully resolved by the numerical integrator. 

Such spurious divergences appear, for example, in the case of the 
non-planar seven-propagator 
two-loop topology with a massive bracket/square bend and all other lines massless, see 
Fig.~\ref{fig:ggtt2diagram}. 
To achieve a convenient representation for this integral (termed $ggtt_2$), 
analytical transformations are performed prior to starting the sector decomposition 
algorithm; this way reducing the number of produced sub-sectors, leading to improved 
numerical behavior. 

The $ggtt_2$ diagram is the most complicated master topology occurring in 
the computation of the light fermionic two-loop corrections 
of $t\bar{t}$ production in the $gg$ channel. Analytic results are 
available, see Ref.~\cite{vonManteuffel:2012je,vonManteuffel:2013uoa}.
Conversely, the master topology containing a massive sub-loop (termed $ggtt_1$, 
see Fig.~\ref{fig:ggtt1diagram}) is not available in analytic form. 
It enters the heavy fermionic corrections to top-quark pair production in the 
gluon fusion channel. 
While the leading pole of $ggtt_2$ is of the order ${\cal O} (1/\eps^{4})$, 
and intermediate expressions during sector decomposition 
show (spurious) pole structures with a higher-than-logarithmic 
degree of divergence,
the integral $ggtt_1$ has only finite contributions. 
The same is true for the all-massive two-loop diagrams $M_7^{\text{P}}$ and $M_7^{\text{NP}}$, 
where "P" refers to 
planar and "NP" to non-planar, see Figs.~\ref{fig:JapPdiagram} 
and \ref{fig:JapNPdiagram}. 
To show the predictive power of the program, the $M_7^{\text{P}}$ diagram is 
shown in a toy mass configuration with thirteen mass scales, a configuration 
which is not even close to being accessible in an analytic calculation. To 
compare with other numerical predictions, the two all-massive 
diagrams are also shown with three different scales involved. 
Lastly, a prediction for the non-planar six-propagator diagram of Fig.~\ref{fig:Bnp6diagram} 
is made, studied in the massless case by Tausk, see Ref.~\cite{Tausk:1999vh}. 

\medskip 

The diagrams $B_6^{\text{NP}}$, $ggtt_1$, $M_7^{\text{P}}$ and $M_7^{\text{NP}}$ 
are evaluated with \secdec{} 2.1 in a fully automated way. Apart from scalar master 
integrals, results for an irreducible tensor integral of rank two for the $ggtt_1$
diagram are given.
In the case of the 
$ggtt_2$ diagram it is advantageous to make some analytical manipulations 
beforehand. The results, presented in Ref.~\cite{Borowka:2013cma} 
were useful as checks in the analytic result presented in 
Ref.~\cite{vonManteuffel:2013uoa}. 

\medskip

The structure of this chapter is as follows: 
in Sec.~\ref{sec:analytprep}, an expression serving as a starting point 
for evaluating the $ggtt_2$ diagram is derived,
and a novel type of transformation is described which can be used 
to reduce the number of sector decompositions and improve 
numerical stability. To check consistency, the expected thresholds 
are derived from the Landau equations, see 
Sec.~\ref{sec:expectedthresholds}. 
In Sec.~\ref{sec:numevaldoubleboxes}, numerical results are 
given for all four-point diagrams previously described, where 
the dependence of the evaluation on the number of mass scales, the singularity 
structure and integrals of higher rank are explored. 
%
%
%
\section{Analytical preparation of the non-planar seven propagator integral $ggtt_2$}
\label{sec:analytprep}
This section explores the possibilities arising from 
a mixed approach, where reformulating the integral 
before sector decomposition algorithm can lead to a large gain 
in numerical efficiency. 
The $ggtt_2$ diagram has four poles in the regulator $\eps$ 
and two spurious linear divergences in the Feynman parameters of the 
type $t_i^{-2-\eps}$ . The available numerical integrators can deal with logarithmic 
divergences efficiently, while the convergence of higher-than-logarithmic 
singularities depends heavily on the kinematics and the number of 
Feynman parameters involved. \\    
After integrating out the $\delta$-constraint, the sub-sector functions of the 
$ggtt_2$ diagram contain six Feynman parameters and subtraction terms 
from two spurious linear divergences. 
The numerical integrator fails convergence in this case.  
Decreasing the number of Feynman parameters or the degree of divergence would 
therefore be beneficial. Methods for the latter are already implemented 
by means of integration by parts, where the exponent of a factorized pole 
is increased. Though a solution for many integrand topologies, the tradeoff in the case of 
$ggtt_2$ are long decomposition times (approximately one week on a multi-core, 
16 GB computer), resulting in an order of $\mathcal{O}(6000)$ 
functions to be integrated. 
Integrating so many expressions leads to computational problems: either large 
cancellations when terms are integrated in isolation, or memory/convergence 
issues when terms are grouped by summation prior to integration.

\medskip

Below, a different approach is presented, reducing the 
number of Feynman parameters, the degree of divergence and the total number of 
functions to be integrated numerically. It is based on Ref.~\cite{Borowka:2013cma}. 
Firstly, a representation where one integration parameter of the $ggtt_2$ diagram 
factorizes naturally is derived, such that it can be integrated out analytically. 
A subsequent remapping and the application of the recently introduced 
backwards transformation~\cite{Borowka:2013cma} to single sectors allows 
for an automated evaluation in 
arbitrary kinematics. The new feature allowing for a treatment of user-defined functions 
in the program \secdec{} was developed for this particular computation and is 
explained in Sec.~\ref{subsec:userdefinedfuncs}. 
It was tested, that an important byproduct of the backwards 
transformation is the reduction in the 
total number of functions by two thirds to be integrated numerically, compared to the initial 
sector decomposition approach and without usage of integration by parts relations. 
%
%
\subsection{Integration in a sub-loop}
\label{subsec:subloopintegration}
%
%
The expression for the scalar integral $ggtt_2$ in momentum space is given by
\begin{align}
\label{eq:gnp}
 \mathcal{G}_{ggtt2}= \left( \frac{1}{\mathrm{i} \pi^{\frac{\mathrm{D}}{2}}}\right)^2 
 \int \frac{\mathrm{d^D}k_1\,\, \mathrm{d^D}k_2}{D_{R_1}D_{R_2}D_{R_3}D_{R_4}D_{B_1}D_{B_2}D_{B_3}}  
\end{align}
where $\mathrm{D}=4-2 \eps$.
The  Feynman propagators $D_{R_i}$ corresponding to the ``rhombus" sub-loop in 
Fig.~\ref{fig:ggtt2diagram} are given by
\begin{subequations}
\begin{align}
  D_{R_1} &= (k_1-k_2)^2+\mathrm{i}\delta \text{ ,}\hspace{15pt}  D_{R_2}= (k_1-k_2+p_2)^2+\mathrm{i}\delta \text{ ,}\\
  D_{R_3} &= (k_2+p_4)^2+\mathrm{i}\delta \text{ ,}\hspace{15pt} D_{R_4}= (k_2+p_1+p_4)^2+\mathrm{i}\delta \text{ ,}
\end{align}
\end{subequations}
where the $p_i$ are the external momenta with $p_3^2=p_4^2=m^2$ and $p_1^2=p_2^2=0$, 
and $k_1$, $k_2$ are the loop momenta. All external momenta are assumed to be ingoing. 
Integrating out the loop momentum $k_2$ first, we are left with an expression containing 
only $k_1$ and external momenta, to be combined with the propagators 
\begin{align}
 D_{B_1} = (k_1-p_3)^2+\mathrm{i}\delta \text{ ,}\hspace{10pt}  D_{B_2}= (k_1+p_4)^2+
 \mathrm{i}\delta \text{ ,}\hspace{10pt} D_{B_3} = k_1^2-m^2+\mathrm{i}\delta \;.
 \label{eq:Dbs}
\end{align}
This procedure is not limited to our particular example, but requires an analytical step 
of introducing a convenient parametrization which can not be found when the sub-loop 
contains massive propagators. 
For the rest of this chapter, the causal $\mathrm{i} \delta$ will be omitted and the 
renormalization scale is set to $\mu_R=1$ for simplicity.

The introduction of Feynman parameters for the one-loop subgraph $\mathcal{I}_{R}$ 
constructed from all propagators containing the loop momentum $k_2$ leads to 
\begin{align}
\non \mathcal{I}_{R} =& \frac{1}{\mathrm{i} \pi^{\mathrm{D}/2}} \int \frac{ \mathrm{d^D}k_2 }
 { D_{R_1}D_{R_2}D_{R_3}D_{R_4}} \\
 =& \Gamma(2+\eps) \int \prod_{i=1}^4\mathrm{d}x_i \;\delta(1-\sum_{j=1}^4 x_j) \;
 \mathcal{F}(\vec{x},k_1)^{-2-\eps} \text{ ,}\label{loop1}
\end{align}
with the second Symanzik polynomial reading 
\begin{align}
\hspace{-25pt} -\mathcal{F}(\vec{x},k_1)= D_{B_1} x_1x_2 + (k_1+p_1+p_4)^2x_1x_3 + 
(k_1+p_2+p_4)^2 x_2x_4 +  D_{B_2} x_3x_4 \text{ .}
\end{align}
During integration of the $\delta$-distribution in Eq.~(\ref{loop1}), the first Symanzik polynomial 
reduces to unity. The substitutions
\begin{align}
x_1 = t_2 (1- t_3) \text{ ,}\hspace{19pt} x_2 = t_1 t_3 \text{ ,}\hspace{19pt} x_3 = (1- t_1) t_3 \text{ ,}
\label{eq:parametrisation}
\end{align}
facilitate a factorization of the parameter $t_3$ which is integrated out analytically. This yields 
\begin{align}
 \mathcal{I}_{R} &= -\frac{2}{\eps}\frac{\Gamma(2+\eps)\Gamma^2(1-\eps)}{\Gamma(1-2\eps)} 
 \int_0^1 \mathrm{d}t_1\int_0^1\mathrm{d}t_2 \, \, \mathcal{\tilde F}(\vec{t},k_1)^{-2-\eps}
\end{align}
with
\begin{align}
-\mathcal{\tilde F}(\vec{t},k_1)= D_{B_1} t_1t_2 + (k_1+p_1+p_4)^2 t_1\bar{t}_2+ 
(k_1+p_2+p_4)^2 \bar{t}_1t_2 +  D_{B_2} \bar{t}_1\bar{t}_2 \text{ ,}
\label{eq:ftilda}
\end{align}
where the shorthand notation $\bar{t}_i=1-t_i$ is introduced. %
The expression for the 1-loop rhombus $\mathcal{I}_{R}$ is combined 
with the remaining $k_1$-dependent propagators, 
treating the expression of Eq.~(\ref{eq:ftilda}) as a fourth propagator with power 
$2+\eps$, to obtain, 
\begin{align}
\nonumber \mathcal{G}_{NP}=&\frac{2}{\eps}\frac{\Gamma(3+2\eps)\Gamma^2(1-\eps)}{\Gamma(1-2\eps)} \int_0^1 \mathrm{d}t_1\int_0^1\mathrm{d}t_2 \,\, \times\label{GNP} \\
 &\prod_{i=1}^4\int_0^1 \mathrm{d}z_i \,z_4^{1+\eps} \,\delta(1-\sum_{j=1}^4 z_j) \,\,\mathcal{F_{NP}}(\vec{z},t_1,t_2)^{-3-2\eps} \,\,\mathcal{U_{NP}}(\vec{z})^{1+3\eps} \text{ }
\end{align}
after integration of $k_1$ and where 
\begin{align}
\mathcal{U_{NP}}(\vec{z})&= \sum_{j=1}^4 z_j \hspace{19pt}\text{ and} \\
\label{eq:Fnp}
\mathcal{F_{NP}}(\vec{z},t_i)&= -s_{12} z_2 z_3 - T z_1 z_4 - S_1 z_2 z_4 -S_2 z_3 z_4 + m^2 z_1 (z_1
+z_4 Q)  \text{ ,}
\end{align}
with
\begin{align}
\nonumber T &= s_{13} \bar{t}_1 t_2 +s_{23} t_1 \bar{t}_2 \text{ ,}\hspace{19pt}  S_1= s_{12} t_1 t_2 \text{ ,}\hspace{19pt} S_2=s_{12} \bar{t}_1 \bar{t}_2 \\
\label{eq:STUdefs}
 Q &=t_1\bar{t}_2+\bar{t}_1t_2\text{ ,}\hspace{19pt} s_{ij} = (p_i+p_j)^2\text{ .}
\end{align}
The full integral $\mathcal{G}_{NP}$ is in total one Feynman parameter short. 
A primary sector decomposition in the newly introduced Feynman parameters 
$z_1,\dots,z_4$ is performed to obtain
\begin{align}
\mathcal{G}_{NP}=&\frac{2}{\eps}\frac{\Gamma(3+2\eps)\Gamma^2(1-\eps)}{\Gamma(1-2\eps)} \int_0^1 \mathrm{d}t_1\int_0^1\mathrm{d}t_2 \sum_{i=1}^4 G_{NP}^i \text{ ,}
\end{align}
with
\begin{subequations}
\begin{align}
G_{NP}^1&=\int_0^1 \mathrm{d}z_2\, \mathrm{d}z_3 \,\mathrm{d}z_4 \,\,\,z_4^{1+\eps}\,\,\,(1+z_2+z_3+z_4)^{1+3 \eps}\,\,\,\mathcal{F}^1(\vec{z},t_i)^{-3-2\eps}  \text{ ,}\\
\mathcal{F}^1(\vec{z},t_i)&= -s_{12} z_2 z_3 - T z_4 - S_1 z_2 z_4 -S_2 z_3 z_4 + m^2 (1 +z_4 Q)  \text{ ,}\\
G_{NP}^2&=\int_0^1 \mathrm{d}z_1 \,\mathrm{d}z_3 \,\mathrm{d}z_4 \,\,\,z_4^{1+\eps}\,\,\,(1+z_1+z_3+z_4)^{1+3 \eps}\,\,\,\mathcal{F}^2(\vec{z},t_i)^{-3-2\eps}  \text{ ,}\\
\mathcal{F}^2(\vec{z},t_i)&= -s_{12} z_3 - T z_1 z_4 - S_1 z_4 -S_2 z_3 z_4 + m^2 z_1 (z_1 +z_4 Q)  \text{ ,}\label{eq:Fsec2}\\
G_{NP}^3&=\int_0^1 \mathrm{d}z_1\, \mathrm{d}z_2 \,\mathrm{d}z_4 \,\,\,z_4^{1+\eps}\,\,\,(1+z_1+z_2+z_4)^{1+3 \eps}\,\,\,\mathcal{F}^3(\vec{z},t_i)^{-3-2\eps}  \text{ ,}\\
\mathcal{F}^3(\vec{z},t_i)&= -s_{12} z_2 - T z_1 z_4 - S_1 z_2 z_4 -S_2 z_4 + m^2 z_1 (z_1 +z_4 Q)  \text{ ,}\\
G_{NP}^4&=\int_0^1 \mathrm{d}z_1\, \mathrm{d}z_2 \,\mathrm{d}z_3 \,\,\,(1+z_1+z_2+z_3)^{1+3 \eps} \,\,\,\mathcal{F}^4(\vec{z},t_i)^{-3-2\eps}  \text{ ,}\\
\mathcal{F}^4(\vec{z},t_i)&= -s_{12} z_2 z_3 - T z_1 - S_1 z_2 -S_2 z_3 + m^2 z_1 (z_1 + Q)  \text{ ,}
\end{align}
\end{subequations}
where the $\delta$-distribution is naturally integrated out. \\
Observing the primary sectors, the first sector $\mathcal{F}^1(\vec{z},t_i)$ is of the 
form $m^2+{\mathrm{func}}(z_i,t_i)$, 
and does not need iterations of the decomposition into further sectors. Primary sector three 
can be remapped to primary sector two by exchanging $z_2 \leftrightarrow z_3$ and 
$S_1 \leftrightarrow S_2$. 
The sectors two and four are therefore the only ones needing further treatment. 
This is a benefit from the prior integration of one sub-loop of the full integral. This treatment 
has a small drawback though: with the introduction of the substitutions in 
Eq.~(\ref{eq:parametrisation}), singularities at the second endpoint are introduced. 
The integrals $G_{NP}^{2,3,4}$ can diverge both at zero and one in $t_1$ and $t_2$. 
With the sector decomposition algorithm, only singularities at zero are factorized automatically. 
Consequently, the singularities located at the upper integration limit are remapped 
to the origin of parameter space by splitting the integration region at $\tfrac{1}{2}$
and transforming the integration variables to remap the integration domain to the 
unit hypercube, see Sec.~\ref{subsec:endpointremapping}. \\
This procedure results in 12 integrals,   
some of which are already finite, such that no subsequent sector decomposition is required.
Other integrals lead to linear divergences of the type $\int_0^1dx\,x^{-2-\eps}$ in 
two Feynman parameters after sector decomposition. These singularities are 
spurious and can be subtracted 
by expanding the Taylor series in the subtraction procedure up to the second term, see 
Sec.~\ref{subsec:extractpoles}.
This procedure is prone to introducing large cancellations between subtraction terms 
and a large number of sectors and therefore challenges numerical stability. 
Avoiding this type of singularity from scratch is therefore a highly desirable goal. \\
The next subsection describes a strategy which can help to reduce the number of 
higher than logarithmic divergences and functions to be integrated numerically. 
\subsection{Backwards transformation}
\label{subsec:backwards}
%
%
The aim of the procedure described in this section is to achieve a transformation of 
potential linear divergences into logarithmic 
divergences as far as possible. 
A different procedure towards this goal based on integration-by-parts identities, 
has been described in Ref.~\cite{Carter:2010hi} and Sec.~\ref{subsec:ibprelations}. 
The latter method however can increase the number of functions 
to be integrated substantially, 
while the method described below in general reduces the number of further iterations
and therefore the number of produced functions.
Yet another method to reduce the number of functions produced during factorization 
has been suggested in Ref.~\cite{Anastasiou:2010pw}, where a non-linear 
transformation in the Feynman parameters aims at a reduction of the exponent 
of the second Symanzik polynomial. Although an attractive idea, it was not beneficial in 
the case of the $ggtt_2$ as further singularities at the upper integration limit are 
introduced with the transformation. A subsequent remapping of the divergences at the 
endpoint one to zero restores, or even deteriorates the original singularity structure with 
spurious linear divergences in two Feynman parameters. 

\medskip

Due to the fact that the iterated decomposition of the integral into sub-sectors 
introduces higher powers in the Feynman parameters, the desire to undo some 
of those decomposition steps before the final integration is 
rather natural. Yet, in most cases this is not possible without the introduction 
of unacceptable new divergences which have to be subtracted before 
integration. 
Although it does not seem beneficial to ``undo'' single sector decomposition 
steps, there may still be a transformation that does a similar trick. 
Such a transformation was introduced and utilized by the author and 
collaborators in Ref.~\cite{Borowka:2013cma}.
It relies on the possibility of 
blowing down an affine $N$-dimensional space as opposed to the 
blowing up used in the sector decomposition approach. 
After performing the blowing down, the splitting of the 
integration region as performed in the 
sector decomposition approach can be applied backwards. 
The original function is thereby split into two again. 
It may at first seem counter-intuitive to achieve a reduction in the number of sectors 
to be integrated, when a splitting of one primary sector using the 
backwards transformation doubles the number of functions. 
However, in spite of these arguments, it turns out that when 
applied in the right way, a reverse splitting of the integration region 
can rearrange the Feynman parameters, such 
that the double linear divergences of primary sectors two and three are 
transformed into logarithmic ones. This leads to an overall reduction in 
the number of functions describing $G_{NP}$ by two thirds. 

\medskip

Below, the necessary preconditions for the application of such a backwards 
transformation and the transformation itself are analyzed in more detail. 
\subsubsection{Preconditions}
%
%
To explain the backwards transformation in more detail, it is 
convenient to recall the sector decomposition and highlighting the 
important aspects necessary for the transformation. 

\medskip

In the sector decomposition algorithm, the integration region is 
split into at least as many parts as there are integration variables. This splitting is done 
such that a clear and definite hierarchy can be observed among the integration parameters. 
As recollection, the example splitting of Sec.~\ref{sec:secdecconcept} reads 
\begin{align}
\non &\int_0^1 \int_0^1 \text{d}x_1 \; \text{d}x_2  \frac{1}{(x_1+x_2)^{2+\eps}} \\
&= \int_0^1 \int_0^1 \text{d}x_1 \; \text{d}x_2  \frac{1}{(x_1+x_2)^{2+\eps}}\; ( \Theta(x_1-x_2) + \Theta(x_2-x_1) ) \text{ ,} \label{eq:secdecback1}
\end{align}
where it is evident that $x_1$ in the first summand on the righthand side 
must always be bigger than $x_2$, otherwise the integral is zero. In the second 
summand of Eq.~(\ref{eq:secdecback1}) the hierarchy between $x_1$ and $x_2$ is 
reversed. 
Both integrals in the second line of Eq.~(\ref{eq:secdecback1}) are of definite 
hierarchy. In the sector decomposition example of Eqs.~(\ref{eq:secdecexample}), 
a blowing up is applied to these two functions, leading to a factorization of 
the previously overlapping singularities. The hierarchy among the Feynman 
parameters is implicitly kept. Assuming one of the resulting functions with 
non-overlapping singularities to be the starting point of the backwards 
transformation, the sector decomposition of Eqs.~(\ref{eq:secdecexample}) 
can be reversed. It reads 
\begin{subequations}
\begin{align}
& \int_0^1  \text{d}x_1  \int_0^1 \text{d} \tilde{x}_2 \; \frac{1}{x_1^{1+\eps} (1 +  \tilde{x}_2)^{2+\eps}} \label{eq:initial}\\
&= \int_0^1 \text{d}x_1 \int_0^{x_1} \text{d}x_2 \; \frac{1}{(x_1+x_2)^{2+\eps}} \label{eq:blowdown} \\
\non &=  \int_0^1  \text{d}x_1 \int_0^1 \text{d}x_2  \; \frac{1}{(x_1+x_2)^{2+\eps}} \; ( \Theta(x_1-x_2) + \Theta(x_2-x_1) ) \\
& \hspace{7pt} - \int_0^1  \text{d}x_2 \int_0^{x_2} \text{d}x_1 \frac{1}{(x_1+x_2)^{2+\eps}} \\
 &= \int_0^1  \text{d}x_1 \int_0^1 \text{d}x_2 \; \frac{1}{(x_1+x_2)^{2+\eps}} 
-  \int_0^1 \text{d}x_2  \int_0^1 \text{d}\tilde{x}_1\,\, \frac{1}{x_2^{1+\eps}(\tilde{x}_1+1)^{2+\eps}}  \text{ .} 
\end{align}%
\end{subequations}%
Although the equations are just rearranged, the implications are different. 
While the factorization of singularities works due to the application of blowup 
sequences, here the prerequisite is the possibility of applying a blowing down. 
Looking at the transition from Eq.~(\ref{eq:initial}) to Eq.~(\ref{eq:blowdown}) more 
thoroughly, one finds
\begin{subequations}
\begin{align}
& \int_0^1 \int_0^1  \text{d}x_1 \; \text{d} \tilde{x}_2 \; \frac{1}{x_1^{1+\eps} 
(1 +  \tilde{x}_2)^{2+\eps}} 
\label{eq:bdownlong} \\
&=  \int_0^1 \int_0^1  \text{d}x_1 \; \text{d} \tilde{x}_2 \; \frac{1}{x_1^{1+\eps} 
(1 +  \tilde{x}_2)^{2+\eps}} \; \Theta (x_1- x_1  \tilde{x}_2)
\label{eq:bdownlong1} \\
&= \int_0^1 \int_0^1  \text{d}x_1 \; \frac{\text{d}x_2}{x_1}  \; \frac{1}{x_1^{1+\eps} (1 + \frac{x_2}{x_1})^{2+\eps}} \; \Theta (x_1- x_2)
\label{eq:bdownlong2} \\
&= \int_0^1 \text{d}x_1 \int_0^{x_1} \text{d}x_2 \; \frac{1}{(x_1+x_2)^{2+\eps}} \text{ .}
\end{align}%
\end{subequations}%
From Eq.~(\ref{eq:bdownlong1}) to Eq.~(\ref{eq:bdownlong2}) the integration 
parameters $x_1$ and $ \tilde{x}_2$ are transformed as
\begin{subequations}
\begin{align}
x_1 &\rightarrow x_1 \text{ ,} \\
x_1 \tilde{x}_2 &\rightarrow x_2 \text{ .}
\end{align}%
\end{subequations}%
Having examined a symmetric and simple example, it should be noted that 
it is not trivial to a priori know that an equality of the type 
between Eqs.~(\ref{eq:bdownlong}) and (\ref{eq:bdownlong1}) indeed holds. 
In general polynomial integrals, the implicit hierarchies among the 
Feynman parameters must 
be made explicit through the introduction of Heaviside $\Theta$ functions. 

\medskip

In the following, a realistic example is presented, uncovering the 
advantages of such a backwards transformation. 
\subsubsection{Application to the $ggtt_2$}
%
%
Returning to the non-planar double box integral $G_{NP}$, the following structure can 
be identified for the function of Eq.~(\ref{eq:Fsec2}) in sector two (and three) after remapping 
\begin{align}
 I= \prod_{i=1}^N \left\{  \int_0^1 \rd t_i \right\} 
 \left[ t_j\left( P(\vec{t}_{jk}) + t_k Q(\vec{t}_{jk}) \right) + R(\vec{t}_{jk}) \right]^{- \alpha} \text{ ,}
\label{eq:secbeforeback}
\end{align}
where $N=5$, $t_j=z_4$ and $t_k=z_1$ and $\alpha > 0$. $P$, $Q$ and $R$ 
are polynomials of arbitrary 
degree in the Feynman 
parameters $\vec{t}_{jk}=(t_1,\dots,\hat{t}_j,\dots,\hat{t}_k,\dots,t_N)$ and 
kinematic invariants. The carets denote those Feynman parameters which are {\it not} part of the 
vector of Feynman parameters. This definition of a vector of Feynman parameters with double 
index $\vec{t}_{jk}$ will be used throughout this subsection. 

\medskip

In Eq.~(\ref{eq:secbeforeback}), all terms multiplied by the Feynman parameter $t_k$ are 
also multiplied by the Feynman parameter $t_j$. An 
explicit hierarchy exists, allowing for a transformation of the 
type $t_j t_k \rightarrow t_{\tilde{k}}$. 
The splitting of the integration region can be performed backwards as 
\begin{subequations}
\begin{align}
I =& \int_0^1  {\rm d}\vec{t}_k \; \rd t_{\tilde{k}} \; \frac{1}{t_j}\left[ t_j P(\vec{t}_{j\tilde{k}}) + t_{\tilde{k}} Q(\vec{t}_{j\tilde{k}}) + R(\vec{t}_{j\tilde{k}})\right]^{- \alpha} \label{eq:secback1} \\
 -& \int_0^1  {\rm d}\vec{t}_{\tilde{j}\tilde{k}} \; \rd t_{\tilde{j}} \; \rd t_{\tilde{k}}
 \; \frac{1}{t_{\tilde{j}}} \; 
 \left[ t_{\tilde{k}} 
 \left( t_{\tilde{j}} P (\vec{t}_{\tilde{j}\tilde{k}}) + Q (\vec{t}_{\tilde{j}\tilde{k}}) \right) + R(\vec{t}_{\tilde{j}\tilde{k}}) 
 \right]^{- \alpha} \text{ .} \label{eq:secback2} 
\end{align}
\end{subequations}
To explain this in more detail, a rearrangement of the terms leads 
to the well-known sector decomposition
\begin{subequations}
\begin{align}
\int_0^1 & {\rm d}\vec{t}_k \; \rd t_{\tilde{k}} \; \frac{1}{t_j} 
\left[ t_j P(\vec{t}_{j\tilde{k}}) + t_{\tilde{k}} Q(\vec{t}_{j\tilde{k}}) + R(\vec{t}_{j\tilde{k}})\right]^{- \alpha} 
[\underbrace{\Theta(t_j-t_{\tilde{k}})}_{(1)} + \underbrace{\Theta(t_{\tilde{k}} - t_j)}_{(2)}]
\label{eq:wayback} \\
=&  \int_0^1 \rd \vec{t} \;
 \left[ t_j\left( P(\vec{t}_{jk}) + t_k Q(\vec{t}_{jk}) \right) + R(\vec{t}_{jk}) \right]^{- \alpha} \\
+&  \int_0^1  {\rm d}\vec{t}_{\tilde{j}\tilde{k}} \; \rd t_{\tilde{j}} \; \rd t_{\tilde{k}}
 \; \frac{1}{t_{\tilde{j}}} \; 
 \left[ t_{\tilde{k}} 
 \left( t_{\tilde{j}} P (\vec{t}_{\tilde{j}\tilde{k}}) + Q (\vec{t}_{\tilde{j}\tilde{k}}) \right) + R(\vec{t}_{\tilde{j}\tilde{k}}) 
 \right]^{- \alpha} \text{ ,}
\end{align}
\end{subequations}
where $t_{\tilde{k}} \rightarrow t_j\,t_k$ was substituted in sector (1) and 
$t_j \rightarrow t_{\tilde{k}} \,t_{\tilde{j}}$ in sector (2). 

\medskip 

The effect of the backwards transformation is twofold: 
The degree of the polynomial in $t_jt_k$ is reduced in Eq.~(\ref{eq:secback1}),  
and in Eq.~(\ref{eq:secback2}) the degree of divergence
in $t_j$ is reduced if $\alpha > 1$. 
It can be beneficial in the reduction of the number of functions as 
compared to the custom sector decomposition procedure, and can be 
particularly advantageous if the factor $Q$ is much simpler than $P$. 
%
%
%

\medskip

After all transformations of this type, the result is a total of 15 functions partly needing an 
iterated sector decomposition. 
Together with the introduction of the new feature of {\it user-defined functions} in \secdec{} 2.1 
described in Sec.~\ref{subsec:userdefinedfuncs}, the computation of the $ggtt_2$ diagram 
is now possible in a reasonable amount of time. The timings are discussed in 
Sec.~\ref{subsec:ggtt2numresults}.
%
%
%
\section{Expected thresholds from the Landau equations}
\label{sec:expectedthresholds}
Before turning to the numerical evaluation of the $ggtt_2$ diagram and 
diverse other non-planar double box integrals, 
it should be analyzed roughly where the thresholds of the $ggtt_2$ diagram 
are expected, to have a measure for the trustworthiness of the result. 
To this end, it is useful to analyze the Landau equations as described 
in Sec.~\ref{sec:landauequations}, although the resulting singularities 
do not necessarily lead to a divergence in the integral.
The first set of equations resulting from applying Eq.~(\ref{eq:landau1}) to $ggtt_2$ reads 
\begin{subequations}
\begin{align}
x_1\, (k_1-k_2)^2 =& 0 \label{eq:landaueq1} \text{ ,}\\
x_2\, (k_1-k_2 + p_2)^2 =& 0  \text{ ,}\\
x_3 \,(k_2+p_4)^2 =& 0  \text{ ,}\\
x_4 \,(k_2+p_1+p_4)^2 =& 0  \text{ ,}\\
x_5 \,(k_1 + p_1 +p_2 +p_4)^2 =& 0  \text{ ,}\\
x_6 \,(k_1+p_4)^2 =& 0  \text{ ,}\\
x_7 \,(k_1^2 -m^2) =& 0 \label{eq:landaueq7} \text{ ,}
\end{align}
\end{subequations}
where the $k_i$ are again the loop momenta, $p_i$ are the external 
momenta and $m$ is the mass appearing in the propagators of the 
$ggtt_2$ diagram. 
The second set of equations resulting from Eq.~(\ref{eq:landau2}) is
\begin{subequations}
\begin{align}
\non I: \quad &x_1 (k_1-k_2)_\mu + x_2 (k_1-k_2 + p_2)_\mu + 
x_5 (k_1 + p_1 +p_2 +p_4)_\mu +\\
& x_6 (k_1+p_4)_\mu + x_7 k_{1\mu}
= 0 \label{eq:landaueq8}  \text{ ,} \\
\non II: \quad  &- x_1 (k_1-k_2)_\mu - x_2 (k_1-k_2 + p_2)_\mu + x_3 (k_2 + p_4)_\mu + \\
& x_4 (k_2+p_1+p_4)_\mu = 0  \label{eq:landaueq9}  \text{ .}
\end{align}
\end{subequations}
Now, Eqs.~(\ref{eq:landaueq1})-(\ref{eq:landaueq7}) are either 
true when the $x_i$ are vanishing or when the scalar products 
composed of the loop momenta $k_i$ and external momenta $p_i$ 
are either equal to a mass, see Eq.~(\ref{eq:landaueq7}) or 
vanish altogether. Now, if one Feynman parameter $x_i$ is 
zero, the graph to be considered is a subgraph of the original 
one because the propagator connecting two vertices was removed. 
To analyze the singularities of the original graph it is therefore 
sufficient to assume $x_i \neq 0 \,\forall \,i$. This means, 
Eqs.~(\ref{eq:landaueq1})-(\ref{eq:landaueq7}) all serve as constraints 
on the scalar products to appear in Eq.~(\ref{eq:landaueq8}) and 
Eq.~(\ref{eq:landaueq9}) when contracting them with the loop $k_i^\mu$ and 
respectively external momentum vectors $p_i^\mu$. Some of these 
are used to express the missing scalar products in terms of other scalar 
products. The resulting equations containing the Landau singularities 
for those sets of kinematic invariants and Feynman parameters for which 
the equations hold, then read
\begin{subequations}
\begin{align}
0 = & \; s_{12} \, (x_1\, x_5 -  x_2 \, x_6 ) \, + x_7\, \Big[ x_1 \,(m^2 - s_{23})\,  + 
 x_2 \,(m^2 - s_{12} - s_{23})  \Big]  \text{ ,} \\
0 = & \; s_{12} \, (x_2\, x_5 -  x_1 \, x_6 ) \, + x_7\, \Big[ x_2 \, (m^2 - s_{13}) \,  + 
x_2 \, (m^2 - s_{12} - s_{13}) \Big]  \text{ ,} 
\end{align}
\begin{align}
0 = & \; s_{12} \,x_2\, (x_1 - x_4)  \text{ ,} \\
0 =&\;  s_{12}\, x_1\, x_2 \label{eq:threshold2}  \text{ ,} \\
0 = & \; s_{12} \, (x_1\, x_3 - x_2\, x_4)  \text{ ,} \\
\non 0 = &\; ( s_{13} - m^2) \,x_2\, x_3 \,  + (s_{23} -m^2) \, x_1\, x_4\, + \\
\non &\; (s_{13} + s_{23}-2\, m^2 )  \Big[x_5 \,(x_1 + x_3 + x_4) \, + x_2 \,(x_4 + x_5) \Big] \,+ \\
&\; (-2\,m^2 ) \,x_7 \, ( x_1 + x_2 + x_3 + x_4) \text{ ,} \\
\non 0 = &\; s_{12}\, x_1\,x_2\, (x_1 + x_2) + ( s_{12} + s_{13} - m^2) \,x_1\, x_2\, x_3 + 
( s_{13} - m^2 )\, x_2^2\, x_3 \,+\\
\non&\;  ( s_{23} - m^2)\, x_4 \, x_1 \,(x_1+ x_2) \, + \\
\non&\; ( s_{13} + s_{23} - 2 \,m^2) \, x_2\,
\Big[x_2\, x_4 + x_5 \, (x_1\, + x_2 +x_3 + x_4) \Big] \,- \\
&\; x_7\, (x_1 + x_2) \Big[ x_2 \,( 3 \,m^2 -s_{13})  + 2\, m^2 (x_1+ x_3) + x_4\,( s_{23} + m^2) \Big] \text{ ,} \\
\non0 = &\; (m^2 - s_{13}) \,x_2\, x_3\, ( x_1 + x_2) - 
 s_{12} \,x_1\, x_2\, (x_1 + x_2 + x_3) \,+ \\
\non &\; x_4\,\Big[ (m^2 - s_{23}) \,x_1\, (x_1 + x_2) + (2 \,m^2 - s_{13} - s_{23})\, x_2^2\Big] \, + \\
&\; 2\, m^2 \, x_7\, (x_1 + x_2) (x_1 + x_2 + x_3 + x_4)
%
%
%
\end{align}
\end{subequations}
Still assuming all Feynman parameters $x_i \neq 0$, a leading Landau 
singularity cannot be detected in the above equations. The graph is 
regulated by its mass parameter. 
Yet, there are many sub-leading Landau singularities. These are given 
when one or more of the above displayed kinematic conditions
are fulfilled. One such Landau singularity appears at $s_{13} = m^2$, when 
the Feynman parameters $x_{i}$ with $i=1,4,5,6,7$ vanish simultaneously. 
Another one appears at $s_{12}=0$ when $x_{j} \to 0$ for $j=3,4,5,7$. 
%
%
\section{Numerical evaluation}
\label{sec:numevaldoubleboxes}
In this section, numerical results for several two-loop four-point functions 
are presented, in the same order as is shown in Fig.~\ref{fig:alldiagschapter6}. 
The results for the analytically prepared $ggtt_2$ diagram are shown first. 
All other results are obtained using the custom \secdec{} setup. 
%
%
\subsection{The $ggtt_2$ diagram}
\label{subsec:ggtt2numresults}
\begin{figure}[htb]
\subfigure[]{	      
\includegraphics[width=0.5\textwidth]{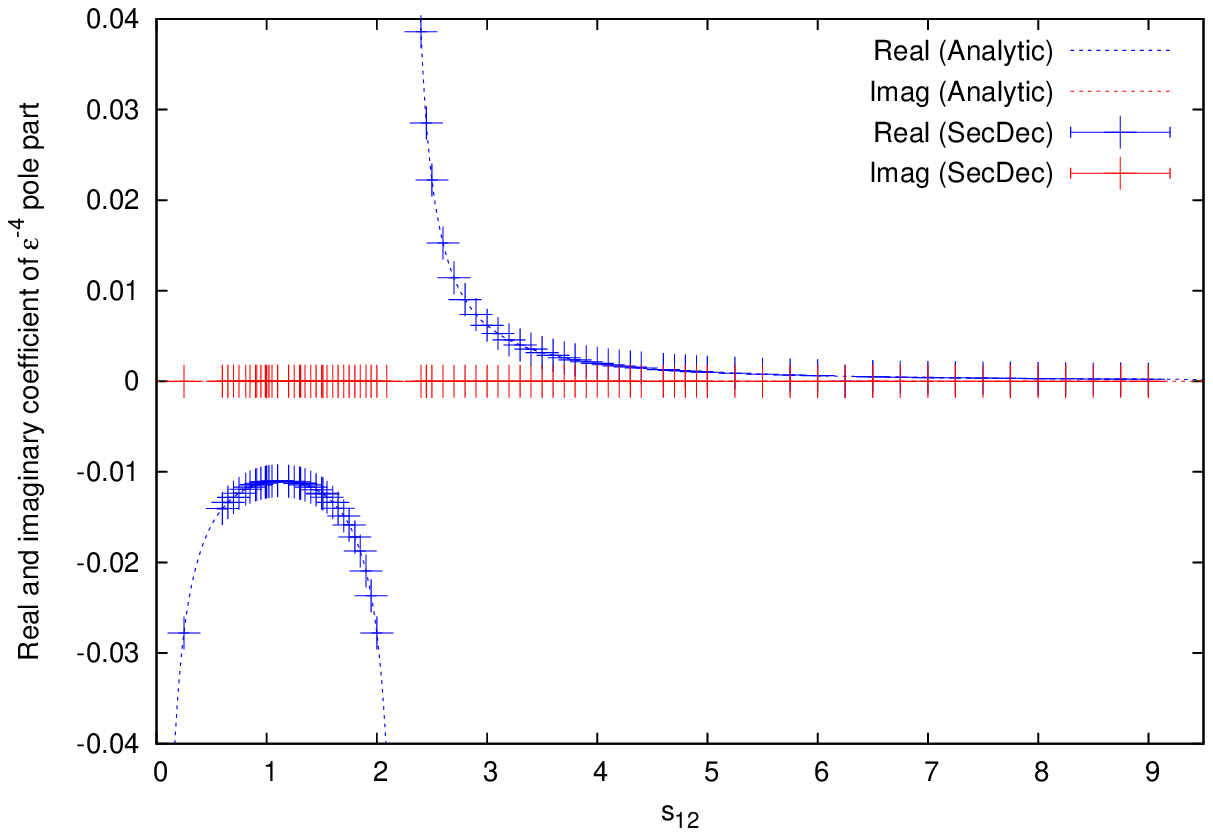}
\label{subfig:leadingggtt2pole}}\hfill
\subfigure[]{
\includegraphics[width=0.5\textwidth]{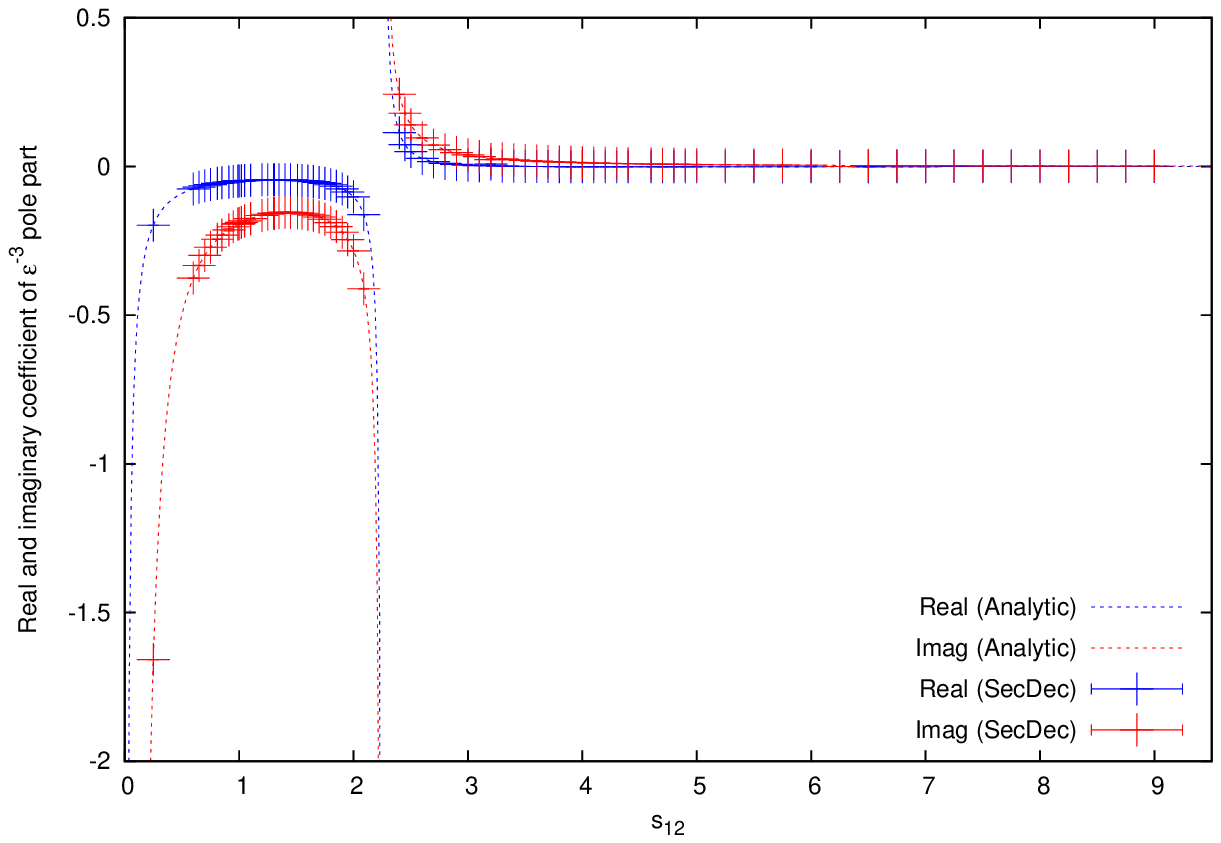}
\label{subfig:subleadingggtt2pole}}
\caption{Comparison of \subref{subfig:leadingggtt2pole} the leading and 
	\subref{subfig:subleadingggtt2pole} the next-to-leading pole coefficients between 
	the analytic result from \cite{AvMACAT} and the \secdec{} result. The real part is 
	shown in blue, the imaginary part in red. As numerical values 
	 $p_3^2=p_4^2=m^2=1$, $s_{23}=-1.25$ is chosen, 
	 assuming four-momentum conservation $s_{13}=2 m^2-s_{12}-s_{23}$. 
	\label{fig:ggtt2_poles43}} 
\end{figure}

\begin{figure}[htb!]
\subfigure[] { \includegraphics[width=0.5\textwidth]{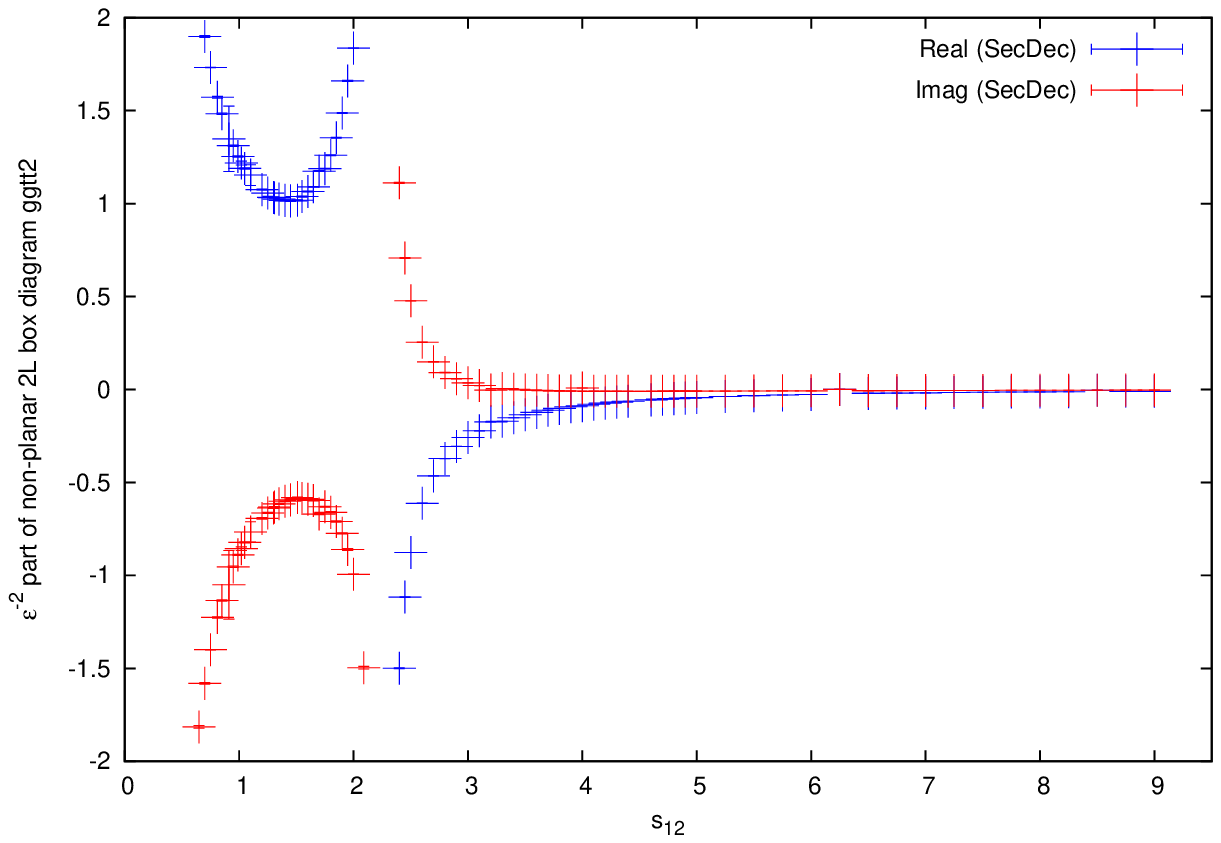}\label{subfig:epsmin2ggtt2pole}}\hfill
\subfigure[] { \includegraphics[width=0.5\textwidth]{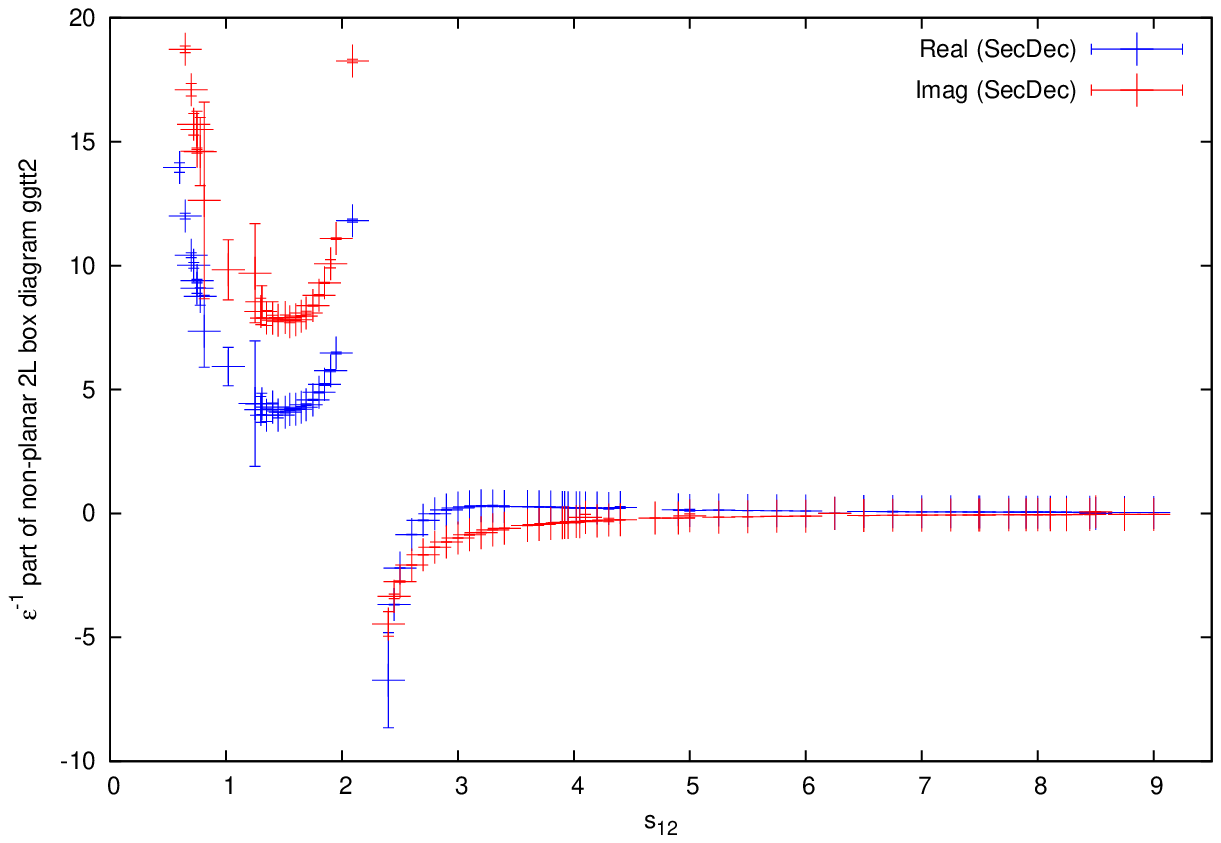}\label{subfig:epsmin1ggtt2pole}}	      
\caption{Results for \subref{subfig:epsmin2ggtt2pole} the $1/\eps^2$ and \subref{subfig:epsmin1ggtt2pole} the $1/\eps$ coefficients of the integral $ ggtt_2$.
The kinematics are the same as in Fig. \ref{fig:ggtt2_poles43}.}
\label{fig:ggtt2_poles21} 
\end{figure}

\begin{figure}[htb!]
\begin{center}
\subfigure[]{
\includegraphics[width=0.66\textwidth]{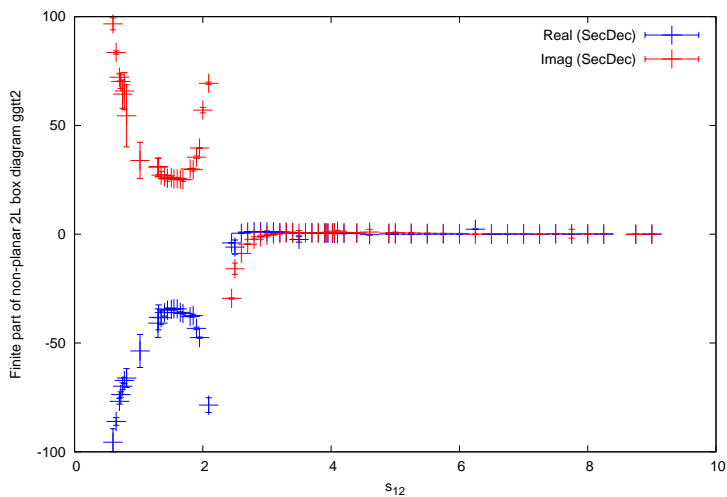}
\label{subfig:finiteggtt2aroundthreshold}} \\
\subfigure[]{
\includegraphics[width=0.66\textwidth]{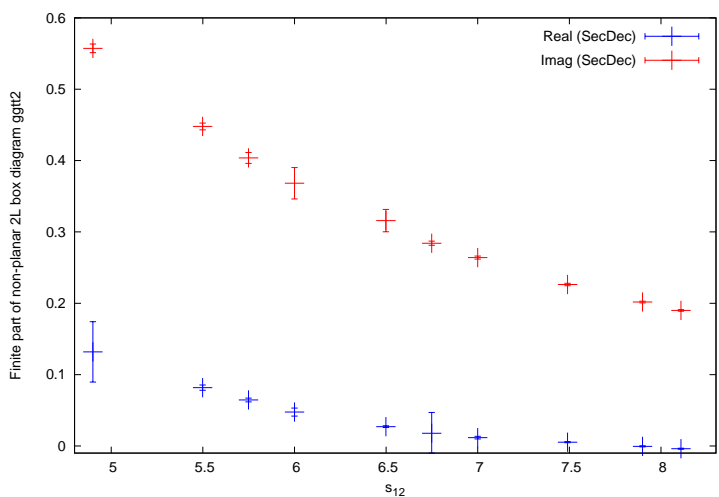} \label{subfig:finiteggtt2beyondthreshold}}
\caption{Results for the finite part of the scalar integral $ ggtt_2$, 
for a larger kinematic range \subref{subfig:finiteggtt2aroundthreshold}, and a 
region further away from threshold \subref{subfig:finiteggtt2beyondthreshold}.
The kinematics are the same as in Fig. \ref{fig:ggtt2_poles43}. 
The vertical bars display the uncertainty of the numerical result.}
\label{fig:ggtt2finite}
\end{center}
\end{figure}

Analytic results for the pole coefficients of the $1/\eps^4$ and $1/\eps^3$ part 
of the diagram $ggtt_2$ shown in Fig.~\ref{fig:ggtt2diagram} have been provided 
in Ref.~\cite{AvMACAT}. Using the mixed analytic and numerical approach 
presented in the work summarized in this thesis, the results of the purely 
analytical pole predictions can be numerically confirmed, see 
Fig.~\ref{fig:ggtt2_poles43}. 
The numerical results for the remaining pole coefficients and the 
finite part of $ggtt_2$ are shown in 
Figs.~\ref{fig:ggtt2_poles21} and \ref{fig:ggtt2finite}. 
For 
Figs.~\ref{fig:ggtt2_poles43} - \ref{fig:ggtt2finite}, an overall 
factor of $-16\,\,\Gamma(1+\eps)^2$ is extracted. %

Two Landau singularities can be observed, one is at $s_{12}=0$ and the 
other at  $s_{13}=m^2$. All other Landau singularities do not 
appear in the plot, as the values for $s_{23}$ and $m^2$ are kept fixed. 

\pagebreak
The numerical result presented here was compared 
to the fully analytical result in two selected phase space 
points~\cite{AvMprivate} for all Laurent coefficients up to 
the finite part, finding agreement within the numerical precision. 
The full prediction obtained from the numerical result of 
Figs.~\ref{fig:ggtt2_poles43} - \ref{fig:ggtt2finite} were 
useful as a cross-check in Ref.~\cite{vonManteuffel:2013uoa}. 
The timings for the leading and next-to-leading pole coefficients of the diagram $ggtt_2$ 
range between fractions of a second and around 20 seconds. 
The coefficients of the $1/\eps^2$ pole take 13 - 300 seconds, while the 
coefficients of the $1/\eps$ pole take between 75 seconds and 50 minutes, depending 
on their distance to thresholds. 
For the finite part, the integration times range from 250 seconds to 67 minutes.
For all Laurent coefficients, a relative accuracy of $5\times 10^{-3}$ has been stipulated,
which was not always reached for the $1/\eps$ and $\eps^0$ coefficients.
It should also be noted that the timings for points close to threshold 
are rather sensitive to the Monte Carlo integration parameters. %
%
%

\clearpage
\subsection{The $ggtt_1$ diagram}
\label{subsec:ggtt1numresults}
\begin{figure}[htb!]
\begin{center}
\subfigure[]{
\includegraphics[width=0.71\textwidth]{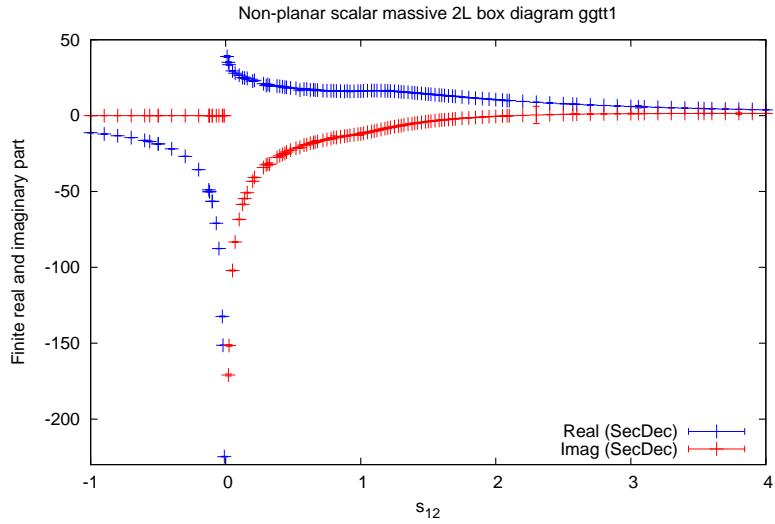} 
\label{subfig:scalarggtt1}}\\
\subfigure[]{
\includegraphics[width=0.71\textwidth]{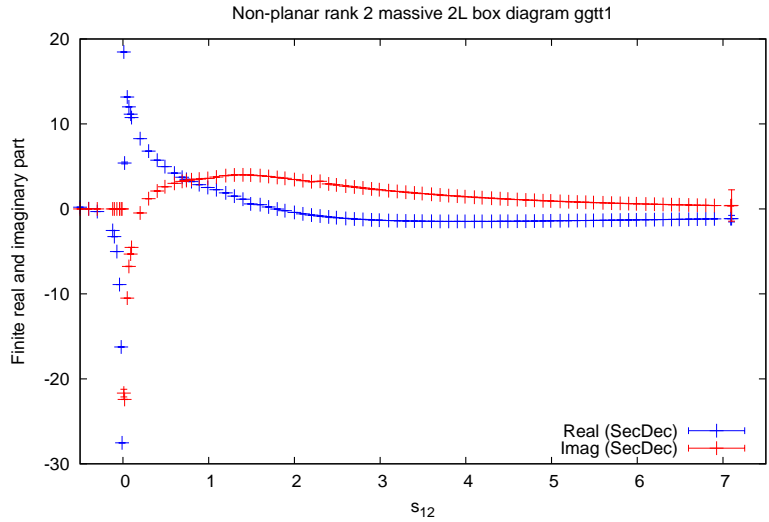} 
\label{subfig:rank2ggtt1}}
\end{center}
\caption{\subref{subfig:scalarggtt1} Results for the scalar integral $ggtt_1$ shown 
in Fig.~\ref{fig:ggtt1diagram}, and \subref{subfig:rank2ggtt1} the corresponding 
rank two tensor integral $ggtt_1$ with the factor 
$k_1\cdot k_2$ in the numerator. The invariant $s_{12}$ is varied, the invariants 
$s_{23}=-1.25$, $m_2=m_1$, $p_3^2=p_4^2=m_1^2=1$ are fixed. 
The uncertainties from numerical integration are 
shown as horizontal markers on the vertical lines. The absence of such markers 
means that the numerical uncertainty is not visible at this scale. }
\label{fig:ggtt1}
 \end{figure}

Numerical results for the diagram $ggtt_1$ with two massive external legs 
are shown for the 
scalar integral and an irreducible rank two tensor integral, 
compare Fig.~\ref{fig:ggtt1diagram} for the corresponding diagram and 
Fig.~\ref{fig:ggtt1} for the numerical results.
 
The integral representation of the diagram $ggtt_1$ is given by
\begin{subequations}
\begin{align}
\label{eq:ggtt1int}
\mathcal{G}_{ggtt_1}= \left( \frac{1}{\mathrm{i} \pi^{\frac{\mathrm{D}}{2}}}\right)^2 
 \int \frac{\mathrm{d^D}k_1\, \mathrm{d^D}k_2}{D_1\ldots D_7} \text{ ,}\\
 \end{align}
with the corresponding Feynman propagators 
 \begin{align}
 & D_1=k_1^2-m_2^2, \quad D_2=(k_1+p_1)^2-m_2^2, \quad D_3=k_2^2-m_2^2, \\
 & D_4=(k_2+p_2)^2-m_2^2, \quad D_5=(k_1-k_2+p_1)^2 \text{,}\\
& D_6=(k_1-k_2-p_2)^2, \quad D_7=(k_1-k_2+p_1+p_3)^2-m_1^2 \text{ ,}
\end{align}
\end{subequations}
where the infinitesimal $\mathrm{i} \delta$ is omitted for brevity, and where 
the convention of all external momenta being ingoing was used. 
The dimension is denoted by $\mathrm{D}$. 
Two external legs 
$p_3$ and $p_4$ are massive and equal, $p_3^2=p_4^2=m_1^2$. 
For the results shown in Fig.~\ref{fig:ggtt1}, the numerical values 
$m_1^2=m_2^2=m^2=1, s_{23}=-1.25, s_{13}=2\,m^2-s_{12}-s_{23}$ were used.
In Fig.~\ref{fig:ggtt1}, the two masses are set to $m_1^2=m_2^2$, 
as this is the topology appearing in the process $gg\to t\bar{t}$ at NNLO 
if the $b$-quarks are assumed to be massless. 
While numerical results for the scalar integral are shown in 
Fig.~\ref{subfig:scalarggtt1}, \subref{subfig:rank2ggtt1} 
corresponds to a rank two tensor integral with the same propagators 
as \subref{subfig:scalarggtt1}, 
with a scalar product of loop momenta $k_1\cdot k_2$, in the numerator. 

\medskip

The timings for one kinematic point for the scalar integral in Fig.~\ref{subfig:scalarggtt1}
range from 11-60 secs for points far from threshold to $27$ minutes
that are very close. In the vicinity of the threshold, the average is around 500 secs. 
A relative accuracy of $10^{-3}$ has been specified for terminating the 
numerical integration, 
while the absolute accuracy has been set to  $10^{-5}$. 
For the tensor integral, the timings are better than for the scalar case, 
as the numerator present in this case smoothes out the integrand. 
A phase-space point far from threshold takes around 5-10 secs, while points 
very close to threshold do not exceed 1 hour 
for the rank 2 tensor integral. 
The results were obtained on a single 8 core Intel i7 machine. 

\bigskip

\bigskip


\pagebreak

\begin{figure}[htb]
\begin{center}
\includegraphics[width=0.8\textwidth]{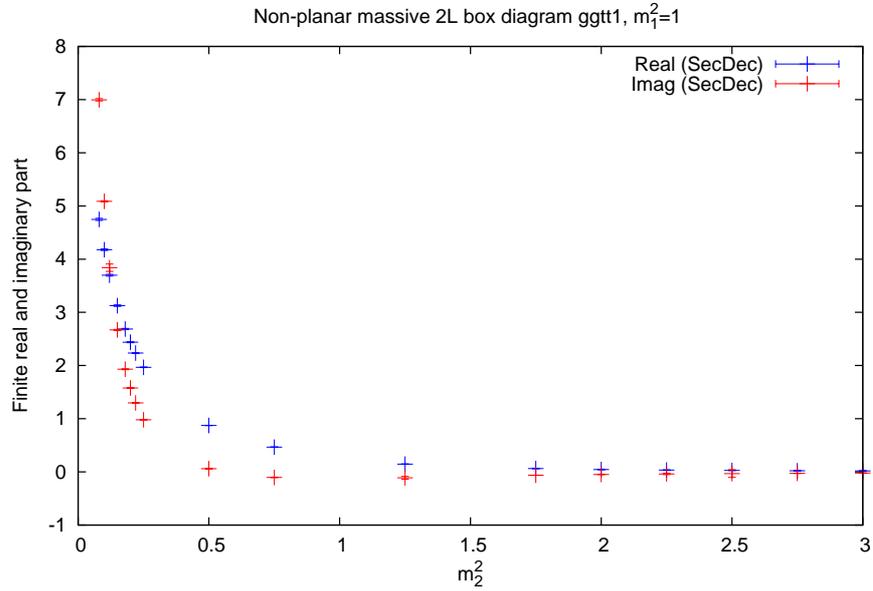} 
\caption{Results for the scalar integral $ggtt_1$ shown in Fig.~\ref{fig:ggtt1diagram}, 
with two different masses. The second mass $m_{2}$ is varied, the other kinematic 
invariants are fixed with the values $s_{12}=5$, $s_{23}=-1.25$ and 
$p_3^2=p_4^2=m_1^2=1$.}
\label{fig:ggtt1m1m2}
\end{center}
 \end{figure}

Numerical results with $m_1\neq m_2$ are shown 
in Fig.~\ref{fig:ggtt1m1m2} to demonstrate that adding another 
mass scale is extremely straightforward with our approach, whereas 
analytical calculations would suffer from enormous additional 
complications.
 \clearpage
\subsection{Planar seven-propagator all massive graph $M_7^{\text{P}}$}
\begin{figure}[htb!]
\begin{center}
\includegraphics[width=0.8\textwidth]{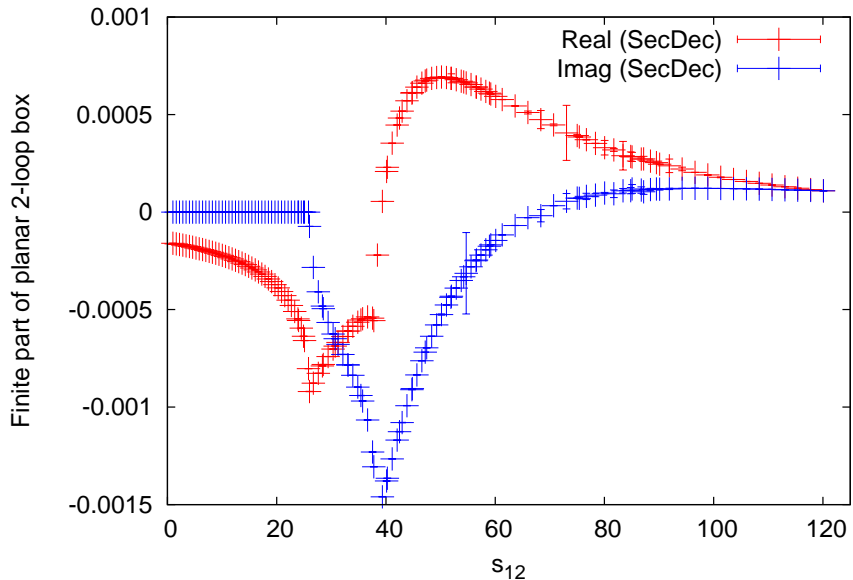}
\end{center}
\caption{Results for the all massive planar seven-propagator graph $M_7^{\text{P}}$ 
with all propagators and external legs massive, using 
$m_1^2=2$, $m_2^2=6$, $m_3^2=7$, $m_4^2=8$, $m_5^2=9$, $m_6^2=10$, 
$m_7^2=12$, $p_1^2=1$, $p_2^2=3$, 
$p_3^2=4$, $p_4^2=5$ and $s_{23}=-0.25$.}
\label{fig:mp7plot}
\end{figure}

To demonstrate the possibility to compute diagrams with arbitrarily many scales, a 
planar seven-propagator toy graph involving the maximal amount of 13 independent 
scales is computed, compare 
Fig.~\ref{fig:JapPdiagram}. The diagram has no poles in the regulator $\eps$. 
All propagators are assumed to have different masses, all external legs 
are chosen to be massive as well. While keeping $s_{23}=-0.25$ fixed, numerical 
values of the finite part of the $M_7^{\text{P}}$ diagram are shown, see 
Fig.~\ref{fig:mp7plot}, where $s_{12}$ and $s_{13}$ are varied using the 
physical constraint $s_{12}+s_{23}+s_{13}=p_1^2+p_2^2+p_3^2+p_4^2$. 

The timings for the numerical integration range between 10 and 180 secs with 
a relative accuracy of 
$10^{-3}$ or an absolute accuracy of $10^{-8}$ if the imaginary part 
is zero, see Sec.~\ref{subsec:program:cubaparameters} for a discussion 
on the relative vs absolute accuracy. \\
These results show that there is in principle no constraint on the number of 
scales involved. 
\subsection{Non-planar seven-propagator all massive graph $M_7^{\text{NP}}$}
\begin{figure}[ht]
\begin{center}
\includegraphics[width=7.cm]{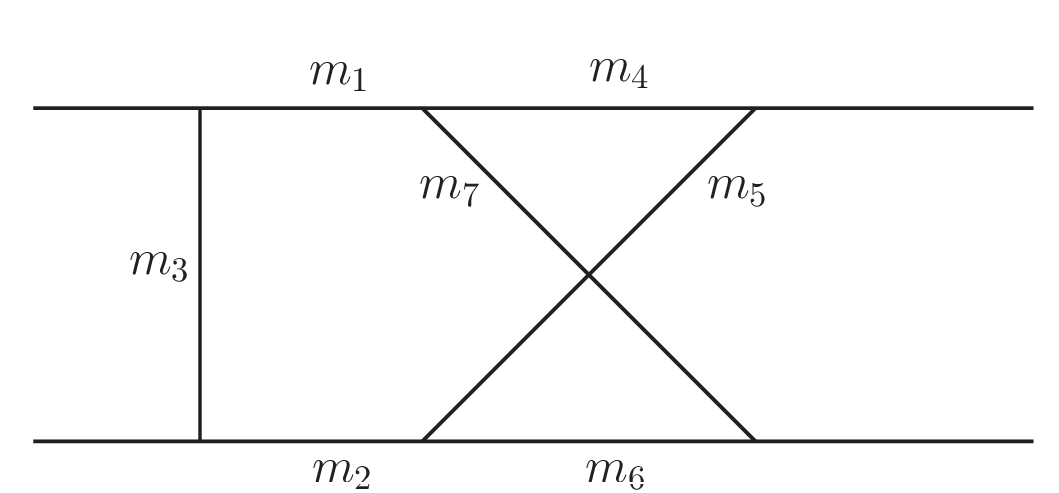}
\end{center}
\caption{Labeling of the masses for the non-planar graph $M_7^{\text{NP}}$.}
\label{fig:JapNP}
\end{figure}

\begin{figure}[ht]
\begin{center}
\includegraphics[width=0.7\textwidth]{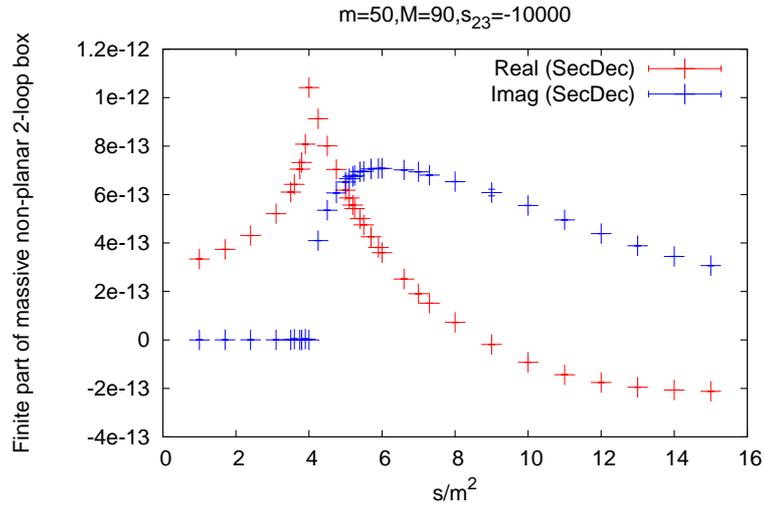}
\end{center}
\caption{Results for the non-planar 7-propagator graph $M_7^{\text{NP}}$ with all 
propagators massive, using $m=50$, $M=90$ and $s_{23}=-10^4$.}
\label{fig:JapNPresults}
\end{figure}

In this example, a seven-propagator non-planar two-loop box integral is considered, 
where all propagators are massive, 
using $m_2=m_4=m_5=m_7=m$,  $m_1=m_3=m_6=M$, $p_1^2=p_2^2=p_3^2=p_4^2=m^2$. 
The labeling is as shown in Fig.~\ref{fig:JapNP}. 
Numerical results for this integral were obtained by Fujimoto 
et al.~\cite{Yuasa:2011ff}, where solutions 
are found by extrapolation in the $i\delta$ parameter. For comparison, 
results produced with \secdec{} 
are shown for the same mass configuration using $m=50, M=90, s_{23}=-10^4$, 
see Fig.~\ref{fig:JapNPresults}. They 
are in agreement with Ref.~\cite{Yuasa:2011ff}.
The computation time for the longest sub-function (for both real and imaginary parts) 
for a relative accuracy of one per mil vary between about 20 secs
for a point far from and about 500 secs close to the threshold.
\subsection{Non-planar six-propagator diagram}
\label{subsec:bnp6}
\begin{figure}[htb]
\begin{center}
\includegraphics[width=0.7\textwidth]{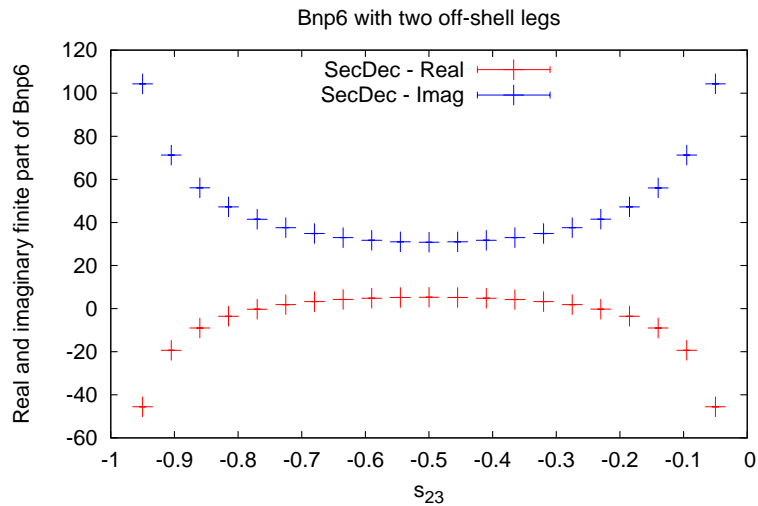}
\caption{Results for the finite part of the graph $B_6^{\text{NP},a}$ with 
massive legs $p_1^2=p_2^2=1$. The error bars are barely seen in the figures as the numerical 
accuracy is about one per mil.}
\label{fig:Bnp6result}
\end{center}
\end{figure}

\begin{figure}[htb!]
\begin{center}
\unitlength=1mm
\includegraphics[width=0.7\textwidth]{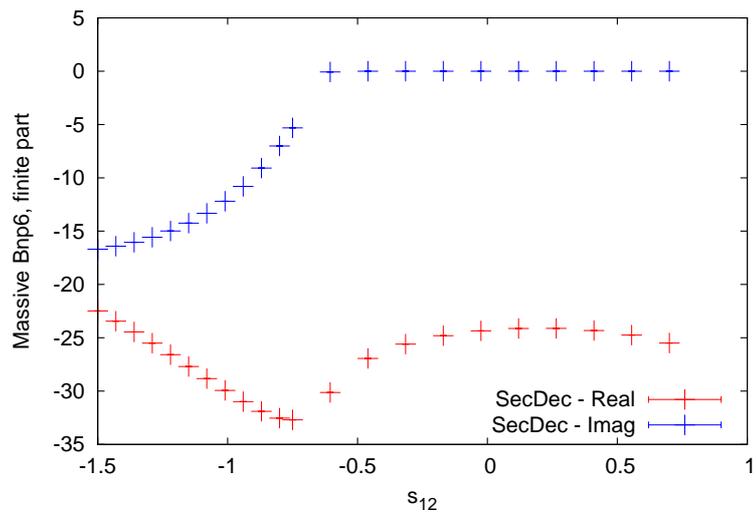}
\end{center}
\caption{The non-planar 6-propagator graph $B_6^{\text{NP},b}$ as a function of $s_{12}$ 
in a region containing a threshold, where $m_1^2=m_2^2=m_5^2=m_6^2=0.25$ and $s_{23}=-0.4$. 
The error bars are barely seen in the figures as the numerical 
accuracy is about one per mil.}
\label{fig:Bnp6threshold}
\end{figure}

First, the non-planar six-propagator two-loop four-point diagram is considered, 
compare Fig.~\ref{fig:Bnp6diagram}. For light-like legs and massless 
propagators, the analytic result has been calculated, see Ref.~\cite{Tausk:1999vh}. 
The name of the graph $B_6^{\text{NP}}$, is adopted from this reference. 
\begin{figure}[htb!]
\begin{center}
\subfigure[$B_6^{\text{NP},a}$]{\includegraphics[width=5.5cm]{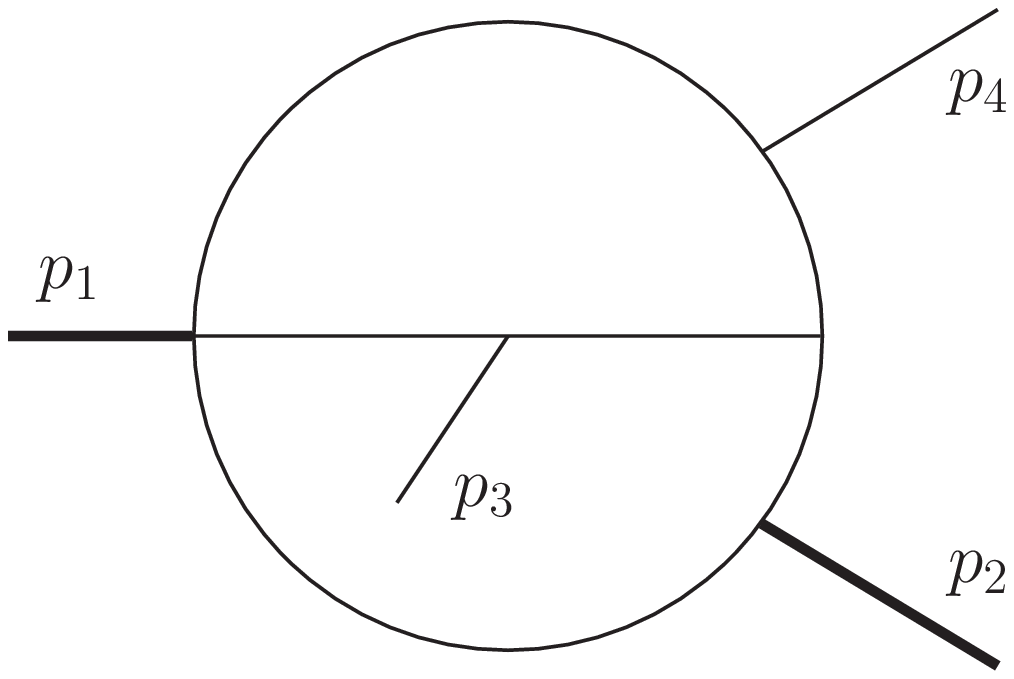}
\label{fig:Bnp6adiagram}} \hspace{30pt}
\subfigure[$B_6^{\text{NP},b}$]{\includegraphics[width=5.5cm]{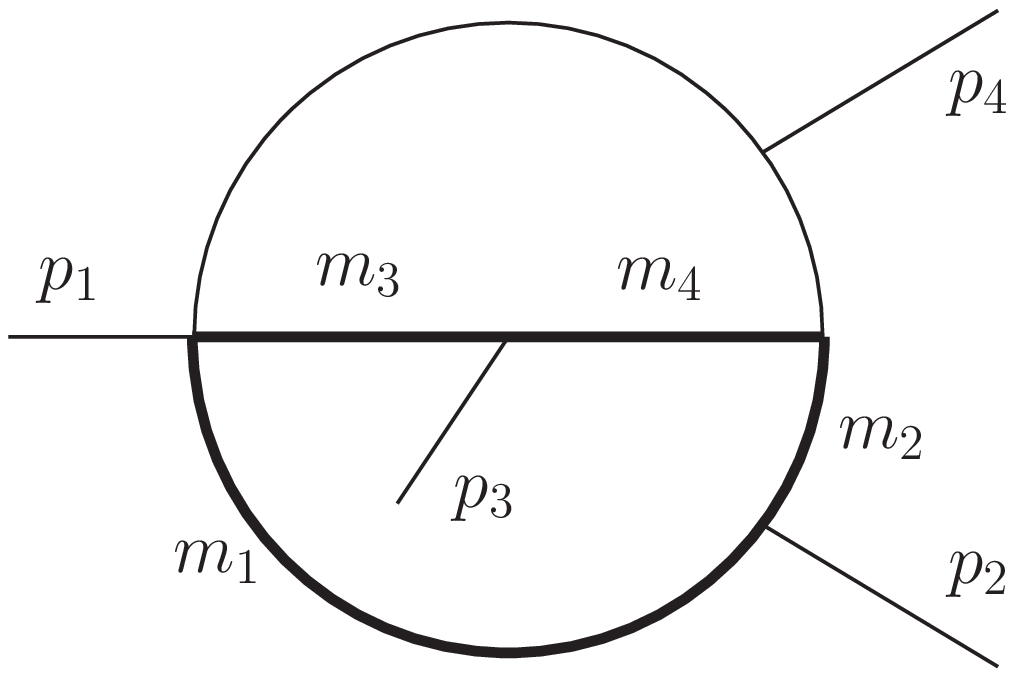}
\label{fig:Bnp6bdiagram}}
\end{center}
\caption{The two non-planar 6-propagator topologies studied here: \subref{fig:Bnp6adiagram} 
$B_6^{\text{NP},a}$ and \subref{fig:Bnp6bdiagram} $B_6^{\text{NP},b}$. The thick lines denote 
massive external legs and propagators, respectively.}
\label{fig:bnptopos} 
\end{figure}
In the following, two different mass configurations of the graph $B_6^{\text{NP}}$ 
are studied \\
\begin{itemize}
\item[$B_6^{\text{NP},a}$:] Two external legs, $p_1^2$ and $p_2^2$, are massive, all 
propagators are massless. The leading pole of this topology is of order $\mathcal{O}(1/\eps^4)$. 
\item[$B_6^{\text{NP},b}$:] All external legs are light-like, four propagators are massive 
with $m_1=m_2=m_3=m_4 \neq 0$, the other two are massless. 
This topology contains poles starting from order $\mathcal{O}(1/\eps)$.
\end{itemize}
For the topology with light-like legs as considered in Ref.~\cite{Tausk:1999vh}, the 
leading pole is of the order $\mathcal{O}(1/\eps^2)$. 
The difference in the pole structure is due to cancellations related to 
the high symmetry of the graph. 
See Figs.~\ref{fig:Bnp6result} and \ref{fig:Bnp6threshold} 
for numerical results of the finite parts of $B_6^{\text{NP},a}$ and 
$B_6^{\text{NP},b}$ and Fig.~\ref{fig:bnptopos} for the corresponding diagrams. 
\\
In accordance with Ref.~\cite{Tausk:1999vh}, an overall prefactor of 
$\Gamma(1+2\eps)\Gamma(1-\eps)^3/\Gamma(1-3\eps)/(1+4\eps)$ has 
been extracted in all numerical results. Also, for all the values given, 
$s_{13}$ is determined by the physical constraint 
$s_{12}+s_{23}+s_{13}=p_1^2+p_2^2$. \\
For Fig.~\ref{fig:Bnp6result}, the numerical value $s_{12}=3$ was adopted 
while scanning over $s_{23}$. The massive external legs are set to 
$p_1^2=p_2^2=1$. In Fig.~\ref{fig:Bnp6threshold}, numerical values for the 
finite part of the $B_6^{\text{NP},b}$ diagram are shown, where the mass 
scale is set to $m=0.5$. The kinematic invariant 
$s_{12}$ is varied, choosing $s_{23}=-0.4$. 
\clearpage
\section{Summary}
%
%
In this chapter the evaluation of diverse non-trivial two-loop diagrams, including 
planar and non-planar topologies with six or seven propagators, was shown. 
It was found that their evaluation with the upgraded version of the program 
\secdec{} is not restricted by the number of involved scales. Complicated 
diagrams hardly or not accessible with analytical techniques can easily 
be computed, proving the program \secdec{} a powerful tool for checks, 
comparisons and predictions. Yet, an evaluation can become 
difficult if the singularity structure is very complicated 
such that spurious linear divergences occur. 
In the example of the massive non-planar two-loop box $ggtt_2$ which exhibits 
such an extremely complicated singularity structure, it has 
proven beneficial to do an analytical preparation of the integral prior to 
numerical integration. A simplification of the functions to be integrated 
numerically can be achieved by a reduction in the number of Feynman 
parameters to integrate over numerically or in the removal of spurious divergences. 
The former was achieved by integration of one Feynman parameter 
in a sub-loop. Towards the latter 
a new type of transformation, introduced by the author and collaborators 
and summarized in this thesis, facilitated a trading of 
linear divergences for logarithmic ones, thereby achieving a 
reduction by two thirds in the total number 
of functions to be integrated.
As was shown, the analytical preparation leads 
to an overall improved numerical convergence. 
Exploiting the newly developed feature of including 
user-defined functions into the \secdec{} setup, an automated 
evaluation of the $ggtt_2$ diagram was possible. 
The examples demonstrated in this chapter may serve as a guideline 
for the evaluation of very complicated integrals, to become of 
importance in future phenomenological applications. 

%% file: application2/application2.tex
\chapter{Neutral MSSM Higgs-boson spectrum at the two-loop level}
\label{chap:application2}%
The momentum-dependent two-loop contributions to the 
neutral ${\cal CP}$-even MSSM Higgs-boson masses are 
computed at order $\mathcal{O}(\alpha_s \alpha_t)$. 
This requires the calculation of two-loop self-energies with a 
proper renormalization at the two-loop level. 
The calculation is performed using the Feynman-diagrammatic 
approach. 
An effective potential approach, 
though leading to compact expressions, 
does not allow for the incorporation of momentum 
dependence. 
%
%

All relevant two-loop self-energy diagrams and 
those one- and two-loop diagrams contributing 
to the renormalization are generated 
using {\sc FeynArts}~\cite{Kublbeck:1990xc,Hahn:2000kx,Hahn:2001rv}. 
From a diagrammatic point of view, the diagrams involved 
in the calculation including the momentum-dependence 
remain the same with respect to the calculation at zero
momentum transfer. This is due to the fact that the diagrams 
are selected by coupling factors. 
The sole difference is in the dependence of the self-energy 
diagrams on the external momentum.  
Yet, this difference is non-trivial because analytical 
expressions involve the evaluation of elliptic integrals 
which can presently not be performed yielding fully analytical results. 
As numerical evaluations are generally more time-consuming, 
the number of diagrams involved needs to be reduced 
considerably, to a minimal set of master integrals.

A reduction of tensor integrals and the evaluation of traces is 
performed with the packages 
{\sc TwoCalc}~\cite{Weiglein:1993hd} and 
{\sc FormCalc}~\cite{Hahn:1998yk,Hahn:2006zy,Agrawal:2012tm}. 
While the package 
{\sc TwoCalc} condenses the two-loop amplitudes to only scalar master 
topologies, {\sc FormCalc} reduces 
all one-loop counter-terms and counter-term insertions to a basis 
of scalar integrals and a small number of tensor coefficients. 
In the reduction, the method of partial fractioning is used 
where applicable. Also the cancelation of denominators after 
taking the derivative with respect to a kinematic invariant is 
exploited, in 
addition to the application of symmetry relations. An important feature of 
{\sc TwoCalc} is benefitting from the extension of the idea of a tensor 
decomposition introduced by Passarino and Veltman~\cite{Passarino:1978jh} 
as a reduction technique at one-loop. In {\sc TwoCalc}, the latter is applied to 
a sub-loop of an integral before the remains of the integral can 
be further decomposed and simplified with the before 
mentioned techniques. 

After studying the one- and two-loop counter-terms renormalizing the 
masses as well as the fields, and assuring a consistent cancellation of all 
divergences, the resulting finite terms 
are evaluated. Where possible, analytic results are used, all 
other integral topologies are computed numerically using 
the program \secdec. 
The resulting self-energy corrections are added to the 
inverse propagator matrix, as discussed in Sec.~\ref{sec:highorderreview},
and the resulting mass shifts to the neutral ${\cal CP}$-even 
Higgs-boson masses are computed. 
The two-loop calculation 
is performed in the $\phi_1$-$\phi_2$ basis. 
The rotation into the physical $h^0$-$H^0$ basis according to 
Eq.~(\ref{eq:physbasis}) is performed afterwards. 
\section{Dominant momentum-dependent two-loop QCD corrections}
To compute the dominant momentum-dependent two-loop 
contributions to the neutral MSSM Higgs-boson spectrum, only those 
self-energy diagrams with couplings 
strictly of the order $\mathcal{O}(\alpha_s \alpha_t)$ are taken into
account.
%
While electro-weak gauge contributions are 
incorporated up to the one-loop level, they are 
assumed negligible at the two-loop level.  
The corrections involving squares in the top Yukawa coupling, Eq.~(\ref{eq:yukdefs}), 
dominate the electro-weak higher-order contributions, 
as $m_t^2 \gg m_{Z,W}^2$. Contributions involving couplings to 
gauge bosons are therefore expected to be relatively small. 

Although the top Yukawa coupling is 
much larger than the bottom Yukawa coupling, $y_t \gg y_b$, 
it might be argued that the scalar bottom (sbottom) quark 
mass could be significantly larger than the bottom 
mass, when supersymmetry is broken. This region of parameter 
space could then lead to 
corrections of similar size with respect to the couplings 
proportional to $y_t^2$.
However, the couplings of the squarks to the Higgs-bosons are 
all proportional to Yukawa couplings, see 
Sec.~\ref{sec:scalarquarktreelevel}. 
They are composed of
the fully supersymmetric F-term and the non-supersymmetric 
soft-breaking contributions, where fully supersymmetric refers to 
the relation $y_{\tilde{t},\tilde{b}}=y_{t,b}$.
The relevant soft-breaking terms are
proportional to the trilinear couplings $A_{t,b}$ and the Standard Model 
Yukawa couplings. Assuming no inverse hierarchy among the trilinear
couplings, the up-type Yukawa terms 
dominate over the down-type ones.
Therefore, all bottom and sbottom contributions can be 
assumed to be negligible. 
%
%
%
\section{Self-energy diagrams}
\label{sec:sediags}
\begin{figure}[htb!]
\begin{center}
\subfigure[Top. 1]{\raisebox{0pt}{\includegraphics[width=0.3\textwidth]{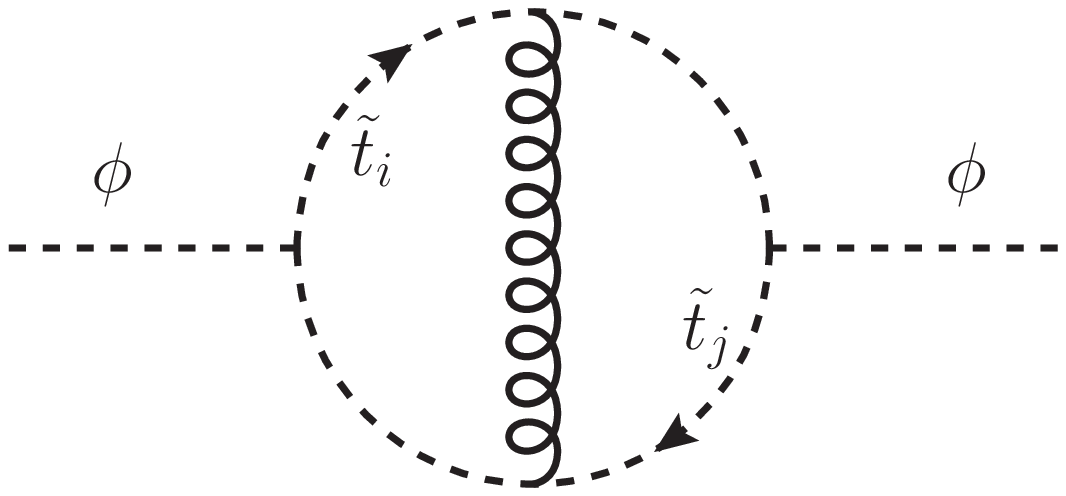}} \label{subfig:setop4}}
\subfigure[Top. 2]{\raisebox{1pt}{\includegraphics[width=0.3\textwidth]{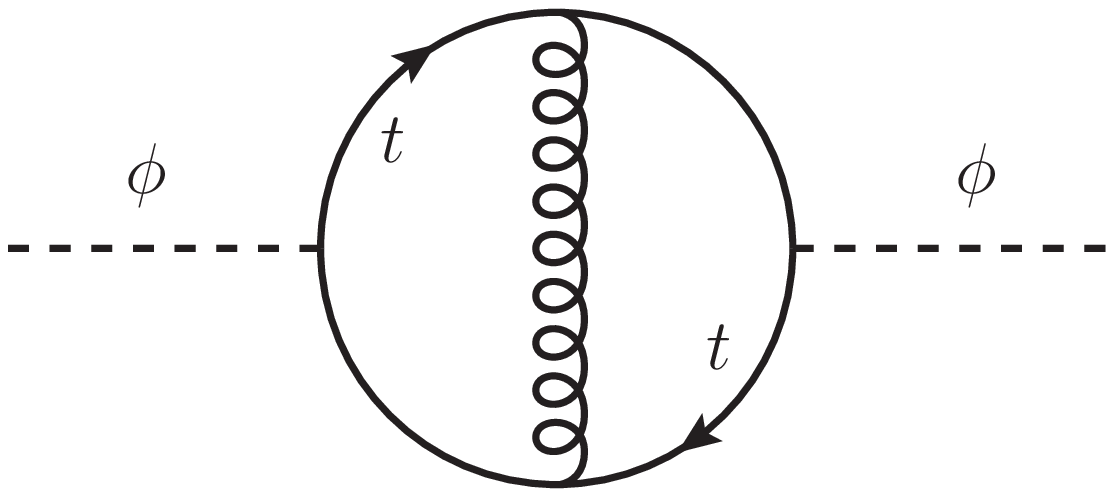}} \label{subfig:setop7}}
\subfigure[Top. 3]{\raisebox{1pt}{\includegraphics[width=0.3\textwidth]{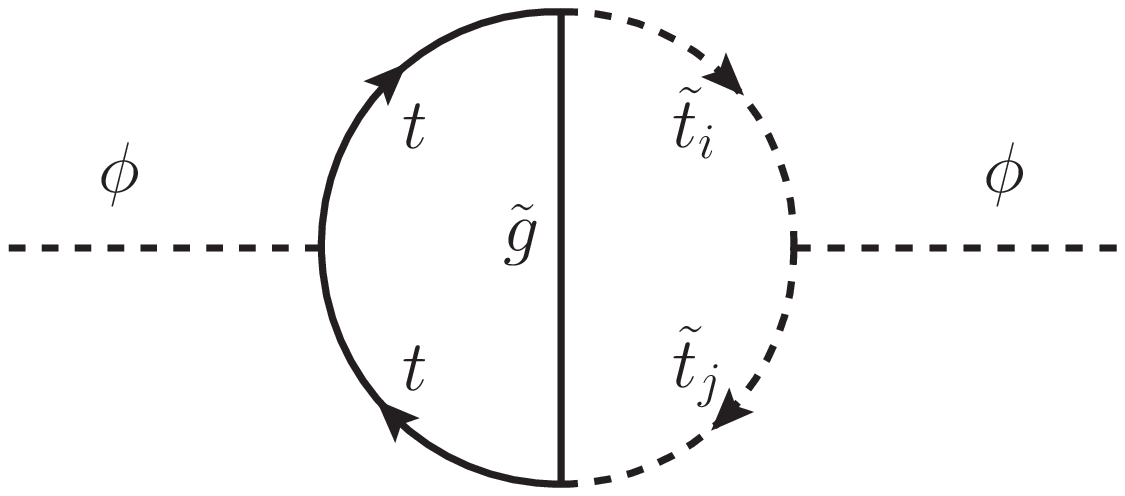}} \label{subfig:setop11}}\\
\subfigure[Top. 4]{\raisebox{0pt}{\includegraphics[width=0.3\textwidth]{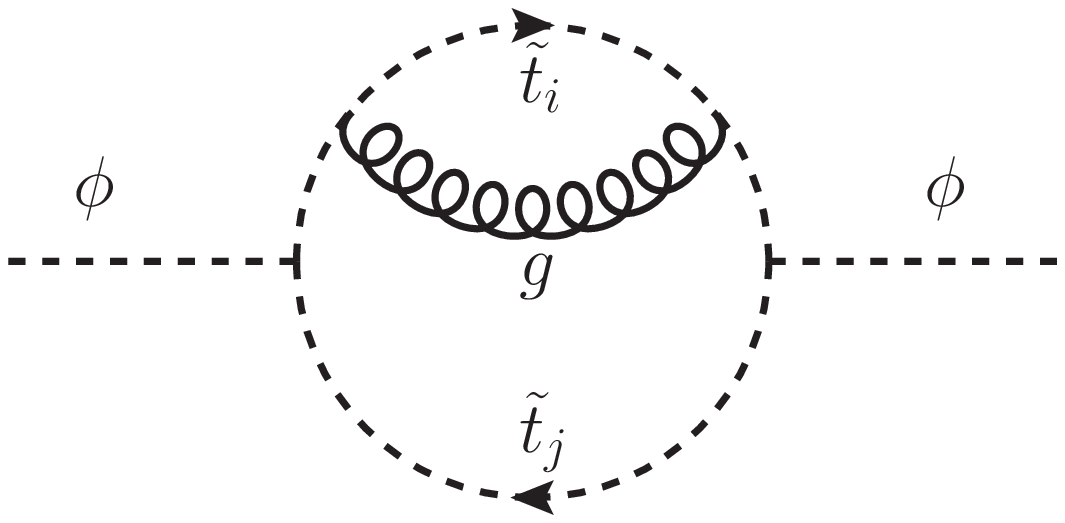}} \label{subfig:setop5}}
\subfigure[Top. 5]{\raisebox{1pt}{\includegraphics[width=0.3\textwidth]{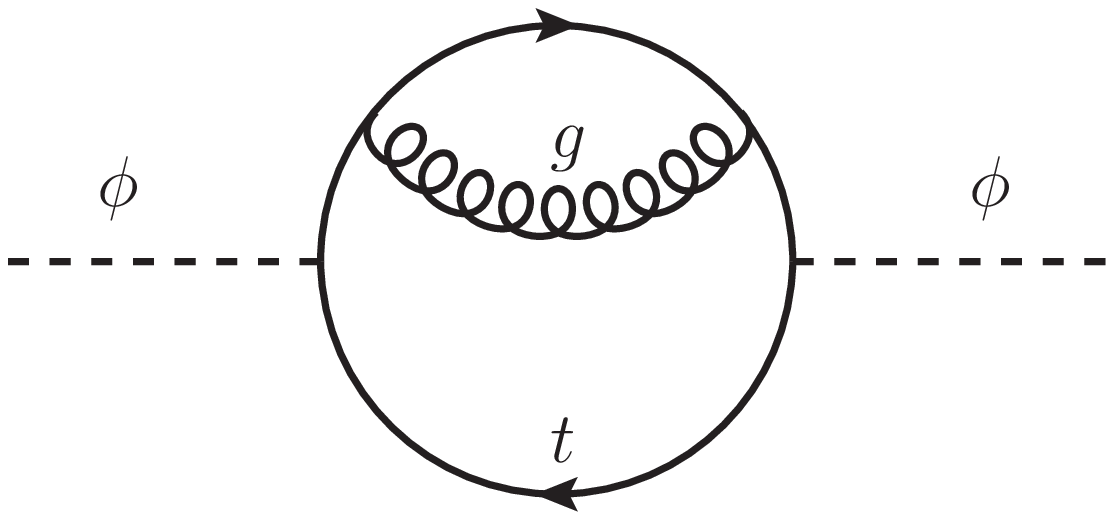}} \label{subfig:setop8}}
\subfigure[Top. 6]{\raisebox{1pt}{\includegraphics[width=0.3\textwidth]{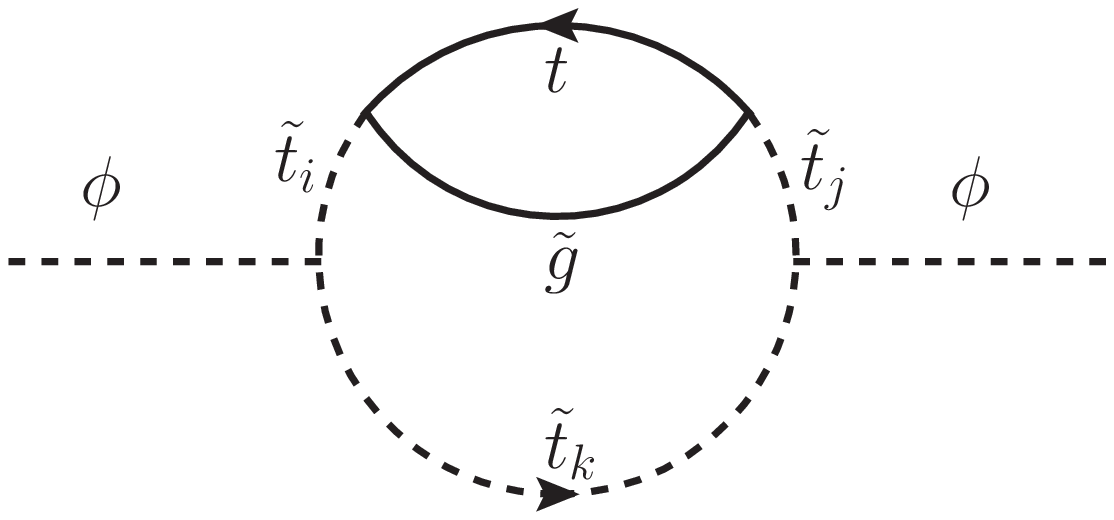}} 
\label{subfig:setop9}}\\\hspace{-20pt}
\subfigure[Top. 7]{\raisebox{0pt}{\includegraphics[width=0.3\textwidth]{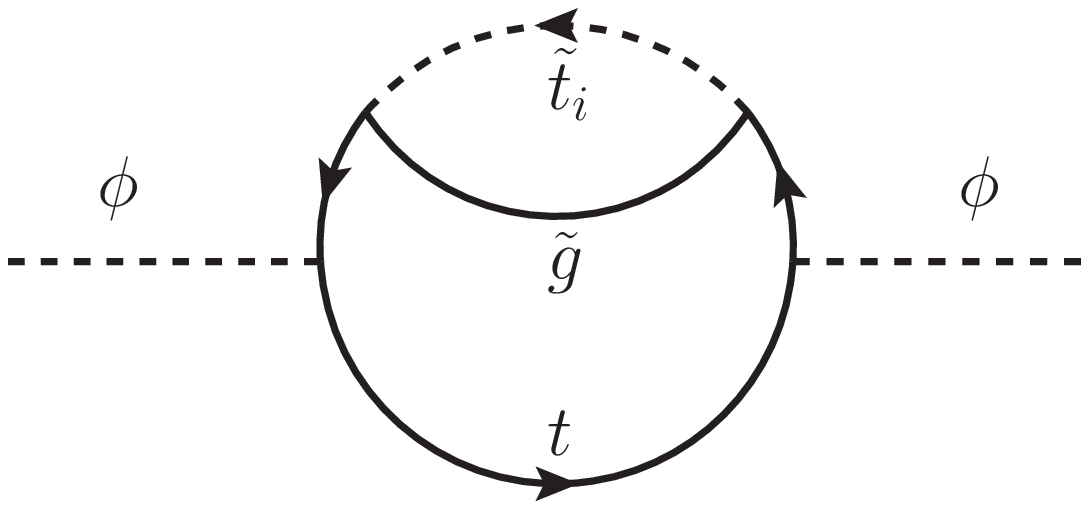}} 
\label{subfig:setop10}}\hspace{19pt}
\subfigure[Top. 8]{\raisebox{1pt}{\includegraphics[width=0.22\textwidth]{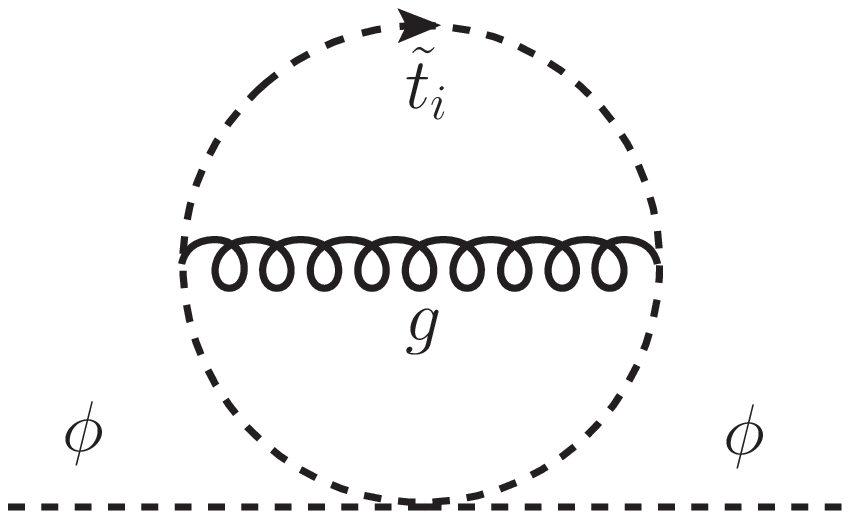}} 
\label{subfig:setop6}}\hspace{34pt}
\subfigure[Top. 9]{\raisebox{0pt}{\includegraphics[width=0.22\textwidth]{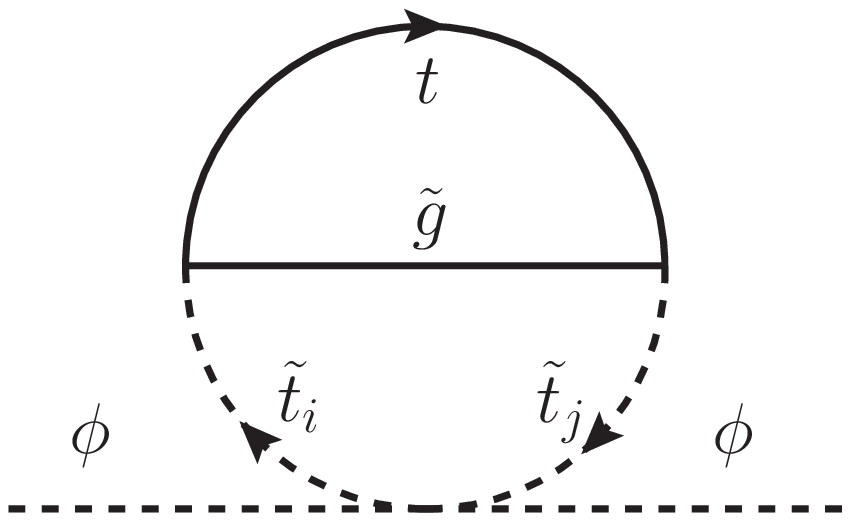}} 
\label{subfig:setop12}}\\\hspace{-2pt}
\subfigure[Top. 10]{\raisebox{0pt}{\includegraphics[width=0.22\textwidth]{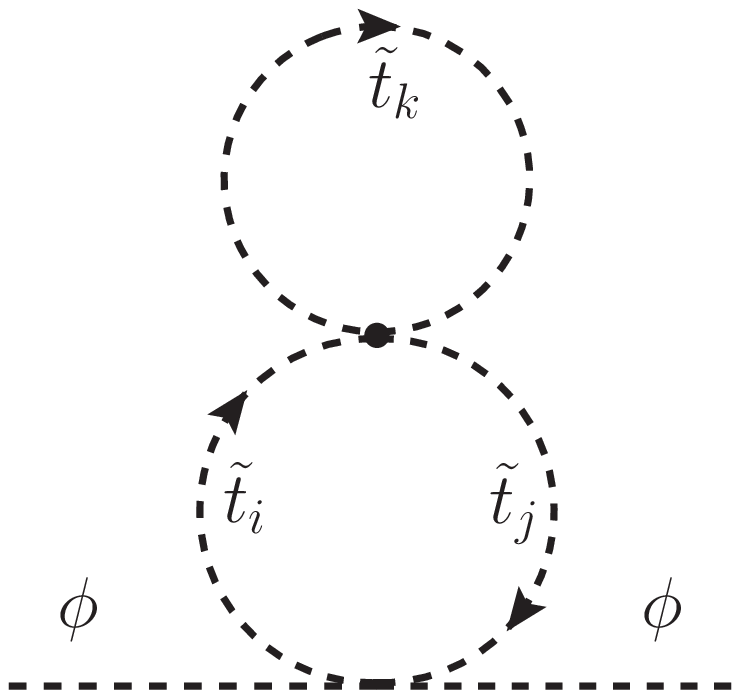}} 
\label{subfig:setop3}}\hspace{2pt}
\subfigure[Top. 11]{\raisebox{0pt}{\includegraphics[width=0.38\textwidth]{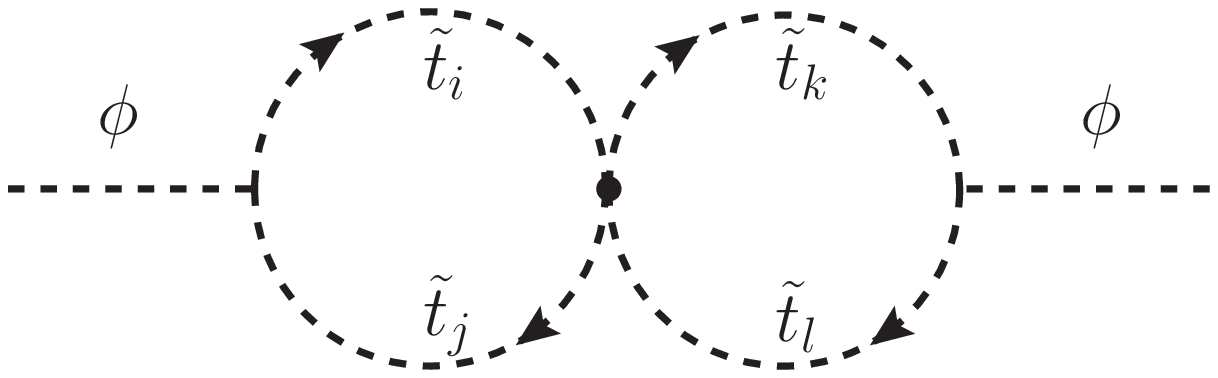}} 
\label{subfig:setop1}}\hspace{1pt}
\subfigure[Top. 12]{\raisebox{2pt}{\includegraphics[width=0.22\textwidth]{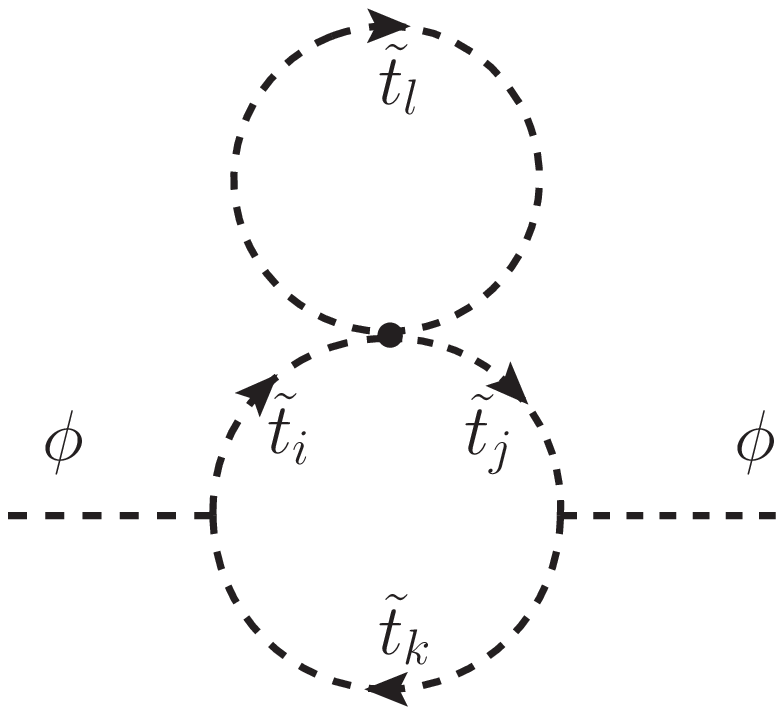}} \label{subfig:setop2}}
\end{center}
\caption{Generic two-loop self-energy diagrams contributing at the order $\mathcal{O}(\alpha_s \alpha_t)$, with $\phi = \phi_1, \phi_2, A^0$ and $i,j,k=1,2$. The 
$A^0$ boson self-energies enter the two-loop counter terms during renormalization. 
Stop particles $\tilde{t}$ carry indices $1$ and $2$, 
in accordance with the Feynman rules.}
\label{fig:selfdiags} 
\end{figure}
Twelve different two-loop topologies contribute 
to the MSSM Higgs-boson self-energy corrections
at the order $\mathcal{O}(\alpha_s\alpha_t)$, see 
Fig.~\ref{fig:selfdiags}. 
They match Eq.~(\ref{eq:sigmaxixj}) 
choosing the scalar fields $\phi_1^0$ and $\phi_2^0$ 
for $x_a$. 
Every vertex of Yukawa type involving a Higgs-boson and quarks or squarks
contains a square root factor of $\alpha_t$. All quark or squark 
interactions with a gluon or gluino and the 4-squark vertices 
contribute with a factor of $\sqrt{\alpha_s}$. Products of these 
coupling factors lead to an overall order of $\alpha_s\alpha_t$ 
for each diagram. 
%
%
The squared one-particle 
irreducible diagrams and reducible two-loop diagrams 
$(\Sigma^{(1)}_{\phi_i \phi_j})^2$ do not contribute 
to the corrections because their amplitudes are not 
of the order $\alpha_s\alpha_t$. 

\pagebreak

\begin{figure}[htb!]
\begin{center}\hspace{31pt}
\subfigure[$T_{134}$]{\raisebox{0pt}{\includegraphics[width=0.129\textwidth]{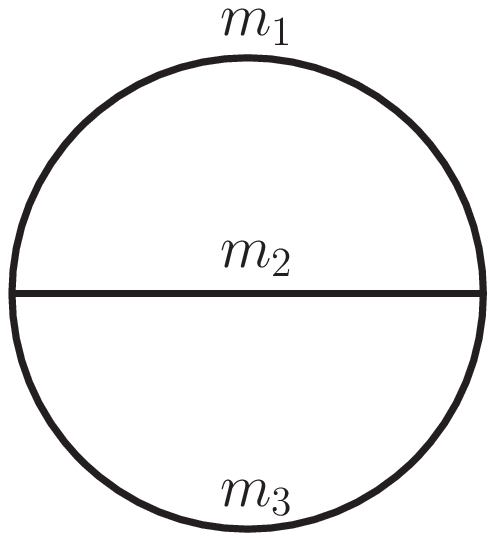}}
\label{subfig:T134}}\hspace{38pt}
\subfigure[$T_{234}$]{\includegraphics[width=0.29\textwidth]{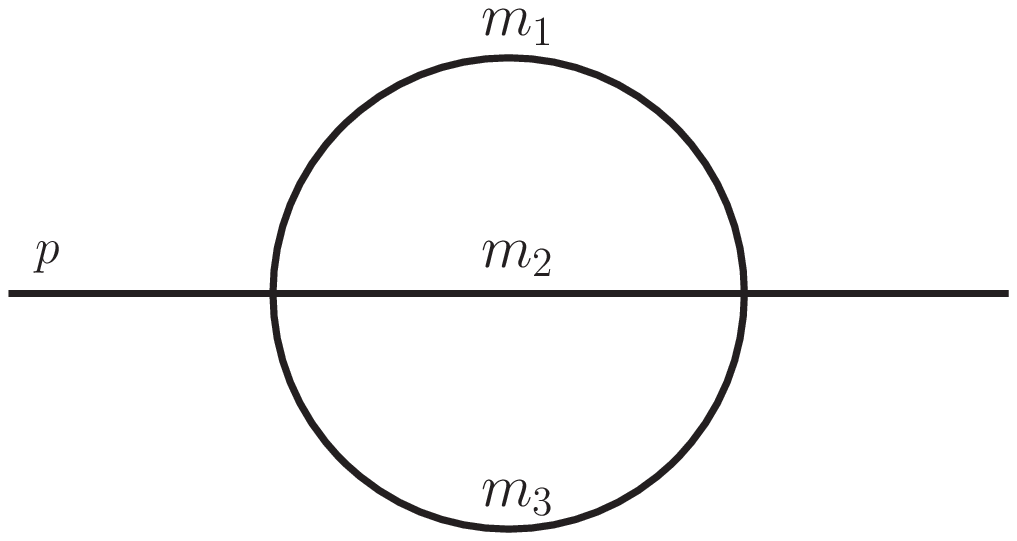}
\label{subfig:T234}}\hspace{18pt}
\subfigure[$T_{1234}$]{\raisebox{0pt}{\includegraphics[width=0.29\textwidth]{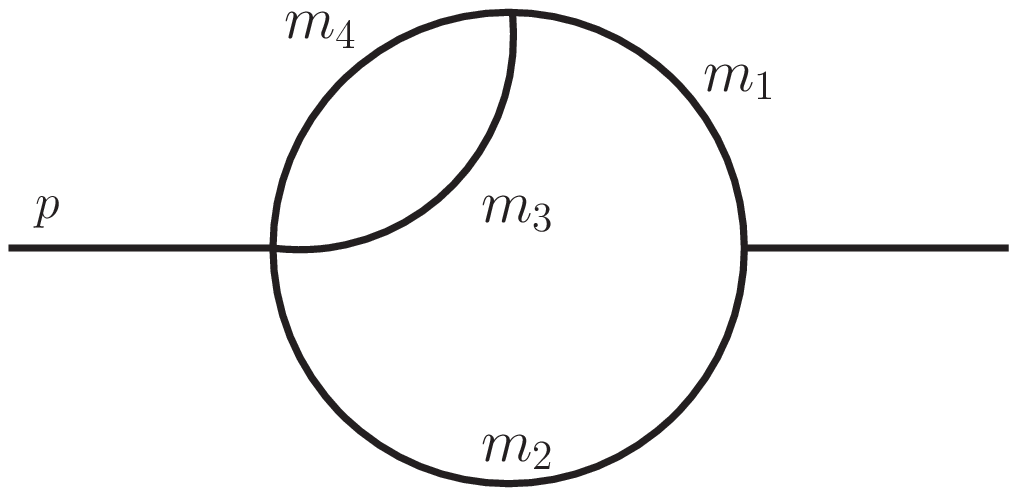}}
\label{subfig:T1234}}\\
\subfigure[$T_{11234}$]{\includegraphics[width=0.29\textwidth]{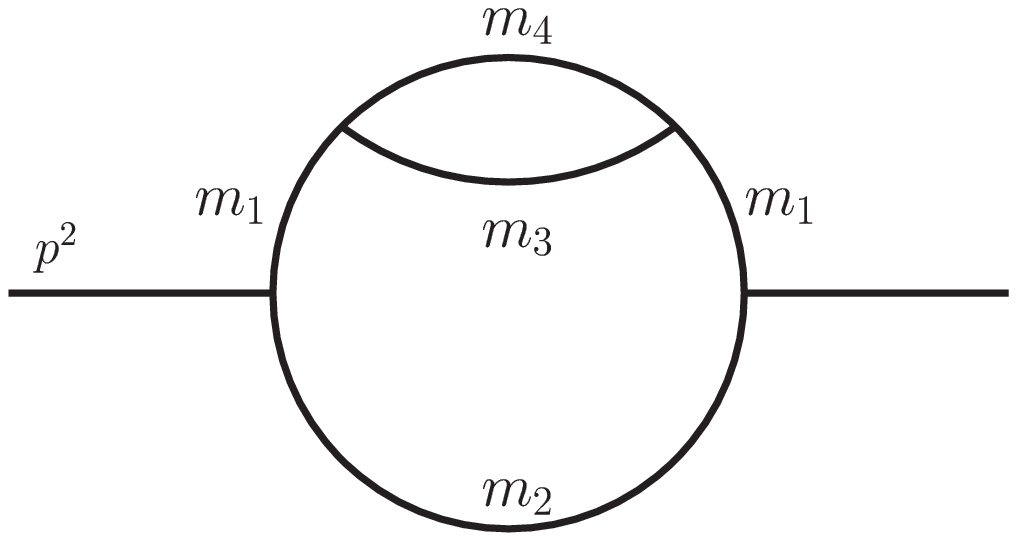}
\label{subfig:T11234}}\hspace{20pt}
\subfigure[$T_{12345}$]{\includegraphics[width=0.29\textwidth]{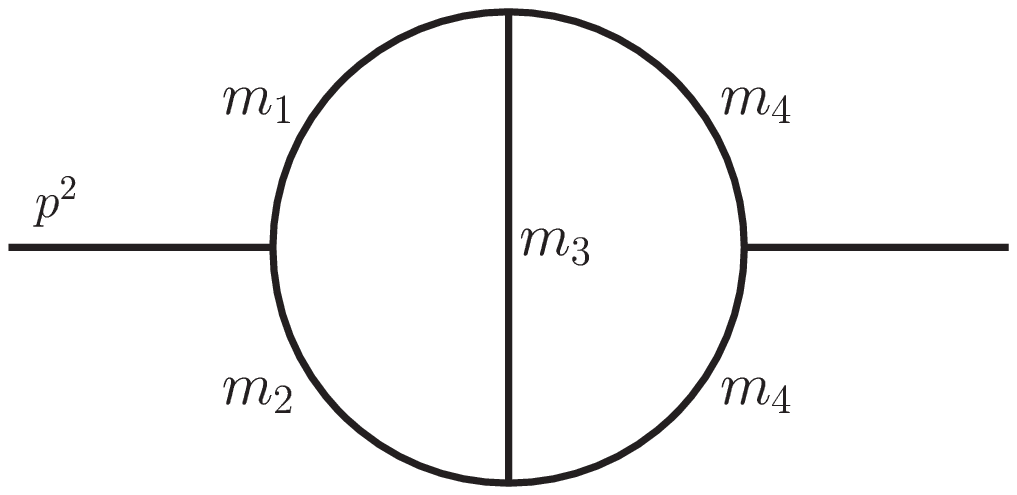}
\label{subfig:T12345}}
\caption{Two-loop master topologies resulting after tensor reduction. 
Some of the internal lines may also be massless.}
\label{fig:Tall}
\end{center}
\end{figure}

After performing a tensor reduction with {\sc TwoCalc}, 
the resulting amplitudes are 
given in terms of a few scalar master integrals. 
Those involve factorizing one-loop 
tadpole $A(m^2)$, one-loop bubble $B(p^2,m_1^2,m_2^2)$ and two-loop 
bubble $T(p^2,\{m_i^2\})$ topologies, compare Fig.~\ref{fig:Tall}. 
While the resulting one-loop integrals can only become UV divergent, the 
two-loop integrals can be ultraviolet and infrared divergent. As the gluons are the 
only massless particles in this 
calculation and appear singly, no infrared singularities 
arise in any of the two-loop diagrams. 
This remains valid 
throughout the whole calculation, even though 
some sub-topologies contain more than one 
massless internal line. In these cases, the integrand
structure is such that an IR singularity does not occur. 
Though IR divergences are absent, additional sub-ultraviolet divergences can arise 
depending on the structure of the integrand. This is true for the diagrams
$T_{1234}$, $T_{11234}$ and $T_{234}$. The integral $T_{12345}$ is 
finite in all mass configurations appearing in this calculation. 
To summarize, the previously shown two-loop self-energies
may contain single and double UV poles. The newly
included momentum-dependence gives rise 
to additional divergent terms.
It is desired to find the correct counter-terms for their 
renormalization.

\medskip

It is beneficial to analyze 
which diagrams carry momentum
dependence, and which contribute with additional
divergent parts. 
First of all, the topologies 8, 9 and 10 of 
Figs.~\ref{subfig:setop6}-\subref{subfig:setop3} do not 
contribute with an additional momentum dependence at the 
two-loop level. 
Taking a diagrammatic point of view, these
diagrams appear as tadpoles with two external legs pinched
to one vertex, and 
all tadpoles are 
independent of the external momentum. 
All other diagrams carry momentum dependence 
in the loop and contain momentum-dependent
single UV divergent pole contributions.
The diagram in Fig.~\ref{subfig:setop2} is the only 
one to contain a single UV pole, all other
diagrams have the double amount. 
Only three diagrams contain momentum-dependent 
double UV divergent terms, namely 
\ref{subfig:setop7}, \subref{subfig:setop8} 
and \subref{subfig:setop10}. 

\medskip

With the additional contributions stemming 
from momentum dependence being exposed, the 
divergent terms can be 
eliminated, before the momentum-dependent contributions 
in the finite self-energies can be 
analyzed.  
%
%
%
%
%
\section{Renormalization}
\label{sec:renormalization}
As mentioned in the previous section, up to two UV poles 
appear in the unrenormalized self-energies. These 
can contain local divergences in a sub-loop. 
To renormalize these, one-loop self-energies with 
counter-term insertions have to cancel those 
divergences arising in a sub-loop of the 
two-loop diagrams. In addition, two-loop counter terms need 
to cancel the rest of the divergences. 
In this calculation, a mixed on-shell and 
$\overline{DR}$ scheme is used. Within the $\overline{DR}$ 
renormalization scheme, conventional dimensional reduction, 
see Sec.~\ref{sec:generalintegraldefinition}, is 
imposed. The bar in $\overline{DR}$ 
denotes that an additional factor of $\gamma_E$ and 
$\text{log}(4\pi)$ stemming from prefactors to 
loop integrals is absorbed into the 
renormalization scale $\mu_r$. 

\medskip

In a renormalized theory, all ultraviolet poles have to 
be canceled and physical quantities need to be finite. 
The proof of the existence of a set of counter-term vertices 
rendering any theory of superficial divergence finite is
summarized in the BPHZ theorem, see 
Refs.~\cite{Bogoliubov:1957gp,Hepp:1966eg,Zimmermann:1968mu}.
While the renormalization of a sub-loop of two-loop self-energies 
is constructed from vertex corrections, 
their two-loop counter terms are determined from the Lagrangian. 
The latter can be derived by looking at 
corrections to the fields and masses.
To this end, each scalar doublet gets a 
multiplicative field renormalization, 
\begin{align}
 \mathcal{H}_a \rightarrow  \mathcal{H}_a \sqrt{Z_{\mathcal{H}_a}} \text{ ,}
\end{align}
where
\begin{align}
 Z_{\mathcal{H}_a} = 1 + \delta Z_{\mathcal{H}_a}^{(1)} + \delta Z_{\mathcal{H}_a}^{(2)} + ... \text{ .}
 \label{eq:fieldrenconstexpansion}
\end{align}
Similarly, the masses get renormalization constants. 
This is 
implemented by adding counter-terms to the 
tree-level mass matrices 
\begin{align}
&M^2_{x} \rightarrow M^2_{x} + \delta M_{x}^{2(1)} + \delta M_{x}^{2(2)} + \dots \text{ ,} \label{eq:masshigherorders}
\end{align}
with $x=\phi^0,\chi^0,\phi^{\pm}$. 
Recapitulating Sec.~\ref{sec:higgstreelevel}, the 
Higgs-boson masses are determined from both, the 
bilinear part of the Lagrangian, 
compare Eq.~(\ref{eq:MSSMmassmatrices}), and
the linear part, see 
Eqs.~(\ref{eq:linVpart})-(\ref{eq:mtilde2}). They contain 
parametric dependences on $m_{A^0}$, $m_Z$, 
$m_W$, the charge $e$, the
electroweak mixing angle $\theta_W$, the angle relating
the vacuum expectation values $\beta$
and the tadpole coefficients $T_1$ and $T_2$, compare 
Eqs.~(\ref{eq:linVpart}), (\ref{eq:vevsdef}) and (\ref{eq:mphi0tree}).
Reformulating the kinetic part of the Lagrangian  
including field renormalization constants yields 
\begin{subequations}
\begin{align}
 \mathcal{L}_{\rm{H_{free}}} =& 
 \sum_{a=1}^2 \partial_\mu \left( \mathcal{H}_a \sqrt{Z_{\mathcal{H}_a}} \right)^\dagger 
 \partial^\mu \mathcal{H}_a \sqrt{Z_{\mathcal{H}_a}} \\
\non =& \frac12 (|\partial_\mu \boldsymbol{\phi}_1^0 \sqrt{Z_{H_1}}|^2 + 
 |\partial_\mu \boldsymbol{\chi}_1^0\sqrt{Z_{H_1}}|^2)+ |\partial_\mu \boldsymbol{\phi}_1^-\sqrt{Z_{H_1}}|^2 \\
 &+ \frac12 (|\partial_\mu \boldsymbol{\phi}_2^0\sqrt{Z_{H_2}}|^2 + |\partial_\mu \boldsymbol{\chi}_2^0\sqrt{Z_{H_2}}|^2) + 
 |\partial_\mu \boldsymbol{\phi}_2^+\sqrt{Z_{H_2}}|^2\text{ .} 
 \label{eq:freeRenLag}
\end{align}
\end{subequations}
Examining those parts of the potential part of the Lagrangian 
contributing to the Higgs-boson masses, 
the following substitutions appear in the formulation 
of UV finite quantities, 
\begin{align}
\hspace{-20pt} 
\mathcal{L}_{\rm{H_{mass}}} = - 
\frac12 \left( 
\begin{matrix} 
x_1 \sqrt{Z_{\mathcal{H}_1}} & x_2 \sqrt{Z_{\mathcal{H}_2}}
\end{matrix} 
\right) (M^2_{x} + \de M_{x}^{2(1)} + \de M_{x}^{2(2)} + ...)\left( 
\begin{matrix} 
x_1 \sqrt{Z_{\mathcal{H}_1}} \\ x_2 \sqrt{Z_{\mathcal{H}_2}}
\end{matrix} 
\right) \text{ ,}
 \label{eq:massperturb}
\end{align}
where the tree-level terms and the one, two and higher loop 
counter-terms can be identified. 
An expansion of the square root becomes necessary when the 
fields $x_1$ and $x_2$ mix, 
\begin{subequations}
\begin{align}
\non x_1 \sqrt{Z_{\mathcal{H}_1}} x_2 \sqrt{Z_{\mathcal{H}_2}} 
=&\,x_1 \;(1+ \frac12 (\delta Z_{\mathcal{H}_1}^{(1)} + 
	\delta Z_{\mathcal{H}_1}^{(2)}) - \frac18 (\delta Z_{\mathcal{H}_1}^{(1)})^2 + ...) \\
&\times \, x_2\; (1+ \frac12 (\delta Z_{\mathcal{H}_2}^{(1)} + 
	\delta Z_{\mathcal{H}_2}^{(2)}) - \frac18 (\delta Z_{\mathcal{H}_2}^{(1)})^2 + ...)\\
\non =& x_1 x_2\; (1+ \frac12 (\delta Z_{\mathcal{H}_1}^{(1)} + \delta Z_{\mathcal{H}_2}^{(1)}) 
+\frac12 ( \delta Z_{\mathcal{H}_1}^{(2)} + \delta Z_{\mathcal{H}_2}^{(2)}) \\
&\hspace{25pt} - \frac18 (\delta Z_{\mathcal{H}_1}^{(1)} - \delta Z_{\mathcal{H}_2}^{(1)})^2 + \dots ) 
%
\label{eq:fieldperturb}
\end{align}
\end{subequations}
where a record of the expansion is only kept until second order. 
As mentioned before, the terms linear in the field 
need to vanish at all orders. 
Looking at these after the application of 
the equations of motion Eq.~(\ref{eq:eoms}), 
the dependence on the fields drops out, meaning 
that there is no field renormalization 
involved here. This is in agreement 
with the fact that tadpoles 
do not carry momentum dependence. Therefore, 
the only counter-term correction needed in 
Eq.~(\ref{eq:linLaghigherorders}) arises from the 
corrections to the tadpoles themselves
\begin{align}
T_a \rightarrow T_a + \delta T_a^{(1)} + \delta T_a^{(2)} + \dots \text{ .} \label{eq:tadhigherorders}
\end{align}
The solution to the equations of motion 
including counter-terms therefore reads
\begin{align}
 T_a + T_a^{(1)} + T_a^{(2)} + \delta T_a^{(1)} + \delta T_a^{(2)} + \ldots \stackrel{\mathrm{!}}{=} 0 \text{ .}
 \label{eq:tadpolects}
\end{align}
\subsection{Two-loop counter terms for the renormalization 
at $\mathcal{O}(\alpha_s\alpha_t)$}
\label{subsec:twoloopcts}
%
%
%
Applying field and parameter renormalization yields 
the renormalized self-energies in 
terms of the scalar fields $\phi_1^0$ and 
$\phi_2^0$ 
\begin{subequations}
\begin{align}
\hat{\Sigma}_{\phi_1^0\phi_1^0}^{(2)}(p^2) &= \Sigma_{\phi_1^0\phi_1^0}^{(2)}(p^2) + \delta \Phi_{1}^{(2)}  
- \delta V_{\phi_1^0\phi_1^0}^{(2)} \\
 \hat{\Sigma}_{\phi_2^0\phi_2^0}^{(2)}(p^2) &= \Sigma_{\phi_2^0\phi_2^0}^{(2)}(p^2) + \delta \Phi_{2}^{(2)} 
- \delta V_{\phi_2^0\phi_2^0}^{(2)} \\
\hat{\Sigma}_{\phi_1^0\phi_2^0}^{(2)}(p^2) &= \Sigma_{\phi_1^0\phi_2^0}^{(2)}(p^2) 
- \delta V_{\phi_1^0\phi_2^0}^{(2)} \text{ .}
\end{align}
\label{eqs:renormalizedselfses}
\end{subequations}
The counter-terms $\delta \Phi_{a}$ arise from the kinetic 
Lagrangian, see Eq.~(\ref{eq:freeRenLag}), and  
$\delta V_{x_ax_b}$ from the potential part of the Lagrangian, 
see Eq.~(\ref{eq:massperturb}). 

\medskip

The counter-terms originating from
perturbative expansions to the free 
fields $\phi_1^0$ and $\phi_2^0$ read
\begin{align}
 \delta \Phi_{a}^{(2)} =p^2 \left( \delta Z_{\mathcal{H}_a}^{(2)} - \frac14 (\delta 
Z_{\mathcal{H}_a}^{(1)})^2 \right) \text{ ,}
\end{align}
as deducible from Eqs.~(\ref{eq:fieldrenconstexpansion}) and~(\ref{eq:freeRenLag}).
In the calculation of the self-energies of the two orders $\alpha_s$ and $\alpha_t$, 
the squared one-loop field renormalization constants do not contribute. They 
would be needed, e.g., in an order ${\cal O} (\alpha_t^2)$ calculation.
This reduces the field renormalization
counter-terms to
\begin{align}
 \delta \Phi_{1}^{(2)} &= p^2 \delta Z_{\mathcal{H}_1^0}^{(2)} \\
 \delta \Phi_{2}^{(2)} &= p^2 \delta Z_{\mathcal{H}_2^0}^{(2)} \text{ .} 
\end{align}

\medskip

Based on the derivation from the beginning of Sec.~\ref{sec:renormalization}, 
the counter-terms arising from the renormalization of the potential 
part of the Lagrangian are given by 
\begin{align}
 \delta V_{\phi_1^0\phi_1^0}^{(2)}  &= 
 M_{\phi_1^0\phi_1^0}^2  \delta Z_{\mathcal{H}_1^0}^{(2)} + 
 \delta  M_{\phi_1^0\phi_1^0}^{2(2)} \\
 \delta V_{\phi_2^0\phi_2^0}^{(2)}  &= 
 M_{\phi_2^0\phi_2^0}^2  \delta Z_{\mathcal{H}_2^0}^{(2)} + 
 \delta  M_{\phi_2^0\phi_2^0}^{2(2)}  \\
 \delta V_{\phi_1^0\phi_2^0}^{(2)}  &= 
 \frac12 M_{\phi_1^0\phi_2^0}^2 ( \delta Z_{\mathcal{H}_1^0}^{(2)} + 
 \delta Z_{\mathcal{H}_2^0}^{(2)} ) + 
 \delta  M_{\phi_1^0\phi_2^0}^{2(2)} \text{ ,}
 \label{eq:2loopcounterterms}
\end{align}
where $\delta  M_{\phi^0}^{2(2)}$ is computed from the Taylor series 
expansion of the mass matrix $M_{\phi^0}^{2}$ in its parameters 
up to second order. 
Recalling 
the mass matrix $M_{\phi^0}^{2}$ of Eq.~(\ref{eq:mphi0tree}), 
the vanishing of the tree-level tadpole parameters due to the 
minimization of the MSSM Higgs-boson potential, 
Eq.~(\ref{eq:minlinpartcond}), 
was already incorporated. 
Tadpole contributions must 
be taken into account at higher orders, 
as their coefficients do not necessarily 
vanish, see Eq.~(\ref{eq:linLaghigherorders}) 
and Eq.~(\ref{eq:tadpolects}). 
Yet, they only enter in the 
renormalization of the momentum independent 
part. 
The mass matrix is then parametrized by 
$m_{A^0}$, $m_Z$, $m_W$, the charge $e$, the
electroweak mixing angle $\theta_W$, the angle relating
the vacuum expectation values $\beta$
and the tadpole coefficients $T_1$ and $T_2$. 
The computation of the Taylor series expansion of the 
mass matrix up to second order 
corresponds to a perturbative expansion 
in these parameters, 
\begin{align}
\label{rMSSM:PhysParamRenorm}
  m_{A^0}^2 &\to m_{A^0}^2 + \delta m_{A^0}^{2(1)} + \delta m_{A^0}^{2(2)},  
& T_1 &\to T_1 + \delta T_1^{(1)} +  \delta T_1^{(2)}, \notag\\ 
  m_Z^2 &\to m_Z^2 + \delta m_Z^{2(1)} + \delta m_Z^{2(2)},  
& T_2 &\to T_2 + \delta T_2^{(1)} +  \delta T_2^{(2)}, \notag \\ 
    \tb & \to \tb \KL 1 + \de\tb^{(1)} + \de\tb^{(2)} \KR\,. & 
\end{align}
With the computation performed in the ``gaugeless'' limit, 
all counter-terms including electroweak gauge particles 
and their contributions 
are discarded.
Furthermore, squares of one-loop renormalization constants 
appearing in $\delta  M_{\phi^0}^{2(2)}$ do not contribute, 
as they are proportional to electroweak coupling factors but they are not 
of the order ${\cal O}(\alpha_s \alpha_t)$. For a full 
explicit presentation of $\delta  M_{\phi^0}^{2(2)}$, see 
e.g. Ref.~\cite{Rzehak:2005zz}. 
The final counter-terms for the potential in the quadratic $\phi_a^0$ terms are 
\begin{subequations}
\begin{align}
\nonumber \delta V_{\phi_1^0\phi_1^0}^{(2)} =& \, \delta m_{A^0}^{2(2)} \,\text{sin}^2 \beta \\
\nonumber & - \delta T_1^{(2)} \, \frac{e}{2\, M_W \sin\theta_W} \;\text{cos}\, \beta \, (1+\text{sin}^2 \beta) \\
\nonumber & + \delta T_2^{(2)} \, \frac{e}{2 \, M_W \sin\theta_W}  \;\text{sin} \, \beta \,\text{cos}^2 \beta \\
\nonumber & + 2 \, \delta \text{tan}\beta^{(2)} \, \text{cos}^2 \beta \,\text{sin}^2 \beta \,m_{A^0}^2 \\
& + \delta Z_{\mathcal{H}_1^0}^{(2)} \, m_{A^0}^{2} \,\text{sin}^2 \beta \text{ ,}
\label{eq:v112Lcounter}\\
\nonumber \delta V_{\phi_2^0\phi_2^0}^{(2)} =& \, \delta m_{A^0}^{2(2)} \,\text{cos}^2 \beta \\
\nonumber & + \delta T_1^{(2)}\, \frac{e}{2 \,M_W \sin\theta_W} \; \text{cos} \,\beta \,\text{sin}^2 \beta \\
\nonumber & - \delta T_2^{(2)} \, \frac{e}{2 \,M_W \sin\theta_W} \;\text{sin} \,\beta \,(1+\text{cos}^2 \beta)\\
\nonumber & - 2\, \delta\text{tan}\beta^{(2)} \,\text{cos}^2 \beta \,\text{sin}^2 \beta \,m_{A^0}^2 \\
& + \delta Z_{\mathcal{H}_2^0}^{(2)}\,  m_{A^0}^{2} \,\text{cos}^2 \beta \text{ ,}
\label{eq:v222Lcounter}\\
\nonumber \delta V_{\phi_1^0\phi_2^0}^{(2)} =& - \delta m_{A^0}^{2(2)}\,\text{sin} \,\beta\,\text{cos}\, \beta \\
\nonumber & - \delta T_1^{(2)} \,\frac{e}{2 \,M_W \sin\theta_W} \;\text{sin}^3 \beta\\
\nonumber & - \delta T_2^{(2)} \,\frac{e}{2 \,M_W \sin\theta_W} \;\text{cos}^3 \beta\\
\nonumber & - \delta\text{tan}\beta^{(2)} \text{cos}\, \beta \, \sin\, \beta \,\cos (2 \beta)\, m_{A^0}^2 \\
& - \frac12 (\delta Z_{\mathcal{H}_1^0}^{(2)}+ \delta Z_{\mathcal{H}_2^0}^{(2)}) \,\text{sin} \beta\,\text{cos} \beta\, m_{A^0}^{2}  \text{ .}
\label{eq:v122Lcounter}
\end{align}
\label{eq:2lctspotentialpart}
\end{subequations}

\begin{figure}[htb]
\begin{center}
\hspace{60pt}
\subfigure[]{\raisebox{0pt}{\hspace{-35pt}\includegraphics[width=0.23\textwidth]{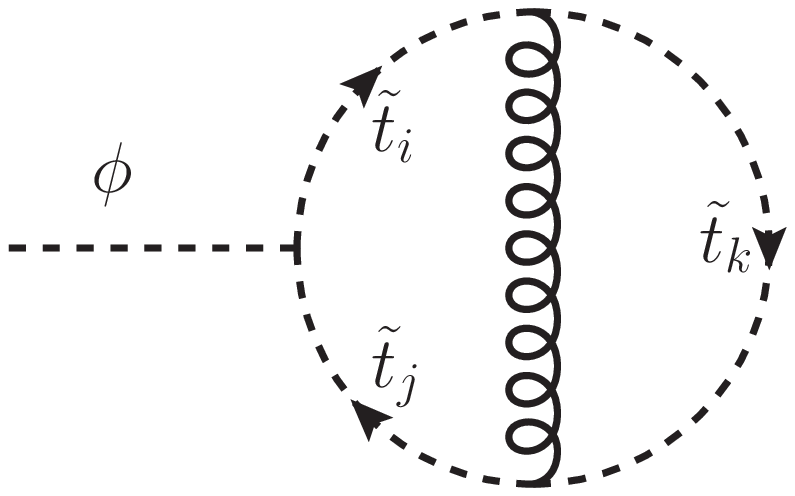}} }\hspace{45pt}
\subfigure[]{\raisebox{0pt}{\hspace{-35pt}\includegraphics[width=0.23\textwidth]{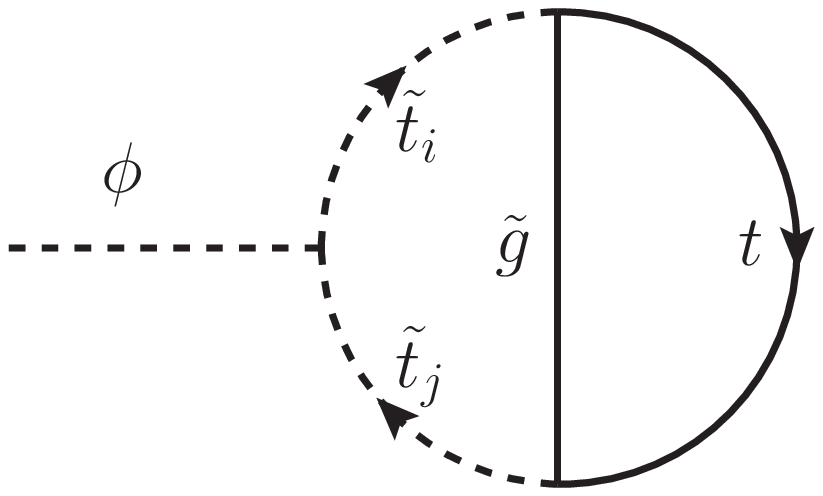}} }\hspace{42pt}
\subfigure[]{\raisebox{5pt}{\hspace{-32pt}\includegraphics[width=0.3\textwidth]{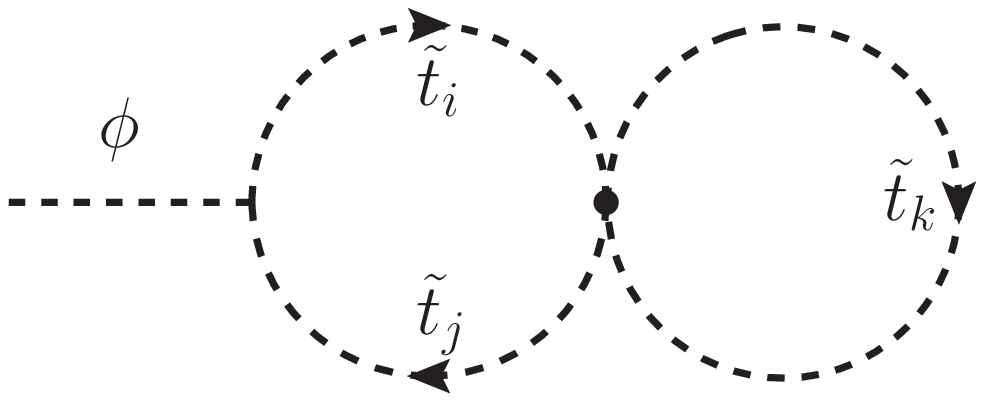}} } 
\subfigure[]{\raisebox{0pt}{\hspace{-35pt}\includegraphics[width=0.23\textwidth]{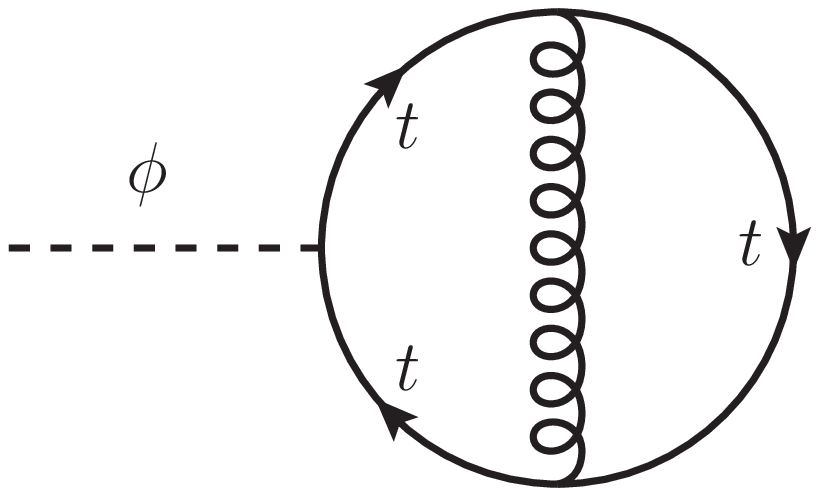}} }\hspace{65pt}
\subfigure[]{\raisebox{0pt}{\hspace{-35pt}\includegraphics[width=0.23\textwidth]{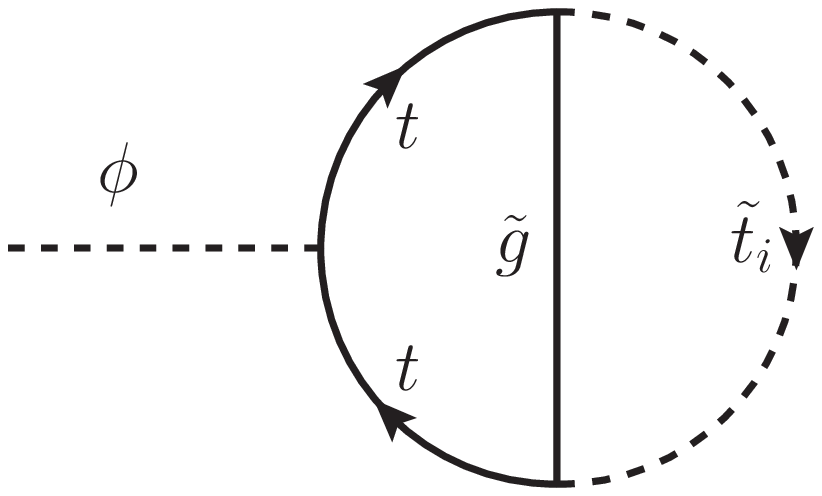}} }
\end{center}
\caption{Two-loop tadpole diagrams contributing in the two-loop counter terms to 
the $\mathcal{O}(\alpha_s \alpha_t)$ self-energies, $\phi = \phi_1, \phi_2$ and $i,j,k=1,2$.}
\label{fig:taddiags} 
\end{figure}

With the derived counter-terms at hand, the renormalization 
constants entering them need to be defined. As 
mentioned previously, these are renormalization scheme dependent. To 
be consistent with previous calculations incorporated in the public 
program {\sc FeynHiggs}, all masses and the tadpole parameters are 
renormalized using the on-shell scheme, and all field contributions 
using the $\overline{DR}$ scheme~\cite{Frank:2002qf,Heinemeyer:2004xw,
Hollik:2003jj,Heinemeyer:2010mm,Fritzsche:2011nr}. 
In the latter, a dependence on the renormalization 
scale $\mu_r$ is exhibited. 
The two-loop renormalization constants entering the mass 
renormalization counter-terms are known for vanishing external 
momentum, see Refs.~\cite{Heinemeyer:1998jw,Heinemeyer:1998kz,
Heinemeyer:1998np,Hempfling:1993qq}. 
Taking the 
momentum dependence into account, 
only the tadpole contributions remain unchanged. 
The tadpole renormalization constants can be deduced from 
Eq.~(\ref{eq:tadpolects}). To meet the condition that 
the sum of all tadpole contributions
must vanish, it must follow 
\begin{align}
  \delta T_a^{(2)} = - T_a^{(2)} \text{ ,}
  \label{eq:tadrenconsts}
\end{align}
adopting an on-shell renormalization. To obtain the 
renormalization constants $\delta T_a^{(2)} $, thus 
two-loop tadpole diagrams have to be computed, matching the
right order in the couplings, compare the diagrams in Fig.~\ref{fig:taddiags}. 

For the renormalization constants $\dZ{\cHe}$, 
$\dZ{\cHz}$ and $\delta\tan\beta$ 
several choices are possible, see the discussion 
in~\cite{Frank:2002qf,Freitas:2002um}. 
As shown there, the most convenient
choice is a \drbar\ renormalization of $\delta\tan\beta$, $\dZ{\cHe}$
and $\dZ{\cHz}$. 
The field renormalization constants can be extracted 
by taking the derivative of the renormalized self-energies, Eqs.~(\ref{eqs:renormalizedselfses}), 
with respect to the external momentum squared, yielding in 
$\overline{DR}$ renormalization 
\begin{align}
 \delta Z_{\mathcal{H}_a}^{(2)} =& \; 
 \text{Re}\left[ 
 \frac{\partial}{\partial p^2} \hat{\Sigma}_{\phi_a^0\phi_a^0}(p^2) - 
 \frac{\partial}{\partial p^2} \Sigma_{\phi_a^0\phi_a^0}(p^2)
 \right]_{p^2=0} \\ 
=& -\text{Re}\left[ \frac{\partial}{\partial p^2} \Sigma_{\phi_a^0\phi_a^0} \right]_{p^2=0}^{\text{div}} \text{ .}
 \label{eq:fieldrenconsts}
\end{align}

The formulation of the field renormalization constants 
in the $\overline{DR}$ scheme has the 
advantage that all ultraviolet divergences of the momentum-dependent integrals contained 
in the $\Sigma_{\phi_a^0\phi_a^0}$ self-energies 
are known analytically. The derivative with respect to $p^2$ can 
therefore be performed in a fully analytical way. 
Adopting an on-shell field renormalization, the derivatives of some 
integrals in the finite part for which closed analytical expressions are 
not available, 
are expected to be less straightforward. 
It could be argued that the calculation can be simplified even further 
adopting a full $\overline{DR}$ scheme, compare 
Refs.~\cite{Martin:2003it,Martin:2004kr,Martin:2005qm}. 
Yet, the 
dependence on the renormalization scale is decreased adopting a 
hybrid renormalization scheme, 
physical effects of higher order in the 
$\overline{DR}$ scheme already appear at the current order in the on-shell scheme, 
and predictions are given with pole masses as input parameters. 
%

Due to the fact that $\tan\beta$ is defined in terms 
of the two vacuum expectation values $v_1$ and $v_2$, Eq.~(\ref{eq:tanb}), 
minimizing the MSSM Higgs-boson potential according to Eq.~(\ref{eq:minlinpartcond}), the 
field renormalization also enters here.
The renormalization of $\tan\beta$ follows from 
\begin{align}
v_a \to \sqrt{Z_a} (v_a + \delta v_1^{(1)} +  \delta v_1^{(2)}) \text{ ,}
\end{align}
yielding in $\overline{DR}$
\begin{align}
\delta\text{tan}\beta^{(2)} = 
\frac12 \, \text{tan}\beta \, (\delta Z_{\mathcal{H}_2}^{(2)} - \delta Z_{\mathcal{H}_1}^{(2)}) \text{ .}
\label{eq:tanbrenconst}
\end{align}
The term in \refeq{eq:tanbrenconst} is in general 
not the proper expression
beyond one-loop order even in the  \drbar\ scheme.
In the approximation with only the top
Yukawa coupling at the two-loop level, 
it is the correct \drbar\ form, see 
Refs.~\cite{Sperling:2013eva,Sperling:2013xqa}.

\medskip

Finally, the two-loop renormalization constant of the $A^0$-boson mass 
is determined in the on-shell scheme as 
\begin{align}
 \delta m_{A^0}^{2(2)}= \text{Re}\left[ \Sigma_{A^0A^0}^{(2)} (p^2\!=\!m_{A^0}^2) \right] \text{ ,}
  \label{eq:marenconst}
\end{align}
in terms of the unrenormalized two-loop $A^0$-boson self-energy $\Sigma_{A^0A^0}^{(2)} $ 
evaluated at the pole mass; 
see Fig.~\ref{fig:selfdiags} for the generic $A^0$-boson self-energy diagrams at 
the two-loop level. 
The appearance of a non-zero momentum in the self-energy goes beyond
the \order{\alt\als} corrections evaluated in 
Refs.~\cite{Heinemeyer:1998jw,Heinemeyer:1998kz,
Heinemeyer:1998np,Hempfling:1993qq}.
Fixing the external momentum of 
the self-energy to the $A^0$-boson mass is necessary 
to cancel additional divergent terms
arising from the two-loop counter-terms involving $\tan\beta$, compare 
Eqs.~(\ref{eq:v112Lcounter})-(\ref{eq:v122Lcounter}). 
The latter contain a dependence on $m_{A^0}^2$, giving rise to 
divergences which do not cancel any divergence arising in 
the neutral ${\cal CP}$-even Higgs-bosons. The latter are independent of 
the $A^0$-boson mass. 
These additional divergences must therefore
be cancelled by $\delta m_{A^0}^{2(2)}$.
\subsection{Renormalization at the sub-loop level}
\label{subsec:subrenorm}
%
%
%
%
With the two-loop counter terms at hand, there is only
one missing piece left in the renormalization 
procedure. 
Filling this gap, those one-loop 
diagrams with counter-term insertions are 
computed, which 
renormalize a sub-loop of the two-loop 
self-energies and of the two-loop tadpole 
diagrams. The latter are needed for the two-loop counter-terms, 
compare Eq.~(\ref{eq:tadrenconsts}). 
The one-loop amplitudes with counter-term insertions 
are generated with the program {\sc FeynArts}~\cite{Kublbeck:1990xc,
Hahn:2000kx,Hahn:2001rv}, using the model file including 
counter-terms from Ref.~\cite{Fritzsche:2013fta}.

\begin{figure}[htb!]
\begin{center}
\hspace{-130pt}
\subfigure[]{\raisebox{0pt}{\includegraphics[width=0.25\textwidth]{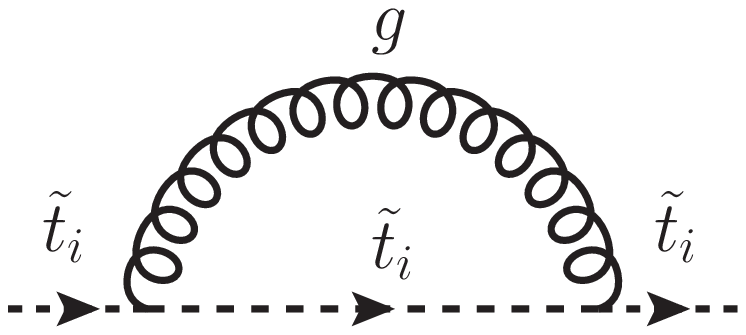}} 
\label{subfig:cttop1}}\hspace{15pt}
\subfigure[]{\raisebox{0pt}{\includegraphics[width=0.25\textwidth]{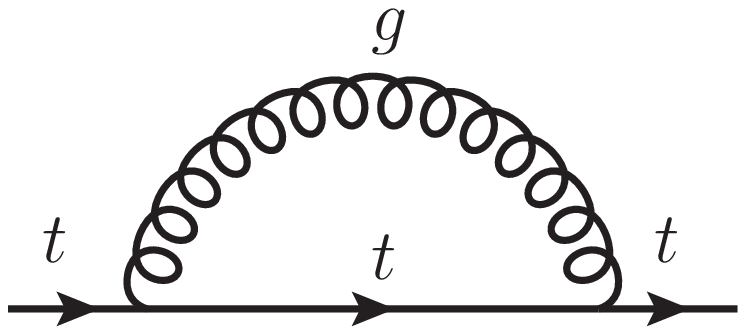}} 
\label{subfig:cttop5}}\\
\subfigure[]{\raisebox{0pt}{\includegraphics[width=0.25\textwidth]{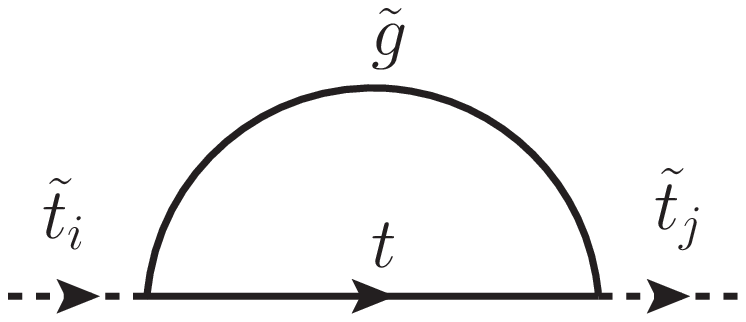}} 
\label{subfig:cttop2}}\hspace{15pt}
\subfigure[]{\raisebox{0pt}{\includegraphics[width=0.25\textwidth]{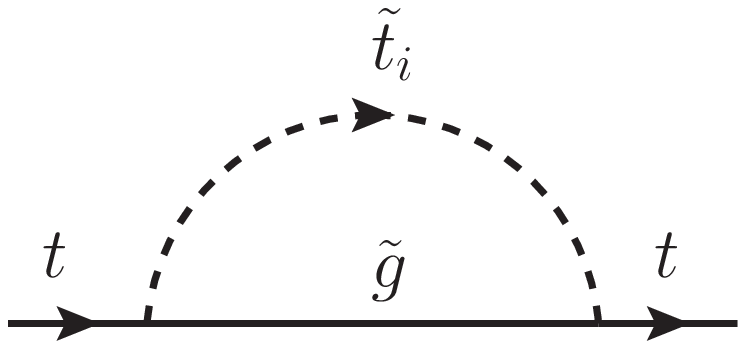}} 
\label{subfig:cttop4}}\hspace{15pt}
\subfigure[]{\raisebox{1pt}{\includegraphics[width=0.25\textwidth]{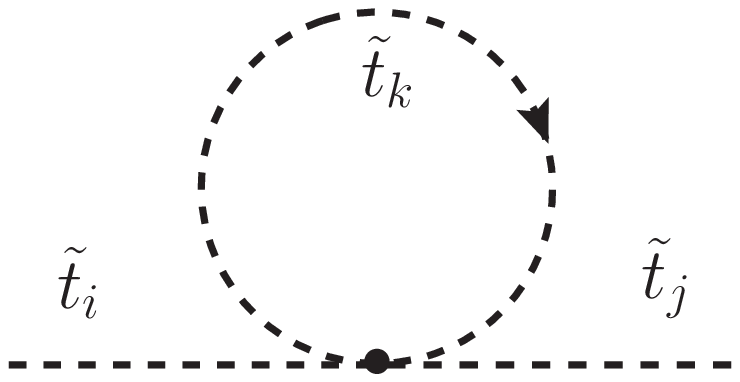}} 
\label{subfig:cttop5}}
\end{center}
\caption{One-loop diagrams for the quark and squark counter-term insertions 
with $i,\,j,\,k=1,2$.}
\label{fig:1Lsquarkdiags} 
\end{figure}

\begin{figure}[htb!]
\begin{center}
\subfigure[]{\raisebox{0pt}{\includegraphics[width=0.3\textwidth]{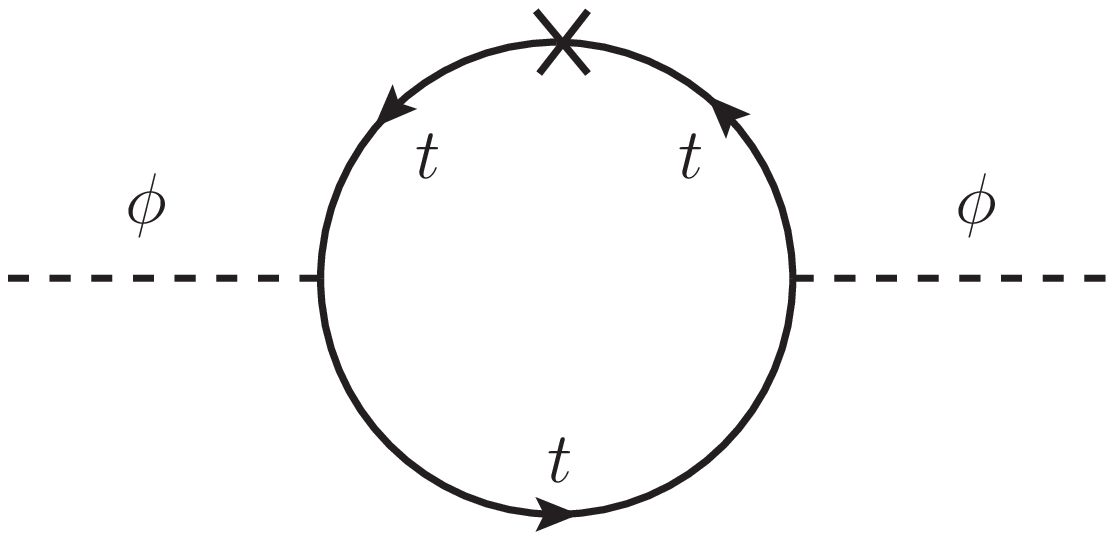}} }
\subfigure[]{\raisebox{0pt}{\includegraphics[width=0.3\textwidth]{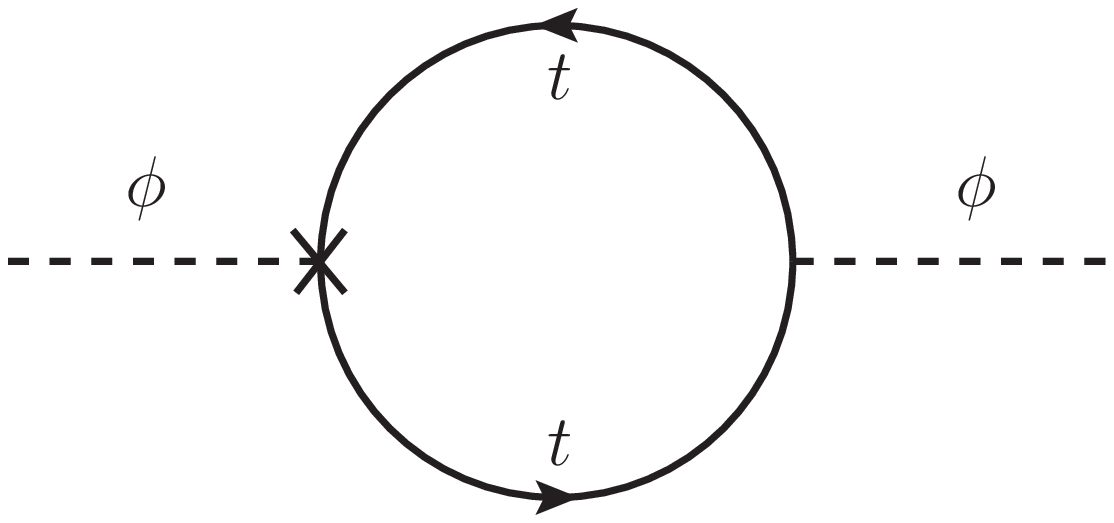}} }
\subfigure[]{\raisebox{-2pt}{\includegraphics[width=0.3\textwidth]{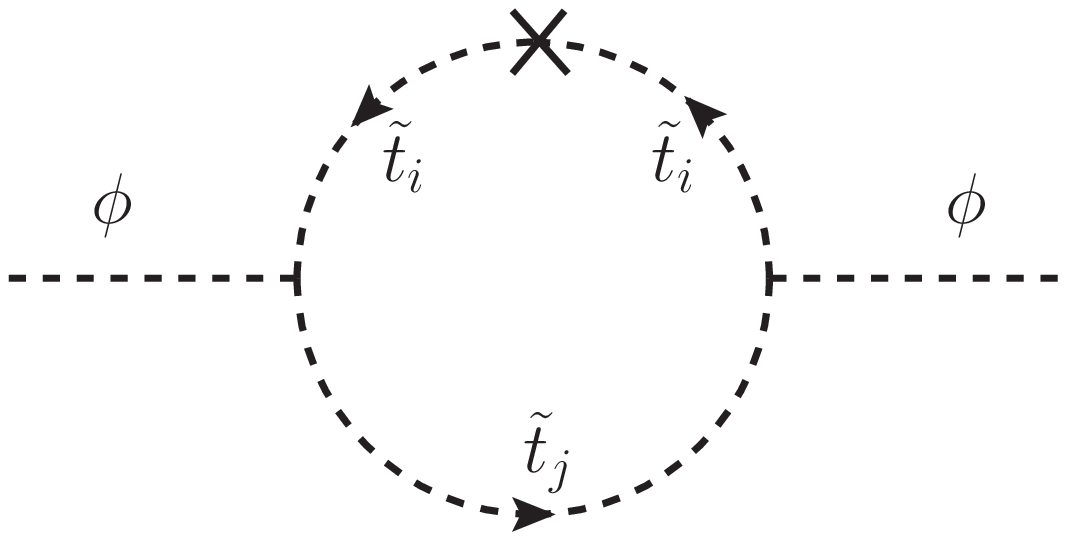}} }\\\hspace{-20pt}
\subfigure[]{\raisebox{0pt}{\includegraphics[width=0.3\textwidth]{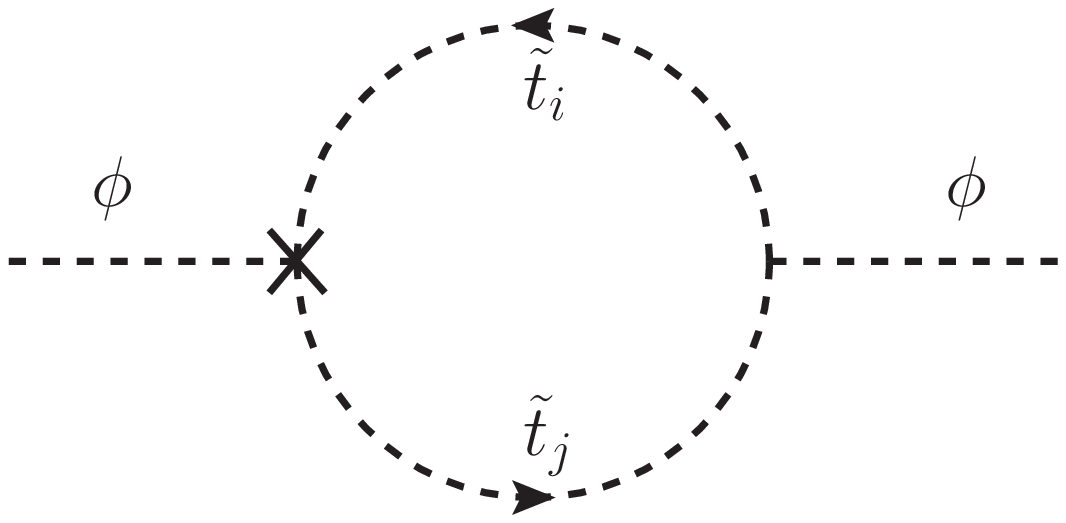}} \label{subfig:oneloopctfigd}}\hspace{16pt}
\subfigure[]{\raisebox{0pt}{\includegraphics[width=0.23\textwidth]{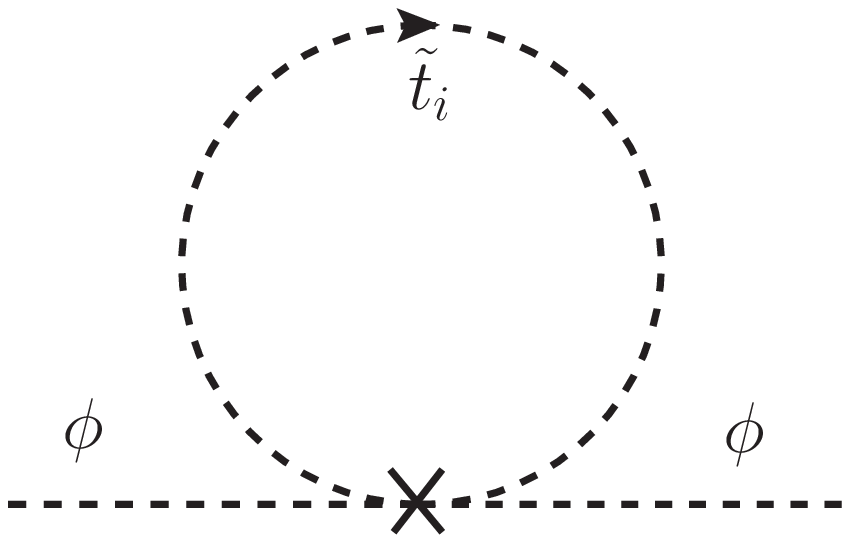}} }\hspace{30pt}
\subfigure[]{\raisebox{0pt}{\includegraphics[width=0.23\textwidth]{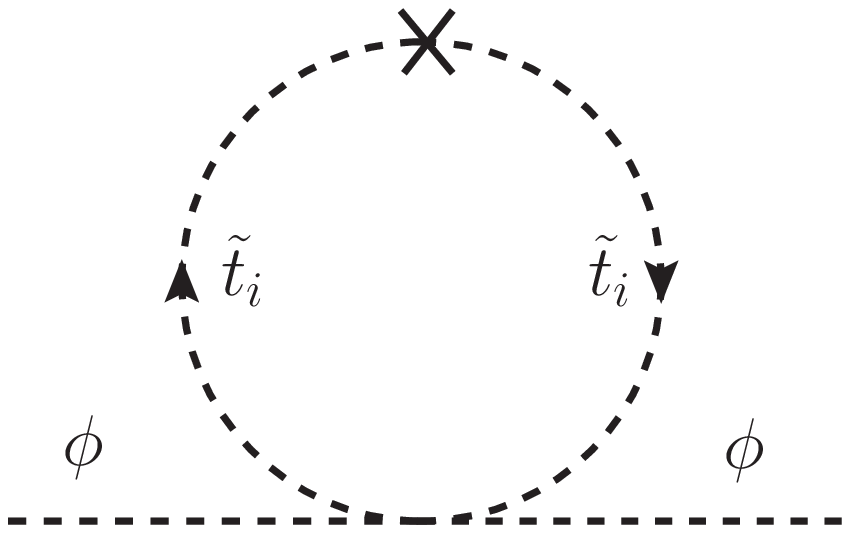}} }
\end{center}
\caption{One-loop counter-term contributions to the sub-loop renormalization of the 
$\mathcal{O}(\alpha_s \alpha_t)$ self-energies, with $\phi = \phi_1, \phi_2, A$ and 
$i,j=1,2$.}
\label{fig:1Lctdiags} 
\end{figure}

\begin{figure}[htb!]
\begin{center}
\subfigure[]{\raisebox{0pt}{\hspace{-35pt}\includegraphics[width=0.23\textwidth]{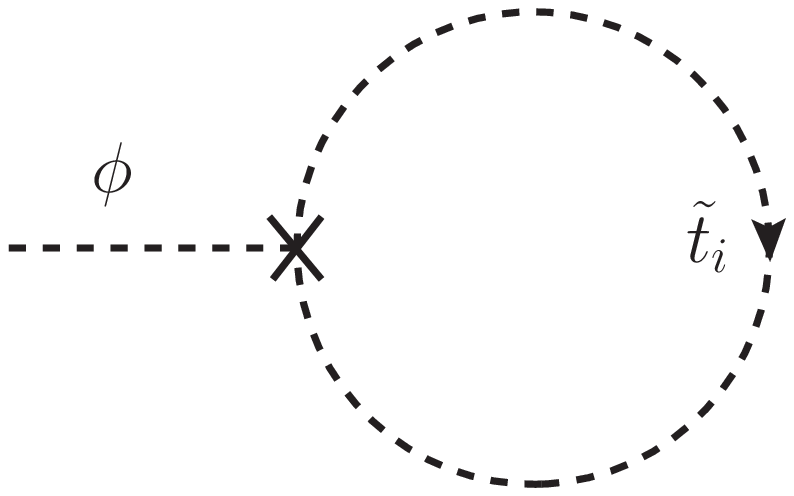}} }\hspace{60pt}
\subfigure[]{\raisebox{0pt}{\hspace{-35pt}\includegraphics[width=0.23\textwidth]{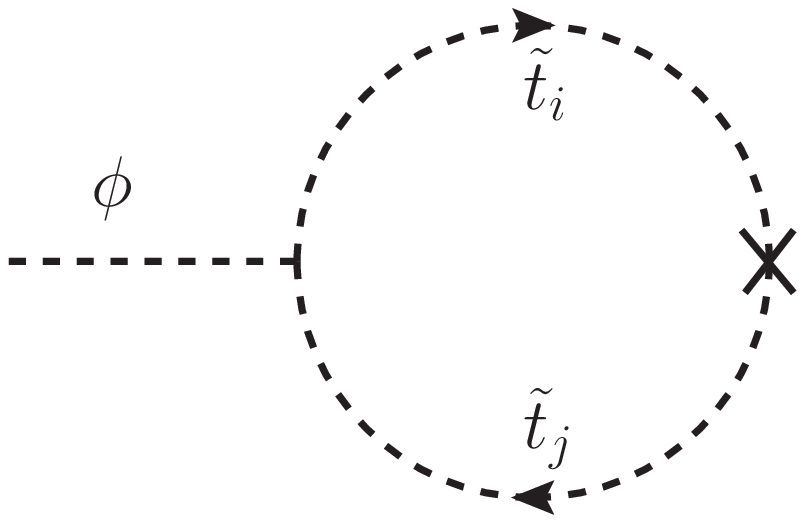}} }\\
\subfigure[]{\raisebox{0pt}{\hspace{-35pt}\includegraphics[width=0.23\textwidth]{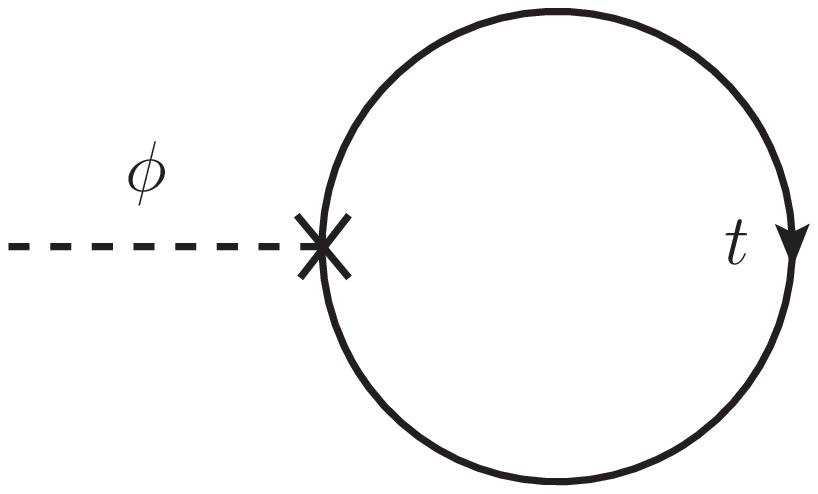}} }\hspace{60pt}
\subfigure[]{\raisebox{0pt}{\hspace{-35pt}\includegraphics[width=0.23\textwidth]{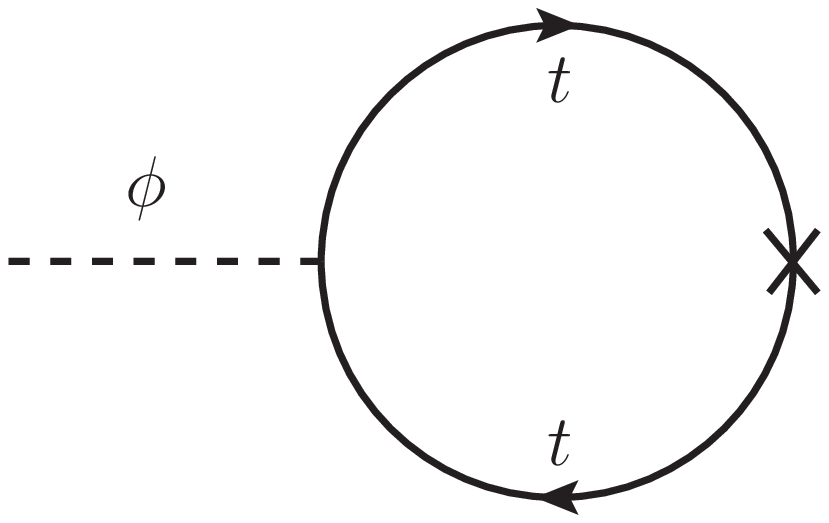}} }
\end{center}
\caption{One-loop counter-term diagrams with insertions, needed in the sub-loop 
renormalization of the two-loop tadpoles, $\phi = \phi_1, \phi_2$ and $i,j=1, 2$.}
\label{fig:tadctdiags} 
\end{figure}

One-loop corrections $x^{(1)}$
to the field $x$ with one-loop counter-term insertions $\delta P^{(1)}$ of a 
given parameter $P$ were already shown in
Eqs.~(\ref{eq:freeRenLag}) and (\ref{eq:massperturb}).
The contributions there, however, are 
of the order $\mathcal{O}(\alpha_t^2)$ and lack
a dependence of $\alpha_s$. 
To find the proper sub-loop renormalization terms, 
the particle interactions proportional to $\alpha_s$, listed in 
Sec.~\ref{sec:scalarquarktreelevel}, need
to be taken into account. They amount to 
five different counter-term 
structures, see 
Fig.~\ref{fig:1Lsquarkdiags}, required for the insertions in 
the one-loop amplitudes of order $\mathcal{O}(\alpha_t)$ in 
Fig.~\ref{fig:1Lctdiags}.%
The latter involve the one-loop neutral
$\cp$-even and $\cp$-odd Higgs-boson amplitudes. 
The $\cp$-odd amplitude is needed for the 
sub-loop renormalization
of the $A^0$-boson self-energy entering the
two-loop counter-terms. 
Furthermore, the tadpole diagrams entering 
the two-loop counter-terms need a proper
sub-loop renormalization. Their 
one-loop amplitudes must be computed as 
well, see the
diagrams in Fig.~\ref{fig:tadctdiags}.
The one-loop diagrams are either of order
$\mathcal{O}(\alpha_t)$ or 
$\mathcal{O}(\sqrt{\alpha_t})$, 
see Figs.~\ref{fig:1Lctdiags} and~\ref{fig:tadctdiags}, respectively.
An additional factor $\sqrt{\alpha_t}$ comes
in with the prefactor to the tadpole
renormalization constants, compare 
Eqs.~(\ref{eq:v112Lcounter})-(\ref{eq:v122Lcounter}).
The program {\sc FormCalc}~\cite{Hahn:1998yk} is used to perform the 
reduction of the one-loop diagrams 
to one-, two-, and three-point master integrals
$A$, $B$ and $C$, respectively. 
In addition, tensor coefficients
to the two- and three-point integrals can arise. 

\medskip

The counter-term insertions 
contain quark and squark field-renormalization components
as well as terms resulting from
the renormalization of the 
tree-level top-quark and squark mass matrices.
%
The renormalization of the top and stop sector at the
one-loop level as discussed below, has been analyzed in detail in 
Refs.~\cite{Heinemeyer:1998np,Hollik:2003jj,
Heinemeyer:2004xw,Rzehak:2005zz,Frank:2006yh,
Heinemeyer:2010mm,Fritzsche:2011nr}. 
One-loop counter-terms to the stop mass 
matrix $M_{\tilde{t}_{12}}^2$
and the fields $\tilde{t}_L$ and $\tilde{t}_R$ read
\begin{subequations}
\begin{align}
 m_{t_i}^2 &\rightarrow m_{\tilde{t}_i}^2 + \delta m_{\tilde{t}_i}^{2(1)} \text{ ,}\\
 \theta_{\tilde{t}} &\rightarrow \theta_{\tilde{t}} + \delta\theta_{\tilde{t}}^{(1)} \text{ ,}\\
 \tilde{t}_{L,R} &\rightarrow \tilde{t}_{L,R} \,\sqrt{Z_{\tilde{t}_{L,R}}} \approx 
 \tilde{t}_{L,R} \,(1+\frac12 \delta Z_{\tilde{t}_{L,R}}^{(1)}) \label{eq:stopfieldren}\text{ .}
\end{align}
\end{subequations}
The stop fields are transformed into the physical 
components $\tilde{t}_1$ and $\tilde{t}_2$ via
\begin{align}
 \begin{pmatrix}
{\tilde{t}}_{1}\\[.2em]{\tilde{t}}_{2} 
\end{pmatrix}
= U_{\tilde{t}}
 \begin{pmatrix}
{\tilde{t}}_{L}\\[.2em]{\tilde{t}}_{R}
\end{pmatrix}\text{ ,}
\end{align}
where $U_{\tilde{t}}$ is the stop mixing matrix defined in 
Eq.~(\ref{eq:utransformtheta}). 
The free-field kinetic terms of the stops are bilinear 
in the fields $\tilde{t}_i$ and
when computing the field renormalization counter-terms 
$\propto | \partial_\mu \tilde{t}_i |^2$; these
receive contributions from both the left-hand and right-hand 
components. To disentangle them, it is beneficial to introduce 
the field renormalization in the mass eigenstate basis 
\begin{align}
\begin{pmatrix}
{\tilde{t}}_{1}\\[.2em]{\tilde{t}}_{2} 
\end{pmatrix}
\rightarrow \mathcal{Z}_{\tilde{t}_{12}}
\begin{pmatrix}
{\tilde{t}}_{1}\\[.2em]{\tilde{t}}_{2}
\end{pmatrix} =
\begin{pmatrix}
1+ \frac12 \delta Z_{\tilde{t}_1}^{(1)} & \frac12 \delta Z_{\tilde{t}_{12}}^{(1)}  \\
\frac12 \delta Z_{\tilde{t}_{21}}^{(1)}  & 1+ \frac12 \delta Z_{\tilde{t}_2}^{(1)} \\
\end{pmatrix}
 \begin{pmatrix}
{\tilde{t}}_{1}\\[.2em]{\tilde{t}}_{2}
\end{pmatrix}\text{ .}
\label{eq:1loopfieldren}
\end{align}
Now, a transformation of coordinates of 
the left- and right-handed field renormalization matrix
can be performed 
\begin{subequations}
\begin{align}
U_{\tilde{t}} &
\begin{pmatrix}
1+ \frac12 \delta Z_{\tilde{t}_L}^{(1)} & 0 \\
0 & 1+ \frac12 \delta Z_{\tilde{t}_R}^{(1)} \\
\end{pmatrix}
 U^T_{\tilde{t}} \\
\hspace{-20pt}  = &
\begin{pmatrix}
1+ \frac12 (\cos^2\theta_{\tilde{t}} \delta Z_{\tilde{t}_L}^{(1)} + 
\sin^2\theta_{\tilde{t}}\delta Z_{\tilde{t}_R}^{(1)}) 
& \frac12 \sin\theta_{\tilde{t}} \cos\theta_{\tilde{t}} 
(\delta Z_{\tilde{t}_R}^{(1)} - \delta Z_{\tilde{t}_L}^{(1)}) \\
\frac12 \sin\theta_{\tilde{t}} \cos\theta_{\tilde{t}} 
(\delta Z_{\tilde{t}_R}^{(1)} - \delta Z_{\tilde{t}_L}^{(1)}) 
& 1+ \frac12 (\sin^2\theta_{\tilde{t}} 
\delta Z_{\tilde{t}_L}^{(1)} + \cos^2\theta_{\tilde{t}}\delta Z_{\tilde{t}_R}^{(1)}) 
\end{pmatrix}\text{ ,}
\end{align}
\end{subequations}
where a comparison of the coefficients with Eq.~(\ref{eq:1loopfieldren})
yields
\begin{subequations}
\begin{align}
\delta Z_{\tilde{t}_1}^{(1)} &= 
\cos^2\theta_{\tilde{t}} \delta Z_{\tilde{t}_L}^{(1)} + \sin^2\theta_{\tilde{t}}\delta Z_{\tilde{t}_R}^{(1)} \text{ ,}\\
\delta Z_{\tilde{t}_2}^{(1)} &=
\sin^2\theta_{\tilde{t}} \delta Z_{\tilde{t}_L}^{(1)} + \cos^2\theta_{\tilde{t}}\delta Z_{\tilde{t}_R}^{(1)} \text{ ,}\\
\delta Z_{\tilde{t}_{12}}^{(1)} &=
\delta Z_{\tilde{t}_{21}}^{(1)} \\&=
\sin\theta_{\tilde{t}} \cos\theta_{\tilde{t}} \,(\delta Z_{\tilde{t}_R}^{(1)} - \delta Z_{\tilde{t}_L}^{(1)})\\&= 
\frac{\sin\theta_{\tilde{t}} \, \cos\theta_{\tilde{t}}}{\cos^2\theta_{\tilde{t}} - \sin^2\theta_{\tilde{t}}} 
\,(\de Z_{\tilde{t}_2}^{(1)} - \de Z_{\tilde{t}_1}^{(1)})
\label{eq:deltaZ12field}
\text{ .}
\end{align}
\end{subequations}
Now, the one-loop free field renormalization terms in
the physical basis read
\begin{subequations}
\begin{align}
 \delta \boldsymbol{\tilde{t}}_1^{(1)} &= p^2 \,\delta Z_{\tilde{t}_{1}}^{(1)} \text{ ,}\\
 \delta \boldsymbol{\tilde{t}}_2^{(1)} &= p^2 \,\delta Z_{\tilde{t}_{2}}^{(1)} \text{ ,}\\
 \delta \boldsymbol{\tilde{t}}_{12}^{(1)} &= p^2 \,\delta Z_{\tilde{t}_{12}}^{(1)} \text{ .}
\label{eq:squarkfreefieldcts}
\end{align}
\end{subequations}
In analogy to the previous sections, the counter-terms 
arising from the squark potential get a field and a mass renormalization 
contribution. 
The one-loop squark potential counter-terms result from 
\begin{align}
\cL_{\tilde{t}, \text{mass}}^{\text{ren}} \supset -
\begin{pmatrix}
{\tilde{t}}_{1}^\dagger & {\tilde{t}}_{2}^\dagger
\end{pmatrix}
\mathcal{Z}^T_{\tilde{t}_{12}}M^2_{\tilde{t}_{12}}\mathcal{Z}_{\tilde{t}_{12}}
\begin{pmatrix}
{\tilde{t}}_{1}\\[.2em]{\tilde{t}}_{2}
\end{pmatrix} \text{ ,}
\end{align}
where $M^2_{\tilde{t}_{12}}$ is defined in Eq.~(\ref{alleqs:msquark12}) 
and $\mathcal{Z}_{\tilde{t}_{12}}$ in Eq.~(\ref{eq:1loopfieldren}).
Expanding them up to the first order they read
\begin{subequations}
\begin{align}
 \delta \boldsymbol{V}_{\tilde{t}_1\tilde{t}_1} &= 
 m_{\tilde{t}_1}^2  \delta Z_{\tilde{t}_1}^{(1)} + 
 \delta  M_{\tilde{t}_1\tilde{t}_1}^{2(1)} \text{ ,}\\
 \delta \boldsymbol{V}_{\tilde{t}_2\tilde{t}_2} &=
 m_{\tilde{t}_2}^2  \delta Z_{\tilde{t}_2}^{(1)} + 
 \delta  M_{\tilde{t}_2\tilde{t}_2}^{2(1)} \text{ ,} \\
 \delta \boldsymbol{V}_{\tilde{t}_1\tilde{t}_2} &=
 \frac12 (m_{\tilde{t}_1}^2 + m_{\tilde{t}_2}^2) \delta Z_{\tilde{t}_{12}}^{(1)} + 
 \delta  M_{\tilde{t}_1\tilde{t}_2}^{2(1)} \text{ .}
\label{eq:squarkpotentialcts}
\end{align}
\end{subequations}
As the field renormalization matrices $\mathcal{Z}_{\tilde{t}_{12}}$ 
are not diagonal, contributions from tree-level masses 
enter in the off-diagonal squark potential counter-terms, 
even though the mass matrix $M^2_{\tilde{t}_{12}}$ is in
diagonal form. 
The mass renormalization counter-terms read
\begin{align}
 \delta  M_{\tilde{t}_{12}}^{2(1)} = U_{\tilde{t}} \;\delta M_{\tilde{t}_{LR}}^{2(1)} \,U^T_{\tilde{t}} 
\end{align}
where the total differential of the matrix $M_{\tilde{t}_{LR}}^{2}$ defines
the one-loop counter-terms. Written out in explicit form, each 
contribution reads as follows, 
\begin{align}
 \delta  M_{\tilde{t}_1\tilde{t}_1}^{2(1)} = \delta m_{\tilde{t}_1}^{2(1)} \text{,}
 \hspace{20pt}
\delta  M_{\tilde{t}_2\tilde{t}_2}^{2(1)} = \delta m_{\tilde{t}_2}^{2(1)} \text{,}
 \hspace{20pt}
\delta  M_{\tilde{t}_1\tilde{t}_2}^{2(1)} =  (m_{\tilde{t}_1}^2 - m_{\tilde{t}_2}^2) \,\delta\theta_{\tilde{t}}^{(1)}
 \text{ .}
\end{align}
Collecting all counter-terms, the renormalized one-loop 
squark self-energies are given by 
\begin{subequations}
\begin{align}
\hat{\Sigma}_{\tilde{t}_1\tilde{t}_1}^{(1)}(p^2) &= \Sigma_{\tilde{t}_1\tilde{t}_1}^{(1)}(p^2) + \delta \boldsymbol{\tilde{t}}_1^{(1)}  
- \delta \boldsymbol{V}_{\tilde{t}_1\tilde{t}_1}^{(1)}\text{ ,} \\
\hat{\Sigma}_{\tilde{t}_2\tilde{t}_2}^{(1)}(p^2) &= \Sigma_{\tilde{t}_2\tilde{t}_2}^{(1)}(p^2) + \delta \boldsymbol{\tilde{t}}_2^{(1)} 
- \delta \boldsymbol{V}_{\tilde{t}_2\tilde{t}_2}^{(1)} \text{ ,} \\
\hat{\Sigma}_{\tilde{t}_1\tilde{t}_2}^{(1)}(p^2) &= \Sigma_{\tilde{t}_1\tilde{t}_2}^{(1)}(p^2) + \delta \boldsymbol{\tilde{t}}_{12}^{(1)}
- \delta \boldsymbol{V}_{\tilde{t}_1\tilde{t}_2}^{(1)} \text{ .}
 \label{eq:1loopselfenergies}
\end{align}
\end{subequations}
While the sum of all field renormalization 
contributions concerning inner fields 
vanishes, the massive contributions
do not. 
All self-energies are renormalized on-shell where the following 
renormalization conditions were used, following Refs.~\cite{Heinemeyer:1998np,Hollik:2003jj,
Heinemeyer:2004xw,
Heinemeyer:2010mm,Fritzsche:2011nr}
\begin{subequations}
\BEA
 \text{Re} [ \hat{\Sigma}_{\tilde{t}_1\tilde{t}_1}^{(1)}(m_{\tilde{t}_1}^2) ] &=& 0 \text{ ,} \\
 \text{Re} [ \hat{\Sigma}_{\tilde{t}_2\tilde{t}_2}^{(1)}(m_{\tilde{t}_2}^2) ] &=& 0 \text{ ,} \\
 \text{Re} [ \hat{\Sigma}_{\tilde{t}_1\tilde{t}_2}^{(1)}(m_{\tilde{t}_2}^2) ] +
 \text{Re} [ \hat{\Sigma}_{\tilde{t}_2\tilde{t}_1}^{(1)}(m_{\tilde{t}_1}^2) ] &=& 0 \text{ ,} \\
 \text{Re} [ \frac{\partial}{\partial p^2}\hat{\Sigma}_{\tilde{t}_1\tilde{t}_1}^{(1)}(m_{\tilde{t}_1}^2) ] &=& 0 \text{ ,} \\
 \text{Re} [ \frac{\partial}{\partial p^2}\hat{\Sigma}_{\tilde{t}_2\tilde{t}_2}^{(1)}(m_{\tilde{t}_2}^2) ] &=& 0 \text{ .} 
\EEA
\end{subequations}
These lead to the following determinations for the renormalization
constants entering the mass counter-terms, 
\begin{align}
\de  m_{\tilde{t}_i}^{2(1)} &= 
\text{Re} [\Si_{\tilde{t}_{i}\tilde{t}_i}^{(1)}(m_{{\tilde{t}}_{i}}^2) ]
\quad\ \text{with} \quad\ i = 1,\,2 \label{eq:dmst}  \text{ ,} \\
\de \theta_{\Stop}^{(1)} &= 
\frac{{\text{Re}}[{\Si}_{\tilde{t}_{1}\tilde{t}_2}^{(1)}(\mste^2)]  +
 {\text{Re}}[{\Si}_{\tilde{t}_{1}\tilde{t}_2}^{(1)}(\mstz^2)]}{2\; (m^2_{\Stop_1} - m^2_{\Stop_2})} \text{ ,} 
 \label{eq:thetarenormalization}
\end{align}
and the field counter terms
\begin{align}
 \de  Z_{\tilde{t}_1}^{(1)} &= 
 - \text{Re} [ \frac{\partial}{\partial p^2}\hat{\Sigma}_{\tilde{t}_1\tilde{t}_1}^{(1)}(m_{\tilde{t}_1}^2) ]  \text{ ,} \\
 \de  Z_{\tilde{t}_2}^{(1)} &= 
 - \text{Re} [ \frac{\partial}{\partial p^2}\hat{\Sigma}_{\tilde{t}_2\tilde{t}_2}^{(1)}(m_{\tilde{t}_2}^2) ]  \text{ .} 
\end{align}
$\de Z_{\tilde{t}_{12}}^{(1)}$ was already defined in Eq.~(\ref{eq:deltaZ12field}).
In addition to the squarks, also the quark sector has to be renormalized.
The top-quark mass is defined on-shell, yielding the one-loop
counter-term
\begin{align}
\de  m_t = \frac{m_t}{2} \;\bigl(
 \text{Re} [{\Si}_t^L (\mt^2) ] 
+ \text{Re} [{\Si}_t^R (\mt^2) ] 
+ 2 \,\text{Re} [ {\Si}_t^S (\mt^2) ] \bigr) ~,
\label{eq:dmt}
\end{align}
where ${\Si}_t^{L,R}$ are the left- and right-handed components of 
the quark self-energy, respectively. The latter can be decomposed
according to its Lorentz structure, 
\begin{align}
{\Si}_{t} (k) &= 
\not\! p {\Si}_t^V (p^2) + 
\not\! p \gamma_5 {\Si}_t^A (p^2) + 
\mt {\Si}_t^S (p^2) \text{ ,}
\label{eq:decomposition}
\end{align}
where it is split into the scalar part of the quark 
self energy ${\Si}_t^S$, the vectorial ${\Si}_t^V$ 
and the axial vectorial part ${\Si}_t^A$.

The renormalization of 
the stop-mixing soft breaking term  $X_{\tilde{t}}$, defined in 
Eq.~(\ref{eq:definitionXt}), reads
\begin{align}
 \de X_{\tilde{t}}^{(1)} = X_{\tilde{t}} 
 \left( 
 \frac{ 1-2\, \sin^2\theta_{\tilde{t}} }{\sin\theta_{\tilde{t}} \cos\theta_{\tilde{t}} } \de \theta_{\Stop}^{(1)}
 + \frac{ \de m_{\tilde{t}_1}^{2(1)} - \de m_{\tilde{t}_2}^{2(1)} }{ m_{\tilde{t}_1}^{2} - m_{\tilde{t}_2}^{2} }
 - \frac{ \de m_{t}^{2(1)} }{ m_{t} }
 \right) \text{ ,}
\end{align}
where the fact that neither $\de \mu^{(1)}$ nor $\de \tanb^{(1)}$ 
have couplings of the order $\mathcal{O}(\alpha_s)$ 
was already taken into account. The renormalization 
constant $ \de X_{\tilde{t}}^{(1)}$ enters in counter-term 
vertex insertions, e.g. in Fig.~\ref{subfig:oneloopctfigd}.
All renormalization \textbf{constants} are independent
of an external momentum, while the 
one-loop diagrams containing the counter-term insertions 
carry momentum dependence. 
%
%
\section{Treatment of the integrals}
\label{sec:treatintegrals}
%
%
In Sec.~\ref{sec:sediags}, some relevant two-loop topologies have
already been introduced diagrammatically, compare Fig.~\ref{fig:Tall} in 
Sec.~\ref{sec:sediags}. 
Below, the following notation
\begin{align}
\nonumber T_{i_1 i_2\dots i_n}&(p^2, m_{i_1}^2,m_{i_2}^2,\dots, m_{i_5}^2) = \left(2 \pi \mu_r \right)^{2(4-D)}\\
\times \iint & \frac{\text{d}^D  q_1}{i \pi^2}\, \frac{\text{d}^D  q_2}{i \pi^2}
\frac{1}{(k_{i_1}^2-m_{i_1}^2+ i \delta)(k_{i_2}^2-m_{i_2}^2+ i \delta)\cdots(k_{i_n}^2-m_{i_n}^2+ i \delta)} \text{ ,}
\label{eq:tints}
\end{align}
for the scalar two-loop integrals is adopted, where $p$ is the 
external momentum, the $q_i$ are the loop momenta, $\mu_r$ is 
the renormalization scale and the $m_i$ denote the masses of the
propagators. 
To comply with $\overline{DR}$ renormalization, 
an additional factor of $\gamma_E$ and 
$\text{log}(4\pi)$ is absorbed into the 
renormalization scale $\mu_r$ as
\begin{align}
 \text{log}(\mu_r) \rightarrow  \text{log}(\mu_r') =
 \text{log}(\mu_r) + \text{log}(4\pi) - \gamma_E \text{ .}
\end{align}
The momenta are continued to $D$ dimensions. 
The number $n$ of indices of the two-loop integral $T_{i_1 i_2 \dots i_n}$ 
corresponds to the number of propagators involved. The indices' 
numbers $i_1,\dots,i_n=1,\dots,5$ label which propagator type 
appears in the integral. There are five different propagator 
types $k_{1},\dots,k_5$, where the $k_i$ read
\begin{align}
\label{eq:kis}
k_1=q_1, \hspace{10pt} k_2=q_1+p, \hspace{10pt} k_3=q_2 -q_1, \hspace{10pt} k_4=q_2, \hspace{10pt} k_5=q_2+p \text{.}
\end{align}

In accordance with Ref.~\cite{'tHooft:1978xw}, the letters $A$, $B$ and $C$ are used
to describe one-loop integrals with one, two and three external legs, respectively. 
Hence, a one-loop tadpole integral $A$ can have one 
momentum-independent propagator of type $k_1$ or $k_4$. 
A one-loop bubble integral $B$ is composed of two propagators of 
type $k_1$ and $k_2$ or $k_4$ and $k_5$.
The three-point vertex integrals $C$ can carry dependence of two different 
external momenta $p_1$ and $p_2$ and have three propagators. 
In this calculation, however, being part of a self-energy computation, they depend on 
one external momentum $p$ only and are therefore 
always reducible to two-point one-loop functions $B$.
\subsection{Analytically known integrals}
\label{subsec:analytintegrals}
%
%
%

A series representation in the regulator $\eps$ 
of all one-loop integrals entering the sub-loop 
renormalization is known in analytical form and a 
variety of different representations is available. 
In this calculation, the approaches of 
Refs.~\cite{'tHooft:1978xw,Nierste:1992wg,
Scharf:1993ds,Berends:1994ed,Bauberger:1994zz,
Denner:1991kt} 
are used. 
At two loops, the resulting integrals from the 
reduction are mostly factorizing one-loop 
diagrams. Also all $\eps$-divergent parts 
and a few finite parts of the two-loop integrals 
appearing in the calculation are known 
from Refs.~\cite{Scharf:1993ds,Berends:1994ed,
Bauberger:1994zz,Bauberger:1994by,
Bauberger:1994hx}. 
Knowing the analytical representations of the 
divergent parts is vital in finding exact 
cancellation of all divergences. 
The full 
analytical cancellation of all divergences
is achieved with the above mentioned integral
representations, taking into consideration 
all counter-terms of Secs.~\ref{subsec:twoloopcts} 
and~\ref{subsec:subrenorm}, with their proper 
renormalization constants. 
The integral representations have all been 
cross-checked intensively with \secdec{} 
and the {\sc Golem95} integral 
library~\cite{Binoth:2008uq,Guillet:2013msa}. 
Especially for the imaginary parts these 
two tools were very useful. 

\medskip

Due to the occurrence of factorizing one-loop
integrals, these must be known to 
first order in $\eps$ as they
can contribute to the finite part through
terms of the type 
\begin{align}
\hspace{-25pt} (\frac{a_1}{\eps} + b_1 + c_1 \eps) 
 (\frac{a_2}{\eps} + b_2 + c_2 \eps) =
 \frac{a_1 a_2}{\eps^2} + \frac{a_1 b_2 + a_2 b_1}{\eps}
 + b_1 b_2 + a_1 c_2 + a_2 c_1 + ... \text{ .}
\end{align}
All necessary parts for the one-loop integrals 
are known analytically~\cite{Nierste:1992wg,
Bauberger:1994zz,Scharf:1993ds,Berends:1994ed}. As 
mentioned in the previous section, three-point 
integrals $C$ result from the tensor reduction in the 
sub-loop renormalization part. 
The $A$ and $B$ integrals are UV 
divergent, the $C$ integrals 
appear as finite integrals only because possible 
IR singularities are regulated by the massive propagators.
Additionally, the third external leg occurs 
in the context of counter term insertions and
therefore always has a
vanishing external momentum, compare 
Figs.~\ref{subfig:cttop1} and \ref{subfig:cttop2}. 
It is due to this fact, that all three-point 
functions can be reduced to derivatives of
two-point functions.
A general three-point function reads 
\begin{align}
\hspace{-20pt} C_0(p_1^2, p_2^2, m_1^2, m_2^2, m_3^2)=& 
\int \text{d}^D q \frac{1}{(q^2-m_1^2)((q+p_1)^2-m_2^2)((q+p_1+p_2)^2-m_3^2)} \text{ .} 
\end{align}
Here, the notation of Ref.~\cite{'tHooft:1978xw} is adopted.
One example of a $C$ integral occurring in the calculation, 
to be reduced to 
the derivative of a two-point function, reads
\begin{subequations}
\begin{align}
C_0(p^2, 0, m_1^2, m_2^2, m_2^2) 
&= \int \text{d}^D q \frac{1}{(q^2-m_1^2)((q+p_1)^2-m_2^2)((q+p_1)^2-m_2^2)}\\
&= \frac{\partial}{\partial m_2^2}  \int\text{d}^D q \frac{1}{(q^2-m_1^2)((q+p_1)^2-m_2^2)}\\
&= \frac{\partial}{\partial m_2^2} B_0(p^2,m_1,m_2) \text{ .}
\end{align}
\end{subequations}
From the derivative in the last line, the finiteness of the 
$C$ integral becomes apparent; the divergent 
part of the two-point function does not depend on 
the masses.

The reduction to purely $A$ and $B$ integrals
is especially useful as derivatives
of all integrals occurring in the self-energies 
are required in the calculation
of the field renormalization constants. 
Products of
$C$ integrals with $A$ and $B$ integrals, entering 
through the counter-term insertions, lead to finite 
three-point function contributions to the derivatives. 
Now, the derivatives of $C$ integrals can be easily 
computed for simpler cases, but for the very general 
case of arbitrary masses in the loop and external 
momentum, analytical results are much more 
involved. For cross checks, the  
derivatives of the $C$ integrals for the simpler
cases could be computed using the algebraic 
output from \secdec{} and integrating the 
expressions analytically. 

Furthermore, the method of partial fractioning can
be exploited, see Ref.~\cite{vanderBij:1983bw}. It 
is used for the reduction of $B$ integrals
to one-point integrals. The method works at
arbitrary loop order. At two loops, e.g., the vacuum
integral $T_{1134}$ function can be reduced as follows  
\begin{align}
\non T_{1134}(& 0, m_1^2,m_2^2,m_3^2,m_4^2) \\
&= \frac{1}{m_1^2-m_2^2} (T_{134}(0,m_1^2,m_3^2,m_4^2)-T_{134}(0,m_2^2,m_3^2,m_4^2)) \text{ ,}
\end{align}
where the first entry in the integrals denotes a 
zero external momentum, in accordance with the 
notation set up in Eq.~(\ref{eq:tints}). 

The additional relation 
\begin{align}
T^{\text{div}}_{1234}(p^2, m^2,m^2,m_3^2, m_4^2) = 
B_0^{\text{fin}}(p^2, m^2, m^2) + \frac{1}{2} \text{ ,}
\label{eq:tdivB0}
\end{align}
is of use, where $T^{\text{div}}_{1234}$ denotes the coefficient 
to the singly divergent $\frac{1}{\eps}$ term in the 
Laurent expansion of the $T_{1234}$ integral in the
regulator $\eps$, and $B_0^{\text{fin}}$ the finite part 
of the $B_0$ integral. 
Relation Eq.~(\ref{eq:tdivB0}) can be found as a direct 
consequence of the 
formulas in the appendix of Ref.~\cite{Scharf:1993ds}. 
Furthermore, the relation
\begin{subequations}
\begin{align}
C_0(0, p^2, m^2, m^2, m^2) 
&= \int \text{d}^D q \frac{1}{(q^2-m^2)^2((q+p)^2-m^2)} \\
&=\frac12 \frac{\partial}{\partial m^2} B_0(p^2, m^2, m^2)  \text{ ,}
\label{eq:C0reducedtoderivB0}
\end{align}
\end{subequations}
deduced from
\begin{subequations}
\begin{align}
 &\frac{\partial}{\partial m^2} B_0(p^2, m^2, m^2) \\
 &= \int \text{d}^D q \frac{1}{(q^2-m^2)((q+p)^2-m^2)^2} + \int \text{d}^D q \frac{1}{(q^2-m^2)^2((q+p)^2-m^2)} \label{eq:dm1b0momshift} \\
 &= 2 \int \text{d}^D q \frac{1}{(q^2-m^2)^2((q+p)^2-m^2)}  \text{ ,}
\end{align}
\end{subequations}
helps achieving a stable cancellation of the UV poles.
A momentum shift was performed in the first integral of 
Eq.~(\ref{eq:dm1b0momshift}). 

The tensor coefficients 
$C_2$ and $B_1$ can be treated by expanding 
their numerators. To give an example, the $C_2$ tensor 
coefficients appearing in this calculation are of the type
\begin{subequations}
\begin{align}
 &C_2(0, p^2, m^2, m^2, m^2) \\
 &= \frac{1}{p^2}\int \text{d}^D q \frac{q\cdot p}{(q^2-m^2)^2((q+p)^2-m^2)} \\
 &= \frac{1}{2 p^2}\int \text{d}^D q \frac{(q+p)^2-q^2-p^2+m^2-m^2}{(q^2-m^2)^2((q+p)^2-m^2)}\\
\nonumber &= \frac{1}{2 p^2}\left( B_0(0,m^2,m^2)-B_0(p^2,m^2,m^2) \right.\\
 &\left. -p^2 C_0(0,p^2,m^2, m^2, m^2)\right) \text{ .}
\end{align}
\end{subequations}
Similarly, the $B_1$ coefficient can be reduced to $B_0$ and $A_0$ integrals.

\medskip

After the cancellation of all divergent parts, the finite
parts must be treated. These maximally include finite terms 
of two-loop integrals and terms up to the order 
$\mathcal{O}(\eps)$ in the one-loop case. 
As is visible from Eq.~(\ref{eq:tdivB0}), the finite part of 
the one-loop bubble is part of a sub-divergence of the 
two-loop integral $T_{1234}$. This fact suggests that one-loop 
terms of order $\mathcal{O}(\eps)$ may contribute to the 
finite parts of some of the two-loop integrals. As a matter of 
fact, $\mathcal{O}(\eps)$ parts of the integral 
\begin{align}
\label{eq:dm1bo}
\frac{\partial}{\partial m_1^2}B_0(p^2, m_1^2,m_2^2) = \int \frac{\text{d}^D  q_1}{\text{i} \pi^2 (2 \pi \mu)^{D-4}} \frac{1}{(k_1^2-m_1^2)^2 (k_2^2-m_2^2)} \text{ ,}
\end{align}
appear in sums and subtractions with the finite part of the integral 
\begin{align}
\label{eq:t11234}
T_{11234}(p^2, m_1^2,m_1^2,m_2^2,m_3^2, m_4^2) = \frac{\partial}{\partial m_1^2} T_{1234}(p^2, m_1^2,m_2^2,m_3^2, m_4^2) \text{ .}
\end{align}
An analytic result of the finite part of the latter is not available. 
As the unknown integrals are computed numerically 
with version 2 of \secdec, analytic and numeric results enter 
the self-energies and their counter-terms. 
A stable evaluation of the analytically available and 
the numerically computed functions is required. 
The analytically available integrals appearing in the 
final result 
are the one-point functions $A_0$, the scalar two-point 
integral $B_0$, the derivative of $B_0$ by one mass, 
Eq.~(\ref{eq:dm1bo}), the two-loop vacuum diagram 
with arbitrary masses $T_{134}$, and the two-loop 
bubble diagram $T_{234}$ with two massive propagators 
of the same mass and one massless propagator. 
Explicit expressions for the integrals are given in 
App.~\ref{app:analytformulae}. 
For the evaluation of the analytical integrals beyond threshold, 
the proper analytical continuation 
into the complex plane is essential. 
Square roots, as well as logarithms, get an imaginary
part as soon as their arguments are negative. 
The correct prescription for the analytical continuation is given 
by the causal $i\delta$ appearing 
in the Feynman propagators. 
For the kinematic invariants, they separately read 
\begin{subequations}
\begin{align}
m^2 &\rightarrow m^2 - i \delta \text{ ,}\\
p^2 &\rightarrow p^2 + i \delta \text{ ,}\\
 s_{ij} &\rightarrow s_{ij} + i \delta \text{ .}
 \label{eq:analytcontinuation}
\end{align}
\end{subequations}
In practice, the infinitesimal quantity $\delta$ must 
be assigned a tiny, but not infinitesimally small 
value. 
This can lead to numerical instabilities 
in the evaluation of an analytical result. 
These can be evaded by explicitly 
choosing a Riemann 
sheet of the square root 
\begin{align}
  \sqrt{a^2 \pm i \delta}= 
  \begin{cases}
  \sqrt{a^2}  & \text{, } a^2 > 0\\
  \pm i\sqrt{-a^2} & \text{, }a^2 < 0 \text{ ,}
\end{cases}
\end{align}
and the logarithm
\begin{align}
   \text{log} (a^2 \pm i \delta) = 
  \begin{cases}
  \text{log} (a^2)  & \text{, } a^2 > 0\\
  \text{log} (-a^2) \pm i \pi  & \text{, }a^2 < 0 \text{ ,}
\end{cases}
\end{align}
as shown, e.g., in Ref.~\cite{Bauberger:1994zz}. 
At the one-loop level, thresholds can be parametrized by 
square roots of the fully symmetric K{\"a}ll{\'e}n function 
\begin{align}
\lambda(x,y,z) = x^2 + y^2 + z^2 - 2 x y - 2 x z - 2 y z \text{ ,}
\end{align}
which frequently appears in the analytic expressions. Its sign 
in different parameter regions is 
\begin{align}
\lambda(x,y,z)
\begin{cases}
  <0 \text{ ,} &  (y + z) - 2 \sqrt{y z} < x < (y + z) + 2 \sqrt{y z}\\
  =0 \text{ ,} & x = (y + z) \pm 2 \sqrt{y z}\\
  >0 \text{ ,}  & \text{else} \\
\end{cases}
\hspace{10pt}\text{ .} 
\end{align}
With this knowledge at hand, all thresholds entering from 
the on-shell renormalization of the sub-loop can be explained as the 
square root of the K{\"a}ll{\'e}n function becomes complex 
for a negative $\lambda(x,y,z)$. 
%
%
\subsection{Analytically unknown integrals}
The evaluation of some of the integrals entering the calculation 
are not known.
This concerns different topologies of the integrals 
$T_{234}$, $T_{1234}$, $T_{11234}$ and $T_{12345}$, 
depicted in Fig.~\ref{fig:Tall}.
While the evaluation is straightforward in the massless cases, 
the all-massive cases are very hard to compute in a full analytical way. 
This is due to the fact that neither the $T_{234}$ nor 
the $T_{12345}$ can be expressed in terms of polylogarithms, 
see Refs.~\cite{Berends:1993ee,Scharf:1991dipl} and references
therein. 
An analytical description of the 
diagram $T_{234}$ in terms of generalized Lauricella 
functions was found in Ref.~\cite{Berends:1993ee}.
The Appell functions appearing in their result 
are extremely hard to evaluate in a full analytical way, 
compare, e.g. Ref.~\cite{TuanKim:1992}. 
Up to date, only one-dimensional
integral representations are available 
for the evaluation of the $T_{234}$ integral with 
arbitrary masses. 
Special cases (large momentum expansion, equation of masses) 
of the $T_{234}$ integral are available, see 
Refs.~\cite{Davydychev:1993pg,Scharf:1993ds,
Berends:1994ed,Laporta:2004rb}. 
The general result for different masses in 
the three propagators is still 
discussed today, among physicists~\cite{Adams:2014vja,
Adams:2013nia,Remiddi:2013joa} and 
mathematicians~\cite{Greynat:2013zqa} alike.
A description of the finite $T_{12345}$ function in 
terms of a double integral
representation~\cite{Kreimer:1991jv}, and
in terms of a single integral representation~\cite{Bauberger:1994hx}, 
is also available. 
Furthermore, 
special cases and asymptotic limits are known, see 
Refs.~\cite{Broadhurst:1987ei,Davydychev:1992mt,
Broadhurst:1993mw,Scharf:1993ds}. 
The numerical evaluation of 
integrals not accessible with purely analytical 
methods can be achieved, benefitting from the automated 
setup of the program \secdec.
\subsubsection{Numerical computation of two-point two-loop integrals}
\label{subsubsec:numericalcomphiggs}
\begin{figure}[htb!]
\subfigure[]{\includegraphics[height=5.15cm]{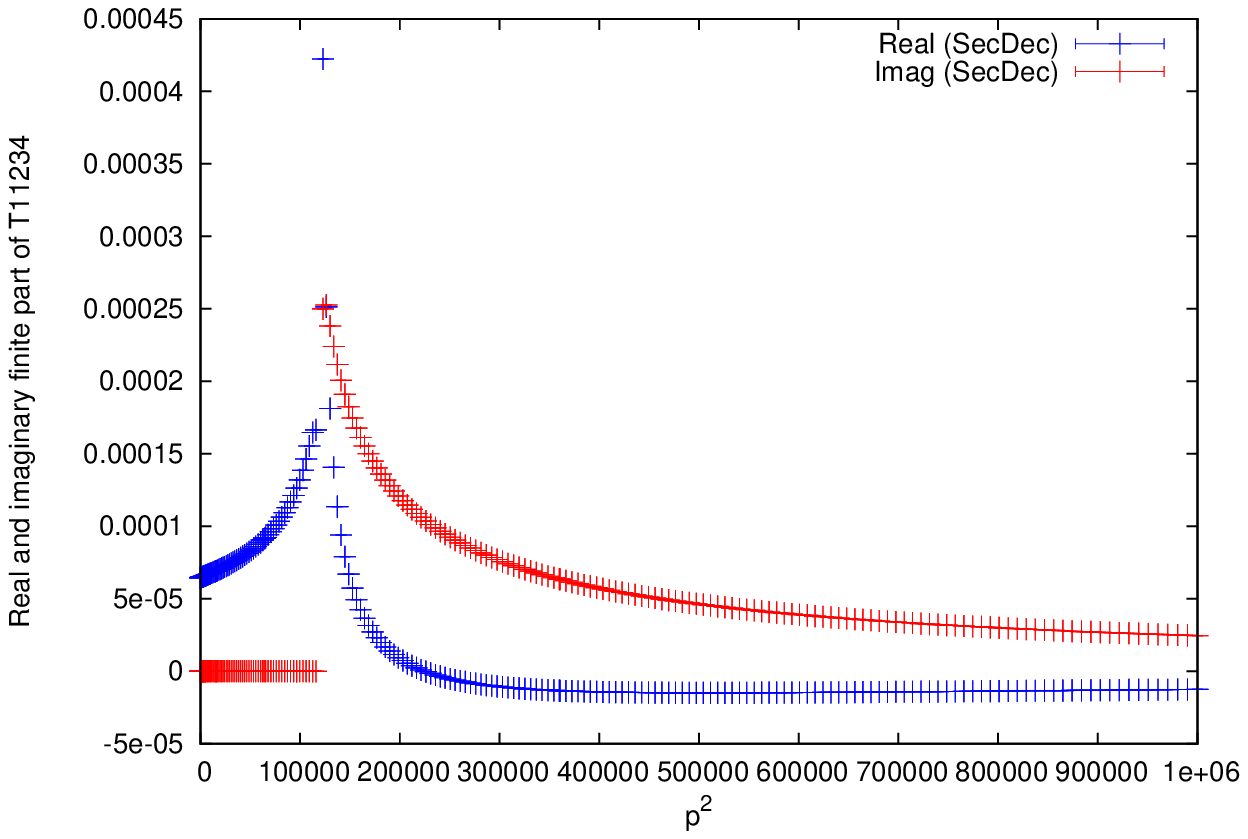} \label{fig:T11234} }\hfill
\subfigure[]{\includegraphics[height=5.15cm]{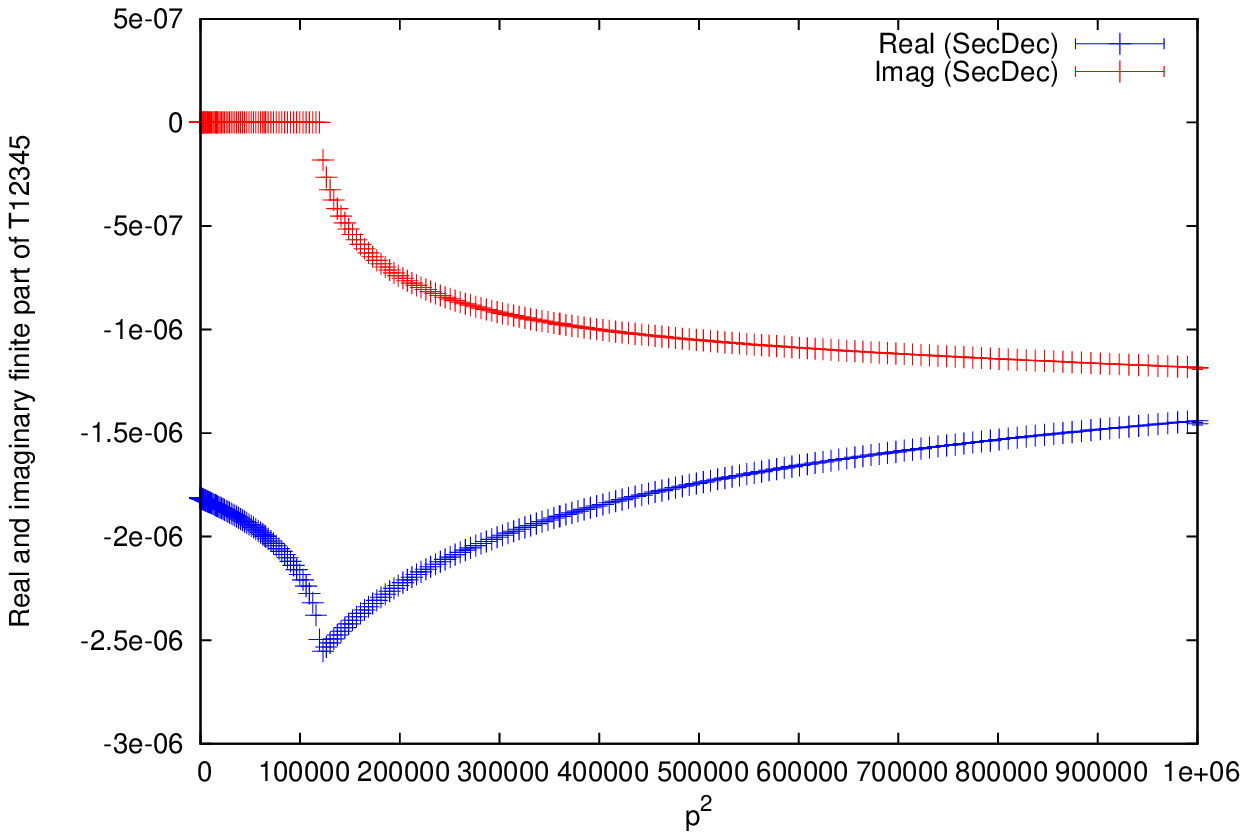} \label{fig:T12345} }
\caption{Two representative numerical \secdec{} results for the integrals of type 
\ref{subfig:T11234} in \subref{fig:T11234}, and of type \ref{subfig:T12345} in 
\subref{fig:T12345}, with three and four different masses, respectively. 
In \subref{fig:T11234}, the values $m_1 = m_2 =173.2$, $m_3 =826.8$ and $m_4=1.5$ 
are used. In \subref{fig:T12345}, the numerical values $m_1 =1173.2$, 
$m_2 =826.8$, $m_3 =1.5$ and $m_4 =173.2$ are chosen. Due to the high 
accuracy of the numerical integration, error bars are barely seen. }
\label{fig:TSecDecInts}
\end{figure}

%
%
The aforementioned four integral topologies $T_{234}$, 
$T_{1234}$, $T_{11234}$ and $T_{12345}$ appear in 34 
different mass configurations. These involve up to four different masses, 
in addition to the mass scale given by the external momentum $p^2$. 
They are all computed with \secdec. 
For the 
whole evaluation, the integrator Divonne contained in the 
{\sc Cuba library}~\cite{Hahn:2004fe,Agrawal:2011tm} is used. The 
integrator uses a deterministic algorithm and can reach very 
accurate results in
integrations of few, but more than one, Feynman parameters.
An additional acceleration was achieved by introducing 
user-defined thresholds to the program \secdec, see 
Secs.~\ref{subsec:program:cubaparameters} 
and~\ref{sec:appendix:usermanual}. This allows the user
to define a lowest threshold condition. Once it is met, \secdec{} 
switches to a deformation of the integration 
contour into the complex plane.
The kinematic values can differ by up to 14 orders of magnitude. 
The evaluation of a single phase space point for the most complicated topology, 
to reach a relative accuracy of at least $10^{-5}$,  
ranges between 0.01 and 100 seconds on an Intel i7 processor, 
where the larger timings are for points very close to a kinematic threshold.
The huge differences in the kinematic
invariants enter, for instance, when choosing a small value for the 
squared external momentum, while testing large squared masses. 
These configurations were of special interest in the performance 
of checks against the results obtained for 
vanishing external momenta. 
For two representative results, see Fig.~\ref{fig:TSecDecInts}. 
The configuration in Fig.~\ref{fig:T11234} has three different mass 
scales where the first two masses can be associated with the top mass 
$m_1 = m_2 =173.2$ GeV, the third mass with one possible value for a 
stop mass $m_3 =826.8$ GeV and the fourth mass to a value for the 
gluino $m_4=1.5$ TeV. 
In Fig.~\ref{fig:T12345}, the numerical values $m_1 =1173.2$ GeV, 
$m_2 =826.8$ GeV, $m_3 =1.5$ TeV and $m_4 =173.2$ GeV are chosen. The
relative accuracy of the plots is beyond $10^{-5}$. 
When evaluating a specific scenario, some integrals needed to 
be evaluated up to a numerical relative accuracy of at least $10^{-5}$ to 
make up for cancellations appearing between analytically 
evaluated integrals and numerical ones. For the evaluation of 
an arbitrary scenario with arbitrary rMSSM parameters, this 
accuracy was therefore demanded for every \secdec{} integral.
%
%
\section{Evaluation of the additional shifts to the Higgs-boson masses}
\label{sec:evaladdshifts}
The calculation is performed in the $\phi_1$-$\phi_2$ basis. 
To be consistent with all other higher-order contributions to the 
Higgs-boson masses incorporated in the public program \fh, 
the renormalized self-energies in the $\phi_1$-$\phi_2$ basis can
be rotated into the physical $h^0$-$H^0$ basis, where the tree-level 
propagator matrix is diagonal, via
\begin{subequations}
\BEA
\ser{H^0H^0}^{(2)}&=& 
\cos^2\!\alpha \,\ser{\Pe\Pe}^{(2)} + 
\sin^2\!\alpha \,\ser{\Pz\Pz}^{(2)} + 
\sin (2 \alpha) \, \ser{\PePz}^{(2)} \text{ ,}\\
\ser{h^0h^0}^{(2)} &=& 
\sin^2\!\alpha \,\ser{\Pe\Pe}^{(2)} + 
\cos^2\!\alpha \,\ser{\Pz\Pz}^{(2)} - 
\sin (2 \alpha) \,   \ser{\PePz}^{(2)} \text{ ,} \\
\ser{h^0H^0}^{(2)} &=& 
\sin \alpha \cos\alpha \,(\ser{\Pz\Pz}^{(2)} - \ser{\Pe\Pe}^{(2)}) + 
\cos (2 \alpha) \,   \,\ser{\PePz}^{(2)} \text{ ,}
\EEA
\label{eq:transformationphi12tohH}%
\end{subequations}%
where $\alpha$ is the tree-level mixing angle and using Eq.~(\ref{eq:physbasis}). 
The former is 
expressible in terms of the parameters $\tan\beta$, 
$m_{A^0}$ and $m_{Z}$, 
see Eq.~(\ref{eq:tan2alphatan2beta}). 
The resulting new contributions to the neutral 
$\cp$-even Higgs-boson 
self-energies, containing all momentum-dependent and additional constant 
terms, are assigned to the differences
\begin{equation}
\label{eq:DeltaSE}
\De\ser{ab}^{(2)}(p^2) = \ser{ab}^{(2)}(p^2) - \tilde\Sigma_{ab}^{(2)}(0)\,,
\qquad
ab = \{H^0H^0,h^0H^0,h^0h^0\}\,.
\end{equation}
Note the tilde (not hat) on $\tilde\Sigma^{(2)}(0)$, which signifies that 
not only the self-energies are evaluated at zero external momentum but
also the corresponding counter-terms,
following Refs.~\cite{Heinemeyer:1998jw,Heinemeyer:1998kz,
Heinemeyer:1998np}.
A finite shift $\De\hat{\Sigma}^{(2)}  (0)$
therefore remains in the limit $p^2\to 0$ 
due to $\de m_{A^0}^{2(2)} = \re\se{A^0A^0}^{(2)}(m_{A^0}^2)$ being computed 
at $p^2=m_{A^0}^2$ 
in $\hat\Sigma^{(2)}$, but at $p^2=0$ in $\tilde\Sigma^{(2)}$, 
as discussed in Sec.~\ref{subsec:twoloopcts}. 

Subtracting the finite shift of $\de m_{A^0}^{2(2)}$, the 
$\De\ser{ab}^{(2)}(p^2)$ in Eq.~(\ref{eq:DeltaSE}) must vanish. 
This limit was tested numerically, see 
Sec.~\ref{subsubsec:numericalcomphiggs}. 
Moreover, the zero momentum limit was checked 
analytically, deriving expressions for the 
vacuum diagrams $T_{1134}$, 
$T_{11134}$ and $T_{11344}$ and using an 
available expression for $T_{134}$. 
The relations were computed from $T_{134}$ using derivatives 
of the integral by the masses and partial fractioning 
as mentioned in Ref.~\cite{vanderBij:1983bw}. Expressions for the
vacuum diagram $T_{134}$ with different mass configurations can be 
found in Refs.~\cite{Berends:1994sa,Davydychev:1992mt}. All 
deduced integrals were checked with \secdec, including the 
three-propagator vacuum bubble $T_{134}$. 
For further comparison, the expression for the $T_{1134}$ integral
of Ref.~\cite{Hoogeveen:1985tf} could be used.

\medskip

The higher-order corrected $\cp$-even Higgs-boson masses in the
MSSM are obtained  from the corresponding propagators
dressed by their self-energies. 
Inserting the fields $h^0$ and $H^0$ in Eq.~(\ref{eq:proppolex}), 
the inverse propagator matrix in the $h^0$-$H^0$ basis 
is given by
\begin{align}
\label{eq:prop}
(\Delta_{\text{Higgs}})^{-1} = -\text{i}
\left( \begin{matrix} 
p^2 - m_{H^0}^2 + \hat{\Sigma}_{H^0H^0}(p^2) & \hat{\Sigma}_{h^0H^0}(p^2)\\ 
\hat{\Sigma}_{h^0H^0}(p^2) & p^2 - m_{h^0}^2 + \hat{\Sigma}_{h^0h^0}(p^2) 
\end{matrix} \right) \text{ .}
\end{align}
The $\cp$-even Higgs boson masses are determined by the
poles of the $h^0$-$H^0$-propagator matrix. 
This is equivalent to solving the equation
\begin{equation}
\left[p^2 - m_{h^0}^2 + \hSi_{h^0h^0}(p^2) \right]
\left[p^2 - m_{H^0}^2 + \hSi_{H^0H^0}(p^2) \right] -
\left[\hSi_{h^0H^0}(p^2)\right]^2 = 0\,~,
\label{eq:proppole}
\end{equation}
yielding the loop-corrected pole masses, $\Mh$ and $\MH$.
%
%
%
%
%
\subsection{Phenomenological motivation for two different scenarios}
%
%
Suitable scenarios to analyze the influence of 
the new self-energies on the Higgs-boson mass shifts should 
cover a range of experimentally allowed parameter space, 
in addition to maximizing the resulting additional shifts. 
It should be noted, that a complete parameter scan over the in 
principle more than one hundred free parameters of the MSSM is 
not feasible and the experimental sensitivity to collectively 
constrain many parameters is not sufficient with present 
experiments. In practice, parameter scans are therefore done 
for a smaller set of parameters with the highest
phenomenological impact on the rMSSM Higgs sector, see e.g. 
Ref.~\cite{Bechtle:2012jw}. 
These reflections result in an $m_h^{\text{max}}$ and a light 
stop scenario, motivated in and following analyses 
from Refs.~\cite{Bechtle:2012jw,Carena:2013qia}. 
%
%
%
%

\medskip

The MSSM input parameters entering the calculation are 
 \begin{align}
\non m_{\tilde{g}},\; m_{\tilde{t}_1},\; m_{\tilde{t}_2},\; \theta_{\tilde{t}},\; m_t,\; \mu,\; m_{A^0},\; \tan\beta \text{ .}
\end{align}
The benchmark scenarios of Ref.~\cite{Carena:2013qia} are 
given in terms of the following set of MSSM input parameters 
 \begin{align}
\non m_{\tilde{g}}, \; \Xt, \; \msusy, \; m_t,\; \mu,\; m_{A^0},\; \tan\beta \text{ .}
\end{align}
In the conversion to the latter parameters, 
the left- and right-handed soft SUSY breaking stop mass
parameters are equated and set to the soft SUSY breaking scale
\begin{align}
 \msusy := \MstL = \MstR \text{ .}
\end{align}
The two stop masses $m_{\tilde{t}_1}$, $m_{\tilde{t}_2}$ 
and the angle $\theta_{\tilde{t}}$ are 
expressed in terms of $\msusy$, 
the soft SUSY breaking mixing parameter $X_{\tilde{t}}$, the
top-quark mass $m_t$ and $m_Z$ using 
Eq.~(\ref{eq:upmassmatrixnicer}) with $\tilde{u}=\tilde{t}$. 
%
%
The value for the top mass $m_t=173.2$ GeV 
is used in both scenarios and taken from the latest combination of all 
top-mass measurements undertaken by the D0 and CDF 
collaborations~\cite{CDF:2013jga}, in agreement with
the top mass resulting from the combination of the latest ATLAS, 
CMS results, see Ref.~\cite{CMS-PAS-TOP-12-001,ATLAS:2014wva}. 
For the computation of the Higgs-boson mass shifts, a value of 
$m_Z = 91.1875$ GeV~\cite{Acciarri:2000kg} is used. \footnote{For 
simplicity, the $Z$ boson mass contribution is not taken into 
account in the Figs.~\ref{fig:se_scenario1reim}, \ref{fig:se_scenario1_madep}, 
\ref{fig:se_scenario2reim} and \ref{fig:se_scenario2_madep}. }

\medskip

Keeping the top and $Z$ boson 
mass fixed, the stop masses can be plotted with respect to the ratio 
$X_{\tilde{t}}/\msusy$. For the limit $ X_{\tilde{t}} \rightarrow 0$, the
squark mass eigenstates are equal. Following the analyses in 
Ref.~\cite{Bechtle:2012jw}, degenerate stop masses are 
excluded, if the observed new particle at the LHC is associated with 
the light Higgs-boson at around $m_h^{obs} \approx 125.7$ GeV. 
Therefore, a maximal mixing of $X_{\tilde{t}}=2\, \msusy$ 
is assumed in all 
scenarios, varying only the SUSY breaking scale $\msusy$ 
between $500$ GeV and $1$ TeV. 

\medskip

The gluino mass parameter $m_{\tilde{g}}$ enters 
the MSSM Higgs-boson mass predictions from two-loop 
order on. It is 
therefore of special interest
to analyze its impact on the MSSM Higgs-boson masses. 
The gluino mass can be indirectly constrained 
from the hitherto non-observation of a second 
neutral Higgs-boson. 

\medskip

The higgsino parameter $\mu$ enters the self-energies 
through the trilinear coupling. 
In all scenarios considered, it is chosen $\mu =200$ GeV, 
in accordance with Ref.~\cite{Carena:2013qia,Bechtle:2012jw} . 

\medskip

The $\cp$-odd $A^0$ boson mass and $\tan \beta$ are left as free 
parameters and can be varied between 
$90 \text{ GeV} \leq m_{A^0}\leq 1 \text{ TeV}$ and 
$1 \leq \tan\beta \leq 60$ respectively, compare 
Ref.~\cite{Bechtle:2012jw}. 
When choosing $m_{A^0}$ to be rather light, the observed
particle $m_h^{obs}$ GeV can be associated 
with the heavy Higgs-boson leaving room for an additional 
lighter state. 
%

\medskip

The corresponding renormalization scale, $\mu_r$, is set to 
$\mu_r = \mt$ in all numerical evaluations. 
The scale uncertainties are expected to be much smaller than the 
parametric uncertainties due to variations of parameters like $\tb,\MA,\mgl,m_{\tilde{t}}$.
\subsection{Renormalized $\mathcal{O}(\alpha_s \alpha_t)$ self-energies}
\subsubsection{Scenario 1: $m_h^{\text{max}}$ scenario}
Scenario~1 is oriented at the \mhmax\ scenario described
in Ref.~\citere{Carena:2013qia}. 
The following values are assigned to the MSSM parameters, 
\begin{align}
\mt &= 173.2\gev,\; \msusy=1\tev,\; \Xt =2\,\msusy\; , \non \\
\mgl &= 1500\gev,\; \mu = 200\gev~, 
\end{align}
leading to stop mass values of
\begin{align}
\mste &= 826.8\gev,\;  \mstz= 1173.2\gev\, .\non
\end{align}

With the introduction of the momentum 
dependence, thresholds occur in the self-energy diagrams 
when the external momentum $p=\sqrt{p^2}$,
in the time-like region, 
is such that a cut of the diagram would correspond to the 
on-shell production of the massive particles 
of the cut propagators, compare the discussion of 
Sec.~\ref{sec:euclidvsphyskinem}.
The resulting imaginary parts will enter in the search for 
the complex poles of the inverse 
propagator matrix of the Higgs-bosons.
Therefore it is interesting to study the 
behavior of the real and imaginary parts of the self-energies.
%
\begin{figure}[htb!]
\centering
\includegraphics[width=0.49\textwidth]{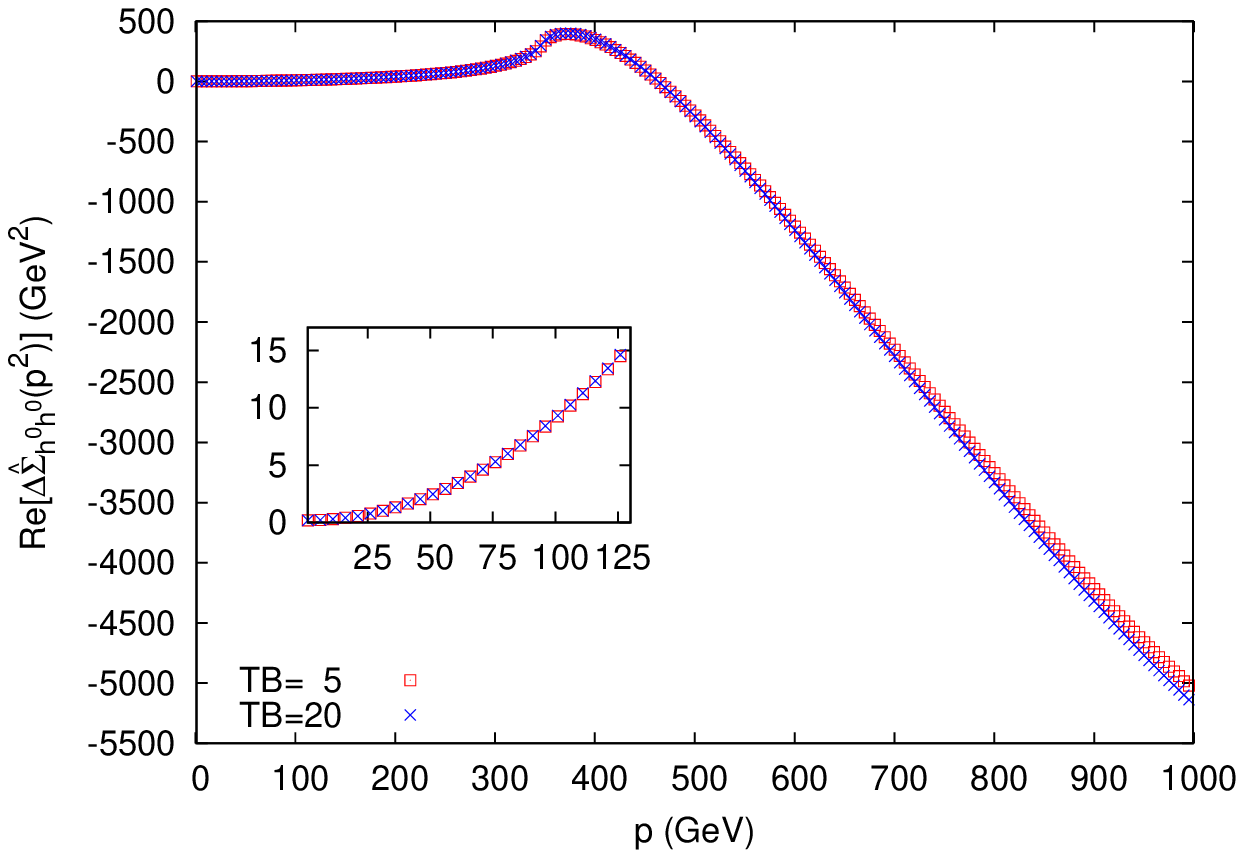}
\includegraphics[width=0.49\textwidth]{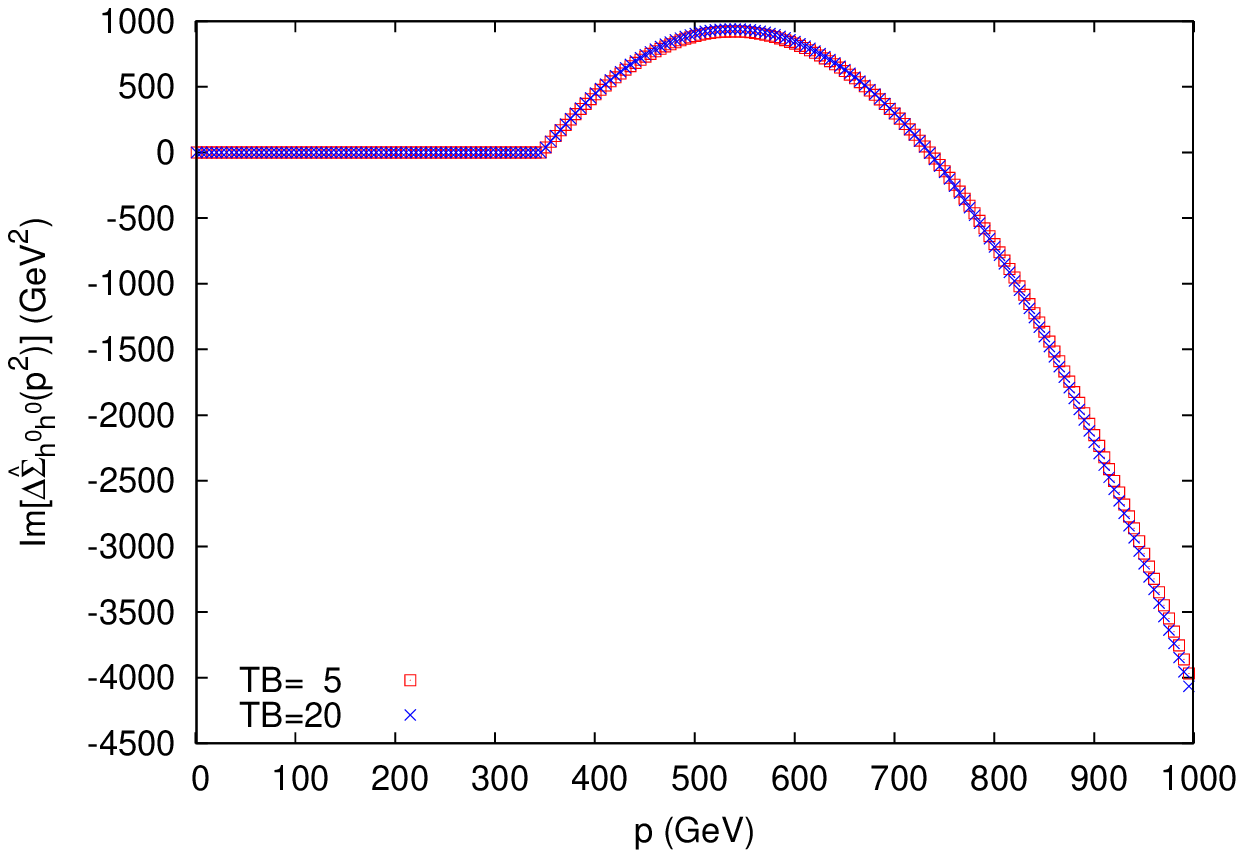}
\\[2em]
\includegraphics[width=0.49\textwidth]{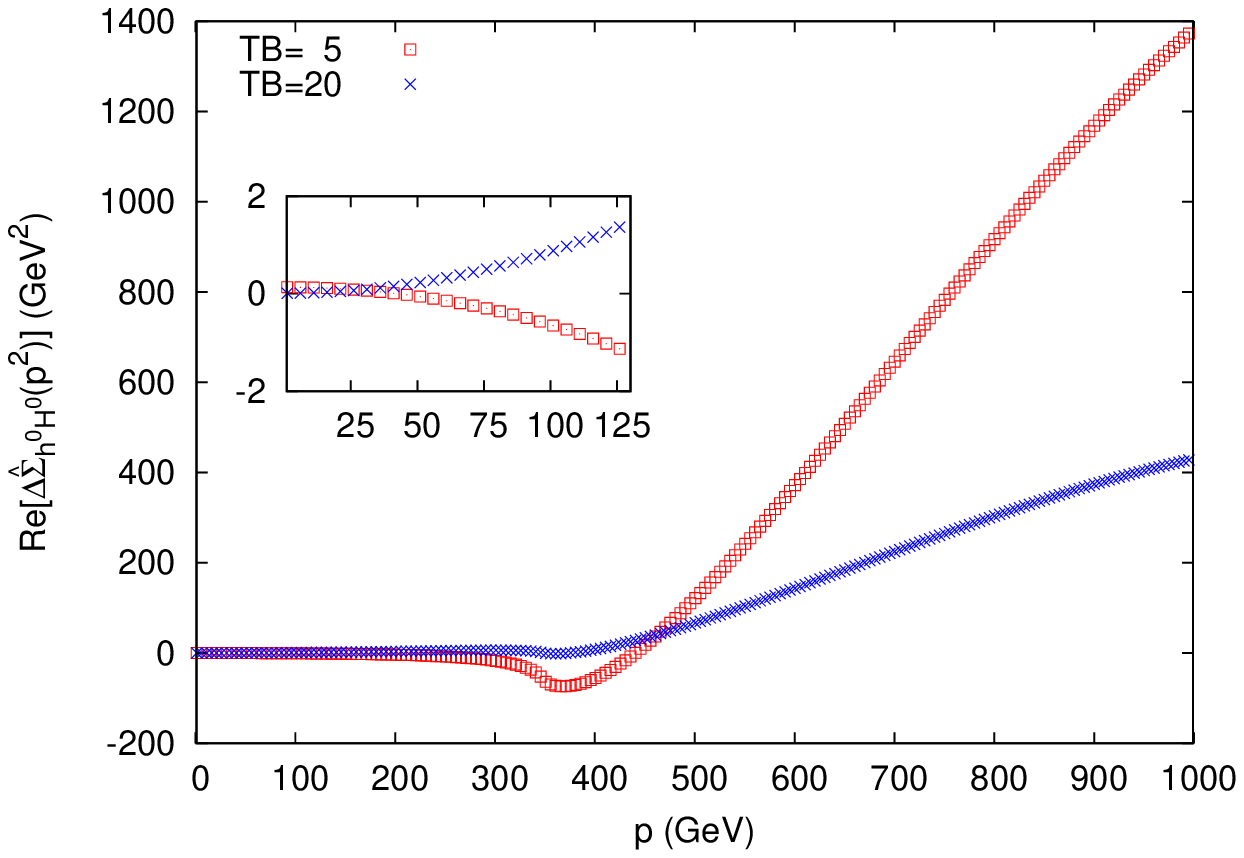}
\includegraphics[width=0.49\textwidth]{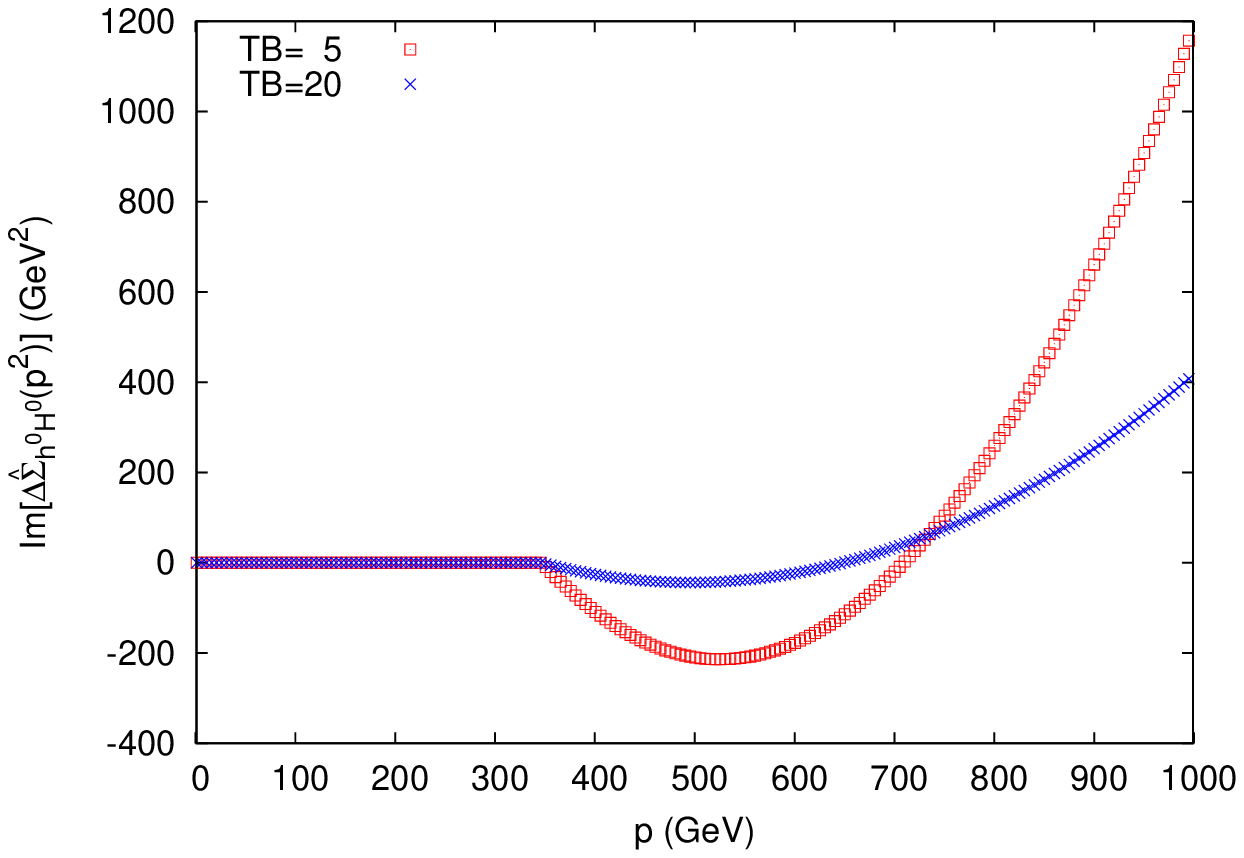}
\\[2em]
\includegraphics[width=0.49\textwidth]{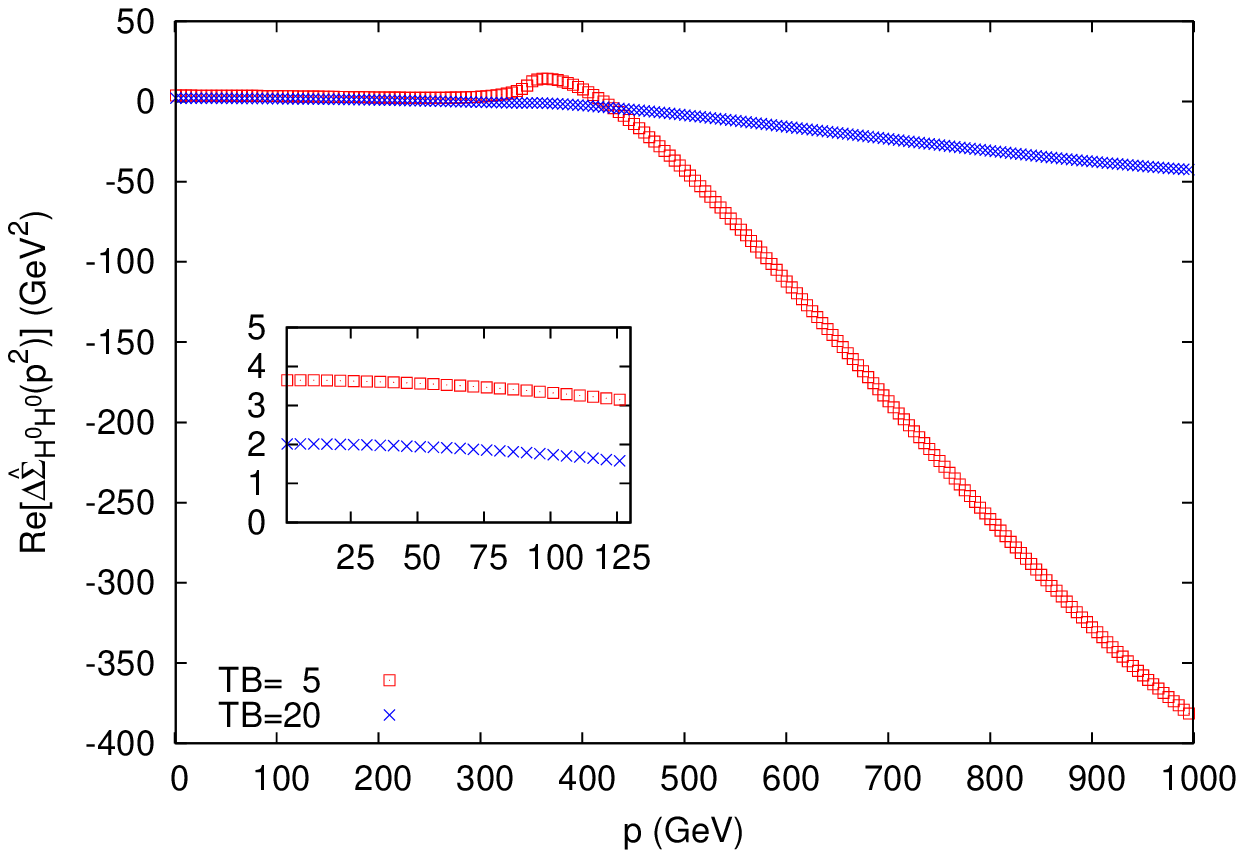}
\includegraphics[width=0.49\textwidth]{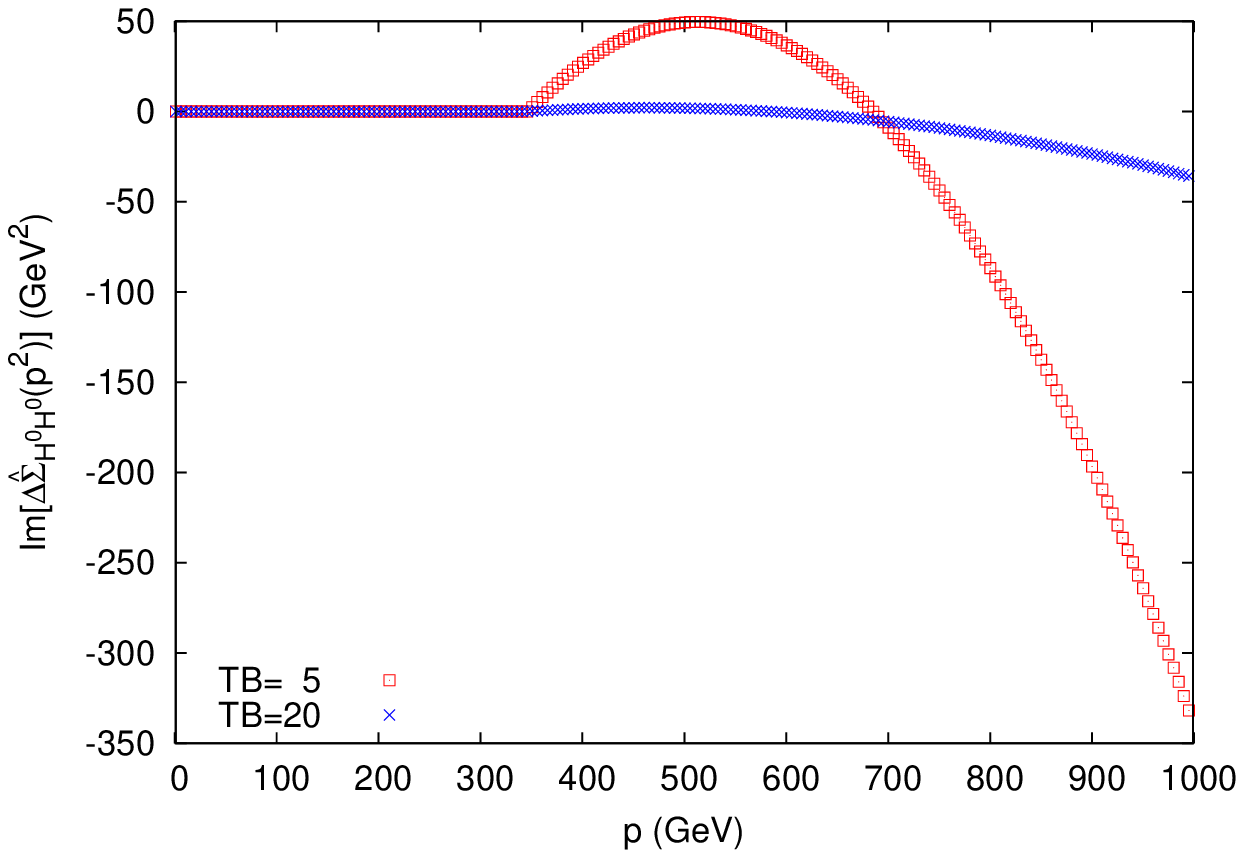}
\\[2em]
\caption{Momentum dependence of the real (left column) and 
imaginary (right column) parts of the two-loop selfenergies 
$\De\hat{\Sigma}_{h^0h^0},\De\hat{\Sigma}_{h^0H^0},\De\hat{\Sigma}_{H^0H^0}$, within
scenario 1,  
for $\tb=5$ (red squares) and $\tb = 20$ (blue crosses) and 
$\MA=250 \gev$. 
One can see that the selfenergies change 
substantially beyond the threshold at $p^2= (2\mt)^2$.}
\label{fig:se_scenario1reim}
\end{figure} 
%
The momentum-dependent 
parts of the renormalized two-loop self-energies are shown 
in the physical basis, \refeq{eq:DeltaSE}, 
for two different values of $\tb$, $\tan\beta=5$ and $\tan\beta=20$, at 
a fixed  $A^0$-boson mass $\MA=250$ GeV, see Fig.~\ref{fig:se_scenario1reim}.
The data points are not connected by a line in order to show
that each numerical point is obtained from a calculation
of the 34 analytically unknown integrals with the program \secdec. 
The inlays in Fig.~\ref{fig:se_scenario1reim} magnify the 
region $p^2\leq (125 \gev)^2$, 
where it can be observed that for $p^2\to 0$, 
the subtracted self-energies are not exactly 
zero.
As mentioned at the beginning of this section, this is due to the fact that the 
on-shell renormalization condition for the $A^0$-boson self-energy 
is defined differently with regard to the calculation without momentum dependence. 
The resulting constant contributions are additionally suppressed by factors 
$\SQb$, $\Sbe\Cb$ and $\CQb$ appearing in the counter-terms 
$\deVez$, $\deVzz$ and $\deVezz$, respectively, see 
Eqs.~(\ref{eq:2lctspotentialpart}). 

\medskip

The imaginary part is independent of
the $A^0$-boson mass, as this mass parameter 
solely appears in the counter-terms of 
\DRbar\ renormalized quantities and 
the $\dMAsqt$ counter-term, where
only the real part contributes. 
Therefore, the imaginary parts do not contain additional 
constant terms, compare Fig.~\ref{fig:se_scenario1reim}.
As to be expected, the imaginary parts are zero below the 
$t\bar{t}$ production threshold at $p=2\,m_t$, 
which results from the fact that the top mass is the 
smallest mass appearing in the loop diagrams. 
Beyond this threshold,
the imaginary parts are nonzero but of the same order of 
magnitude as the real parts. 
{}From these observations, the mass shifts in the region 
below the first threshold at $p=2\,m_t$ are expected not to be large.

\clearpage
\begin{figure}[htb!]
\centering
\includegraphics[width=0.6\textwidth]{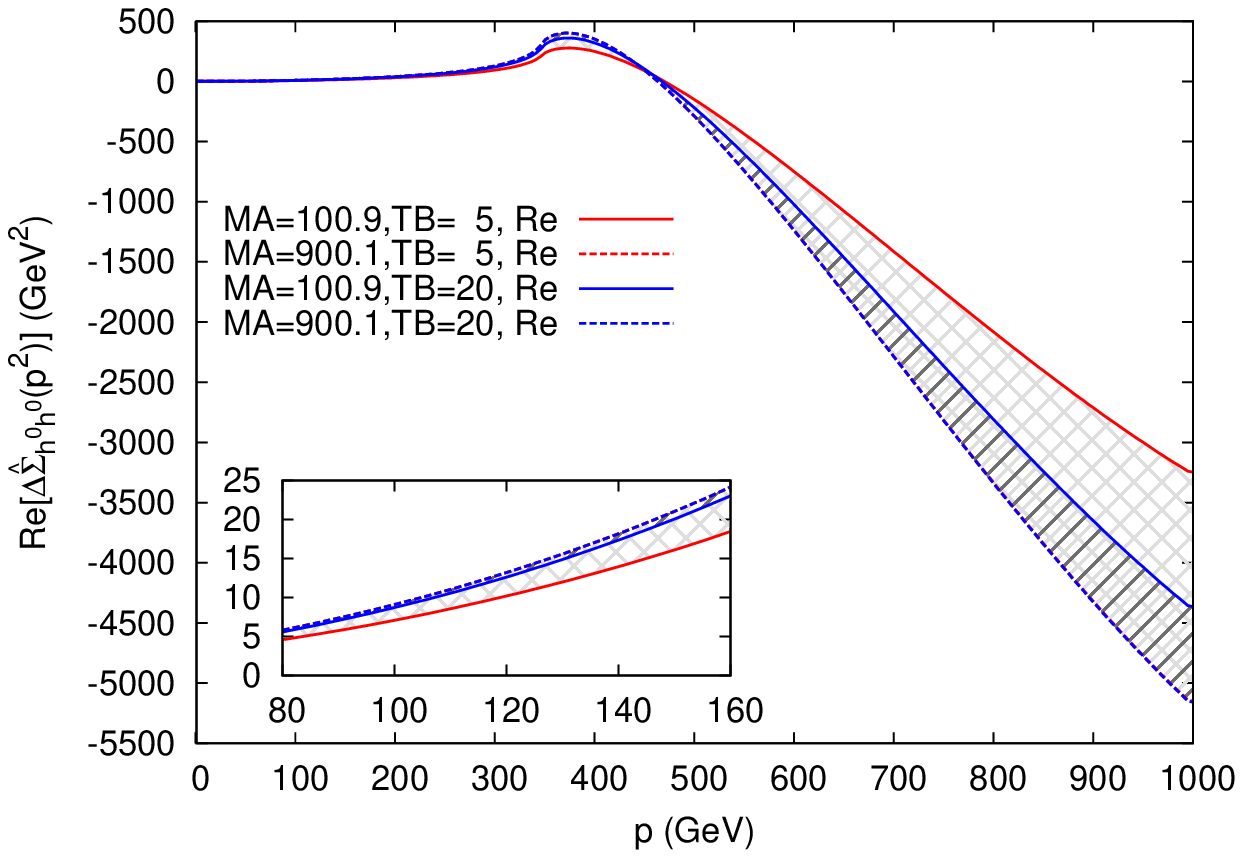}\\
\includegraphics[width=0.6\textwidth]{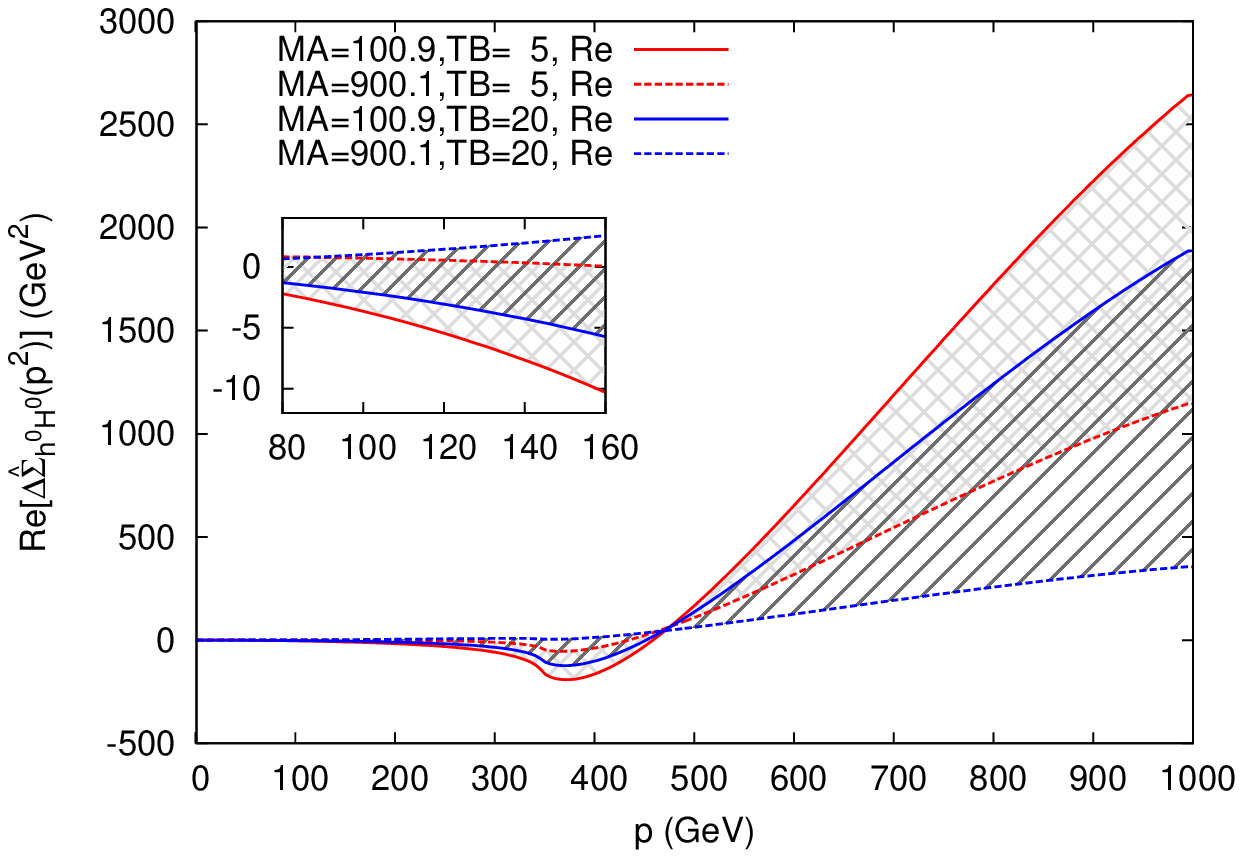}\\
\includegraphics[width=0.6\textwidth]{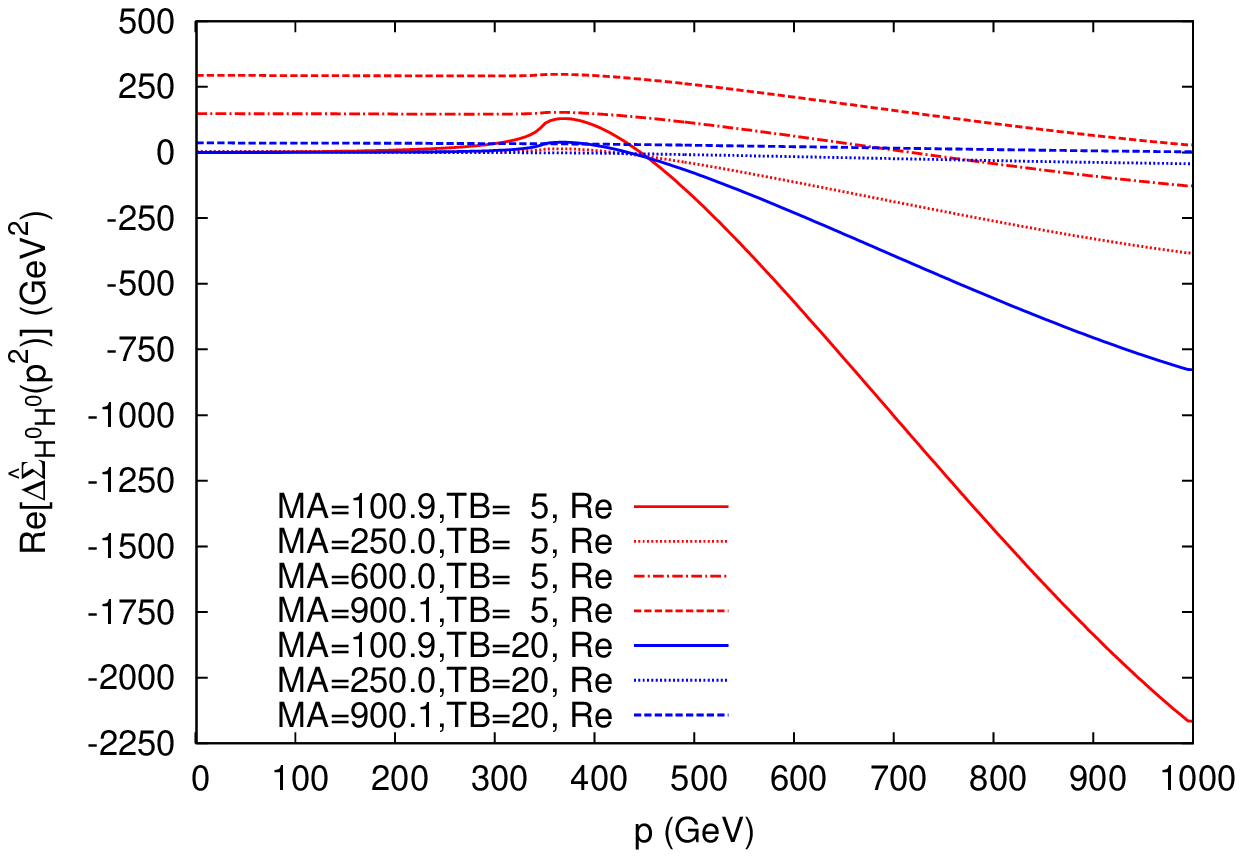}
\caption{Momentum dependence of the real part of the two-loop self-energies 
$\De\hat{\Sigma}_{h^0h^0}$, $\De\hat{\Sigma}_{h^0H^0}$, 
$\De\hat{\Sigma}_{H^0H^0}$, within scenario~1, 
for two different  values of $\tb$ and a range of $\MA$ values.
}
\label{fig:se_scenario1_madep}
\end{figure} 
%

\clearpage
Similar results, now including a variation of $\MA$ are shown 
in~\reffi{fig:se_scenario1_madep}. 
In the upper plot for $\De\ser{h^0h^0}$ and in the middle plot for $\De\ser{h^0H^0}$
the solid lines depict $\MA \sim 100 \gev$, 
while the dashed lines are for $\MA \sim 900 \gev$.
In these plots the light shading covers the range for $\tb = 5$, while
the dark shading for $\tb = 20$.
In the lower plot for $\De\ser{H^0H^0}$ results for 
$\MA \sim 100, 250, 600, 900 \gev$ are shown as solid, dotted, dot-dashed, dashed
lines, respectively (and shading has been omitted).
For $\De\ser{h^0h^0}$ at low values of the momentum~$p$ only a 
small variation with $\MA$ can
be observed. For $p$ and $\MA$ large, the 
contributions to the self-energy are bigger. In $\De\ser{h^0H^0}$
larger effects are observed at smaller $\MA$ for both, small and 
large~$p$ values. For $\De\ser{H^0H^0}$, on the other hand, at low~$p$ values,
large effects can be observed for large $\MA$ due to the aforementioned
counter-term contribution $\sim \dMAsqt = \re\se{A^0A^0}^{(2)}(\MA^2)$. At
large~$p$, as before, small $\MA$ values give a more sizable contribution.
%
%
%
\subsubsection{Scenario 2: Light stop scenario}
Scenario~2 is oriented at the ``light-stop scenario'' of
\citere{Carena:2013qia}%
\footnote{
While the original scenario in \citere{Carena:2013qia} is challenged by
recent scalar-top searches at ATLAS and CMS, a small modification in the
gaugino-mass parameters (which play no or only a very minor role here)
to $M_1 = 340 \gev$, $M_2 = \mu = 400 \gev$ 
leads to a SUSY spectrum that is very difficult to test at the LHC.
}. %
The following values are assigned to the MSSM parameters
\begin{align}
\mt &= 173.2\gev,\; \msusy= 0.5 \tev,\; \Xt =2\,\msusy\; , \non \\
\mgl &= 1600\gev,\; \mu = 200\gev~, 
\end{align}
leading to stop mass values of
\begin{align}
\mste &= 326.8 \gev,\;  \mstz= 673.2 \gev\, .\non
\end{align}

\begin{figure}[htb!]
\centering
\includegraphics[width=0.49\textwidth]{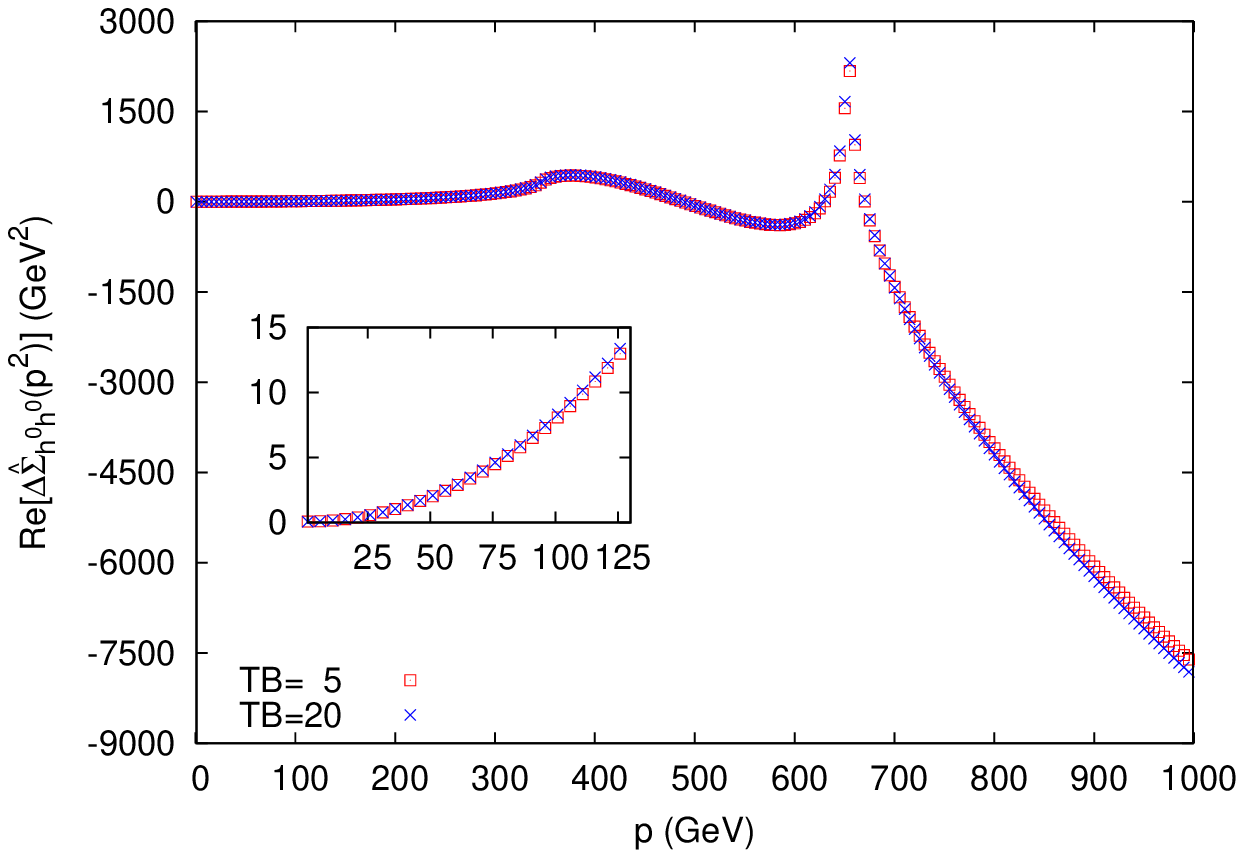}
\includegraphics[width=0.49\textwidth]{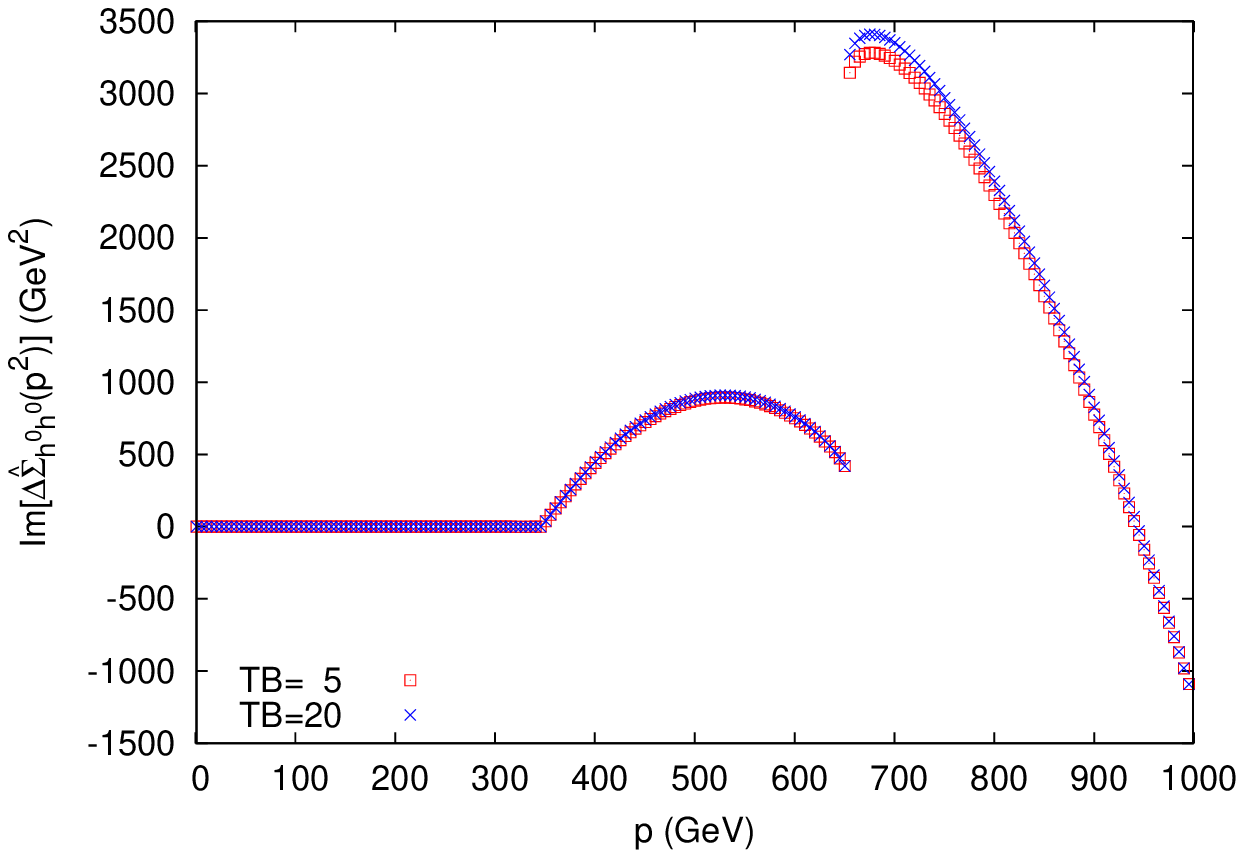}
\\[2em]
\includegraphics[width=0.49\textwidth]{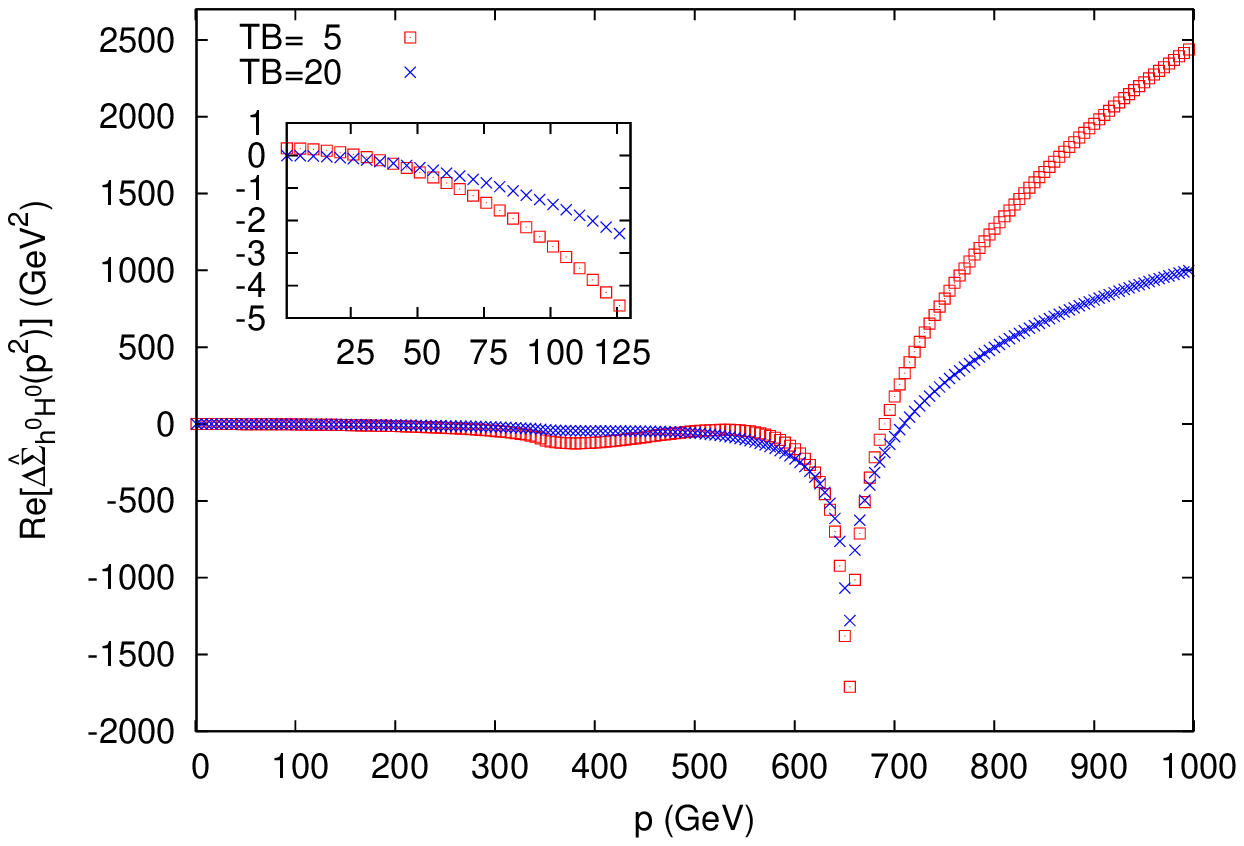}
\includegraphics[width=0.49\textwidth]{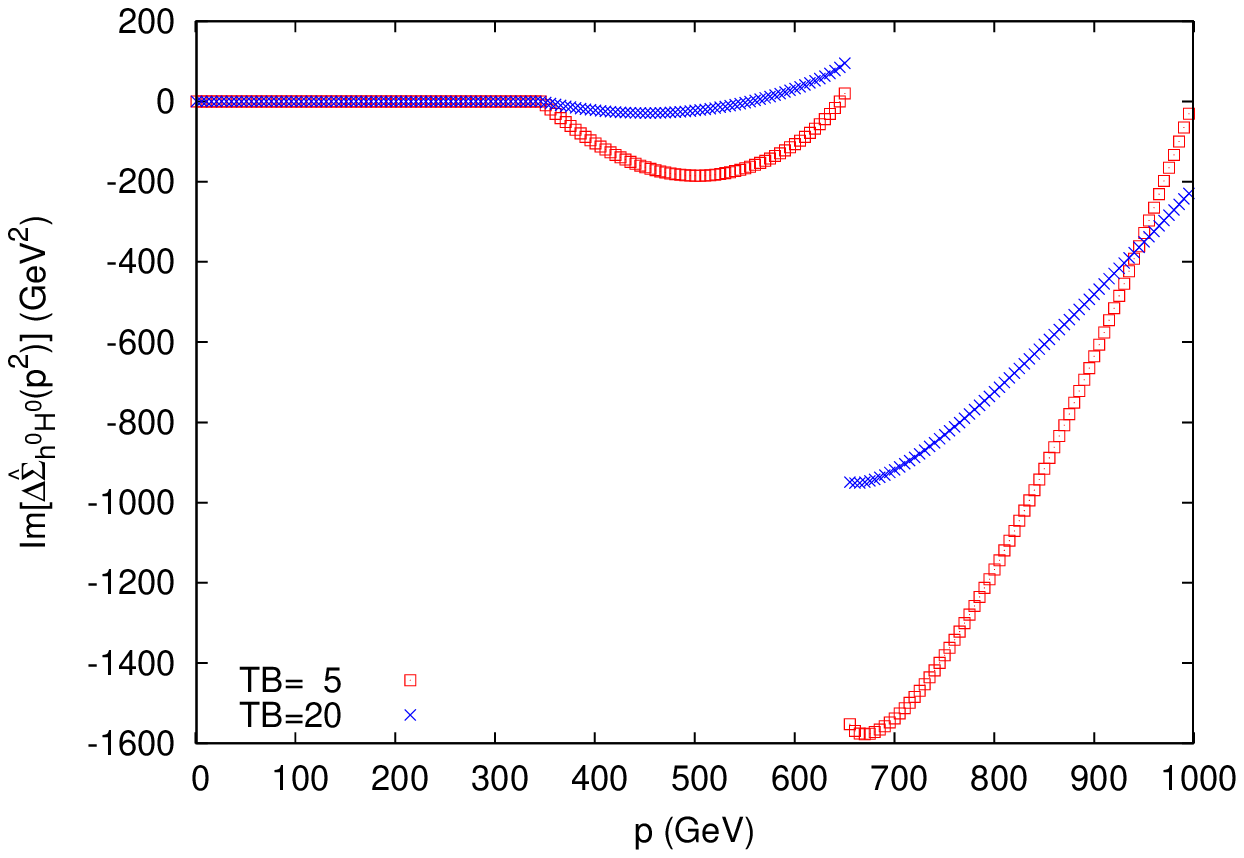}
\\[2em]
\includegraphics[width=0.49\textwidth]{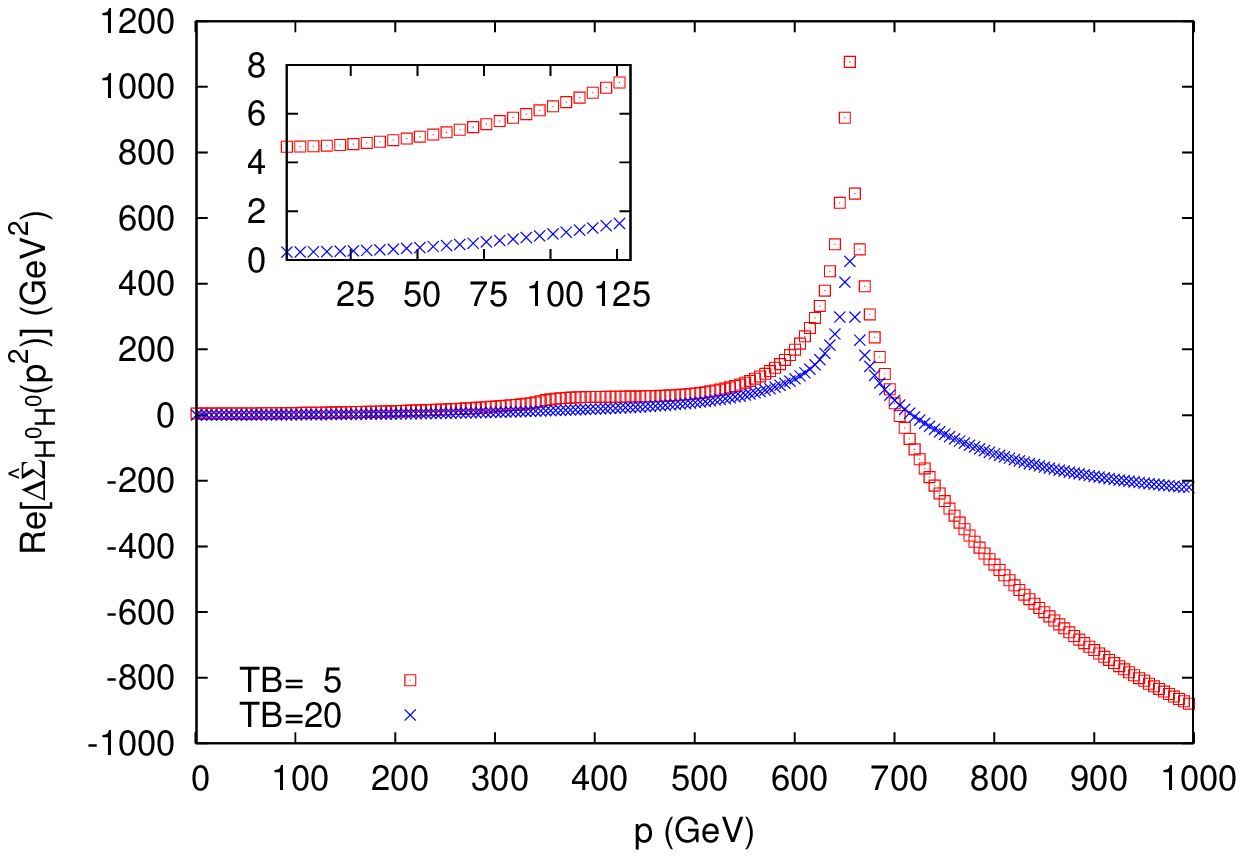}
\includegraphics[width=0.49\textwidth]{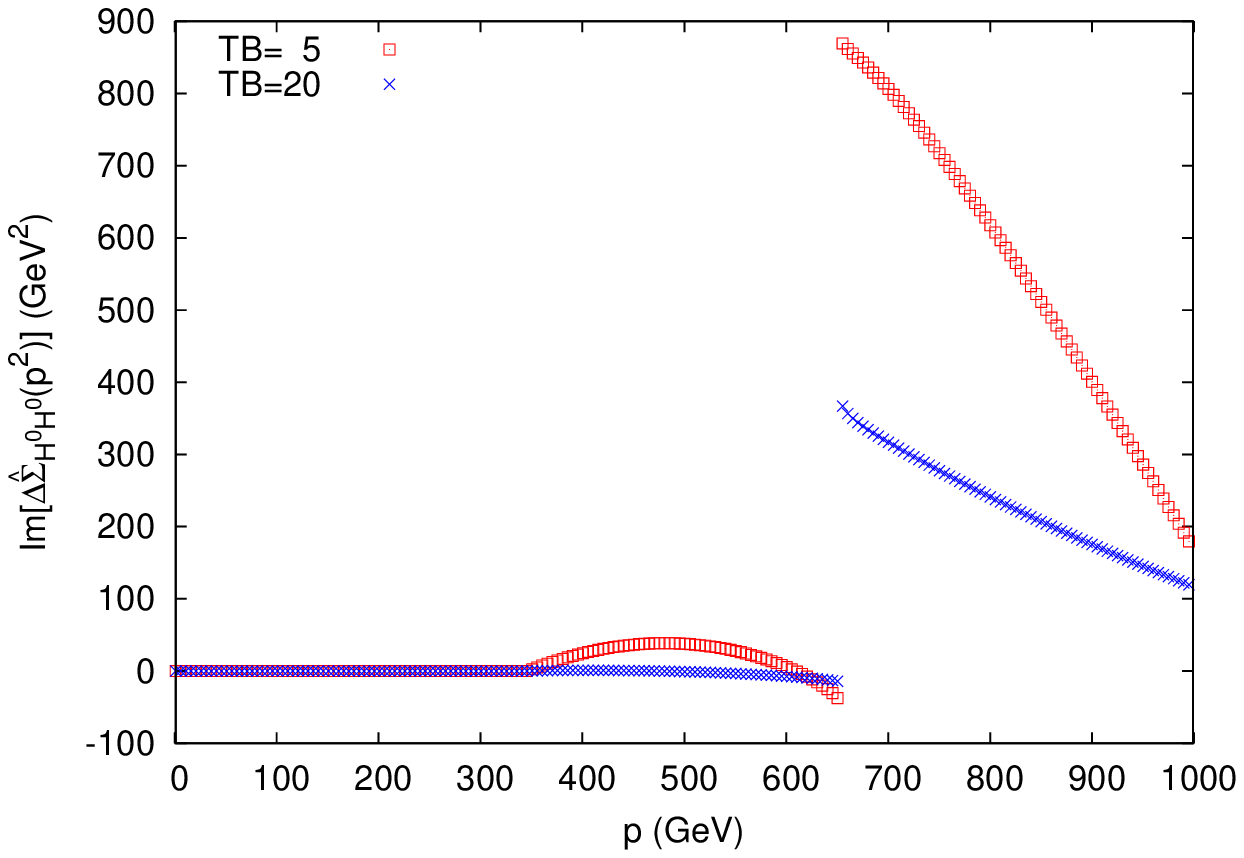}
\\[2em]
\caption{Momentum dependence of the real and imaginary parts of the
  two-loop self-energies 
$\De\hat{\Sigma}_{h^0h^0}$, $\De\hat{\Sigma}_{h^0H^0}$, $\De\hat{\Sigma}_{H^0H^0}$ within 
scenario 2, with $\tb=5,20$ and $\MA=250 \gev$ with the same color
coding as in \reffi{fig:se_scenario1reim}.}
\label{fig:se_scenario2reim}
\end{figure}

Scenario~2 is analyzed with the same set of plots shown for scenario~1.
The effects of the new momentum-dependent two-loop contributions on the
renormalized Higgs-boson self-energies, $\De\ser{ab}(p^2)$, are shown in
\reffi{fig:se_scenario2reim}. As before, separate results are shown 
for the real and imaginary parts of the self-energies. 
An additional threshold beyond the top-mass threshold appears at 
$p = 2\, \mste$. 
Analogously to scenario~1, the largest contributions in the 
region below $200\gev$ arise in the real part of 
$\De\hat{\Sigma}_{h^0h^0}$ amounting to about  $15\gev^2$ at $p=125\gev$, 
where the dependence on the value of $\tanb$ is rather weak.
The imaginary part equals the one of scenario~1 up to the 
$p = 2\, \mste$ threshold. The discontinuity at the latter enters 
through the integral involving the derivative of the $B_0$ function with 
respect to $\mste^2$, 
$\frac{\partial}{\partial \mste^2} B_0(p^2,\mste^2,\mste^2) $, see 
Sec.~\ref{subsec:analytintegrals} and Eq.~(\ref{eq:C0reducedtoderivB0}) 
therein. 

\begin{figure}[htb!]
\centering
\includegraphics[width=0.6\textwidth]{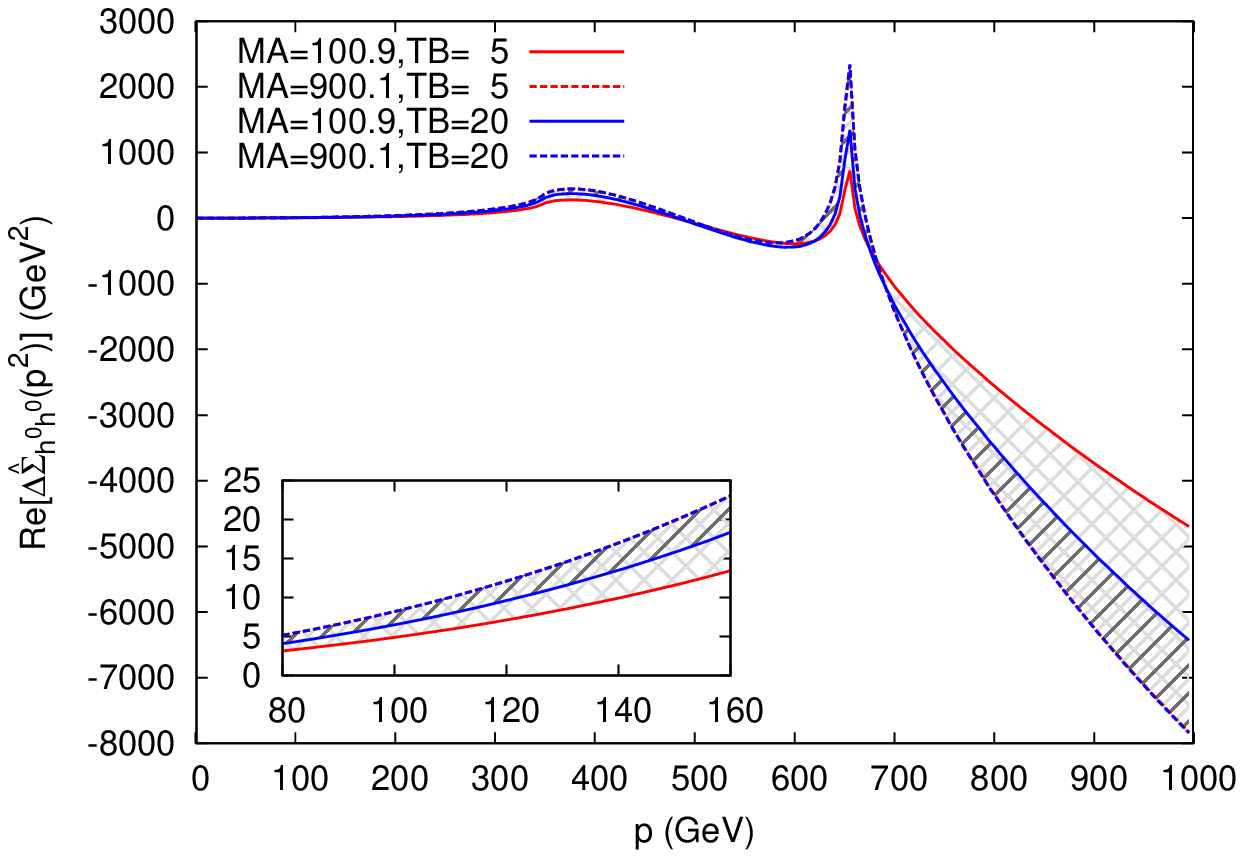}\\
\includegraphics[width=0.6\textwidth]{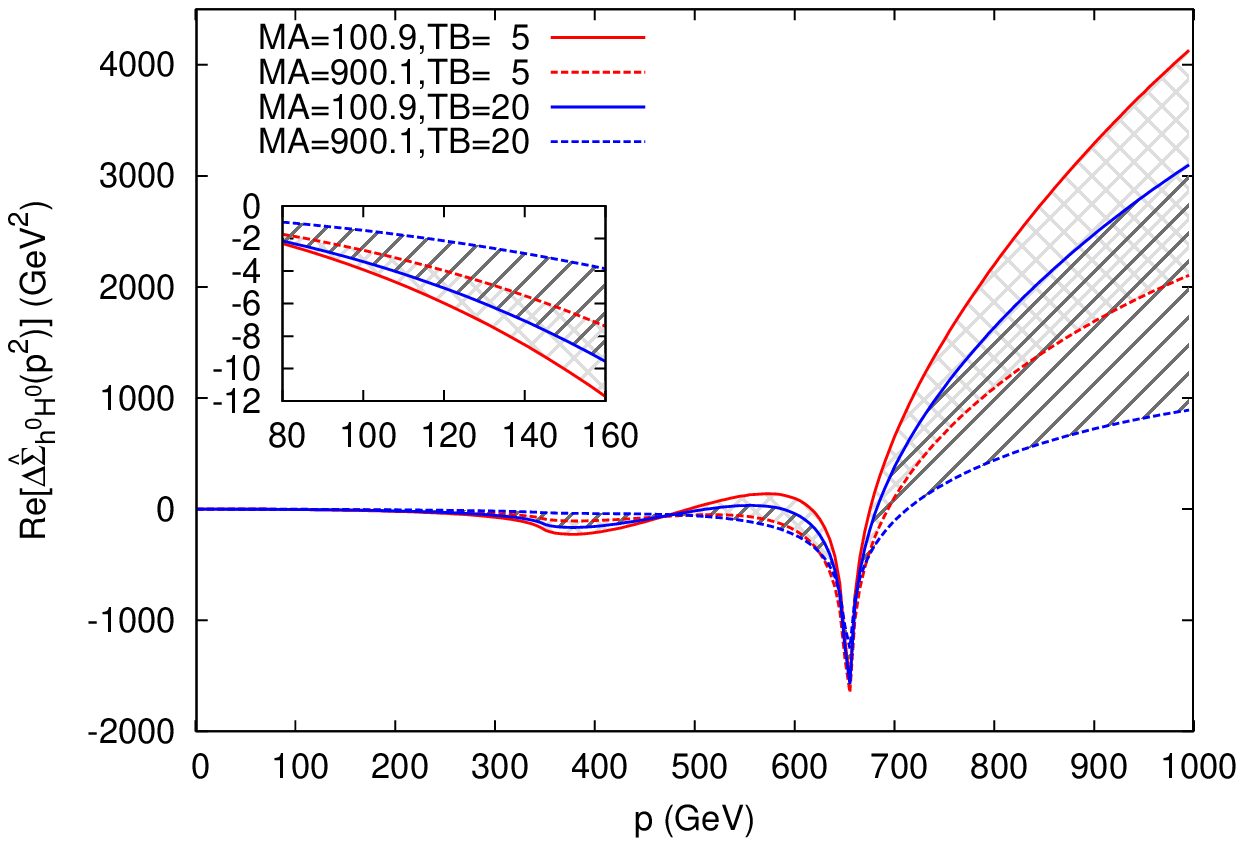}\\
\includegraphics[width=0.6\textwidth]{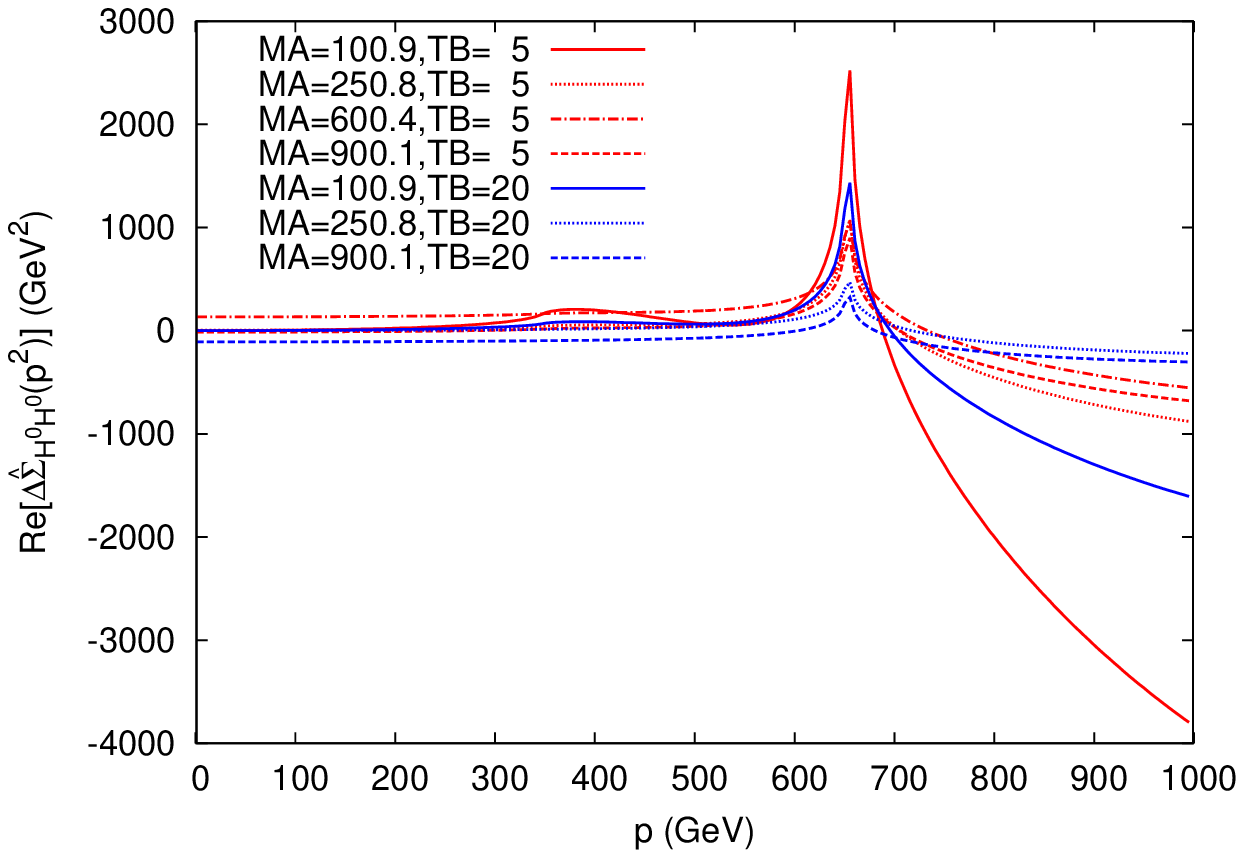}
\caption{Momentum dependence of the real parts of the two-loop self-energies 
$\De\hat{\Sigma}_{h^0h^0}$, $\De\hat{\Sigma}_{h^0H^0}$, $\De\hat{\Sigma}_{H^0H^0}$ in scenario~2 
for two different  values of $\tb$ and various values of $\MA$ (see text).}
\label{fig:se_scenario2_madep}
\end{figure}

%
The dependence of $\De\ser{ab}(p^2)$ on $\MA$ is shown in
\reffi{fig:se_scenario2_madep}, using the same line styles as in
\reffi{fig:se_scenario1_madep}. The curves show the same qualitative 
behavior as in \reffi{fig:se_scenario2reim}, exhibiting again  
the new threshold at $p = 2\, \mste$. 
In general, outside the threshold region
the effects in scenario~2 are slightly smaller than in scenario~1. 

\clearpage

%
%
\subsection{Neutral $\mathcal{CP}$-even Higgs-boson mass corrections}
The numerical evaluation to derive the physical masses for $h^0,H^0$ as the
poles (real parts) of the dressed propagators proceeds on the 
basis of~\refeq{eq:proppole} in an iterative way. 
\begin{itemize}
\item[$\bullet$]
In a first step, the squared masses $M_{h,0}^2, M_{H,0}^2$ are determined 
by solving \refeq{eq:proppole} excluding the new terms $\De\hat\Sigma_{ab}^{(2)}(p^2)$
from the self-energies. The masses $M_{h,0}, M_{H,0}$ are computed 
based on the higher-order contributions of 
Refs.~\cite{Heinemeyer:1998yj,Hahn:2009zz,
Heinemeyer:1998np,Degrassi:2002fi,Frank:2006yh,Hahn:2013ria}. 
\item[$\bullet$]
In a second step, the shifts $\De\hat\Sigma^{(2)}_{ab}(M_{h,0}^2) \equiv c_{ab}^h$ and
$\De\hat\Sigma^{(2)}_{ab}(M_{H,0}^2) \equiv c_{ab}^H$ are calculated and 
added as constants to the self-energies in \refeq{eq:proppole},
$\ser{ab}(p^2) \to \ser{ab}(p^2) + c_{ab}^{h(H)}$.
\item[$\bullet$]
In the third step, \refeq{eq:proppole} is solved again, now including the
constant shifts $c_{ab}^{h(H)}$ in the self-energies, to deliver the 
refined masses $\Mh$ (with $c_{ab}^h)$ and $\MH$ (with $c_{ab}^H$). 
\end{itemize}
This procedure can be repeated for improving the accuracy; numerically
it turns out that going beyond the first iteration yields 
only marginal changes. 
Below, results are shown for the mass shifts
\begin{align}
\De M_h = M_h - M_{h,0}, \quad \De M_H = M_H - M_{H,0} \text{ .}
\end{align} 
%
The mass shifts, in particular $\De\Mh$ for the light
$\cp$-even Higgs-boson, can directly be compared with the current
experimental uncertainty as well as with the anticipated future ILC
accuracy~\cite{Baer:2013cma} of 
\begin{align}
\de\Mh^{\rm exp, ILC} \lsim 0.05 \gev~.
\label{dMhILC}
\end{align}
The results are obtained for two different scenarios, 
varying parameters like $\tb,\MA,\mgl$, and illustrate the impact 
of these parameters via the new two-loop corrections on
the neutral $\cp$-even Higgs-boson masses, $\Mh$ and $\MH$. 
\subsubsection{Implementation in the program \fh}
\label{sec:fh}
The corrections of \refeq{eq:DeltaSE} are incorporated in \fh{}\footnote{The embedding 
of these corrections into the \fh{} code was performed with 
substantial support by Thomas Hahn.} by 
the following recipe, which is more general and in principle
applicable also to the case of the complex MSSM with $\cp$-violation.
\begin{enumerate}
\item Determine the Higgs-boson masses $M_{h_{i},0}$ 
without the momentum-dependent terms of \refeq{eq:DeltaSE}; 
the index $i=1,\dots,4$ enumerates 
the masses of $h, H, A, H^\pm$ in the real MSSM. 
This is done by invoking the \fh\ mass-finder.

\item Compute the shifts 
 $c_{ab}^{h_k} = \De\ser{ab}^{(2)} (M_{h_k,0}^2)$  with $a,b,h_k = {h,H}$.

\item Run \fh' mass-finder again including the $c_{ab}^{h_k}$
as constant shifts in the self-energies
to determine the refined Higgs masses $M_{h}$ and $M_{H}$.
\end{enumerate}
This procedure could conceivably be iterated until full self-consistency 
is reached; yet the resulting mass improvements turn out to be too small 
to justify extra CPU time.

\bigskip
On the technical side, an interface for an external program to 
\fh{} was added, which exports relevant model parameters to the external program's 
environment, currently
\begin{tabbing}
\texttt{FHscalefactor} \qquad\= ren. scale multiplicator, 
\qquad\qquad \= 	
\texttt{FHTB}	\qquad	\= $\tan\beta$, \\
         \texttt{FHAlfasMT}  \> $\alpha_s(\mt)$, \>
      \texttt{FHGF}	\qquad	\= $G_F$, \\
      \texttt{FHMHiggs2$i$} \qquad\> $M_{h_i,0}^2$, $i = 1\dots 4$
      , \>
     \texttt{FHMSt$i$}		\> $m_{\tilde t,i}$, $i = 1, 2$, \\
     \texttt{FH\{Re,Im\}USt1$i$} \> $U_{\tilde t,1i}$, $i = 1, 2$, \>
       \texttt{FHMGl}		\> $m_{\tilde g}$ , \\
     \texttt{FH\{Re,Im\}MUE}	\> $\mu$ , \>
     \texttt{FHMA0}		\> $M_A$, 
\end{tabbing}
where the $U_{\tilde t,1i}$ denote the elements of the stop mixing 
matrix, $\alpha_s(\mt)$ the running strong coupling 
at the scale $\mt$, and $G_F$  
the Fermi constant. The renormalization 
scale is defined within \fh\
as $\mu_r = m_t \cdot \texttt{FHscalefactor}$. 
Invocation of the external program is switched on by providing its path 
in the environment variable \texttt{FHEXTSE}.  The program is executed 
from inside a temporary directory which is afterwards removed.

\smallskip
The output (stdout) is scanned for lines of the form 
`\textit{se}@$m$ $c_r$ $c_i$'
which specify the correction $c_r + \mathrm{i} c_i$ [with
$c_r = {\rm Re}(c_{ab}^{h_k}), \; c_i = {\rm Im}(c_{ab}^{h_k})$]
to self-energy 
\textit{se} in the computation of mass $m$, where $m$ is one of 
\texttt{Mh0}, \texttt{MHH}, \texttt{MA0}, \texttt{MHp}, and \textit{se} 
is one of \texttt{h0h0}, \texttt{HHHH}, \texttt{A0A0}, \texttt{HmHp}, 
\texttt{h0HH}, \texttt{h0A0}, \texttt{HHA0}, \texttt{G0G0}, 
\texttt{h0G0}, \texttt{HHG0}, \texttt{A0G0}, \texttt{GmGp}, 
\texttt{HmGp}, \texttt{F1F1}, \texttt{F2F2}, \texttt{F1F2}.  The latter 
three, if given, substitute
\begin{subequations}
\begin{align}
\mathtt{HHHH} &= \cos^2\alpha\,\mathtt{F1F1} +
                 \sin^2\alpha\,\mathtt{F2F2} +
                 \sin 2\alpha\,\mathtt{F1F2}\,, \\
\mathtt{h0h0} &= \sin^2\alpha\,\mathtt{F1F1} + 
                 \cos^2\alpha\,\mathtt{F2F2} -
                 \sin 2\alpha\,\mathtt{F1F2}\,, \\
\mathtt{h0HH} &= \cos 2\alpha\,\mathtt{F1F2} +
                 \tfrac 12\sin 2\alpha\,(\mathtt{F2F2} - \mathtt{F1F1})\,,
\end{align}
\end{subequations}
in accordance with Eq.~(\ref{eq:transformationphi12tohH}).
Self-energies not given are assumed zero.

\smallskip
The zero-momentum contributions $\tilde\Sigma_{ab}^{(2)}(0)$ with $ab = 
\{H^0H^0,h^0H^0,h^0h^0\}$, defined in Eq.~(\ref{eq:DeltaSE}) are 
subtracted if the output of the external program contains one or more of 
`\texttt{sub asat}', `\texttt{sub atat}', `\texttt{sub asab}', 
`\texttt{sub atab}' for the $\alpha_s\alpha_t$, $\alpha_t^2$, 
$\alpha_s\alpha_b$, and $\alpha_t\alpha_b$ contributions, respectively.
All other lines in the output are ignored.

%
%
%
%
\subsubsection{Scenario 1: $m_h^{\text{max}}$ scenario}
\begin{figure}[ht!]
\centering
\includegraphics[width=0.7\textwidth]{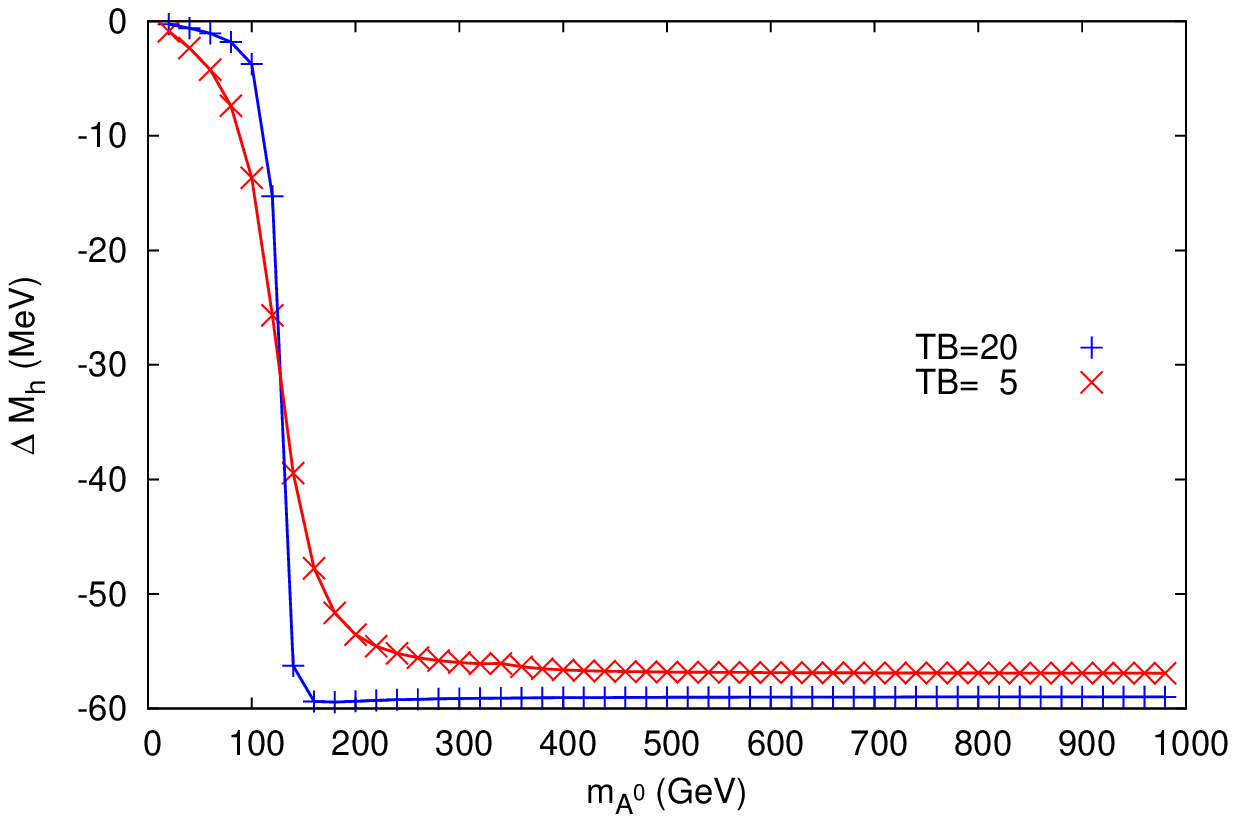}\\
\includegraphics[width=0.7\textwidth]{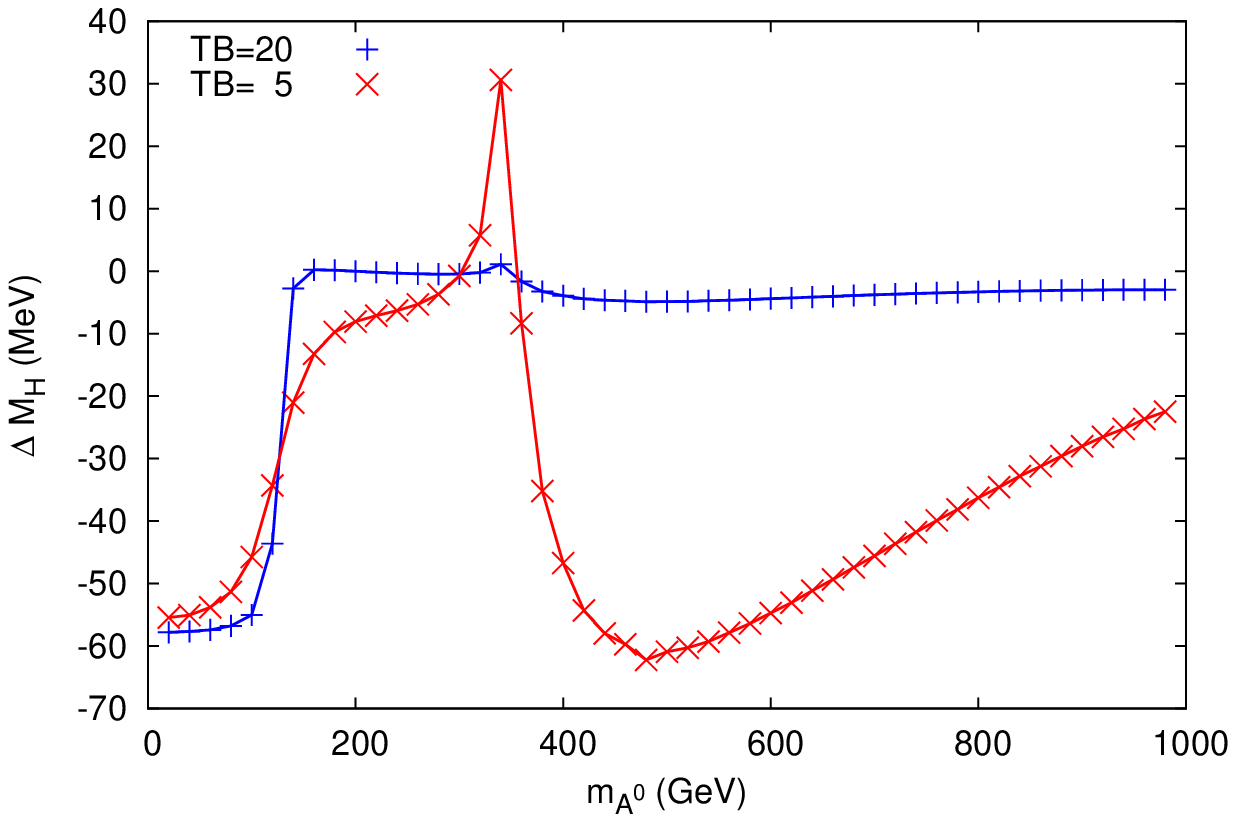}
\caption{Variation of the mass shifts $\Delta\Mh,\Delta\MH$ with the
  $A^0$-boson mass $\MA$ within scenario~1, 
for $\tb=5$ (blue) and $\tb = 20$ (red). The small peak in $\Delta\MH$
originates from a threshold at $2\,\mt$.} 
\label{fig:shiftswithma}
\end{figure}

The effects of the newly computed momentum-dependent two-loop 
corrections on the Higgs-boson masses $M_{h,H}$ via the mass shifts
$\De\Mh$ and $\De\MH$ are now studied. 
In \reffi{fig:shiftswithma}, $\De\Mh$ (upper plot) and $\De\MH$ (lower plot) are 
shown as a function of $\MA$ for $\tb = 5$ (blue) 
and $\tb = 20$ (red). 
For $\MA \gsim 200 \gev$, the additional shifts from momentum-dependence 
of up to $\Delta\Mh \sim - 60 \mev$
are of the size of the expected future experimental
precision, see \refeq{dMhILC}. 
The contribution to the heavy $\cp$-even Higgs-boson mass 
is of similar order of magnitude. Around the threshold at $\MA=2\mt$, 
the heavy Higgs-boson mass is shifted upwards. 

\medskip

\begin{figure}[htb!]
\centering
\includegraphics[width=0.7\textwidth]{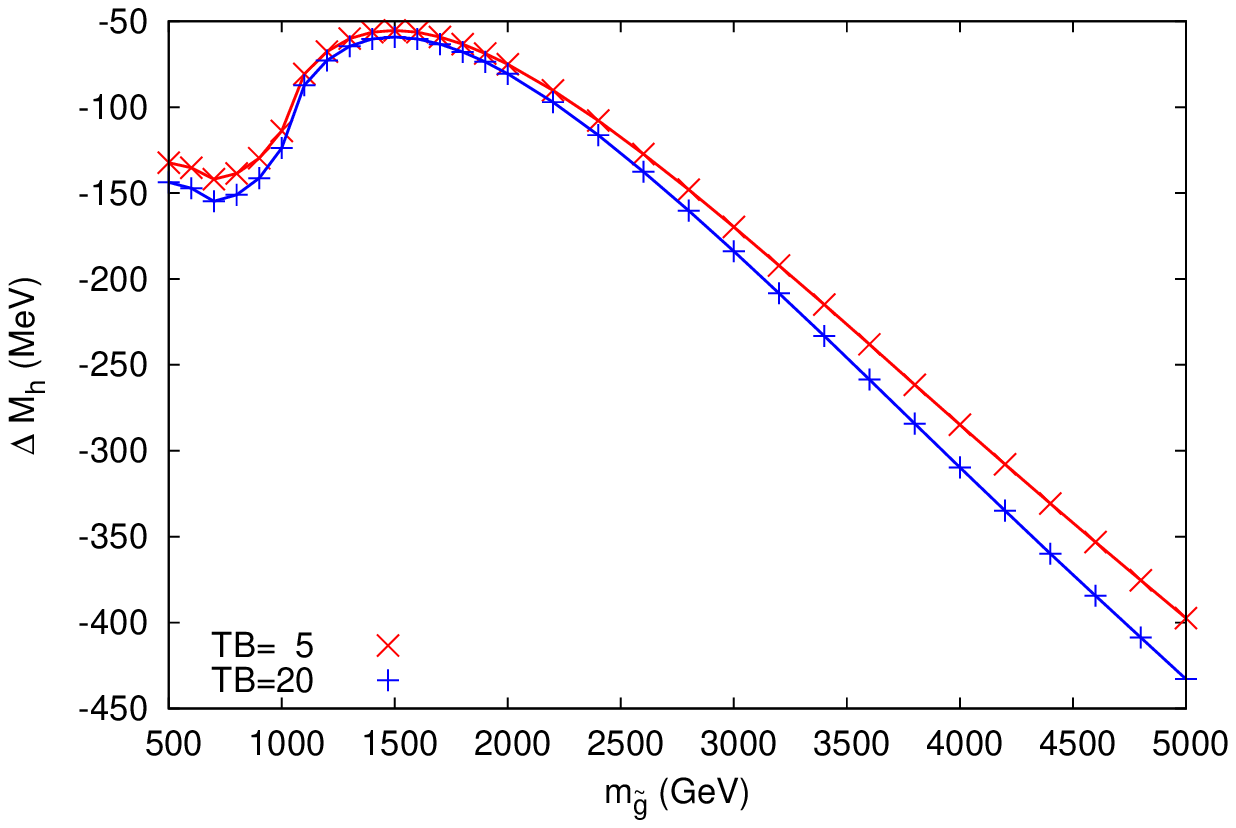}
\includegraphics[width=0.7\textwidth]{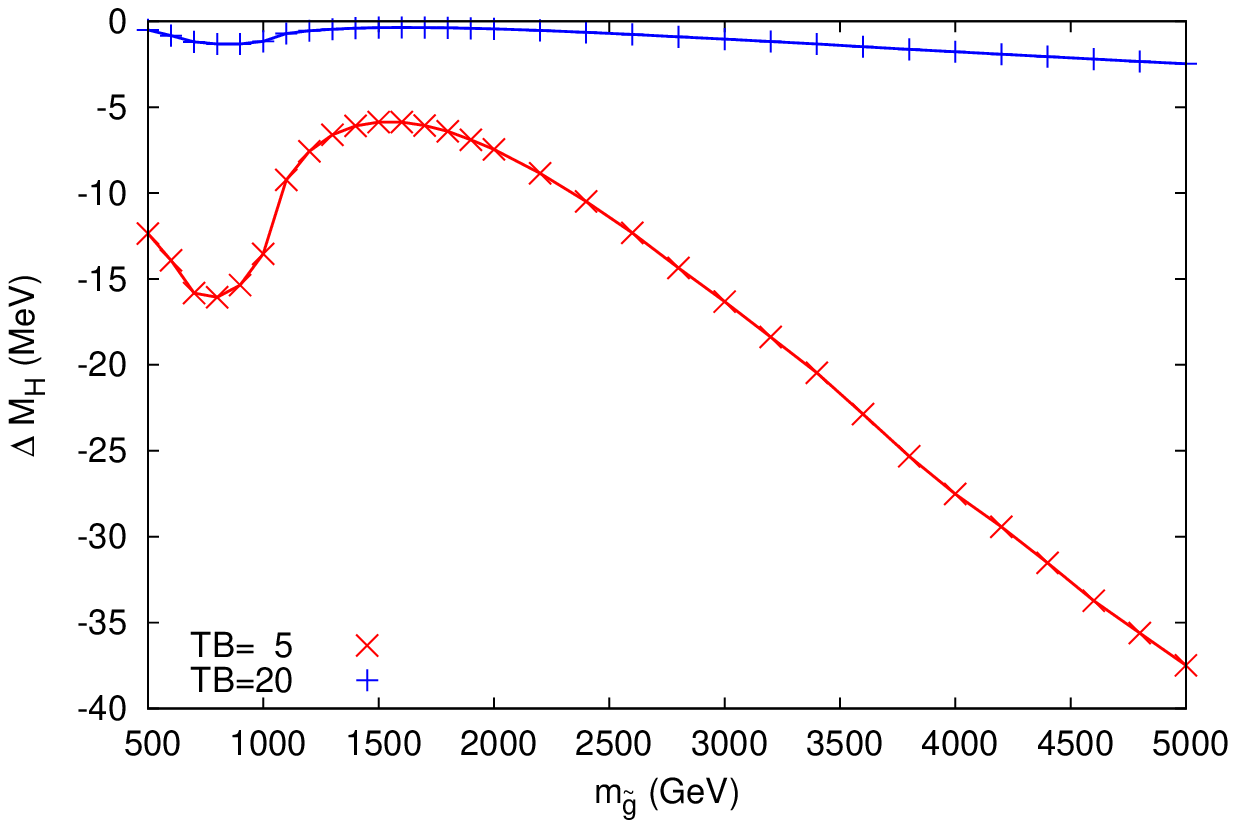}
\caption{Variation of the mass shifts $\Delta\Mh,\Delta\MH$ with the
  gluino mass, within scenario 1, for two different  values of
  $\tb=5,20$ and $\MA = 250 \gev$.
}
\label{fig:variationmgluino}
\end{figure}

Finally, the dependence of $\Mh$ and
$\MH$ on the gluino mass $\mgl$ is examined. 
The results are shown in~\reffi{fig:variationmgluino} for $\De\Mh$ (upper plot) and $\De\MH$
(lower plot) for $\MA = 250 \gev$,
with the same color coding as in \reffi{fig:shiftswithma}.
In the upper
plot one can observe that the effects are particularly small for the
default value of $\mgl$ in scenario~1.
More sizeable shifts occur for larger gluino masses,   
by more than $-400 \mev$ for $\mgl \gsim 4 \tev$, 
reaching thus the level of
the current experimental accuracy in the Higgs-boson mass 
determination.
The corrections to $\MH$, for the given value of $\MA = 250 \gev$
do not exceed $-50 \mev$ in the considered $\mgl$ range.

%
%
\subsubsection{Scenario 2: Light stop scenario} 
\begin{figure}[htb!]
\centering
\includegraphics[width=0.7\textwidth]{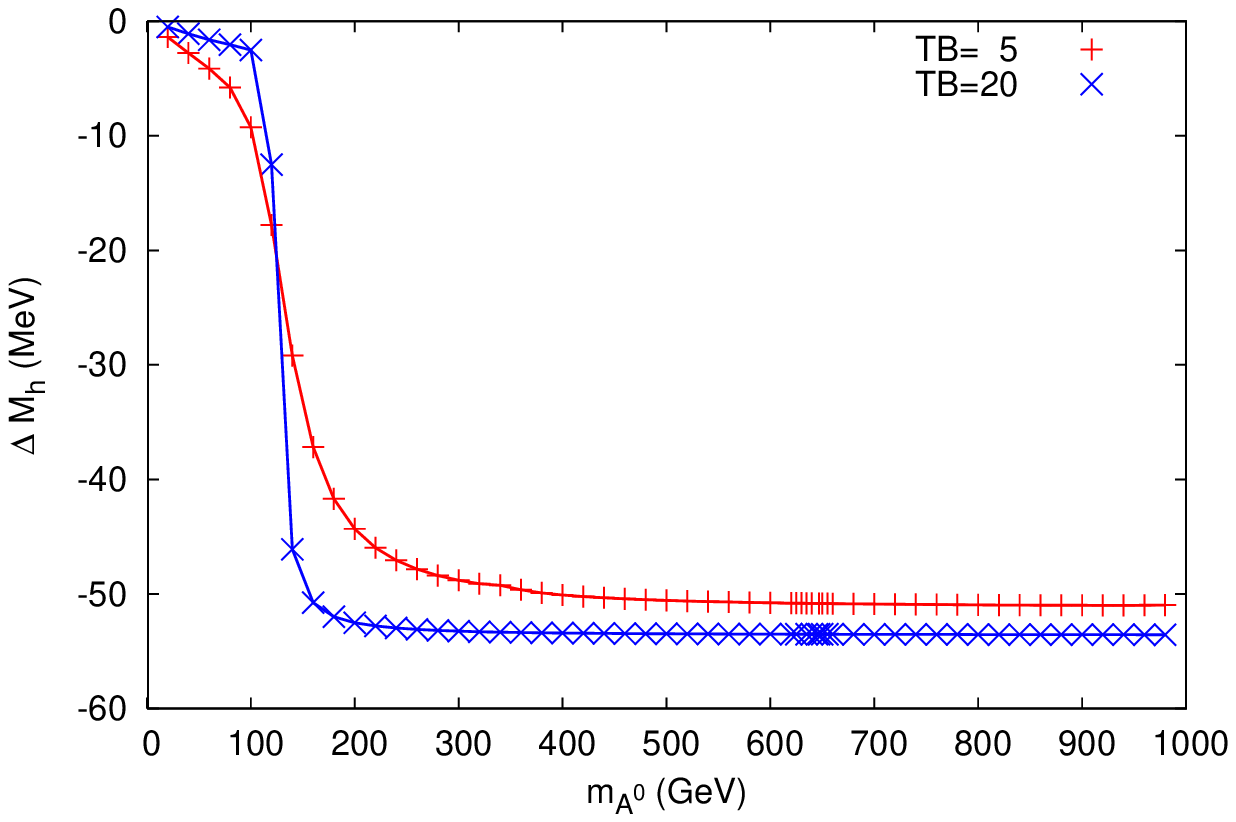}\\
\includegraphics[width=0.7\textwidth]{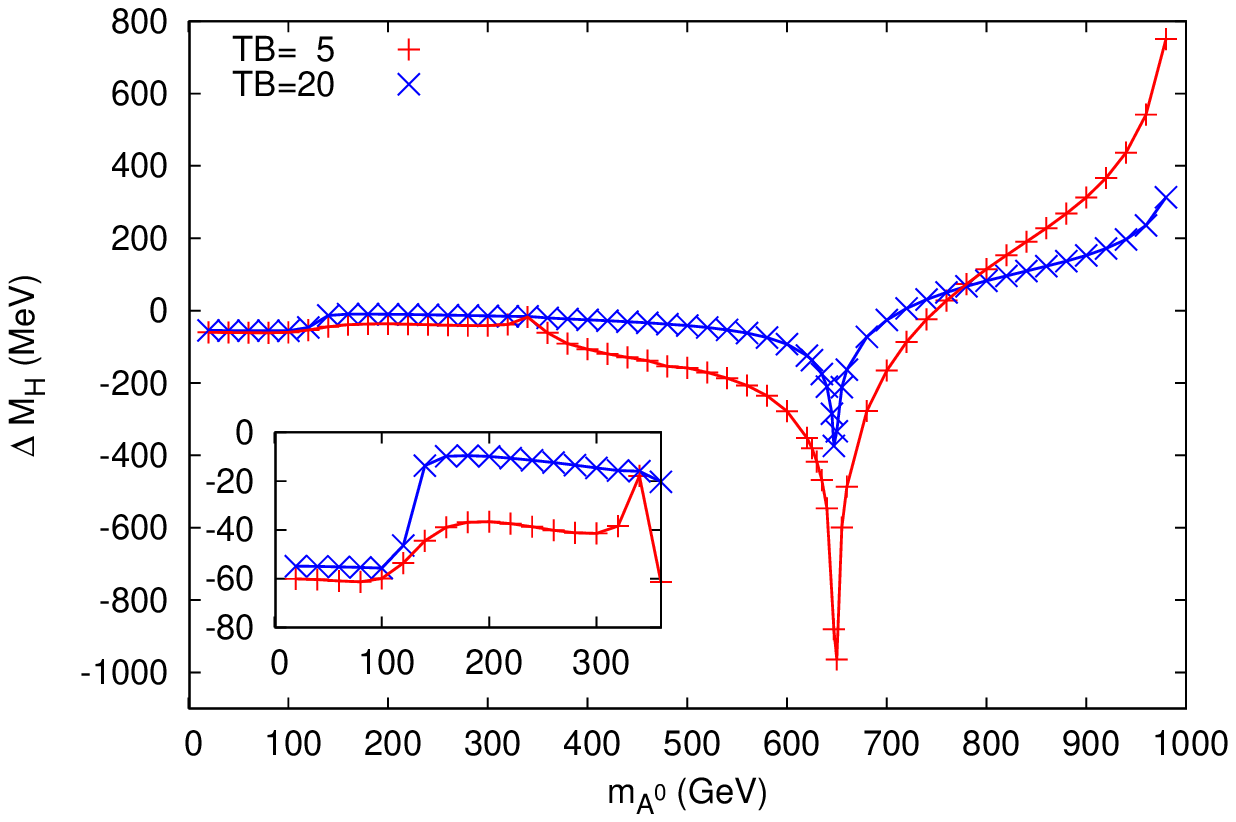}
\caption{Variation of the mass shifts $\Delta\Mh,\Delta\MH$ with  the $A^0$-boson mass $\MA$ within scenario~2,
for two different  values of $\tb=5,20$. 
}
\label{fig:scen2shiftswithma}
\end{figure}

The effects on the physical neutral $\cp$-even Higgs-boson masses are now 
analyzed within scenario~2. 
The results are shown for $\De\Mh$ (upper plot) and 
$\De\MH$ (lower plot) as a function of $\MA$
(with the same line styles as in \reffi{fig:shiftswithma}), 
see \reffi{fig:scen2shiftswithma}. As can
be expected from the previous figures, below the $p = 2\,\mt$ threshold, 
the effects on $\Mh$ and $\MH$ 
are in general slightly smaller in scenario~2 than in scenario~1, where
$\De\mh$ still reaches the anticipated ILC accuracy, see
\refeq{dMhILC}. Above the $p = 2\,\mt$ threshold, the variation of 
$\De\MH$ is larger than in scenario~1 for $\tb=5$. 
Around the threshold $p = 2\,\mste$ a shift in the 
GeV range towards lower masses is found, reaching the level of about~0.2\%. 
Beyond this threshold, the heavy Higgs-boson mass 
can be shifted towards higher masses. The latter shift is 
also roughly in the GeV range.

\medskip

\begin{figure}[htb!]
\centering
\includegraphics[width=0.7\textwidth]{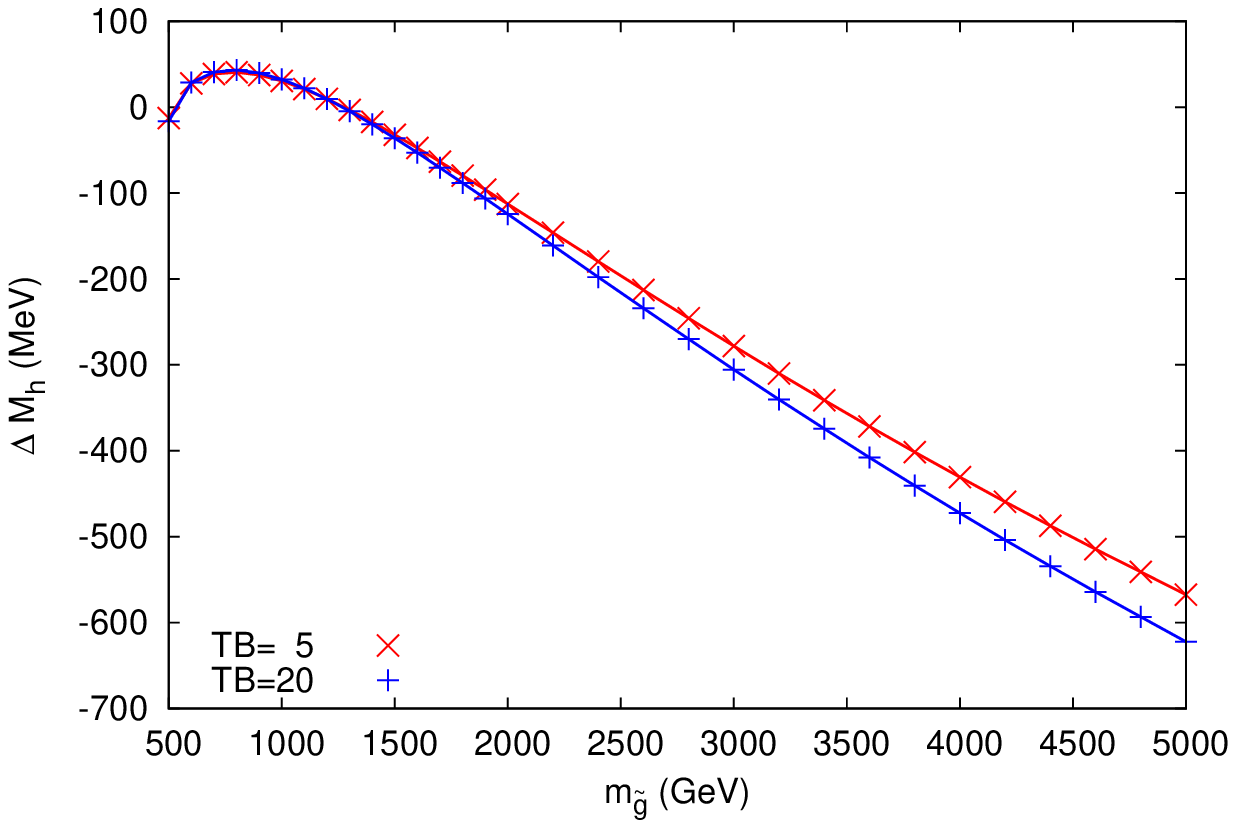}
\includegraphics[width=0.7\textwidth]{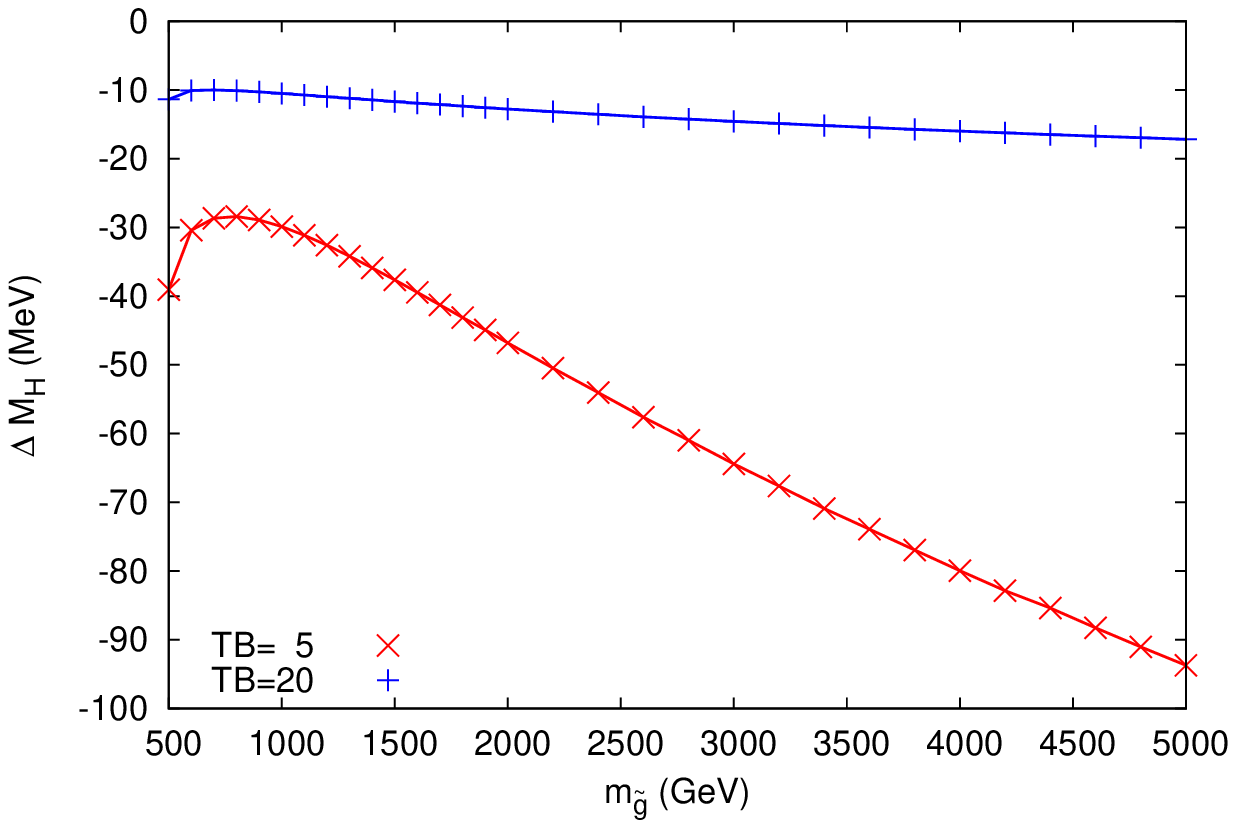}
\caption{Variation of the mass shifts $\Delta\Mh,\Delta\MH$ with the
  gluino mass, within scenario 2, for two different  values of
  $\tb=5,20$ and $\MA = 250 \gev$.
}
\label{fig:scen2variationmgluino}
\end{figure}

Finally the dependence on $\mgl$ is examined, see 
\reffi{fig:scen2variationmgluino}. In the upper plot $\De\Mh$ is shown 
for $\tb = 5$ and $\tb = 20$, where both values yield very similar
results. As in scenario~1, ``accidentally'' small values of $\De\Mh$ are
found around $\mgl \sim 1600 \gev$. For larger gluino mass values the
shifts induced by the new momentum-dependent two-loop corrections exceed
$- 500 \mev$ and are thus larger than the current experimental
uncertainty. The results for $\De\MH$ are shown in the lower plot. While
they are roughly twice as large as in scenario~1, they do not exceed
$-100 \mev$.
%
%
%

\clearpage

\section{Summary and Perspectives}
The calculation and respective results 
for the leading momentum-dependent
\order{\als\alt} contributions to the masses of neutral 
$\cp$-even Higgs-bosons in the MSSM were 
presented. 
They are obtained by calculating
the corresponding contributions to the dressed Higgs-boson propagators 
in the Feynman-diagrammatic approach. 

The required two-loop self-energy diagrams 
and one-loop counter-terms 
with counter-term insertions are generated using {\sc FeynArts}, 
and reduced to a
set of basic integrals with the help of  {\sc TwoCalc} and {\sc FormCalc}.

The mass and field renormalization is performed 
adopting a mixed
on-shell/\DRbar\ renormalization scheme. 
More precisely, the field renormalization part is renormalized 
in the \DRbar\ scheme, the on-shell scheme is used for the remaining 
mass renormalization. This scheme choice has been 
beneficial in other contexts~\cite{Frank:2002qf,Heinemeyer:2004xw,
Hollik:2003jj,Heinemeyer:2010mm,Fritzsche:2011nr} and proved 
constructive in this calculation as well.  
Furthermore, the on-shell renormalization of the two-loop $A^0$-boson mass 
counter-term lead to an additional physical shift in the self-energies which 
has not been found in 
the calculation at zero momentum. 
The scheme choice is in contrast to 
Ref.~\cite{Martin:2003it,Martin:2004kr}, where 
the same contribution was computed using a full \DRbar\ scheme.  

The renormalized momentum-dependent two-loop 
Higgs-boson self-energies 
contain analytically inaccessible two-loop integrals. 
These are computed numerically using the program \secdec.
The new momentum-dependent contributions are 
incorporated in the public program \fh, including an 
interface to \secdec. 
The analysis of the mass shifts is performed with the 
upgraded version of \fh. 

The numerical analysis showed that the effects on the light
$\cp$-even Higgs boson mass, $\Mh$, depend strongly 
on the value of the gluino mass, $\mgl$.
For values of $\mgl \sim 1.5 \tev$ corrections 
of about $ -50 \mev$ are found, at the level of the anticipated
future ILC accuracy. If the gluino mass is assumed 
very large, $\mgl \gtrsim 4 \tev$, the corrections are 
substantially larger and at the level of the current experimental 
accuracy~\cite{Gennai:2007ys}. 
The shifts in the heavy $\cp$-even 
Higgs-boson mass, resulting from the incorporation of the 
momentum dependence are mostly
below current and future anticipated accuracies. 
Only close to thresholds, e.g.\ around $p = 2\,\mste$, the 
corrections are larger but do not exceed 0.2\%. 

\medskip

The evaluation times for the computation of the 
${\cp}$-even Higgs-boson masses including the new 
momentum-dependent two-loop contributions
strongly depend on the performance 
of \secdec{}. They range 
between one minute for a point far, 
and maximally one hour for a point very close 
to a threshold. 
Being able to compute multi-loop 
multi-scale integrals, 
the program \secdec{} reaches far beyond 
the applicability to two-loop 
two-point functions which are required for the 
self-energy corrections of the MSSM Higgs-boson 
masses. Yet, it is due to its universal applicability that 
a program specifically tailored to the desired 
two-loop two-point functions may perform better in terms 
of evaluation times. 
One such alternative could be a package provided 
by Bauberger~\cite{Bauberger:1994zz}, which contains 
the desired loop integrals in terms of one-parameter 
integral representations. 

\medskip

With the computation of the renormalized $\cp$-odd 
Higgs-boson self-energy at hand, it could be 
interesting to extend the analysis of the mass shifts 
resulting from momentum dependence to the 
charged Higgs-bosons. Furthermore, it could be 
interesting to compute the 
effective couplings to gauge bosons.

As a further improvement on the precision of the 
prediction of the Higgs-boson masses, the 
contributions involving couplings to the electroweak 
gauge bosons can be included.  
Moreover, the computation of the 
leading QCD corrections at the three-loop order 
to be incorporated into \fh 
can be considered. 

%% file: conclusion/conclusion.tex
\chapter{Conclusions}
\label{chap:conclusion}%
The upgrade of the program \secdec{} towards 
the automated computation of multi-loop multi-scale Feynman 
graphs 
in the physical region including thresholds 
was presented, thereby lifting the restriction  
to the Euclidean region. 

The program allows for an automated algebraic factorization of 
dimensionally regulated singularities and a numerical 
evaluation of the resulting pole coefficients. 
These can be multi-loop Feynman integrals with up to 
several mass scales and with in principle no limitation on the tensor rank, or 
more general parametric functions 
with singularities only at the endpoints of the integration 
region. 
Additionally, the program was enhanced by 
allowing for the automated evaluation of user-defined 
functions in the physical region. 

The extension to physical kinematics was achieved 
by the implementation of an automated analytical 
continuation of the integrand. An algorithm to find the 
optimal according deformation of the integration contour 
was further developed, allowing for a stable evaluation 
of integrals over large regions of values for the kinematic 
invariants. 

\medskip

Provided with the upgraded version of the program, a plethora of 
new applications are feasible. 
Two applications were shown in this thesis. 

\medskip

In the first application, massive 
planar and non-planar six- and seven-propagator integrals 
with four external legs were computed. 
It was shown that the method is in principle 
independent of the number of involved scales by 
providing numerical results for a planar four-point 
seven-propagator diagram with internal and external 
lines all massive. 

Furthermore, and in contrast to analytical 
evaluation techniques, adding 
massive lines to a topology proved 
beneficial in terms of evaluation times. 
This was shown for two of the most complicated massive 
non-planar 
seven-propagator double box integrals, entering in the 
NNLO prediction for top-quark pair production. 
Differing in the massive sub-loop topology, the 
singularity structure of the diagram involved in the 
heavy fermionic corrections is significantly simpler 
than the diagram entering the light fermionic 
corrections. While the custom \secdec{} setup 
can be used in the computation of the former, 
the latter diagram challenges the automated 
setup of the program. 
Therefore, an analytical preparation of this diagram prior 
to its treatment in \secdec{} was explored, leading to an 
improved numerical behavior.  

A systematic improvement of the numerical 
convergence can be achieved through a 
reduction in the number of integration parameters involved and 
the elimination of spurious divergences. 
The analytical preparation of the massive non-planar box 
diagram entering the light fermionic corrections 
follows these aims, reducing the number 
of involved parameters by 
integration of one Feynman parameter 
in a sub-loop of the integral. 
%
Hereby, 
singularities are mapped to both ends of the integration 
region and spurious linear singularities appear in pairs. 
While a remapping cures the 
occurrence of singularities at both endpoints 
of the integration region, a newly introduced backwards 
transformation serves in distributing the linear 
divergences more evenly among the Feynman 
parameters, thereby achieving a total reduction by two thirds 
in the number of functions to be integrated 
numerically.

\medskip

In its second application, the new features of the program are 
demonstrated in the computation of the  
the leading momentum-dependent two-loop QCD corrections 
to the masses of neutral 
$\cp$-even Higgs-bosons in the MSSM. 
These are obtained by calculating
the corresponding contributions to the dressed Higgs-boson propagators 
using the Feynman-diagrammatic approach and adopting a mixed
on-shell/\DRbar\ renormalization scheme. 

A revised two-loop mass and field 
renormalization has to be carried 
out for the mass of the neutral Higgs-bosons to cancel the 
additional divergences arising from incorporating the momentum 
dependence. 
An additional shift with respect to previous calculations 
of the order $\order{\als\alt}$ at zero momentum transfer 
appears from evaluating the ${\cal CP}$-odd Higgs-boson mass
counter-term at its pole mass, $m_{A^0}^2$. 

The effect of the new momentum-dependent
two-loop corrections on the predictions for
the $\cp$-even Higgs boson masses were 
analyzed numerically. 
The program \secdec{} is used in the evaluation of 
the finite parts of analytically unaccessible two-loop 
integrals with several mass scales. 

The obtained mass shifts of the light
$\cp$-even Higgs-boson mass 
exhibit an overall strong dependence on the 
mass of the gluino. 
For values of $\mgl \sim 1.5 \tev$ corrections 
of about $ -50 \mev$ are found, at the level of the anticipated
future ILC accuracy. For very large gluino masses, 
$\mgl \gtrsim 4 \tev$, on the other hand, substantially larger 
corrections are found, at the level of the current experimental 
accuracy at the LHC. 
Additional shifts in the heavy $\cp$-even 
Higgs-boson mass, are mostly
below current and future anticipated accuracies. 
Only close to
thresholds, e.g.\ around $p = 2\,\mste$, the corrections
are larger but do not exceed 0,2\%. 
The new momentum-dependent two-loop contributions 
have been incorporated into the program \fh.

\medskip

In conclusion, the leading momentum-dependent two-loop 
QCD corrections to the neutral $\cp$-even Higgs-boson 
masses should be taken into
account in precision analyses interpreting the discovered 
scalar particle as a Higgs-boson in the MSSM.

\medskip

Especially in this last application of the upgraded version 
of the program \secdec{} within this thesis, 
it was shown that the program can be used in the 
computation of phenomenological quantities. 
Yet, for an application where Monte Carlo sampling of an amplitude 
at millions of phase-space points is required, the
speed in the evaluation has to be improved. 

Nonetheless, in its new version \secdec{} has 
proven an invaluable 
tool in independent checks to analytical 
calculations of very complicated two-loop 
topologies involving masses and multiple 
scales. 

Based on the observation that the numerical 
evaluation in \secdec{} is 
rather limited in the cases of diagrams with very complicated 
singularity structures than limited by the 
number of mass scales involved, this numerical 
method forms a vital counter part to purely analytical 
approaches. 
Its extremely high potential 
needs to be exploited further in the future. 
Considering the fact that the method is very 
suitable for intense parallelization, the program 
has the potential to be further applicable in a multitude of 
higher-order corrections in quantum field
theories. 

%
%
%
%
%
%
%
%
%

%% file: appendix/appendix.tex
\chapter{Appendix}
\label{sec:appendix}%
\section{Analytical formulae}
\label{app:analytformulae}
All following formulae are based on 
Refs.~\cite{'tHooft:1978xw,vanderBij:1983bw,Nierste:1992wg,
Scharf:1993ds,Berends:1994ed,Bauberger:1994zz,
Bauberger:1994by,Bauberger:1994hx} and were 
either recalculated analytically or checked numerically 
with \secdec{} or Golem95~\cite{Binoth:2008uq,Guillet:2013msa}. 

\medskip

The prefactor at loop order $L$ reads
\begin{align}
P^L = \left( \frac{(2 \pi \mu_r)^{(4-D)}}{i \pi^2} \right)^L \text{ ,}
\end{align}
where the dimension $D=4- 2\, \eps$ contains the dimensional regulator $\eps$. 
The fully symmetric K{\"a}ll{\'e}n function reads
\begin{align}
\lambda(x,y,z) =x^2 + y^2 + z^2 - 2 xy - 2xz - 2yz \text{ .}
\end{align}
The following definitions facilitate numerical 
stability of analytic results. They are based 
on work of, e.g., Refs.~\cite{Bauberger:1994zz,Binoth:2005ff}.
\begin{align}
&\sqrt{\lambda(p^2,m_1^2, m_2^2)} =
\begin{cases}
\sqrt{(p^2-m_{12}^+)(p^2-m_{12}^-)}, 
  &  p^2 < m_{12}^+ \text{ and } p^2 \leq m_{12}^- ,\\
  i \sqrt{(p^2-m_{12}^- )(m_{12}^+-p^2)},  
  & p^2 < m_{12}^+ \text{ and } p^2 >m_{12}^- \text{ ,} \\
    \sqrt{(p^2-m_{12}^+)(p^2-m_{12}^-)},
  & p^2 \geq m_{12}^+ \text{ .}
\end{cases} 
\end{align}
where
\begin{align}
&m_{12}^+ = (m_1 + m_2)^2 \text{ ,} \\
&m_{12}^- = (m_1 - m_2)^2 \text{ .}
\end{align}
Furthermore, the Riemann sheet of the logarithm can be chosen 
explicitly as
\begin{align}
\text{log}(x)^+ = \log (x + i \delta) = 
\begin{cases}
  \text{log}(x),  
  & x > 0 \text{ ,}\\
  \text{log}(-x) + i \pi, 
  & x < 0 \text{ ,}
\end{cases}
\end{align}
or 
\begin{align}
\text{log}(x)^- = \log (x - i \delta) = 
\begin{cases}
  \text{log}(x),  
  & x > 0 \text{ ,}\\
  \text{log}(-x) - i \pi, 
  & x < 0 \text{ .}
\end{cases}
\end{align}
Similarly, two choices for an analytical continuation of the dilogarithm read
\begin{align}
\text{Li}_2(x)^+ = \text{Li}_2(x + i \delta) \text{ ,}\\
\text{Li}_2(x)^- = \text{Li}_2(x - i \delta) \text{ .} 
\end{align}
Furthermore, the following expressions can be abbreviated 
\begin{align}
\text{L}_m(m^2) = \gamma_E + \text{log}\left( \frac{m^2}{4 \pi \mu_r} \right)^- \text{ ,}\\
\text{L}_p(p^2) = \gamma_E + \text{log}\left( \frac{-p^2}{4 \pi \mu_r} \right)^- \text{ .}
\end{align}
\subsection{One-loop representations}
\subsubsection{One-loop tadpole}
The one-loop one-point function reads 
\begin{align}
A (m^2) =& P \int \text{d}^D q \frac{1}{(q^2-m^2)} \\
	=& - m^2  \left( \frac{m^2}{4 \pi \mu_r} \right)^{-\eps} \;
	\Gamma \left( \eps - 1 \right) \\
=& \frac{m^2}{\eps} + m^2 (1- \text{L}_m(m^2) )+ \mathcal{O}(\eps) \text{ .}
\end{align}
%
%
\subsubsection{One-loop bubble}
The one-loop two-point function is defined as follows 
\begin{align}
B (p^2, m_1^2, m_2^2) =& P \int \text{d}^D q \frac{1}{(q^2-m_1^2)((q+p_1)^2-m_2^2)} \\
 =& \frac{1}{\eps} +B^{\text{fin}} (p^2, m_1^2, m_2^2) + \mathcal{O}(\eps) \text{ .}
\end{align}
Special cases of the finite part read
\begin{align}
B^{\text{fin}} (0, m^2, 0) = B^{\text{fin}} (0, 0, m^2)  \text{ ,} 
\end{align}
\begin{align}
B^{\text{fin}} (0, 0, m^2) = 1- \text{L}_m(m^2)  \text{ ,}
\end{align}
\begin{align}
B^{\text{fin}} (0, m^2, m^2) = -  \text{L}_m(m^2)  \text{ ,}
\end{align}
\begin{align}
B^{\text{fin}} (0, m_1^2, m_2^2) = \frac{ A^{\text{fin}} (m_1^2) - A^{\text{fin}} (m_2^2) }{m_1^2 - m_2^2}  \text{ ,}
\end{align}
\begin{align}
B^{\text{fin}} (p^2, 0, 0) = 2 - \text{L}_p(p^2)  \text{ ,}
\end{align}
\begin{align}
B^{\text{fin}} (m^2, 0, m^2) = 2 -  \gamma_E - \text{log}\left( \frac{m^2}{4 \pi \mu_r} \right)^+  \text{ ,}
\end{align}
\begin{align}
B^{\text{fin}} (p^2, 0, m^2) =& B^{\text{fin}} (p^2, m^2, 0)  \text{ ,}\\
 =& 2 - \frac{m^2}{p^2} \text{L}_m(m^2) 
+ \frac{m^2-p^2}{p^2} \text{L}_m(m^2-p^2)  \text{ ,}
\end{align}
For a general representation of the finite part $B^{\text{fin}} (p^2, m_1^2, m_2^2)$, see 
Ref.~\cite{Scharf:1993ds}. 
For the coefficient of order ${\cal O} (\eps)$, see Refs.~\cite{Bauberger:1994zz,
Nierste:1992wg,Scharf:1993ds}. 
\subsubsection{Derivative of the one-loop bubble}
The derivative of the one-loop two-point function with respect to the first 
mass reads
\begin{align}
\frac{\partial}{\partial m_1^2}B (p^2, m_1^2, m_2^2) =& 
\partial_{m_1^2}B (p^2, m_1^2, m_2^2) \\
=& P \int \text{d}^D q \frac{1}{(q^2-m_1^2)^2((q+p_1)^2-m_2^2)} \\
=& \partial_{m_1^2}B^{\text{fin}} (p^2, m_1^2, m_2^2) + \mathcal{O}(\eps)  \text{ .}
\end{align}
It is finite as the divergent terms are independent of the masses. 
Special cases read 
\begin{align}
\partial_{m_1^2}B^{\text{fin}} (0, m_1^2, 0) =& - \frac{1}{m_1^2}  \text{ ,}\\
\partial_{m_1^2}B^{\text{fin}} (0, m^2, m^2) =& - \frac{1}{2 m_1^2}  \text{ ,}\\
\partial_{m_1^2}B^{\text{fin}} (0, m_1^2, m_2^2) =& \frac{1}{m_2^2 - m_1^2} + 
\frac{m_2^2}{(m_1^2-m_2^2)^2} \log\left( \frac{m_1}{m_2^2} \right)  \text{ ,}\\
\partial_{m_1^2}B^{\text{fin}} (p^2, m_1^2, 0) = & \frac{1}{p^2} \log \left( \frac{m_1^2-p^2}{m_1^2} \right)^- \text{ .}
\end{align}
For a representation of the general case with arbitrary masses, see Ref.~\cite{Bauberger:1994zz,Nierste:1992wg}. 
\subsection{Two-loop representations}
\subsubsection{Two-loop vacuum diagram}
The two-loop three-propagator vacuum integral corresponding to the diagram 
in Fig.~\ref{subfig:T134} reads
\begin{align}
T_{134}(m_{1}^2,m_{2}^2,m_{3}^2) =& P^2 \iint  
\frac{\text{d}^D  q_1 \text{d}^D  q_2}{(k_{1}^2-m_{1}^2+ i \delta)(k_{3}^2-m_{2}^2+ i \delta)(k_{4}^2-m_{3}^2+ i \delta)} \\
\non =& \frac{1}{\eps^2} \, T_{134}^{\text{div2}}(m_{1}^2,m_{2}^2,m_{3}^2) + 
 \frac{1}{\eps} \, T_{134}^{\text{div}}(m_{1}^2,m_{2}^2,m_{3}^2) \,+ \\
 & T_{134}^{\text{fin}}(m_{1}^2,m_{2}^2,m_{3}^2) + \mathcal{O}(\eps) \\
\non  =&\frac{1}{2 \eps^2} (m_1^2 + m_2^2 + m_3^2) + 
\frac{1}{ \eps} \,\Big[\, \frac32 \,(m_1^2 + m_2^2 + m_3^2) \,- \\
\non & m_1^2\, \text{L}_m(m_1^2) -
m_2^2 \,\text{L}_m(m_2^2) - m_3^2\, \text{L}_m(m_3^2)\, \Big] + \\
& T_{134}^{\text{fin}}(m_{1}^2,m_{2}^2,m_{3}^2) + \mathcal{O}(\eps) \text{ .}
\end{align}
A representation of the finite part 
$T_{134}^{\text{fin}}(m_{1}^2,m_{2}^2,m_{3}^2)$ is included in 
Ref.~\cite{Berends:1994ed}.
\subsubsection{Two-loop two-point three-propagator (sunrise) diagram}
The two-loop two-point three-propagator function corresponding to the diagram 
in Fig.~\ref{subfig:T234} reads
\begin{align}
\hspace{-20pt} T_{234}(p^2,m_{1}^2,m_{2}^2,m_{3}^2) =P^2 \iint  
\frac{\text{d}^D  q_1 \text{d}^D  q_2}{(k_{2}^2-m_{1}^2+ i \delta)(k_{3}^2-m_{2}^2+ i \delta)(k_{4}^2-m_{3}^2+ i \delta)} \text{ .}
\end{align}
The special case of one massless propagator reads 
\begin{align}
T_{234}(p^2,m_{1}^2,m_{2}^2,0) =& \frac{1}{2 \eps^2} (m_1^2 + m_2^2 ) +
\frac{1}{ \eps} \, ( T_{134}^{\text{fin}}(m_{1}^2,m_{2}^2,0) - \frac{1}{4} p^2 ) \,+ \\
& T_{234}^{\text{fin}}(p^2,m_{1}^2,m_{2}^2,0)
+ \mathcal{O}(\eps) \text{ .}
\end{align}
A representation of the finite part $T_{234}^{\text{fin}}(p^2,m_{1}^2,m_{2}^2,0)$ is 
included in Ref.~\cite{Berends:1994ed}.
\subsubsection{Two-loop two-point four-propagator diagram}
The two-loop two-point four-propagator function corresponding to the diagram 
in Fig.~\ref{subfig:T1234} reads
\begin{align}
\non &T_{1234}(p^2,m_{1}^2,m_{2}^2,m_{3}^2,m_{4}^2) \\
&=P^2 \iint  \frac{ \text{d}^D  q_1 \text{d}^D  q_2}
{(k_{1}^2-m_{1}^2+ i \delta)(k_{2}^2-m_{2}^2+ i \delta)(k_{3}^2-m_{3}^2+ i \delta)(k_{4}^2-m_{4}^2+ i \delta)} \text{ .}
\end{align}
The divergent part can be expressed as
\begin{align}
\non &T_{1234}(p^2,m_{1}^2,m_{2}^2,m_{3}^2,m_{4}^2) \\
&= \frac{1}{2 \eps^2} + 
\frac{1}{\eps} \left( B^{\text{fin}}(p^2,m_{1}^2,m_{2}^2 ) + \frac12 \right) \,
+ T_{1234}^{\text{fin}}(p^2,m_{1}^2,m_{2}^2,m_{3}^2,m_{4}^2) + \mathcal{O}(\eps) \text{ .}
\end{align}
A result for the finite part is not available in analytical form. 
\subsubsection{Two-loop two-point five-propagator diagram}
The two-loop two-point five-propagator function corresponding to the diagram 
in Fig.~\ref{subfig:T11234} reads
\begin{align}
&T_{11234}(p^2,m_{1}^2,m_{2}^2,m_{3}^2,m_{4}^2) \\
&=\frac{\partial}{\partial m_1^2} T_{1234}(p^2,m_{1}^2,m_{2}^2,m_{3}^2,m_{4}^2)\\
 &= P^2 \iint  \frac{ \text{d}^D  q_1 \text{d}^D  q_2}
 {(k_{1}^2-m_{1}^2+ i \delta)^2 (k_{2}^2-m_{2}^2+ i \delta)(k_{3}^2-m_{3}^2+ i \delta)(k_{4}^2-m_{4}^2+ i \delta)} \text{ .}
\end{align}
The divergent part can be expressed as
\begin{align}
\non &T_{11234}(p^2,m_{1}^2,m_{2}^2,m_{3}^2,m_{4}^2) \\
&= \frac{1}{\eps} \partial_{m_1^2} B^{\text{fin}}(p^2,m_{1}^2,m_{2}^2 ) \,+ 
 T_{11234}^{\text{fin}}(p^2,m_{1}^2,m_{2}^2,m_{3}^2,m_{4}^2) + \mathcal{O}(\eps) \text{ .}
\end{align}
A result for the finite part is not available in analytical form. 
\section{\secdec{} User Manual}
\label{sec:appendix:usermanual}
%
%
%
%
\subsection{Installation}
\label{subsec:install}
The program can be downloaded from {\tt http://secdec.hepforge.org}.
\medskip \\
Unpacking the tar archive via `{\it  tar xzvf SecDec-2.x.tar.gz}'  
will create a directory called {\tt SecDec-2.x} 
with the subdirectories as described in the previous section. Changing 
to the directory {\tt SecDec-2.x}, the program is installed by 
running `{\it ./install}'. 
\medskip \\
Prerequisites are Mathematica, version 6 or above, Perl (installed by default on 
most Unix/Linux systems), a Fortran compiler (e.g. gfortran, ifort) or a C$^{++}$ 
compiler if the C$^{++}$ option is used.
\medskip \\
In order to use the program, the user only has to edit the two files {\tt param*.input} and 
{\tt template*.m}. \secdec{} has three different setups, the user might be interested in. 
This is the setup to compute standard loop integrals, termed `Loop setup' in the following, 
generalized parametric functions, termed `General setup' in what follows, and 
and functions with a similar structure as loop integrals, referred to as `User-defined setup'. 

\subsubsection{Loop setup}
\begin{itemize}
\item {\tt paramloop.input}: (text file, Perl readable format)\\
In this file the user needs to specify paths, the type of integrand, 
the desired order in $\eps$, the output format, the parameters and kinematic 
values for numerical integration, the parameters for contour 
deformation and further options.
\item {\tt templateloop.m}: (Mathematica syntax) \\
Here enters the specification of the loop momenta and propagators, 
resp. of the topology; optionally a numerator different from 1, non-standard 
propagator powers, parameters to be split in the middle 
of the integration region and the space-time dimension.
\end{itemize}
\subsubsection{General setup}
\begin{itemize}
\item {\tt param.input}: (text file, Perl readable format)\\
In this file the user needs to specify paths, the type of integrand, the 
symbols and dummy functions utilized in the template file, 
the desired order in $\eps$, the output format, the parameters and values 
for numerical integration and further options.
\item {\tt Template.m}: (Mathematica syntax) \\
Here enters the specification of the integration variables, the factors of the 
integrand, variables to be split in the middle 
of the integration region and the space-time dimension.
\end{itemize}
\subsubsection{User-defined setup}
\begin{itemize}
\item {\tt paramuserdefined.input}: (text file, Perl readable format)\\
In this file the user needs to specify paths, the type of integrand, 
the desired order in $\eps$, the output format, the parameters and kinematic 
values for numerical integration, the parameters for contour 
deformation and further options.
\item {\tt templateuserdefined.m}: (Mathematica syntax) \\
Here enters the specification of the user-defined 
functions, the list of powers of the original propagators (optional), 
the rank of the integrand, the variables to be split in the 
middle of the integration region and the space-time dimension. 
As the file is read in by Mathematica, additional functions needed 
to evaluate the user-defined functions can be included without 
further specification in the parameter file. 
\end{itemize}
\subsection{Operation}
\label{subsec:operation}
\begin{enumerate}
\item Change to the subdirectory {\tt loop} for the calculation of 
a loop or a user-defined integral or to the subdirectory 
{\tt general} to evaluate a more general parameter integral.
\item Copy the files {\tt param*.input} and {\tt template*.m} to 
create your own parameter and template files  {\tt myparamfile.input} and {\tt mytemplatefile.m}, 
respectively. These two files serve to define the integrand and the 
parameters for the numerical integration.
\item Set the desired parameters in {\tt myparamfile.input} and specify the 
integrand in {\tt mytemplatefile.m}.
\item Issue the command `{\it ./launch -p myparamfile.input -t mytemplatefile.m}' 
in the shell. If you run the command with an additional `{\it -u}' 
the user-defined setup is used. \\
If you omit the option `{\it -p myparamfile.input}', the file {\tt param.input}, \\
{\tt paramloop.input} or {\tt paramuserdefined.input} will be taken as default, 
depending on the current directory (either 
{\tt general} or {\tt loop}) and whether the user-defined setup was chosen by adding the `{\it -u}'. 
Likewise, if you omit the option `{\it -t mytemplatefile.m}', 
the file {\tt Template.m}, {\tt templateloop.m} or {\tt templateuserdefined.m} 
will be taken as default.
If your files {\tt myparamfile.input, mytemplatefile.m} are in a different directory, say, 
{\tt myworkingdir}, 
use the option {\it -d myworkingdir}. The shell command then reads 
`{\it ./launch -d myworkingdir -p myparamfile.input -t mytemplatefile.m}', 
 executed from the directory {\tt SecDec/general} or
 {\tt SecDec/loop}. \\
\item Collect the results. Depending on whether you have used a single machine or 
submitted the jobs to a cluster, the following actions will be performed:
 \begin{itemize}
\item If the calculations are done sequentially on a single machine, 
    the results will be collected automatically (via {\tt results.pl}, {\tt resultsloop.pl} or \\
    {\tt resultsuserdefined.pl} called by {\tt launch}).
    The output file will be displayed with the text editor specified in the 
    {\tt myparamfile.input}.
\item If the jobs have been submitted to a cluster,    
	execute  the command 
	{\it ./results*.pl [-d myworkingdir -p myparamfile]} 
	when all jobs have finished. 
	This will write the final results to files in the {\tt graph} subdirectory
	specified in the input file.
\end{itemize}
\item After the calculation and the collection of the results is completed, 
you can use the shell command {\it ./launchclean[graph]}
to remove obsolete files.
\end{enumerate}
It should be mentioned that the code starts working on the most complicated 
pole structure, which also takes longest. 
When the jobs expected to take longest are submitted to a cluster first, the 
time the user has to wait for the results is minimized. 
\subsection{Program input parameters}
\label{subsec:programinputparameters}
The user manual is for loop diagrams; the input files in the subdirectory {\tt general}
to compute more general parametric functions are very similar.

\medskip

For the computation of an arbitrary loop integral the user should switch to the 
directory {\tt loop}, copy the files {\tt paramloop.input} and {\tt templateloop.m} 
and rename them arbitrarily to arrive at {\tt myparamloop.input}  and 
{\tt mytemplateloop.m}, respectively. 
In the file {\tt myparamloop.input} which is written in Perl readable format, the 
user then needs to name the graph to be 
computed and specify its number of propagators, external legs and loops.
Furthermore, the kinematic invariants $s_{ij}$, $p^2$ and $m^2$ need to be given 
numeric values. Apart from these definitions only one further 
flag ({\it cutconstruct=0 or 1}) needs to be set which decides 
how the user chooses to define the propagators of the 
diagram at hand in the {\tt mytemplateloop.m} file, compare 
Sec.~\ref{subsec:cutconstruct} for further explanations. All other 
parameters do not need to be specified, default values will be chosen. 

\medskip

\underline{The following parameters can be specified: }
\begin{description}
\item[subdir]
specifies the name of the subdirectory to which the graph should be written to. 
If it does not exist yet, it will be created. The specified {\it subdir} contains 
the directory specified in {\it outputdir}. 
\item[outputdir]
The name for the desired output directory can be given here by 
specifying the full path to the desired output directory. 
If {\it outputdir} is not specified, the default directory for the output will have the graph name 
(see below) appended to the directory {\it subdir}.

The output directory will contain all the files produced during the decomposition, subtraction, 
expansion and numerical integration, and the results. 
The output of the decomposition into sectors is found in the outputdir directly. 
The functions from subtraction and expansion and the respective 
files for numerical integration are 
found in subdirectories. The latter are named by the pole structure 
and contain subdirectories named after the order in $\eps$ to which 
the Laurent coefficients contained in these folders contribute.
\item[graph]
The name of the diagram or parametric function to be computed is specified here. 
The graph name can contain underscores and numbers, but should not contain commas.
\item[propagators]
Here, the number of propagators the diagram has is specified. This specification is 
mandatory in the computation of loop integrals using the automated setup. When 
utilizing the user-defined setup, the number of propagators only needs to be 
specified if the exponent of the two Symanzik polynomials should be computed 
in an automated way. 
\item[legs]
The number of external legs the diagram has is specified here (mandatory).
\item[loops]
The number of loops the diagram has is specified here (mandatory).
\item[cutconstruct]
If the graph to be computed corresponds to a scalar integral, 
the integrand (${\cal F}$ and $\cal U$) can be constructed via topological cuts.
In this case set {\it cutconstruct=1}, the default is =0. 
If cutconstruct is switched on, the input for the graph structure (*.m file) 
is just a list of labels connecting vertices, 
as explained in Secs.~\ref{subsec:cutconstruct} and \ref{subsec:graphm}.
\item[epsord]
The order to which the Laurent series in $\eps$ should be expanded, starting from $\eps^{-maxpole}$, can be specified here. 
The default is {\it epsord=0} where the Laurent series is cut after finite part $\eps^0$. 
If epsord is set to a negative value, only the pole coefficients up to this order are computed.
\item[prefactorflag]
Possible values for the {\it prefactorflag} are 0 (default), 1 and 2. 
\begin{itemize}
\item 0: The default prefactor $(-1)^{N}\,\Gamma[N-Nloops*Dim/2]$ is factored out of the numerical result.
\item 1: The default prefactor $(-1)^{N}\,\Gamma[N-Nloops*Dim/2]$ is included in the numerical result.
\item 2: Give the desired prefactor in {\it prefactor=} to be factored out in the final result. 
\end{itemize}
\item[prefactor]
If option 2 has been chosen in the {\it prefactorflag}, write down the desired {\it prefactor} in Mathematica syntax. In combination with options 0 or 1 in the {\it prefactorflag} this entry will be ignored. 
Use {\it Nn}, {\it Nloops} and {\it Dim} to denote the number of propagators, loops and dimension ({\it Dim=4-2*eps} by default).
\item[IBPflag]
Set {\it IBPflag=0} if the integration by parts option should not be used and {\it=1} if it should be used. 
{\it IBPflag=2} is designed to use IBP relations when it is more efficient to do so. 

Using the integrations by parts method takes more time in the subtraction and 
expansion step and generally 
results in more functions for numerical integration. 
However, it can be useful if (spurious) linear poles of the type $x^{-2-b\eps}$ are found 
in the decomposition, 
as it reduces the power of $x$ in the denominator.
\item[compiler]
Choose a Fortran compiler (tested with gfortran, ifort, g77) if {\it language=Fortran}. 
Left blank, the default is gfortran.
\item[exeflag]
The {\it exeflag} can be used to execute the program in steps within one 
calculation. 
\begin{itemize}
\item 0: The iterated sector decomposition is done and the scripts to do the subtraction, the expansion in epsilon, the creation of the Fortran/C$^{++}$ files and to launch the numerical integration are created (scripts batch* in the subdirectory {\it graph}) but not run. This can be useful if a cluster is available to run each pole structure on a different node.
\item 1: In addition to the steps done in 0, the subtraction and epsilon expansion is performed and the resulting functions are written to Fortran/C$^{++}$ files. 
\item 2: In addition to the steps done in 1, all the files needed for the numerical integration are created.
\item 3: In addition to the steps done in 2, the compilation of the Fortran/C$^{++}$ files is launched to make the executables. 
\item 4: In addition to the steps done in 3, the executables are run, either by batch submission or locally.
\end{itemize}
\item[clusterflag]
The {\it clusterflag} determines how jobs are submitted. Setting {\it clusterflag=0} (default) the jobs will run on
a single machine, setting it {\it =1} the jobs will run on a cluster (a batch system to submit jobs).
\item[batchsystem]
If a cluster is used ({\it clusterflag=1}), this flag should be set to 0 to use the setup 
for the PBS (Portable batch system). If the flag is set to 1
a user-defined setup is activated. Currently this is the submission via {\tt condor}, but it 
can be adapted to other batch systems by editing {\tt perlsrc/makejob.pm}. 
\item[maxjobs]
When using a cluster, specify the maximum number of jobs allowed in the queue here.
\item[maxcput]
Specify the estimated maximal CPU time (in hours). 
This option is only used to send a job to a particular queue on a batch system. 
\item[pointname]
The name of the point to calculate is specified here. 
It should be either blank or a string and is useful to label the result files in case of different 
runs for different numerical values of the Mandelstam variables, masses etc.
\item[sij]
The values for Mandelstam invariants $s_{ij}=(p_i+p_j)^2$ in numbers are specified here (mandatory). 
The $s_{ij}$ should be $\leq 0$ in the Euclidean region. 
\item[pi2]
Massive external legs $p_1^2$, $p_2^2$,... are specified here (mandatory). 
$p_i^2$ should be $\leq 0$ in the Euclidean region. Light-like external legs 
must be specified in the {\it onshell=} conditions in 
the {\tt mytemplatefile.m}, see Sec.~\ref{subsec:graphm}. 
\item[ms2]
Specify the masses of propagators $m_1^2$, $m_2^2$,... 
using the notation ms[i] for  $m_i^2$ (mandatory). The masses should 
not be complex numbers. 
\item[integrator]
The program for numerical integration can be chosen here. 
BASES ({\it integrator=0}) can only be used in the Fortran 
version. Vegas ({\it integrator=1}), Suave ({\it integrator=2}), 
Divonne ({\it integrator=3}, default) and Cuhre ({\it integrator=4}) 
are part of the \textsc{Cuba} library and can be used in both the Fortran and 
the C$^{++}$ version. 
In practice, Divonne usually gives the fastest results when using the C$^{++}$ version. 
In the following we therefore concentrate on the adjustment of the parameters 
needed for numerical 
integration using Divonne. For more details about the \textsc{Cuba} parameters, the reader 
is referred to Ref.~\cite{Hahn:2004fe}.
\item[cubapath]
The path to the \textsc{Cuba} library can be specified here. The 
default directory is {\it [your path to SecDec]/Cuba-3.2}. 
\textsc{Cuba}-3.2 uses parallel processing during the numerical 
evaluation of the integral. The older version (\textsc{Cuba}-2.1) is 
still supported and can be used.
\item[maxeval]
Separated by commas and starting with the lowest order coefficient in $\eps$, 
specify the maximal number of evaluations to be used by the numerical integrator for each order in $\eps$. 
If {\it maxeval} is not equal to {\it mineval}, the maximal number of evaluations does not have to be reached.
\item[mineval]
Separated by commas and starting with the lowest order coefficient in $\eps$, 
specify the number of evaluations which should at least be done before the numerical integrator returns a result. 
The default is 0. 
\item[epsrel]
Separated by commas and starting with the lowest order coefficient in $\eps$, 
specify the desired relative accuracy for the numerical evaluation.
\item[epsabs]
Separated by commas and starting with the lowest order coefficient in $\eps$, 
specify the desired absolute accuracy for the numerical evaluation. 
These values are particularly important when either the real or the imaginary 
part of an integral is close to zero. Note, {\it epsabs=} must be 
chosen smaller than the resulting values for the integral. 
\item[cubaflags]
Set the cuba verbosity flags. The default is 2 which means, the \textsc{Cuba} input parameters and other useful 
information, e.g. about numerical convergence, are echoed during the 
numerical integration.
\item[key1]
Separated by commas and starting with the lowest order coefficient in $\eps$, 
specify {\it key1} which determines the sampling to be used for the partitioning phase in Divonne. 
With a positive {\it key1}, a Korobov quasi-random sample of {\it key1} points is used.
A {\it key1} of about 1000 (default) usually is a good choice. 
\item[key2]
Separated by commas and starting with the lowest order coefficient in $\eps$, 
specify {\it key2} which determines the sampling to be used for the final integration phase in Divonne. 
With a positive {\it key2}, a Korobov quasi-random sample is used. 
The default is {\it key2=1} which means, the number of points needed to reach the prescribed accuracy is estimated 
by Divonne. 
\item[key3]
Separated by commas and starting with the lowest order coefficient in $\eps$, 
specify the {\it key3} to be used for the refinement phase in Divonne. 
Setting {\it key3=1} (default), each subregion is split once more.
\item[maxpass]
Separated by commas and starting with the lowest order coefficient in $\eps$, 
specify how good the convergence has to be during the partitioning phase until the program passes 
on to the main integration phase. 
A {\it maxpass} of 3 (default) is usually sufficient to get a quick and good result.  
\item[border]
Separated by commas and starting with the lowest order coefficient in $\eps$, 
specify the border for the numerical integration. 
The points in the interval $[0,border]$ and $[1-border,1]$ are not included in the integration 
but are extrapolated from a few points of the excluded range. This can be useful 
if the integrand is known to be peaked close to endpoints of the integration variables.
\item[maxchisq]
Separated by commas and starting with the lowest order coefficient in $\eps$, 
specify the maximally allowed $\chi^2$ at the end of the numerical integration. 
\item[mindeviation]
Separated by commas and starting with the lowest order coefficient in $\eps$, 
specify the deviation two sample averages in one region can show without being treated any further.
\end{description}

\bigskip

\underline{These parameters are advanced options:}
\begin{description}
\item[primarysectors]
Specify a list of primary sectors to be treated here. 
If left blank, {\it primarysectors} defaults to all, i.e. {\it primarysectors=1,...,N} will 
be assumed, where $N$ is the number of  
propagators. This option is useful if a diagram has symmetries such that some primary 
sectors yield the same result. 
\item[multiplicities]
Specify the {\it multiplicities} of the primary sectors listed above. List the {\it multiplicities} in same order as 
the corresponding sectors above. If left blank, a default multiplicity of $1$ is set for 
each primary sector.
\item[infinitesectors] 
The alternative heuristic sector decomposition strategy, 
compare \\
Sec.~\ref{subsec:heuristicstrategies}, is activated when 
specifying, separated by commas, those primary sectors which should get a 
pre-decomposition. Writing \\
{\it infinitesectors=2,3} \\
results in the application of the alternative sector decomposition 
strategy to primary sectors 2 and 3.
{\it infinitesectors} must be left empty for the default strategy to be applied to all 
primary sectors. 
\item[togetherflag]
This flag defines whether to integrate subsets of functions for each 
pole order separately {\it togetherflag=0} (default) 
or to sum all functions for a certain pole order prior to integration 
{\it togetherflag=1}. The latter will allow 
cancellations between different functions and thus give a more 
realistic error, but should not be used for complicated 
diagrams where the individual functions are very large.
\item[grouping]
Even though {\it togetherflag=0} is chosen, it can be beneficial to first sum a few 
functions before integrating them. 
Choosing a value for the grouping which is unequal to zero, 
defines how many bytes a summed function may have. 
The number of bytes is set by {\it grouping=\#bytes}. 
Setting {\it grouping=0}, all functions {\tt f*.f} resp. {\tt f*.cc} are integrated 
separately. In practice, a grouping=0 has proven to lead to faster 
convergence and more accurate results. 
When considering integrals where large cancellations among 
the different functions occur, the grouping value should be chosen $\neq0$. 
The log files {\tt *results*.log} in the results directory contain 
the results from the individual sub-sector integration. They can be 
viewed to spot cancellations between the individual functions. 
\item[editor]
Choose here which editor should be used to display the result. 
If {\it editor=none} is set, the full result will not be displayed in an editor 
window at the end of the calculation.
\item[language]
The choice between Fortran and C$^{++}$ can be made in 
the {\tt myparam.input} file by choosing 
either {\it language=Cpp} (default) or {\it language=Fortran}.
For diagrams with purely Euclidean kinematics, both languages can 
be chosen. 
In combination with {\it contourdef=True}, {\it language=Cpp} must be 
used, as the inclusion of an imaginary part in the result is implemented 
in C$^{++}$ only. 
\item[rescale]
If all invariants are very small or very large it is beneficial to rescale 
them to reach faster convergence during numerical integration. The 
rescaling (scaling out the largest invariant
in the numerical integration part) can be switched on with {\it rescale=1} 
and switched off when set to 0 (default). 
If switched on, it is not  possible to set explicit values for any 
non-zero invariant in the {\it onshell=} conditions in the Mathematica 
template file {\tt mytemplate.m}.
\item[contourdef]
The contour deformation can be switched on or off, by choosing 
{\it contourdef=True/False} in the input file {\tt myparamloop.input}. 
For multi-scale problems, respectively diagrams with non-Euclidean kinematics, 
set {\it contourdef=True} (default is False). 
In this case, a deformation of the integration contour in the form of 
Eq.~(\ref{eq:condef}) is done. 
In addition to the functions {\tt f*.cc} to be integrated, auxiliary files 
{\tt g*.cc} are produced which serve to
optimize the deformation for each integrand function.
\item[lambda]
The initial value for $\lambda$ can be set for the deformation 
of Eq.~(\ref{eq:condef}) by assigning a value to {\it lambda=}. 
The program takes the $\lambda$ value given by the user in the 
{\tt myparam.input} file as a starting point. During the checks 
listed in Sec.~\ref{subsec:program:contourdef}, the 
appropriate value for $\lambda$ is found automatically by the program. 
The user should pick an initial value which is 
rather too large than too small. 
A too large initial value for $\lambda$ can easily be accommodated to 
have the right size during the computation, while a too small initial value 
can not be increased anymore. 
{\it lambda=3.0} usually serves as a good initial value. 
Without any knowledge about the characteristics of the 
integrand, {\it lambda=1.0} is a good choice. 
If the diagram contains mostly massless propagators and light-like legs, 
it can be useful to choose the initial $\lambda$ larger (e.g. {\it lambda=5.0}), in order to 
compensate for cases where the remainders of the IR 
subtraction lead to large cancellations for $x_i\to 0$. 
For diagrams with mostly massive propagators the initial 
lambda can be chosen smaller, e.g. {\it lambda=0.1}.    
\item[smalldefs]
If the integrand is expected to be oscillatory and hence sensitive to small changes in the 
deformation parameter $\lambda$, {\it smalldefs} should be set to 1 (default is 0). 
If switched on, the argument of each sub-sector function ${\cal F}$ is minimized.
\item[largedefs]
If the integrand is expected to  have (integrable) endpoint singularities 
at $x_j=0$ or 1, the deformation should be maximized. If {\it largedefs=1}, the 
program maximizes the deformation.
The default is {\it largedefs=0}. 
\item[optlamevals]
The number of pre-samples to determine the optimal contour 
deformation parameter $\lambda$ can be chosen by assigning a number to 
{\it optlamevals=} in the {\tt myparam.input}. The default value is 4000.
\end{description}
\subsection{Input for the definition of the integrand}
\label{subsec:graphm}
The Mathematica input file should be called {\tt *.m}. 
The following parameters can be specified in Mathematica readable format 
\begin{description}
\item[momlist]
If {\it cutconstruct=0} is set in the input file, specify the names of the loop momenta here.
\item[proplist]
Specify the diagram topology here (mandatory). The syntax for {\it cutconstruct=1} is 
described in Section \ref{subsec:graphtopo}. 
If {\it cutconstruct=0} has been chosen, the propagators have to be given explicitly.
An example propagator list could be \\
{\it proplist=\{k\^\,2-ms[1],(k+p1)\^\,2-ms[1]\}} \\
with the loop momentum $k$, 
the propagator mass $m_1^2$ and external momentum $p_1$. 
\item[numerator]
If present, specify the numerator of the integrand here. 
If not given, a {\it numerator=\{1\}} is assumed. Please note that the option 
{\it cutconstruct=1} is not available in combination with numerator functions. 
\item[powerlist]
The propagator powers must be specified. If all propagators are raised to the power 
one, an example syntax reads 
{\it powerlist=Table[1,\{i,Length[proplist]\}]; }.
\item[onshell]
Specify replacements for kinematic invariants here. 
The specification of light-like external 
legs is of specific importance in the generation of the correct 
integrand topology. 
These can be assigned by writing \\
{\it onshell=\{ ssp[1] -> 0\}}, \\
where {\it ssp[1]} denotes the first external momentum squared, $p_1^2$.
The kinematic invariants can be assigned other specific values, e.g. {\it ssp[1] -> 0.25}. 
Furthermore, relations among the invariants can 
be set (e.g. {\it ssp[1]$\to$sp[1,3]}).
This option can not be used in combination with {\it rescale=1}.
\item[Dim]
Set the space-time dimension. The default is {\it Dim=4-2*eps}. The symbol 
for the regulator, {\it eps}, must be kept.
\item[threshold] This option works in combination with C$^{++}$ and 
{\it contourdef=True}. Specify a kinematical threshold condition above which, an 
imaginary part is expected. For the calculation below this threshold, the 
imaginary part is set to zero and the contour deformation parameter $\lambda$ is 
decreased to a very small value. Set {\it threshold=none}, remove or comment the line
out if the threshold option should not be used (default). Usage of constants and 
kinematic invariants {\it ms[i], ssp[i]} and {\it sp[i,j]} is allowed. 
An example syntax reads {\it threshold = sp[1,2] > 0;}.
\item[splitlist]
The integration region of those integrals over Feynman 
parameters $t_{j_i}$, specified in a 
{\it splitlist=\{ $j_1$, $j_2$, \dots, $j_n$ \}} 
in the {\tt mytemplatefile.m} is split at $1/2$ and the resulting 
two integrals are remapped to the unit interval. 
The procedure follows the explanations of Sec.~\ref{subsec:endpointremapping}. 
\end{description}
\subsection{Topology based construction of an integrand}
\label{subsec:graphtopo} 
The implementation of the topology based construction of an 
integrand in the program \secdec{} is such that the 
user only has to label the external momenta $p_i$, the vertices $i$ and 
the masses $ms[k]$ of a graph. It is selected by choosing 
{\it cutconstruct=1} in the input file. 
If an external momentum $p_i$ is  
part of a vertex, this vertex needs to carry the label $i$. 
The labeling of vertices containing only internal lines is arbitrary.
In the {\tt mytemplate.m} file, the user has to specify the {\it proplist} as a 
list of entries of the form 
$\{ms[k],\{i,j\}\}$, where $ms[k]$ is the mass squared of the 
propagator connecting vertex $i$ and vertex $j$.
The mass label $k$ must correspond the the $k$th entry of the list of masses given in 
{\tt paramloop.input}. While $k$ needs to be the number labeling 
the masses, $ms[k]$ (with $k$ being an integer) can be left symbolic 
until numerical integration. If the mass is zero, the user has to put 
 $\{0,\{i,j\}\}$, because this changes the singularity structure 
 at the level of the decomposition into sectors. 
More examples can be found in the Mathematica template files 
{\tt templateP126.m, templateBnp6*.m, templateJapNP.m, templateggtt*.m}
in the subdirectory {\tt loop/demos}. 
The original form of specifying the 
propagators by their momenta and including numerators different from one, 
as done in {\secdec}\,1.0~\cite{Carter:2010hi}, is still operational. 
To use the latter option, the flag {\it cutconstruct=} must be set to zero. 
\subsection{Utilization of the user-defined setup}
\label{subsec:graphuserdef}
When the functions $\mathcal U$ and $\mathcal F$ are already known and 
the $\delta$-distribution is integrated out, a user-defined setup can 
be considered. The integrand could then be defined in 
the {\tt mytemplateuserdefined.m} file. First, the two known Symanzik 
polynomials, the numerator and the rank of the integral are 
written to the template file. 
Taking the diagram $P_{126}$ as an example, compare Sec.~\ref{subsec:p126}, 
the input in the {\tt mytemplateuserdefined.m} file 
reads \\
~\\
{\it U}[z\_] := z[1]*z[2] + z[2]*z[3] + z[1]*z[4] + z[2]*z[4] + z[3]*z[4] + z[1]*z[5] +                                        
 z[2]*z[5] + z[3]*z[5] + z[1]*z[6] + z[2]*z[6] + z[3]*z[6] \\
 ~\\
{\it F}[z\_] := -(sp[1, 2]*z[1]*z[2]*z[3]) - sp[1, 2]*z[1]*z[3]*z[4] -                                                         
 sp[1, 2]*z[1]*z[3]*z[5] - sp[1, 2]*z[2]*z[3]*z[5] -                                                                  
 sp[1, 2]*z[1]*z[2]*z[6] - sp[1, 2]*z[1]*z[3]*z[6] -                                                                  
 sp[1, 2]*z[1]*z[5]*z[6] - sp[1, 2]*z[2]*z[5]*z[6] -                                                                  
 sp[1, 2]*z[3]*z[5]*z[6] + (ms[1]*z[1] + ms[1]*z[2] + ms[1]*z[3])*                                                    
  (z[1]*z[2] + z[2]*z[3] + z[1]*z[4] + z[2]*z[4] + z[3]*z[4] + z[1]*z[5] +                                            
   z[2]*z[5] + z[3]*z[5] + z[1]*z[6] + z[2]*z[6] + z[3]*z[6]) \\
   ~\\
{\it Num} = 1; \\
{\it rank} = 0; \\
~\\
The naming of these functions is arbitrary and only needed for 
a clearer presentation of the functions in the {\it functionlist}, to be 
explained now. The diagram $P_{126}$ has 
six propagators and is therefore assumed to have six primary 
sectors after having integrated out the $\delta$-distribution. The  
primary sectors are counted in the first entry of each function. In 
a more general application, the first entry just lists those functions 
of equal exponents in the Symanzik polynomials 
which should be grouped together. The second entry of each function 
in the {\it functionlist} gives the exponents of each Feynman parameter, 
starting with the exponent of $t_1$. The total number of Feynman 
parameters occurring here (and to be integrated over afterwards) is 5 
because the $\delta$-distribution was already integrated out. 
The third entry in the list is the function, corresponding to the 
first Symanzik polynomial in the first primary sector, followed by 
its exponent and a flag whether the function needs further sector 
decomposition or not. Here, $B$ denotes that further decomposition 
is necessary while $A$ means that the function is already fully 
decomposed into sectors. \\
~\\
{\it functionlist} = \{\\
{\small \{1, \{0,0,0,0,0\}, 
\{\{(U[t]/.t[1]->1)/.t[6]->t[1], $XU$, $B$\}, \{(F[t]/.t[1]->1)/.t[6]->t[1], $XF$, $B$\}, Num\}, \\
\{2, \{0,0,0,0,0\}, 
\{\{(U[t]/.t[2]->1)/.t[6]->t[2], $XU$, $B$\}, \{(F[t]/.t[2]->1)/.t[6]->t[2], $XF$, $B$\}, Num\}, \\
\{3, \{0,0,0,0,0\}, 
\{\{(U[t]/.t[3]->1)/.t[6]->t[3], $XU$, $B$\}, \{(F[t]/.t[3]->1)/.t[6]->t[3], $XF$, $B$\}, Num\}, \\
\{4, \{0,0,0,0,0\}, 
\{\{(U[t]/.t[4]->1)/.t[6]->t[4], $XU$, $B$\}, \{(F[t]/.t[4]->1)/.t[6]->t[4], $XF$, $B$\}, Num\}, \\
\{5, \{0,0,0,0,0\}, 
\{\{(U[t]/.t[5]->1)/.t[6]->t[5], $XU$, $B$\}, \{(F[t]/.t[5]->1)/.t[6]->t[5], $XF$, $B$\}, Num\}, \\
\{6, \{0,0,0,0,0\}, 
\{\{U[t]/.t[6]->1, $XU$, $B$\}, \{F[t]/.t[6]->1, $XF$, $B$\}, Num\}\}; }\\
~\\
Choosing $XU$ and $XF$, the exponents of the functions $\cal U$ and $\cal F$, 
respectively, are computed automatically by the program using the information 
about the number of loops, the number of propagators, their powers, the rank 
and the space-time dimension. 
To run this example, from the {\tt loop} directory, issue the command
`{\it ./launch -p paramuserdefined.input -t templateuserdefined.m -u}'. 
The demo files {\tt paramuserdefined.m} and {\tt templateuserdefined.m} in the 
{\tt loop} directory come with the code. 
\subsection{Looping over ranges of parameters}
\label{subsec:appendixmultinumerics}
A looping over ranges of parameters in \secdec{} is  
put into practice using the perl script {\tt multinumerics*}, where the {\tt *} 
stands either for {\tt loop.pl}, {\tt userdefined.pl} or just {\tt .pl}, depending on 
whether a standard loop, a user-defined integral or a more general 
parametric integral should be computed for ranges of parameters. 
The sets of parameters to be evaluated are specified in a text file
{\tt mymultiparam.input} in {\tt myworkingdir}, to be read by the 
{\tt multinumerics*}-script. 

The following information must be contained in this textfile:
\begin{itemize}
\item {\it paramfile = myparam.input}: Specify the name of the parameter file 
containing the graph info.
\item {\it pointname = my\_prefix} (optional): Specify a name which is used as prefix to 
each kinematical point. To distinguish different kinematical points, each gets 
a different name using the prefix and sequential number, e.g. {\it
my\_prefix1} for the first point, {\it my\_prefix2} for the second and so on. 
\item {\it lines = $a$} (optional): Specify the number $a$ of points you 
wish to calculate - if omitted all points (listed in separate lines) will 
be calculated.
\item {\it xplot = $b$} (optional): If this option is set, a tab-separated data file is written with 
a variable of choice in the first column, the numeric result for the real part in the second 
column, the uncertainty in the third column, the numeric result for the 
imaginary part in the fourth column and its uncertainty in the fifth. The integer $b$ 
defines the number of the column in the {\tt multiparam.input} file containing the 
values which should be stored in the first column of the data file (default is 1). 
The resulting data files can directly be used for producing plots with, e.g., gnuplot. 
\end{itemize}
The subsequent specifications are different for parametric integrals and 
loop (or similar) integrals. \\
~\\
\underline{{\tt multinumericsloop.pl} and {\tt multinumericsuserdefined.pl}:} \\
~\\
When computing either a standard loop or a user-defined integral, 
the number of values given for $s_{ij}$, $p_{i}^2$ and  $m_i^2$ need to 
be specified by \\
{\it numsij=} \\
{\it numpi2=} \\
{\it numms2= }. \\
Numerical values of the parameters for each point to calculate
need to ensue these definitions. 
Examples come with the code of the program, one is found 
in \\{\tt loop/demos/multiparam.input}. 
The perl script {\tt helpmulti.pl} can be used to generate 
{\tt multiparam.input} files automatically, to avoid typing 
large sets of numerical values.
\\
~\\
\underline{{\tt multinumerics.pl}:} \\
~\\
When computing a more general parametric function, depending, 
for instance, on the two symbols $a$ and $b$ (defined as such in the 
{\tt param.input} file), values for these can either be specified 
explicitly in the {\tt multiparam.input} file as \\
{\it values1=0.1,0.2,0.4\\
values2=0.1,0.3,0.6} \;,\\
where {\it values1} is linked to the first symbol 
specified in the {\tt param.input} file, {\it values2} to the 
second symbol and so on.
Or, if the user desires to calculate the integrand for values of parameters at
incremental steps, the following syntax applies \\
{\it minvals=0.1,0.1\\
maxvals=0.3,0.7\\
stepvals=0.1,0.2} .\\
This input would calculate each combination of $a=0.1,0.2,0.3$ and 
$b=0.1,0.3,0.5,0.7$, where $a$ and $b$ are specified first and second 
in the list of symbols in the {\tt param.input} file.
Further examples can be found in {\tt general/demos/multiparam.input}.

\medskip

Before executing the script {\tt multinumerics*}, the functions 
generated by Mathematica must already be in place. The simplest 
way to guarantee this is to run the {\tt launch} script, choosing 
{\it exeflag=1} in the {\tt myparamfile.input} and subsequently 
issue the command \\
{\it `./multinumerics* [-d myworkingdir -p multiparamfile]'} .\\
In the single-machine mode ({\it clusterflag=0})
all integrations are performed sequentially, in the batch mode, 
they are run in parallel. \\
Running the script again with the additional argument ``1'' as \\
{\it `./multinumerics* [-d myworkingdir -p multiparamfile] 1'} \,,\\
all results are collated before writing the output as {\tt *.out} 
files into the graph directory specified in the 
{\tt myparamfile.input}. If specified in the {\tt multiparam.input} 
file, an additional {\tt *.gpdat} file is written, containing the 
results of all computed points.\\
The script generates a parameter file for each numerical point 
calculated. To remove such intermediate parameter files, issue the 
command \\
{\it `./multinumerics* [-d myworkingdir -p multiparamfile] 2'}. \\ 
This should only be done after the results have been collected.
\subsection{Leaving functions implicit during the algebraic part}
\label{subsec:appendixdummy}
To use this option, the Mathematica template file 
can contain a function which is left undefined, but needs to be listed 
with the option {\it dummys=} in the {\tt myparam.input} file. 
If symbolic parameters are used in addition, these
do not need to be listed as arguments of the implicit function. \\
Once the template and parameter files are set up, the functions need to be defined
explicitly so that they can be used in the calculation. The
simplest way to do this is to prepare a Mathematica syntax file for each
implicit function, and place them in the output directory specified as
 {\it outputdir=} in the {\tt myparam.input} file. For a function 
named {\it dum1} of two variables, defined as 
$dum1(x_1,x_2)=1+x_1+x_2$, the following lines would need to be 
inserted in a file {\tt dum1.m} named after the dummy function \\
$intvars=\{z_1,z_2\};\\dum1=1+z_1+z_2;$\\
where $z_1,z_2$ can be replaced by any variable name you wish, as long as they
are used consistently in {\tt dum1.m}. 
Once these Mathematica files are in place, 
issue the command\\ 
`{\it createdummyfortran.pl [-d myworkingdir -p myparamfile]}' . \\
This triggers the generation of the necessary Fortran files for the user-defined 
dummy functions. The files are stored in the same subdirectory as the originals. \\
It is equally possible to write these Fortran files oneself instead of
having them generated by the program, although the automated 
procedure is recommended. 
An example of this can be found in the directory {\tt general/demos}, 
comprised in the files {\tt paramdummy.input}, {\tt templatedummy.m} and 
in the directory {\tt /testdummy}.

%% file: borowka_thesis.bbl
\begin{thebibliography}{100}
\expandafter\ifx\csname url\endcsname\relax
  \def\url#1{\texttt{#1}}\fi
\expandafter\ifx\csname urlprefix\endcsname\relax\def\urlprefix{URL }\fi
\expandafter\ifx\csname href\endcsname\relax
  \def\href#1#2{#2} \def\path#1{#1}\fi

\bibitem{Borowka:2012yc}
S.~Borowka, J.~Carter, G.~Heinrich, {Numerical Evaluation of Multi-Loop
  Integrals for Arbitrary Kinematics with SecDec 2.0}, Comput.Phys.Commun. 184
  (2013) 396--408.
\newblock \href {http://arxiv.org/abs/1204.4152} {\path{arXiv:1204.4152}},
  \href {http://dx.doi.org/10.1016/j.cpc.2012.09.020}
  {\path{doi:10.1016/j.cpc.2012.09.020}}.

\bibitem{Borowka:2013cma}
S.~Borowka, G.~Heinrich, {Massive non-planar two-loop four-point integrals with
  SecDec 2.1}, Comput.Phys.Commun. 184 (2013) 2552--2561.
\newblock \href {http://arxiv.org/abs/1303.1157} {\path{arXiv:1303.1157}},
  \href {http://dx.doi.org/10.1016/j.cpc.2013.05.022}
  {\path{doi:10.1016/j.cpc.2013.05.022}}.

\bibitem{Borowka:2014wla}
S.~Borowka, T.~Hahn, S.~Heinemeyer, G.~Heinrich, W.~Hollik, {Momentum-dependent
  two-loop QCD corrections to the neutral Higgs-boson masses in the MSSM},
  Eur.Phys.J. C74~(8) (2014) 2994.
\newblock \href {http://arxiv.org/abs/1404.7074} {\path{arXiv:1404.7074}},
  \href {http://dx.doi.org/10.1140/epjc/s10052-014-2994-0}
  {\path{doi:10.1140/epjc/s10052-014-2994-0}}.

\bibitem{Borowka:2012ii}
S.~Borowka, J.~Carter, G.~Heinrich, {SecDec: A tool for numerical multi-loop
  calculations}, J.Phys.Conf.Ser. 368 (2012) 012051.
\newblock \href {http://arxiv.org/abs/1206.4908} {\path{arXiv:1206.4908}},
  \href {http://dx.doi.org/10.1088/1742-6596/366/1/012051}
  {\path{doi:10.1088/1742-6596/366/1/012051}}.

\bibitem{Borowka:2012rt}
S.~Borowka, G.~Heinrich, {Numerical evaluation of massive multi-loop integrals
  with SecDec}, PoS LL2012 (2012) 038.
\newblock \href {http://arxiv.org/abs/1209.6345} {\path{arXiv:1209.6345}}.

\bibitem{Borowka:2013lda}
S.~Borowka, G.~Heinrich, {Numerical multi-loop calculations with SecDec}, C
  13-05-16.
\newblock \href {http://arxiv.org/abs/1309.3492} {\path{arXiv:1309.3492}}.

\bibitem{Borowka:2013uea}
S.~Borowka, G.~Heinrich, {Two-loop applications of the program SecDec}, PoS
  RADCOR2013 (2014) 009.
\newblock \href {http://arxiv.org/abs/1311.6476} {\path{arXiv:1311.6476}}.

\bibitem{Glashow:1961tr}
S.~Glashow, {Partial Symmetries of Weak Interactions}, Nucl.Phys. 22 (1961)
  579--588.
\newblock \href {http://dx.doi.org/10.1016/0029-5582(61)90469-2}
  {\path{doi:10.1016/0029-5582(61)90469-2}}.

\bibitem{Weinberg:1967tq}
S.~Weinberg, {A Model of Leptons}, Phys.Rev.Lett. 19 (1967) 1264--1266.
\newblock \href {http://dx.doi.org/10.1103/PhysRevLett.19.1264}
  {\path{doi:10.1103/PhysRevLett.19.1264}}.

\bibitem{Salam:1968rm}
A.~Salam, {Weak and Electromagnetic Interactions}, Conf.Proc. C680519 (1968)
  367--377.

\bibitem{Fritzsch:1973pi}
H.~Fritzsch, M.~Gell-Mann, H.~Leutwyler, {Advantages of the Color Octet Gluon
  Picture}, Phys.Lett. B47 (1973) 365--368.
\newblock \href {http://dx.doi.org/10.1016/0370-2693(73)90625-4}
  {\path{doi:10.1016/0370-2693(73)90625-4}}.

\bibitem{Gross:1973ju}
D.~Gross, F.~Wilczek, {Asymptotically Free Gauge Theories. 1}, Phys.Rev. D8
  (1973) 3633--3652.
\newblock \href {http://dx.doi.org/10.1103/PhysRevD.8.3633}
  {\path{doi:10.1103/PhysRevD.8.3633}}.

\bibitem{Gross:1973id}
D.~J. Gross, F.~Wilczek, {Ultraviolet Behavior of Nonabelian Gauge Theories},
  Phys.Rev.Lett. 30 (1973) 1343--1346.
\newblock \href {http://dx.doi.org/10.1103/PhysRevLett.30.1343}
  {\path{doi:10.1103/PhysRevLett.30.1343}}.

\bibitem{Politzer:1973fx}
H.~D. Politzer, {Reliable Perturbative Results for Strong Interactions?},
  Phys.Rev.Lett. 30 (1973) 1346--1349.
\newblock \href {http://dx.doi.org/10.1103/PhysRevLett.30.1346}
  {\path{doi:10.1103/PhysRevLett.30.1346}}.

\bibitem{Guralnik:1964eu}
G.~Guralnik, C.~Hagen, T.~Kibble, {Global Conservation Laws and Massless
  Particles}, Phys.Rev.Lett. 13 (1964) 585--587.
\newblock \href {http://dx.doi.org/10.1103/PhysRevLett.13.585}
  {\path{doi:10.1103/PhysRevLett.13.585}}.

\bibitem{Englert:1964et}
F.~Englert, R.~Brout, {Broken Symmetry and the Mass of Gauge Vector Mesons},
  Phys.Rev.Lett. 13 (1964) 321--323.
\newblock \href {http://dx.doi.org/10.1103/PhysRevLett.13.321}
  {\path{doi:10.1103/PhysRevLett.13.321}}.

\bibitem{Higgs:1964ia}
P.~W. Higgs, {Broken symmetries, massless particles and gauge fields},
  Phys.Lett. 12 (1964) 132--133.
\newblock \href {http://dx.doi.org/10.1016/0031-9163(64)91136-9}
  {\path{doi:10.1016/0031-9163(64)91136-9}}.

\bibitem{Higgs:1966ev}
P.~W. Higgs, {Spontaneous Symmetry Breakdown without Massless Bosons},
  Phys.Rev. 145 (1966) 1156--1163.
\newblock \href {http://dx.doi.org/10.1103/PhysRev.145.1156}
  {\path{doi:10.1103/PhysRev.145.1156}}.

\bibitem{Aad:2012tfa}
G.~Aad, et~al., {Observation of a new particle in the search for the Standard
  Model Higgs boson with the ATLAS detector at the LHC}, Phys.Lett. B716 (2012)
  1--29.
\newblock \href {http://arxiv.org/abs/1207.7214} {\path{arXiv:1207.7214}},
  \href {http://dx.doi.org/10.1016/j.physletb.2012.08.020}
  {\path{doi:10.1016/j.physletb.2012.08.020}}.

\bibitem{Chatrchyan:2012ufa}
S.~Chatrchyan, et~al., {Observation of a new boson at a mass of 125 GeV with
  the CMS experiment at the LHC}, Phys.Lett. B716 (2012) 30--61.
\newblock \href {http://arxiv.org/abs/1207.7235} {\path{arXiv:1207.7235}},
  \href {http://dx.doi.org/10.1016/j.physletb.2012.08.021}
  {\path{doi:10.1016/j.physletb.2012.08.021}}.

\bibitem{Aad:2013xqa}
G.~Aad, et~al., {Evidence for the spin-0 nature of the Higgs boson using ATLAS
  data}, Phys.Lett. B726 (2013) 120--144.
\newblock \href {http://arxiv.org/abs/1307.1432} {\path{arXiv:1307.1432}},
  \href {http://dx.doi.org/10.1016/j.physletb.2013.08.026}
  {\path{doi:10.1016/j.physletb.2013.08.026}}.

\bibitem{ATLAS-CONF-2014-009}
{Updated coupling measurements of the Higgs boson with the ATLAS detector using
  up to 25 fb$^{-1}$ of proton-proton collision data}, Tech. Rep.
  ATLAS-CONF-2014-009, CERN, Geneva (Mar 2014).

\bibitem{Chatrchyan:2013mxa}
S.~Chatrchyan, et~al., {Measurement of the properties of a Higgs boson in the
  four-lepton final state }\href {http://arxiv.org/abs/1312.5353}
  {\path{arXiv:1312.5353}}.

\bibitem{Chatrchyan:2013iaa}
S.~Chatrchyan, et~al., {Measurement of Higgs boson production and properties in
  the WW decay channel with leptonic final states}, JHEP 1401 (2014) 096.
\newblock \href {http://arxiv.org/abs/1312.1129} {\path{arXiv:1312.1129}},
  \href {http://dx.doi.org/10.1007/JHEP01(2014)096}
  {\path{doi:10.1007/JHEP01(2014)096}}.

\bibitem{Begeman:1991iy}
K.~Begeman, A.~Broeils, R.~Sanders, {Extended rotation curves of spiral
  galaxies: Dark haloes and modified dynamics}, Mon.Not.Roy.Astron.Soc. 249
  (1991) 523.

\bibitem{Clowe:2006eq}
D.~Clowe, M.~Bradac, A.~H. Gonzalez, M.~Markevitch, S.~W. Randall, et~al., {A
  direct empirical proof of the existence of dark matter}, Astrophys.J. 648
  (2006) L109--L113.
\newblock \href {http://arxiv.org/abs/astro-ph/0608407}
  {\path{arXiv:astro-ph/0608407}}, \href {http://dx.doi.org/10.1086/508162}
  {\path{doi:10.1086/508162}}.

\bibitem{Ade:2013zuv}
P.~Ade, et~al., {Planck 2013 results. XVI. Cosmological parameters }\href
  {http://arxiv.org/abs/1303.5076} {\path{arXiv:1303.5076}}.

\bibitem{Cabibbo:1979ay}
N.~Cabibbo, L.~Maiani, G.~Parisi, R.~Petronzio, {Bounds on the Fermions and
  Higgs Boson Masses in Grand Unified Theories}, Nucl.Phys. B158 (1979)
  295--305.
\newblock \href {http://dx.doi.org/10.1016/0550-3213(79)90167-6}
  {\path{doi:10.1016/0550-3213(79)90167-6}}.

\bibitem{Hung:1979dn}
P.~Q. Hung, {Vacuum Instability and New Constraints on Fermion Masses},
  Phys.Rev.Lett. 42 (1979) 873.
\newblock \href {http://dx.doi.org/10.1103/PhysRevLett.42.873}
  {\path{doi:10.1103/PhysRevLett.42.873}}.

\bibitem{Lindner:1985uk}
M.~Lindner, {Implications of Triviality for the Standard Model}, Z.Phys. C31
  (1986) 295.
\newblock \href {http://dx.doi.org/10.1007/BF01479540}
  {\path{doi:10.1007/BF01479540}}.

\bibitem{Degrassi:2012ry}
G.~Degrassi, S.~Di~Vita, J.~Elias-Miro, J.~R. Espinosa, G.~F. Giudice, et~al.,
  {Higgs mass and vacuum stability in the Standard Model at NNLO}, JHEP 1208
  (2012) 098.
\newblock \href {http://arxiv.org/abs/1205.6497} {\path{arXiv:1205.6497}},
  \href {http://dx.doi.org/10.1007/JHEP08(2012)098}
  {\path{doi:10.1007/JHEP08(2012)098}}.

\bibitem{Weinberg:1979bn}
S.~Weinberg, {Implications of Dynamical Symmetry Breaking: An Addendum},
  Phys.Rev. D19 (1979) 1277--1280.
\newblock \href {http://dx.doi.org/10.1103/PhysRevD.19.1277}
  {\path{doi:10.1103/PhysRevD.19.1277}}.

\bibitem{Gildener:1976ih}
E.~Gildener, S.~Weinberg, {Symmetry Breaking and Scalar Bosons}, Phys.Rev. D13
  (1976) 3333.
\newblock \href {http://dx.doi.org/10.1103/PhysRevD.13.3333}
  {\path{doi:10.1103/PhysRevD.13.3333}}.

\bibitem{Cao:2013cfa}
J.~Cao, Y.~He, P.~Wu, M.~Zhang, J.~Zhu, {Higgs Phenomenology in the Minimal
  Dilaton Model after Run I of the LHC}, JHEP 1401 (2014) 150.
\newblock \href {http://arxiv.org/abs/1311.6661} {\path{arXiv:1311.6661}},
  \href {http://dx.doi.org/10.1007/JHEP01(2014)150}
  {\path{doi:10.1007/JHEP01(2014)150}}.

\bibitem{Kaplan:1983fs}
D.~B. Kaplan, H.~Georgi, {SU(2) x U(1) Breaking by Vacuum Misalignment},
  Phys.Lett. B136 (1984) 183.
\newblock \href {http://dx.doi.org/10.1016/0370-2693(84)91177-8}
  {\path{doi:10.1016/0370-2693(84)91177-8}}.

\bibitem{Giudice:2007fh}
G.~Giudice, C.~Grojean, A.~Pomarol, R.~Rattazzi, {The Strongly-Interacting
  Light Higgs}, JHEP 0706 (2007) 045.
\newblock \href {http://arxiv.org/abs/hep-ph/0703164}
  {\path{arXiv:hep-ph/0703164}}, \href
  {http://dx.doi.org/10.1088/1126-6708/2007/06/045}
  {\path{doi:10.1088/1126-6708/2007/06/045}}.

\bibitem{Nilles:1983ge}
H.~P. Nilles, {Supersymmetry, Supergravity and Particle Physics}, Phys.Rept.
  110 (1984) 1--162.
\newblock \href {http://dx.doi.org/10.1016/0370-1573(84)90008-5}
  {\path{doi:10.1016/0370-1573(84)90008-5}}.

\bibitem{Haber:1984rc}
H.~E. Haber, G.~L. Kane, {The Search for Supersymmetry: Probing Physics Beyond
  the Standard Model}, Phys.Rept. 117 (1985) 75--263.
\newblock \href {http://dx.doi.org/10.1016/0370-1573(85)90051-1}
  {\path{doi:10.1016/0370-1573(85)90051-1}}.

\bibitem{Barbieri:1987xf}
R.~Barbieri, {Looking Beyond the Standard Model: The Supersymmetric Option},
  Riv.Nuovo Cim. 11N4 (1988) 1--45.
\newblock \href {http://dx.doi.org/10.1007/BF02725953}
  {\path{doi:10.1007/BF02725953}}.

\bibitem{Ramond:1971gb}
P.~Ramond, {Dual Theory for Free Fermions}, Phys.Rev. D3 (1971) 2415--2418.
\newblock \href {http://dx.doi.org/10.1103/PhysRevD.3.2415}
  {\path{doi:10.1103/PhysRevD.3.2415}}.

\bibitem{Neveu:1971rx}
A.~Neveu, J.~Schwarz, {Factorizable dual model of pions}, Nucl.Phys. B31 (1971)
  86--112.
\newblock \href {http://dx.doi.org/10.1016/0550-3213(71)90448-2}
  {\path{doi:10.1016/0550-3213(71)90448-2}}.

\bibitem{Gervais:1971ji}
J.-L. Gervais, B.~Sakita, {Field Theory Interpretation of Supergauges in Dual
  Models}, Nucl.Phys. B34 (1971) 632--639.
\newblock \href {http://dx.doi.org/10.1016/0550-3213(71)90351-8}
  {\path{doi:10.1016/0550-3213(71)90351-8}}.

\bibitem{Golfand:1971iw}
Y.~Golfand, E.~Likhtman, {Extension of the Algebra of Poincare Group Generators
  and Violation of p Invariance}, JETP Lett. 13 (1971) 323--326.

\bibitem{Volkov:1973ix}
D.~Volkov, V.~Akulov, {Is the Neutrino a Goldstone Particle?}, Phys.Lett. B46
  (1973) 109--110.
\newblock \href {http://dx.doi.org/10.1016/0370-2693(73)90490-5}
  {\path{doi:10.1016/0370-2693(73)90490-5}}.

\bibitem{Wess:1973kz}
J.~Wess, B.~Zumino, {A Lagrangian Model Invariant Under Supergauge
  Transformations}, Phys.Lett. B49 (1974) 52.
\newblock \href {http://dx.doi.org/10.1016/0370-2693(74)90578-4}
  {\path{doi:10.1016/0370-2693(74)90578-4}}.

\bibitem{Wess:1974tw}
J.~Wess, B.~Zumino, {Supergauge Transformations in Four-Dimensions}, Nucl.Phys.
  B70 (1974) 39--50.
\newblock \href {http://dx.doi.org/10.1016/0550-3213(74)90355-1}
  {\path{doi:10.1016/0550-3213(74)90355-1}}.

\bibitem{Haag:1974qh}
R.~Haag, J.~T. Lopuszanski, M.~Sohnius, {All Possible Generators of
  Supersymmetries of the s Matrix}, Nucl.Phys. B88 (1975) 257.
\newblock \href {http://dx.doi.org/10.1016/0550-3213(75)90279-5}
  {\path{doi:10.1016/0550-3213(75)90279-5}}.

\bibitem{Coleman:1967ad}
S.~R. Coleman, J.~Mandula, {All Possible Symmetries of the S Matrix}, Phys.Rev.
  159 (1967) 1251--1256.
\newblock \href {http://dx.doi.org/10.1103/PhysRev.159.1251}
  {\path{doi:10.1103/PhysRev.159.1251}}.

\bibitem{Fayet:1974pd}
P.~Fayet, {Supergauge Invariant Extension of the Higgs Mechanism and a Model
  for the electron and Its Neutrino}, Nucl.Phys. B90 (1975) 104--124.
\newblock \href {http://dx.doi.org/10.1016/0550-3213(75)90636-7}
  {\path{doi:10.1016/0550-3213(75)90636-7}}.

\bibitem{Witten:1981nf}
E.~Witten, {Dynamical Breaking of Supersymmetry}, Nucl.Phys. B188 (1981) 513.
\newblock \href {http://dx.doi.org/10.1016/0550-3213(81)90006-7}
  {\path{doi:10.1016/0550-3213(81)90006-7}}.

\bibitem{Dimopoulos:1981zb}
S.~Dimopoulos, H.~Georgi, {Softly Broken Supersymmetry and SU(5)}, Nucl.Phys.
  B193 (1981) 150.
\newblock \href {http://dx.doi.org/10.1016/0550-3213(81)90522-8}
  {\path{doi:10.1016/0550-3213(81)90522-8}}.

\bibitem{Sakai:1981gr}
N.~Sakai, {Naturalness in Supersymmetric Guts}, Z.Phys. C11 (1981) 153.
\newblock \href {http://dx.doi.org/10.1007/BF01573998}
  {\path{doi:10.1007/BF01573998}}.

\bibitem{Inoue:1982ej}
K.~Inoue, A.~Kakuto, H.~Komatsu, S.~Takeshita, {Low-Energy Parameters and
  Particle Masses in a Supersymmetric Grand Unified Model}, Prog.Theor.Phys. 67
  (1982) 1889.
\newblock \href {http://dx.doi.org/10.1143/PTP.67.1889}
  {\path{doi:10.1143/PTP.67.1889}}.

\bibitem{Inoue:1982pi}
K.~Inoue, A.~Kakuto, H.~Komatsu, S.~Takeshita, {Aspects of Grand Unified Models
  with Softly Broken Supersymmetry}, Prog.Theor.Phys. 68 (1982) 927.
\newblock \href {http://dx.doi.org/10.1143/PTP.68.927}
  {\path{doi:10.1143/PTP.68.927}}.

\bibitem{Inoue:1983pp}
K.~Inoue, A.~Kakuto, H.~Komatsu, S.~Takeshita, {Renormalization of
  Supersymmetry Breaking Parameters Revisited}, Prog.Theor.Phys. 71 (1984) 413.
\newblock \href {http://dx.doi.org/10.1143/PTP.71.413}
  {\path{doi:10.1143/PTP.71.413}}.

\bibitem{Glashow:1976nt}
S.~L. Glashow, S.~Weinberg, {Natural Conservation Laws for Neutral Currents},
  Phys.Rev. D15 (1977) 1958.
\newblock \href {http://dx.doi.org/10.1103/PhysRevD.15.1958}
  {\path{doi:10.1103/PhysRevD.15.1958}}.

\bibitem{Butterworth:2014efa}
J.~Butterworth, G.~Dissertori, S.~Dittmaier, D.~de~Florian, N.~Glover, et~al.,
  {Les Houches 2013: Physics at TeV Colliders: Standard Model Working Group
  Report }\href {http://arxiv.org/abs/1405.1067} {\path{arXiv:1405.1067}}.

\bibitem{Czakon:2013goa}
M.~Czakon, P.~Fiedler, A.~Mitov, {Total Top-Quark Pair-Production Cross Section
  at Hadron Colliders Through $O(\alpha_s^4)$}, Phys.Rev.Lett. 110~(25) (2013)
  252004.
\newblock \href {http://arxiv.org/abs/1303.6254} {\path{arXiv:1303.6254}},
  \href {http://dx.doi.org/10.1103/PhysRevLett.110.252004}
  {\path{doi:10.1103/PhysRevLett.110.252004}}.

\bibitem{Korner:2008bn}
J.~Korner, Z.~Merebashvili, M.~Rogal, {NNLO $O(\alpha_s^{4})$ results for heavy
  quark pair production in quark-antiquark collisions: The One-loop squared
  contributions}, Phys.Rev. D77 (2008) 094011.
\newblock \href {http://arxiv.org/abs/0802.0106} {\path{arXiv:0802.0106}},
  \href {http://dx.doi.org/10.1103/PhysRevD.77.094011,
  10.1103/PhysRevD.85.119904} {\path{doi:10.1103/PhysRevD.77.094011,
  10.1103/PhysRevD.85.119904}}.

\bibitem{Bonciani:2008az}
R.~Bonciani, A.~Ferroglia, T.~Gehrmann, D.~Maitre, C.~Studerus, {Two-Loop
  Fermionic Corrections to Heavy-Quark Pair Production: The Quark-Antiquark
  Channel}, JHEP 0807 (2008) 129.
\newblock \href {http://arxiv.org/abs/0806.2301} {\path{arXiv:0806.2301}},
  \href {http://dx.doi.org/10.1088/1126-6708/2008/07/129}
  {\path{doi:10.1088/1126-6708/2008/07/129}}.

\bibitem{Bonciani:2009nb}
R.~Bonciani, A.~Ferroglia, T.~Gehrmann, C.~Studerus, {Two-Loop Planar
  Corrections to Heavy-Quark Pair Production in the Quark-Antiquark Channel},
  JHEP 0908 (2009) 067.
\newblock \href {http://arxiv.org/abs/0906.3671} {\path{arXiv:0906.3671}},
  \href {http://dx.doi.org/10.1088/1126-6708/2009/08/067}
  {\path{doi:10.1088/1126-6708/2009/08/067}}.

\bibitem{Bonciani:2010mn}
R.~Bonciani, A.~Ferroglia, T.~Gehrmann, A.~Manteuffel, C.~Studerus, {Two-Loop
  Leading Color Corrections to Heavy-Quark Pair Production in the Gluon Fusion
  Channel}, JHEP 1101 (2011) 102.
\newblock \href {http://arxiv.org/abs/1011.6661} {\path{arXiv:1011.6661}},
  \href {http://dx.doi.org/10.1007/JHEP01(2011)102}
  {\path{doi:10.1007/JHEP01(2011)102}}.

\bibitem{Aliev:2010zk}
M.~Aliev, H.~Lacker, U.~Langenfeld, S.~Moch, P.~Uwer, et~al., {HATHOR: HAdronic
  Top and Heavy quarks crOss section calculatoR}, Comput.Phys.Commun. 182
  (2011) 1034--1046.
\newblock \href {http://arxiv.org/abs/1007.1327} {\path{arXiv:1007.1327}},
  \href {http://dx.doi.org/10.1016/j.cpc.2010.12.040}
  {\path{doi:10.1016/j.cpc.2010.12.040}}.

\bibitem{Abelof:2011ap}
G.~Abelof, A.~Gehrmann-De~Ridder, {Double real radiation corrections to
  $t\bar{t}$ production at the LHC: the all-fermion processes}, JHEP 1204
  (2012) 076.
\newblock \href {http://arxiv.org/abs/1112.4736} {\path{arXiv:1112.4736}},
  \href {http://dx.doi.org/10.1007/JHEP04(2012)076}
  {\path{doi:10.1007/JHEP04(2012)076}}.

\bibitem{Czakon:2011ve}
M.~Czakon, {Double-real radiation in hadronic top quark pair production as a
  proof of a certain concept}, Nucl.Phys. B849 (2011) 250--295.
\newblock \href {http://arxiv.org/abs/1101.0642} {\path{arXiv:1101.0642}},
  \href {http://dx.doi.org/10.1016/j.nuclphysb.2011.03.020}
  {\path{doi:10.1016/j.nuclphysb.2011.03.020}}.

\bibitem{Czakon:2012pz}
M.~Czakon, A.~Mitov, {NNLO corrections to top pair production at hadron
  colliders: the quark-gluon reaction}, JHEP 1301 (2013) 080.
\newblock \href {http://arxiv.org/abs/1210.6832} {\path{arXiv:1210.6832}},
  \href {http://dx.doi.org/10.1007/JHEP01(2013)080}
  {\path{doi:10.1007/JHEP01(2013)080}}.

\bibitem{Czakon:2012zr}
M.~Czakon, A.~Mitov, {NNLO corrections to top-pair production at hadron
  colliders: the all-fermionic scattering channels}, JHEP 1212 (2012) 054.
\newblock \href {http://arxiv.org/abs/1207.0236} {\path{arXiv:1207.0236}},
  \href {http://dx.doi.org/10.1007/JHEP12(2012)054}
  {\path{doi:10.1007/JHEP12(2012)054}}.

\bibitem{Baernreuther:2012ws}
P.~Baernreuther, M.~Czakon, A.~Mitov, {Percent Level Precision Physics at the
  Tevatron: First Genuine NNLO QCD Corrections to $q \bar{q} \to t \bar{t} +
  X$}, Phys.Rev.Lett. 109 (2012) 132001.
\newblock \href {http://arxiv.org/abs/1204.5201} {\path{arXiv:1204.5201}},
  \href {http://dx.doi.org/10.1103/PhysRevLett.109.132001}
  {\path{doi:10.1103/PhysRevLett.109.132001}}.

\bibitem{vonManteuffel:2012je}
A.~von Manteuffel, C.~Studerus, {Top quark pairs at two loops and Reduze 2 }{To
  appear in the proceedings of the conference ``Loops and Legs in Quantum Field
  Theory", 2012}.
\newblock \href {http://arxiv.org/abs/1210.1436} {\path{arXiv:1210.1436}}.

\bibitem{Abelof:2012he}
G.~Abelof, A.~Gehrmann-De~Ridder, O.~Dekkers, {Antenna subtraction with massive
  fermions at NNLO: Double real initial-final configurations}, JHEP 1212 (2012)
  107.
\newblock \href {http://arxiv.org/abs/1210.5059} {\path{arXiv:1210.5059}},
  \href {http://dx.doi.org/10.1007/JHEP12(2012)107}
  {\path{doi:10.1007/JHEP12(2012)107}}.

\bibitem{Abelof:2012rv}
G.~Abelof, A.~Gehrmann-De~Ridder, {Double real radiation corrections to
  $t\bar{t}$ production at the LHC: the $gg\rightarrow t\bar{t}q\bar{q}$
  channel}, JHEP 1211 (2012) 074.
\newblock \href {http://arxiv.org/abs/1207.6546} {\path{arXiv:1207.6546}},
  \href {http://dx.doi.org/10.1007/JHEP11(2012)074}
  {\path{doi:10.1007/JHEP11(2012)074}}.

\bibitem{Czakon:2013vfa}
M.~Czakon, A.~Mitov, {Precision top pair production at hadron colliders},
  J.Phys.Conf.Ser. 452 (2013) 012026.
\newblock \href {http://arxiv.org/abs/1303.0693} {\path{arXiv:1303.0693}},
  \href {http://dx.doi.org/10.1088/1742-6596/452/1/012026}
  {\path{doi:10.1088/1742-6596/452/1/012026}}.

\bibitem{Moch:2012mk}
S.~Moch, P.~Uwer, A.~Vogt, {On top-pair hadro-production at
  next-to-next-to-leading order}, Phys.Lett. B714 (2012) 48--54.
\newblock \href {http://arxiv.org/abs/1203.6282} {\path{arXiv:1203.6282}},
  \href {http://dx.doi.org/10.1016/j.physletb.2012.06.048}
  {\path{doi:10.1016/j.physletb.2012.06.048}}.

\bibitem{Cacciari:2011hy}
M.~Cacciari, M.~Czakon, M.~Mangano, A.~Mitov, P.~Nason, {Top-pair production at
  hadron colliders with next-to-next-to-leading logarithmic soft-gluon
  resummation}, Phys.Lett. B710 (2012) 612--622.
\newblock \href {http://arxiv.org/abs/1111.5869} {\path{arXiv:1111.5869}},
  \href {http://dx.doi.org/10.1016/j.physletb.2012.03.013}
  {\path{doi:10.1016/j.physletb.2012.03.013}}.

\bibitem{Czakon:2011xx}
M.~Czakon, A.~Mitov, {Top++: A Program for the Calculation of the Top-Pair
  Cross-Section at Hadron Colliders }\href {http://arxiv.org/abs/1112.5675}
  {\path{arXiv:1112.5675}}.

\bibitem{Beneke:2011mq}
M.~Beneke, P.~Falgari, S.~Klein, C.~Schwinn, {Hadronic top-quark pair
  production with NNLL threshold resummation}, Nucl.Phys. B855 (2012) 695--741.
\newblock \href {http://arxiv.org/abs/1109.1536} {\path{arXiv:1109.1536}},
  \href {http://dx.doi.org/10.1016/j.nuclphysb.2011.10.021}
  {\path{doi:10.1016/j.nuclphysb.2011.10.021}}.

\bibitem{Beneke:2012wb}
M.~Beneke, P.~Falgari, S.~Klein, J.~Piclum, C.~Schwinn, et~al., {Inclusive
  Top-Pair Production Phenomenology with TOPIXS}, JHEP 1207 (2012) 194.
\newblock \href {http://arxiv.org/abs/1206.2454} {\path{arXiv:1206.2454}},
  \href {http://dx.doi.org/10.1007/JHEP07(2012)194}
  {\path{doi:10.1007/JHEP07(2012)194}}.

\bibitem{Ahrens:2011px}
V.~Ahrens, A.~Ferroglia, M.~Neubert, B.~D. Pecjak, L.~L. Yang, {Precision
  predictions for the t+t(bar) production cross section at hadron colliders},
  Phys.Lett. B703 (2011) 135--141.
\newblock \href {http://arxiv.org/abs/1105.5824} {\path{arXiv:1105.5824}},
  \href {http://dx.doi.org/10.1016/j.physletb.2011.07.058}
  {\path{doi:10.1016/j.physletb.2011.07.058}}.

\bibitem{Kidonakis:2010dk}
N.~Kidonakis, {Next-to-next-to-leading soft-gluon corrections for the top quark
  cross section and transverse momentum distribution}, Phys.Rev. D82 (2010)
  114030.
\newblock \href {http://arxiv.org/abs/1009.4935} {\path{arXiv:1009.4935}},
  \href {http://dx.doi.org/10.1103/PhysRevD.82.114030}
  {\path{doi:10.1103/PhysRevD.82.114030}}.

\bibitem{Bonciani:2013ywa}
R.~Bonciani, A.~Ferroglia, T.~Gehrmann, A.~von Manteuffel, C.~Studerus,
  {Light-quark two-loop corrections to heavy-quark pair production in the gluon
  fusion channel}, JHEP 1312 (2013) 038.
\newblock \href {http://arxiv.org/abs/1309.4450} {\path{arXiv:1309.4450}},
  \href {http://dx.doi.org/10.1007/JHEP12(2013)038}
  {\path{doi:10.1007/JHEP12(2013)038}}.

\bibitem{vonManteuffel:2013uoa}
A.~von Manteuffel, C.~Studerus, {Massive planar and non-planar double box
  integrals for light $N_{f}$ contributions to $gg \to t\bar{t}$}, JHEP 1310
  (2013) 037.
\newblock \href {http://arxiv.org/abs/1306.3504} {\path{arXiv:1306.3504}},
  \href {http://dx.doi.org/10.1007/JHEP10(2013)037}
  {\path{doi:10.1007/JHEP10(2013)037}}.

\bibitem{Gunion:1984yn}
J.~Gunion, H.~E. Haber, {Higgs Bosons in Supersymmetric Models. 1.}, Nucl.Phys.
  B272 (1986) 1.
\newblock \href {http://dx.doi.org/10.1016/0550-3213(86)90340-8}
  {\path{doi:10.1016/0550-3213(86)90340-8}}.

\bibitem{Gunion:1986nh}
J.~Gunion, H.~E. Haber, {Higgs Bosons in Supersymmetric Models. 2. Implications
  for Phenomenology}, Nucl.Phys. B278 (1986) 449.
\newblock \href {http://dx.doi.org/10.1016/0550-3213(86)90050-7}
  {\path{doi:10.1016/0550-3213(86)90050-7}}.

\bibitem{Fayet:1976cr}
P.~Fayet, S.~Ferrara, {Supersymmetry}, Phys.Rept. 32 (1977) 249--334.
\newblock \href {http://dx.doi.org/10.1016/0370-1573(77)90066-7}
  {\path{doi:10.1016/0370-1573(77)90066-7}}.

\bibitem{Gunion:1989we}
J.~F. Gunion, H.~E. Haber, G.~L. Kane, S.~Dawson, {The Higgs Hunter's Guide},
  Vol.~80, 2000.

\bibitem{Rzehak:2005zz}
H.~A. Rzehak, {Two-loop contributions to the supersymmetric Higgs sector},
  Ph.D. thesis (2005).

\bibitem{Faddeev:1967fc}
L.~Faddeev, V.~Popov, {Feynman Diagrams for the Yang-Mills Field}, Phys.Lett.
  B25 (1967) 29--30.
\newblock \href {http://dx.doi.org/10.1016/0370-2693(67)90067-6}
  {\path{doi:10.1016/0370-2693(67)90067-6}}.

\bibitem{Girardello:1981wz}
L.~Girardello, M.~T. Grisaru, {Soft Breaking of Supersymmetry}, Nucl.Phys. B194
  (1982) 65.
\newblock \href {http://dx.doi.org/10.1016/0550-3213(82)90512-0}
  {\path{doi:10.1016/0550-3213(82)90512-0}}.

\bibitem{Carena:2001bg}
M.~S. Carena, H.~E. Haber, H.~E. Logan, S.~Mrenna, {Distinguishing a MSSM Higgs
  boson from the SM Higgs boson at a linear collider}, Phys.Rev. D65 (2002)
  055005.
\newblock \href {http://arxiv.org/abs/hep-ph/0106116}
  {\path{arXiv:hep-ph/0106116}}, \href
  {http://dx.doi.org/10.1103/PhysRevD.65.055005, 10.1103/PhysRevD.65.099902}
  {\path{doi:10.1103/PhysRevD.65.055005, 10.1103/PhysRevD.65.099902}}.

\bibitem{Ferrara:1974pu}
S.~Ferrara, B.~Zumino, {Supergauge Invariant Yang-Mills Theories}, Nucl.Phys.
  B79 (1974) 413.
\newblock \href {http://dx.doi.org/10.1016/0550-3213(74)90559-8}
  {\path{doi:10.1016/0550-3213(74)90559-8}}.

\bibitem{Drees:2004jm}
M.~Drees, R.~Godbole, P.~Roy, {Theory and phenomenology of sparticles: An
  account of four-dimensional N=1 supersymmetry in high energy physics }, 2004.

\bibitem{Jackiw:1974cv}
R.~Jackiw, {Functional evaluation of the effective potential}, Phys.Rev. D9
  (1974) 1686.
\newblock \href {http://dx.doi.org/10.1103/PhysRevD.9.1686}
  {\path{doi:10.1103/PhysRevD.9.1686}}.

\bibitem{Peskin:1995ev}
M.~E. Peskin, D.~V. Schroeder, {An Introduction to quantum field theory }.

\bibitem{Ellis:1990nz}
J.~R. Ellis, G.~Ridolfi, F.~Zwirner, {Radiative corrections to the masses of
  supersymmetric Higgs bosons}, Phys.Lett. B257 (1991) 83--91.
\newblock \href {http://dx.doi.org/10.1016/0370-2693(91)90863-L}
  {\path{doi:10.1016/0370-2693(91)90863-L}}.

\bibitem{Haber:1990aw}
H.~E. Haber, R.~Hempfling, {Can the mass of the lightest Higgs boson of the
  minimal supersymmetric model be larger than m(Z)?}, Phys.Rev.Lett. 66 (1991)
  1815--1818.
\newblock \href {http://dx.doi.org/10.1103/PhysRevLett.66.1815}
  {\path{doi:10.1103/PhysRevLett.66.1815}}.

\bibitem{Okada:1990vk}
Y.~Okada, M.~Yamaguchi, T.~Yanagida, {Upper bound of the lightest Higgs boson
  mass in the minimal supersymmetric standard model}, Prog.Theor.Phys. 85
  (1991) 1--6.
\newblock \href {http://dx.doi.org/10.1143/PTP.85.1}
  {\path{doi:10.1143/PTP.85.1}}.

\bibitem{Brignole:1992uf}
A.~Brignole, {Radiative corrections to the supersymmetric neutral Higgs boson
  masses}, Phys.Lett. B281 (1992) 284--294.
\newblock \href {http://dx.doi.org/10.1016/0370-2693(92)91142-V}
  {\path{doi:10.1016/0370-2693(92)91142-V}}.

\bibitem{Chankowski:1992ej}
P.~H. Chankowski, S.~Pokorski, J.~Rosiek, {One loop corrections to the
  supersymmetric Higgs boson couplings and LEP phenomenology}, Phys.Lett. B286
  (1992) 307--314.
\newblock \href {http://dx.doi.org/10.1016/0370-2693(92)91780-D}
  {\path{doi:10.1016/0370-2693(92)91780-D}}.

\bibitem{Chankowski:1992er}
P.~H. Chankowski, S.~Pokorski, J.~Rosiek, {Complete on-shell renormalization
  scheme for the minimal supersymmetric Higgs sector}, Nucl.Phys. B423 (1994)
  437--496.
\newblock \href {http://arxiv.org/abs/hep-ph/9303309}
  {\path{arXiv:hep-ph/9303309}}, \href
  {http://dx.doi.org/10.1016/0550-3213(94)90141-4}
  {\path{doi:10.1016/0550-3213(94)90141-4}}.

\bibitem{Dabelstein:1994hb}
A.~Dabelstein, {The One loop renormalization of the MSSM Higgs sector and its
  application to the neutral scalar Higgs masses}, Z.Phys. C67 (1995) 495--512.
\newblock \href {http://arxiv.org/abs/hep-ph/9409375}
  {\path{arXiv:hep-ph/9409375}}, \href {http://dx.doi.org/10.1007/BF01624592}
  {\path{doi:10.1007/BF01624592}}.

\bibitem{Dabelstein:1995js}
A.~Dabelstein, {Fermionic decays of neutral MSSM Higgs bosons at the one loop
  level}, Nucl.Phys. B456 (1995) 25--56.
\newblock \href {http://arxiv.org/abs/hep-ph/9503443}
  {\path{arXiv:hep-ph/9503443}}, \href
  {http://dx.doi.org/10.1016/0550-3213(95)00523-2}
  {\path{doi:10.1016/0550-3213(95)00523-2}}.

\bibitem{Martin:2002iu}
S.~P. Martin, {Two loop effective potential for the minimal supersymmetric
  standard model}, Phys.Rev. D66 (2002) 096001.
\newblock \href {http://arxiv.org/abs/hep-ph/0206136}
  {\path{arXiv:hep-ph/0206136}}, \href
  {http://dx.doi.org/10.1103/PhysRevD.66.096001}
  {\path{doi:10.1103/PhysRevD.66.096001}}.

\bibitem{Martin:2002wn}
S.~P. Martin, {Complete two loop effective potential approximation to the
  lightest Higgs scalar boson mass in supersymmetry}, Phys.Rev. D67 (2003)
  095012.
\newblock \href {http://arxiv.org/abs/hep-ph/0211366}
  {\path{arXiv:hep-ph/0211366}}, \href
  {http://dx.doi.org/10.1103/PhysRevD.67.095012}
  {\path{doi:10.1103/PhysRevD.67.095012}}.

\bibitem{Kodaira:1993yt}
J.~Kodaira, Y.~Yasui, K.~Sasaki, {The Mass of the lightest supersymmetric Higgs
  boson beyond the leading logarithm approximation}, Phys.Rev. D50 (1994)
  7035--7041.
\newblock \href {http://arxiv.org/abs/hep-ph/9311366}
  {\path{arXiv:hep-ph/9311366}}, \href
  {http://dx.doi.org/10.1103/PhysRevD.50.7035}
  {\path{doi:10.1103/PhysRevD.50.7035}}.

\bibitem{Gladyshev:1994iw}
A.~Gladyshev, D.~Kazakov, {Renormalization group improved radiative corrections
  to the supersymmetric Higgs boson masses}, Mod.Phys.Lett. A10 (1995)
  3129--3138.
\newblock \href {http://arxiv.org/abs/hep-ph/9411209}
  {\path{arXiv:hep-ph/9411209}}, \href
  {http://dx.doi.org/10.1142/S0217732395003288}
  {\path{doi:10.1142/S0217732395003288}}.

\bibitem{Carena:1995bx}
M.~S. Carena, J.~Espinosa, M.~Quiros, C.~Wagner, {Analytical expressions for
  radiatively corrected Higgs masses and couplings in the MSSM}, Phys.Lett.
  B355 (1995) 209--221.
\newblock \href {http://arxiv.org/abs/hep-ph/9504316}
  {\path{arXiv:hep-ph/9504316}}, \href
  {http://dx.doi.org/10.1016/0370-2693(95)00694-G}
  {\path{doi:10.1016/0370-2693(95)00694-G}}.

\bibitem{Casas:1994us}
J.~Casas, J.~Espinosa, M.~Quiros, A.~Riotto, {The Lightest Higgs boson mass in
  the minimal supersymmetric standard model}, Nucl.Phys. B436 (1995) 3--29.
\newblock \href {http://arxiv.org/abs/hep-ph/9407389}
  {\path{arXiv:hep-ph/9407389}}, \href
  {http://dx.doi.org/10.1016/0550-3213(94)00508-C}
  {\path{doi:10.1016/0550-3213(94)00508-C}}.

\bibitem{Carena:1995wu}
M.~S. Carena, M.~Quiros, C.~Wagner, {Effective potential methods and the Higgs
  mass spectrum in the MSSM}, Nucl.Phys. B461 (1996) 407--436.
\newblock \href {http://arxiv.org/abs/hep-ph/9508343}
  {\path{arXiv:hep-ph/9508343}}, \href
  {http://dx.doi.org/10.1016/0550-3213(95)00665-6}
  {\path{doi:10.1016/0550-3213(95)00665-6}}.

\bibitem{Hempfling:1993qq}
R.~Hempfling, A.~H. Hoang, {Two loop radiative corrections to the upper limit
  of the lightest Higgs boson mass in the minimal supersymmetric model},
  Phys.Lett. B331 (1994) 99--106.
\newblock \href {http://arxiv.org/abs/hep-ph/9401219}
  {\path{arXiv:hep-ph/9401219}}, \href
  {http://dx.doi.org/10.1016/0370-2693(94)90948-2}
  {\path{doi:10.1016/0370-2693(94)90948-2}}.

\bibitem{Zhang:1998bm}
R.-J. Zhang, {Two loop effective potential calculation of the lightest CP even
  Higgs boson mass in the MSSM}, Phys.Lett. B447 (1999) 89--97.
\newblock \href {http://arxiv.org/abs/hep-ph/9808299}
  {\path{arXiv:hep-ph/9808299}}, \href
  {http://dx.doi.org/10.1016/S0370-2693(98)01575-5}
  {\path{doi:10.1016/S0370-2693(98)01575-5}}.

\bibitem{Espinosa:1999zm}
J.~R. Espinosa, R.-J. Zhang, {MSSM lightest CP even Higgs boson mass to
  O(alpha(s) alpha(t)): The Effective potential approach}, JHEP 0003 (2000)
  026.
\newblock \href {http://arxiv.org/abs/hep-ph/9912236}
  {\path{arXiv:hep-ph/9912236}}, \href
  {http://dx.doi.org/10.1088/1126-6708/2000/03/026}
  {\path{doi:10.1088/1126-6708/2000/03/026}}.

\bibitem{Espinosa:2000df}
J.~R. Espinosa, R.-J. Zhang, {Complete two loop dominant corrections to the
  mass of the lightest CP even Higgs boson in the minimal supersymmetric
  standard model}, Nucl.Phys. B586 (2000) 3--38.
\newblock \href {http://arxiv.org/abs/hep-ph/0003246}
  {\path{arXiv:hep-ph/0003246}}, \href
  {http://dx.doi.org/10.1016/S0550-3213(00)00421-1}
  {\path{doi:10.1016/S0550-3213(00)00421-1}}.

\bibitem{Degrassi:2001yf}
G.~Degrassi, P.~Slavich, F.~Zwirner, {On the neutral Higgs boson masses in the
  MSSM for arbitrary stop mixing}, Nucl.Phys. B611 (2001) 403--422.
\newblock \href {http://arxiv.org/abs/hep-ph/0105096}
  {\path{arXiv:hep-ph/0105096}}, \href
  {http://dx.doi.org/10.1016/S0550-3213(01)00343-1}
  {\path{doi:10.1016/S0550-3213(01)00343-1}}.

\bibitem{Brignole:2001jy}
A.~Brignole, G.~Degrassi, P.~Slavich, F.~Zwirner, {On the O(alpha(t)**2) two
  loop corrections to the neutral Higgs boson masses in the MSSM}, Nucl.Phys.
  B631 (2002) 195--218.
\newblock \href {http://arxiv.org/abs/hep-ph/0112177}
  {\path{arXiv:hep-ph/0112177}}, \href
  {http://dx.doi.org/10.1016/S0550-3213(02)00184-0}
  {\path{doi:10.1016/S0550-3213(02)00184-0}}.

\bibitem{Brignole:2002bz}
A.~Brignole, G.~Degrassi, P.~Slavich, F.~Zwirner, {On the two loop sbottom
  corrections to the neutral Higgs boson masses in the MSSM}, Nucl.Phys. B643
  (2002) 79--92.
\newblock \href {http://arxiv.org/abs/hep-ph/0206101}
  {\path{arXiv:hep-ph/0206101}}, \href
  {http://dx.doi.org/10.1016/S0550-3213(02)00748-4}
  {\path{doi:10.1016/S0550-3213(02)00748-4}}.

\bibitem{Dedes:2003km}
A.~Dedes, G.~Degrassi, P.~Slavich, {On the two loop Yukawa corrections to the
  MSSM Higgs boson masses at large tan beta}, Nucl.Phys. B672 (2003) 144--162.
\newblock \href {http://arxiv.org/abs/hep-ph/0305127}
  {\path{arXiv:hep-ph/0305127}}, \href
  {http://dx.doi.org/10.1016/j.nuclphysb.2003.08.033}
  {\path{doi:10.1016/j.nuclphysb.2003.08.033}}.

\bibitem{Heinemeyer:1998jw}
S.~Heinemeyer, W.~Hollik, G.~Weiglein, {QCD corrections to the masses of the
  neutral CP - even Higgs bosons in the MSSM}, Phys.Rev. D58 (1998) 091701.
\newblock \href {http://arxiv.org/abs/hep-ph/9803277}
  {\path{arXiv:hep-ph/9803277}}, \href
  {http://dx.doi.org/10.1103/PhysRevD.58.091701}
  {\path{doi:10.1103/PhysRevD.58.091701}}.

\bibitem{Heinemeyer:1998kz}
S.~Heinemeyer, W.~Hollik, G.~Weiglein, {Precise prediction for the mass of the
  lightest Higgs boson in the MSSM}, Phys.Lett. B440 (1998) 296--304.
\newblock \href {http://arxiv.org/abs/hep-ph/9807423}
  {\path{arXiv:hep-ph/9807423}}, \href
  {http://dx.doi.org/10.1016/S0370-2693(98)01116-2}
  {\path{doi:10.1016/S0370-2693(98)01116-2}}.

\bibitem{Heinemeyer:1998np}
S.~Heinemeyer, W.~Hollik, G.~Weiglein, {The Masses of the neutral CP - even
  Higgs bosons in the MSSM: Accurate analysis at the two loop level},
  Eur.Phys.J. C9 (1999) 343--366.
\newblock \href {http://arxiv.org/abs/hep-ph/9812472}
  {\path{arXiv:hep-ph/9812472}}, \href
  {http://dx.doi.org/10.1007/s100529900006} {\path{doi:10.1007/s100529900006}}.

\bibitem{Heinemeyer:1999be}
S.~Heinemeyer, W.~Hollik, G.~Weiglein, {The Mass of the lightest MSSM Higgs
  boson: A Compact analytical expression at the two loop level}, Phys.Lett.
  B455 (1999) 179--191.
\newblock \href {http://arxiv.org/abs/hep-ph/9903404}
  {\path{arXiv:hep-ph/9903404}}, \href
  {http://dx.doi.org/10.1016/S0370-2693(99)00417-7}
  {\path{doi:10.1016/S0370-2693(99)00417-7}}.

\bibitem{Espinosa:1991fc}
J.~Espinosa, M.~Quiros, {Two loop radiative corrections to the mass of the
  lightest Higgs boson in supersymmetric standard models}, Phys.Lett. B266
  (1991) 389--396.
\newblock \href {http://dx.doi.org/10.1016/0370-2693(91)91056-2}
  {\path{doi:10.1016/0370-2693(91)91056-2}}.

\bibitem{Hempfling:1993kv}
R.~Hempfling, {Yukawa coupling unification with supersymmetric threshold
  corrections}, Phys.Rev. D49 (1994) 6168--6172.
\newblock \href {http://dx.doi.org/10.1103/PhysRevD.49.6168}
  {\path{doi:10.1103/PhysRevD.49.6168}}.

\bibitem{Hall:1993gn}
L.~J. Hall, R.~Rattazzi, U.~Sarid, {The Top quark mass in supersymmetric SO(10)
  unification}, Phys.Rev. D50 (1994) 7048--7065.
\newblock \href {http://arxiv.org/abs/hep-ph/9306309}
  {\path{arXiv:hep-ph/9306309}}, \href
  {http://dx.doi.org/10.1103/PhysRevD.50.7048}
  {\path{doi:10.1103/PhysRevD.50.7048}}.

\bibitem{Carena:1994bv}
M.~S. Carena, M.~Olechowski, S.~Pokorski, C.~Wagner, {Electroweak symmetry
  breaking and bottom - top Yukawa unification}, Nucl.Phys. B426 (1994)
  269--300.
\newblock \href {http://arxiv.org/abs/hep-ph/9402253}
  {\path{arXiv:hep-ph/9402253}}, \href
  {http://dx.doi.org/10.1016/0550-3213(94)90313-1}
  {\path{doi:10.1016/0550-3213(94)90313-1}}.

\bibitem{Carena:1999py}
M.~S. Carena, D.~Garcia, U.~Nierste, C.~E. Wagner, {Effective Lagrangian for
  the $\bar{t} b H^{+}$ interaction in the MSSM and charged Higgs
  phenomenology}, Nucl.Phys. B577 (2000) 88--120.
\newblock \href {http://arxiv.org/abs/hep-ph/9912516}
  {\path{arXiv:hep-ph/9912516}}, \href
  {http://dx.doi.org/10.1016/S0550-3213(00)00146-2}
  {\path{doi:10.1016/S0550-3213(00)00146-2}}.

\bibitem{Espinosa:2001mm}
J.~Espinosa, I.~Navarro, {Radiative corrections to the Higgs boson mass for a
  hierarchical stop spectrum}, Nucl.Phys. B615 (2001) 82--116.
\newblock \href {http://arxiv.org/abs/hep-ph/0104047}
  {\path{arXiv:hep-ph/0104047}}, \href
  {http://dx.doi.org/10.1016/S0550-3213(01)00429-1}
  {\path{doi:10.1016/S0550-3213(01)00429-1}}.

\bibitem{Carena:2000dp}
M.~S. Carena, H.~Haber, S.~Heinemeyer, W.~Hollik, C.~Wagner, et~al.,
  {Reconciling the two loop diagrammatic and effective field theory
  computations of the mass of the lightest CP - even Higgs boson in the MSSM},
  Nucl.Phys. B580 (2000) 29--57.
\newblock \href {http://arxiv.org/abs/hep-ph/0001002}
  {\path{arXiv:hep-ph/0001002}}, \href
  {http://dx.doi.org/10.1016/S0550-3213(00)00212-1}
  {\path{doi:10.1016/S0550-3213(00)00212-1}}.

\bibitem{Degrassi:2002fi}
G.~Degrassi, S.~Heinemeyer, W.~Hollik, P.~Slavich, G.~Weiglein, {Towards high
  precision predictions for the MSSM Higgs sector}, Eur.Phys.J. C28 (2003)
  133--143.
\newblock \href {http://arxiv.org/abs/hep-ph/0212020}
  {\path{arXiv:hep-ph/0212020}}, \href
  {http://dx.doi.org/10.1140/epjc/s2003-01152-2}
  {\path{doi:10.1140/epjc/s2003-01152-2}}.

\bibitem{Martin:2007pg}
S.~P. Martin, {Three-loop corrections to the lightest Higgs scalar boson mass
  in supersymmetry}, Phys.Rev. D75 (2007) 055005.
\newblock \href {http://arxiv.org/abs/hep-ph/0701051}
  {\path{arXiv:hep-ph/0701051}}, \href
  {http://dx.doi.org/10.1103/PhysRevD.75.055005}
  {\path{doi:10.1103/PhysRevD.75.055005}}.

\bibitem{Harlander:2008ju}
R.~Harlander, P.~Kant, L.~Mihaila, M.~Steinhauser, {Higgs boson mass in
  supersymmetry to three loops}, Phys.Rev.Lett. 100 (2008) 191602.
\newblock \href {http://arxiv.org/abs/0803.0672} {\path{arXiv:0803.0672}},
  \href {http://dx.doi.org/10.1103/PhysRevLett.101.039901,
  10.1103/PhysRevLett.100.191602} {\path{doi:10.1103/PhysRevLett.101.039901,
  10.1103/PhysRevLett.100.191602}}.

\bibitem{Kant:2010tf}
P.~Kant, R.~Harlander, L.~Mihaila, M.~Steinhauser, {Light MSSM Higgs boson mass
  to three-loop accuracy}, JHEP 1008 (2010) 104.
\newblock \href {http://arxiv.org/abs/1005.5709} {\path{arXiv:1005.5709}},
  \href {http://dx.doi.org/10.1007/JHEP08(2010)104}
  {\path{doi:10.1007/JHEP08(2010)104}}.

\bibitem{Hahn:2013ria}
T.~Hahn, S.~Heinemeyer, W.~Hollik, H.~Rzehak, G.~Weiglein, {High-precision
  predictions for the light CP-even Higgs Boson Mass of the MSSM }\href
  {http://arxiv.org/abs/1312.4937} {\path{arXiv:1312.4937}}.

\bibitem{Draper:2013oza}
P.~Draper, G.~Lee, C.~E.~M. Wagner, {Precise Estimates of the Higgs Mass in
  Heavy SUSY}, Phys.Rev. D89 (2014) 055023.
\newblock \href {http://arxiv.org/abs/1312.5743} {\path{arXiv:1312.5743}},
  \href {http://dx.doi.org/10.1103/PhysRevD.89.055023}
  {\path{doi:10.1103/PhysRevD.89.055023}}.

\bibitem{Heinemeyer:2004ms}
S.~Heinemeyer, {MSSM Higgs physics at higher orders}, Int.J.Mod.Phys. A21
  (2006) 2659--2772.
\newblock \href {http://arxiv.org/abs/hep-ph/0407244}
  {\path{arXiv:hep-ph/0407244}}, \href
  {http://dx.doi.org/10.1142/S0217751X06031028}
  {\path{doi:10.1142/S0217751X06031028}}.

\bibitem{Frank:2006yh}
M.~Frank, T.~Hahn, S.~Heinemeyer, W.~Hollik, H.~Rzehak, et~al., {The Higgs
  Boson Masses and Mixings of the Complex MSSM in the Feynman-Diagrammatic
  Approach}, JHEP 0702 (2007) 047.
\newblock \href {http://arxiv.org/abs/hep-ph/0611326}
  {\path{arXiv:hep-ph/0611326}}, \href
  {http://dx.doi.org/10.1088/1126-6708/2007/02/047}
  {\path{doi:10.1088/1126-6708/2007/02/047}}.

\bibitem{Martin:2003it}
S.~P. Martin, {Two loop scalar self energies in a general renormalizable theory
  at leading order in gauge couplings}, Phys.Rev. D70 (2004) 016005.
\newblock \href {http://arxiv.org/abs/hep-ph/0312092}
  {\path{arXiv:hep-ph/0312092}}, \href
  {http://dx.doi.org/10.1103/PhysRevD.70.016005}
  {\path{doi:10.1103/PhysRevD.70.016005}}.

\bibitem{Martin:2004kr}
S.~P. Martin, {Strong and Yukawa two-loop contributions to Higgs scalar boson
  self-energies and pole masses in supersymmetry}, Phys.Rev. D71 (2005) 016012.
\newblock \href {http://arxiv.org/abs/hep-ph/0405022}
  {\path{arXiv:hep-ph/0405022}}, \href
  {http://dx.doi.org/10.1103/PhysRevD.71.016012}
  {\path{doi:10.1103/PhysRevD.71.016012}}.

\bibitem{Martin:2005qm}
S.~P. Martin, D.~G. Robertson, {TSIL: A Program for the calculation of two-loop
  self-energy integrals}, Comput.Phys.Commun. 174 (2006) 133--151.
\newblock \href {http://arxiv.org/abs/hep-ph/0501132}
  {\path{arXiv:hep-ph/0501132}}, \href
  {http://dx.doi.org/10.1016/j.cpc.2005.08.005}
  {\path{doi:10.1016/j.cpc.2005.08.005}}.

\bibitem{Heinemeyer:1998yj}
S.~Heinemeyer, W.~Hollik, G.~Weiglein, {FeynHiggs: A Program for the
  calculation of the masses of the neutral CP even Higgs bosons in the MSSM},
  Comput.Phys.Commun. 124 (2000) 76--89.
\newblock \href {http://arxiv.org/abs/hep-ph/9812320}
  {\path{arXiv:hep-ph/9812320}}, \href
  {http://dx.doi.org/10.1016/S0010-4655(99)00364-1}
  {\path{doi:10.1016/S0010-4655(99)00364-1}}.

\bibitem{Frank:2002qf}
M.~Frank, S.~Heinemeyer, W.~Hollik, G.~Weiglein, {FeynHiggs1.2: Hybrid MS-bar /
  on-shell renormalization for the CP even Higgs boson sector in the MSSM
  }\href {http://arxiv.org/abs/hep-ph/0202166} {\path{arXiv:hep-ph/0202166}}.

\bibitem{Hahn:2009zz}
T.~Hahn, S.~Heinemeyer, W.~Hollik, H.~Rzehak, G.~Weiglein, {FeynHiggs: A
  program for the calculation of MSSM Higgs-boson observables - Version 2.6.5},
  Comput.Phys.Commun. 180 (2009) 1426--1427.
\newblock \href {http://dx.doi.org/10.1016/j.cpc.2009.02.014}
  {\path{doi:10.1016/j.cpc.2009.02.014}}.

\bibitem{Hahn:2010te}
T.~Hahn, S.~Heinemeyer, W.~Hollik, H.~Rzehak, G.~Weiglein, {FeynHiggs 2.7},
  Nucl.Phys.Proc.Suppl. 205-206 (2010) 152--157.
\newblock \href {http://arxiv.org/abs/1007.0956} {\path{arXiv:1007.0956}},
  \href {http://dx.doi.org/10.1016/j.nuclphysbps.2010.08.035}
  {\path{doi:10.1016/j.nuclphysbps.2010.08.035}}.

\bibitem{Bogoliubov:1957gp}
N.~Bogoliubov, O.~a. Parasiuk, {On the Multiplication of the causal function in
  the quantum theory of fields}, Acta Math. 97 (1957) 227--266.
\newblock \href {http://dx.doi.org/10.1007/BF02392399}
  {\path{doi:10.1007/BF02392399}}.

\bibitem{Hepp:1966eg}
K.~Hepp, Proof of the {B}ogolyubov-{P}arasiuk theorem on renormalization,
  Commun. Math. Phys. 2 (1966) 301--326.

\bibitem{Zimmermann:1968mu}
W.~Zimmermann, {The power counting theorem for minkowski metric},
  Commun.Math.Phys. 11 (1968) 1--8.
\newblock \href {http://dx.doi.org/10.1007/BF01654298}
  {\path{doi:10.1007/BF01654298}}.

\bibitem{Frixione:1995ms}
S.~Frixione, Z.~Kunszt, A.~Signer, {Three jet cross-sections to next-to-leading
  order}, Nucl.Phys. B467 (1996) 399--442.
\newblock \href {http://arxiv.org/abs/hep-ph/9512328}
  {\path{arXiv:hep-ph/9512328}}, \href
  {http://dx.doi.org/10.1016/0550-3213(96)00110-1}
  {\path{doi:10.1016/0550-3213(96)00110-1}}.

\bibitem{Frixione:1997np}
S.~Frixione, {A General approach to jet cross-sections in QCD}, Nucl.Phys. B507
  (1997) 295--314.
\newblock \href {http://arxiv.org/abs/hep-ph/9706545}
  {\path{arXiv:hep-ph/9706545}}, \href
  {http://dx.doi.org/10.1016/S0550-3213(97)00574-9}
  {\path{doi:10.1016/S0550-3213(97)00574-9}}.

\bibitem{Catani:1996vz}
S.~Catani, M.~Seymour, {A General algorithm for calculating jet cross-sections
  in NLO QCD}, Nucl.Phys. B485 (1997) 291--419.
\newblock \href {http://arxiv.org/abs/hep-ph/9605323}
  {\path{arXiv:hep-ph/9605323}}, \href
  {http://dx.doi.org/10.1016/S0550-3213(96)00589-5}
  {\path{doi:10.1016/S0550-3213(96)00589-5}}.

\bibitem{Catani:1998nv}
S.~Catani, M.~Grazzini, {Collinear factorization and splitting functions for
  next-to-next-to-leading order QCD calculations}, Phys.Lett. B446 (1999)
  143--152.
\newblock \href {http://arxiv.org/abs/hep-ph/9810389}
  {\path{arXiv:hep-ph/9810389}}, \href
  {http://dx.doi.org/10.1016/S0370-2693(98)01513-5}
  {\path{doi:10.1016/S0370-2693(98)01513-5}}.

\bibitem{Catani:1999ss}
S.~Catani, M.~Grazzini, {Infrared factorization of tree level QCD amplitudes at
  the next-to-next-to-leading order and beyond}, Nucl.Phys. B570 (2000)
  287--325.
\newblock \href {http://arxiv.org/abs/hep-ph/9908523}
  {\path{arXiv:hep-ph/9908523}}, \href
  {http://dx.doi.org/10.1016/S0550-3213(99)00778-6}
  {\path{doi:10.1016/S0550-3213(99)00778-6}}.

\bibitem{Binoth:2000ps}
T.~Binoth, G.~Heinrich, An automatized algorithm to compute infrared divergent
  multi-loop integrals, Nucl. Phys. B585 (2000) 741--759.
\newblock \href {http://arxiv.org/abs/hep-ph/0004013}
  {\path{arXiv:hep-ph/0004013}}.

\bibitem{Heinrich:2008si}
G.~Heinrich, {Sector Decomposition}, Int. J. Mod. Phys. A23 (2008) 1457--1486.
\newblock \href {http://arxiv.org/abs/0803.4177} {\path{arXiv:0803.4177}},
  \href {http://dx.doi.org/10.1142/S0217751X08040263}
  {\path{doi:10.1142/S0217751X08040263}}.

\bibitem{Heinrich:2002rc}
G.~Heinrich, {A numerical method for NNLO calculations}, Nucl.Phys.Proc.Suppl.
  116 (2003) 368--372.
\newblock \href {http://arxiv.org/abs/hep-ph/0211144}
  {\path{arXiv:hep-ph/0211144}}, \href
  {http://dx.doi.org/10.1016/S0920-5632(03)80201-3}
  {\path{doi:10.1016/S0920-5632(03)80201-3}}.

\bibitem{GehrmannDeRidder:2003bm}
A.~Gehrmann-De~Ridder, T.~Gehrmann, G.~Heinrich, {Four particle phase space
  integrals in massless QCD}, Nucl.Phys. B682 (2004) 265--288.
\newblock \href {http://arxiv.org/abs/hep-ph/0311276}
  {\path{arXiv:hep-ph/0311276}}, \href
  {http://dx.doi.org/10.1016/j.nuclphysb.2004.01.023}
  {\path{doi:10.1016/j.nuclphysb.2004.01.023}}.

\bibitem{Anastasiou:2003gr}
C.~Anastasiou, K.~Melnikov, F.~Petriello, {A new method for real radiation at
  NNLO}, Phys.Rev. D69 (2004) 076010.
\newblock \href {http://arxiv.org/abs/hep-ph/0311311}
  {\path{arXiv:hep-ph/0311311}}, \href
  {http://dx.doi.org/10.1103/PhysRevD.69.076010}
  {\path{doi:10.1103/PhysRevD.69.076010}}.

\bibitem{Binoth:2004jv}
T.~Binoth, G.~Heinrich, {Numerical evaluation of phase space integrals by
  sector decomposition}, Nucl.Phys. B693 (2004) 134--148.
\newblock \href {http://arxiv.org/abs/hep-ph/0402265}
  {\path{arXiv:hep-ph/0402265}}, \href
  {http://dx.doi.org/10.1016/j.nuclphysb.2004.06.005}
  {\path{doi:10.1016/j.nuclphysb.2004.06.005}}.

\bibitem{Anastasiou:2004qd}
C.~Anastasiou, K.~Melnikov, F.~Petriello, {Real radiation at NNLO: $e^+ e^- \to
  2$ jets through O(alpha**2(s))}, Phys.Rev.Lett. 93 (2004) 032002.
\newblock \href {http://arxiv.org/abs/hep-ph/0402280}
  {\path{arXiv:hep-ph/0402280}}, \href
  {http://dx.doi.org/10.1103/PhysRevLett.93.032002}
  {\path{doi:10.1103/PhysRevLett.93.032002}}.

\bibitem{Czakon:2010td}
M.~Czakon, {A novel subtraction scheme for double-real radiation at NNLO},
  Phys.Lett. B693 (2010) 259--268.
\newblock \href {http://arxiv.org/abs/1005.0274} {\path{arXiv:1005.0274}},
  \href {http://dx.doi.org/10.1016/j.physletb.2010.08.036}
  {\path{doi:10.1016/j.physletb.2010.08.036}}.

\bibitem{Boughezal:2011jf}
R.~Boughezal, K.~Melnikov, F.~Petriello, {A subtraction scheme for NNLO
  computations}, Phys.Rev. D85 (2012) 034025.
\newblock \href {http://arxiv.org/abs/1111.7041} {\path{arXiv:1111.7041}},
  \href {http://dx.doi.org/10.1103/PhysRevD.85.034025}
  {\path{doi:10.1103/PhysRevD.85.034025}}.

\bibitem{Kosower:1997zr}
D.~A. Kosower, {Antenna factorization of gauge theory amplitudes}, Phys.Rev.
  D57 (1998) 5410--5416.
\newblock \href {http://arxiv.org/abs/hep-ph/9710213}
  {\path{arXiv:hep-ph/9710213}}, \href
  {http://dx.doi.org/10.1103/PhysRevD.57.5410}
  {\path{doi:10.1103/PhysRevD.57.5410}}.

\bibitem{Campbell:1998nn}
J.~M. Campbell, M.~Cullen, E.~N. Glover, {Four jet event shapes in electron -
  positron annihilation}, Eur.Phys.J. C9 (1999) 245--265.
\newblock \href {http://arxiv.org/abs/hep-ph/9809429}
  {\path{arXiv:hep-ph/9809429}}, \href
  {http://dx.doi.org/10.1007/s100529900034} {\path{doi:10.1007/s100529900034}}.

\bibitem{Kosower:2003bh}
D.~A. Kosower, {Antenna factorization in strongly ordered limits}, Phys.Rev.
  D71 (2005) 045016.
\newblock \href {http://arxiv.org/abs/hep-ph/0311272}
  {\path{arXiv:hep-ph/0311272}}, \href
  {http://dx.doi.org/10.1103/PhysRevD.71.045016}
  {\path{doi:10.1103/PhysRevD.71.045016}}.

\bibitem{GehrmannDeRidder:2004tv}
A.~Gehrmann-De~Ridder, T.~Gehrmann, E.~N. Glover, {Infrared structure of $e^+
  e^- \to 2$ jets at NNLO}, Nucl.Phys. B691 (2004) 195--222.
\newblock \href {http://arxiv.org/abs/hep-ph/0403057}
  {\path{arXiv:hep-ph/0403057}}, \href
  {http://dx.doi.org/10.1016/j.nuclphysb.2004.05.017}
  {\path{doi:10.1016/j.nuclphysb.2004.05.017}}.

\bibitem{GehrmannDeRidder:2005cm}
A.~Gehrmann-De~Ridder, T.~Gehrmann, E.~N. Glover, {Antenna subtraction at
  NNLO}, JHEP 0509 (2005) 056.
\newblock \href {http://arxiv.org/abs/hep-ph/0505111}
  {\path{arXiv:hep-ph/0505111}}, \href
  {http://dx.doi.org/10.1088/1126-6708/2005/09/056}
  {\path{doi:10.1088/1126-6708/2005/09/056}}.

\bibitem{GehrmannDeRidder:2007bj}
A.~Gehrmann-De~Ridder, T.~Gehrmann, E.~Glover, G.~Heinrich, {Second-order QCD
  corrections to the thrust distribution}, Phys.Rev.Lett. 99 (2007) 132002.
\newblock \href {http://arxiv.org/abs/0707.1285} {\path{arXiv:0707.1285}},
  \href {http://dx.doi.org/10.1103/PhysRevLett.99.132002}
  {\path{doi:10.1103/PhysRevLett.99.132002}}.

\bibitem{Currie:2013vh}
J.~Currie, E.~Glover, S.~Wells, {Infrared Structure at NNLO Using Antenna
  Subtraction}, JHEP 1304 (2013) 066.
\newblock \href {http://arxiv.org/abs/1301.4693} {\path{arXiv:1301.4693}},
  \href {http://dx.doi.org/10.1007/JHEP04(2013)066}
  {\path{doi:10.1007/JHEP04(2013)066}}.

\bibitem{Somogyi:2006cz}
G.~Somogyi, Z.~Trocsanyi, {A New subtraction scheme for computing QCD jet cross
  sections at next-to-leading order accuracy }\href
  {http://arxiv.org/abs/hep-ph/0609041} {\path{arXiv:hep-ph/0609041}}.

\bibitem{Somogyi:2006da}
G.~Somogyi, Z.~Trocsanyi, V.~Del~Duca, {A Subtraction scheme for computing QCD
  jet cross sections at NNLO: Regularization of doubly-real emissions}, JHEP
  0701 (2007) 070.
\newblock \href {http://arxiv.org/abs/hep-ph/0609042}
  {\path{arXiv:hep-ph/0609042}}, \href
  {http://dx.doi.org/10.1088/1126-6708/2007/01/070}
  {\path{doi:10.1088/1126-6708/2007/01/070}}.

\bibitem{Alwall:2014hca}
J.~Alwall, R.~Frederix, S.~Frixione, V.~Hirschi, F.~Maltoni, et~al., {The
  automated computation of tree-level and next-to-leading order differential
  cross sections, and their matching to parton shower simulations }\href
  {http://arxiv.org/abs/1405.0301} {\path{arXiv:1405.0301}}.

\bibitem{Berger:2008sj}
C.~Berger, Z.~Bern, L.~Dixon, F.~Febres~Cordero, D.~Forde, et~al., {An
  Automated Implementation of On-Shell Methods for One-Loop Amplitudes},
  Phys.Rev. D78 (2008) 036003.
\newblock \href {http://arxiv.org/abs/0803.4180} {\path{arXiv:0803.4180}},
  \href {http://dx.doi.org/10.1103/PhysRevD.78.036003}
  {\path{doi:10.1103/PhysRevD.78.036003}}.

\bibitem{Hahn:2010zi}
T.~Hahn, {Feynman Diagram Calculations with FeynArts, FormCalc, and LoopTools},
  PoS ACAT2010 (2010) 078.
\newblock \href {http://arxiv.org/abs/1006.2231} {\path{arXiv:1006.2231}}.

\bibitem{Cullen:2014yla}
G.~Cullen, H.~van Deurzen, N.~Greiner, G.~Heinrich, G.~Luisoni, et~al.,
  {GoSam-2.0: a tool for automated one-loop calculations within the Standard
  Model and beyond }\href {http://arxiv.org/abs/1404.7096}
  {\path{arXiv:1404.7096}}.

\bibitem{Bevilacqua:2011xh}
G.~Bevilacqua, M.~Czakon, M.~Garzelli, A.~van Hameren, A.~Kardos, et~al.,
  {HELAC-NLO}, Comput.Phys.Commun. 184 (2013) 986--997.
\newblock \href {http://arxiv.org/abs/1110.1499} {\path{arXiv:1110.1499}},
  \href {http://dx.doi.org/10.1016/j.cpc.2012.10.033}
  {\path{doi:10.1016/j.cpc.2012.10.033}}.

\bibitem{Bahr:2008pv}
M.~Bahr, S.~Gieseke, M.~Gigg, D.~Grellscheid, K.~Hamilton, et~al., {Herwig++
  Physics and Manual}, Eur.Phys.J. C58 (2008) 639--707.
\newblock \href {http://arxiv.org/abs/0803.0883} {\path{arXiv:0803.0883}},
  \href {http://dx.doi.org/10.1140/epjc/s10052-008-0798-9}
  {\path{doi:10.1140/epjc/s10052-008-0798-9}}.

\bibitem{Platzer:2011bc}
S.~Platzer, S.~Gieseke, {Dipole Showers and Automated NLO Matching in
  Herwig++}, Eur.Phys.J. C72 (2012) 2187.
\newblock \href {http://arxiv.org/abs/1109.6256} {\path{arXiv:1109.6256}},
  \href {http://dx.doi.org/10.1140/epjc/s10052-012-2187-7}
  {\path{doi:10.1140/epjc/s10052-012-2187-7}}.

\bibitem{Campbell:2011bn}
J.~M. Campbell, R.~K. Ellis, C.~Williams, {Vector boson pair production at the
  LHC}, JHEP 1107 (2011) 018.
\newblock \href {http://arxiv.org/abs/1105.0020} {\path{arXiv:1105.0020}},
  \href {http://dx.doi.org/10.1007/JHEP07(2011)018}
  {\path{doi:10.1007/JHEP07(2011)018}}.

\bibitem{Badger:2012pg}
S.~Badger, B.~Biedermann, P.~Uwer, V.~Yundin, {Numerical evaluation of virtual
  corrections to multi-jet production in massless QCD}, Comput.Phys.Commun. 184
  (2013) 1981--1998.
\newblock \href {http://arxiv.org/abs/1209.0100} {\path{arXiv:1209.0100}},
  \href {http://dx.doi.org/10.1016/j.cpc.2013.03.018}
  {\path{doi:10.1016/j.cpc.2013.03.018}}.

\bibitem{Cascioli:2011va}
F.~Cascioli, P.~Maierhofer, S.~Pozzorini, {Scattering Amplitudes with Open
  Loops}, Phys.Rev.Lett. 108 (2012) 111601.
\newblock \href {http://arxiv.org/abs/1111.5206} {\path{arXiv:1111.5206}},
  \href {http://dx.doi.org/10.1103/PhysRevLett.108.111601}
  {\path{doi:10.1103/PhysRevLett.108.111601}}.

\bibitem{Alioli:2010xd}
S.~Alioli, P.~Nason, C.~Oleari, E.~Re, {A general framework for implementing
  NLO calculations in shower Monte Carlo programs: the POWHEG BOX}, JHEP 1006
  (2010) 043.
\newblock \href {http://arxiv.org/abs/1002.2581} {\path{arXiv:1002.2581}},
  \href {http://dx.doi.org/10.1007/JHEP06(2010)043}
  {\path{doi:10.1007/JHEP06(2010)043}}.

\bibitem{Actis:2012qn}
S.~Actis, A.~Denner, L.~Hofer, A.~Scharf, S.~Uccirati, {Recursive generation of
  one-loop amplitudes in the Standard Model}, JHEP 1304 (2013) 037.
\newblock \href {http://arxiv.org/abs/1211.6316} {\path{arXiv:1211.6316}},
  \href {http://dx.doi.org/10.1007/JHEP04(2013)037}
  {\path{doi:10.1007/JHEP04(2013)037}}.

\bibitem{Gleisberg:2008ta}
T.~Gleisberg, S.~Hoeche, F.~Krauss, M.~Schonherr, S.~Schumann, et~al., {Event
  generation with SHERPA 1.1}, JHEP 0902 (2009) 007.
\newblock \href {http://arxiv.org/abs/0811.4622} {\path{arXiv:0811.4622}},
  \href {http://dx.doi.org/10.1088/1126-6708/2009/02/007}
  {\path{doi:10.1088/1126-6708/2009/02/007}}.

\bibitem{Arnold:2011wj}
K.~Arnold, J.~Bellm, G.~Bozzi, M.~Brieg, F.~Campanario, et~al., {VBFNLO: A
  Parton Level Monte Carlo for Processes with Electroweak Bosons -- Manual for
  Version 2.5.0 }\href {http://arxiv.org/abs/1107.4038}
  {\path{arXiv:1107.4038}}.

\bibitem{Kublbeck:1990xc}
J.~Kublbeck, M.~Bohm, A.~Denner, {Feyn Arts: Computer Algebraic Generation of
  Feynman Graphs and Amplitudes}, Comput.Phys.Commun. 60 (1990) 165--180.
\newblock \href {http://dx.doi.org/10.1016/0010-4655(90)90001-H}
  {\path{doi:10.1016/0010-4655(90)90001-H}}.

\bibitem{Hahn:2000kx}
T.~Hahn, {Generating Feynman diagrams and amplitudes with FeynArts 3},
  Comput.Phys.Commun. 140 (2001) 418--431.
\newblock \href {http://arxiv.org/abs/hep-ph/0012260}
  {\path{arXiv:hep-ph/0012260}}, \href
  {http://dx.doi.org/10.1016/S0010-4655(01)00290-9}
  {\path{doi:10.1016/S0010-4655(01)00290-9}}.

\bibitem{Nogueira:1991ex}
P.~Nogueira, {Automatic Feynman graph generation}, J.Comput.Phys. 105 (1993)
  279--289.
\newblock \href {http://dx.doi.org/10.1006/jcph.1993.1074}
  {\path{doi:10.1006/jcph.1993.1074}}.

\bibitem{Hahn:1998yk}
T.~Hahn, M.~Perez-Victoria, {Automatized one loop calculations in
  four-dimensions and D-dimensions}, Comput.Phys.Commun. 118 (1999) 153--165.
\newblock \href {http://arxiv.org/abs/hep-ph/9807565}
  {\path{arXiv:hep-ph/9807565}}, \href
  {http://dx.doi.org/10.1016/S0010-4655(98)00173-8}
  {\path{doi:10.1016/S0010-4655(98)00173-8}}.

\bibitem{Ellis:2007qk}
R.~K. Ellis, G.~Zanderighi, {Scalar one-loop integrals for QCD}, JHEP 0802
  (2008) 002.
\newblock \href {http://arxiv.org/abs/0712.1851} {\path{arXiv:0712.1851}},
  \href {http://dx.doi.org/10.1088/1126-6708/2008/02/002}
  {\path{doi:10.1088/1126-6708/2008/02/002}}.

\bibitem{Binoth:2008uq}
T.~Binoth, J.-P. Guillet, G.~Heinrich, E.~Pilon, T.~Reiter, {Golem95: A
  Numerical program to calculate one-loop tensor integrals with up to six
  external legs}, Comput.Phys.Commun. 180 (2009) 2317--2330.
\newblock \href {http://arxiv.org/abs/0810.0992} {\path{arXiv:0810.0992}},
  \href {http://dx.doi.org/10.1016/j.cpc.2009.06.024}
  {\path{doi:10.1016/j.cpc.2009.06.024}}.

\bibitem{vanHameren:2010cp}
A.~van Hameren, {OneLOop: For the evaluation of one-loop scalar functions},
  Comput.Phys.Commun. 182 (2011) 2427--2438.
\newblock \href {http://arxiv.org/abs/1007.4716} {\path{arXiv:1007.4716}},
  \href {http://dx.doi.org/10.1016/j.cpc.2011.06.011}
  {\path{doi:10.1016/j.cpc.2011.06.011}}.

\bibitem{'tHooft:1972fi}
G.~'t~Hooft, M.~Veltman, {Regularization and Renormalization of Gauge Fields},
  Nucl.Phys. B44 (1972) 189--213.
\newblock \href {http://dx.doi.org/10.1016/0550-3213(72)90279-9}
  {\path{doi:10.1016/0550-3213(72)90279-9}}.

\bibitem{Tkachov:1981wb}
F.~Tkachov, {A Theorem on Analytical Calculability of Four Loop Renormalization
  Group Functions}, Phys.Lett. B100 (1981) 65--68.
\newblock \href {http://dx.doi.org/10.1016/0370-2693(81)90288-4}
  {\path{doi:10.1016/0370-2693(81)90288-4}}.

\bibitem{Chetyrkin:1981qh}
K.~Chetyrkin, F.~Tkachov, {Integration by Parts: The Algorithm to Calculate
  beta Functions in 4 Loops}, Nucl.Phys. B192 (1981) 159--204.
\newblock \href {http://dx.doi.org/10.1016/0550-3213(81)90199-1}
  {\path{doi:10.1016/0550-3213(81)90199-1}}.

\bibitem{Gehrmann:1999as}
T.~Gehrmann, E.~Remiddi, {Differential equations for two loop four point
  functions}, Nucl.Phys. B580 (2000) 485--518.
\newblock \href {http://arxiv.org/abs/hep-ph/9912329}
  {\path{arXiv:hep-ph/9912329}}, \href
  {http://dx.doi.org/10.1016/S0550-3213(00)00223-6}
  {\path{doi:10.1016/S0550-3213(00)00223-6}}.

\bibitem{Bardin:1999ak}
D.~Y. Bardin, G.~Passarino, {The standard model in the making: Precision study
  of the electroweak interactions }.

\bibitem{Hollik:2005nn}
W.~Hollik, D.~Stockinger, {MSSM Higgs-boson mass predictions and two-loop
  non-supersymmetric counterterms}, Phys.Lett. B634 (2006) 63--68.
\newblock \href {http://arxiv.org/abs/hep-ph/0509298}
  {\path{arXiv:hep-ph/0509298}}, \href
  {http://dx.doi.org/10.1016/j.physletb.2006.01.030}
  {\path{doi:10.1016/j.physletb.2006.01.030}}.

\bibitem{Siegel:1979wq}
W.~Siegel, {Supersymmetric Dimensional Regularization via Dimensional
  Reduction}, Phys.Lett. B84 (1979) 193.
\newblock \href {http://dx.doi.org/10.1016/0370-2693(79)90282-X}
  {\path{doi:10.1016/0370-2693(79)90282-X}}.

\bibitem{Capper:1979ns}
D.~Capper, D.~Jones, P.~van Nieuwenhuizen, {Regularization by Dimensional
  Reduction of Supersymmetric and Nonsupersymmetric Gauge Theories}, Nucl.Phys.
  B167 (1980) 479.
\newblock \href {http://dx.doi.org/10.1016/0550-3213(80)90244-8}
  {\path{doi:10.1016/0550-3213(80)90244-8}}.

\bibitem{Tarasov:1996br}
O.~Tarasov, {Connection between Feynman integrals having different values of
  the space-time dimension}, Phys.Rev. D54 (1996) 6479--6490.
\newblock \href {http://arxiv.org/abs/hep-th/9606018}
  {\path{arXiv:hep-th/9606018}}, \href
  {http://dx.doi.org/10.1103/PhysRevD.54.6479}
  {\path{doi:10.1103/PhysRevD.54.6479}}.

\bibitem{Smirnov:2006ry}
V.~A. Smirnov, {Feynman integral calculus }, Springer, 2006.

\bibitem{Smirnov:1999gc}
V.~A. Smirnov, {Analytical result for dimensionally regularized massless on
  shell double box}, Phys.Lett. B460 (1999) 397--404.
\newblock \href {http://arxiv.org/abs/hep-ph/9905323}
  {\path{arXiv:hep-ph/9905323}}, \href
  {http://dx.doi.org/10.1016/S0370-2693(99)00777-7}
  {\path{doi:10.1016/S0370-2693(99)00777-7}}.

\bibitem{Tausk:1999vh}
J.~Tausk, {Nonplanar massless two loop Feynman diagrams with four on-shell
  legs}, Phys.Lett. B469 (1999) 225--234.
\newblock \href {http://arxiv.org/abs/hep-ph/9909506}
  {\path{arXiv:hep-ph/9909506}}, \href
  {http://dx.doi.org/10.1016/S0370-2693(99)01277-0}
  {\path{doi:10.1016/S0370-2693(99)01277-0}}.

\bibitem{Czakon:2005rk}
M.~Czakon, {Automatized analytic continuation of Mellin-Barnes integrals},
  Comput.Phys.Commun. 175 (2006) 559--571.
\newblock \href {http://arxiv.org/abs/hep-ph/0511200}
  {\path{arXiv:hep-ph/0511200}}, \href
  {http://dx.doi.org/10.1016/j.cpc.2006.07.002}
  {\path{doi:10.1016/j.cpc.2006.07.002}}.

\bibitem{Smirnov:2009up}
A.~Smirnov, V.~Smirnov, {On the Resolution of Singularities of Multiple
  Mellin-Barnes Integrals}, Eur.Phys.J. C62 (2009) 445--449.
\newblock \href {http://arxiv.org/abs/0901.0386} {\path{arXiv:0901.0386}},
  \href {http://dx.doi.org/10.1140/epjc/s10052-009-1039-6}
  {\path{doi:10.1140/epjc/s10052-009-1039-6}}.

\bibitem{Gluza:2010rn}
J.~Gluza, K.~Kajda, T.~Riemann, V.~Yundin, {Numerical Evaluation of Tensor
  Feynman Integrals in Euclidean Kinematics}, Eur.Phys.J. C71 (2011) 1516.
\newblock \href {http://arxiv.org/abs/1010.1667} {\path{arXiv:1010.1667}},
  \href {http://dx.doi.org/10.1140/epjc/s10052-010-1516-y}
  {\path{doi:10.1140/epjc/s10052-010-1516-y}}.

\bibitem{Anastasiou:2005cb}
C.~Anastasiou, A.~Daleo, {Numerical evaluation of loop integrals}, JHEP 0610
  (2006) 031.
\newblock \href {http://arxiv.org/abs/hep-ph/0511176}
  {\path{arXiv:hep-ph/0511176}}, \href
  {http://dx.doi.org/10.1088/1126-6708/2006/10/031}
  {\path{doi:10.1088/1126-6708/2006/10/031}}.

\bibitem{Freitas:2010nx}
A.~Freitas, Y.-C. Huang, {On the Numerical Evaluation of Loop Integrals With
  Mellin-Barnes Representations}, JHEP 1004 (2010) 074.
\newblock \href {http://arxiv.org/abs/1001.3243} {\path{arXiv:1001.3243}},
  \href {http://dx.doi.org/10.1007/JHEP04(2010)074}
  {\path{doi:10.1007/JHEP04(2010)074}}.

\bibitem{Smirnov:2013eza}
A.~V. Smirnov, {FIESTA 3: cluster-parallelizable multiloop numerical
  calculations in physical regions }\href {http://arxiv.org/abs/1312.3186}
  {\path{arXiv:1312.3186}}.

\bibitem{Pilipp:2008ef}
V.~Pilipp, {Semi-numerical power expansion of Feynman integrals}, JHEP 0809
  (2008) 135.
\newblock \href {http://arxiv.org/abs/0808.2555} {\path{arXiv:0808.2555}},
  \href {http://dx.doi.org/10.1088/1126-6708/2008/09/135}
  {\path{doi:10.1088/1126-6708/2008/09/135}}.

\bibitem{Kotikov:1990kg}
A.~Kotikov, {Differential equations method: New technique for massive Feynman
  diagrams calculation}, Phys.Lett. B254 (1991) 158--164.
\newblock \href {http://dx.doi.org/10.1016/0370-2693(91)90413-K}
  {\path{doi:10.1016/0370-2693(91)90413-K}}.

\bibitem{Remiddi:1997ny}
E.~Remiddi, {Differential equations for Feynman graph amplitudes}, Nuovo Cim.
  A110 (1997) 1435--1452.
\newblock \href {http://arxiv.org/abs/hep-th/9711188}
  {\path{arXiv:hep-th/9711188}}.

\bibitem{Caffo:1998yd}
M.~Caffo, H.~Czyz, S.~Laporta, E.~Remiddi, {Master equations for master
  amplitudes}, Acta Phys.Polon. B29 (1998) 2627--2635.
\newblock \href {http://arxiv.org/abs/hep-th/9807119}
  {\path{arXiv:hep-th/9807119}}.

\bibitem{Remiddi:1999ew}
E.~Remiddi, J.~Vermaseren, {Harmonic polylogarithms}, Int.J.Mod.Phys. A15
  (2000) 725--754.
\newblock \href {http://arxiv.org/abs/hep-ph/9905237}
  {\path{arXiv:hep-ph/9905237}}, \href
  {http://dx.doi.org/10.1142/S0217751X00000367}
  {\path{doi:10.1142/S0217751X00000367}}.

\bibitem{Goncharov:1998kja}
A.~B. Goncharov, {Multiple polylogarithms, cyclotomy and modular complexes},
  Math.Res.Lett. 5 (1998) 497--516.
\newblock \href {http://arxiv.org/abs/1105.2076} {\path{arXiv:1105.2076}},
  \href {http://dx.doi.org/10.4310/MRL.1998.v5.n4.a7}
  {\path{doi:10.4310/MRL.1998.v5.n4.a7}}.

\bibitem{Bonciani:2003cj}
R.~Bonciani, A.~Ferroglia, P.~Mastrolia, E.~Remiddi, J.~van~der Bij, {Planar
  box diagram for the (N(F) = 1) two loop QED virtual corrections to Bhabha
  scattering}, Nucl.Phys. B681 (2004) 261--291.
\newblock \href {http://arxiv.org/abs/hep-ph/0310333}
  {\path{arXiv:hep-ph/0310333}}, \href
  {http://dx.doi.org/10.1016/j.nuclphysb.2004.08.003}
  {\path{doi:10.1016/j.nuclphysb.2004.08.003}}.

\bibitem{0728.11062}
D.~Zagier, {Polylogarithms, Dedekind zeta functions, and the algebraic K-theory
  of fields.}, {
  Prog. Math.} 89 (1991) 391--430.

\bibitem{0863.19004}
A.~Goncharov, {Geometry of configurations, polylogarithms, and motivic
  cohomology.}, Adv. Math. 114~(2) (1995) 197--318.
\newblock \href {http://dx.doi.org/10.1006/aima.1995.1045}
  {\path{doi:10.1006/aima.1995.1045}}.

\bibitem{2009arXiv0908.2238G}
A.~B. {Goncharov}, {A simple construction of Grassmannian polylogarithms},
  ArXiv e-prints\href {http://arxiv.org/abs/0908.2238}
  {\path{arXiv:0908.2238}}.

\bibitem{Goncharov:2010jf}
A.~B. Goncharov, M.~Spradlin, C.~Vergu, A.~Volovich, {Classical Polylogarithms
  for Amplitudes and Wilson Loops}, Phys.Rev.Lett. 105 (2010) 151605.
\newblock \href {http://arxiv.org/abs/1006.5703} {\path{arXiv:1006.5703}},
  \href {http://dx.doi.org/10.1103/PhysRevLett.105.151605}
  {\path{doi:10.1103/PhysRevLett.105.151605}}.

\bibitem{Alday:2010jz}
L.~F. Alday, {Some analytic results for two-loop scattering amplitudes}, JHEP
  1107 (2011) 080.
\newblock \href {http://arxiv.org/abs/1009.1110} {\path{arXiv:1009.1110}},
  \href {http://dx.doi.org/10.1007/JHEP07(2011)080}
  {\path{doi:10.1007/JHEP07(2011)080}}.

\bibitem{Gaiotto:2011dt}
D.~Gaiotto, J.~Maldacena, A.~Sever, P.~Vieira, {Pulling the straps of
  polygons}, JHEP 1112 (2011) 011.
\newblock \href {http://arxiv.org/abs/1102.0062} {\path{arXiv:1102.0062}},
  \href {http://dx.doi.org/10.1007/JHEP12(2011)011}
  {\path{doi:10.1007/JHEP12(2011)011}}.

\bibitem{DelDuca:2011ne}
V.~Del~Duca, C.~Duhr, V.~A. Smirnov, {The massless hexagon integral in D = 6
  dimensions}, Phys.Lett. B703 (2011) 363--365.
\newblock \href {http://arxiv.org/abs/1104.2781} {\path{arXiv:1104.2781}},
  \href {http://dx.doi.org/10.1016/j.physletb.2011.07.079}
  {\path{doi:10.1016/j.physletb.2011.07.079}}.

\bibitem{Dixon:2011ng}
L.~J. Dixon, J.~M. Drummond, J.~M. Henn, {The one-loop six-dimensional hexagon
  integral and its relation to MHV amplitudes in N=4 SYM}, JHEP 1106 (2011)
  100.
\newblock \href {http://arxiv.org/abs/1104.2787} {\path{arXiv:1104.2787}},
  \href {http://dx.doi.org/10.1007/JHEP06(2011)100}
  {\path{doi:10.1007/JHEP06(2011)100}}.

\bibitem{DelDuca:2011jm}
V.~Del~Duca, C.~Duhr, V.~A. Smirnov, {The One-Loop One-Mass Hexagon Integral in
  D=6 Dimensions}, JHEP 1107 (2011) 064.
\newblock \href {http://arxiv.org/abs/1105.1333} {\path{arXiv:1105.1333}},
  \href {http://dx.doi.org/10.1007/JHEP07(2011)064}
  {\path{doi:10.1007/JHEP07(2011)064}}.

\bibitem{DelDuca:2011wh}
V.~Del~Duca, L.~J. Dixon, J.~M. Drummond, C.~Duhr, J.~M. Henn, et~al., {The
  one-loop six-dimensional hexagon integral with three massive corners},
  Phys.Rev. D84 (2011) 045017.
\newblock \href {http://arxiv.org/abs/1105.2011} {\path{arXiv:1105.2011}},
  \href {http://dx.doi.org/10.1103/PhysRevD.84.045017}
  {\path{doi:10.1103/PhysRevD.84.045017}}.

\bibitem{CaronHuot:2011ky}
S.~Caron-Huot, {Superconformal symmetry and two-loop amplitudes in planar N=4
  super Yang-Mills}, JHEP 1112 (2011) 066.
\newblock \href {http://arxiv.org/abs/1105.5606} {\path{arXiv:1105.5606}},
  \href {http://dx.doi.org/10.1007/JHEP12(2011)066}
  {\path{doi:10.1007/JHEP12(2011)066}}.

\bibitem{Dixon:2011pw}
L.~J. Dixon, J.~M. Drummond, J.~M. Henn, {Bootstrapping the three-loop
  hexagon}, JHEP 1111 (2011) 023.
\newblock \href {http://arxiv.org/abs/1108.4461} {\path{arXiv:1108.4461}},
  \href {http://dx.doi.org/10.1007/JHEP11(2011)023}
  {\path{doi:10.1007/JHEP11(2011)023}}.

\bibitem{Heslop:2011hv}
P.~Heslop, V.~V. Khoze, {Wilson Loops @ 3-Loops in Special Kinematics}, JHEP
  1111 (2011) 152.
\newblock \href {http://arxiv.org/abs/1109.0058} {\path{arXiv:1109.0058}},
  \href {http://dx.doi.org/10.1007/JHEP11(2011)152}
  {\path{doi:10.1007/JHEP11(2011)152}}.

\bibitem{Brandhuber:2012vm}
A.~Brandhuber, G.~Travaglini, G.~Yang, {Analytic two-loop form factors in N=4
  SYM}, JHEP 1205 (2012) 082.
\newblock \href {http://arxiv.org/abs/1201.4170} {\path{arXiv:1201.4170}},
  \href {http://dx.doi.org/10.1007/JHEP05(2012)082}
  {\path{doi:10.1007/JHEP05(2012)082}}.

\bibitem{Buehler:2011ev}
S.~Buehler, C.~Duhr, {CHAPLIN - Complex Harmonic Polylogarithms in Fortran
  }\href {http://arxiv.org/abs/1106.5739} {\path{arXiv:1106.5739}}.

\bibitem{Duhr:2011zq}
C.~Duhr, H.~Gangl, J.~R. Rhodes, {From polygons and symbols to polylogarithmic
  functions}, JHEP 1210 (2012) 075.
\newblock \href {http://arxiv.org/abs/1110.0458} {\path{arXiv:1110.0458}},
  \href {http://dx.doi.org/10.1007/JHEP10(2012)075}
  {\path{doi:10.1007/JHEP10(2012)075}}.

\bibitem{Duhr:2012fh}
C.~Duhr, {Hopf algebras, coproducts and symbols: an application to Higgs boson
  amplitudes}, JHEP 1208 (2012) 043.
\newblock \href {http://arxiv.org/abs/1203.0454} {\path{arXiv:1203.0454}},
  \href {http://dx.doi.org/10.1007/JHEP08(2012)043}
  {\path{doi:10.1007/JHEP08(2012)043}}.

\bibitem{Gehrmann:2013vga}
T.~Gehrmann, L.~Tancredi, E.~Weihs, {Two-loop QCD helicity amplitudes for $g\,g
  \to Z\,g$ and $g\,g \to Z\,\gamma $}, JHEP 1304 (2013) 101.
\newblock \href {http://arxiv.org/abs/1302.2630} {\path{arXiv:1302.2630}},
  \href {http://dx.doi.org/10.1007/JHEP04(2013)101}
  {\path{doi:10.1007/JHEP04(2013)101}}.

\bibitem{Anastasiou:2013srw}
C.~Anastasiou, C.~Duhr, F.~Dulat, B.~Mistlberger, {Soft triple-real radiation
  for Higgs production at N3LO}, JHEP 1307 (2013) 003.
\newblock \href {http://arxiv.org/abs/1302.4379} {\path{arXiv:1302.4379}},
  \href {http://dx.doi.org/10.1007/JHEP07(2013)003}
  {\path{doi:10.1007/JHEP07(2013)003}}.

\bibitem{Gehrmann:2013cxs}
T.~Gehrmann, L.~Tancredi, E.~Weihs, {Two-loop master integrals for $q \bar{q}
  \to VV$: the planar topologies}, JHEP 1308 (2013) 070.
\newblock \href {http://arxiv.org/abs/1306.6344} {\path{arXiv:1306.6344}},
  \href {http://dx.doi.org/10.1007/JHEP08(2013)070}
  {\path{doi:10.1007/JHEP08(2013)070}}.

\bibitem{Abreu:2014cla}
S.~Abreu, R.~Britto, C.~Duhr, E.~Gardi, {From multiple unitarity cuts to the
  coproduct of Feynman integrals }\href {http://arxiv.org/abs/1401.3546}
  {\path{arXiv:1401.3546}}.

\bibitem{Gehrmann:2014bfa}
T.~Gehrmann, A.~von Manteuffel, L.~Tancredi, E.~Weihs, {The Two-Loop Master
  Integrals for $q \bar{q} \to V V$ }\href {http://arxiv.org/abs/1404.4853}
  {\path{arXiv:1404.4853}}.

\bibitem{Goncharov:2002math}
A.~B. {Goncharov}, {Galois symmetries of fundamental groupoids and
  noncommutative geometry}, ArXiv Mathematics e-prints\href
  {http://arxiv.org/abs/math/0208144} {\path{arXiv:math/0208144}}.

\bibitem{Brown:1102.1310B}
F.~{Brown}, {On the decomposition of motivic multiple zeta values}, ArXiv
  e-prints\href {http://arxiv.org/abs/1102.1310} {\path{arXiv:1102.1310}}.

\bibitem{Henn:2013pwa}
J.~M. Henn, {Multiloop integrals in dimensional regularization made simple},
  Phys.Rev.Lett. 110~(25) (2013) 251601.
\newblock \href {http://arxiv.org/abs/1304.1806} {\path{arXiv:1304.1806}},
  \href {http://dx.doi.org/10.1103/PhysRevLett.110.251601}
  {\path{doi:10.1103/PhysRevLett.110.251601}}.

\bibitem{Henn:2013tua}
J.~M. Henn, A.~V. Smirnov, V.~A. Smirnov, {Analytic results for planar
  three-loop four-point integrals from a Knizhnik-Zamolodchikov equation}, JHEP
  1307 (2013) 128.
\newblock \href {http://arxiv.org/abs/1306.2799} {\path{arXiv:1306.2799}},
  \href {http://dx.doi.org/10.1007/JHEP07(2013)128}
  {\path{doi:10.1007/JHEP07(2013)128}}.

\bibitem{Henn:2013woa}
J.~M. Henn, V.~A. Smirnov, {Analytic results for two-loop master integrals for
  Bhabha scattering I}, JHEP 1311 (2013) 041.
\newblock \href {http://arxiv.org/abs/1307.4083} {\path{arXiv:1307.4083}},
  \href {http://dx.doi.org/10.1007/JHEP11(2013)041}
  {\path{doi:10.1007/JHEP11(2013)041}}.

\bibitem{Henn:2013nsa}
J.~M. Henn, A.~V. Smirnov, V.~A. Smirnov, {Evaluating single-scale and/or
  non-planar diagrams by differential equations}, JHEP 1403 (2014) 088.
\newblock \href {http://arxiv.org/abs/1312.2588} {\path{arXiv:1312.2588}},
  \href {http://dx.doi.org/10.1007/JHEP03(2014)088}
  {\path{doi:10.1007/JHEP03(2014)088}}.

\bibitem{Argeri:2014qva}
M.~Argeri, S.~Di~Vita, P.~Mastrolia, E.~Mirabella, J.~Schlenk, et~al., {Magnus
  and Dyson Series for Master Integrals}, JHEP 1403 (2014) 082.
\newblock \href {http://arxiv.org/abs/1401.2979} {\path{arXiv:1401.2979}},
  \href {http://dx.doi.org/10.1007/JHEP03(2014)082}
  {\path{doi:10.1007/JHEP03(2014)082}}.

\bibitem{Henn:2014lfa}
J.~M. Henn, K.~Melnikov, V.~A. Smirnov, {Two-loop planar master integrals for
  the production of off-shell vector bosons in hadron collisions }\href
  {http://arxiv.org/abs/1402.7078} {\path{arXiv:1402.7078}}.

\bibitem{Caron-Huot:2014lda}
S.~Caron-Huot, J.~M. Henn, {Iterative structure of finite loop integrals }\href
  {http://arxiv.org/abs/1404.2922} {\path{arXiv:1404.2922}}.

\bibitem{Bern:1993kr}
Z.~Bern, L.~J. Dixon, D.~A. Kosower, {Dimensionally regulated pentagon
  integrals}, Nucl.Phys. B412 (1994) 751--816.
\newblock \href {http://arxiv.org/abs/hep-ph/9306240}
  {\path{arXiv:hep-ph/9306240}}, \href
  {http://dx.doi.org/10.1016/0550-3213(94)90398-0}
  {\path{doi:10.1016/0550-3213(94)90398-0}}.

\bibitem{Tarasov:1996bz}
O.~Tarasov, {A New approach to the momentum expansion of multiloop Feynman
  diagrams}, Nucl.Phys. B480 (1996) 397--412.
\newblock \href {http://arxiv.org/abs/hep-ph/9606238}
  {\path{arXiv:hep-ph/9606238}}, \href
  {http://dx.doi.org/10.1016/S0550-3213(96)00466-X}
  {\path{doi:10.1016/S0550-3213(96)00466-X}}.

\bibitem{Campbell:1996zw}
J.~M. Campbell, E.~N. Glover, D.~Miller, {One loop tensor integrals in
  dimensional regularization}, Nucl.Phys. B498 (1997) 397--442.
\newblock \href {http://arxiv.org/abs/hep-ph/9612413}
  {\path{arXiv:hep-ph/9612413}}, \href
  {http://dx.doi.org/10.1016/S0550-3213(97)00268-X}
  {\path{doi:10.1016/S0550-3213(97)00268-X}}.

\bibitem{Tarasov:1998nx}
O.~Tarasov, {Reduction of Feynman graph amplitudes to a minimal set of basic
  integrals}, Acta Phys.Polon. B29 (1998) 2655.
\newblock \href {http://arxiv.org/abs/hep-ph/9812250}
  {\path{arXiv:hep-ph/9812250}}.

\bibitem{Fleischer:1999hq}
J.~Fleischer, F.~Jegerlehner, O.~Tarasov, {Algebraic reduction of one loop
  Feynman graph amplitudes}, Nucl.Phys. B566 (2000) 423--440.
\newblock \href {http://arxiv.org/abs/hep-ph/9907327}
  {\path{arXiv:hep-ph/9907327}}, \href
  {http://dx.doi.org/10.1016/S0550-3213(99)00678-1}
  {\path{doi:10.1016/S0550-3213(99)00678-1}}.

\bibitem{Halliday:1987an}
I.~Halliday, R.~Ricotta, {Negative dimensional integrals. 1. Feynman graphs},
  Phys.Lett. B193 (1987) 241.
\newblock \href {http://dx.doi.org/10.1016/0370-2693(87)91229-9}
  {\path{doi:10.1016/0370-2693(87)91229-9}}.

\bibitem{Dunne:1987am}
G.~V. Dunne, I.~Halliday, {Negative Dimensional Integration. 2. Path Integrals
  and Fermionic Equivalence}, Phys.Lett. B193 (1987) 247.
\newblock \href {http://dx.doi.org/10.1016/0370-2693(87)91230-5}
  {\path{doi:10.1016/0370-2693(87)91230-5}}.

\bibitem{Dunne:1987qb}
G.~V. Dunne, I.~Halliday, {Negative dimensional oscillators}, Nucl.Phys. B308
  (1988) 589.
\newblock \href {http://dx.doi.org/10.1016/0550-3213(88)90579-2}
  {\path{doi:10.1016/0550-3213(88)90579-2}}.

\bibitem{Ricotta:1989ia}
R.~M. Ricotta, {Negative dimensions in quantum field theory }.

\bibitem{Broadhurst:1987tv}
D.~J. Broadhurst, {Two loop negative dimensional integration}, Phys.Lett. B197
  (1987) 179.
\newblock \href {http://dx.doi.org/10.1016/0370-2693(87)90364-9}
  {\path{doi:10.1016/0370-2693(87)90364-9}}.

\bibitem{Anastasiou:1999ui}
C.~Anastasiou, E.~N. Glover, C.~Oleari, {Scalar one loop integrals using the
  negative dimension approach}, Nucl.Phys. B572 (2000) 307--360.
\newblock \href {http://arxiv.org/abs/hep-ph/9907494}
  {\path{arXiv:hep-ph/9907494}}, \href
  {http://dx.doi.org/10.1016/S0550-3213(99)00637-9}
  {\path{doi:10.1016/S0550-3213(99)00637-9}}.

\bibitem{Landau:1959fi}
L.~D. Landau, {On analytic properties of vertex parts in quantum field theory},
  Nucl. Phys. 13 (1959) 181--192.

\bibitem{Cutkosky:1960sp}
R.~Cutkosky, {Singularities and discontinuities of Feynman amplitudes},
  J.Math.Phys. 1 (1960) 429--433.
\newblock \href {http://dx.doi.org/10.1063/1.1703676}
  {\path{doi:10.1063/1.1703676}}.

\bibitem{'tHooft:1973pz}
G.~'t~Hooft, M.~Veltman, {DIAGRAMMAR}, NATO Adv.Study Inst.Ser.B Phys. 4 (1974)
  177--322.

\bibitem{Remiddi:1981hn}
E.~Remiddi, {Dispersion Relations for Feynman Graphs}, Helv.Phys.Acta 54 (1982)
  364.

\bibitem{Brown:2004svm}
F.~Brown, {Single-valued multiple polylogarithms in one variable}, C. R. Acad.
  Sci. Paris Ser. I 338.

\bibitem{Brown:2009ta}
F.~C. Brown, {On the periods of some Feynman integrals }\href
  {http://arxiv.org/abs/0910.0114} {\path{arXiv:0910.0114}}.

\bibitem{Brown:2009qja}
F.~C. Brown, {Multiple zeta values and periods of moduli spaces M 0 ,n ( R )},
  Annales Sci.Ecole Norm.Sup. 42 (2009) 371.
\newblock \href {http://arxiv.org/abs/math/0606419}
  {\path{arXiv:math/0606419}}.

\bibitem{Brown:2008um}
F.~Brown, {The Massless higher-loop two-point function}, Commun.Math.Phys. 287
  (2009) 925--958.
\newblock \href {http://arxiv.org/abs/0804.1660} {\path{arXiv:0804.1660}},
  \href {http://dx.doi.org/10.1007/s00220-009-0740-5}
  {\path{doi:10.1007/s00220-009-0740-5}}.

\bibitem{Bogner:2013tia}
C.~Bogner, M.~Luders, {Multiple polylogarithms and linearly reducible Feynman
  graphs }\href {http://arxiv.org/abs/1302.6215} {\path{arXiv:1302.6215}}.

\bibitem{Panzer:2013cha}
E.~Panzer, {On the analytic computation of massless propagators in dimensional
  regularization}\href {http://arxiv.org/abs/1305.2161}
  {\path{arXiv:1305.2161}}, \href
  {http://dx.doi.org/10.1016/j.nuclphysb.2013.05.025}
  {\path{doi:10.1016/j.nuclphysb.2013.05.025}}.

\bibitem{Panzer:2014gra}
E.~Panzer, {On hyperlogarithms and Feynman integrals with divergences and many
  scales}, JHEP 1403 (2014) 071.
\newblock \href {http://arxiv.org/abs/1401.4361} {\path{arXiv:1401.4361}},
  \href {http://dx.doi.org/10.1007/JHEP03(2014)071}
  {\path{doi:10.1007/JHEP03(2014)071}}.

\bibitem{Panzer:2014caa}
E.~Panzer, {Algorithms for the symbolic integration of hyperlogarithms with
  applications to Feynman integrals }\href {http://arxiv.org/abs/1403.3385}
  {\path{arXiv:1403.3385}}.

\bibitem{Passarino:2001wv}
G.~Passarino, {An Approach toward the numerical evaluation of multiloop Feynman
  diagrams}, Nucl.Phys. B619 (2001) 257--312.
\newblock \href {http://arxiv.org/abs/hep-ph/0108252}
  {\path{arXiv:hep-ph/0108252}}, \href
  {http://dx.doi.org/10.1016/S0550-3213(01)00528-4}
  {\path{doi:10.1016/S0550-3213(01)00528-4}}.

\bibitem{Passarino:2001jd}
G.~Passarino, S.~Uccirati, {Algebraic numerical evaluation of Feynman diagrams:
  Two loop selfenergies}, Nucl.Phys. B629 (2002) 97--187.
\newblock \href {http://arxiv.org/abs/hep-ph/0112004}
  {\path{arXiv:hep-ph/0112004}}, \href
  {http://dx.doi.org/10.1016/S0550-3213(02)00138-4}
  {\path{doi:10.1016/S0550-3213(02)00138-4}}.

\bibitem{Passarino:2006gv}
G.~Passarino, S.~Uccirati, {Two-loop vertices in quantum field theory: Infrared
  and collinear divergent configurations}, Nucl.Phys. B747 (2006) 113--189.
\newblock \href {http://arxiv.org/abs/hep-ph/0603121}
  {\path{arXiv:hep-ph/0603121}}, \href
  {http://dx.doi.org/10.1016/j.nuclphysb.2006.04.014}
  {\path{doi:10.1016/j.nuclphysb.2006.04.014}}.

\bibitem{Passarino:2007fp}
G.~Passarino, C.~Sturm, S.~Uccirati, {Complete Two-Loop Corrections to $H \to
  \gamma \gamma$}, Phys.Lett. B655 (2007) 298--306.
\newblock \href {http://arxiv.org/abs/0707.1401} {\path{arXiv:0707.1401}},
  \href {http://dx.doi.org/10.1016/j.physletb.2007.09.002}
  {\path{doi:10.1016/j.physletb.2007.09.002}}.

\bibitem{Actis:2008ts}
S.~Actis, G.~Passarino, C.~Sturm, S.~Uccirati, {NNLO Computational Techniques:
  The Cases $H \to \gamma \gamma$ and $H \to g g$}, Nucl.Phys. B811 (2009)
  182--273.
\newblock \href {http://arxiv.org/abs/0809.3667} {\path{arXiv:0809.3667}},
  \href {http://dx.doi.org/10.1016/j.nuclphysb.2008.11.024}
  {\path{doi:10.1016/j.nuclphysb.2008.11.024}}.

\bibitem{Anastasiou:2007qb}
C.~Anastasiou, S.~Beerli, A.~Daleo, {Evaluating multi-loop Feynman diagrams
  with infrared and threshold singularities numerically}, JHEP 0705 (2007) 071.
\newblock \href {http://arxiv.org/abs/hep-ph/0703282}
  {\path{arXiv:hep-ph/0703282}}, \href
  {http://dx.doi.org/10.1088/1126-6708/2007/05/071}
  {\path{doi:10.1088/1126-6708/2007/05/071}}.

\bibitem{Beerli:2008zz}
S.~Beerli, {A New method for evaluating two-loop Feynman integrals and its
  application to Higgs production}, Ph.D. thesis (2008).

\bibitem{Anastasiou:2008rm}
C.~Anastasiou, S.~Beerli, A.~Daleo, {The Two-loop QCD amplitude $gg \to h,H$ in
  the Minimal Supersymmetric Standard Model}, Phys.Rev.Lett. 100 (2008) 241806.
\newblock \href {http://arxiv.org/abs/0803.3065} {\path{arXiv:0803.3065}},
  \href {http://dx.doi.org/10.1103/PhysRevLett.100.241806}
  {\path{doi:10.1103/PhysRevLett.100.241806}}.

\bibitem{Freitas:2012iu}
A.~Freitas, {Numerical evaluation of multi-loop integrals using subtraction
  terms}, JHEP 1207 (2012) 132.
\newblock \href {http://arxiv.org/abs/1205.3515} {\path{arXiv:1205.3515}},
  \href {http://dx.doi.org/10.1007/JHEP07(2012)132, 10.1007/JHEP09(2012)129}
  {\path{doi:10.1007/JHEP07(2012)132, 10.1007/JHEP09(2012)129}}.

\bibitem{Boughezal:2013uia}
R.~Boughezal, F.~Caola, K.~Melnikov, F.~Petriello, M.~Schulze, {Higgs boson
  production in association with a jet at next-to-next-to-leading order in
  perturbative QCD}, JHEP 1306 (2013) 072.
\newblock \href {http://arxiv.org/abs/1302.6216} {\path{arXiv:1302.6216}},
  \href {http://dx.doi.org/10.1007/JHEP06(2013)072}
  {\path{doi:10.1007/JHEP06(2013)072}}.

\bibitem{Soper:1999xk}
D.~E. Soper, Techniques for {QCD} calculations by numerical integration, Phys.
  Rev. D62 (2000) 014009.
\newblock \href {http://arxiv.org/abs/hep-ph/9910292}
  {\path{arXiv:hep-ph/9910292}}.

\bibitem{Binoth:2005ff}
T.~Binoth, J.~P. Guillet, G.~Heinrich, E.~Pilon, C.~Schubert, An algebraic /
  numerical formalism for one-loop multi-leg amplitudes, JHEP 10 (2005) 015.
\newblock \href {http://arxiv.org/abs/hep-ph/0504267}
  {\path{arXiv:hep-ph/0504267}}.

\bibitem{Nagy:2006xy}
Z.~Nagy, D.~E. Soper, Numerical integration of one-loop {F}eynman diagrams for
  {N}-photon amplitudes, Phys. Rev. D74 (2006) 093006.
\newblock \href {http://arxiv.org/abs/hep-ph/0610028}
  {\path{arXiv:hep-ph/0610028}}.

\bibitem{Roth:1996pd}
M.~Roth, A.~Denner, High-energy approximation of one-loop {F}eynman integrals,
  Nucl. Phys. B479 (1996) 495--514.
\newblock \href {http://arxiv.org/abs/hep-ph/9605420}
  {\path{arXiv:hep-ph/9605420}}.

\bibitem{Hartshorne:1977}
R.~{Hartshorne}, {Algebraic geometry.}, 1977.

\bibitem{Bogner:2007cr}
C.~Bogner, S.~Weinzierl, {Resolution of singularities for multi-loop
  integrals}, Comput.Phys.Commun. 178 (2008) 596--610.
\newblock \href {http://arxiv.org/abs/0709.4092} {\path{arXiv:0709.4092}},
  \href {http://dx.doi.org/10.1016/j.cpc.2007.11.012}
  {\path{doi:10.1016/j.cpc.2007.11.012}}.

\bibitem{Smirnov:2008py}
A.~Smirnov, M.~Tentyukov, {Feynman Integral Evaluation by a Sector
  decomposiTion Approach (FIESTA)}, Comput.Phys.Commun. 180 (2009) 735--746.
\newblock \href {http://arxiv.org/abs/0807.4129} {\path{arXiv:0807.4129}},
  \href {http://dx.doi.org/10.1016/j.cpc.2008.11.006}
  {\path{doi:10.1016/j.cpc.2008.11.006}}.

\bibitem{Hironaka1964}
H.~Hironaka, \href{http://www.jstor.org/stable/1970486}{Resolution of
  singularities of an algebraic variety over a field of characteristic zero:
  I}, Annals of Mathematics 79~(1) (1964) pp. 109--203.
\newline\urlprefix\url{http://www.jstor.org/stable/1970486}

\bibitem{1245.14052}
D.~Zeillinger, {A short solution to Hironaka's polyhedra game. }.

\bibitem{0531.14009}
M.~Spivakovsky, {A solution to Hironaka's polyhedra game.}

\bibitem{1059.14022}
S.~Encinas, H.~Hauser, {Strong resolution of singularities in characteristic
  zero.}, Comment. Math. Helv. 77~(4) (2002) 821--845.
\newblock \href {http://dx.doi.org/10.1007/PL00012443}
  {\path{doi:10.1007/PL00012443}}.

\bibitem{Bogner:2008ry}
C.~Bogner, S.~Weinzierl, {Blowing up Feynman integrals}, Nucl.Phys.Proc.Suppl.
  183 (2008) 256--261.
\newblock \href {http://arxiv.org/abs/0806.4307} {\path{arXiv:0806.4307}},
  \href {http://dx.doi.org/10.1016/j.nuclphysbps.2008.09.113}
  {\path{doi:10.1016/j.nuclphysbps.2008.09.113}}.

\bibitem{Speer:1975dc}
E.~Speer, {Ultraviolet and Infrared Singularity Structure of Generic Feynman
  Amplitudes}, Annales Poincare Phys.Theor. 23 (1975) 1--21.

\bibitem{Smirnov:2008aw}
A.~Smirnov, V.~Smirnov, {Hepp and Speer Sectors within Modern Strategies of
  Sector Decomposition}, JHEP 0905 (2009) 004.
\newblock \href {http://arxiv.org/abs/0812.4700} {\path{arXiv:0812.4700}},
  \href {http://dx.doi.org/10.1088/1126-6708/2009/05/004}
  {\path{doi:10.1088/1126-6708/2009/05/004}}.

\bibitem{Smirnov:2009pb}
A.~Smirnov, V.~Smirnov, M.~Tentyukov, {FIESTA 2: Parallelizeable multiloop
  numerical calculations}, Comput.Phys.Commun. 182 (2011) 790--803.
\newblock \href {http://arxiv.org/abs/0912.0158} {\path{arXiv:0912.0158}},
  \href {http://dx.doi.org/10.1016/j.cpc.2010.11.025}
  {\path{doi:10.1016/j.cpc.2010.11.025}}.

\bibitem{Kaneko:2009qx}
T.~Kaneko, T.~Ueda, {A Geometric method of sector decomposition},
  Comput.Phys.Commun. 181 (2010) 1352--1361.
\newblock \href {http://arxiv.org/abs/0908.2897} {\path{arXiv:0908.2897}},
  \href {http://dx.doi.org/10.1016/j.cpc.2010.04.001}
  {\path{doi:10.1016/j.cpc.2010.04.001}}.

\bibitem{Kaneko:2010kj}
T.~Kaneko, T.~Ueda, {Sector Decomposition Via Computational Geometry}, PoS
  ACAT2010 (2010) 082.
\newblock \href {http://arxiv.org/abs/1004.5490} {\path{arXiv:1004.5490}}.

\bibitem{Ueda:2009xx}
T.~Ueda, J.~Fujimoto, {New implementation of the sector decomposition on FORM},
  PoS ACAT08 (2008) 120.
\newblock \href {http://arxiv.org/abs/0902.2656} {\path{arXiv:0902.2656}}.

\bibitem{Barber96thequickhull}
C.~B. Barber, D.~P. Dobkin, H.~Huhdanpaa, The quickhull algorithm for convex
  hulls, ACM Transactions on mathematical software 22~(4) (1996) 469--483.

\bibitem{ELOP}
R.~J. Eden, P.~V. Landshoff, D.~I. Olive, J.~C. Polkinghorne, The Analytic
  S-Matrix, Cambridge University Press, 1966.

\bibitem{Nakanishi}
N.~Nakanishi, Graph Theory and {F}eynman Integrals, Gordon and Breach, New
  York, 1971.

\bibitem{Bjorken:1965zz}
J.~D. Bjorken, S.~D. Drell, {Relativistic quantum fields }.

\bibitem{Carter:2010hi}
J.~Carter, G.~Heinrich, {SecDec: A general program for sector decomposition},
  Comput.Phys.Commun. 182 (2011) 1566--1581.
\newblock \href {http://arxiv.org/abs/1011.5493} {\path{arXiv:1011.5493}},
  \href {http://dx.doi.org/10.1016/j.cpc.2011.03.026}
  {\path{doi:10.1016/j.cpc.2011.03.026}}.

\bibitem{Hahn:2004fe}
T.~Hahn, {CUBA: A library for multidimensional numerical integration}, Comput.
  Phys. Commun. 168 (2005) 78--95.
\newblock \href {http://arxiv.org/abs/hep-ph/0404043}
  {\path{arXiv:hep-ph/0404043}}, \href
  {http://dx.doi.org/10.1016/j.cpc.2005.01.010}
  {\path{doi:10.1016/j.cpc.2005.01.010}}.

\bibitem{Agrawal:2011tm}
S.~Agrawal, T.~Hahn, E.~Mirabella, {FormCalc 7}, J.Phys.Conf.Ser. 368 (2012)
  012054.
\newblock \href {http://arxiv.org/abs/1112.0124} {\path{arXiv:1112.0124}},
  \href {http://dx.doi.org/10.1088/1742-6596/368/1/012054}
  {\path{doi:10.1088/1742-6596/368/1/012054}}.

\bibitem{Carter:2011uza}
J.~P. Carter, {Higher Order Corrections in Perturbative Quantum Field Theory
  via Sector Decomposition}, Ph.D. thesis.

\bibitem{Kawabata:1995th}
S.~Kawabata, {A New version of the multidimensional integration and event
  generation package BASES/SPRING}, Comp. Phys. Commun. 88 (1995) 309--326.
\newblock \href {http://dx.doi.org/10.1016/0010-4655(95)00028-E}
  {\path{doi:10.1016/0010-4655(95)00028-E}}.

\bibitem{Wolfram}
{Mathematica, Copyright by Wolfram Research}.

\bibitem{Robodoc}
F.~Slothouber, et~al., {ROBODoc 4.99.40},
  http://www.xs4all.nl/~rfsber/Robo/robodoc.html.

\bibitem{Davydychev:2003mv}
A.~I. Davydychev, M.~Y. Kalmykov, {Massive Feynman diagrams and inverse
  binomial sums}, Nucl.Phys. B699 (2004) 3--64.
\newblock \href {http://arxiv.org/abs/hep-th/0303162}
  {\path{arXiv:hep-th/0303162}}, \href
  {http://dx.doi.org/10.1016/j.nuclphysb.2004.08.020}
  {\path{doi:10.1016/j.nuclphysb.2004.08.020}}.

\bibitem{Ferroglia:2003yj}
A.~Ferroglia, M.~Passera, G.~Passarino, S.~Uccirati, {Two loop vertices in
  quantum field theory: Infrared convergent scalar configurations}, Nucl.Phys.
  B680 (2004) 199--270.
\newblock \href {http://arxiv.org/abs/hep-ph/0311186}
  {\path{arXiv:hep-ph/0311186}}, \href
  {http://dx.doi.org/10.1016/j.nuclphysb.2003.12.016}
  {\path{doi:10.1016/j.nuclphysb.2003.12.016}}.

\bibitem{Bonciani:2003hc}
R.~Bonciani, P.~Mastrolia, E.~Remiddi, {Master integrals for the two loop QCD
  virtual corrections to the forward backward asymmetry}, Nucl.Phys. B690
  (2004) 138--176.
\newblock \href {http://arxiv.org/abs/hep-ph/0311145}
  {\path{arXiv:hep-ph/0311145}}, \href
  {http://dx.doi.org/10.1016/j.nuclphysb.2004.04.011}
  {\path{doi:10.1016/j.nuclphysb.2004.04.011}}.

\bibitem{Anastasiou:2010pw}
C.~Anastasiou, F.~Herzog, A.~Lazopoulos, {On the factorization of overlapping
  singularities at NNLO}, JHEP 1103 (2011) 038.
\newblock \href {http://arxiv.org/abs/1011.4867} {\path{arXiv:1011.4867}},
  \href {http://dx.doi.org/10.1007/JHEP03(2011)038}
  {\path{doi:10.1007/JHEP03(2011)038}}.

\bibitem{AvMACAT}
A.~von Manteuffel, {{Talk given at the conference ACAT 2011, Uxbridge, London,
  September 2011 }}.

\bibitem{AvMprivate}
A.~von Manteuffel, {private communication}.

\bibitem{Yuasa:2011ff}
F.~Yuasa, E.~de~Doncker, N.~Hamaguchi, T.~Ishikawa, K.~Kato, et~al., {Numerical
  Computation of Two-loop Box Diagrams with Masses }\href
  {http://arxiv.org/abs/1112.0637} {\path{arXiv:1112.0637}}.

\bibitem{Hahn:2001rv}
T.~Hahn, C.~Schappacher, {The Implementation of the minimal supersymmetric
  standard model in FeynArts and FormCalc}, Comput.Phys.Commun. 143 (2002)
  54--68.
\newblock \href {http://arxiv.org/abs/hep-ph/0105349}
  {\path{arXiv:hep-ph/0105349}}, \href
  {http://dx.doi.org/10.1016/S0010-4655(01)00436-2}
  {\path{doi:10.1016/S0010-4655(01)00436-2}}.

\bibitem{Weiglein:1993hd}
G.~Weiglein, R.~Scharf, M.~Bohm, {Reduction of general two loop selfenergies to
  standard scalar integrals}, Nucl.Phys. B416 (1994) 606--644.
\newblock \href {http://arxiv.org/abs/hep-ph/9310358}
  {\path{arXiv:hep-ph/9310358}}, \href
  {http://dx.doi.org/10.1016/0550-3213(94)90325-5}
  {\path{doi:10.1016/0550-3213(94)90325-5}}.

\bibitem{Hahn:2006zy}
T.~Hahn, {A Mathematica interface for FormCalc-generated code},
  Comput.Phys.Commun. 178 (2008) 217--221.
\newblock \href {http://arxiv.org/abs/hep-ph/0611273}
  {\path{arXiv:hep-ph/0611273}}, \href
  {http://dx.doi.org/10.1016/j.cpc.2007.09.004}
  {\path{doi:10.1016/j.cpc.2007.09.004}}.

\bibitem{Agrawal:2012tm}
S.~Agrawal, T.~Hahn, E.~Mirabella, {FormCalc 7}, J.Phys.Conf.Ser. 368 (2012)
  012054.
\newblock \href {http://arxiv.org/abs/1112.0124} {\path{arXiv:1112.0124}},
  \href {http://dx.doi.org/10.1088/1742-6596/368/1/012054}
  {\path{doi:10.1088/1742-6596/368/1/012054}}.

\bibitem{Passarino:1978jh}
G.~Passarino, M.~Veltman, {One Loop Corrections for e+ e- Annihilation Into mu+
  mu- in the Weinberg Model}, Nucl.Phys. B160 (1979) 151.
\newblock \href {http://dx.doi.org/10.1016/0550-3213(79)90234-7}
  {\path{doi:10.1016/0550-3213(79)90234-7}}.

\bibitem{Heinemeyer:2004xw}
S.~Heinemeyer, W.~Hollik, H.~Rzehak, G.~Weiglein, {High-precision predictions
  for the MSSM Higgs sector at O(alpha(b) alpha(s))}, Eur.Phys.J. C39 (2005)
  465--481.
\newblock \href {http://arxiv.org/abs/hep-ph/0411114}
  {\path{arXiv:hep-ph/0411114}}, \href
  {http://dx.doi.org/10.1140/epjc/s2005-02112-6}
  {\path{doi:10.1140/epjc/s2005-02112-6}}.

\bibitem{Hollik:2003jj}
W.~Hollik, H.~Rzehak, {The Sfermion mass spectrum of the MSSM at the one loop
  level}, Eur.Phys.J. C32 (2003) 127--133.
\newblock \href {http://arxiv.org/abs/hep-ph/0305328}
  {\path{arXiv:hep-ph/0305328}}, \href
  {http://dx.doi.org/10.1140/epjc/s2003-01387-9}
  {\path{doi:10.1140/epjc/s2003-01387-9}}.

\bibitem{Heinemeyer:2010mm}
S.~Heinemeyer, H.~Rzehak, C.~Schappacher, {Proposals for Bottom Quark/Squark
  Renormalization in the Complex MSSM}, Phys.Rev. D82 (2010) 075010.
\newblock \href {http://arxiv.org/abs/1007.0689} {\path{arXiv:1007.0689}},
  \href {http://dx.doi.org/10.1103/PhysRevD.82.075010}
  {\path{doi:10.1103/PhysRevD.82.075010}}.

\bibitem{Fritzsche:2011nr}
T.~Fritzsche, S.~Heinemeyer, H.~Rzehak, C.~Schappacher, {Heavy Scalar Top Quark
  Decays in the Complex MSSM: A Full One-Loop Analysis}, Phys.Rev. D86 (2012)
  035014.
\newblock \href {http://arxiv.org/abs/1111.7289} {\path{arXiv:1111.7289}},
  \href {http://dx.doi.org/10.1103/PhysRevD.86.035014}
  {\path{doi:10.1103/PhysRevD.86.035014}}.

\bibitem{Freitas:2002um}
A.~Freitas, D.~Stoeckinger, {Gauge dependence and renormalization of tan beta
  in the MSSM}, Phys.Rev. D66 (2002) 095014.
\newblock \href {http://arxiv.org/abs/hep-ph/0205281}
  {\path{arXiv:hep-ph/0205281}}, \href
  {http://dx.doi.org/10.1103/PhysRevD.66.095014}
  {\path{doi:10.1103/PhysRevD.66.095014}}.

\bibitem{Sperling:2013eva}
M.~Sperling, D.~Stöckinger, A.~Voigt, {Renormalization of vacuum expectation
  values in spontaneously broken gauge theories}, JHEP 1307 (2013) 132.
\newblock \href {http://arxiv.org/abs/1305.1548} {\path{arXiv:1305.1548}},
  \href {http://dx.doi.org/10.1007/JHEP07(2013)132}
  {\path{doi:10.1007/JHEP07(2013)132}}.

\bibitem{Sperling:2013xqa}
M.~Sperling, D.~Stoeckinger, A.~Voigt, {Renormalization of vacuum expectation
  values in spontaneously broken gauge theories: Two-loop results}, JHEP 1401
  (2014) 068.
\newblock \href {http://arxiv.org/abs/1310.7629} {\path{arXiv:1310.7629}},
  \href {http://dx.doi.org/10.1007/JHEP01(2014)068}
  {\path{doi:10.1007/JHEP01(2014)068}}.

\bibitem{Fritzsche:2013fta}
T.~Fritzsche, T.~Hahn, S.~Heinemeyer, F.~von~der Pahlen, H.~Rzehak, et~al.,
  {The Implementation of the Renormalized Complex MSSM in FeynArts and FormCalc
  }\href {http://arxiv.org/abs/1309.1692} {\path{arXiv:1309.1692}}.

\bibitem{'tHooft:1978xw}
G.~'t~Hooft, M.~Veltman, {Scalar One Loop Integrals}, Nucl.Phys. B153 (1979)
  365--401.
\newblock \href {http://dx.doi.org/10.1016/0550-3213(79)90605-9}
  {\path{doi:10.1016/0550-3213(79)90605-9}}.

\bibitem{Nierste:1992wg}
U.~Nierste, D.~Muller, M.~Bohm, {Two loop relevant parts of D-dimensional
  massive scalar one loop integrals}, Z.Phys. C57 (1993) 605--614.
\newblock \href {http://dx.doi.org/10.1007/BF01561479}
  {\path{doi:10.1007/BF01561479}}.

\bibitem{Scharf:1993ds}
R.~Scharf, J.~Tausk, {Scalar two loop integrals for gauge boson selfenergy
  diagrams with a massless fermion loop}, Nucl.Phys. B412 (1994) 523--552.
\newblock \href {http://dx.doi.org/10.1016/0550-3213(94)90391-3}
  {\path{doi:10.1016/0550-3213(94)90391-3}}.

\bibitem{Berends:1994ed}
F.~A. Berends, J.~Tausk, {On the numerical evaluation of scalar two loop
  selfenergy diagrams}, Nucl.Phys. B421 (1994) 456--470.
\newblock \href {http://dx.doi.org/10.1016/0550-3213(94)90336-0}
  {\path{doi:10.1016/0550-3213(94)90336-0}}.

\bibitem{Bauberger:1994zz}
S.~Bauberger, {Massive scalar two-loop self-energy integrals}, Master's thesis
  (1994).

\bibitem{Denner:1991kt}
A.~Denner, {Techniques for calculation of electroweak radiative corrections at
  the one loop level and results for W physics at LEP-200}, Fortsch.Phys. 41
  (1993) 307--420.
\newblock \href {http://arxiv.org/abs/0709.1075} {\path{arXiv:0709.1075}},
  \href {http://dx.doi.org/10.1002/prop.2190410402}
  {\path{doi:10.1002/prop.2190410402}}.

\bibitem{Bauberger:1994by}
S.~Bauberger, F.~A. Berends, M.~Bohm, M.~Buza, {Analytical and numerical
  methods for massive two loop selfenergy diagrams}, Nucl.Phys. B434 (1995)
  383--407.
\newblock \href {http://arxiv.org/abs/hep-ph/9409388}
  {\path{arXiv:hep-ph/9409388}}, \href
  {http://dx.doi.org/10.1016/0550-3213(94)00475-T}
  {\path{doi:10.1016/0550-3213(94)00475-T}}.

\bibitem{Bauberger:1994hx}
S.~Bauberger, M.~Bohm, {Simple one-dimensional integral representations for two
  loop selfenergies: The Master diagram}, Nucl.Phys. B445 (1995) 25--48.
\newblock \href {http://arxiv.org/abs/hep-ph/9501201}
  {\path{arXiv:hep-ph/9501201}}, \href
  {http://dx.doi.org/10.1016/0550-3213(95)00199-3}
  {\path{doi:10.1016/0550-3213(95)00199-3}}.

\bibitem{Guillet:2013msa}
J.~P. Guillet, G.~Heinrich, J.~von Soden-Fraunhofen, {Tools for NLO automation:
  extension of the golem95C integral library }\href
  {http://arxiv.org/abs/1312.3887} {\path{arXiv:1312.3887}}.

\bibitem{vanderBij:1983bw}
J.~van~der Bij, M.~Veltman, {Two Loop Large Higgs Mass Correction to the rho
  Parameter}, Nucl.Phys. B231 (1984) 205.
\newblock \href {http://dx.doi.org/10.1016/0550-3213(84)90284-0}
  {\path{doi:10.1016/0550-3213(84)90284-0}}.

\bibitem{Berends:1993ee}
F.~A. Berends, M.~Buza, M.~Bohm, R.~Scharf, {Closed expressions for specific
  massive multiloop selfenergy integrals}, Z.Phys. C63 (1994) 227--234.
\newblock \href {http://dx.doi.org/10.1007/BF01411014}
  {\path{doi:10.1007/BF01411014}}.

\bibitem{Scharf:1991dipl}
R.~Scharf, {University of W{\"u}rzburg}, Master's thesis (1991).

\bibitem{TuanKim:1992}
V.~K. Tuan, R.~G. Buschman, {Integral representations of generalized Lauricella
  hypergeometric functions}, International Journal of Mathematics and
  Mathematical Sciences 15 number 4 (1992) 653--657.
\newblock \href {http://dx.doi.org/10.1155/S0161171292000863}
  {\path{doi:10.1155/S0161171292000863}}.

\bibitem{Davydychev:1993pg}
A.~I. Davydychev, V.~A. Smirnov, J.~Tausk, {Large momentum expansion of two
  loop selfenergy diagrams with arbitrary masses}, Nucl.Phys. B410 (1993)
  325--342.
\newblock \href {http://arxiv.org/abs/hep-ph/9307371}
  {\path{arXiv:hep-ph/9307371}}, \href
  {http://dx.doi.org/10.1016/0550-3213(93)90436-S}
  {\path{doi:10.1016/0550-3213(93)90436-S}}.

\bibitem{Laporta:2004rb}
S.~Laporta, E.~Remiddi, {Analytic treatment of the two loop equal mass sunrise
  graph}, Nucl.Phys. B704 (2005) 349--386.
\newblock \href {http://arxiv.org/abs/hep-ph/0406160}
  {\path{arXiv:hep-ph/0406160}}, \href
  {http://dx.doi.org/10.1016/j.nuclphysb.2004.10.044}
  {\path{doi:10.1016/j.nuclphysb.2004.10.044}}.

\bibitem{Adams:2014vja}
L.~Adams, C.~Bogner, S.~Weinzierl, {The two-loop sunrise graph with arbitrary
  masses in terms of elliptic dilogarithms }\href
  {http://arxiv.org/abs/1405.5640} {\path{arXiv:1405.5640}}.

\bibitem{Adams:2013nia}
L.~Adams, C.~Bogner, S.~Weinzierl, {The two-loop sunrise graph with arbitrary
  masses }\href {http://arxiv.org/abs/1302.7004} {\path{arXiv:1302.7004}}.

\bibitem{Remiddi:2013joa}
E.~Remiddi, L.~Tancredi, {Schouten identities for Feynman graph amplitudes; The
  Master Integrals for the two-loop massive sunrise graph}, Nucl.Phys. B880
  (2014) 343--377.
\newblock \href {http://arxiv.org/abs/1311.3342} {\path{arXiv:1311.3342}},
  \href {http://dx.doi.org/10.1016/j.nuclphysb.2014.01.009}
  {\path{doi:10.1016/j.nuclphysb.2014.01.009}}.

\bibitem{Greynat:2013zqa}
D.~Greynat, J.~Sesma, G.~Vulvert, {Epsilon expansion of Appell and Kamp\'e de
  F\'eriet functions }\href {http://arxiv.org/abs/1310.7700}
  {\path{arXiv:1310.7700}}.

\bibitem{Kreimer:1991jv}
D.~Kreimer, {The Master two loop two point function: The General case},
  Phys.Lett. B273 (1991) 277--281.
\newblock \href {http://dx.doi.org/10.1016/0370-2693(91)91684-N}
  {\path{doi:10.1016/0370-2693(91)91684-N}}.

\bibitem{Broadhurst:1987ei}
D.~J. Broadhurst, {The Master Two Loop Diagram With Masses}, Z.Phys. C47 (1990)
  115--124.
\newblock \href {http://dx.doi.org/10.1007/BF01551921}
  {\path{doi:10.1007/BF01551921}}.

\bibitem{Davydychev:1992mt}
A.~I. Davydychev, J.~Tausk, {Two loop selfenergy diagrams with different masses
  and the momentum expansion}, Nucl.Phys. B397 (1993) 123--142.
\newblock \href {http://dx.doi.org/10.1016/0550-3213(93)90338-P}
  {\path{doi:10.1016/0550-3213(93)90338-P}}.

\bibitem{Broadhurst:1993mw}
D.~J. Broadhurst, J.~Fleischer, O.~Tarasov, {Two loop two point functions with
  masses: Asymptotic expansions and Taylor series, in any dimension}, Z.Phys.
  C60 (1993) 287--302.
\newblock \href {http://arxiv.org/abs/hep-ph/9304303}
  {\path{arXiv:hep-ph/9304303}}, \href {http://dx.doi.org/10.1007/BF01474625}
  {\path{doi:10.1007/BF01474625}}.

\bibitem{Berends:1994sa}
F.~A. Berends, A.~I. Davydychev, V.~A. Smirnov, J.~Tausk, {Zero threshold
  expansion of two loop selfenergy diagrams}, Nucl.Phys. B439 (1995) 536--560.
\newblock \href {http://arxiv.org/abs/hep-ph/9410232}
  {\path{arXiv:hep-ph/9410232}}, \href
  {http://dx.doi.org/10.1016/0550-3213(95)00018-N}
  {\path{doi:10.1016/0550-3213(95)00018-N}}.

\bibitem{Hoogeveen:1985tf}
F.~Hoogeveen, {The Influence of a Heavy Fermion Doublet on Higgs Boson
  Production via the Gluon Fusion Mechanism}, Nucl.Phys. B259 (1985) 19.
\newblock \href {http://dx.doi.org/10.1016/0550-3213(85)90295-0}
  {\path{doi:10.1016/0550-3213(85)90295-0}}.

\bibitem{Bechtle:2012jw}
P.~Bechtle, S.~Heinemeyer, O.~Stal, T.~Stefaniak, G.~Weiglein, et~al., {MSSM
  Interpretations of the LHC Discovery: Light or Heavy Higgs?}, Eur.Phys.J. C73
  (2013) 2354.
\newblock \href {http://arxiv.org/abs/1211.1955} {\path{arXiv:1211.1955}},
  \href {http://dx.doi.org/10.1140/epjc/s10052-013-2354-5}
  {\path{doi:10.1140/epjc/s10052-013-2354-5}}.

\bibitem{Carena:2013qia}
M.~Carena, S.~Heinemeyer, O.~Stal, C.~Wagner, G.~Weiglein, {MSSM Higgs Boson
  Searches at the LHC: Benchmark Scenarios after the Discovery of a Higgs-like
  Particle}, Eur. Phys. J. C73 (2013) 2552.
\newblock \href {http://arxiv.org/abs/1302.7033} {\path{arXiv:1302.7033}},
  \href {http://dx.doi.org/10.1140/epjc/s10052-013-2552-1}
  {\path{doi:10.1140/epjc/s10052-013-2552-1}}.

\bibitem{CDF:2013jga}
M.~Muether, CDF, {Combination of CDF and DO results on the mass of the top
  quark using up to 8.7 fb$^{-1}$ at the Tevatron }\href
  {http://arxiv.org/abs/1305.3929} {\path{arXiv:1305.3929}}.

\bibitem{CMS-PAS-TOP-12-001}
{LHC Combination: Top mass}, Tech. Rep. CMS-PAS-TOP-12-001, CERN, Geneva
  (2012).

\bibitem{ATLAS:2014wva}
{First combination of Tevatron and LHC measurements of the top-quark mass}\href
  {http://arxiv.org/abs/1403.4427} {\path{arXiv:1403.4427}}.

\bibitem{Acciarri:2000kg}
M.~Acciarri, et~al., {Determination of $\gamma / Z$ interference in $e^{+}
  e^{-}$ annihilation at LEP}, Phys.Lett. B489 (2000) 93--101.
\newblock \href {http://arxiv.org/abs/hep-ex/0007006}
  {\path{arXiv:hep-ex/0007006}}, \href
  {http://dx.doi.org/10.1016/S0370-2693(00)00889-3}
  {\path{doi:10.1016/S0370-2693(00)00889-3}}.

\bibitem{Baer:2013cma}
H.~Baer, T.~Barklow, K.~Fujii, Y.~Gao, A.~Hoang, et~al., {The International
  Linear Collider Technical Design Report - Volume 2: Physics }\href
  {http://arxiv.org/abs/1306.6352} {\path{arXiv:1306.6352}}.

\bibitem{Gennai:2007ys}
S.~Gennai, S.~Heinemeyer, A.~Kalinowski, R.~Kinnunen, S.~Lehti, et~al., {Search
  for heavy neutral MSSM Higgs bosons with CMS: Reach and Higgs-mass
  precision}, Eur.Phys.J. C52 (2007) 383--395.
\newblock \href {http://arxiv.org/abs/0704.0619} {\path{arXiv:0704.0619}},
  \href {http://dx.doi.org/10.1140/epjc/s10052-007-0398-0}
  {\path{doi:10.1140/epjc/s10052-007-0398-0}}.

\end{thebibliography}
